%% file: main.tex
\documentclass[twocolumn,twoside,aps]{revtex4}

\usepackage{microtype}

\usepackage{amsmath}
\usepackage{amssymb}

\usepackage{graphicx}
\usepackage[caption=false]{subfig}
\usepackage{tikz}

\usepackage{array}

\newcommand{\ket}[1]{\left|#1\right\rangle}
\newcommand{\bra}[1]{\left\langle#1\right|}

\begin{document}

\title{Checking the error correction strength of arbitrary surface code
logical gates}

\author{Thomas J. Milburn, Austin G. Fowler}

\affiliation{The Melbourne Quantum Computer Science Group, \\ Centre for
Quantum Computation and Communication Technology, \\ School of Physics, the
University of Melbourne, Victoria 3010, Australia}

\date{\today}

\begin{abstract}
Topologically quantum error corrected logical gates are complex. Chains of
errors can form in space and time and diagonally in spacetime. It is highly
nontrivial to determine whether a given logical gate is free of low weight
combinations of errors leading to failure. We report a new tool Nestcheck
capable of analyzing an arbitrary topological computation and determining the
minimum number of errors required to cause failure.
\end{abstract}

\maketitle

\section{Introduction}
\label{sec:Introduction}

Constructing a quantum device for simulating quantum systems was first
suggested by Feynman in 1982 \cite{feynman1982spw}. The specific device he
proposed, however, was not software programmable. The first proposal for a
programmable quantum computer was given by Deutsch in 1985
\cite{deutsch1985qtt}. The paradigm of quantum computation was thus born, but
it had yet to acquire motivation. This came in 1994 with the advent of Shor's
factoring algorithm \cite{shor1994afq}.

A concern raised early within the quantum computation community was that of
its infeasability due to decoherence \cite{berthiaume1994tso}. Although this
concern has yet to be eliminated, it has been considerably lessened. With the
invention of quantum error correction in 1995 \cite{shor1995sfr,
steane1996ecc} and the proof of the threshold theorem in 1996
\cite{knill1996taf, aharonov1997ftq} there is now a focus on constructing
quantum error correction codes.

Topological approaches to quantum error correction are particularly promising
\cite{bravyi1998qco, dennis2002tqm, raussendorf2006aft, raussendorf2007tft,
fowler2009tcs}. One such class is surface codes. The specific surface code we
consider in this paper is explained well in \cite{fowler2012sct}. The
elementary systems of the surface code are physical qubits (two level quantum
systems) \cite{schumacher1995qc}, and the information elements are logical
qubits (encoded over the physical qubits). This code allows efficient
implementation of quantum algorithms and its physical requirements are modest:
a 2D lattice of qubits, nearest neighbor interactions, parallelizability, and
gate error rates around $1\%$ \cite{wang2011scq, fowler2012tpc}.

During computation, various gates are continually performed on the physical
qubits with each gate having an error rate. These errors can form chains in
space and time and diagonally in spacetime. Determining the error correction
strength of a proposal for a surface code logical gate, which may consist of
many of physical operations, can thus be somewhat complex. It can only
reliably be accomplished via directly simulating the logical gate.

We report a new tool for this task: Nestcheck. Using Nestcheck we are able to
analyse an arbitrary surface code logical gate and determine the minimum
number of errors required to cause failure.

\section{The surface code}
\label{sec:The surface code}

Group theoretic concepts provide a convenient formalism for describing both
the logical qubits and gates of the surface code \cite{gottesman1997sca,
gottesman1998tof}. Homology also lends itself to our subject
\cite{raussendorf2007tft, raussendorf2007ftq}, but encasing our discussion in
the terms of homology might render it inaccessible. The following is an
extremely brief account of the surface code.

Consider a set of $n$ qubits, $\mathcal{Q}$. Decoherence of
$\ket{\mathcal{Q}}$ is due to $\mathcal{Q}$ interacting with an unmonitored
bath, $\mathcal{B}$. This may be written $U (\ket{\mathcal{Q}} \ket{B})$,
where $U$ is a unitary operator. The effective reduced density operator is
then
\begin{gather}
		\rho_\mathcal{Q} = \text{tr}_B ( U \ket{\mathcal{Q}} \ket{B}
\bra{\mathcal{Q}} \bra{B} U^\dag ) = \sum_{E \in \mathcal{E}} E
\ket{\mathcal{Q}} \bra{\mathcal{Q}} E^\dag ,\label{eq:Topological quantum
error correction codes.1}
\end{gather}
where $\mathcal{E}$ are Krauss operators. We call $\mathcal{E}$ the errors
and often relax the Krauss normalization condition \cite{kaye2007ait}. By
Eq.~\ref{eq:Topological quantum error correction codes.1} and the linearity
of quantum mechanics, in order to correct decoherence we need only correct a
basis for $\mathcal{E}$ \cite{nielsen2000qca}. A convenient choice is the
complete pauli basis, $\mathcal{G} = \{1, -1, i, -1\} \times \{I, X, Y,
Z\}^{\otimes n}$. This is a group under multiplication.

The stabilizer of $\ket{\mathcal{Q}}$ is the set of operators $\mathcal{S}
\subset \mathcal{G}$ such that $\mathcal{S} \ket{\mathcal{Q}} =
\ket{\mathcal{Q}}$. Every element of $\mathcal{G}$ either commutes or
anticommutes with each other element. Consider some $S \in \hat{\mathcal{S}}$
(a generating set for $\mathcal{S}$) and applying an operator $G \in
\mathcal{G}$: If $[G, S] = 0$ then $S G \ket{\mathcal{Q}} = G
\ket{\mathcal{Q}}$ and measuring $S$ yields $1$, whereas if $\{G, S\} = 0$
then $S G \ket{\mathcal{Q}} = - G \ket{\mathcal{Q}}$ and measuring $S$ yields
$-1$.

The surface code is a stabilizer code: we enforce a certain set of operators
to be $\hat{\mathcal{S}}$, a set of stabilizer generators. We construct it
such that every sufficiently low-weight and distinct error anticommutes with a
unique subset of the phase factor $1$ version of $\hat{\mathcal{S}}$. At
regular intervals we measure the phase factor $1$ version of
$\hat{\mathcal{S}}$, each called a round of error correction. This yields an
ordered set of $1$s and $-1$s, which we call a syndrome. We call an element of
a syndrome a syndrome result. If one syndrome differs from another then there
has been either measurement or physical qubit error on some number of physical
qubits. Assuming only those low-weight errors mentioned above, we can
determine from syndrome changes the errors that caused them.

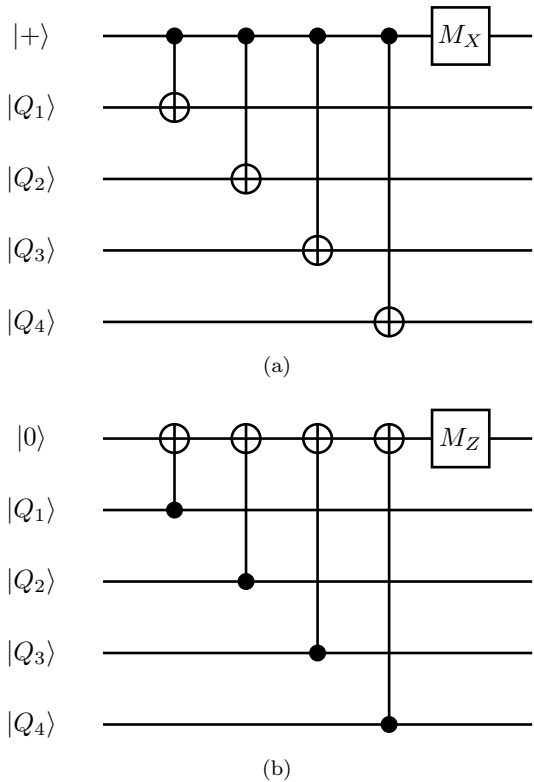
\begin{figure}[t!]
	\begin{center}
		\subfloat[]{%
			{\normalsize%
				\input{pic_circuits.tex}
			}
		}\\
		\subfloat[]{%
			{\normalsize%
				\input{pic_circuits2.tex}
			}
		}
		\caption{Quantum circuits for measuring (a) $XXXX$ and (b) $ZZZZ$.}
		\label{fig:The surface code.2}
	\end{center}
\end{figure}

Concerning the surface code specifically, elements of the phase factor $1$
version of $\hat{\mathcal{S}}$ have the form $XXXX$, $XXX$ and $XX$, and
$ZZZZ$, $ZZZ$ and $ZZ$. Measuring the phase factor $1$ version of
$\hat{\mathcal{S}}$ requires ancilla qubits \cite{divincenzo1997fte}, which we
call syndrome qubits. See Fig.~\ref{fig:The surface code.2}. We call the
qubits in which logical information is stored data qubits. We separate
$\hat{\mathcal{S}}$ into two generating sets: primal, the $X$-operators, and
dual, the $Z$-operators. Primal and dual stabilizer generators are used to
correct $Z$- and $X$-errors respectively. This is sufficient because an
$I$-error is trivial and $Y = \alpha X Z$.

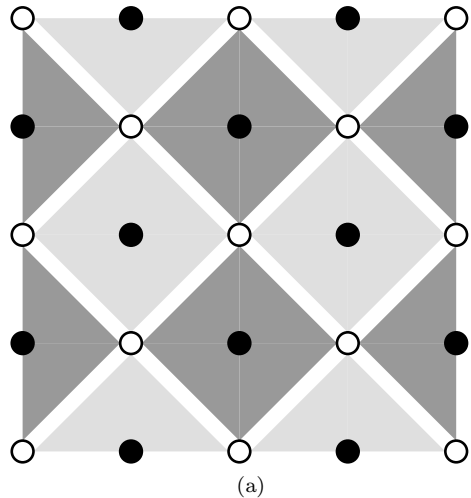
\begin{figure}[t!]
	\begin{center}
		\subfloat[]{%
			{\normalsize%
				\input{pic_smallestplate.tex}
			}
		}
		\caption{A surface code plate protecting one logical qubit. Filled and
empty circles denote syndrome and data qubits respectively. Darkly and lightly
shaded shapes represent primal ($X$) and dual ($Z$) stabilizer generators
respectively.}
		\label{fig:The surface code.1}
	\end{center}
\end{figure}

Consider Fig.~\ref{fig:The surface code.1}. This is a surface code plate
protecting one logical qubit. If a $Z$-error acts on the center data qubit
then the primal stabilizer generators adjacent to it reverse phase causing two
syndrome result changes. In larger surface code plates chains of errors can
form, the stabilizer generators at the ends of which reverse phase. Chains can
undetectably connect to boundaries. In Fig.~\ref{fig:The surface code.1} the
shortest undetectable chain of errors has weight three. For example, an
$X$-error on the center-left, center and center-right data qubit. We say
`distance three', write $d = 3$, and this is a measure of the strength of a
surface code logical gate.

The normalizer of $\hat{\mathcal{S}}$ is the set of operators $\mathcal{N}$
such that $(\forall N \in \mathcal{N}) [N, \hat{\mathcal{S}}] = 0$. This is
the set of logical operators. By defining particular physical operators as
particular logical operators we define basis logical states. For example,
in Fig.~\ref{fig:The surface code.1} a natural choice for $X_\text{L}$
(logical $X$) is any chain of $X$-operators from the left edge to the right
that commutes with all dual ($Z$) stabilizers.

The specific error correction of the surface code proceeds via pairing
syndrome results to each other and to boundaries. This is because every
correctible error may be decomposed into a set of error chains each of which
causes either two syndrome result changes, or one due to the error chain
connecting to a boundary. Note that an error chain that connects to two
boundaries commutes $\hat{\mathcal{S}}$ and is hence uncorrectible. According
that shorter error chains are more likely than longer ones, syndrome results
should be paired using a minimum weight matching algorithm. Autotune, the tool
we use for this task and which is breifly discussed in
section~\ref{sec:Nestcheck}, uses Edmond's minimum weight perfect matching
algorithm \cite{edmonds1965pta, edmonds1965mma, fowler2011tpc,
fowler2012tpc}.

\section{Nestcheck}
\label{sec:Nestcheck}

A single round of measuring the stabilizers of Fig.~\ref{fig:The surface
code.1} does not change the encoded data. In other words, $I_\text{L}$ is
applied. Less trivial logical gates involve rounds of error correction during
which data qubits may be initialized, measured, swapped, and so on.
Furthermore, we are not so fortunate that errors can only occur on data qubits
between rounds of error correction: errors can occur on any qubit at any time,
even measurements are prone to error.

\begin{table*}[t!]
	\begin{center}
		\subfloat[]{%
			\begin{tabular*}{\linewidth}{l@{\extracolsep{\fill}}c>{\raggedleft\arraybackslash}p{0.8\linewidth}}
				\hline\hline
				Command & Figure & Action \\
				\hline
				\texttt{ACTIVE} & \begin{tikzpicture}
					\draw [line width=1pt] (0,0) circle (2pt);
				\end{tikzpicture}
				& Either $\text{SWAP}$-application due to a diagonally
neighboring data qubit having a \texttt{SWAP\symbol{95}NW},
\texttt{SWAP\symbol{95}NE}, \texttt{SWAP\symbol{95}SW} or
\texttt{SWAP\symbol{95}SE} command, or $\Lambda(X)$-application due to
measurement of neighboring stabilizer generators. \\
				\texttt{INACTIVE} & & Consider data qubit as nonexistent. \\
				\texttt{ADD\symbol{95}X} & \begin{tikzpicture}
					\draw [line width=1pt] (0,0) circle (2pt);
					\draw (0,0) node [above right] {$|+\rangle$};
				\end{tikzpicture}
				& $\ket{+}$-initialization followed
by $\Lambda(X)$-application due to measurement of neighboring stabilizer
generators. \\
				\texttt{ADD\symbol{95}Z} & \begin{tikzpicture}
					\draw [line width=1pt] (0,0) circle (2pt);
					\draw (0,0) node [above right] {$|0\rangle$};
				\end{tikzpicture}
				& $\ket{0}$-initialization followed
by $\Lambda(X)$-application due to measurement of neighboring stabilizer
generators. \\
				\texttt{TRIM\symbol{95}X} & \begin{tikzpicture}
					\draw [line width=1pt] (0,0) circle (2pt);
					\draw (0,0) node [above right] {$M_X$};
				\end{tikzpicture}
				& $\Lambda(X)$-application due to measurement of neighboring
stabilizer generators followed by $X$-basis-measurement. \\
				\texttt{TRIM\symbol{95}Z} & \begin{tikzpicture}
					\draw [line width=1pt] (0,0) circle (2pt);
					\draw (0,0) node [above right] {$M_Z$};
				\end{tikzpicture}
				& $\Lambda(X)$-application due to measurement of neighboring
stabilizer generators followed by $Z$-basis-measurement. \\
				\texttt{HAD} & \begin{tikzpicture}
					\draw [line width=1pt] (0,0) circle (2pt);
					\draw (0,0) node [above right] {$H$};
				\end{tikzpicture}
				& $\Lambda(X)$-application due to measurement of
neighboring stabilizer generators followed by $H$-application. \\
				\texttt{SWAP\symbol{95}NW} & \begin{tikzpicture}
					\draw [line width=1pt] (0,0) circle (2pt);
					\draw [line width=1pt,->] (-2pt,2pt) -- (-14pt,14pt);
				\end{tikzpicture}
				& $\text{SWAP}$-application of this
data qubit and the syndrome qubit to the north followed by
$\text{SWAP}$-application of the syndrome qubit to the north and the data
qubit to the north-west. \\
				\texttt{SWAP\symbol{95}NE} & \begin{tikzpicture}
					\draw [line width=1pt] (0,0) circle (2pt);
					\draw [line width=1pt,->] (2pt,2pt) -- (14pt,14pt);
				\end{tikzpicture}
				& $\text{SWAP}$-application of this
data qubit and the syndrome qubit to the north followed by
$\text{SWAP}$-application of the syndrome qubit to the north and the data
qubit to the north-east. \\
				\texttt{SWAP\symbol{95}SW} & \begin{tikzpicture}
					\draw [line width=1pt] (0,0) circle (2pt);
					\draw [line width=1pt,->] (-2pt,-2pt) -- (-14pt,-14pt);
				\end{tikzpicture}& $\text{SWAP}$-application of this
data qubit and the syndrome qubit to the south followed by
$\text{SWAP}$-application of the syndrome qubit to the south and the data
qubit to the south-west. \\
				\texttt{SWAP\symbol{95}SE} & \begin{tikzpicture}
					\draw [line width=1pt] (0,0) circle (2pt);
					\draw [line width=1pt,->] (2pt,-2pt) -- (14pt,-14pt);
				\end{tikzpicture}& $\text{SWAP}$-application of this
data qubit and the syndrome qubit to the south followed by
$\text{SWAP}$-application of the syndrome qubit to the south and the data
qubit to the south-east. \\
				\hline\hline
			\end{tabular*}
		}\\
		\subfloat[]{%
			\begin{tabular*}{\linewidth}{l@{\extracolsep{\fill}}c>{\raggedleft\arraybackslash}p{0.8\linewidth}}
				\hline\hline
				Command & Figure & Action \\
				\hline
				\texttt{ACTIVE} & \begin{tikzpicture}
					\draw [line width=1pt,fill=black] (0,0) circle (2pt);
				\end{tikzpicture}
				& Either use syndrome qubit as an ancilla for stabilizer
generator measurement where the data qubits included are those neighboring
data qubits that have neither the command \texttt{INACTIVE}, nor
\texttt{SWAP\symbol{95}NW}, \texttt{SWAP\symbol{95}NE},
\texttt{SWAP\symbol{95}SW} nor \texttt{SWAP\symbol{95}SE}, or
$\text{SWAP}$-application due to a neighboring data qubit having one of the
latter four commands. \\
				\texttt{INACTIVE} & & Consider syndrome qubit as nonexistent. \\
				\hline\hline
			\end{tabular*}
		}
		\caption{(a) Possible data qubit commands. (b) Possible syndrome qubit
commands. Note that in figures stabilizers are represented as per
Fig.~\ref{fig:The surface code.1} for convenience. This is not required in our
set of commands because in any lattice of qubits we index the qubits with $i$
and $j$ such that if $i + j \text{ mod } 2 = 0$ we have a syndrome qubit and if
also $i \text{ mod } 2 = 0$ we have a dual syndrome qubit.}
		\label{tab:Nestcheck.1}
	\end{center}
\end{table*}

In order to determine the error correction strength of a surface code logical
gate we first construct a primal and a dual graph. The vertices of such a
graph are the spacetime locations of potential syndrome result changes and the
edges correspond to potential connections between syndrome result changes due
to a single error. We construct this graph using the tool Autotune
\cite{fowler2012tca}.

In the Autotune lexicon, our graph of vertices and edges is a nest of balls
and sticks. From section~\ref{sec:The surface code} the data of error
correction are the various measurements made during computation. Accordingly,
Autotune creates a nest by processing many sets of measurements. A set is
located in spacetime and has the property that if its measurements multiply to
$-1$, as opposed to $1$, then its location is that of a syndrome result
change. This implies either measurement error or phase-reversal of the
associated stabilizer generator and is called a detection event. In a
nutshell: qubits are simulated, during which course all possible errors are
generated and propagated, measurements of these qubits are placed in sets, all
possible detection events are formed to create balls, and pairs of detection
events generated by the same error are connected by a stick. We thus obtain
our nest of balls and sticks. Note that the proximity of sets to temporal and
spatial boundaries must be specified by the user so that unpaired detection
events generate a stick to this boundary specifically.

The user must place every measurement in either two sets, or one set and a
boundary set. Deciding when to create a set and in which set a particular
measurement must be placed depends on the physical gate sequence the user is
applying. This physical gate sequence can be rather complex: a surface code
computation typically involves introduction, deformation and elimination of
boundaries, measurements in various bases, and single and double physical
qubit gates. We desire a short list of commands to be given locally to
physical qubits such that an arbitrary surface code computation can be
specified. Table~\ref{tab:Nestcheck.1} details the list we use. Note that
initializations specific to state-injection have not been included, but will
be in future work. 

\begin{table}[t!]
	\begin{center}
		\begin{tabular*}{\linewidth}{c@{\extracolsep{\fill}}r}
			\hline\hline
			Figure & Description \\
			\hline
			\begin{tikzpicture}
				\draw [line width=1pt] (0,0) -- (0,20pt);
			\end{tikzpicture}
			& $I_\text{L}$-application. \\
			\begin{tikzpicture}
				\draw [line width=1pt] (0,0) -- (0,20pt);
				\draw [line width=1pt,fill=white] (-2pt,8pt) -- (-2pt,12pt) -- (2pt,12pt) -- (2pt,8pt) -- cycle;
			\end{tikzpicture}
			& $\ket{0}$-initialization. \\
			\begin{tikzpicture}
				\draw [line width=1pt] (0,0) -- (0,20pt);
				\draw [line width=1pt,fill=white] (-2pt,10pt) -- (0,12pt) -- (2pt, 10pt) -- (0,8pt) -- cycle;
			\end{tikzpicture}
			& $H$-application. \\
			\begin{tikzpicture}
				\draw [line width=1pt] (0,0) -- (0,20pt);
				\draw [line width=1pt,fill=white] (0,10pt) circle (2pt);
			\end{tikzpicture}
			& $Z$-basis-measurement. \\
			\begin{tikzpicture}
				\draw [line width=1pt] (0,0) -- (0,20pt);
				\draw [line width=1pt] (20pt,0) -- (20pt,20pt);
				\draw [line width=1pt,fill=black] (0,10pt) circle (2pt);
				\draw [line width=1pt] (20pt,10pt) circle (4pt);
				\draw [line width=1pt] (0,10pt) -- (24pt,10pt);
			\end{tikzpicture}
			& $\Lambda(X)$-application (left qubit is the control). \\
			\begin{tikzpicture}
				\draw [line width=1pt] (0,0) -- (0,20pt);
				\draw [line width=1pt] (20pt,0) -- (20pt,20pt);
				\draw [line width=1pt] (0,10pt) -- (20pt,10pt);
				\draw [line width=1pt] (-4pt,14pt) -- (4pt,6pt);
				\draw [line width=1pt] (-4pt,6pt) -- (4pt,14pt);
				\draw [line width=1pt] (16pt,14pt) -- (24pt,6pt);
				\draw [line width=1pt] (16pt,6pt) -- (24pt,14pt);
			\end{tikzpicture}
			& $\text{SWAP}$-application. \\
			\begin{tikzpicture}
				\draw [line width=1pt,dashed,rounded corners] (-8pt,-8pt) -- (-8pt,8pt) -- (8pt,8pt) -- (8pt,-8pt) -- cycle;
			\end{tikzpicture}
			& A set, whose measurements are those enclosed. \\
			\begin{tikzpicture}
				\draw [line width=1pt,dotted] (0,10pt) circle (8pt);
			\end{tikzpicture}
			& The measurement enclosed is in a boundary set. \\
			\hline\hline
		\end{tabular*}
		\caption{Notation scheme for parts of figures~\ref{fig:Nestcheck.1}
to~\ref{fig:Nestcheck.5} that display physical gate sequences specified by
frames.}
		\label{tab:Nestcheck.2}
	\end{center}
\end{table}

\begin{figure}[t!]
	\begin{center}
		\subfloat[]{%
			{\normalsize
				\input{pic_isoaddprimal.tex}
			}
		}
		\subfloat[]{%
			{\normalsize
				\input{pic_isoaddprimal2.tex}
			}
		}\\
		\subfloat[]{%
			{\normalsize
				\input{pic_addX.tex}
			}
		}
		\subfloat[]{%
			{\normalsize
				\input{pic_addX2.tex}
			}
		}
		\caption{Each pair (a)-(c) and (b)-(d) is a program consisting of a
an $I_\text{L}$-frame followed by a \texttt{ADD\symbol{95}X}-frame. (a) and
(b) are isometric views with upwards world lines: See
table~\ref{tab:Nestcheck.2} for the notation scheme. (c) and (d) are
representations of the program frames: Read upwards. See
table~\ref{tab:Nestcheck.1} for the notation scheme.}
		\label{fig:Nestcheck.2}
	\end{center}
\end{figure}
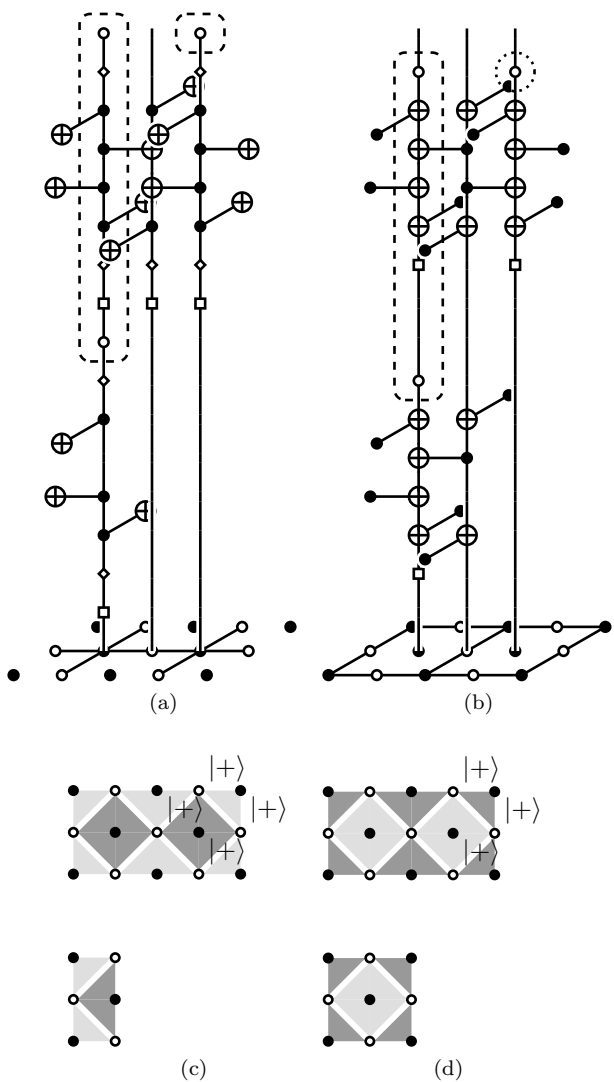

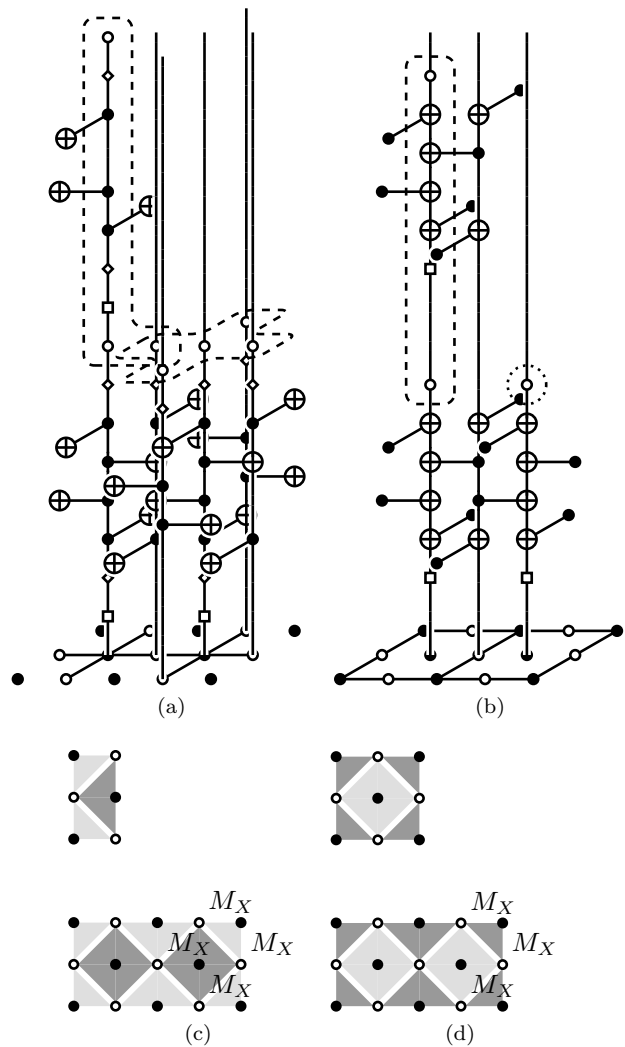
\begin{figure}[t!]
	\begin{center}
		\subfloat[]{%
			{\normalsize
				\input{pic_isotrimprimal.tex}
			}
		}
		\subfloat[]{%
			{\normalsize
				\input{pic_isotrimprimal2.tex}
			}
		}\\
		\subfloat[]{%
			{\normalsize
				\input{pic_trimX.tex}
			}
		}
		\subfloat[]{%
			{\normalsize
				\input{pic_trimX2.tex}
			}
		}
		\caption{Each pair (a)-(c) and (b)-(d) is a program consisting of
an \texttt{TRIM\symbol{95}X}-frame followed by a $I_\text{L}$-frame. (a) and
(b) are isometric views with upwards world lines: See
table~\ref{tab:Nestcheck.2} for the notation scheme. (c) and (d) are
representations of the program frames: Read upwards. See
table~\ref{tab:Nestcheck.1} for the notation scheme.}
		\label{fig:Nestcheck.3}
	\end{center}
\end{figure}

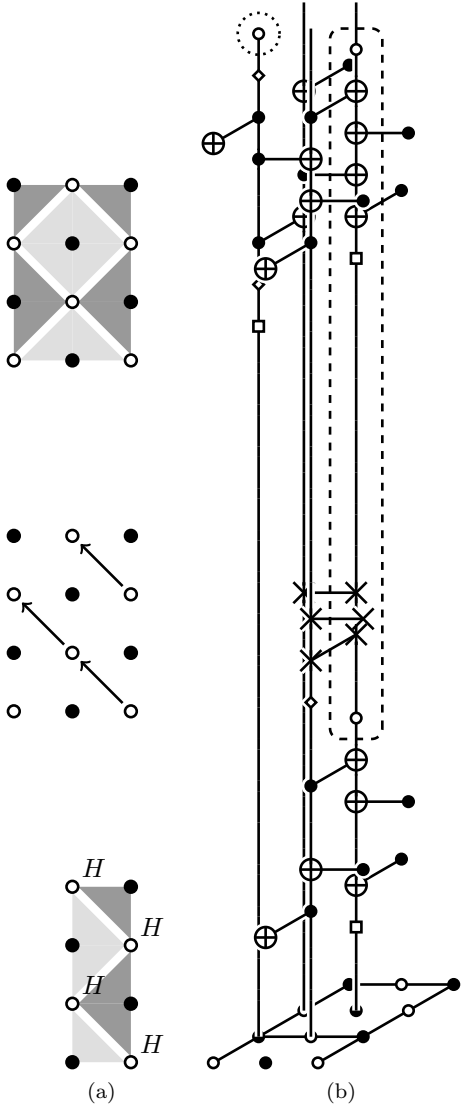
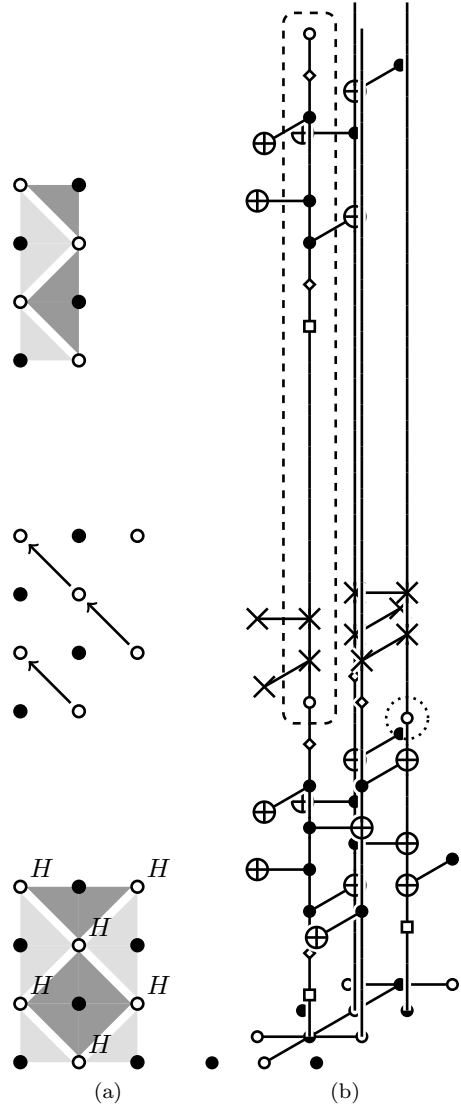
\begin{figure}[t!]
	\begin{center}
		\subfloat[]{%
			{\normalsize
				\input{pic_swap.tex}
			}
		}
		\subfloat[]{%
			{\normalsize
				\input{pic_isoswap.tex}
			}
		}
		\caption{A program consisting of a \texttt{HAD}-frame followed by a
\texttt{SWAP\symbol{95}NW}-frame followed by an $I_\text{L}$-frame. Note that
these frames are to be understood as possibly extending an arbitrary distance
in the north, west, east and south directions, lest data qubits appear from
and dissapear to nowhere (see Fig.~\ref{fig:Distance three logical
hadamard.2}(d) of appendix~\ref{appx:Distance three logical hadamard} for an
example of a complete swapping). (a) Representation of the program frames:
Read upwards. See table~\ref{tab:Nestcheck.1} for the notation scheme. (b)
Isometric view with upwards world lines: See table~\ref{tab:Nestcheck.2} for
the notation scheme.}
		\label{fig:Nestcheck.4}
	\end{center}
\end{figure}

We construct a surface code computation by writing a program. A program is a
series of frames. A frame specifies a single round of error correction and is
a list of physical qubits each with two data: a command and a boundary
specification. We require boundary specifications for while Autotune can
discern when a boundary set is required, it cannot discern which boundary set.
Figs.~\ref{fig:Nestcheck.1} to~\ref{fig:Nestcheck.5} display small programs
and elucidate how these specify gate sequences and sets. From these figures
the logic of creating and placing measurements in sets can in part be infered.
We expand on this below.

There is some arbitrariness in how we use frames to specify sets. In sets
containing two measurements, the measurements often have different times. The
two measurements are specified by different frames and so the set is specified
by these two frames conjointly. It is thus somewhat arbitrary which frame we
choose to specify the boundary set that this set may connect to. We choose the
frame executed at a later time.

It is convenient to introduce some terminology for discussing frames: If every
data qubit has either the command \texttt{INACTIVE} or \texttt{ACTIVE} we call
the frame an $I_\text{L}$-frame. Otherwise, we call the frame by the commands
of the data qubits that are not \texttt{INACTIVE} or \texttt{ACTIVE}. For
example, if some data qubit has the command \texttt{HAD} we call the frame a
\texttt{HAD}-frame. Obviously, this is not the best terminology for frames in
general: what do we call a frame in which some qubits have the command
\texttt{ADD\symbol{95}X} while others have the command
\texttt{ADD\symbol{95}Z}? Nevertheless, it serves us well in the discussion of
boundary sets below.

In order to handle boundary sets, nonrepeating $I_\text{L}$-frames must often
intersperse the frames of a program. By `nonrepeating $I_\text{L}$-frame' we
mean an $I_\text{L}$-frame in which more boundaries are specified than
necessary were we to continually execute it. This is best illustrated by
examining programs in which various frames are followed by
$I_\text{L}$-frames. There are four cases:

\begin{figure}[t!]
	\begin{center}
		\subfloat[]{%
			{\normalsize
				\input{pic_identity.tex}
			}
		}
		\subfloat[]{%
			{\normalsize
				\input{pic_isostandardplate.tex}
			}
		}
		\caption{A program consisting of two $I_\text{L}$-frames. (a)
Representation of the program frames: Read upwards. See
table~\ref{tab:Nestcheck.1} for the notation scheme. (b) Isometric view with
upwards world lines: See table~\ref{tab:Nestcheck.2} for the notation scheme.}
		\label{fig:Nestcheck.1}
	\end{center}
\end{figure}
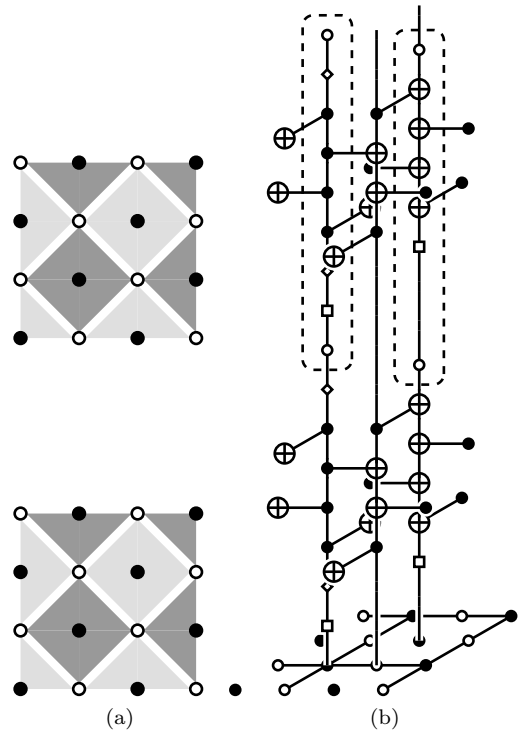

(i) An $I_\text{L}$-frame followed by $I_\text{L}$-frames: Obviously,
of the $I_\text{L}$-frames following the $I_\text{L}$-frame none needs to be
nonrepeating. See Fig.~\ref{fig:Nestcheck.1}.

(ii) An \texttt{ADD\symbol{95}X}- or \texttt{ADD\symbol{95}Z}-frame
followed by $I_\text{L}$-frames: Consider specifically the
\texttt{ADD\symbol{95}X}-frame shown in Fig.~\ref{fig:Nestcheck.2}(d).
Syndrome qubits are activated in this frame. Since we are
$\ket{+}$-initialising data qubits, the first syndrome result of a dual
syndrome qubit activated in this frame is random and must be placed in a
boundary set and a set that connects to this boundary. The user must specify
which boundary set in the \texttt{ADD\symbol{95}X}-frame, and while Autotune
can discern that the connection mentioned is required it cannot discern which
boundary set must be connected to. The user must therefore specify the
connection in the directly following $I_\text{L}$-frame. Such boundary
information is not required in a repeating $I_\text{L}$-frame. Thus the
$I_\text{L}$-frame that directly follows the \texttt{ADD\symbol{95}X}- or
\texttt{ADD\symbol{95}Z}-frame needs to be nonrepeating, all other following
$I_\text{L}$-frames can be repeating.

(iii) A \texttt{TRIM\symbol{95}X}- or \texttt{TRIM\symbol{95}Z}-frame
followed by $I_\text{L}$-frames: Consider specifically the
\texttt{TRIM\symbol{95}X}-frame in Fig.~\ref{fig:Nestcheck.3}(d). In the
directly following frame, syndrome qubits are deactivated. We cannot use the
$X$-basis-measurements of data qubits to construct dual ($Z$) syndrome
results. The final syndrome result of a dual syndrome qubit deactivated in the
directly following $I_\text{L}$-frame must hence be placed in a boundary set
and a set that connects to this boundary. Both of these sets are specified by
the \texttt{TRIM\symbol{95}X}-frame. Thus none of the $I_\text{L}$-frames that
follows the \texttt{TRIM\symbol{95}X}- or \texttt{TRIM\symbol{95}Z}-frame
needs to be nonrepeating.

(iv) A \texttt{HAD}-frame followed by a \texttt{SWAP\symbol{95}NW}-,
\texttt{SWAP\symbol{95}NE}-, \texttt{SWAP\symbol{95}SW}- or
\texttt{SWAP\symbol{95}SE}-frame followed by $I_\text{L}$-frames: Consider
specifically the \texttt{SWAP\symbol{95}NW}-frame in
Fig.~\ref{fig:Nestcheck.4}(b). In the directly following frame, syndrome
qubits are activated. Ideally, the first syndrome result of each syndrome
qubit activated in the directly following frame should be matched with that
of the previous \texttt{HAD}-frame diagonally offset from it. This would
involve, however, converting a primal nest into a dual nest and vice versa,
which would in practive be quite a bit of effort code. Perhaps in future
research we shall concern ourselves with this, but for the moment we use a
shortcut. We simply consider activated syndromes as entirely new. The first
syndrome result of each syndrome qubit activated in the directly following
frame is then considered random and must be placed in a boundary set and a set
that connects to this boundary. This situation is similar to case~(i) but one
frame onwards. Thus two of the $I_\text{L}$-frames that directly follow the
\texttt{SWAP\symbol{95}NW}-, \texttt{SWAP\symbol{95}NE}-,
\texttt{SWAP\symbol{95}SW}- or \texttt{SWAP\symbol{95}SE}-frame need to be
nonrepeating.

\begin{figure}[t!]
	\begin{center}
		\subfloat[]{%
			{\normalsize
				\input{pic_swap2.tex}
			}
		}
		\subfloat[]{%
			{\normalsize
				\input{pic_isoswap2.tex}
			}
		}
		\caption{A program consisting of a \texttt{HAD}-frame followed by a
\texttt{SWAP\symbol{95}NW}-frame followed by an $I_\text{L}$-frame. Note that
these frames are to be understood as possibly extending an arbitrary distance
in the north, west, east and south directions, lest data qubits appear from
and dissapear to nowhere (see Fig.~\ref{fig:Distance three logical
hadamard.2}(d) of appendix~\ref{appx:Distance three logical hadamard} for an
example of a complete swapping). (a) Representation of the program frames:
Read upwards. See table~\ref{tab:Nestcheck.1} for the notation scheme. (b)
Isometric view with upwards world lines: See table~\ref{tab:Nestcheck.2} for
the notation scheme.}
		\label{fig:Nestcheck.5}
	\end{center}
\end{figure}

A further note is necessary on case~(iv): The \texttt{HAD}-frame requires more
boundaries than one might na\"{i}vely expect to be specified. Consider
specifically the \texttt{SWAP\symbol{95}NW}-frame in
Fig.~\ref{fig:Nestcheck.5}(b). In this frame syndrome qubits are deactivated.
Since we are using a shortcut, the final syndrome result of each syndrome
qubit deactivated in this frame must be placed in a boundary set and a set
that connects to this set. Both of these sets are specified by the previous
frame, the \texttt{HAD}-frame. Thus the \texttt{HAD}-frame must include
boundary information to this effect.

An indispensible tool used in debugging the logic for handling sets is a
simple Blender visualizer we have developed \cite{fowler2012tca}. This tool
creates a 3D blender model of a nest so that we can see whether the correct
boundaries have been specified and the correct connections exist.
Fig.~\ref{fig:Nestcheck.6} is such a model of a section of the primal nest for
$d=3$ $H_\text{L}$ as set out in appendix~\ref{appx:Distance three logical
hadamard}.

\begin{figure}[t!]
	\begin{center}
		\includegraphics[width=0.8\linewidth]{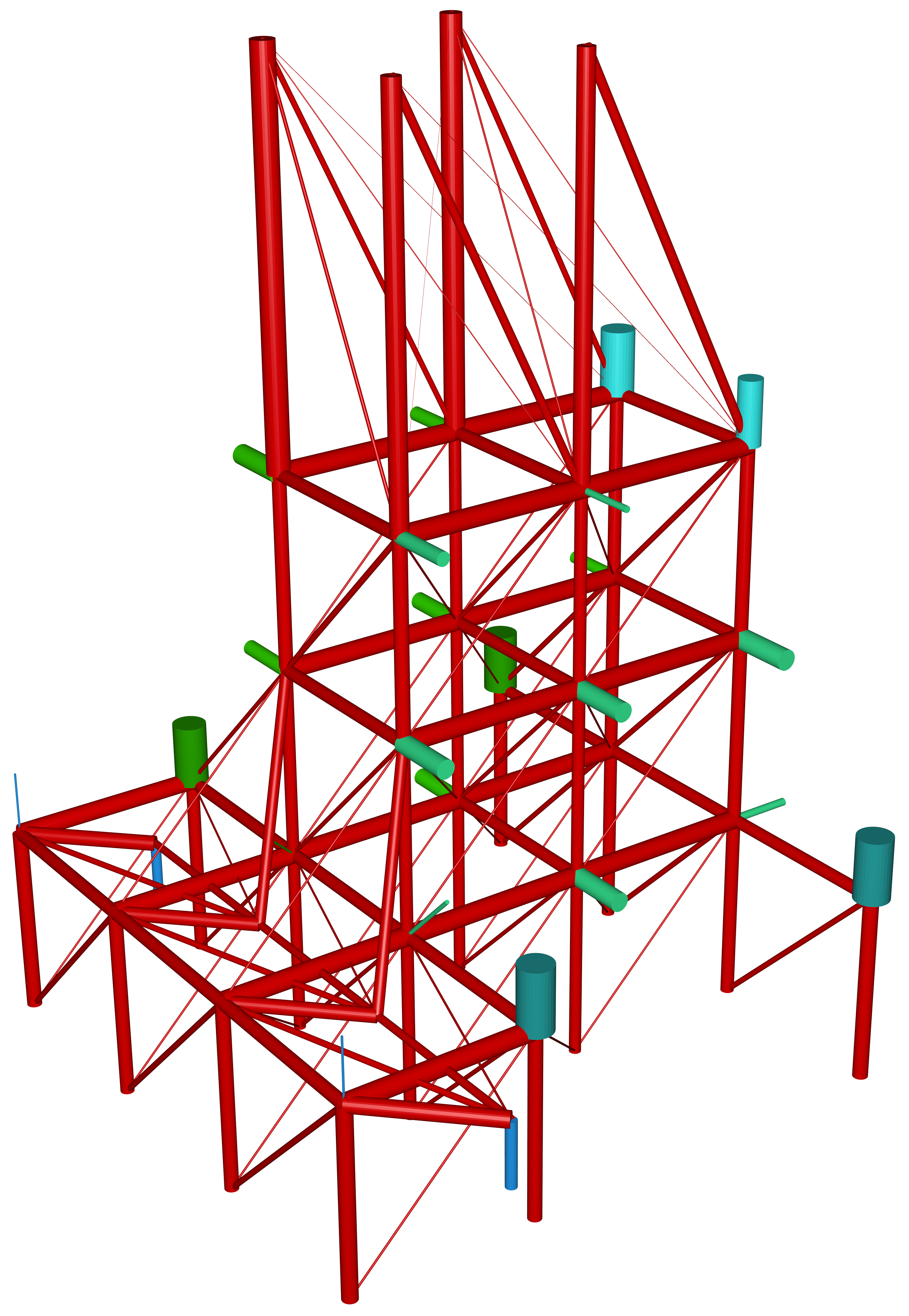}
		\caption{Color online. Blender model of a section of the primal nest
for $d=3$ $H_\text{L}$ as presented in appendix~\ref{appx:Distance three logical
hadamard}. Thicker cylinders (sticks) represent more probable connections
between potential syndrome result changes (balls) due to a single error. Red
cylinders represent sticks between sets, differently colored cylinders
represent sticks between a set and a boundary set. In this nest section we see
trimming and the beginning of swapping. There are more sets than usual in a
trimming time-slice, hence the offset cylinder intersection points.}
		\label{fig:Nestcheck.6}
	\end{center}
\end{figure}

With the primal and dual nests built for a particular surface code logical
gate we then need to find the shortest chain between any two boundaries and
the smallest topologically nontrivial ring. The length of the smaller of these
two is the distance of this surface code logical gate and so the error
correction strength.

Finding the shortest chain between two boundaries is simple. We loop over
pairs of boundaries and perform a breadth first search between them.

As to finding the smallest topologically nontrivial ring the situation is not
quite so undemanding and we have not yet implemented such a search. This is
not to invalidate Nestcheck, however, for rings are generally easy to see.
Nevertheless, without this ability Nestcheck cannot be considered complete.

In future work we plan to search for a set of smallest topologically distinct
and nontrivial rings. An outline of our prospective method is thus: The `hole'
around which a topologically nontrivial ring in a nest of one type can exist
is itself a ring, or at least a closed structure \cite{fowler2012abt}, of the
other type. This latter ring can be specified by the user. The task is then
reduced to, for each input ring of a certain type, looping over an
appropriately constructed set of rings in the nest of the other type and
checking whether each interlocks with the input ring. The conceptually
principal part of this algorithm is a method for determining whether a given
primal ring interlocks with a given dual ring.

Nestcheck has been tested extensively on a wide variety of physical gate
sequences. Furthermore, as an example, we have run Nestcheck on a proposal for
$d=3$ $H_\text{L}$ (see appendix~\ref{appx:Distance three logical hadamard}).
Nestcheck found that this proposal is $d=3$ as purported.

\section{Conclusion}
\label{sec:Conclusion}

We have reported a new tool Nestcheck capable of analysing an arbitrary
surface code computation and determining the minimum number of errors required
to cause failure. 3D topological cluster states may also be analysed since
the data that Nestcheck analyses is created using Autotune. In future
research, surface code computations consisting of many qubits shall be
constructed, for which it is highly nontrivial to check the strength of error
correction. Nestcheck shall therefore be of utility in this endeavour.
Furthermore, using the framework of Nestcheck (programs, frames, and so on),
we are able to easily specify a complex surface code computation thus enabling
simulation of such to be relatively quickly and painlessly written.

\begin{acknowledgments}
This research was conducted by the Australian Research Council Centre of
Excellence for Quantum Computation and Communication Technology (project
number CE110001027), with support from the United States National Security
Agency and the United States Army Research Office under contract number
W911NF-08-1-0527. Supported by the Intelligence Advanced Research Projects
Activity (IARPA) via Department of Interior National Business Center contract
number D11PC20166. The United States Government is authorized to reproduce and
distribute reprints for governmental purposes notwithstanding any copyright
annotation thereon. Disclaimer: The views and conclusions contained herein are
those of the authors and should not be interpreted as necessarily representing
the official policies or endorsements, either expressed or implied, of IARPA,
DoI/NBC, or the United States Government.
\end{acknowledgments}

\appendix

\section{Full program for a distance three plate}
\label{appx:Distance three plate with boundary specifications}

\begin{figure}[t!]
	\begin{center}
		\subfloat[]{%
			{\normalsize%
				\input{pic_d3_3.tex}
			}
		}\\
		\subfloat[]{%
			{\normalsize%
				\input{pic_d3_2.tex}
			}
		}\\
		\subfloat[]{%
			{\normalsize%
				\input{pic_d3_1.tex}
			}
		}
		\caption{Full program for simulating the surface code plate of
Fig.~\ref{fig:The surface code.1}. Read (a), (b) then (c). See 
table~\ref{tab:Nestcheck.1} for the notation scheme. Arrows represent boundary
specifications: solid and dashed for primal and dual respectively, and bent
and straight for temporal and spatial respectively. We see three primal
boundaries: a temporal one, one along the top and one along the bottom. We see
two dual boundaries: one along the left and another along the right. Note that
in general spatiotemporal boundary specifications are possible.}
		\label{fig:Full program for a distance three plate.1}
	\end{center}
\end{figure}
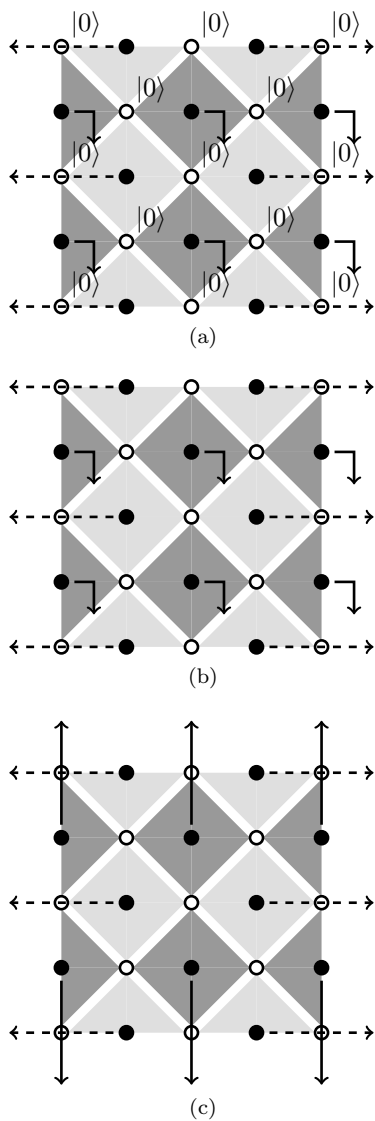

Fig.~\ref{fig:Full program for a distance three plate.1} details an entire
program for simulating the surface code plate of Fig.~\ref{fig:The surface
code.1}. Boundary specifications have been included. The final frame,
Fig.~\ref{fig:Full program for a distance three plate.1}(c), is
repeating in the sense of section~\ref{sec:Nestcheck}.

\section{Distance three logical hadamard}
\label{appx:Distance three logical hadamard}

A scheme for $d=7$ $H_\text{L}$ is presented in \cite{fowler2012los}. We adapt
this to $d=3$. Besides nonrepeating $I_\text{L}$-frames (see
section~\ref{sec:Nestcheck}, Figs.~\ref{fig:Distance three logical hadamard.1}
to~\ref{fig:Distance three logical hadamard.3} present this adaptation.

\section{Checking the logic of specific surface code logical gates}
\label{appx:Checking the logic of specific surface code logical gates}

There would be little point in checking the error correction strength of a
proposal for surface code logical gate without checking whether it is indeed
the logical gate purported. Future research shall potentially yield a general
tool for this task \cite{fowler2012abt}, but for the moment we resort to
specific methods.

Excepting initializations specific to state-injection, the physical gates of
an arbitary surface code computation are in the clifford group. Algorithms for
efficiently simulating clifford group quantum circuits have been developed and
implemented, and are available for download. We currently use Graphsim
\cite{anders2006fso}. Constructing a Graphsim simulation is simple: we declare
a qubit register and then call our gate sequence.

To view the state of the qubit register at a point during simulation there are
two methods. We may either print a table of stabilizers, or simply create a
duplicate of the qubit register and measure all the qubits in an appropriate
basis. Since we are checking rather than investigating logic we choose the
latter method.

Take $H_\text{L}$ as an example. Here we must check whether the gate maps
$\ket{0}$, $\ket{1}$, $\ket{+}$ and $\ket{-}$ to $\ket{+}$, $\ket{-}$,
$\ket{0}$ and $\ket{1}$ respectively. Consider the case $\ket{0} \mapsto
\ket{+}$: We initialize the logical qubit to $\ket{0}$ and then apply the gate
sequence of $H_\text{L}$, during which course byproduct-operators
\cite{fowler2009tcs} accrue. Then we measure in the $X$-basis each qubit to
which $X_\text{L}$ is nontrivial and check whether the multiple of these
measurements with those associated with the appropriate byproduct-operators is
$1$. Success is declared if so, otherwise failure.

This method of checking logic is stochastic. In order to reliably check using
this method, then, we must apply it a large number of times. If failure is
declared even once then the proposal fails entirely.

\bibliography{References}

\begin{figure*}[f]
	\begin{center}
		\subfloat[]{%
			{\normalsize
				\input{001ex2.tex}
			}
		}
		\subfloat[]{%
			{\normalsize
				\input{002ex2.tex}
			}
		}\\
		\subfloat[]{%
			{\normalsize
				\input{003ex2.tex}
			}
		}
		\subfloat[]{%
			{\normalsize
				\input{004ex2.tex}
			}
		}
	\end{center}
	\caption{Frames for simulating $d=3$ $H_\text{L}$. See
table~\ref{tab:Nestcheck.1} for the notation scheme.}
	\label{fig:Distance three logical hadamard.1}
\end{figure*}

\begin{figure*}[f]
	\begin{center}
		\subfloat[]{%
			{\normalsize
				\input{005ex2.tex}
			}
		}
		\subfloat[]{%
			{\normalsize
				\input{006ex2.tex}
			}
		}\\
		\subfloat[]{%
			{\normalsize
				\input{007ex2.tex}
			}
		}
		\subfloat[]{%
			{\normalsize
				\input{008ex2.tex}
			}
		}
	\end{center}
	\caption{Frames for simulating $d=3$ $H_\text{L}$. See
table~\ref{tab:Nestcheck.1} for the notation scheme.}
	\label{fig:Distance three logical hadamard.2}
\end{figure*}
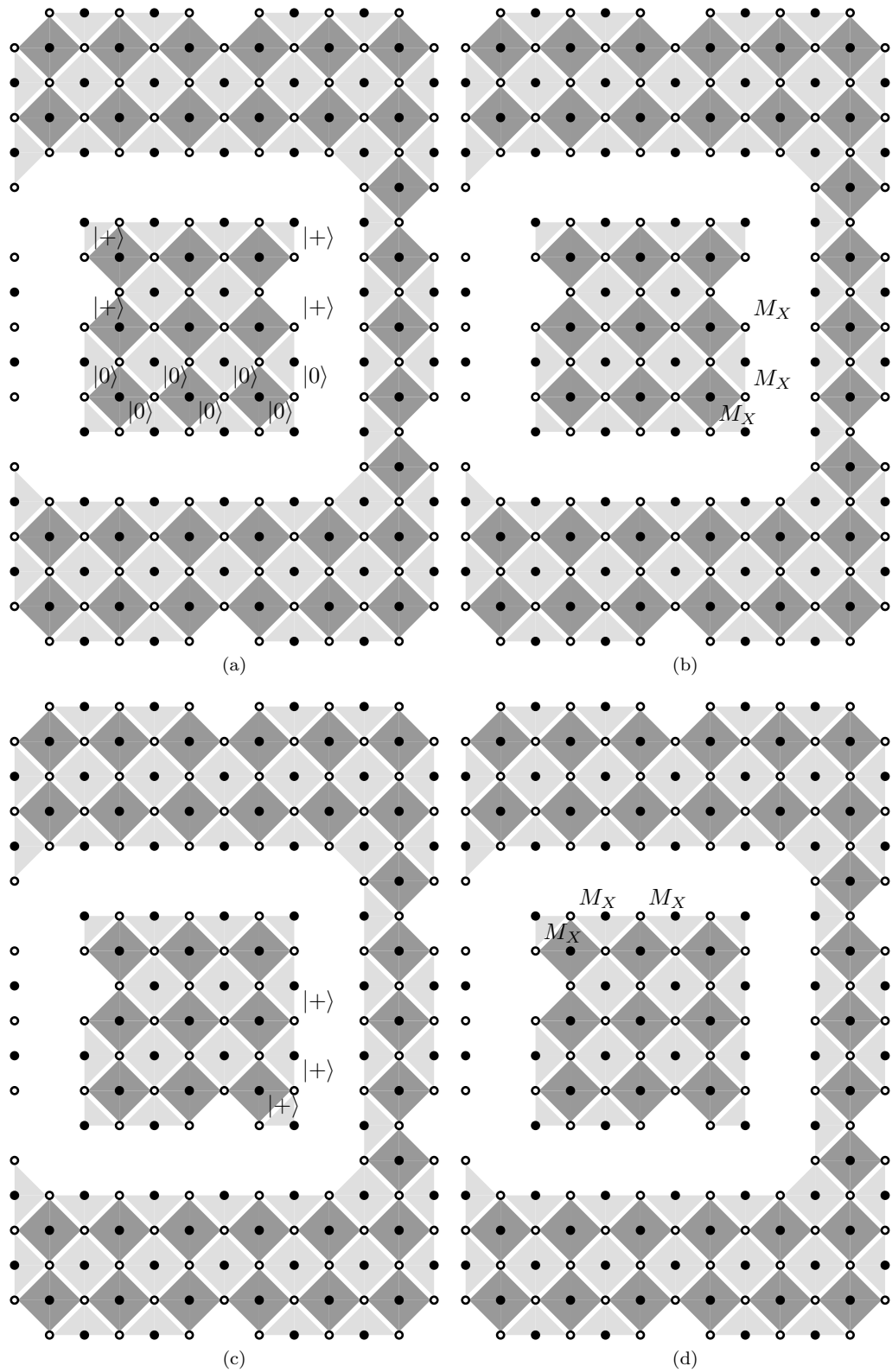

\begin{figure*}[f]
	\begin{center}
		\subfloat[]{%
			{\normalsize
				\input{009ex2.tex}
			}
		}
		\subfloat[]{%
			{\normalsize
				\input{010ex2.tex}
			}
		}
	\end{center}
	\caption{Frames for simulating $d=3$ $H_\text{L}$. See
table~\ref{tab:Nestcheck.1} for the notation scheme.}
	\label{fig:Distance three logical hadamard.3}
\end{figure*}

\end{document}

%% file: pic_circuits.tex
\begin{tikzpicture}[x=0.11\linewidth,y=0.11\linewidth]
	\draw (0,0) node {$\ket{Q_4}$};
	\draw (0,1) node {$\ket{Q_3}$};
	\draw (0,2) node {$\ket{Q_2}$};
	\draw (0,3) node {$\ket{Q_1}$};
	\draw (0,4) node {$\ket{+}$};
	\draw [line width=1pt] (1,0) -- (7,0);
	\draw [line width=1pt] (1,1) -- (7,1);
	\draw [line width=1pt] (1,2) -- (7,2);
	\draw [line width=1pt] (1,3) -- (7,3);
	\draw [line width=1pt] (1,4) -- (7,4);
	\draw [line width=1pt,fill=black] (2,4) circle (0.1);
	\draw [line width=1pt,fill=black] (3,4) circle (0.1);
	\draw [line width=1pt,fill=black] (4,4) circle (0.1);
	\draw [line width=1pt,fill=black] (5,4) circle (0.1);
	\draw [line width=1pt] (2,4) -- (2,2.8);
	\draw [line width=1pt] (3,4) -- (3,1.8);
	\draw [line width=1pt] (4,4) -- (4,0.8);
	\draw [line width=1pt] (5,4) -- (5,-0.2);
	\draw [line width=1pt] (2,3) circle (0.2);
	\draw [line width=1pt] (3,2) circle (0.2);
	\draw [line width=1pt] (4,1) circle (0.2);
	\draw [line width=1pt] (5,0) circle (0.2);
	\draw [line width=1pt,fill=white] (5.6,3.6) -- (5.6,4.4) -- (6.4,4.4) -- (6.4,3.6) -- cycle;
	\draw (6,4) node {$M_X$};
\end{tikzpicture}

%% file: pic_circuits2.tex
\begin{tikzpicture}[x=0.11\linewidth,y=0.11\linewidth]
	\draw (0,0) node {$\ket{Q_4}$};
	\draw (0,1) node {$\ket{Q_3}$};
	\draw (0,2) node {$\ket{Q_2}$};
	\draw (0,3) node {$\ket{Q_1}$};
	\draw (0,4) node {$\ket{0}$};
	\draw [line width=1pt] (1,0) -- (7,0);
	\draw [line width=1pt] (1,1) -- (7,1);
	\draw [line width=1pt] (1,2) -- (7,2);
	\draw [line width=1pt] (1,3) -- (7,3);
	\draw [line width=1pt] (1,4) -- (7,4);
	\draw [line width=1pt] (2,4) circle (0.2);
	\draw [line width=1pt] (3,4) circle (0.2);
	\draw [line width=1pt] (4,4) circle (0.2);
	\draw [line width=1pt] (5,4) circle (0.2);
	\draw [line width=1pt] (2,4.2) -- (2,3);
	\draw [line width=1pt] (3,4.2) -- (3,2);
	\draw [line width=1pt] (4,4.2) -- (4,1);
	\draw [line width=1pt] (5,4.2) -- (5,0);
	\draw [line width=1pt,fill=black] (2,3) circle (0.1);
	\draw [line width=1pt,fill=black] (3,2) circle (0.1);
	\draw [line width=1pt,fill=black] (4,1) circle (0.1);
	\draw [line width=1pt,fill=black] (5,0) circle (0.1);
	\draw [line width=1pt,fill=white] (5.6,3.6) -- (5.6,4.4) -- (6.4,4.4) -- (6.4,3.6) -- cycle;
	\draw (6,4) node {$M_Z$};
\end{tikzpicture}

%% file: pic_smallestplate.tex
\begin{tikzpicture}[x=0.16666666*\the\linewidth,y=0.16666666*\the\linewidth]
	\fill[gray!25] (3.000000,5.000000) -- (3.900000,5.000000) -- (3.000000,4.100000) -- cycle;
	\fill[gray!25] (3.000000,5.000000) -- (3.000000,4.100000) -- (2.100000,5.000000) -- cycle;
	\fill[gray!25] (5.000000,5.000000) -- (5.900000,5.000000) -- (5.000000,4.100000) -- cycle;
	\fill[gray!25] (5.000000,5.000000) -- (5.000000,4.100000) -- (4.100000,5.000000) -- cycle;
	\fill[gray!80] (2.000000,4.000000) -- (2.000000,4.900000) -- (2.900000,4.000000) -- cycle;
	\fill[gray!80] (2.000000,4.000000) -- (2.900000,4.000000) -- (2.000000,3.100000) -- cycle;
	\fill[gray!80] (4.000000,4.000000) -- (3.100000,4.000000) -- (4.000000,4.900000) -- cycle;
	\fill[gray!80] (4.000000,4.000000) -- (4.000000,4.900000) -- (4.900000,4.000000) -- cycle;
	\fill[gray!80] (4.000000,4.000000) -- (4.900000,4.000000) -- (4.000000,3.100000) -- cycle;
	\fill[gray!80] (4.000000,4.000000) -- (4.000000,3.100000) -- (3.100000,4.000000) -- cycle;
	\fill[gray!80] (6.000000,4.000000) -- (5.100000,4.000000) -- (6.000000,4.900000) -- cycle;
	\fill[gray!80] (6.000000,4.000000) -- (6.000000,3.100000) -- (5.100000,4.000000) -- cycle;
	\fill[gray!25] (3.000000,3.000000) -- (2.100000,3.000000) -- (3.000000,3.900000) -- cycle;
	\fill[gray!25] (3.000000,3.000000) -- (3.000000,3.900000) -- (3.900000,3.000000) -- cycle;
	\fill[gray!25] (3.000000,3.000000) -- (3.900000,3.000000) -- (3.000000,2.100000) -- cycle;
	\fill[gray!25] (3.000000,3.000000) -- (3.000000,2.100000) -- (2.100000,3.000000) -- cycle;
	\fill[gray!25] (5.000000,3.000000) -- (4.100000,3.000000) -- (5.000000,3.900000) -- cycle;
	\fill[gray!25] (5.000000,3.000000) -- (5.000000,3.900000) -- (5.900000,3.000000) -- cycle;
	\fill[gray!25] (5.000000,3.000000) -- (5.900000,3.000000) -- (5.000000,2.100000) -- cycle;
	\fill[gray!25] (5.000000,3.000000) -- (5.000000,2.100000) -- (4.100000,3.000000) -- cycle;
	\fill[gray!80] (2.000000,2.000000) -- (2.000000,2.900000) -- (2.900000,2.000000) -- cycle;
	\fill[gray!80] (2.000000,2.000000) -- (2.900000,2.000000) -- (2.000000,1.100000) -- cycle;
	\fill[gray!80] (4.000000,2.000000) -- (3.100000,2.000000) -- (4.000000,2.900000) -- cycle;
	\fill[gray!80] (4.000000,2.000000) -- (4.000000,2.900000) -- (4.900000,2.000000) -- cycle;
	\fill[gray!80] (4.000000,2.000000) -- (4.900000,2.000000) -- (4.000000,1.100000) -- cycle;
	\fill[gray!80] (4.000000,2.000000) -- (4.000000,1.100000) -- (3.100000,2.000000) -- cycle;
	\fill[gray!80] (6.000000,2.000000) -- (5.100000,2.000000) -- (6.000000,2.900000) -- cycle;
	\fill[gray!80] (6.000000,2.000000) -- (6.000000,1.100000) -- (5.100000,2.000000) -- cycle;
	\fill[gray!25] (3.000000,1.000000) -- (2.100000,1.000000) -- (3.000000,1.900000) -- cycle;
	\fill[gray!25] (3.000000,1.000000) -- (3.000000,1.900000) -- (3.900000,1.000000) -- cycle;
	\fill[gray!25] (5.000000,1.000000) -- (4.100000,1.000000) -- (5.000000,1.900000) -- cycle;
	\fill[gray!25] (5.000000,1.000000) -- (5.000000,1.900000) -- (5.900000,1.000000) -- cycle;
	\draw[line width=1pt] (2.000000,5.000000) circle (0.100000);
	\draw[line width=1pt,fill=black] (3.000000,5.000000) circle (0.100000);
	\draw[line width=1pt] (4.000000,5.000000) circle (0.100000);
	\draw[line width=1pt,fill=black] (5.000000,5.000000) circle (0.100000);
	\draw[line width=1pt] (6.000000,5.000000) circle (0.100000);
	\draw[line width=1pt,fill=black] (2.000000,4.000000) circle (0.100000);
	\draw[line width=1pt] (3.000000,4.000000) circle (0.100000);
	\draw[line width=1pt,fill=black] (4.000000,4.000000) circle (0.100000);
	\draw[line width=1pt] (5.000000,4.000000) circle (0.100000);
	\draw[line width=1pt,fill=black] (6.000000,4.000000) circle (0.100000);
	\draw[line width=1pt] (2.000000,3.000000) circle (0.100000);
	\draw[line width=1pt,fill=black] (3.000000,3.000000) circle (0.100000);
	\draw[line width=1pt] (4.000000,3.000000) circle (0.100000);
	\draw[line width=1pt,fill=black] (5.000000,3.000000) circle (0.100000);
	\draw[line width=1pt] (6.000000,3.000000) circle (0.100000);
	\draw[line width=1pt,fill=black] (2.000000,2.000000) circle (0.100000);
	\draw[line width=1pt] (3.000000,2.000000) circle (0.100000);
	\draw[line width=1pt,fill=black] (4.000000,2.000000) circle (0.100000);
	\draw[line width=1pt] (5.000000,2.000000) circle (0.100000);
	\draw[line width=1pt,fill=black] (6.000000,2.000000) circle (0.100000);
	\draw[line width=1pt] (2.000000,1.000000) circle (0.100000);
	\draw[line width=1pt,fill=black] (3.000000,1.000000) circle (0.100000);
	\draw[line width=1pt] (4.000000,1.000000) circle (0.100000);
	\draw[line width=1pt,fill=black] (5.000000,1.000000) circle (0.100000);
	\draw[line width=1pt] (6.000000,1.000000) circle (0.100000);
\end{tikzpicture}

%% file: pic_identity.tex
\begin{tikzpicture}[x=0.090000\linewidth,y=0.090000\linewidth]
	\fill[gray!80] (3.000000,11.000000) -- (3.900000,11.000000) -- (3.000000,10.100000) -- cycle;
	\fill[gray!80] (3.000000,11.000000) -- (3.000000,10.100000) -- (2.100000,11.000000) -- cycle;
	\fill[gray!80] (5.000000,11.000000) -- (5.000000,10.100000) -- (4.100000,11.000000) -- cycle;
	\fill[gray!25] (2.000000,10.000000) -- (2.000000,10.900000) -- (2.900000,10.000000) -- cycle;
	\fill[gray!25] (2.000000,10.000000) -- (2.900000,10.000000) -- (2.000000,9.100000) -- cycle;
	\fill[gray!25] (4.000000,10.000000) -- (3.100000,10.000000) -- (4.000000,10.900000) -- cycle;
	\fill[gray!25] (4.000000,10.000000) -- (4.000000,10.900000) -- (4.900000,10.000000) -- cycle;
	\fill[gray!25] (4.000000,10.000000) -- (4.900000,10.000000) -- (4.000000,9.100000) -- cycle;
	\fill[gray!25] (4.000000,10.000000) -- (4.000000,9.100000) -- (3.100000,10.000000) -- cycle;
	\fill[gray!80] (3.000000,9.000000) -- (2.100000,9.000000) -- (3.000000,9.900000) -- cycle;
	\fill[gray!80] (3.000000,9.000000) -- (3.000000,9.900000) -- (3.900000,9.000000) -- cycle;
	\fill[gray!80] (3.000000,9.000000) -- (3.900000,9.000000) -- (3.000000,8.100000) -- cycle;
	\fill[gray!80] (3.000000,9.000000) -- (3.000000,8.100000) -- (2.100000,9.000000) -- cycle;
	\fill[gray!80] (5.000000,9.000000) -- (4.100000,9.000000) -- (5.000000,9.900000) -- cycle;
	\fill[gray!80] (5.000000,9.000000) -- (5.000000,8.100000) -- (4.100000,9.000000) -- cycle;
	\fill[gray!25] (2.000000,8.000000) -- (2.000000,8.900000) -- (2.900000,8.000000) -- cycle;
	\fill[gray!25] (4.000000,8.000000) -- (3.100000,8.000000) -- (4.000000,8.900000) -- cycle;
	\fill[gray!25] (4.000000,8.000000) -- (4.000000,8.900000) -- (4.900000,8.000000) -- cycle;
	\fill[gray!80] (3.000000,5.000000) -- (3.900000,5.000000) -- (3.000000,4.100000) -- cycle;
	\fill[gray!80] (3.000000,5.000000) -- (3.000000,4.100000) -- (2.100000,5.000000) -- cycle;
	\fill[gray!80] (5.000000,5.000000) -- (5.000000,4.100000) -- (4.100000,5.000000) -- cycle;
	\fill[gray!25] (2.000000,4.000000) -- (2.000000,4.900000) -- (2.900000,4.000000) -- cycle;
	\fill[gray!25] (2.000000,4.000000) -- (2.900000,4.000000) -- (2.000000,3.100000) -- cycle;
	\fill[gray!25] (4.000000,4.000000) -- (3.100000,4.000000) -- (4.000000,4.900000) -- cycle;
	\fill[gray!25] (4.000000,4.000000) -- (4.000000,4.900000) -- (4.900000,4.000000) -- cycle;
	\fill[gray!25] (4.000000,4.000000) -- (4.900000,4.000000) -- (4.000000,3.100000) -- cycle;
	\fill[gray!25] (4.000000,4.000000) -- (4.000000,3.100000) -- (3.100000,4.000000) -- cycle;
	\fill[gray!80] (3.000000,3.000000) -- (2.100000,3.000000) -- (3.000000,3.900000) -- cycle;
	\fill[gray!80] (3.000000,3.000000) -- (3.000000,3.900000) -- (3.900000,3.000000) -- cycle;
	\fill[gray!80] (3.000000,3.000000) -- (3.900000,3.000000) -- (3.000000,2.100000) -- cycle;
	\fill[gray!80] (3.000000,3.000000) -- (3.000000,2.100000) -- (2.100000,3.000000) -- cycle;
	\fill[gray!80] (5.000000,3.000000) -- (4.100000,3.000000) -- (5.000000,3.900000) -- cycle;
	\fill[gray!80] (5.000000,3.000000) -- (5.000000,2.100000) -- (4.100000,3.000000) -- cycle;
	\fill[gray!25] (2.000000,2.000000) -- (2.000000,2.900000) -- (2.900000,2.000000) -- cycle;
	\fill[gray!25] (4.000000,2.000000) -- (3.100000,2.000000) -- (4.000000,2.900000) -- cycle;
	\fill[gray!25] (4.000000,2.000000) -- (4.000000,2.900000) -- (4.900000,2.000000) -- cycle;
	\draw[line width=1pt] (2.000000,11.000000) circle (0.100000);
	\draw[line width=1pt,fill=black] (3.000000,11.000000) circle (0.100000);
	\draw[line width=1pt] (4.000000,11.000000) circle (0.100000);
	\draw[line width=1pt,fill=black] (5.000000,11.000000) circle (0.100000);
	\draw[line width=1pt,fill=black] (2.000000,10.000000) circle (0.100000);
	\draw[line width=1pt] (3.000000,10.000000) circle (0.100000);
	\draw[line width=1pt,fill=black] (4.000000,10.000000) circle (0.100000);
	\draw[line width=1pt] (5.000000,10.000000) circle (0.100000);
	\draw[line width=1pt] (2.000000,9.000000) circle (0.100000);
	\draw[line width=1pt,fill=black] (3.000000,9.000000) circle (0.100000);
	\draw[line width=1pt] (4.000000,9.000000) circle (0.100000);
	\draw[line width=1pt,fill=black] (5.000000,9.000000) circle (0.100000);
	\draw[line width=1pt,fill=black] (2.000000,8.000000) circle (0.100000);
	\draw[line width=1pt] (3.000000,8.000000) circle (0.100000);
	\draw[line width=1pt,fill=black] (4.000000,8.000000) circle (0.100000);
	\draw[line width=1pt] (5.000000,8.000000) circle (0.100000);
	\draw[line width=1pt] (2.000000,5.000000) circle (0.100000);
	\draw[line width=1pt,fill=black] (3.000000,5.000000) circle (0.100000);
	\draw[line width=1pt] (4.000000,5.000000) circle (0.100000);
	\draw[line width=1pt,fill=black] (5.000000,5.000000) circle (0.100000);
	\draw[line width=1pt,fill=black] (2.000000,4.000000) circle (0.100000);
	\draw[line width=1pt] (3.000000,4.000000) circle (0.100000);
	\draw[line width=1pt,fill=black] (4.000000,4.000000) circle (0.100000);
	\draw[line width=1pt] (5.000000,4.000000) circle (0.100000);
	\draw[line width=1pt] (2.000000,3.000000) circle (0.100000);
	\draw[line width=1pt,fill=black] (3.000000,3.000000) circle (0.100000);
	\draw[line width=1pt] (4.000000,3.000000) circle (0.100000);
	\draw[line width=1pt,fill=black] (5.000000,3.000000) circle (0.100000);
	\draw[line width=1pt,fill=black] (2.000000,2.000000) circle (0.100000);
	\draw[line width=1pt] (3.000000,2.000000) circle (0.100000);
	\draw[line width=1pt,fill=black] (4.000000,2.000000) circle (0.100000);
	\draw[line width=1pt] (5.000000,2.000000) circle (0.100000);
\end{tikzpicture}

%% file: pic_isostandardplate.tex
\begin{tikzpicture}[x=0.075780\linewidth,y=0.075780\linewidth]
	\draw[line width=1pt,fill=black] (0.000000,0.000000) circle (0.100000);
	\draw[line width=1pt] (0.866025,0.500000) circle (0.100000);
	\draw[line width=1pt,fill=black] (1.732051,1.000000) circle (0.100000);
	\draw[line width=1pt] (2.598076,1.500000) circle (0.100000);
	\draw[line width=1pt] (1.000000,0.000000) circle (0.100000);
	\draw[line width=1pt] (0.966025,0.500000) -- (1.766025,0.500000);
	\draw[line width=1pt] (1.779423,0.450000) -- (1.086603,0.050000);
	\draw[line width=1pt,fill=black] (1.866025,0.500000) circle (0.100000);
	\draw[line width=1pt] (2.645448,0.950000) -- (1.952628,0.550000);
	\draw[line width=1pt] (2.732051,1.000000) circle (0.100000);
	\draw[line width=1pt] (2.698076,1.500000) -- (3.498076,1.500000);
	\draw[line width=1pt] (3.511474,1.450000) -- (2.818653,1.050000);
	\draw[line width=1pt,fill=black] (3.598076,1.500000) circle (0.100000);
	\draw[line width=1pt,fill=black] (2.000000,0.000000) circle (0.100000);
	\draw[line width=1pt] (1.966025,0.500000) -- (2.766025,0.500000);
	\draw[line width=1pt] (2.866025,0.500000) circle (0.100000);
	\draw[line width=1pt,fill=black] (3.732051,1.000000) circle (0.100000);
	\draw[line width=1pt] (3.698076,1.500000) -- (4.498076,1.500000);
	\draw[line width=1pt] (4.598076,1.500000) circle (0.100000);
	\draw[line width=1pt] (3.000000,0.000000) circle (0.100000);
	\draw[line width=1pt] (2.966025,0.500000) -- (3.766025,0.500000);
	\draw[line width=1pt] (3.779423,0.450000) -- (3.086603,0.050000);
	\draw[line width=1pt,fill=black] (3.866025,0.500000) circle (0.100000);
	\draw[line width=1pt] (4.645448,0.950000) -- (3.952628,0.550000);
	\draw[line width=1pt] (4.732051,1.000000) circle (0.100000);
	\draw[line width=1pt] (4.698076,1.500000) -- (5.498076,1.500000);
	\draw[line width=1pt] (5.511474,1.450000) -- (4.818653,1.050000);
	\draw[line width=1pt,fill=black] (5.598076,1.500000) circle (0.100000);
	\draw[line width=3pt,draw=white] (3.732051,1.000000) -- (3.732051,1.700000);
	\draw[line width=3pt,draw=white] (3.732051,1.700000) -- (3.732051,1.900000);
	\draw[line width=3pt,draw=white] (3.732051,1.900000) -- (3.732051,2.500000);
	\draw[line width=3pt,fill=white,draw=white] (3.632051,2.500000) -- (3.632051,2.700000) -- (3.832051,2.700000) -- (3.832051,2.500000) -- cycle;
	\draw[line width=3pt,draw=white] (3.732051,2.700000) -- (3.732051,3.300000);
	\draw[line width=3pt,draw=white,fill=white] (3.732051,3.400000) circle (0.200000);
	\draw[line width=3pt,draw=white] (3.532051,3.400000) -- (3.932051,3.400000);
	\draw[line width=3pt,draw=white] (3.732051,3.200000) -- (3.732051,3.600000);
	\draw[line width=3pt,draw=white] (3.732051,3.400000) -- (4.598076,3.900000);
	\draw[line width=3pt,draw=white,fill=white] (4.598076,3.900000) circle (0.100000);
	\draw[line width=3pt,draw=white] (3.732051,3.500000) -- (3.732051,4.100000);
	\draw[line width=3pt,draw=white,fill=white] (3.732051,4.200000) circle (0.200000);
	\draw[line width=3pt,draw=white] (3.532051,4.200000) -- (3.932051,4.200000);
	\draw[line width=3pt,draw=white] (3.732051,4.000000) -- (3.732051,4.400000);
	\draw[line width=3pt,draw=white] (3.732051,4.200000) -- (2.732051,4.200000);
	\draw[line width=3pt,draw=white,fill=white] (2.732051,4.200000) circle (0.100000);
	\draw[line width=3pt,draw=white] (3.732051,4.300000) -- (3.732051,4.900000);
	\draw[line width=3pt,draw=white,fill=white] (3.732051,5.000000) circle (0.200000);
	\draw[line width=3pt,draw=white] (3.532051,5.000000) -- (3.932051,5.000000);
	\draw[line width=3pt,draw=white] (3.732051,4.800000) -- (3.732051,5.200000);
	\draw[line width=3pt,draw=white] (3.732051,5.000000) -- (4.732051,5.000000);
	\draw[line width=3pt,draw=white,fill=white] (4.732051,5.000000) circle (0.100000);
	\draw[line width=3pt,draw=white] (3.732051,5.100000) -- (3.732051,5.700000);
	\draw[line width=3pt,draw=white,fill=white] (3.732051,5.800000) circle (0.200000);
	\draw[line width=3pt,draw=white] (3.532051,5.800000) -- (3.932051,5.800000);
	\draw[line width=3pt,draw=white] (3.732051,5.600000) -- (3.732051,6.000000);
	\draw[line width=3pt,draw=white] (3.732051,5.800000) -- (2.866025,5.300000);
	\draw[line width=3pt,draw=white,fill=white] (2.866025,5.300000) circle (0.100000);
	\draw[line width=3pt,draw=white] (3.732051,5.900000) -- (3.732051,6.500000);
	\draw[line width=3pt,fill=white,draw=white] (3.732051,6.600000) circle (0.100000);
	\draw[line width=3pt,draw=white] (3.732051,6.700000) -- (3.732051,7.300000);
	\draw[line width=3pt,draw=white] (3.732051,7.300000) -- (3.732051,7.500000);
	\draw[line width=3pt,draw=white] (3.732051,7.500000) -- (3.732051,8.100000);
	\draw[line width=3pt,draw=white] (3.732051,8.100000) -- (3.732051,8.300000);
	\draw[line width=3pt,draw=white] (3.732051,8.300000) -- (3.732051,8.900000);
	\draw[line width=3pt,fill=white,draw=white] (3.632051,8.900000) -- (3.632051,9.100000) -- (3.832051,9.100000) -- (3.832051,8.900000) -- cycle;
	\draw[line width=3pt,draw=white] (3.732051,9.100000) -- (3.732051,9.700000);
	\draw[line width=3pt,draw=white,fill=white] (3.732051,9.800000) circle (0.200000);
	\draw[line width=3pt,draw=white] (3.532051,9.800000) -- (3.932051,9.800000);
	\draw[line width=3pt,draw=white] (3.732051,9.600000) -- (3.732051,10.000000);
	\draw[line width=3pt,draw=white] (3.732051,9.800000) -- (4.598076,10.300000);
	\draw[line width=3pt,draw=white,fill=white] (4.598076,10.300000) circle (0.100000);
	\draw[line width=3pt,draw=white] (3.732051,9.900000) -- (3.732051,10.500000);
	\draw[line width=3pt,draw=white,fill=white] (3.732051,10.600000) circle (0.200000);
	\draw[line width=3pt,draw=white] (3.532051,10.600000) -- (3.932051,10.600000);
	\draw[line width=3pt,draw=white] (3.732051,10.400000) -- (3.732051,10.800000);
	\draw[line width=3pt,draw=white] (3.732051,10.600000) -- (2.732051,10.600000);
	\draw[line width=3pt,draw=white,fill=white] (2.732051,10.600000) circle (0.100000);
	\draw[line width=3pt,draw=white] (3.732051,10.700000) -- (3.732051,11.300000);
	\draw[line width=3pt,draw=white,fill=white] (3.732051,11.400000) circle (0.200000);
	\draw[line width=3pt,draw=white] (3.532051,11.400000) -- (3.932051,11.400000);
	\draw[line width=3pt,draw=white] (3.732051,11.200000) -- (3.732051,11.600000);
	\draw[line width=3pt,draw=white] (3.732051,11.400000) -- (4.732051,11.400000);
	\draw[line width=3pt,draw=white,fill=white] (4.732051,11.400000) circle (0.100000);
	\draw[line width=3pt,draw=white] (3.732051,11.500000) -- (3.732051,12.100000);
	\draw[line width=3pt,draw=white,fill=white] (3.732051,12.200000) circle (0.200000);
	\draw[line width=3pt,draw=white] (3.532051,12.200000) -- (3.932051,12.200000);
	\draw[line width=3pt,draw=white] (3.732051,12.000000) -- (3.732051,12.400000);
	\draw[line width=3pt,draw=white] (3.732051,12.200000) -- (2.866025,11.700000);
	\draw[line width=3pt,draw=white,fill=white] (2.866025,11.700000) circle (0.100000);
	\draw[line width=3pt,draw=white] (3.732051,12.300000) -- (3.732051,12.900000);
	\draw[line width=3pt,fill=white,draw=white] (3.732051,13.000000) circle (0.100000);
	\draw[line width=3pt,draw=white] (3.732051,13.100000) -- (3.732051,13.700000);
	\draw[line width=3pt,draw=white] (3.732051,13.700000) -- (3.732051,13.900000);
	\draw[line width=1pt] (3.732051,1.000000) -- (3.732051,1.700000);
	\draw[line width=1pt] (3.732051,1.700000) -- (3.732051,1.900000);
	\draw[line width=1pt] (3.732051,1.900000) -- (3.732051,2.500000);
	\draw[line width=1pt,fill=white] (3.632051,2.500000) -- (3.632051,2.700000) -- (3.832051,2.700000) -- (3.832051,2.500000) -- cycle;
	\draw[line width=1pt] (3.732051,2.700000) -- (3.732051,3.300000);
	\draw[line width=1pt] (3.732051,3.400000) -- (4.598076,3.900000);
	\draw[line width=1pt,fill=white] (3.732051,3.400000) circle (0.200000);
	\draw[line width=1pt] (3.532051,3.400000) -- (3.932051,3.400000);
	\draw[line width=1pt] (3.732051,3.200000) -- (3.732051,3.600000);
	\draw[line width=1pt,fill=black] (4.598076,3.900000) circle (0.100000);
	\draw[line width=1pt] (3.732051,3.500000) -- (3.732051,4.100000);
	\draw[line width=1pt] (3.732051,4.200000) -- (2.732051,4.200000);
	\draw[line width=1pt,fill=white] (3.732051,4.200000) circle (0.200000);
	\draw[line width=1pt] (3.532051,4.200000) -- (3.932051,4.200000);
	\draw[line width=1pt] (3.732051,4.000000) -- (3.732051,4.400000);
	\draw[line width=1pt,fill=black] (2.732051,4.200000) circle (0.100000);
	\draw[line width=1pt] (3.732051,4.300000) -- (3.732051,4.900000);
	\draw[line width=1pt] (3.732051,5.000000) -- (4.732051,5.000000);
	\draw[line width=1pt,fill=white] (3.732051,5.000000) circle (0.200000);
	\draw[line width=1pt] (3.532051,5.000000) -- (3.932051,5.000000);
	\draw[line width=1pt] (3.732051,4.800000) -- (3.732051,5.200000);
	\draw[line width=1pt,fill=black] (4.732051,5.000000) circle (0.100000);
	\draw[line width=1pt] (3.732051,5.100000) -- (3.732051,5.700000);
	\draw[line width=1pt] (3.732051,5.800000) -- (2.866025,5.300000);
	\draw[line width=1pt,fill=white] (3.732051,5.800000) circle (0.200000);
	\draw[line width=1pt] (3.532051,5.800000) -- (3.932051,5.800000);
	\draw[line width=1pt] (3.732051,5.600000) -- (3.732051,6.000000);
	\draw[line width=1pt,fill=black] (2.866025,5.300000) circle (0.100000);
	\draw[line width=1pt] (2.866025,5.400000) -- +(0.0,2.0pt);
	\draw[line width=1pt] (2.866025,5.200000) -- +(0.0,-2.0pt);
	\draw[line width=1pt] (3.732051,5.900000) -- (3.732051,6.500000);
	\draw[line width=1pt,fill=white] (3.732051,6.600000) circle (0.100000);
	\draw[line width=1pt] (3.732051,6.700000) -- (3.732051,7.300000);
	\draw[line width=1pt] (3.732051,7.300000) -- (3.732051,7.500000);
	\draw[line width=1pt] (3.732051,7.500000) -- (3.732051,8.100000);
	\draw[line width=1pt] (3.732051,8.100000) -- (3.732051,8.300000);
	\draw[line width=1pt] (3.732051,8.300000) -- (3.732051,8.900000);
	\draw[line width=1pt,fill=white] (3.632051,8.900000) -- (3.632051,9.100000) -- (3.832051,9.100000) -- (3.832051,8.900000) -- cycle;
	\draw[line width=1pt] (3.732051,9.100000) -- (3.732051,9.700000);
	\draw[line width=1pt] (3.732051,9.800000) -- (4.598076,10.300000);
	\draw[line width=1pt,fill=white] (3.732051,9.800000) circle (0.200000);
	\draw[line width=1pt] (3.532051,9.800000) -- (3.932051,9.800000);
	\draw[line width=1pt] (3.732051,9.600000) -- (3.732051,10.000000);
	\draw[line width=1pt,fill=black] (4.598076,10.300000) circle (0.100000);
	\draw[line width=1pt] (3.732051,9.900000) -- (3.732051,10.500000);
	\draw[line width=1pt] (3.732051,10.600000) -- (2.732051,10.600000);
	\draw[line width=1pt,fill=white] (3.732051,10.600000) circle (0.200000);
	\draw[line width=1pt] (3.532051,10.600000) -- (3.932051,10.600000);
	\draw[line width=1pt] (3.732051,10.400000) -- (3.732051,10.800000);
	\draw[line width=1pt,fill=black] (2.732051,10.600000) circle (0.100000);
	\draw[line width=1pt] (3.732051,10.700000) -- (3.732051,11.300000);
	\draw[line width=1pt] (3.732051,11.400000) -- (4.732051,11.400000);
	\draw[line width=1pt,fill=white] (3.732051,11.400000) circle (0.200000);
	\draw[line width=1pt] (3.532051,11.400000) -- (3.932051,11.400000);
	\draw[line width=1pt] (3.732051,11.200000) -- (3.732051,11.600000);
	\draw[line width=1pt,fill=black] (4.732051,11.400000) circle (0.100000);
	\draw[line width=1pt] (3.732051,11.500000) -- (3.732051,12.100000);
	\draw[line width=1pt] (3.732051,12.200000) -- (2.866025,11.700000);
	\draw[line width=1pt,fill=white] (3.732051,12.200000) circle (0.200000);
	\draw[line width=1pt] (3.532051,12.200000) -- (3.932051,12.200000);
	\draw[line width=1pt] (3.732051,12.000000) -- (3.732051,12.400000);
	\draw[line width=1pt,fill=black] (2.866025,11.700000) circle (0.100000);
	\draw[line width=1pt] (2.866025,11.800000) -- +(0.0,2.0pt);
	\draw[line width=1pt] (2.866025,11.600000) -- +(0.0,-2.0pt);
	\draw[line width=1pt] (3.732051,12.300000) -- (3.732051,12.900000);
	\draw[line width=1pt,fill=white] (3.732051,13.000000) circle (0.100000);
	\draw[line width=1pt] (3.732051,13.100000) -- (3.732051,13.700000);
	\draw[line width=1pt] (3.732051,13.700000) -- (3.732051,13.900000);
	\draw[line width=3pt,draw=white] (1.866025,0.500000) -- (1.866025,1.200000);
	\draw[line width=3pt,fill=white,draw=white] (1.766025,1.200000) -- (1.766025,1.400000) -- (1.966025,1.400000) -- (1.966025,1.200000) -- cycle;
	\draw[line width=3pt,draw=white] (1.866025,1.400000) -- (1.866025,2.000000);
	\draw[line width=3pt,fill=white,draw=white] (1.766025,2.100000) -- (1.866025,2.200000) -- (1.966025,2.100000) -- (1.866025,2.000000) -- cycle;
	\draw[line width=3pt,draw=white] (1.866025,2.200000) -- (1.866025,2.800000);
	\draw[line width=3pt,draw=white,fill=white] (1.866025,2.900000) circle (0.100000);
	\draw[line width=3pt,draw=white] (1.866025,2.900000) -- (2.732051,3.400000);
	\draw[line width=3pt,draw=white,fill=white] (2.732051,3.400000) circle (0.200000);
	\draw[line width=3pt,draw=white] (2.532051,3.400000) -- (2.932051,3.400000);
	\draw[line width=3pt,draw=white] (2.732051,3.200000) -- (2.732051,3.600000);
	\draw[line width=3pt,draw=white] (1.866025,3.000000) -- (1.866025,3.600000);
	\draw[line width=3pt,draw=white,fill=white] (1.866025,3.700000) circle (0.100000);
	\draw[line width=3pt,draw=white] (1.866025,3.700000) -- (0.866025,3.700000);
	\draw[line width=3pt,draw=white,fill=white] (0.866025,3.700000) circle (0.200000);
	\draw[line width=3pt,draw=white] (0.666025,3.700000) -- (1.066025,3.700000);
	\draw[line width=3pt,draw=white] (0.866025,3.500000) -- (0.866025,3.900000);
	\draw[line width=3pt,draw=white] (1.866025,3.800000) -- (1.866025,4.400000);
	\draw[line width=3pt,draw=white,fill=white] (1.866025,4.500000) circle (0.100000);
	\draw[line width=3pt,draw=white] (1.866025,4.500000) -- (2.866025,4.500000);
	\draw[line width=3pt,draw=white,fill=white] (2.866025,4.500000) circle (0.200000);
	\draw[line width=3pt,draw=white] (2.666025,4.500000) -- (3.066025,4.500000);
	\draw[line width=3pt,draw=white] (2.866025,4.300000) -- (2.866025,4.700000);
	\draw[line width=3pt,draw=white] (1.866025,4.600000) -- (1.866025,5.200000);
	\draw[line width=3pt,draw=white,fill=white] (1.866025,5.300000) circle (0.100000);
	\draw[line width=3pt,draw=white] (1.866025,5.300000) -- (1.000000,4.800000);
	\draw[line width=3pt,draw=white,fill=white] (1.000000,4.800000) circle (0.200000);
	\draw[line width=3pt,draw=white] (0.800000,4.800000) -- (1.200000,4.800000);
	\draw[line width=3pt,draw=white] (1.000000,4.600000) -- (1.000000,5.000000);
	\draw[line width=3pt,draw=white] (1.866025,5.400000) -- (1.866025,6.000000);
	\draw[line width=3pt,fill=white,draw=white] (1.766025,6.100000) -- (1.866025,6.200000) -- (1.966025,6.100000) -- (1.866025,6.000000) -- cycle;
	\draw[line width=3pt,draw=white] (1.866025,6.200000) -- (1.866025,6.800000);
	\draw[line width=3pt,fill=white,draw=white] (1.866025,6.900000) circle (0.100000);
	\draw[line width=3pt,draw=white] (1.866025,7.000000) -- (1.866025,7.600000);
	\draw[line width=3pt,fill=white,draw=white] (1.766025,7.600000) -- (1.766025,7.800000) -- (1.966025,7.800000) -- (1.966025,7.600000) -- cycle;
	\draw[line width=3pt,draw=white] (1.866025,7.800000) -- (1.866025,8.400000);
	\draw[line width=3pt,fill=white,draw=white] (1.766025,8.500000) -- (1.866025,8.600000) -- (1.966025,8.500000) -- (1.866025,8.400000) -- cycle;
	\draw[line width=3pt,draw=white] (1.866025,8.600000) -- (1.866025,9.200000);
	\draw[line width=3pt,draw=white,fill=white] (1.866025,9.300000) circle (0.100000);
	\draw[line width=3pt,draw=white] (1.866025,9.300000) -- (2.732051,9.800000);
	\draw[line width=3pt,draw=white,fill=white] (2.732051,9.800000) circle (0.200000);
	\draw[line width=3pt,draw=white] (2.532051,9.800000) -- (2.932051,9.800000);
	\draw[line width=3pt,draw=white] (2.732051,9.600000) -- (2.732051,10.000000);
	\draw[line width=3pt,draw=white] (1.866025,9.400000) -- (1.866025,10.000000);
	\draw[line width=3pt,draw=white,fill=white] (1.866025,10.100000) circle (0.100000);
	\draw[line width=3pt,draw=white] (1.866025,10.100000) -- (0.866025,10.100000);
	\draw[line width=3pt,draw=white,fill=white] (0.866025,10.100000) circle (0.200000);
	\draw[line width=3pt,draw=white] (0.666025,10.100000) -- (1.066025,10.100000);
	\draw[line width=3pt,draw=white] (0.866025,9.900000) -- (0.866025,10.300000);
	\draw[line width=3pt,draw=white] (1.866025,10.200000) -- (1.866025,10.800000);
	\draw[line width=3pt,draw=white,fill=white] (1.866025,10.900000) circle (0.100000);
	\draw[line width=3pt,draw=white] (1.866025,10.900000) -- (2.866025,10.900000);
	\draw[line width=3pt,draw=white,fill=white] (2.866025,10.900000) circle (0.200000);
	\draw[line width=3pt,draw=white] (2.666025,10.900000) -- (3.066025,10.900000);
	\draw[line width=3pt,draw=white] (2.866025,10.700000) -- (2.866025,11.100000);
	\draw[line width=3pt,draw=white] (1.866025,11.000000) -- (1.866025,11.600000);
	\draw[line width=3pt,draw=white,fill=white] (1.866025,11.700000) circle (0.100000);
	\draw[line width=3pt,draw=white] (1.866025,11.700000) -- (1.000000,11.200000);
	\draw[line width=3pt,draw=white,fill=white] (1.000000,11.200000) circle (0.200000);
	\draw[line width=3pt,draw=white] (0.800000,11.200000) -- (1.200000,11.200000);
	\draw[line width=3pt,draw=white] (1.000000,11.000000) -- (1.000000,11.400000);
	\draw[line width=3pt,draw=white] (1.866025,11.800000) -- (1.866025,12.400000);
	\draw[line width=3pt,fill=white,draw=white] (1.766025,12.500000) -- (1.866025,12.600000) -- (1.966025,12.500000) -- (1.866025,12.400000) -- cycle;
	\draw[line width=3pt,draw=white] (1.866025,12.600000) -- (1.866025,13.200000);
	\draw[line width=3pt,fill=white,draw=white] (1.866025,13.300000) circle (0.100000);
	\draw[line width=1pt] (1.866025,0.500000) -- (1.866025,1.200000);
	\draw[line width=1pt,fill=white] (1.766025,1.200000) -- (1.766025,1.400000) -- (1.966025,1.400000) -- (1.966025,1.200000) -- cycle;
	\draw[line width=1pt] (1.866025,1.400000) -- (1.866025,2.000000);
	\draw[line width=1pt,fill=white] (1.766025,2.100000) -- (1.866025,2.200000) -- (1.966025,2.100000) -- (1.866025,2.000000) -- cycle;
	\draw[line width=1pt] (1.866025,2.200000) -- (1.866025,2.800000);
	\draw[line width=1pt] (1.866025,2.900000) -- (2.732051,3.400000);
	\draw[line width=1pt,fill=black] (1.866025,2.900000) circle (0.100000);
	\draw[line width=1pt,fill=white] (2.732051,3.400000) circle (0.200000);
	\draw[line width=1pt] (2.532051,3.400000) -- (2.932051,3.400000);
	\draw[line width=1pt] (2.732051,3.200000) -- (2.732051,3.600000);
	\draw[line width=1pt] (1.866025,3.000000) -- (1.866025,3.600000);
	\draw[line width=1pt] (1.866025,3.700000) -- (0.866025,3.700000);
	\draw[line width=1pt,fill=black] (1.866025,3.700000) circle (0.100000);
	\draw[line width=1pt,fill=white] (0.866025,3.700000) circle (0.200000);
	\draw[line width=1pt] (0.666025,3.700000) -- (1.066025,3.700000);
	\draw[line width=1pt] (0.866025,3.500000) -- (0.866025,3.900000);
	\draw[line width=1pt] (1.866025,3.800000) -- (1.866025,4.400000);
	\draw[line width=1pt] (1.866025,4.500000) -- (2.866025,4.500000);
	\draw[line width=1pt,fill=black] (1.866025,4.500000) circle (0.100000);
	\draw[line width=1pt,fill=white] (2.866025,4.500000) circle (0.200000);
	\draw[line width=1pt] (2.666025,4.500000) -- (3.066025,4.500000);
	\draw[line width=1pt] (2.866025,4.300000) -- (2.866025,4.700000);
	\draw[line width=1pt] (2.866025,4.700000) -- +(0.0,2.0pt);
	\draw[line width=1pt] (2.866025,4.300000) -- +(0.0,-2.0pt);
	\draw[line width=1pt] (1.866025,4.600000) -- (1.866025,5.200000);
	\draw[line width=1pt] (1.866025,5.300000) -- (1.000000,4.800000);
	\draw[line width=1pt,fill=black] (1.866025,5.300000) circle (0.100000);
	\draw[line width=1pt,fill=white] (1.000000,4.800000) circle (0.200000);
	\draw[line width=1pt] (0.800000,4.800000) -- (1.200000,4.800000);
	\draw[line width=1pt] (1.000000,4.600000) -- (1.000000,5.000000);
	\draw[line width=1pt] (1.866025,5.400000) -- (1.866025,6.000000);
	\draw[line width=1pt,fill=white] (1.766025,6.100000) -- (1.866025,6.200000) -- (1.966025,6.100000) -- (1.866025,6.000000) -- cycle;
	\draw[line width=1pt] (1.866025,6.200000) -- (1.866025,6.800000);
	\draw[line width=1pt,fill=white] (1.866025,6.900000) circle (0.100000);
	\draw[line width=1pt] (1.866025,7.000000) -- (1.866025,7.600000);
	\draw[line width=1pt,fill=white] (1.766025,7.600000) -- (1.766025,7.800000) -- (1.966025,7.800000) -- (1.966025,7.600000) -- cycle;
	\draw[line width=1pt] (1.866025,7.800000) -- (1.866025,8.400000);
	\draw[line width=1pt,fill=white] (1.766025,8.500000) -- (1.866025,8.600000) -- (1.966025,8.500000) -- (1.866025,8.400000) -- cycle;
	\draw[line width=1pt] (1.866025,8.600000) -- (1.866025,9.200000);
	\draw[line width=1pt] (1.866025,9.300000) -- (2.732051,9.800000);
	\draw[line width=1pt,fill=black] (1.866025,9.300000) circle (0.100000);
	\draw[line width=1pt,fill=white] (2.732051,9.800000) circle (0.200000);
	\draw[line width=1pt] (2.532051,9.800000) -- (2.932051,9.800000);
	\draw[line width=1pt] (2.732051,9.600000) -- (2.732051,10.000000);
	\draw[line width=1pt] (1.866025,9.400000) -- (1.866025,10.000000);
	\draw[line width=1pt] (1.866025,10.100000) -- (0.866025,10.100000);
	\draw[line width=1pt,fill=black] (1.866025,10.100000) circle (0.100000);
	\draw[line width=1pt,fill=white] (0.866025,10.100000) circle (0.200000);
	\draw[line width=1pt] (0.666025,10.100000) -- (1.066025,10.100000);
	\draw[line width=1pt] (0.866025,9.900000) -- (0.866025,10.300000);
	\draw[line width=1pt] (1.866025,10.200000) -- (1.866025,10.800000);
	\draw[line width=1pt] (1.866025,10.900000) -- (2.866025,10.900000);
	\draw[line width=1pt,fill=black] (1.866025,10.900000) circle (0.100000);
	\draw[line width=1pt,fill=white] (2.866025,10.900000) circle (0.200000);
	\draw[line width=1pt] (2.666025,10.900000) -- (3.066025,10.900000);
	\draw[line width=1pt] (2.866025,10.700000) -- (2.866025,11.100000);
	\draw[line width=1pt] (2.866025,11.100000) -- +(0.0,2.0pt);
	\draw[line width=1pt] (2.866025,10.700000) -- +(0.0,-2.0pt);
	\draw[line width=1pt] (1.866025,11.000000) -- (1.866025,11.600000);
	\draw[line width=1pt] (1.866025,11.700000) -- (1.000000,11.200000);
	\draw[line width=1pt,fill=black] (1.866025,11.700000) circle (0.100000);
	\draw[line width=1pt,fill=white] (1.000000,11.200000) circle (0.200000);
	\draw[line width=1pt] (0.800000,11.200000) -- (1.200000,11.200000);
	\draw[line width=1pt] (1.000000,11.000000) -- (1.000000,11.400000);
	\draw[line width=1pt] (1.866025,11.800000) -- (1.866025,12.400000);
	\draw[line width=1pt,fill=white] (1.766025,12.500000) -- (1.866025,12.600000) -- (1.966025,12.500000) -- (1.866025,12.400000) -- cycle;
	\draw[line width=1pt] (1.866025,12.600000) -- (1.866025,13.200000);
	\draw[line width=1pt,fill=white] (1.866025,13.300000) circle (0.100000);
	\draw[line width=3pt,draw=white] (2.866025,0.500000) -- (2.866025,1.200000);
	\draw[line width=3pt,draw=white] (2.866025,1.200000) -- (2.866025,1.400000);
	\draw[line width=3pt,draw=white] (2.866025,1.400000) -- (2.866025,2.000000);
	\draw[line width=3pt,draw=white] (2.866025,2.000000) -- (2.866025,2.200000);
	\draw[line width=3pt,draw=white] (2.866025,2.200000) -- (2.866025,2.800000);
	\draw[line width=3pt,draw=white,fill=white] (2.866025,2.900000) circle (0.100000);
	\draw[line width=3pt,draw=white] (2.866025,2.900000) -- (2.000000,2.400000);
	\draw[line width=3pt,draw=white,fill=white] (2.000000,2.400000) circle (0.200000);
	\draw[line width=3pt,draw=white] (1.800000,2.400000) -- (2.200000,2.400000);
	\draw[line width=3pt,draw=white] (2.000000,2.200000) -- (2.000000,2.600000);
	\draw[line width=3pt,draw=white] (2.866025,3.000000) -- (2.866025,3.600000);
	\draw[line width=3pt,draw=white,fill=white] (2.866025,3.700000) circle (0.200000);
	\draw[line width=3pt,draw=white] (2.666025,3.700000) -- (3.066025,3.700000);
	\draw[line width=3pt,draw=white] (2.866025,3.500000) -- (2.866025,3.900000);
	\draw[line width=3pt,draw=white] (2.866025,3.700000) -- (3.866025,3.700000);
	\draw[line width=3pt,draw=white,fill=white] (3.866025,3.700000) circle (0.100000);
	\draw[line width=3pt,draw=white] (2.866025,3.800000) -- (2.866025,4.400000);
	\draw[line width=3pt,draw=white,fill=white] (2.866025,4.500000) circle (0.200000);
	\draw[line width=3pt,draw=white] (2.666025,4.500000) -- (3.066025,4.500000);
	\draw[line width=3pt,draw=white] (2.866025,4.300000) -- (2.866025,4.700000);
	\draw[line width=3pt,draw=white] (2.866025,4.500000) -- (1.866025,4.500000);
	\draw[line width=3pt,draw=white,fill=white] (1.866025,4.500000) circle (0.100000);
	\draw[line width=3pt,draw=white] (2.866025,4.600000) -- (2.866025,5.200000);
	\draw[line width=3pt,draw=white,fill=white] (2.866025,5.300000) circle (0.100000);
	\draw[line width=3pt,draw=white] (2.866025,5.300000) -- (3.732051,5.800000);
	\draw[line width=3pt,draw=white,fill=white] (3.732051,5.800000) circle (0.200000);
	\draw[line width=3pt,draw=white] (3.532051,5.800000) -- (3.932051,5.800000);
	\draw[line width=3pt,draw=white] (3.732051,5.600000) -- (3.732051,6.000000);
	\draw[line width=3pt,draw=white] (2.866025,5.400000) -- (2.866025,6.000000);
	\draw[line width=3pt,draw=white] (2.866025,6.000000) -- (2.866025,6.200000);
	\draw[line width=3pt,draw=white] (2.866025,6.200000) -- (2.866025,6.800000);
	\draw[line width=3pt,draw=white] (2.866025,6.800000) -- (2.866025,7.000000);
	\draw[line width=3pt,draw=white] (2.866025,7.000000) -- (2.866025,7.600000);
	\draw[line width=3pt,draw=white] (2.866025,7.600000) -- (2.866025,7.800000);
	\draw[line width=3pt,draw=white] (2.866025,7.800000) -- (2.866025,8.400000);
	\draw[line width=3pt,draw=white] (2.866025,8.400000) -- (2.866025,8.600000);
	\draw[line width=3pt,draw=white] (2.866025,8.600000) -- (2.866025,9.200000);
	\draw[line width=3pt,draw=white,fill=white] (2.866025,9.300000) circle (0.100000);
	\draw[line width=3pt,draw=white] (2.866025,9.300000) -- (2.000000,8.800000);
	\draw[line width=3pt,draw=white,fill=white] (2.000000,8.800000) circle (0.200000);
	\draw[line width=3pt,draw=white] (1.800000,8.800000) -- (2.200000,8.800000);
	\draw[line width=3pt,draw=white] (2.000000,8.600000) -- (2.000000,9.000000);
	\draw[line width=3pt,draw=white] (2.866025,9.400000) -- (2.866025,10.000000);
	\draw[line width=3pt,draw=white,fill=white] (2.866025,10.100000) circle (0.200000);
	\draw[line width=3pt,draw=white] (2.666025,10.100000) -- (3.066025,10.100000);
	\draw[line width=3pt,draw=white] (2.866025,9.900000) -- (2.866025,10.300000);
	\draw[line width=3pt,draw=white] (2.866025,10.100000) -- (3.866025,10.100000);
	\draw[line width=3pt,draw=white,fill=white] (3.866025,10.100000) circle (0.100000);
	\draw[line width=3pt,draw=white] (2.866025,10.200000) -- (2.866025,10.800000);
	\draw[line width=3pt,draw=white,fill=white] (2.866025,10.900000) circle (0.200000);
	\draw[line width=3pt,draw=white] (2.666025,10.900000) -- (3.066025,10.900000);
	\draw[line width=3pt,draw=white] (2.866025,10.700000) -- (2.866025,11.100000);
	\draw[line width=3pt,draw=white] (2.866025,10.900000) -- (1.866025,10.900000);
	\draw[line width=3pt,draw=white,fill=white] (1.866025,10.900000) circle (0.100000);
	\draw[line width=3pt,draw=white] (2.866025,11.000000) -- (2.866025,11.600000);
	\draw[line width=3pt,draw=white,fill=white] (2.866025,11.700000) circle (0.100000);
	\draw[line width=3pt,draw=white] (2.866025,11.700000) -- (3.732051,12.200000);
	\draw[line width=3pt,draw=white,fill=white] (3.732051,12.200000) circle (0.200000);
	\draw[line width=3pt,draw=white] (3.532051,12.200000) -- (3.932051,12.200000);
	\draw[line width=3pt,draw=white] (3.732051,12.000000) -- (3.732051,12.400000);
	\draw[line width=3pt,draw=white] (2.866025,11.800000) -- (2.866025,12.400000);
	\draw[line width=3pt,draw=white] (2.866025,12.400000) -- (2.866025,12.600000);
	\draw[line width=3pt,draw=white] (2.866025,12.600000) -- (2.866025,13.200000);
	\draw[line width=3pt,draw=white] (2.866025,13.200000) -- (2.866025,13.400000);
	\draw[line width=1pt] (2.866025,0.500000) -- (2.866025,1.200000);
	\draw[line width=1pt] (2.866025,1.200000) -- (2.866025,1.400000);
	\draw[line width=1pt] (2.866025,1.400000) -- (2.866025,2.000000);
	\draw[line width=1pt] (2.866025,2.000000) -- (2.866025,2.200000);
	\draw[line width=1pt] (2.866025,2.200000) -- (2.866025,2.800000);
	\draw[line width=1pt] (2.866025,2.900000) -- (2.000000,2.400000);
	\draw[line width=1pt,fill=black] (2.866025,2.900000) circle (0.100000);
	\draw[line width=1pt,fill=white] (2.000000,2.400000) circle (0.200000);
	\draw[line width=1pt] (1.800000,2.400000) -- (2.200000,2.400000);
	\draw[line width=1pt] (2.000000,2.200000) -- (2.000000,2.600000);
	\draw[line width=1pt] (2.866025,3.000000) -- (2.866025,3.600000);
	\draw[line width=1pt] (2.866025,3.700000) -- (3.866025,3.700000);
	\draw[line width=1pt,fill=white] (2.866025,3.700000) circle (0.200000);
	\draw[line width=1pt] (2.666025,3.700000) -- (3.066025,3.700000);
	\draw[line width=1pt] (2.866025,3.500000) -- (2.866025,3.900000);
	\draw[line width=1pt,fill=black] (3.866025,3.700000) circle (0.100000);
	\draw[line width=1pt] (2.866025,3.800000) -- (2.866025,4.400000);
	\draw[line width=1pt] (2.866025,4.500000) -- (1.866025,4.500000);
	\draw[line width=1pt,fill=white] (2.866025,4.500000) circle (0.200000);
	\draw[line width=1pt] (2.666025,4.500000) -- (3.066025,4.500000);
	\draw[line width=1pt] (2.866025,4.300000) -- (2.866025,4.700000);
	\draw[line width=1pt,fill=black] (1.866025,4.500000) circle (0.100000);
	\draw[line width=1pt] (1.866025,4.600000) -- +(0.0,2.0pt);
	\draw[line width=1pt] (1.866025,4.400000) -- +(0.0,-2.0pt);
	\draw[line width=1pt] (2.866025,4.600000) -- (2.866025,5.200000);
	\draw[line width=1pt] (2.866025,5.300000) -- (3.732051,5.800000);
	\draw[line width=1pt,fill=black] (2.866025,5.300000) circle (0.100000);
	\draw[line width=1pt,fill=white] (3.732051,5.800000) circle (0.200000);
	\draw[line width=1pt] (3.532051,5.800000) -- (3.932051,5.800000);
	\draw[line width=1pt] (3.732051,5.600000) -- (3.732051,6.000000);
	\draw[line width=1pt] (3.732051,6.000000) -- +(0.0,2.0pt);
	\draw[line width=1pt] (3.732051,5.600000) -- +(0.0,-2.0pt);
	\draw[line width=1pt] (2.866025,5.400000) -- (2.866025,6.000000);
	\draw[line width=1pt] (2.866025,6.000000) -- (2.866025,6.200000);
	\draw[line width=1pt] (2.866025,6.200000) -- (2.866025,6.800000);
	\draw[line width=1pt] (2.866025,6.800000) -- (2.866025,7.000000);
	\draw[line width=1pt] (2.866025,7.000000) -- (2.866025,7.600000);
	\draw[line width=1pt] (2.866025,7.600000) -- (2.866025,7.800000);
	\draw[line width=1pt] (2.866025,7.800000) -- (2.866025,8.400000);
	\draw[line width=1pt] (2.866025,8.400000) -- (2.866025,8.600000);
	\draw[line width=1pt] (2.866025,8.600000) -- (2.866025,9.200000);
	\draw[line width=1pt] (2.866025,9.300000) -- (2.000000,8.800000);
	\draw[line width=1pt,fill=black] (2.866025,9.300000) circle (0.100000);
	\draw[line width=1pt,fill=white] (2.000000,8.800000) circle (0.200000);
	\draw[line width=1pt] (1.800000,8.800000) -- (2.200000,8.800000);
	\draw[line width=1pt] (2.000000,8.600000) -- (2.000000,9.000000);
	\draw[line width=1pt] (2.866025,9.400000) -- (2.866025,10.000000);
	\draw[line width=1pt] (2.866025,10.100000) -- (3.866025,10.100000);
	\draw[line width=1pt,fill=white] (2.866025,10.100000) circle (0.200000);
	\draw[line width=1pt] (2.666025,10.100000) -- (3.066025,10.100000);
	\draw[line width=1pt] (2.866025,9.900000) -- (2.866025,10.300000);
	\draw[line width=1pt,fill=black] (3.866025,10.100000) circle (0.100000);
	\draw[line width=1pt] (2.866025,10.200000) -- (2.866025,10.800000);
	\draw[line width=1pt] (2.866025,10.900000) -- (1.866025,10.900000);
	\draw[line width=1pt,fill=white] (2.866025,10.900000) circle (0.200000);
	\draw[line width=1pt] (2.666025,10.900000) -- (3.066025,10.900000);
	\draw[line width=1pt] (2.866025,10.700000) -- (2.866025,11.100000);
	\draw[line width=1pt,fill=black] (1.866025,10.900000) circle (0.100000);
	\draw[line width=1pt] (1.866025,11.000000) -- +(0.0,2.0pt);
	\draw[line width=1pt] (1.866025,10.800000) -- +(0.0,-2.0pt);
	\draw[line width=1pt] (2.866025,11.000000) -- (2.866025,11.600000);
	\draw[line width=1pt] (2.866025,11.700000) -- (3.732051,12.200000);
	\draw[line width=1pt,fill=black] (2.866025,11.700000) circle (0.100000);
	\draw[line width=1pt,fill=white] (3.732051,12.200000) circle (0.200000);
	\draw[line width=1pt] (3.532051,12.200000) -- (3.932051,12.200000);
	\draw[line width=1pt] (3.732051,12.000000) -- (3.732051,12.400000);
	\draw[line width=1pt] (3.732051,12.400000) -- +(0.0,2.0pt);
	\draw[line width=1pt] (3.732051,12.000000) -- +(0.0,-2.0pt);
	\draw[line width=1pt] (2.866025,11.800000) -- (2.866025,12.400000);
	\draw[line width=1pt] (2.866025,12.400000) -- (2.866025,12.600000);
	\draw[line width=1pt] (2.866025,12.600000) -- (2.866025,13.200000);
	\draw[line width=1pt] (2.866025,13.200000) -- (2.866025,13.400000);
	\draw[line width=1pt,rounded corners,dashed] (2.366025,6.500000) -- (2.366025,13.700000) -- (1.366025,13.700000) -- (1.366025,6.500000) -- cycle;
	\draw[line width=1pt,rounded corners,dashed] (4.232051,6.200000) -- (4.232051,13.400000) -- (3.232051,13.400000) -- (3.232051,6.200000) -- cycle;
\end{tikzpicture}

%% file: pic_isoaddprimal.tex
\begin{tikzpicture}[x=0.074272\linewidth,y=0.074272\linewidth]
	\draw[line width=1pt,fill=black] (0.000000,0.000000) circle (0.100000);
	\draw[line width=1pt] (0.866025,0.500000) circle (0.100000);
	\draw[line width=1pt,fill=black] (1.732051,1.000000) circle (0.100000);
	\draw[line width=1pt] (1.000000,0.000000) circle (0.100000);
	\draw[line width=1pt] (0.966025,0.500000) -- (1.766025,0.500000);
	\draw[line width=1pt] (1.779423,0.450000) -- (1.086603,0.050000);
	\draw[line width=1pt,fill=black] (1.866025,0.500000) circle (0.100000);
	\draw[line width=1pt] (2.645448,0.950000) -- (1.952628,0.550000);
	\draw[line width=1pt] (2.732051,1.000000) circle (0.100000);
	\draw[line width=1pt,fill=black] (2.000000,0.000000) circle (0.100000);
	\draw[line width=1pt] (1.966025,0.500000) -- (2.766025,0.500000);
	\draw[line width=1pt] (2.866025,0.500000) circle (0.100000);
	\draw[line width=1pt,fill=black] (3.732051,1.000000) circle (0.100000);
	\draw[line width=1pt] (3.000000,0.000000) circle (0.100000);
	\draw[line width=1pt] (2.966025,0.500000) -- (3.766025,0.500000);
	\draw[line width=1pt] (3.779423,0.450000) -- (3.086603,0.050000);
	\draw[line width=1pt,fill=black] (3.866025,0.500000) circle (0.100000);
	\draw[line width=1pt] (4.645448,0.950000) -- (3.952628,0.550000);
	\draw[line width=1pt] (4.732051,1.000000) circle (0.100000);
	\draw[line width=1pt,fill=black] (4.000000,0.000000) circle (0.100000);
	\draw[line width=1pt] (3.966025,0.500000) -- (4.766025,0.500000);
	\draw[line width=1pt] (4.866025,0.500000) circle (0.100000);
	\draw[line width=1pt,fill=black] (5.732051,1.000000) circle (0.100000);
	\draw[line width=3pt,draw=white] (1.866025,0.500000) -- (1.866025,1.200000);
	\draw[line width=3pt,fill=white,draw=white] (1.766025,1.200000) -- (1.766025,1.400000) -- (1.966025,1.400000) -- (1.966025,1.200000) -- cycle;
	\draw[line width=3pt,draw=white] (1.866025,1.400000) -- (1.866025,2.000000);
	\draw[line width=3pt,fill=white,draw=white] (1.766025,2.100000) -- (1.866025,2.200000) -- (1.966025,2.100000) -- (1.866025,2.000000) -- cycle;
	\draw[line width=3pt,draw=white] (1.866025,2.200000) -- (1.866025,2.800000);
	\draw[line width=3pt,draw=white,fill=white] (1.866025,2.900000) circle (0.100000);
	\draw[line width=3pt,draw=white] (1.866025,2.900000) -- (2.732051,3.400000);
	\draw[line width=3pt,draw=white,fill=white] (2.732051,3.400000) circle (0.200000);
	\draw[line width=3pt,draw=white] (2.532051,3.400000) -- (2.932051,3.400000);
	\draw[line width=3pt,draw=white] (2.732051,3.200000) -- (2.732051,3.600000);
	\draw[line width=3pt,draw=white] (1.866025,3.000000) -- (1.866025,3.600000);
	\draw[line width=3pt,draw=white,fill=white] (1.866025,3.700000) circle (0.100000);
	\draw[line width=3pt,draw=white] (1.866025,3.700000) -- (0.866025,3.700000);
	\draw[line width=3pt,draw=white,fill=white] (0.866025,3.700000) circle (0.200000);
	\draw[line width=3pt,draw=white] (0.666025,3.700000) -- (1.066025,3.700000);
	\draw[line width=3pt,draw=white] (0.866025,3.500000) -- (0.866025,3.900000);
	\draw[line width=3pt,draw=white] (1.866025,3.800000) -- (1.866025,4.400000);
	\draw[line width=3pt,draw=white] (1.866025,4.400000) -- (1.866025,4.600000);
	\draw[line width=3pt,draw=white] (1.866025,4.600000) -- (1.866025,5.200000);
	\draw[line width=3pt,draw=white,fill=white] (1.866025,5.300000) circle (0.100000);
	\draw[line width=3pt,draw=white] (1.866025,5.300000) -- (1.000000,4.800000);
	\draw[line width=3pt,draw=white,fill=white] (1.000000,4.800000) circle (0.200000);
	\draw[line width=3pt,draw=white] (0.800000,4.800000) -- (1.200000,4.800000);
	\draw[line width=3pt,draw=white] (1.000000,4.600000) -- (1.000000,5.000000);
	\draw[line width=3pt,draw=white] (1.866025,5.400000) -- (1.866025,6.000000);
	\draw[line width=3pt,fill=white,draw=white] (1.766025,6.100000) -- (1.866025,6.200000) -- (1.966025,6.100000) -- (1.866025,6.000000) -- cycle;
	\draw[line width=3pt,draw=white] (1.866025,6.200000) -- (1.866025,6.800000);
	\draw[line width=3pt,fill=white,draw=white] (1.866025,6.900000) circle (0.100000);
	\draw[line width=3pt,draw=white] (1.866025,7.000000) -- (1.866025,7.600000);
	\draw[line width=3pt,fill=white,draw=white] (1.766025,7.600000) -- (1.766025,7.800000) -- (1.966025,7.800000) -- (1.966025,7.600000) -- cycle;
	\draw[line width=3pt,draw=white] (1.866025,7.800000) -- (1.866025,8.400000);
	\draw[line width=3pt,fill=white,draw=white] (1.766025,8.500000) -- (1.866025,8.600000) -- (1.966025,8.500000) -- (1.866025,8.400000) -- cycle;
	\draw[line width=3pt,draw=white] (1.866025,8.600000) -- (1.866025,9.200000);
	\draw[line width=3pt,draw=white,fill=white] (1.866025,9.300000) circle (0.100000);
	\draw[line width=3pt,draw=white] (1.866025,9.300000) -- (2.732051,9.800000);
	\draw[line width=3pt,draw=white,fill=white] (2.732051,9.800000) circle (0.200000);
	\draw[line width=3pt,draw=white] (2.532051,9.800000) -- (2.932051,9.800000);
	\draw[line width=3pt,draw=white] (2.732051,9.600000) -- (2.732051,10.000000);
	\draw[line width=3pt,draw=white] (1.866025,9.400000) -- (1.866025,10.000000);
	\draw[line width=3pt,draw=white,fill=white] (1.866025,10.100000) circle (0.100000);
	\draw[line width=3pt,draw=white] (1.866025,10.100000) -- (0.866025,10.100000);
	\draw[line width=3pt,draw=white,fill=white] (0.866025,10.100000) circle (0.200000);
	\draw[line width=3pt,draw=white] (0.666025,10.100000) -- (1.066025,10.100000);
	\draw[line width=3pt,draw=white] (0.866025,9.900000) -- (0.866025,10.300000);
	\draw[line width=3pt,draw=white] (1.866025,10.200000) -- (1.866025,10.800000);
	\draw[line width=3pt,draw=white] (1.866025,10.800000) -- (1.866025,11.000000);
	\draw[line width=3pt,draw=white] (1.866025,11.000000) -- (1.866025,11.600000);
	\draw[line width=3pt,draw=white,fill=white] (1.866025,11.700000) circle (0.100000);
	\draw[line width=3pt,draw=white] (1.866025,11.700000) -- (1.000000,11.200000);
	\draw[line width=3pt,draw=white,fill=white] (1.000000,11.200000) circle (0.200000);
	\draw[line width=3pt,draw=white] (0.800000,11.200000) -- (1.200000,11.200000);
	\draw[line width=3pt,draw=white] (1.000000,11.000000) -- (1.000000,11.400000);
	\draw[line width=3pt,draw=white] (1.866025,11.800000) -- (1.866025,12.400000);
	\draw[line width=3pt,fill=white,draw=white] (1.766025,12.500000) -- (1.866025,12.600000) -- (1.966025,12.500000) -- (1.866025,12.400000) -- cycle;
	\draw[line width=3pt,draw=white] (1.866025,12.600000) -- (1.866025,13.200000);
	\draw[line width=3pt,fill=white,draw=white] (1.866025,13.300000) circle (0.100000);
	\draw[line width=1pt] (1.866025,0.500000) -- (1.866025,1.200000);
	\draw[line width=1pt,fill=white] (1.766025,1.200000) -- (1.766025,1.400000) -- (1.966025,1.400000) -- (1.966025,1.200000) -- cycle;
	\draw[line width=1pt] (1.866025,1.400000) -- (1.866025,2.000000);
	\draw[line width=1pt,fill=white] (1.766025,2.100000) -- (1.866025,2.200000) -- (1.966025,2.100000) -- (1.866025,2.000000) -- cycle;
	\draw[line width=1pt] (1.866025,2.200000) -- (1.866025,2.800000);
	\draw[line width=1pt] (1.866025,2.900000) -- (2.732051,3.400000);
	\draw[line width=1pt,fill=black] (1.866025,2.900000) circle (0.100000);
	\draw[line width=1pt,fill=white] (2.732051,3.400000) circle (0.200000);
	\draw[line width=1pt] (2.532051,3.400000) -- (2.932051,3.400000);
	\draw[line width=1pt] (2.732051,3.200000) -- (2.732051,3.600000);
	\draw[line width=1pt] (1.866025,3.000000) -- (1.866025,3.600000);
	\draw[line width=1pt] (1.866025,3.700000) -- (0.866025,3.700000);
	\draw[line width=1pt,fill=black] (1.866025,3.700000) circle (0.100000);
	\draw[line width=1pt,fill=white] (0.866025,3.700000) circle (0.200000);
	\draw[line width=1pt] (0.666025,3.700000) -- (1.066025,3.700000);
	\draw[line width=1pt] (0.866025,3.500000) -- (0.866025,3.900000);
	\draw[line width=1pt] (1.866025,3.800000) -- (1.866025,4.400000);
	\draw[line width=1pt] (1.866025,4.400000) -- (1.866025,4.600000);
	\draw[line width=1pt] (1.866025,4.600000) -- (1.866025,5.200000);
	\draw[line width=1pt] (1.866025,5.300000) -- (1.000000,4.800000);
	\draw[line width=1pt,fill=black] (1.866025,5.300000) circle (0.100000);
	\draw[line width=1pt,fill=white] (1.000000,4.800000) circle (0.200000);
	\draw[line width=1pt] (0.800000,4.800000) -- (1.200000,4.800000);
	\draw[line width=1pt] (1.000000,4.600000) -- (1.000000,5.000000);
	\draw[line width=1pt] (1.866025,5.400000) -- (1.866025,6.000000);
	\draw[line width=1pt,fill=white] (1.766025,6.100000) -- (1.866025,6.200000) -- (1.966025,6.100000) -- (1.866025,6.000000) -- cycle;
	\draw[line width=1pt] (1.866025,6.200000) -- (1.866025,6.800000);
	\draw[line width=1pt,fill=white] (1.866025,6.900000) circle (0.100000);
	\draw[line width=1pt] (1.866025,7.000000) -- (1.866025,7.600000);
	\draw[line width=1pt,fill=white] (1.766025,7.600000) -- (1.766025,7.800000) -- (1.966025,7.800000) -- (1.966025,7.600000) -- cycle;
	\draw[line width=1pt] (1.866025,7.800000) -- (1.866025,8.400000);
	\draw[line width=1pt,fill=white] (1.766025,8.500000) -- (1.866025,8.600000) -- (1.966025,8.500000) -- (1.866025,8.400000) -- cycle;
	\draw[line width=1pt] (1.866025,8.600000) -- (1.866025,9.200000);
	\draw[line width=1pt] (1.866025,9.300000) -- (2.732051,9.800000);
	\draw[line width=1pt,fill=black] (1.866025,9.300000) circle (0.100000);
	\draw[line width=1pt,fill=white] (2.732051,9.800000) circle (0.200000);
	\draw[line width=1pt] (2.532051,9.800000) -- (2.932051,9.800000);
	\draw[line width=1pt] (2.732051,9.600000) -- (2.732051,10.000000);
	\draw[line width=1pt] (1.866025,9.400000) -- (1.866025,10.000000);
	\draw[line width=1pt] (1.866025,10.100000) -- (0.866025,10.100000);
	\draw[line width=1pt,fill=black] (1.866025,10.100000) circle (0.100000);
	\draw[line width=1pt,fill=white] (0.866025,10.100000) circle (0.200000);
	\draw[line width=1pt] (0.666025,10.100000) -- (1.066025,10.100000);
	\draw[line width=1pt] (0.866025,9.900000) -- (0.866025,10.300000);
	\draw[line width=1pt] (1.866025,10.200000) -- (1.866025,10.800000);
	\draw[line width=1pt] (1.866025,10.800000) -- (1.866025,11.000000);
	\draw[line width=1pt] (1.866025,11.000000) -- (1.866025,11.600000);
	\draw[line width=1pt] (1.866025,11.700000) -- (1.000000,11.200000);
	\draw[line width=1pt,fill=black] (1.866025,11.700000) circle (0.100000);
	\draw[line width=1pt,fill=white] (1.000000,11.200000) circle (0.200000);
	\draw[line width=1pt] (0.800000,11.200000) -- (1.200000,11.200000);
	\draw[line width=1pt] (1.000000,11.000000) -- (1.000000,11.400000);
	\draw[line width=1pt] (1.866025,11.800000) -- (1.866025,12.400000);
	\draw[line width=1pt,fill=white] (1.766025,12.500000) -- (1.866025,12.600000) -- (1.966025,12.500000) -- (1.866025,12.400000) -- cycle;
	\draw[line width=1pt] (1.866025,12.600000) -- (1.866025,13.200000);
	\draw[line width=1pt,fill=white] (1.866025,13.300000) circle (0.100000);
	\draw[line width=3pt,draw=white] (2.866025,0.500000) -- (2.866025,1.200000);
	\draw[line width=3pt,draw=white] (2.866025,1.200000) -- (2.866025,1.400000);
	\draw[line width=3pt,draw=white] (2.866025,1.400000) -- (2.866025,2.000000);
	\draw[line width=3pt,draw=white] (2.866025,2.000000) -- (2.866025,2.200000);
	\draw[line width=3pt,draw=white] (2.866025,2.200000) -- (2.866025,2.800000);
	\draw[line width=3pt,draw=white] (2.866025,2.800000) -- (2.866025,3.000000);
	\draw[line width=3pt,draw=white] (2.866025,3.000000) -- (2.866025,3.600000);
	\draw[line width=3pt,draw=white] (2.866025,3.600000) -- (2.866025,3.800000);
	\draw[line width=3pt,draw=white] (2.866025,3.800000) -- (2.866025,4.400000);
	\draw[line width=3pt,draw=white] (2.866025,4.400000) -- (2.866025,4.600000);
	\draw[line width=3pt,draw=white] (2.866025,4.600000) -- (2.866025,5.200000);
	\draw[line width=3pt,draw=white] (2.866025,5.200000) -- (2.866025,5.400000);
	\draw[line width=3pt,draw=white] (2.866025,5.400000) -- (2.866025,6.000000);
	\draw[line width=3pt,draw=white] (2.866025,6.000000) -- (2.866025,6.200000);
	\draw[line width=3pt,draw=white] (2.866025,6.200000) -- (2.866025,6.800000);
	\draw[line width=3pt,draw=white] (2.866025,6.800000) -- (2.866025,7.000000);
	\draw[line width=3pt,draw=white] (2.866025,7.000000) -- (2.866025,7.600000);
	\draw[line width=3pt,fill=white,draw=white] (2.766025,7.600000) -- (2.766025,7.800000) -- (2.966025,7.800000) -- (2.966025,7.600000) -- cycle;
	\draw[line width=3pt,draw=white] (2.866025,7.800000) -- (2.866025,8.400000);
	\draw[line width=3pt,fill=white,draw=white] (2.766025,8.500000) -- (2.866025,8.600000) -- (2.966025,8.500000) -- (2.866025,8.400000) -- cycle;
	\draw[line width=3pt,draw=white] (2.866025,8.600000) -- (2.866025,9.200000);
	\draw[line width=3pt,draw=white,fill=white] (2.866025,9.300000) circle (0.100000);
	\draw[line width=3pt,draw=white] (2.866025,9.300000) -- (2.000000,8.800000);
	\draw[line width=3pt,draw=white,fill=white] (2.000000,8.800000) circle (0.200000);
	\draw[line width=3pt,draw=white] (1.800000,8.800000) -- (2.200000,8.800000);
	\draw[line width=3pt,draw=white] (2.000000,8.600000) -- (2.000000,9.000000);
	\draw[line width=3pt,draw=white] (2.866025,9.400000) -- (2.866025,10.000000);
	\draw[line width=3pt,draw=white,fill=white] (2.866025,10.100000) circle (0.200000);
	\draw[line width=3pt,draw=white] (2.666025,10.100000) -- (3.066025,10.100000);
	\draw[line width=3pt,draw=white] (2.866025,9.900000) -- (2.866025,10.300000);
	\draw[line width=3pt,draw=white] (2.866025,10.100000) -- (3.866025,10.100000);
	\draw[line width=3pt,draw=white,fill=white] (3.866025,10.100000) circle (0.100000);
	\draw[line width=3pt,draw=white] (2.866025,10.200000) -- (2.866025,10.800000);
	\draw[line width=3pt,draw=white,fill=white] (2.866025,10.900000) circle (0.200000);
	\draw[line width=3pt,draw=white] (2.666025,10.900000) -- (3.066025,10.900000);
	\draw[line width=3pt,draw=white] (2.866025,10.700000) -- (2.866025,11.100000);
	\draw[line width=3pt,draw=white] (2.866025,10.900000) -- (1.866025,10.900000);
	\draw[line width=3pt,draw=white,fill=white] (1.866025,10.900000) circle (0.100000);
	\draw[line width=3pt,draw=white] (2.866025,11.000000) -- (2.866025,11.600000);
	\draw[line width=3pt,draw=white,fill=white] (2.866025,11.700000) circle (0.100000);
	\draw[line width=3pt,draw=white] (2.866025,11.700000) -- (3.732051,12.200000);
	\draw[line width=3pt,draw=white,fill=white] (3.732051,12.200000) circle (0.200000);
	\draw[line width=3pt,draw=white] (3.532051,12.200000) -- (3.932051,12.200000);
	\draw[line width=3pt,draw=white] (3.732051,12.000000) -- (3.732051,12.400000);
	\draw[line width=3pt,draw=white] (2.866025,11.800000) -- (2.866025,12.400000);
	\draw[line width=3pt,draw=white] (2.866025,12.400000) -- (2.866025,12.600000);
	\draw[line width=3pt,draw=white] (2.866025,12.600000) -- (2.866025,13.200000);
	\draw[line width=3pt,draw=white] (2.866025,13.200000) -- (2.866025,13.400000);
	\draw[line width=1pt] (2.866025,0.500000) -- (2.866025,1.200000);
	\draw[line width=1pt] (2.866025,1.200000) -- (2.866025,1.400000);
	\draw[line width=1pt] (2.866025,1.400000) -- (2.866025,2.000000);
	\draw[line width=1pt] (2.866025,2.000000) -- (2.866025,2.200000);
	\draw[line width=1pt] (2.866025,2.200000) -- (2.866025,2.800000);
	\draw[line width=1pt] (2.866025,2.800000) -- (2.866025,3.000000);
	\draw[line width=1pt] (2.866025,3.000000) -- (2.866025,3.600000);
	\draw[line width=1pt] (2.866025,3.600000) -- (2.866025,3.800000);
	\draw[line width=1pt] (2.866025,3.800000) -- (2.866025,4.400000);
	\draw[line width=1pt] (2.866025,4.400000) -- (2.866025,4.600000);
	\draw[line width=1pt] (2.866025,4.600000) -- (2.866025,5.200000);
	\draw[line width=1pt] (2.866025,5.200000) -- (2.866025,5.400000);
	\draw[line width=1pt] (2.866025,5.400000) -- (2.866025,6.000000);
	\draw[line width=1pt] (2.866025,6.000000) -- (2.866025,6.200000);
	\draw[line width=1pt] (2.866025,6.200000) -- (2.866025,6.800000);
	\draw[line width=1pt] (2.866025,6.800000) -- (2.866025,7.000000);
	\draw[line width=1pt] (2.866025,7.000000) -- (2.866025,7.600000);
	\draw[line width=1pt,fill=white] (2.766025,7.600000) -- (2.766025,7.800000) -- (2.966025,7.800000) -- (2.966025,7.600000) -- cycle;
	\draw[line width=1pt] (2.866025,7.800000) -- (2.866025,8.400000);
	\draw[line width=1pt,fill=white] (2.766025,8.500000) -- (2.866025,8.600000) -- (2.966025,8.500000) -- (2.866025,8.400000) -- cycle;
	\draw[line width=1pt] (2.866025,8.600000) -- (2.866025,9.200000);
	\draw[line width=1pt] (2.866025,9.300000) -- (2.000000,8.800000);
	\draw[line width=1pt,fill=black] (2.866025,9.300000) circle (0.100000);
	\draw[line width=1pt,fill=white] (2.000000,8.800000) circle (0.200000);
	\draw[line width=1pt] (1.800000,8.800000) -- (2.200000,8.800000);
	\draw[line width=1pt] (2.000000,8.600000) -- (2.000000,9.000000);
	\draw[line width=1pt] (2.866025,9.400000) -- (2.866025,10.000000);
	\draw[line width=1pt] (2.866025,10.100000) -- (3.866025,10.100000);
	\draw[line width=1pt,fill=white] (2.866025,10.100000) circle (0.200000);
	\draw[line width=1pt] (2.666025,10.100000) -- (3.066025,10.100000);
	\draw[line width=1pt] (2.866025,9.900000) -- (2.866025,10.300000);
	\draw[line width=1pt,fill=black] (3.866025,10.100000) circle (0.100000);
	\draw[line width=1pt] (3.866025,10.200000) -- +(0.0,2.0pt);
	\draw[line width=1pt] (3.866025,10.000000) -- +(0.0,-2.0pt);
	\draw[line width=1pt] (2.866025,10.200000) -- (2.866025,10.800000);
	\draw[line width=1pt] (2.866025,10.900000) -- (1.866025,10.900000);
	\draw[line width=1pt,fill=white] (2.866025,10.900000) circle (0.200000);
	\draw[line width=1pt] (2.666025,10.900000) -- (3.066025,10.900000);
	\draw[line width=1pt] (2.866025,10.700000) -- (2.866025,11.100000);
	\draw[line width=1pt,fill=black] (1.866025,10.900000) circle (0.100000);
	\draw[line width=1pt] (1.866025,11.000000) -- +(0.0,2.0pt);
	\draw[line width=1pt] (1.866025,10.800000) -- +(0.0,-2.0pt);
	\draw[line width=1pt] (2.866025,11.000000) -- (2.866025,11.600000);
	\draw[line width=1pt] (2.866025,11.700000) -- (3.732051,12.200000);
	\draw[line width=1pt,fill=black] (2.866025,11.700000) circle (0.100000);
	\draw[line width=1pt,fill=white] (3.732051,12.200000) circle (0.200000);
	\draw[line width=1pt] (3.532051,12.200000) -- (3.932051,12.200000);
	\draw[line width=1pt] (3.732051,12.000000) -- (3.732051,12.400000);
	\draw[line width=1pt] (2.866025,11.800000) -- (2.866025,12.400000);
	\draw[line width=1pt] (2.866025,12.400000) -- (2.866025,12.600000);
	\draw[line width=1pt] (2.866025,12.600000) -- (2.866025,13.200000);
	\draw[line width=1pt] (2.866025,13.200000) -- (2.866025,13.400000);
	\draw[line width=3pt,draw=white] (3.866025,0.500000) -- (3.866025,1.200000);
	\draw[line width=3pt,draw=white] (3.866025,1.200000) -- (3.866025,1.400000);
	\draw[line width=3pt,draw=white] (3.866025,1.400000) -- (3.866025,2.000000);
	\draw[line width=3pt,draw=white] (3.866025,2.000000) -- (3.866025,2.200000);
	\draw[line width=3pt,draw=white] (3.866025,2.200000) -- (3.866025,2.800000);
	\draw[line width=3pt,draw=white] (3.866025,2.800000) -- (3.866025,3.000000);
	\draw[line width=3pt,draw=white] (3.866025,3.000000) -- (3.866025,3.600000);
	\draw[line width=3pt,draw=white] (3.866025,3.600000) -- (3.866025,3.800000);
	\draw[line width=3pt,draw=white] (3.866025,3.800000) -- (3.866025,4.400000);
	\draw[line width=3pt,draw=white] (3.866025,4.400000) -- (3.866025,4.600000);
	\draw[line width=3pt,draw=white] (3.866025,4.600000) -- (3.866025,5.200000);
	\draw[line width=3pt,draw=white] (3.866025,5.200000) -- (3.866025,5.400000);
	\draw[line width=3pt,draw=white] (3.866025,5.400000) -- (3.866025,6.000000);
	\draw[line width=3pt,draw=white] (3.866025,6.000000) -- (3.866025,6.200000);
	\draw[line width=3pt,draw=white] (3.866025,6.200000) -- (3.866025,6.800000);
	\draw[line width=3pt,draw=white] (3.866025,6.800000) -- (3.866025,7.000000);
	\draw[line width=3pt,draw=white] (3.866025,7.000000) -- (3.866025,7.600000);
	\draw[line width=3pt,fill=white,draw=white] (3.766025,7.600000) -- (3.766025,7.800000) -- (3.966025,7.800000) -- (3.966025,7.600000) -- cycle;
	\draw[line width=3pt,draw=white] (3.866025,7.800000) -- (3.866025,8.400000);
	\draw[line width=3pt,fill=white,draw=white] (3.766025,8.500000) -- (3.866025,8.600000) -- (3.966025,8.500000) -- (3.866025,8.400000) -- cycle;
	\draw[line width=3pt,draw=white] (3.866025,8.600000) -- (3.866025,9.200000);
	\draw[line width=3pt,draw=white,fill=white] (3.866025,9.300000) circle (0.100000);
	\draw[line width=3pt,draw=white] (3.866025,9.300000) -- (4.732051,9.800000);
	\draw[line width=3pt,draw=white,fill=white] (4.732051,9.800000) circle (0.200000);
	\draw[line width=3pt,draw=white] (4.532051,9.800000) -- (4.932051,9.800000);
	\draw[line width=3pt,draw=white] (4.732051,9.600000) -- (4.732051,10.000000);
	\draw[line width=3pt,draw=white] (3.866025,9.400000) -- (3.866025,10.000000);
	\draw[line width=3pt,draw=white,fill=white] (3.866025,10.100000) circle (0.100000);
	\draw[line width=3pt,draw=white] (3.866025,10.100000) -- (2.866025,10.100000);
	\draw[line width=3pt,draw=white,fill=white] (2.866025,10.100000) circle (0.200000);
	\draw[line width=3pt,draw=white] (2.666025,10.100000) -- (3.066025,10.100000);
	\draw[line width=3pt,draw=white] (2.866025,9.900000) -- (2.866025,10.300000);
	\draw[line width=3pt,draw=white] (3.866025,10.200000) -- (3.866025,10.800000);
	\draw[line width=3pt,draw=white,fill=white] (3.866025,10.900000) circle (0.100000);
	\draw[line width=3pt,draw=white] (3.866025,10.900000) -- (4.866025,10.900000);
	\draw[line width=3pt,draw=white,fill=white] (4.866025,10.900000) circle (0.200000);
	\draw[line width=3pt,draw=white] (4.666025,10.900000) -- (5.066025,10.900000);
	\draw[line width=3pt,draw=white] (4.866025,10.700000) -- (4.866025,11.100000);
	\draw[line width=3pt,draw=white] (3.866025,11.000000) -- (3.866025,11.600000);
	\draw[line width=3pt,draw=white,fill=white] (3.866025,11.700000) circle (0.100000);
	\draw[line width=3pt,draw=white] (3.866025,11.700000) -- (3.000000,11.200000);
	\draw[line width=3pt,draw=white,fill=white] (3.000000,11.200000) circle (0.200000);
	\draw[line width=3pt,draw=white] (2.800000,11.200000) -- (3.200000,11.200000);
	\draw[line width=3pt,draw=white] (3.000000,11.000000) -- (3.000000,11.400000);
	\draw[line width=3pt,draw=white] (3.866025,11.800000) -- (3.866025,12.400000);
	\draw[line width=3pt,fill=white,draw=white] (3.766025,12.500000) -- (3.866025,12.600000) -- (3.966025,12.500000) -- (3.866025,12.400000) -- cycle;
	\draw[line width=3pt,draw=white] (3.866025,12.600000) -- (3.866025,13.200000);
	\draw[line width=3pt,fill=white,draw=white] (3.866025,13.300000) circle (0.100000);
	\draw[line width=1pt] (3.866025,0.500000) -- (3.866025,1.200000);
	\draw[line width=1pt] (3.866025,1.200000) -- (3.866025,1.400000);
	\draw[line width=1pt] (3.866025,1.400000) -- (3.866025,2.000000);
	\draw[line width=1pt] (3.866025,2.000000) -- (3.866025,2.200000);
	\draw[line width=1pt] (3.866025,2.200000) -- (3.866025,2.800000);
	\draw[line width=1pt] (3.866025,2.800000) -- (3.866025,3.000000);
	\draw[line width=1pt] (3.866025,3.000000) -- (3.866025,3.600000);
	\draw[line width=1pt] (3.866025,3.600000) -- (3.866025,3.800000);
	\draw[line width=1pt] (3.866025,3.800000) -- (3.866025,4.400000);
	\draw[line width=1pt] (3.866025,4.400000) -- (3.866025,4.600000);
	\draw[line width=1pt] (3.866025,4.600000) -- (3.866025,5.200000);
	\draw[line width=1pt] (3.866025,5.200000) -- (3.866025,5.400000);
	\draw[line width=1pt] (3.866025,5.400000) -- (3.866025,6.000000);
	\draw[line width=1pt] (3.866025,6.000000) -- (3.866025,6.200000);
	\draw[line width=1pt] (3.866025,6.200000) -- (3.866025,6.800000);
	\draw[line width=1pt] (3.866025,6.800000) -- (3.866025,7.000000);
	\draw[line width=1pt] (3.866025,7.000000) -- (3.866025,7.600000);
	\draw[line width=1pt,fill=white] (3.766025,7.600000) -- (3.766025,7.800000) -- (3.966025,7.800000) -- (3.966025,7.600000) -- cycle;
	\draw[line width=1pt] (3.866025,7.800000) -- (3.866025,8.400000);
	\draw[line width=1pt,fill=white] (3.766025,8.500000) -- (3.866025,8.600000) -- (3.966025,8.500000) -- (3.866025,8.400000) -- cycle;
	\draw[line width=1pt] (3.866025,8.600000) -- (3.866025,9.200000);
	\draw[line width=1pt] (3.866025,9.300000) -- (4.732051,9.800000);
	\draw[line width=1pt,fill=black] (3.866025,9.300000) circle (0.100000);
	\draw[line width=1pt,fill=white] (4.732051,9.800000) circle (0.200000);
	\draw[line width=1pt] (4.532051,9.800000) -- (4.932051,9.800000);
	\draw[line width=1pt] (4.732051,9.600000) -- (4.732051,10.000000);
	\draw[line width=1pt] (3.866025,9.400000) -- (3.866025,10.000000);
	\draw[line width=1pt] (3.866025,10.100000) -- (2.866025,10.100000);
	\draw[line width=1pt,fill=black] (3.866025,10.100000) circle (0.100000);
	\draw[line width=1pt,fill=white] (2.866025,10.100000) circle (0.200000);
	\draw[line width=1pt] (2.666025,10.100000) -- (3.066025,10.100000);
	\draw[line width=1pt] (2.866025,9.900000) -- (2.866025,10.300000);
	\draw[line width=1pt] (2.866025,10.300000) -- +(0.0,2.0pt);
	\draw[line width=1pt] (2.866025,9.900000) -- +(0.0,-2.0pt);
	\draw[line width=1pt] (3.866025,10.200000) -- (3.866025,10.800000);
	\draw[line width=1pt] (3.866025,10.900000) -- (4.866025,10.900000);
	\draw[line width=1pt,fill=black] (3.866025,10.900000) circle (0.100000);
	\draw[line width=1pt,fill=white] (4.866025,10.900000) circle (0.200000);
	\draw[line width=1pt] (4.666025,10.900000) -- (5.066025,10.900000);
	\draw[line width=1pt] (4.866025,10.700000) -- (4.866025,11.100000);
	\draw[line width=1pt] (3.866025,11.000000) -- (3.866025,11.600000);
	\draw[line width=1pt] (3.866025,11.700000) -- (3.000000,11.200000);
	\draw[line width=1pt,fill=black] (3.866025,11.700000) circle (0.100000);
	\draw[line width=1pt,fill=white] (3.000000,11.200000) circle (0.200000);
	\draw[line width=1pt] (2.800000,11.200000) -- (3.200000,11.200000);
	\draw[line width=1pt] (3.000000,11.000000) -- (3.000000,11.400000);
	\draw[line width=1pt] (3.866025,11.800000) -- (3.866025,12.400000);
	\draw[line width=1pt,fill=white] (3.766025,12.500000) -- (3.866025,12.600000) -- (3.966025,12.500000) -- (3.866025,12.400000) -- cycle;
	\draw[line width=1pt] (3.866025,12.600000) -- (3.866025,13.200000);
	\draw[line width=1pt,fill=white] (3.866025,13.300000) circle (0.100000);
	\draw[line width=1pt,rounded corners,dashed] (1.366025,6.500000) -- (2.366025,6.500000) -- (2.366025,13.700000) -- (1.366025,13.700000) -- cycle;
	\draw[line width=1pt,rounded corners,dashed] (3.366025,13.700000) -- (4.366025,13.700000) -- (4.366025,12.900000) -- (3.366025,12.900000) -- cycle;
\end{tikzpicture}

%% file: pic_isoaddprimal2.tex
\begin{tikzpicture}[x=0.074272\linewidth,y=0.074272\linewidth]
	\draw[line width=1pt,fill=black] (0.000000,0.000000) circle (0.100000);
	\draw[line width=1pt] (0.779423,0.450000) -- (0.086603,0.050000);
	\draw[line width=1pt] (0.866025,0.500000) circle (0.100000);
	\draw[line width=1pt] (1.645448,0.950000) -- (0.952628,0.550000);
	\draw[line width=1pt,fill=black] (1.732051,1.000000) circle (0.100000);
	\draw[line width=1pt] (0.100000,0.000000) -- (0.900000,0.000000);
	\draw[line width=1pt] (1.000000,0.000000) circle (0.100000);
	\draw[line width=1pt,fill=black] (1.866025,0.500000) circle (0.100000);
	\draw[line width=1pt] (1.832051,1.000000) -- (2.632051,1.000000);
	\draw[line width=1pt] (2.732051,1.000000) circle (0.100000);
	\draw[line width=1pt] (1.100000,0.000000) -- (1.900000,0.000000);
	\draw[line width=1pt,fill=black] (2.000000,0.000000) circle (0.100000);
	\draw[line width=1pt] (2.779423,0.450000) -- (2.086603,0.050000);
	\draw[line width=1pt] (2.866025,0.500000) circle (0.100000);
	\draw[line width=1pt] (2.832051,1.000000) -- (3.632051,1.000000);
	\draw[line width=1pt] (3.645448,0.950000) -- (2.952628,0.550000);
	\draw[line width=1pt,fill=black] (3.732051,1.000000) circle (0.100000);
	\draw[line width=1pt] (2.100000,0.000000) -- (2.900000,0.000000);
	\draw[line width=1pt] (3.000000,0.000000) circle (0.100000);
	\draw[line width=1pt,fill=black] (3.866025,0.500000) circle (0.100000);
	\draw[line width=1pt] (3.832051,1.000000) -- (4.632051,1.000000);
	\draw[line width=1pt] (4.732051,1.000000) circle (0.100000);
	\draw[line width=1pt] (3.100000,0.000000) -- (3.900000,0.000000);
	\draw[line width=1pt,fill=black] (4.000000,0.000000) circle (0.100000);
	\draw[line width=1pt] (4.779423,0.450000) -- (4.086603,0.050000);
	\draw[line width=1pt] (4.866025,0.500000) circle (0.100000);
	\draw[line width=1pt] (4.832051,1.000000) -- (5.632051,1.000000);
	\draw[line width=1pt] (5.645448,0.950000) -- (4.952628,0.550000);
	\draw[line width=1pt,fill=black] (5.732051,1.000000) circle (0.100000);
	\draw[line width=3pt,draw=white] (1.866025,0.500000) -- (1.866025,1.200000);
	\draw[line width=3pt,draw=white] (1.866025,1.200000) -- (1.866025,1.400000);
	\draw[line width=3pt,draw=white] (1.866025,1.400000) -- (1.866025,2.000000);
	\draw[line width=3pt,fill=white,draw=white] (1.766025,2.000000) -- (1.766025,2.200000) -- (1.966025,2.200000) -- (1.966025,2.000000) -- cycle;
	\draw[line width=3pt,draw=white] (1.866025,2.200000) -- (1.866025,2.800000);
	\draw[line width=3pt,draw=white,fill=white] (1.866025,2.900000) circle (0.200000);
	\draw[line width=3pt,draw=white] (1.666025,2.900000) -- (2.066025,2.900000);
	\draw[line width=3pt,draw=white] (1.866025,2.700000) -- (1.866025,3.100000);
	\draw[line width=3pt,draw=white] (1.866025,2.900000) -- (2.732051,3.400000);
	\draw[line width=3pt,draw=white,fill=white] (2.732051,3.400000) circle (0.100000);
	\draw[line width=3pt,draw=white] (1.866025,3.000000) -- (1.866025,3.600000);
	\draw[line width=3pt,draw=white,fill=white] (1.866025,3.700000) circle (0.200000);
	\draw[line width=3pt,draw=white] (1.666025,3.700000) -- (2.066025,3.700000);
	\draw[line width=3pt,draw=white] (1.866025,3.500000) -- (1.866025,3.900000);
	\draw[line width=3pt,draw=white] (1.866025,3.700000) -- (0.866025,3.700000);
	\draw[line width=3pt,draw=white,fill=white] (0.866025,3.700000) circle (0.100000);
	\draw[line width=3pt,draw=white] (1.866025,3.800000) -- (1.866025,4.400000);
	\draw[line width=3pt,draw=white,fill=white] (1.866025,4.500000) circle (0.200000);
	\draw[line width=3pt,draw=white] (1.666025,4.500000) -- (2.066025,4.500000);
	\draw[line width=3pt,draw=white] (1.866025,4.300000) -- (1.866025,4.700000);
	\draw[line width=3pt,draw=white] (1.866025,4.500000) -- (2.866025,4.500000);
	\draw[line width=3pt,draw=white,fill=white] (2.866025,4.500000) circle (0.100000);
	\draw[line width=3pt,draw=white] (1.866025,4.600000) -- (1.866025,5.200000);
	\draw[line width=3pt,draw=white,fill=white] (1.866025,5.300000) circle (0.200000);
	\draw[line width=3pt,draw=white] (1.666025,5.300000) -- (2.066025,5.300000);
	\draw[line width=3pt,draw=white] (1.866025,5.100000) -- (1.866025,5.500000);
	\draw[line width=3pt,draw=white] (1.866025,5.300000) -- (1.000000,4.800000);
	\draw[line width=3pt,draw=white,fill=white] (1.000000,4.800000) circle (0.100000);
	\draw[line width=3pt,draw=white] (1.866025,5.400000) -- (1.866025,6.000000);
	\draw[line width=3pt,fill=white,draw=white] (1.866025,6.100000) circle (0.100000);
	\draw[line width=3pt,draw=white] (1.866025,6.200000) -- (1.866025,6.800000);
	\draw[line width=3pt,draw=white] (1.866025,6.800000) -- (1.866025,7.000000);
	\draw[line width=3pt,draw=white] (1.866025,7.000000) -- (1.866025,7.600000);
	\draw[line width=3pt,draw=white] (1.866025,7.600000) -- (1.866025,7.800000);
	\draw[line width=3pt,draw=white] (1.866025,7.800000) -- (1.866025,8.400000);
	\draw[line width=3pt,fill=white,draw=white] (1.766025,8.400000) -- (1.766025,8.600000) -- (1.966025,8.600000) -- (1.966025,8.400000) -- cycle;
	\draw[line width=3pt,draw=white] (1.866025,8.600000) -- (1.866025,9.200000);
	\draw[line width=3pt,draw=white,fill=white] (1.866025,9.300000) circle (0.200000);
	\draw[line width=3pt,draw=white] (1.666025,9.300000) -- (2.066025,9.300000);
	\draw[line width=3pt,draw=white] (1.866025,9.100000) -- (1.866025,9.500000);
	\draw[line width=3pt,draw=white] (1.866025,9.300000) -- (2.732051,9.800000);
	\draw[line width=3pt,draw=white,fill=white] (2.732051,9.800000) circle (0.100000);
	\draw[line width=3pt,draw=white] (1.866025,9.400000) -- (1.866025,10.000000);
	\draw[line width=3pt,draw=white,fill=white] (1.866025,10.100000) circle (0.200000);
	\draw[line width=3pt,draw=white] (1.666025,10.100000) -- (2.066025,10.100000);
	\draw[line width=3pt,draw=white] (1.866025,9.900000) -- (1.866025,10.300000);
	\draw[line width=3pt,draw=white] (1.866025,10.100000) -- (0.866025,10.100000);
	\draw[line width=3pt,draw=white,fill=white] (0.866025,10.100000) circle (0.100000);
	\draw[line width=3pt,draw=white] (1.866025,10.200000) -- (1.866025,10.800000);
	\draw[line width=3pt,draw=white,fill=white] (1.866025,10.900000) circle (0.200000);
	\draw[line width=3pt,draw=white] (1.666025,10.900000) -- (2.066025,10.900000);
	\draw[line width=3pt,draw=white] (1.866025,10.700000) -- (1.866025,11.100000);
	\draw[line width=3pt,draw=white] (1.866025,10.900000) -- (2.866025,10.900000);
	\draw[line width=3pt,draw=white,fill=white] (2.866025,10.900000) circle (0.100000);
	\draw[line width=3pt,draw=white] (1.866025,11.000000) -- (1.866025,11.600000);
	\draw[line width=3pt,draw=white,fill=white] (1.866025,11.700000) circle (0.200000);
	\draw[line width=3pt,draw=white] (1.666025,11.700000) -- (2.066025,11.700000);
	\draw[line width=3pt,draw=white] (1.866025,11.500000) -- (1.866025,11.900000);
	\draw[line width=3pt,draw=white] (1.866025,11.700000) -- (1.000000,11.200000);
	\draw[line width=3pt,draw=white,fill=white] (1.000000,11.200000) circle (0.100000);
	\draw[line width=3pt,draw=white] (1.866025,11.800000) -- (1.866025,12.400000);
	\draw[line width=3pt,fill=white,draw=white] (1.866025,12.500000) circle (0.100000);
	\draw[line width=3pt,draw=white] (1.866025,12.600000) -- (1.866025,13.200000);
	\draw[line width=3pt,draw=white] (1.866025,13.200000) -- (1.866025,13.400000);
	\draw[line width=1pt] (1.866025,0.500000) -- (1.866025,1.200000);
	\draw[line width=1pt] (1.866025,1.200000) -- (1.866025,1.400000);
	\draw[line width=1pt] (1.866025,1.400000) -- (1.866025,2.000000);
	\draw[line width=1pt,fill=white] (1.766025,2.000000) -- (1.766025,2.200000) -- (1.966025,2.200000) -- (1.966025,2.000000) -- cycle;
	\draw[line width=1pt] (1.866025,2.200000) -- (1.866025,2.800000);
	\draw[line width=1pt] (1.866025,2.900000) -- (2.732051,3.400000);
	\draw[line width=1pt,fill=white] (1.866025,2.900000) circle (0.200000);
	\draw[line width=1pt] (1.666025,2.900000) -- (2.066025,2.900000);
	\draw[line width=1pt] (1.866025,2.700000) -- (1.866025,3.100000);
	\draw[line width=1pt,fill=black] (2.732051,3.400000) circle (0.100000);
	\draw[line width=1pt] (1.866025,3.000000) -- (1.866025,3.600000);
	\draw[line width=1pt] (1.866025,3.700000) -- (0.866025,3.700000);
	\draw[line width=1pt,fill=white] (1.866025,3.700000) circle (0.200000);
	\draw[line width=1pt] (1.666025,3.700000) -- (2.066025,3.700000);
	\draw[line width=1pt] (1.866025,3.500000) -- (1.866025,3.900000);
	\draw[line width=1pt,fill=black] (0.866025,3.700000) circle (0.100000);
	\draw[line width=1pt] (1.866025,3.800000) -- (1.866025,4.400000);
	\draw[line width=1pt] (1.866025,4.500000) -- (2.866025,4.500000);
	\draw[line width=1pt,fill=white] (1.866025,4.500000) circle (0.200000);
	\draw[line width=1pt] (1.666025,4.500000) -- (2.066025,4.500000);
	\draw[line width=1pt] (1.866025,4.300000) -- (1.866025,4.700000);
	\draw[line width=1pt,fill=black] (2.866025,4.500000) circle (0.100000);
	\draw[line width=1pt] (2.866025,4.600000) -- +(0.0,2.0pt);
	\draw[line width=1pt] (2.866025,4.400000) -- +(0.0,-2.0pt);
	\draw[line width=1pt] (1.866025,4.600000) -- (1.866025,5.200000);
	\draw[line width=1pt] (1.866025,5.300000) -- (1.000000,4.800000);
	\draw[line width=1pt,fill=white] (1.866025,5.300000) circle (0.200000);
	\draw[line width=1pt] (1.666025,5.300000) -- (2.066025,5.300000);
	\draw[line width=1pt] (1.866025,5.100000) -- (1.866025,5.500000);
	\draw[line width=1pt,fill=black] (1.000000,4.800000) circle (0.100000);
	\draw[line width=1pt] (1.866025,5.400000) -- (1.866025,6.000000);
	\draw[line width=1pt,fill=white] (1.866025,6.100000) circle (0.100000);
	\draw[line width=1pt] (1.866025,6.200000) -- (1.866025,6.800000);
	\draw[line width=1pt] (1.866025,6.800000) -- (1.866025,7.000000);
	\draw[line width=1pt] (1.866025,7.000000) -- (1.866025,7.600000);
	\draw[line width=1pt] (1.866025,7.600000) -- (1.866025,7.800000);
	\draw[line width=1pt] (1.866025,7.800000) -- (1.866025,8.400000);
	\draw[line width=1pt,fill=white] (1.766025,8.400000) -- (1.766025,8.600000) -- (1.966025,8.600000) -- (1.966025,8.400000) -- cycle;
	\draw[line width=1pt] (1.866025,8.600000) -- (1.866025,9.200000);
	\draw[line width=1pt] (1.866025,9.300000) -- (2.732051,9.800000);
	\draw[line width=1pt,fill=white] (1.866025,9.300000) circle (0.200000);
	\draw[line width=1pt] (1.666025,9.300000) -- (2.066025,9.300000);
	\draw[line width=1pt] (1.866025,9.100000) -- (1.866025,9.500000);
	\draw[line width=1pt,fill=black] (2.732051,9.800000) circle (0.100000);
	\draw[line width=1pt] (1.866025,9.400000) -- (1.866025,10.000000);
	\draw[line width=1pt] (1.866025,10.100000) -- (0.866025,10.100000);
	\draw[line width=1pt,fill=white] (1.866025,10.100000) circle (0.200000);
	\draw[line width=1pt] (1.666025,10.100000) -- (2.066025,10.100000);
	\draw[line width=1pt] (1.866025,9.900000) -- (1.866025,10.300000);
	\draw[line width=1pt,fill=black] (0.866025,10.100000) circle (0.100000);
	\draw[line width=1pt] (1.866025,10.200000) -- (1.866025,10.800000);
	\draw[line width=1pt] (1.866025,10.900000) -- (2.866025,10.900000);
	\draw[line width=1pt,fill=white] (1.866025,10.900000) circle (0.200000);
	\draw[line width=1pt] (1.666025,10.900000) -- (2.066025,10.900000);
	\draw[line width=1pt] (1.866025,10.700000) -- (1.866025,11.100000);
	\draw[line width=1pt,fill=black] (2.866025,10.900000) circle (0.100000);
	\draw[line width=1pt] (2.866025,11.000000) -- +(0.0,2.0pt);
	\draw[line width=1pt] (2.866025,10.800000) -- +(0.0,-2.0pt);
	\draw[line width=1pt] (1.866025,11.000000) -- (1.866025,11.600000);
	\draw[line width=1pt] (1.866025,11.700000) -- (1.000000,11.200000);
	\draw[line width=1pt,fill=white] (1.866025,11.700000) circle (0.200000);
	\draw[line width=1pt] (1.666025,11.700000) -- (2.066025,11.700000);
	\draw[line width=1pt] (1.866025,11.500000) -- (1.866025,11.900000);
	\draw[line width=1pt,fill=black] (1.000000,11.200000) circle (0.100000);
	\draw[line width=1pt] (1.866025,11.800000) -- (1.866025,12.400000);
	\draw[line width=1pt,fill=white] (1.866025,12.500000) circle (0.100000);
	\draw[line width=1pt] (1.866025,12.600000) -- (1.866025,13.200000);
	\draw[line width=1pt] (1.866025,13.200000) -- (1.866025,13.400000);
	\draw[line width=3pt,draw=white] (2.866025,0.500000) -- (2.866025,1.200000);
	\draw[line width=3pt,draw=white] (2.866025,1.200000) -- (2.866025,1.400000);
	\draw[line width=3pt,draw=white] (2.866025,1.400000) -- (2.866025,2.000000);
	\draw[line width=3pt,draw=white] (2.866025,2.000000) -- (2.866025,2.200000);
	\draw[line width=3pt,draw=white] (2.866025,2.200000) -- (2.866025,2.800000);
	\draw[line width=3pt,draw=white,fill=white] (2.866025,2.900000) circle (0.200000);
	\draw[line width=3pt,draw=white] (2.666025,2.900000) -- (3.066025,2.900000);
	\draw[line width=3pt,draw=white] (2.866025,2.700000) -- (2.866025,3.100000);
	\draw[line width=3pt,draw=white] (2.866025,2.900000) -- (2.000000,2.400000);
	\draw[line width=3pt,draw=white,fill=white] (2.000000,2.400000) circle (0.100000);
	\draw[line width=3pt,draw=white] (2.866025,3.000000) -- (2.866025,3.600000);
	\draw[line width=3pt,draw=white] (2.866025,3.600000) -- (2.866025,3.800000);
	\draw[line width=3pt,draw=white] (2.866025,3.800000) -- (2.866025,4.400000);
	\draw[line width=3pt,draw=white,fill=white] (2.866025,4.500000) circle (0.100000);
	\draw[line width=3pt,draw=white] (2.866025,4.500000) -- (1.866025,4.500000);
	\draw[line width=3pt,draw=white,fill=white] (1.866025,4.500000) circle (0.200000);
	\draw[line width=3pt,draw=white] (1.666025,4.500000) -- (2.066025,4.500000);
	\draw[line width=3pt,draw=white] (1.866025,4.300000) -- (1.866025,4.700000);
	\draw[line width=3pt,draw=white] (2.866025,4.600000) -- (2.866025,5.200000);
	\draw[line width=3pt,draw=white,fill=white] (2.866025,5.300000) circle (0.200000);
	\draw[line width=3pt,draw=white] (2.666025,5.300000) -- (3.066025,5.300000);
	\draw[line width=3pt,draw=white] (2.866025,5.100000) -- (2.866025,5.500000);
	\draw[line width=3pt,draw=white] (2.866025,5.300000) -- (3.732051,5.800000);
	\draw[line width=3pt,draw=white,fill=white] (3.732051,5.800000) circle (0.100000);
	\draw[line width=3pt,draw=white] (2.866025,5.400000) -- (2.866025,6.000000);
	\draw[line width=3pt,draw=white] (2.866025,6.000000) -- (2.866025,6.200000);
	\draw[line width=3pt,draw=white] (2.866025,6.200000) -- (2.866025,6.800000);
	\draw[line width=3pt,draw=white] (2.866025,6.800000) -- (2.866025,7.000000);
	\draw[line width=3pt,draw=white] (2.866025,7.000000) -- (2.866025,7.600000);
	\draw[line width=3pt,draw=white] (2.866025,7.600000) -- (2.866025,7.800000);
	\draw[line width=3pt,draw=white] (2.866025,7.800000) -- (2.866025,8.400000);
	\draw[line width=3pt,draw=white] (2.866025,8.400000) -- (2.866025,8.600000);
	\draw[line width=3pt,draw=white] (2.866025,8.600000) -- (2.866025,9.200000);
	\draw[line width=3pt,draw=white,fill=white] (2.866025,9.300000) circle (0.200000);
	\draw[line width=3pt,draw=white] (2.666025,9.300000) -- (3.066025,9.300000);
	\draw[line width=3pt,draw=white] (2.866025,9.100000) -- (2.866025,9.500000);
	\draw[line width=3pt,draw=white] (2.866025,9.300000) -- (2.000000,8.800000);
	\draw[line width=3pt,draw=white,fill=white] (2.000000,8.800000) circle (0.100000);
	\draw[line width=3pt,draw=white] (2.866025,9.400000) -- (2.866025,10.000000);
	\draw[line width=3pt,draw=white,fill=white] (2.866025,10.100000) circle (0.100000);
	\draw[line width=3pt,draw=white] (2.866025,10.100000) -- (3.866025,10.100000);
	\draw[line width=3pt,draw=white,fill=white] (3.866025,10.100000) circle (0.200000);
	\draw[line width=3pt,draw=white] (3.666025,10.100000) -- (4.066025,10.100000);
	\draw[line width=3pt,draw=white] (3.866025,9.900000) -- (3.866025,10.300000);
	\draw[line width=3pt,draw=white] (2.866025,10.200000) -- (2.866025,10.800000);
	\draw[line width=3pt,draw=white,fill=white] (2.866025,10.900000) circle (0.100000);
	\draw[line width=3pt,draw=white] (2.866025,10.900000) -- (1.866025,10.900000);
	\draw[line width=3pt,draw=white,fill=white] (1.866025,10.900000) circle (0.200000);
	\draw[line width=3pt,draw=white] (1.666025,10.900000) -- (2.066025,10.900000);
	\draw[line width=3pt,draw=white] (1.866025,10.700000) -- (1.866025,11.100000);
	\draw[line width=3pt,draw=white] (2.866025,11.000000) -- (2.866025,11.600000);
	\draw[line width=3pt,draw=white,fill=white] (2.866025,11.700000) circle (0.200000);
	\draw[line width=3pt,draw=white] (2.666025,11.700000) -- (3.066025,11.700000);
	\draw[line width=3pt,draw=white] (2.866025,11.500000) -- (2.866025,11.900000);
	\draw[line width=3pt,draw=white] (2.866025,11.700000) -- (3.732051,12.200000);
	\draw[line width=3pt,draw=white,fill=white] (3.732051,12.200000) circle (0.100000);
	\draw[line width=3pt,draw=white] (2.866025,11.800000) -- (2.866025,12.400000);
	\draw[line width=3pt,draw=white] (2.866025,12.400000) -- (2.866025,12.600000);
	\draw[line width=3pt,draw=white] (2.866025,12.600000) -- (2.866025,13.200000);
	\draw[line width=3pt,draw=white] (2.866025,13.200000) -- (2.866025,13.400000);
	\draw[line width=1pt] (2.866025,0.500000) -- (2.866025,1.200000);
	\draw[line width=1pt] (2.866025,1.200000) -- (2.866025,1.400000);
	\draw[line width=1pt] (2.866025,1.400000) -- (2.866025,2.000000);
	\draw[line width=1pt] (2.866025,2.000000) -- (2.866025,2.200000);
	\draw[line width=1pt] (2.866025,2.200000) -- (2.866025,2.800000);
	\draw[line width=1pt] (2.866025,2.900000) -- (2.000000,2.400000);
	\draw[line width=1pt,fill=white] (2.866025,2.900000) circle (0.200000);
	\draw[line width=1pt] (2.666025,2.900000) -- (3.066025,2.900000);
	\draw[line width=1pt] (2.866025,2.700000) -- (2.866025,3.100000);
	\draw[line width=1pt,fill=black] (2.000000,2.400000) circle (0.100000);
	\draw[line width=1pt] (2.866025,3.000000) -- (2.866025,3.600000);
	\draw[line width=1pt] (2.866025,3.600000) -- (2.866025,3.800000);
	\draw[line width=1pt] (2.866025,3.800000) -- (2.866025,4.400000);
	\draw[line width=1pt] (2.866025,4.500000) -- (1.866025,4.500000);
	\draw[line width=1pt,fill=black] (2.866025,4.500000) circle (0.100000);
	\draw[line width=1pt,fill=white] (1.866025,4.500000) circle (0.200000);
	\draw[line width=1pt] (1.666025,4.500000) -- (2.066025,4.500000);
	\draw[line width=1pt] (1.866025,4.300000) -- (1.866025,4.700000);
	\draw[line width=1pt] (1.866025,4.700000) -- +(0.0,2.0pt);
	\draw[line width=1pt] (1.866025,4.300000) -- +(0.0,-2.0pt);
	\draw[line width=1pt] (2.866025,4.600000) -- (2.866025,5.200000);
	\draw[line width=1pt] (2.866025,5.300000) -- (3.732051,5.800000);
	\draw[line width=1pt,fill=white] (2.866025,5.300000) circle (0.200000);
	\draw[line width=1pt] (2.666025,5.300000) -- (3.066025,5.300000);
	\draw[line width=1pt] (2.866025,5.100000) -- (2.866025,5.500000);
	\draw[line width=1pt,fill=black] (3.732051,5.800000) circle (0.100000);
	\draw[line width=1pt] (2.866025,5.400000) -- (2.866025,6.000000);
	\draw[line width=1pt] (2.866025,6.000000) -- (2.866025,6.200000);
	\draw[line width=1pt] (2.866025,6.200000) -- (2.866025,6.800000);
	\draw[line width=1pt] (2.866025,6.800000) -- (2.866025,7.000000);
	\draw[line width=1pt] (2.866025,7.000000) -- (2.866025,7.600000);
	\draw[line width=1pt] (2.866025,7.600000) -- (2.866025,7.800000);
	\draw[line width=1pt] (2.866025,7.800000) -- (2.866025,8.400000);
	\draw[line width=1pt] (2.866025,8.400000) -- (2.866025,8.600000);
	\draw[line width=1pt] (2.866025,8.600000) -- (2.866025,9.200000);
	\draw[line width=1pt] (2.866025,9.300000) -- (2.000000,8.800000);
	\draw[line width=1pt,fill=white] (2.866025,9.300000) circle (0.200000);
	\draw[line width=1pt] (2.666025,9.300000) -- (3.066025,9.300000);
	\draw[line width=1pt] (2.866025,9.100000) -- (2.866025,9.500000);
	\draw[line width=1pt,fill=black] (2.000000,8.800000) circle (0.100000);
	\draw[line width=1pt] (2.866025,9.400000) -- (2.866025,10.000000);
	\draw[line width=1pt] (2.866025,10.100000) -- (3.866025,10.100000);
	\draw[line width=1pt,fill=black] (2.866025,10.100000) circle (0.100000);
	\draw[line width=1pt,fill=white] (3.866025,10.100000) circle (0.200000);
	\draw[line width=1pt] (3.666025,10.100000) -- (4.066025,10.100000);
	\draw[line width=1pt] (3.866025,9.900000) -- (3.866025,10.300000);
	\draw[line width=1pt] (3.866025,10.300000) -- +(0.0,2.0pt);
	\draw[line width=1pt] (3.866025,9.900000) -- +(0.0,-2.0pt);
	\draw[line width=1pt] (2.866025,10.200000) -- (2.866025,10.800000);
	\draw[line width=1pt] (2.866025,10.900000) -- (1.866025,10.900000);
	\draw[line width=1pt,fill=black] (2.866025,10.900000) circle (0.100000);
	\draw[line width=1pt,fill=white] (1.866025,10.900000) circle (0.200000);
	\draw[line width=1pt] (1.666025,10.900000) -- (2.066025,10.900000);
	\draw[line width=1pt] (1.866025,10.700000) -- (1.866025,11.100000);
	\draw[line width=1pt] (1.866025,11.100000) -- +(0.0,2.0pt);
	\draw[line width=1pt] (1.866025,10.700000) -- +(0.0,-2.0pt);
	\draw[line width=1pt] (2.866025,11.000000) -- (2.866025,11.600000);
	\draw[line width=1pt] (2.866025,11.700000) -- (3.732051,12.200000);
	\draw[line width=1pt,fill=white] (2.866025,11.700000) circle (0.200000);
	\draw[line width=1pt] (2.666025,11.700000) -- (3.066025,11.700000);
	\draw[line width=1pt] (2.866025,11.500000) -- (2.866025,11.900000);
	\draw[line width=1pt,fill=black] (3.732051,12.200000) circle (0.100000);
	\draw[line width=1pt] (2.866025,11.800000) -- (2.866025,12.400000);
	\draw[line width=1pt] (2.866025,12.400000) -- (2.866025,12.600000);
	\draw[line width=1pt] (2.866025,12.600000) -- (2.866025,13.200000);
	\draw[line width=1pt] (2.866025,13.200000) -- (2.866025,13.400000);
	\draw[line width=3pt,draw=white] (3.866025,0.500000) -- (3.866025,1.200000);
	\draw[line width=3pt,draw=white] (3.866025,1.200000) -- (3.866025,1.400000);
	\draw[line width=3pt,draw=white] (3.866025,1.400000) -- (3.866025,2.000000);
	\draw[line width=3pt,draw=white] (3.866025,2.000000) -- (3.866025,2.200000);
	\draw[line width=3pt,draw=white] (3.866025,2.200000) -- (3.866025,2.800000);
	\draw[line width=3pt,draw=white] (3.866025,2.800000) -- (3.866025,3.000000);
	\draw[line width=3pt,draw=white] (3.866025,3.000000) -- (3.866025,3.600000);
	\draw[line width=3pt,draw=white] (3.866025,3.600000) -- (3.866025,3.800000);
	\draw[line width=3pt,draw=white] (3.866025,3.800000) -- (3.866025,4.400000);
	\draw[line width=3pt,draw=white] (3.866025,4.400000) -- (3.866025,4.600000);
	\draw[line width=3pt,draw=white] (3.866025,4.600000) -- (3.866025,5.200000);
	\draw[line width=3pt,draw=white] (3.866025,5.200000) -- (3.866025,5.400000);
	\draw[line width=3pt,draw=white] (3.866025,5.400000) -- (3.866025,6.000000);
	\draw[line width=3pt,draw=white] (3.866025,6.000000) -- (3.866025,6.200000);
	\draw[line width=3pt,draw=white] (3.866025,6.200000) -- (3.866025,6.800000);
	\draw[line width=3pt,draw=white] (3.866025,6.800000) -- (3.866025,7.000000);
	\draw[line width=3pt,draw=white] (3.866025,7.000000) -- (3.866025,7.600000);
	\draw[line width=3pt,draw=white] (3.866025,7.600000) -- (3.866025,7.800000);
	\draw[line width=3pt,draw=white] (3.866025,7.800000) -- (3.866025,8.400000);
	\draw[line width=3pt,fill=white,draw=white] (3.766025,8.400000) -- (3.766025,8.600000) -- (3.966025,8.600000) -- (3.966025,8.400000) -- cycle;
	\draw[line width=3pt,draw=white] (3.866025,8.600000) -- (3.866025,9.200000);
	\draw[line width=3pt,draw=white,fill=white] (3.866025,9.300000) circle (0.200000);
	\draw[line width=3pt,draw=white] (3.666025,9.300000) -- (4.066025,9.300000);
	\draw[line width=3pt,draw=white] (3.866025,9.100000) -- (3.866025,9.500000);
	\draw[line width=3pt,draw=white] (3.866025,9.300000) -- (4.732051,9.800000);
	\draw[line width=3pt,draw=white,fill=white] (4.732051,9.800000) circle (0.100000);
	\draw[line width=3pt,draw=white] (3.866025,9.400000) -- (3.866025,10.000000);
	\draw[line width=3pt,draw=white,fill=white] (3.866025,10.100000) circle (0.200000);
	\draw[line width=3pt,draw=white] (3.666025,10.100000) -- (4.066025,10.100000);
	\draw[line width=3pt,draw=white] (3.866025,9.900000) -- (3.866025,10.300000);
	\draw[line width=3pt,draw=white] (3.866025,10.100000) -- (2.866025,10.100000);
	\draw[line width=3pt,draw=white,fill=white] (2.866025,10.100000) circle (0.100000);
	\draw[line width=3pt,draw=white] (3.866025,10.200000) -- (3.866025,10.800000);
	\draw[line width=3pt,draw=white,fill=white] (3.866025,10.900000) circle (0.200000);
	\draw[line width=3pt,draw=white] (3.666025,10.900000) -- (4.066025,10.900000);
	\draw[line width=3pt,draw=white] (3.866025,10.700000) -- (3.866025,11.100000);
	\draw[line width=3pt,draw=white] (3.866025,10.900000) -- (4.866025,10.900000);
	\draw[line width=3pt,draw=white,fill=white] (4.866025,10.900000) circle (0.100000);
	\draw[line width=3pt,draw=white] (3.866025,11.000000) -- (3.866025,11.600000);
	\draw[line width=3pt,draw=white,fill=white] (3.866025,11.700000) circle (0.200000);
	\draw[line width=3pt,draw=white] (3.666025,11.700000) -- (4.066025,11.700000);
	\draw[line width=3pt,draw=white] (3.866025,11.500000) -- (3.866025,11.900000);
	\draw[line width=3pt,draw=white] (3.866025,11.700000) -- (3.000000,11.200000);
	\draw[line width=3pt,draw=white,fill=white] (3.000000,11.200000) circle (0.100000);
	\draw[line width=3pt,draw=white] (3.866025,11.800000) -- (3.866025,12.400000);
	\draw[line width=3pt,fill=white,draw=white] (3.866025,12.500000) circle (0.100000);
	\draw[line width=3pt,draw=white] (3.866025,12.600000) -- (3.866025,13.200000);
	\draw[line width=3pt,draw=white] (3.866025,13.200000) -- (3.866025,13.400000);
	\draw[line width=1pt] (3.866025,0.500000) -- (3.866025,1.200000);
	\draw[line width=1pt] (3.866025,1.200000) -- (3.866025,1.400000);
	\draw[line width=1pt] (3.866025,1.400000) -- (3.866025,2.000000);
	\draw[line width=1pt] (3.866025,2.000000) -- (3.866025,2.200000);
	\draw[line width=1pt] (3.866025,2.200000) -- (3.866025,2.800000);
	\draw[line width=1pt] (3.866025,2.800000) -- (3.866025,3.000000);
	\draw[line width=1pt] (3.866025,3.000000) -- (3.866025,3.600000);
	\draw[line width=1pt] (3.866025,3.600000) -- (3.866025,3.800000);
	\draw[line width=1pt] (3.866025,3.800000) -- (3.866025,4.400000);
	\draw[line width=1pt] (3.866025,4.400000) -- (3.866025,4.600000);
	\draw[line width=1pt] (3.866025,4.600000) -- (3.866025,5.200000);
	\draw[line width=1pt] (3.866025,5.200000) -- (3.866025,5.400000);
	\draw[line width=1pt] (3.866025,5.400000) -- (3.866025,6.000000);
	\draw[line width=1pt] (3.866025,6.000000) -- (3.866025,6.200000);
	\draw[line width=1pt] (3.866025,6.200000) -- (3.866025,6.800000);
	\draw[line width=1pt] (3.866025,6.800000) -- (3.866025,7.000000);
	\draw[line width=1pt] (3.866025,7.000000) -- (3.866025,7.600000);
	\draw[line width=1pt] (3.866025,7.600000) -- (3.866025,7.800000);
	\draw[line width=1pt] (3.866025,7.800000) -- (3.866025,8.400000);
	\draw[line width=1pt,fill=white] (3.766025,8.400000) -- (3.766025,8.600000) -- (3.966025,8.600000) -- (3.966025,8.400000) -- cycle;
	\draw[line width=1pt] (3.866025,8.600000) -- (3.866025,9.200000);
	\draw[line width=1pt] (3.866025,9.300000) -- (4.732051,9.800000);
	\draw[line width=1pt,fill=white] (3.866025,9.300000) circle (0.200000);
	\draw[line width=1pt] (3.666025,9.300000) -- (4.066025,9.300000);
	\draw[line width=1pt] (3.866025,9.100000) -- (3.866025,9.500000);
	\draw[line width=1pt,fill=black] (4.732051,9.800000) circle (0.100000);
	\draw[line width=1pt] (3.866025,9.400000) -- (3.866025,10.000000);
	\draw[line width=1pt] (3.866025,10.100000) -- (2.866025,10.100000);
	\draw[line width=1pt,fill=white] (3.866025,10.100000) circle (0.200000);
	\draw[line width=1pt] (3.666025,10.100000) -- (4.066025,10.100000);
	\draw[line width=1pt] (3.866025,9.900000) -- (3.866025,10.300000);
	\draw[line width=1pt,fill=black] (2.866025,10.100000) circle (0.100000);
	\draw[line width=1pt] (2.866025,10.200000) -- +(0.0,2.0pt);
	\draw[line width=1pt] (2.866025,10.000000) -- +(0.0,-2.0pt);
	\draw[line width=1pt] (3.866025,10.200000) -- (3.866025,10.800000);
	\draw[line width=1pt] (3.866025,10.900000) -- (4.866025,10.900000);
	\draw[line width=1pt,fill=white] (3.866025,10.900000) circle (0.200000);
	\draw[line width=1pt] (3.666025,10.900000) -- (4.066025,10.900000);
	\draw[line width=1pt] (3.866025,10.700000) -- (3.866025,11.100000);
	\draw[line width=1pt,fill=black] (4.866025,10.900000) circle (0.100000);
	\draw[line width=1pt] (3.866025,11.000000) -- (3.866025,11.600000);
	\draw[line width=1pt] (3.866025,11.700000) -- (3.000000,11.200000);
	\draw[line width=1pt,fill=white] (3.866025,11.700000) circle (0.200000);
	\draw[line width=1pt] (3.666025,11.700000) -- (4.066025,11.700000);
	\draw[line width=1pt] (3.866025,11.500000) -- (3.866025,11.900000);
	\draw[line width=1pt,fill=black] (3.000000,11.200000) circle (0.100000);
	\draw[line width=1pt] (3.866025,11.800000) -- (3.866025,12.400000);
	\draw[line width=1pt,fill=white] (3.866025,12.500000) circle (0.100000);
	\draw[line width=1pt] (3.866025,12.600000) -- (3.866025,13.200000);
	\draw[line width=1pt] (3.866025,13.200000) -- (3.866025,13.400000);
	\draw[line width=1pt,rounded corners,dashed] (1.366025,5.700000) -- (2.366025,5.700000) -- (2.366025,12.900000) -- (1.366025,12.900000) -- cycle;
	\draw[line width=1pt,dotted] (3.866025,12.500000) circle (0.400000);
\end{tikzpicture}

%% file: pic_addX.tex
\begin{tikzpicture}[x=0.064286\linewidth,y=0.064286\linewidth]
	\fill[gray!25] (2.000000,8.000000) -- (2.900000,8.000000) -- (2.000000,7.100000) -- cycle;
	\fill[gray!25] (4.000000,8.000000) -- (4.900000,8.000000) -- (4.000000,7.100000) -- cycle;
	\fill[gray!25] (4.000000,8.000000) -- (4.000000,7.100000) -- (3.100000,8.000000) -- cycle;
	\fill[gray!25] (6.000000,8.000000) -- (6.000000,7.100000) -- (5.100000,8.000000) -- cycle;
	\fill[gray!80] (3.000000,7.000000) -- (2.100000,7.000000) -- (3.000000,7.900000) -- cycle;
	\fill[gray!80] (3.000000,7.000000) -- (3.000000,7.900000) -- (3.900000,7.000000) -- cycle;
	\fill[gray!80] (3.000000,7.000000) -- (3.900000,7.000000) -- (3.000000,6.100000) -- cycle;
	\fill[gray!80] (3.000000,7.000000) -- (3.000000,6.100000) -- (2.100000,7.000000) -- cycle;
	\fill[gray!80] (5.000000,7.000000) -- (4.100000,7.000000) -- (5.000000,7.900000) -- cycle;
	\fill[gray!80] (5.000000,7.000000) -- (5.000000,7.900000) -- (5.900000,7.000000) -- cycle;
	\fill[gray!80] (5.000000,7.000000) -- (5.900000,7.000000) -- (5.000000,6.100000) -- cycle;
	\fill[gray!80] (5.000000,7.000000) -- (5.000000,6.100000) -- (4.100000,7.000000) -- cycle;
	\fill[gray!25] (2.000000,6.000000) -- (2.000000,6.900000) -- (2.900000,6.000000) -- cycle;
	\fill[gray!25] (4.000000,6.000000) -- (3.100000,6.000000) -- (4.000000,6.900000) -- cycle;
	\fill[gray!25] (4.000000,6.000000) -- (4.000000,6.900000) -- (4.900000,6.000000) -- cycle;
	\fill[gray!25] (6.000000,6.000000) -- (5.100000,6.000000) -- (6.000000,6.900000) -- cycle;
	\fill[gray!25] (2.000000,4.000000) -- (2.900000,4.000000) -- (2.000000,3.100000) -- cycle;
	\fill[gray!80] (3.000000,3.000000) -- (2.100000,3.000000) -- (3.000000,3.900000) -- cycle;
	\fill[gray!80] (3.000000,3.000000) -- (3.000000,2.100000) -- (2.100000,3.000000) -- cycle;
	\fill[gray!25] (2.000000,2.000000) -- (2.000000,2.900000) -- (2.900000,2.000000) -- cycle;
	\draw[line width=1pt,fill=black] (2.000000,8.000000) circle (0.100000);
	\draw[line width=1pt] (3.000000,8.000000) circle (0.100000);
	\draw[line width=1pt,fill=black] (4.000000,8.000000) circle (0.100000);
	\draw (5.000000,8.000000) node [above right] {$|+\rangle$};
grestore
	\draw[line width=1pt] (5.000000,8.000000) circle (0.100000);
	\draw[line width=1pt,fill=black] (6.000000,8.000000) circle (0.100000);
	\draw[line width=1pt] (2.000000,7.000000) circle (0.100000);
	\draw[line width=1pt,fill=black] (3.000000,7.000000) circle (0.100000);
	\draw (4.000000,7.000000) node [above right] {$|+\rangle$};
grestore
	\draw[line width=1pt] (4.000000,7.000000) circle (0.100000);
	\draw[line width=1pt,fill=black] (5.000000,7.000000) circle (0.100000);
	\draw (6.000000,7.000000) node [above right] {$|+\rangle$};
grestore
	\draw[line width=1pt] (6.000000,7.000000) circle (0.100000);
	\draw[line width=1pt,fill=black] (2.000000,6.000000) circle (0.100000);
	\draw[line width=1pt] (3.000000,6.000000) circle (0.100000);
	\draw[line width=1pt,fill=black] (4.000000,6.000000) circle (0.100000);
	\draw (5.000000,6.000000) node [above right] {$|+\rangle$};
grestore
	\draw[line width=1pt] (5.000000,6.000000) circle (0.100000);
	\draw[line width=1pt,fill=black] (6.000000,6.000000) circle (0.100000);
	\draw[line width=1pt,fill=black] (2.000000,4.000000) circle (0.100000);
	\draw[line width=1pt] (3.000000,4.000000) circle (0.100000);
	\draw[line width=1pt] (2.000000,3.000000) circle (0.100000);
	\draw[line width=1pt,fill=black] (3.000000,3.000000) circle (0.100000);
	\draw[line width=1pt,fill=black] (2.000000,2.000000) circle (0.100000);
	\draw[line width=1pt] (3.000000,2.000000) circle (0.100000);
\end{tikzpicture}

%% file: pic_addX2.tex
\begin{tikzpicture}[x=0.064000\linewidth,y=0.064000\linewidth]
	\fill[gray!80] (1.000000,7.000000) -- (1.900000,7.000000) -- (1.000000,6.100000) -- cycle;
	\fill[gray!80] (3.000000,7.000000) -- (3.900000,7.000000) -- (3.000000,6.100000) -- cycle;
	\fill[gray!80] (3.000000,7.000000) -- (3.000000,6.100000) -- (2.100000,7.000000) -- cycle;
	\fill[gray!80] (5.000000,7.000000) -- (5.000000,6.100000) -- (4.100000,7.000000) -- cycle;
	\fill[gray!25] (2.000000,6.000000) -- (1.100000,6.000000) -- (2.000000,6.900000) -- cycle;
	\fill[gray!25] (2.000000,6.000000) -- (2.000000,6.900000) -- (2.900000,6.000000) -- cycle;
	\fill[gray!25] (2.000000,6.000000) -- (2.900000,6.000000) -- (2.000000,5.100000) -- cycle;
	\fill[gray!25] (2.000000,6.000000) -- (2.000000,5.100000) -- (1.100000,6.000000) -- cycle;
	\fill[gray!25] (4.000000,6.000000) -- (3.100000,6.000000) -- (4.000000,6.900000) -- cycle;
	\fill[gray!25] (4.000000,6.000000) -- (4.000000,6.900000) -- (4.900000,6.000000) -- cycle;
	\fill[gray!25] (4.000000,6.000000) -- (4.900000,6.000000) -- (4.000000,5.100000) -- cycle;
	\fill[gray!25] (4.000000,6.000000) -- (4.000000,5.100000) -- (3.100000,6.000000) -- cycle;
	\fill[gray!80] (1.000000,5.000000) -- (1.000000,5.900000) -- (1.900000,5.000000) -- cycle;
	\fill[gray!80] (3.000000,5.000000) -- (2.100000,5.000000) -- (3.000000,5.900000) -- cycle;
	\fill[gray!80] (3.000000,5.000000) -- (3.000000,5.900000) -- (3.900000,5.000000) -- cycle;
	\fill[gray!80] (5.000000,5.000000) -- (4.100000,5.000000) -- (5.000000,5.900000) -- cycle;
	\fill[gray!80] (1.000000,3.000000) -- (1.900000,3.000000) -- (1.000000,2.100000) -- cycle;
	\fill[gray!80] (3.000000,3.000000) -- (3.000000,2.100000) -- (2.100000,3.000000) -- cycle;
	\fill[gray!25] (2.000000,2.000000) -- (1.100000,2.000000) -- (2.000000,2.900000) -- cycle;
	\fill[gray!25] (2.000000,2.000000) -- (2.000000,2.900000) -- (2.900000,2.000000) -- cycle;
	\fill[gray!25] (2.000000,2.000000) -- (2.900000,2.000000) -- (2.000000,1.100000) -- cycle;
	\fill[gray!25] (2.000000,2.000000) -- (2.000000,1.100000) -- (1.100000,2.000000) -- cycle;
	\fill[gray!80] (1.000000,1.000000) -- (1.000000,1.900000) -- (1.900000,1.000000) -- cycle;
	\fill[gray!80] (3.000000,1.000000) -- (2.100000,1.000000) -- (3.000000,1.900000) -- cycle;
	\draw[line width=1pt,fill=black] (1.000000,7.000000) circle (0.100000);
	\draw[line width=1pt] (2.000000,7.000000) circle (0.100000);
	\draw[line width=1pt,fill=black] (3.000000,7.000000) circle (0.100000);
	\draw (4.000000,7.000000) node [above right] {$|+\rangle$};
grestore
	\draw[line width=1pt] (4.000000,7.000000) circle (0.100000);
	\draw[line width=1pt,fill=black] (5.000000,7.000000) circle (0.100000);
	\draw[line width=1pt] (1.000000,6.000000) circle (0.100000);
	\draw[line width=1pt,fill=black] (2.000000,6.000000) circle (0.100000);
	\draw[line width=1pt] (3.000000,6.000000) circle (0.100000);
	\draw[line width=1pt,fill=black] (4.000000,6.000000) circle (0.100000);
	\draw (5.000000,6.000000) node [above right] {$|+\rangle$};
grestore
	\draw[line width=1pt] (5.000000,6.000000) circle (0.100000);
	\draw[line width=1pt,fill=black] (1.000000,5.000000) circle (0.100000);
	\draw[line width=1pt] (2.000000,5.000000) circle (0.100000);
	\draw[line width=1pt,fill=black] (3.000000,5.000000) circle (0.100000);
	\draw (4.000000,5.000000) node [above right] {$|+\rangle$};
grestore
	\draw[line width=1pt] (4.000000,5.000000) circle (0.100000);
	\draw[line width=1pt,fill=black] (5.000000,5.000000) circle (0.100000);
	\draw[line width=1pt,fill=black] (1.000000,3.000000) circle (0.100000);
	\draw[line width=1pt] (2.000000,3.000000) circle (0.100000);
	\draw[line width=1pt,fill=black] (3.000000,3.000000) circle (0.100000);
	\draw[line width=1pt] (1.000000,2.000000) circle (0.100000);
	\draw[line width=1pt,fill=black] (2.000000,2.000000) circle (0.100000);
	\draw[line width=1pt] (3.000000,2.000000) circle (0.100000);
	\draw[line width=1pt,fill=black] (1.000000,1.000000) circle (0.100000);
	\draw[line width=1pt] (2.000000,1.000000) circle (0.100000);
	\draw[line width=1pt,fill=black] (3.000000,1.000000) circle (0.100000);
\end{tikzpicture}

%% file: pic_isotrimprimal.tex
\begin{tikzpicture}[x=0.074272\linewidth,y=0.074272\linewidth]
	\draw[line width=1pt,fill=black] (0.000000,0.000000) circle (0.100000);
	\draw[line width=1pt] (0.866025,0.500000) circle (0.100000);
	\draw[line width=1pt,fill=black] (1.732051,1.000000) circle (0.100000);
	\draw[line width=1pt] (1.000000,0.000000) circle (0.100000);
	\draw[line width=1pt] (0.966025,0.500000) -- (1.766025,0.500000);
	\draw[line width=1pt] (1.779423,0.450000) -- (1.086603,0.050000);
	\draw[line width=1pt,fill=black] (1.866025,0.500000) circle (0.100000);
	\draw[line width=1pt] (2.645448,0.950000) -- (1.952628,0.550000);
	\draw[line width=1pt] (2.732051,1.000000) circle (0.100000);
	\draw[line width=1pt,fill=black] (2.000000,0.000000) circle (0.100000);
	\draw[line width=1pt] (1.966025,0.500000) -- (2.766025,0.500000);
	\draw[line width=1pt] (2.866025,0.500000) circle (0.100000);
	\draw[line width=1pt,fill=black] (3.732051,1.000000) circle (0.100000);
	\draw[line width=1pt] (3.000000,0.000000) circle (0.100000);
	\draw[line width=1pt] (2.966025,0.500000) -- (3.766025,0.500000);
	\draw[line width=1pt] (3.779423,0.450000) -- (3.086603,0.050000);
	\draw[line width=1pt,fill=black] (3.866025,0.500000) circle (0.100000);
	\draw[line width=1pt] (4.645448,0.950000) -- (3.952628,0.550000);
	\draw[line width=1pt] (4.732051,1.000000) circle (0.100000);
	\draw[line width=1pt,fill=black] (4.000000,0.000000) circle (0.100000);
	\draw[line width=1pt] (3.966025,0.500000) -- (4.766025,0.500000);
	\draw[line width=1pt] (4.866025,0.500000) circle (0.100000);
	\draw[line width=1pt,fill=black] (5.732051,1.000000) circle (0.100000);
	\draw[line width=3pt,draw=white] (4.732051,1.000000) -- (4.732051,1.700000);
	\draw[line width=3pt,draw=white] (4.732051,1.700000) -- (4.732051,1.900000);
	\draw[line width=3pt,draw=white] (4.732051,1.900000) -- (4.732051,2.500000);
	\draw[line width=3pt,draw=white] (4.732051,2.500000) -- (4.732051,2.700000);
	\draw[line width=3pt,draw=white] (4.732051,2.700000) -- (4.732051,3.300000);
	\draw[line width=3pt,draw=white,fill=white] (4.732051,3.400000) circle (0.200000);
	\draw[line width=3pt,draw=white] (4.532051,3.400000) -- (4.932051,3.400000);
	\draw[line width=3pt,draw=white] (4.732051,3.200000) -- (4.732051,3.600000);
	\draw[line width=3pt,draw=white] (4.732051,3.400000) -- (3.866025,2.900000);
	\draw[line width=3pt,draw=white,fill=white] (3.866025,2.900000) circle (0.100000);
	\draw[line width=3pt,draw=white] (4.732051,3.500000) -- (4.732051,4.100000);
	\draw[line width=3pt,draw=white,fill=white] (4.732051,4.200000) circle (0.100000);
	\draw[line width=3pt,draw=white] (4.732051,4.200000) -- (5.732051,4.200000);
	\draw[line width=3pt,draw=white,fill=white] (5.732051,4.200000) circle (0.200000);
	\draw[line width=3pt,draw=white] (5.532051,4.200000) -- (5.932051,4.200000);
	\draw[line width=3pt,draw=white] (5.732051,4.000000) -- (5.732051,4.400000);
	\draw[line width=3pt,draw=white] (4.732051,4.300000) -- (4.732051,4.900000);
	\draw[line width=3pt,draw=white,fill=white] (4.732051,5.000000) circle (0.100000);
	\draw[line width=3pt,draw=white] (4.732051,5.000000) -- (3.732051,5.000000);
	\draw[line width=3pt,draw=white,fill=white] (3.732051,5.000000) circle (0.200000);
	\draw[line width=3pt,draw=white] (3.532051,5.000000) -- (3.932051,5.000000);
	\draw[line width=3pt,draw=white] (3.732051,4.800000) -- (3.732051,5.200000);
	\draw[line width=3pt,draw=white] (4.732051,5.100000) -- (4.732051,5.700000);
	\draw[line width=3pt,draw=white] (4.732051,5.700000) -- (4.732051,5.900000);
	\draw[line width=3pt,draw=white] (4.732051,5.900000) -- (4.732051,6.500000);
	\draw[line width=3pt,fill=white,draw=white] (4.632051,6.600000) -- (4.732051,6.700000) -- (4.832051,6.600000) -- (4.732051,6.500000) -- cycle;
	\draw[line width=3pt,draw=white] (4.732051,6.700000) -- (4.732051,7.300000);
	\draw[line width=3pt,fill=white,draw=white] (4.732051,7.400000) circle (0.100000);
	\draw[line width=3pt,draw=white] (4.732051,7.500000) -- (4.732051,8.100000);
	\draw[line width=3pt,draw=white] (4.732051,8.100000) -- (4.732051,8.300000);
	\draw[line width=3pt,draw=white] (4.732051,8.300000) -- (4.732051,8.900000);
	\draw[line width=3pt,draw=white] (4.732051,8.900000) -- (4.732051,9.100000);
	\draw[line width=3pt,draw=white] (4.732051,9.100000) -- (4.732051,9.700000);
	\draw[line width=3pt,draw=white] (4.732051,9.700000) -- (4.732051,9.900000);
	\draw[line width=3pt,draw=white] (4.732051,9.900000) -- (4.732051,10.500000);
	\draw[line width=3pt,draw=white] (4.732051,10.500000) -- (4.732051,10.700000);
	\draw[line width=3pt,draw=white] (4.732051,10.700000) -- (4.732051,11.300000);
	\draw[line width=3pt,draw=white] (4.732051,11.300000) -- (4.732051,11.500000);
	\draw[line width=3pt,draw=white] (4.732051,11.500000) -- (4.732051,12.100000);
	\draw[line width=3pt,draw=white] (4.732051,12.100000) -- (4.732051,12.300000);
	\draw[line width=3pt,draw=white] (4.732051,12.300000) -- (4.732051,12.900000);
	\draw[line width=3pt,draw=white] (4.732051,12.900000) -- (4.732051,13.100000);
	\draw[line width=3pt,draw=white] (4.732051,13.100000) -- (4.732051,13.700000);
	\draw[line width=3pt,draw=white] (4.732051,13.700000) -- (4.732051,13.900000);
	\draw[line width=1pt] (4.732051,1.000000) -- (4.732051,1.700000);
	\draw[line width=1pt] (4.732051,1.700000) -- (4.732051,1.900000);
	\draw[line width=1pt] (4.732051,1.900000) -- (4.732051,2.500000);
	\draw[line width=1pt] (4.732051,2.500000) -- (4.732051,2.700000);
	\draw[line width=1pt] (4.732051,2.700000) -- (4.732051,3.300000);
	\draw[line width=1pt] (4.732051,3.400000) -- (3.866025,2.900000);
	\draw[line width=1pt,fill=white] (4.732051,3.400000) circle (0.200000);
	\draw[line width=1pt] (4.532051,3.400000) -- (4.932051,3.400000);
	\draw[line width=1pt] (4.732051,3.200000) -- (4.732051,3.600000);
	\draw[line width=1pt,fill=black] (3.866025,2.900000) circle (0.100000);
	\draw[line width=1pt] (3.866025,3.000000) -- +(0.0,2.0pt);
	\draw[line width=1pt] (3.866025,2.800000) -- +(0.0,-2.0pt);
	\draw[line width=1pt] (4.732051,3.500000) -- (4.732051,4.100000);
	\draw[line width=1pt] (4.732051,4.200000) -- (5.732051,4.200000);
	\draw[line width=1pt,fill=black] (4.732051,4.200000) circle (0.100000);
	\draw[line width=1pt,fill=white] (5.732051,4.200000) circle (0.200000);
	\draw[line width=1pt] (5.532051,4.200000) -- (5.932051,4.200000);
	\draw[line width=1pt] (5.732051,4.000000) -- (5.732051,4.400000);
	\draw[line width=1pt] (4.732051,4.300000) -- (4.732051,4.900000);
	\draw[line width=1pt] (4.732051,5.000000) -- (3.732051,5.000000);
	\draw[line width=1pt,fill=black] (4.732051,5.000000) circle (0.100000);
	\draw[line width=1pt,fill=white] (3.732051,5.000000) circle (0.200000);
	\draw[line width=1pt] (3.532051,5.000000) -- (3.932051,5.000000);
	\draw[line width=1pt] (3.732051,4.800000) -- (3.732051,5.200000);
	\draw[line width=1pt] (4.732051,5.100000) -- (4.732051,5.700000);
	\draw[line width=1pt] (4.732051,5.700000) -- (4.732051,5.900000);
	\draw[line width=1pt] (4.732051,5.900000) -- (4.732051,6.500000);
	\draw[line width=1pt,fill=white] (4.632051,6.600000) -- (4.732051,6.700000) -- (4.832051,6.600000) -- (4.732051,6.500000) -- cycle;
	\draw[line width=1pt] (4.732051,6.700000) -- (4.732051,7.300000);
	\draw[line width=1pt,fill=white] (4.732051,7.400000) circle (0.100000);
	\draw[line width=1pt] (4.732051,7.500000) -- (4.732051,8.100000);
	\draw[line width=1pt] (4.732051,8.100000) -- (4.732051,8.300000);
	\draw[line width=1pt] (4.732051,8.300000) -- (4.732051,8.900000);
	\draw[line width=1pt] (4.732051,8.900000) -- (4.732051,9.100000);
	\draw[line width=1pt] (4.732051,9.100000) -- (4.732051,9.700000);
	\draw[line width=1pt] (4.732051,9.700000) -- (4.732051,9.900000);
	\draw[line width=1pt] (4.732051,9.900000) -- (4.732051,10.500000);
	\draw[line width=1pt] (4.732051,10.500000) -- (4.732051,10.700000);
	\draw[line width=1pt] (4.732051,10.700000) -- (4.732051,11.300000);
	\draw[line width=1pt] (4.732051,11.300000) -- (4.732051,11.500000);
	\draw[line width=1pt] (4.732051,11.500000) -- (4.732051,12.100000);
	\draw[line width=1pt] (4.732051,12.100000) -- (4.732051,12.300000);
	\draw[line width=1pt] (4.732051,12.300000) -- (4.732051,12.900000);
	\draw[line width=1pt] (4.732051,12.900000) -- (4.732051,13.100000);
	\draw[line width=1pt] (4.732051,13.100000) -- (4.732051,13.700000);
	\draw[line width=1pt] (4.732051,13.700000) -- (4.732051,13.900000);
	\draw[line width=3pt,draw=white] (1.866025,0.500000) -- (1.866025,1.200000);
	\draw[line width=3pt,fill=white,draw=white] (1.766025,1.200000) -- (1.766025,1.400000) -- (1.966025,1.400000) -- (1.966025,1.200000) -- cycle;
	\draw[line width=3pt,draw=white] (1.866025,1.400000) -- (1.866025,2.000000);
	\draw[line width=3pt,fill=white,draw=white] (1.766025,2.100000) -- (1.866025,2.200000) -- (1.966025,2.100000) -- (1.866025,2.000000) -- cycle;
	\draw[line width=3pt,draw=white] (1.866025,2.200000) -- (1.866025,2.800000);
	\draw[line width=3pt,draw=white,fill=white] (1.866025,2.900000) circle (0.100000);
	\draw[line width=3pt,draw=white] (1.866025,2.900000) -- (2.732051,3.400000);
	\draw[line width=3pt,draw=white,fill=white] (2.732051,3.400000) circle (0.200000);
	\draw[line width=3pt,draw=white] (2.532051,3.400000) -- (2.932051,3.400000);
	\draw[line width=3pt,draw=white] (2.732051,3.200000) -- (2.732051,3.600000);
	\draw[line width=3pt,draw=white] (1.866025,3.000000) -- (1.866025,3.600000);
	\draw[line width=3pt,draw=white,fill=white] (1.866025,3.700000) circle (0.100000);
	\draw[line width=3pt,draw=white] (1.866025,3.700000) -- (0.866025,3.700000);
	\draw[line width=3pt,draw=white,fill=white] (0.866025,3.700000) circle (0.200000);
	\draw[line width=3pt,draw=white] (0.666025,3.700000) -- (1.066025,3.700000);
	\draw[line width=3pt,draw=white] (0.866025,3.500000) -- (0.866025,3.900000);
	\draw[line width=3pt,draw=white] (1.866025,3.800000) -- (1.866025,4.400000);
	\draw[line width=3pt,draw=white,fill=white] (1.866025,4.500000) circle (0.100000);
	\draw[line width=3pt,draw=white] (1.866025,4.500000) -- (2.866025,4.500000);
	\draw[line width=3pt,draw=white,fill=white] (2.866025,4.500000) circle (0.200000);
	\draw[line width=3pt,draw=white] (2.666025,4.500000) -- (3.066025,4.500000);
	\draw[line width=3pt,draw=white] (2.866025,4.300000) -- (2.866025,4.700000);
	\draw[line width=3pt,draw=white] (1.866025,4.600000) -- (1.866025,5.200000);
	\draw[line width=3pt,draw=white,fill=white] (1.866025,5.300000) circle (0.100000);
	\draw[line width=3pt,draw=white] (1.866025,5.300000) -- (1.000000,4.800000);
	\draw[line width=3pt,draw=white,fill=white] (1.000000,4.800000) circle (0.200000);
	\draw[line width=3pt,draw=white] (0.800000,4.800000) -- (1.200000,4.800000);
	\draw[line width=3pt,draw=white] (1.000000,4.600000) -- (1.000000,5.000000);
	\draw[line width=3pt,draw=white] (1.866025,5.400000) -- (1.866025,6.000000);
	\draw[line width=3pt,fill=white,draw=white] (1.766025,6.100000) -- (1.866025,6.200000) -- (1.966025,6.100000) -- (1.866025,6.000000) -- cycle;
	\draw[line width=3pt,draw=white] (1.866025,6.200000) -- (1.866025,6.800000);
	\draw[line width=3pt,fill=white,draw=white] (1.866025,6.900000) circle (0.100000);
	\draw[line width=3pt,draw=white] (1.866025,7.000000) -- (1.866025,7.600000);
	\draw[line width=3pt,fill=white,draw=white] (1.766025,7.600000) -- (1.766025,7.800000) -- (1.966025,7.800000) -- (1.966025,7.600000) -- cycle;
	\draw[line width=3pt,draw=white] (1.866025,7.800000) -- (1.866025,8.400000);
	\draw[line width=3pt,fill=white,draw=white] (1.766025,8.500000) -- (1.866025,8.600000) -- (1.966025,8.500000) -- (1.866025,8.400000) -- cycle;
	\draw[line width=3pt,draw=white] (1.866025,8.600000) -- (1.866025,9.200000);
	\draw[line width=3pt,draw=white,fill=white] (1.866025,9.300000) circle (0.100000);
	\draw[line width=3pt,draw=white] (1.866025,9.300000) -- (2.732051,9.800000);
	\draw[line width=3pt,draw=white,fill=white] (2.732051,9.800000) circle (0.200000);
	\draw[line width=3pt,draw=white] (2.532051,9.800000) -- (2.932051,9.800000);
	\draw[line width=3pt,draw=white] (2.732051,9.600000) -- (2.732051,10.000000);
	\draw[line width=3pt,draw=white] (1.866025,9.400000) -- (1.866025,10.000000);
	\draw[line width=3pt,draw=white,fill=white] (1.866025,10.100000) circle (0.100000);
	\draw[line width=3pt,draw=white] (1.866025,10.100000) -- (0.866025,10.100000);
	\draw[line width=3pt,draw=white,fill=white] (0.866025,10.100000) circle (0.200000);
	\draw[line width=3pt,draw=white] (0.666025,10.100000) -- (1.066025,10.100000);
	\draw[line width=3pt,draw=white] (0.866025,9.900000) -- (0.866025,10.300000);
	\draw[line width=3pt,draw=white] (1.866025,10.200000) -- (1.866025,10.800000);
	\draw[line width=3pt,draw=white] (1.866025,10.800000) -- (1.866025,11.000000);
	\draw[line width=3pt,draw=white] (1.866025,11.000000) -- (1.866025,11.600000);
	\draw[line width=3pt,draw=white,fill=white] (1.866025,11.700000) circle (0.100000);
	\draw[line width=3pt,draw=white] (1.866025,11.700000) -- (1.000000,11.200000);
	\draw[line width=3pt,draw=white,fill=white] (1.000000,11.200000) circle (0.200000);
	\draw[line width=3pt,draw=white] (0.800000,11.200000) -- (1.200000,11.200000);
	\draw[line width=3pt,draw=white] (1.000000,11.000000) -- (1.000000,11.400000);
	\draw[line width=3pt,draw=white] (1.866025,11.800000) -- (1.866025,12.400000);
	\draw[line width=3pt,fill=white,draw=white] (1.766025,12.500000) -- (1.866025,12.600000) -- (1.966025,12.500000) -- (1.866025,12.400000) -- cycle;
	\draw[line width=3pt,draw=white] (1.866025,12.600000) -- (1.866025,13.200000);
	\draw[line width=3pt,fill=white,draw=white] (1.866025,13.300000) circle (0.100000);
	\draw[line width=1pt] (1.866025,0.500000) -- (1.866025,1.200000);
	\draw[line width=1pt,fill=white] (1.766025,1.200000) -- (1.766025,1.400000) -- (1.966025,1.400000) -- (1.966025,1.200000) -- cycle;
	\draw[line width=1pt] (1.866025,1.400000) -- (1.866025,2.000000);
	\draw[line width=1pt,fill=white] (1.766025,2.100000) -- (1.866025,2.200000) -- (1.966025,2.100000) -- (1.866025,2.000000) -- cycle;
	\draw[line width=1pt] (1.866025,2.200000) -- (1.866025,2.800000);
	\draw[line width=1pt] (1.866025,2.900000) -- (2.732051,3.400000);
	\draw[line width=1pt,fill=black] (1.866025,2.900000) circle (0.100000);
	\draw[line width=1pt,fill=white] (2.732051,3.400000) circle (0.200000);
	\draw[line width=1pt] (2.532051,3.400000) -- (2.932051,3.400000);
	\draw[line width=1pt] (2.732051,3.200000) -- (2.732051,3.600000);
	\draw[line width=1pt] (1.866025,3.000000) -- (1.866025,3.600000);
	\draw[line width=1pt] (1.866025,3.700000) -- (0.866025,3.700000);
	\draw[line width=1pt,fill=black] (1.866025,3.700000) circle (0.100000);
	\draw[line width=1pt,fill=white] (0.866025,3.700000) circle (0.200000);
	\draw[line width=1pt] (0.666025,3.700000) -- (1.066025,3.700000);
	\draw[line width=1pt] (0.866025,3.500000) -- (0.866025,3.900000);
	\draw[line width=1pt] (1.866025,3.800000) -- (1.866025,4.400000);
	\draw[line width=1pt] (1.866025,4.500000) -- (2.866025,4.500000);
	\draw[line width=1pt,fill=black] (1.866025,4.500000) circle (0.100000);
	\draw[line width=1pt,fill=white] (2.866025,4.500000) circle (0.200000);
	\draw[line width=1pt] (2.666025,4.500000) -- (3.066025,4.500000);
	\draw[line width=1pt] (2.866025,4.300000) -- (2.866025,4.700000);
	\draw[line width=1pt] (2.866025,4.700000) -- +(0.0,2.0pt);
	\draw[line width=1pt] (2.866025,4.300000) -- +(0.0,-2.0pt);
	\draw[line width=1pt] (1.866025,4.600000) -- (1.866025,5.200000);
	\draw[line width=1pt] (1.866025,5.300000) -- (1.000000,4.800000);
	\draw[line width=1pt,fill=black] (1.866025,5.300000) circle (0.100000);
	\draw[line width=1pt,fill=white] (1.000000,4.800000) circle (0.200000);
	\draw[line width=1pt] (0.800000,4.800000) -- (1.200000,4.800000);
	\draw[line width=1pt] (1.000000,4.600000) -- (1.000000,5.000000);
	\draw[line width=1pt] (1.866025,5.400000) -- (1.866025,6.000000);
	\draw[line width=1pt,fill=white] (1.766025,6.100000) -- (1.866025,6.200000) -- (1.966025,6.100000) -- (1.866025,6.000000) -- cycle;
	\draw[line width=1pt] (1.866025,6.200000) -- (1.866025,6.800000);
	\draw[line width=1pt,fill=white] (1.866025,6.900000) circle (0.100000);
	\draw[line width=1pt] (1.866025,7.000000) -- (1.866025,7.600000);
	\draw[line width=1pt,fill=white] (1.766025,7.600000) -- (1.766025,7.800000) -- (1.966025,7.800000) -- (1.966025,7.600000) -- cycle;
	\draw[line width=1pt] (1.866025,7.800000) -- (1.866025,8.400000);
	\draw[line width=1pt,fill=white] (1.766025,8.500000) -- (1.866025,8.600000) -- (1.966025,8.500000) -- (1.866025,8.400000) -- cycle;
	\draw[line width=1pt] (1.866025,8.600000) -- (1.866025,9.200000);
	\draw[line width=1pt] (1.866025,9.300000) -- (2.732051,9.800000);
	\draw[line width=1pt,fill=black] (1.866025,9.300000) circle (0.100000);
	\draw[line width=1pt,fill=white] (2.732051,9.800000) circle (0.200000);
	\draw[line width=1pt] (2.532051,9.800000) -- (2.932051,9.800000);
	\draw[line width=1pt] (2.732051,9.600000) -- (2.732051,10.000000);
	\draw[line width=1pt] (1.866025,9.400000) -- (1.866025,10.000000);
	\draw[line width=1pt] (1.866025,10.100000) -- (0.866025,10.100000);
	\draw[line width=1pt,fill=black] (1.866025,10.100000) circle (0.100000);
	\draw[line width=1pt,fill=white] (0.866025,10.100000) circle (0.200000);
	\draw[line width=1pt] (0.666025,10.100000) -- (1.066025,10.100000);
	\draw[line width=1pt] (0.866025,9.900000) -- (0.866025,10.300000);
	\draw[line width=1pt] (1.866025,10.200000) -- (1.866025,10.800000);
	\draw[line width=1pt] (1.866025,10.800000) -- (1.866025,11.000000);
	\draw[line width=1pt] (1.866025,11.000000) -- (1.866025,11.600000);
	\draw[line width=1pt] (1.866025,11.700000) -- (1.000000,11.200000);
	\draw[line width=1pt,fill=black] (1.866025,11.700000) circle (0.100000);
	\draw[line width=1pt,fill=white] (1.000000,11.200000) circle (0.200000);
	\draw[line width=1pt] (0.800000,11.200000) -- (1.200000,11.200000);
	\draw[line width=1pt] (1.000000,11.000000) -- (1.000000,11.400000);
	\draw[line width=1pt] (1.866025,11.800000) -- (1.866025,12.400000);
	\draw[line width=1pt,fill=white] (1.766025,12.500000) -- (1.866025,12.600000) -- (1.966025,12.500000) -- (1.866025,12.400000) -- cycle;
	\draw[line width=1pt] (1.866025,12.600000) -- (1.866025,13.200000);
	\draw[line width=1pt,fill=white] (1.866025,13.300000) circle (0.100000);
	\draw[line width=3pt,draw=white] (2.866025,0.500000) -- (2.866025,1.200000);
	\draw[line width=3pt,draw=white] (2.866025,1.200000) -- (2.866025,1.400000);
	\draw[line width=3pt,draw=white] (2.866025,1.400000) -- (2.866025,2.000000);
	\draw[line width=3pt,draw=white] (2.866025,2.000000) -- (2.866025,2.200000);
	\draw[line width=3pt,draw=white] (2.866025,2.200000) -- (2.866025,2.800000);
	\draw[line width=3pt,draw=white,fill=white] (2.866025,2.900000) circle (0.100000);
	\draw[line width=3pt,draw=white] (2.866025,2.900000) -- (2.000000,2.400000);
	\draw[line width=3pt,draw=white,fill=white] (2.000000,2.400000) circle (0.200000);
	\draw[line width=3pt,draw=white] (1.800000,2.400000) -- (2.200000,2.400000);
	\draw[line width=3pt,draw=white] (2.000000,2.200000) -- (2.000000,2.600000);
	\draw[line width=3pt,draw=white] (2.866025,3.000000) -- (2.866025,3.600000);
	\draw[line width=3pt,draw=white,fill=white] (2.866025,3.700000) circle (0.200000);
	\draw[line width=3pt,draw=white] (2.666025,3.700000) -- (3.066025,3.700000);
	\draw[line width=3pt,draw=white] (2.866025,3.500000) -- (2.866025,3.900000);
	\draw[line width=3pt,draw=white] (2.866025,3.700000) -- (3.866025,3.700000);
	\draw[line width=3pt,draw=white,fill=white] (3.866025,3.700000) circle (0.100000);
	\draw[line width=3pt,draw=white] (2.866025,3.800000) -- (2.866025,4.400000);
	\draw[line width=3pt,draw=white,fill=white] (2.866025,4.500000) circle (0.200000);
	\draw[line width=3pt,draw=white] (2.666025,4.500000) -- (3.066025,4.500000);
	\draw[line width=3pt,draw=white] (2.866025,4.300000) -- (2.866025,4.700000);
	\draw[line width=3pt,draw=white] (2.866025,4.500000) -- (1.866025,4.500000);
	\draw[line width=3pt,draw=white,fill=white] (1.866025,4.500000) circle (0.100000);
	\draw[line width=3pt,draw=white] (2.866025,4.600000) -- (2.866025,5.200000);
	\draw[line width=3pt,draw=white,fill=white] (2.866025,5.300000) circle (0.100000);
	\draw[line width=3pt,draw=white] (2.866025,5.300000) -- (3.732051,5.800000);
	\draw[line width=3pt,draw=white,fill=white] (3.732051,5.800000) circle (0.200000);
	\draw[line width=3pt,draw=white] (3.532051,5.800000) -- (3.932051,5.800000);
	\draw[line width=3pt,draw=white] (3.732051,5.600000) -- (3.732051,6.000000);
	\draw[line width=3pt,draw=white] (2.866025,5.400000) -- (2.866025,6.000000);
	\draw[line width=3pt,fill=white,draw=white] (2.766025,6.100000) -- (2.866025,6.200000) -- (2.966025,6.100000) -- (2.866025,6.000000) -- cycle;
	\draw[line width=3pt,draw=white] (2.866025,6.200000) -- (2.866025,6.800000);
	\draw[line width=3pt,fill=white,draw=white] (2.866025,6.900000) circle (0.100000);
	\draw[line width=3pt,draw=white] (2.866025,7.000000) -- (2.866025,7.600000);
	\draw[line width=3pt,draw=white] (2.866025,7.600000) -- (2.866025,7.800000);
	\draw[line width=3pt,draw=white] (2.866025,7.800000) -- (2.866025,8.400000);
	\draw[line width=3pt,draw=white] (2.866025,8.400000) -- (2.866025,8.600000);
	\draw[line width=3pt,draw=white] (2.866025,8.600000) -- (2.866025,9.200000);
	\draw[line width=3pt,draw=white] (2.866025,9.200000) -- (2.866025,9.400000);
	\draw[line width=3pt,draw=white] (2.866025,9.400000) -- (2.866025,10.000000);
	\draw[line width=3pt,draw=white] (2.866025,10.000000) -- (2.866025,10.200000);
	\draw[line width=3pt,draw=white] (2.866025,10.200000) -- (2.866025,10.800000);
	\draw[line width=3pt,draw=white] (2.866025,10.800000) -- (2.866025,11.000000);
	\draw[line width=3pt,draw=white] (2.866025,11.000000) -- (2.866025,11.600000);
	\draw[line width=3pt,draw=white] (2.866025,11.600000) -- (2.866025,11.800000);
	\draw[line width=3pt,draw=white] (2.866025,11.800000) -- (2.866025,12.400000);
	\draw[line width=3pt,draw=white] (2.866025,12.400000) -- (2.866025,12.600000);
	\draw[line width=3pt,draw=white] (2.866025,12.600000) -- (2.866025,13.200000);
	\draw[line width=3pt,draw=white] (2.866025,13.200000) -- (2.866025,13.400000);
	\draw[line width=1pt] (2.866025,0.500000) -- (2.866025,1.200000);
	\draw[line width=1pt] (2.866025,1.200000) -- (2.866025,1.400000);
	\draw[line width=1pt] (2.866025,1.400000) -- (2.866025,2.000000);
	\draw[line width=1pt] (2.866025,2.000000) -- (2.866025,2.200000);
	\draw[line width=1pt] (2.866025,2.200000) -- (2.866025,2.800000);
	\draw[line width=1pt] (2.866025,2.900000) -- (2.000000,2.400000);
	\draw[line width=1pt,fill=black] (2.866025,2.900000) circle (0.100000);
	\draw[line width=1pt,fill=white] (2.000000,2.400000) circle (0.200000);
	\draw[line width=1pt] (1.800000,2.400000) -- (2.200000,2.400000);
	\draw[line width=1pt] (2.000000,2.200000) -- (2.000000,2.600000);
	\draw[line width=1pt] (2.866025,3.000000) -- (2.866025,3.600000);
	\draw[line width=1pt] (2.866025,3.700000) -- (3.866025,3.700000);
	\draw[line width=1pt,fill=white] (2.866025,3.700000) circle (0.200000);
	\draw[line width=1pt] (2.666025,3.700000) -- (3.066025,3.700000);
	\draw[line width=1pt] (2.866025,3.500000) -- (2.866025,3.900000);
	\draw[line width=1pt,fill=black] (3.866025,3.700000) circle (0.100000);
	\draw[line width=1pt] (3.866025,3.800000) -- +(0.0,2.0pt);
	\draw[line width=1pt] (3.866025,3.600000) -- +(0.0,-2.0pt);
	\draw[line width=1pt] (2.866025,3.800000) -- (2.866025,4.400000);
	\draw[line width=1pt] (2.866025,4.500000) -- (1.866025,4.500000);
	\draw[line width=1pt,fill=white] (2.866025,4.500000) circle (0.200000);
	\draw[line width=1pt] (2.666025,4.500000) -- (3.066025,4.500000);
	\draw[line width=1pt] (2.866025,4.300000) -- (2.866025,4.700000);
	\draw[line width=1pt,fill=black] (1.866025,4.500000) circle (0.100000);
	\draw[line width=1pt] (1.866025,4.600000) -- +(0.0,2.0pt);
	\draw[line width=1pt] (1.866025,4.400000) -- +(0.0,-2.0pt);
	\draw[line width=1pt] (2.866025,4.600000) -- (2.866025,5.200000);
	\draw[line width=1pt] (2.866025,5.300000) -- (3.732051,5.800000);
	\draw[line width=1pt,fill=black] (2.866025,5.300000) circle (0.100000);
	\draw[line width=1pt,fill=white] (3.732051,5.800000) circle (0.200000);
	\draw[line width=1pt] (3.532051,5.800000) -- (3.932051,5.800000);
	\draw[line width=1pt] (3.732051,5.600000) -- (3.732051,6.000000);
	\draw[line width=1pt] (2.866025,5.400000) -- (2.866025,6.000000);
	\draw[line width=1pt,fill=white] (2.766025,6.100000) -- (2.866025,6.200000) -- (2.966025,6.100000) -- (2.866025,6.000000) -- cycle;
	\draw[line width=1pt] (2.866025,6.200000) -- (2.866025,6.800000);
	\draw[line width=1pt,fill=white] (2.866025,6.900000) circle (0.100000);
	\draw[line width=1pt] (2.866025,7.000000) -- (2.866025,7.600000);
	\draw[line width=1pt] (2.866025,7.600000) -- (2.866025,7.800000);
	\draw[line width=1pt] (2.866025,7.800000) -- (2.866025,8.400000);
	\draw[line width=1pt] (2.866025,8.400000) -- (2.866025,8.600000);
	\draw[line width=1pt] (2.866025,8.600000) -- (2.866025,9.200000);
	\draw[line width=1pt] (2.866025,9.200000) -- (2.866025,9.400000);
	\draw[line width=1pt] (2.866025,9.400000) -- (2.866025,10.000000);
	\draw[line width=1pt] (2.866025,10.000000) -- (2.866025,10.200000);
	\draw[line width=1pt] (2.866025,10.200000) -- (2.866025,10.800000);
	\draw[line width=1pt] (2.866025,10.800000) -- (2.866025,11.000000);
	\draw[line width=1pt] (2.866025,11.000000) -- (2.866025,11.600000);
	\draw[line width=1pt] (2.866025,11.600000) -- (2.866025,11.800000);
	\draw[line width=1pt] (2.866025,11.800000) -- (2.866025,12.400000);
	\draw[line width=1pt] (2.866025,12.400000) -- (2.866025,12.600000);
	\draw[line width=1pt] (2.866025,12.600000) -- (2.866025,13.200000);
	\draw[line width=1pt] (2.866025,13.200000) -- (2.866025,13.400000);
	\draw[line width=3pt,draw=white] (3.866025,0.500000) -- (3.866025,1.200000);
	\draw[line width=3pt,fill=white,draw=white] (3.766025,1.200000) -- (3.766025,1.400000) -- (3.966025,1.400000) -- (3.966025,1.200000) -- cycle;
	\draw[line width=3pt,draw=white] (3.866025,1.400000) -- (3.866025,2.000000);
	\draw[line width=3pt,fill=white,draw=white] (3.766025,2.100000) -- (3.866025,2.200000) -- (3.966025,2.100000) -- (3.866025,2.000000) -- cycle;
	\draw[line width=3pt,draw=white] (3.866025,2.200000) -- (3.866025,2.800000);
	\draw[line width=3pt,draw=white,fill=white] (3.866025,2.900000) circle (0.100000);
	\draw[line width=3pt,draw=white] (3.866025,2.900000) -- (4.732051,3.400000);
	\draw[line width=3pt,draw=white,fill=white] (4.732051,3.400000) circle (0.200000);
	\draw[line width=3pt,draw=white] (4.532051,3.400000) -- (4.932051,3.400000);
	\draw[line width=3pt,draw=white] (4.732051,3.200000) -- (4.732051,3.600000);
	\draw[line width=3pt,draw=white] (3.866025,3.000000) -- (3.866025,3.600000);
	\draw[line width=3pt,draw=white,fill=white] (3.866025,3.700000) circle (0.100000);
	\draw[line width=3pt,draw=white] (3.866025,3.700000) -- (2.866025,3.700000);
	\draw[line width=3pt,draw=white,fill=white] (2.866025,3.700000) circle (0.200000);
	\draw[line width=3pt,draw=white] (2.666025,3.700000) -- (3.066025,3.700000);
	\draw[line width=3pt,draw=white] (2.866025,3.500000) -- (2.866025,3.900000);
	\draw[line width=3pt,draw=white] (3.866025,3.800000) -- (3.866025,4.400000);
	\draw[line width=3pt,draw=white,fill=white] (3.866025,4.500000) circle (0.100000);
	\draw[line width=3pt,draw=white] (3.866025,4.500000) -- (4.866025,4.500000);
	\draw[line width=3pt,draw=white,fill=white] (4.866025,4.500000) circle (0.200000);
	\draw[line width=3pt,draw=white] (4.666025,4.500000) -- (5.066025,4.500000);
	\draw[line width=3pt,draw=white] (4.866025,4.300000) -- (4.866025,4.700000);
	\draw[line width=3pt,draw=white] (3.866025,4.600000) -- (3.866025,5.200000);
	\draw[line width=3pt,draw=white,fill=white] (3.866025,5.300000) circle (0.100000);
	\draw[line width=3pt,draw=white] (3.866025,5.300000) -- (3.000000,4.800000);
	\draw[line width=3pt,draw=white,fill=white] (3.000000,4.800000) circle (0.200000);
	\draw[line width=3pt,draw=white] (2.800000,4.800000) -- (3.200000,4.800000);
	\draw[line width=3pt,draw=white] (3.000000,4.600000) -- (3.000000,5.000000);
	\draw[line width=3pt,draw=white] (3.866025,5.400000) -- (3.866025,6.000000);
	\draw[line width=3pt,fill=white,draw=white] (3.766025,6.100000) -- (3.866025,6.200000) -- (3.966025,6.100000) -- (3.866025,6.000000) -- cycle;
	\draw[line width=3pt,draw=white] (3.866025,6.200000) -- (3.866025,6.800000);
	\draw[line width=3pt,fill=white,draw=white] (3.866025,6.900000) circle (0.100000);
	\draw[line width=3pt,draw=white] (3.866025,7.000000) -- (3.866025,7.600000);
	\draw[line width=3pt,draw=white] (3.866025,7.600000) -- (3.866025,7.800000);
	\draw[line width=3pt,draw=white] (3.866025,7.800000) -- (3.866025,8.400000);
	\draw[line width=3pt,draw=white] (3.866025,8.400000) -- (3.866025,8.600000);
	\draw[line width=3pt,draw=white] (3.866025,8.600000) -- (3.866025,9.200000);
	\draw[line width=3pt,draw=white] (3.866025,9.200000) -- (3.866025,9.400000);
	\draw[line width=3pt,draw=white] (3.866025,9.400000) -- (3.866025,10.000000);
	\draw[line width=3pt,draw=white] (3.866025,10.000000) -- (3.866025,10.200000);
	\draw[line width=3pt,draw=white] (3.866025,10.200000) -- (3.866025,10.800000);
	\draw[line width=3pt,draw=white] (3.866025,10.800000) -- (3.866025,11.000000);
	\draw[line width=3pt,draw=white] (3.866025,11.000000) -- (3.866025,11.600000);
	\draw[line width=3pt,draw=white] (3.866025,11.600000) -- (3.866025,11.800000);
	\draw[line width=3pt,draw=white] (3.866025,11.800000) -- (3.866025,12.400000);
	\draw[line width=3pt,draw=white] (3.866025,12.400000) -- (3.866025,12.600000);
	\draw[line width=3pt,draw=white] (3.866025,12.600000) -- (3.866025,13.200000);
	\draw[line width=3pt,draw=white] (3.866025,13.200000) -- (3.866025,13.400000);
	\draw[line width=1pt] (3.866025,0.500000) -- (3.866025,1.200000);
	\draw[line width=1pt,fill=white] (3.766025,1.200000) -- (3.766025,1.400000) -- (3.966025,1.400000) -- (3.966025,1.200000) -- cycle;
	\draw[line width=1pt] (3.866025,1.400000) -- (3.866025,2.000000);
	\draw[line width=1pt,fill=white] (3.766025,2.100000) -- (3.866025,2.200000) -- (3.966025,2.100000) -- (3.866025,2.000000) -- cycle;
	\draw[line width=1pt] (3.866025,2.200000) -- (3.866025,2.800000);
	\draw[line width=1pt] (3.866025,2.900000) -- (4.732051,3.400000);
	\draw[line width=1pt,fill=black] (3.866025,2.900000) circle (0.100000);
	\draw[line width=1pt,fill=white] (4.732051,3.400000) circle (0.200000);
	\draw[line width=1pt] (4.532051,3.400000) -- (4.932051,3.400000);
	\draw[line width=1pt] (4.732051,3.200000) -- (4.732051,3.600000);
	\draw[line width=1pt] (4.732051,3.600000) -- +(0.0,2.0pt);
	\draw[line width=1pt] (4.732051,3.200000) -- +(0.0,-2.0pt);
	\draw[line width=1pt] (3.866025,3.000000) -- (3.866025,3.600000);
	\draw[line width=1pt] (3.866025,3.700000) -- (2.866025,3.700000);
	\draw[line width=1pt,fill=black] (3.866025,3.700000) circle (0.100000);
	\draw[line width=1pt,fill=white] (2.866025,3.700000) circle (0.200000);
	\draw[line width=1pt] (2.666025,3.700000) -- (3.066025,3.700000);
	\draw[line width=1pt] (2.866025,3.500000) -- (2.866025,3.900000);
	\draw[line width=1pt] (2.866025,3.900000) -- +(0.0,2.0pt);
	\draw[line width=1pt] (2.866025,3.500000) -- +(0.0,-2.0pt);
	\draw[line width=1pt] (3.866025,3.800000) -- (3.866025,4.400000);
	\draw[line width=1pt] (3.866025,4.500000) -- (4.866025,4.500000);
	\draw[line width=1pt,fill=black] (3.866025,4.500000) circle (0.100000);
	\draw[line width=1pt,fill=white] (4.866025,4.500000) circle (0.200000);
	\draw[line width=1pt] (4.666025,4.500000) -- (5.066025,4.500000);
	\draw[line width=1pt] (4.866025,4.300000) -- (4.866025,4.700000);
	\draw[line width=1pt] (4.866025,4.700000) -- +(0.0,2.0pt);
	\draw[line width=1pt] (4.866025,4.300000) -- +(0.0,-2.0pt);
	\draw[line width=1pt] (3.866025,4.600000) -- (3.866025,5.200000);
	\draw[line width=1pt] (3.866025,5.300000) -- (3.000000,4.800000);
	\draw[line width=1pt,fill=black] (3.866025,5.300000) circle (0.100000);
	\draw[line width=1pt,fill=white] (3.000000,4.800000) circle (0.200000);
	\draw[line width=1pt] (2.800000,4.800000) -- (3.200000,4.800000);
	\draw[line width=1pt] (3.000000,4.600000) -- (3.000000,5.000000);
	\draw[line width=1pt] (3.000000,5.000000) -- +(0.0,2.0pt);
	\draw[line width=1pt] (3.000000,4.600000) -- +(0.0,-2.0pt);
	\draw[line width=1pt] (3.866025,5.400000) -- (3.866025,6.000000);
	\draw[line width=1pt,fill=white] (3.766025,6.100000) -- (3.866025,6.200000) -- (3.966025,6.100000) -- (3.866025,6.000000) -- cycle;
	\draw[line width=1pt] (3.866025,6.200000) -- (3.866025,6.800000);
	\draw[line width=1pt,fill=white] (3.866025,6.900000) circle (0.100000);
	\draw[line width=1pt] (3.866025,7.000000) -- (3.866025,7.600000);
	\draw[line width=1pt] (3.866025,7.600000) -- (3.866025,7.800000);
	\draw[line width=1pt] (3.866025,7.800000) -- (3.866025,8.400000);
	\draw[line width=1pt] (3.866025,8.400000) -- (3.866025,8.600000);
	\draw[line width=1pt] (3.866025,8.600000) -- (3.866025,9.200000);
	\draw[line width=1pt] (3.866025,9.200000) -- (3.866025,9.400000);
	\draw[line width=1pt] (3.866025,9.400000) -- (3.866025,10.000000);
	\draw[line width=1pt] (3.866025,10.000000) -- (3.866025,10.200000);
	\draw[line width=1pt] (3.866025,10.200000) -- (3.866025,10.800000);
	\draw[line width=1pt] (3.866025,10.800000) -- (3.866025,11.000000);
	\draw[line width=1pt] (3.866025,11.000000) -- (3.866025,11.600000);
	\draw[line width=1pt] (3.866025,11.600000) -- (3.866025,11.800000);
	\draw[line width=1pt] (3.866025,11.800000) -- (3.866025,12.400000);
	\draw[line width=1pt] (3.866025,12.400000) -- (3.866025,12.600000);
	\draw[line width=1pt] (3.866025,12.600000) -- (3.866025,13.200000);
	\draw[line width=1pt] (3.866025,13.200000) -- (3.866025,13.400000);
	\draw[line width=3pt,draw=white] (4.866025,0.500000) -- (4.866025,1.200000);
	\draw[line width=3pt,draw=white] (4.866025,1.200000) -- (4.866025,1.400000);
	\draw[line width=3pt,draw=white] (4.866025,1.400000) -- (4.866025,2.000000);
	\draw[line width=3pt,draw=white] (4.866025,2.000000) -- (4.866025,2.200000);
	\draw[line width=3pt,draw=white] (4.866025,2.200000) -- (4.866025,2.800000);
	\draw[line width=3pt,draw=white,fill=white] (4.866025,2.900000) circle (0.100000);
	\draw[line width=3pt,draw=white] (4.866025,2.900000) -- (4.000000,2.400000);
	\draw[line width=3pt,draw=white,fill=white] (4.000000,2.400000) circle (0.200000);
	\draw[line width=3pt,draw=white] (3.800000,2.400000) -- (4.200000,2.400000);
	\draw[line width=3pt,draw=white] (4.000000,2.200000) -- (4.000000,2.600000);
	\draw[line width=3pt,draw=white] (4.866025,3.000000) -- (4.866025,3.600000);
	\draw[line width=3pt,draw=white] (4.866025,3.600000) -- (4.866025,3.800000);
	\draw[line width=3pt,draw=white] (4.866025,3.800000) -- (4.866025,4.400000);
	\draw[line width=3pt,draw=white,fill=white] (4.866025,4.500000) circle (0.200000);
	\draw[line width=3pt,draw=white] (4.666025,4.500000) -- (5.066025,4.500000);
	\draw[line width=3pt,draw=white] (4.866025,4.300000) -- (4.866025,4.700000);
	\draw[line width=3pt,draw=white] (4.866025,4.500000) -- (3.866025,4.500000);
	\draw[line width=3pt,draw=white,fill=white] (3.866025,4.500000) circle (0.100000);
	\draw[line width=3pt,draw=white] (4.866025,4.600000) -- (4.866025,5.200000);
	\draw[line width=3pt,draw=white,fill=white] (4.866025,5.300000) circle (0.100000);
	\draw[line width=3pt,draw=white] (4.866025,5.300000) -- (5.732051,5.800000);
	\draw[line width=3pt,draw=white,fill=white] (5.732051,5.800000) circle (0.200000);
	\draw[line width=3pt,draw=white] (5.532051,5.800000) -- (5.932051,5.800000);
	\draw[line width=3pt,draw=white] (5.732051,5.600000) -- (5.732051,6.000000);
	\draw[line width=3pt,draw=white] (4.866025,5.400000) -- (4.866025,6.000000);
	\draw[line width=3pt,fill=white,draw=white] (4.766025,6.100000) -- (4.866025,6.200000) -- (4.966025,6.100000) -- (4.866025,6.000000) -- cycle;
	\draw[line width=3pt,draw=white] (4.866025,6.200000) -- (4.866025,6.800000);
	\draw[line width=3pt,fill=white,draw=white] (4.866025,6.900000) circle (0.100000);
	\draw[line width=3pt,draw=white] (4.866025,7.000000) -- (4.866025,7.600000);
	\draw[line width=3pt,draw=white] (4.866025,7.600000) -- (4.866025,7.800000);
	\draw[line width=3pt,draw=white] (4.866025,7.800000) -- (4.866025,8.400000);
	\draw[line width=3pt,draw=white] (4.866025,8.400000) -- (4.866025,8.600000);
	\draw[line width=3pt,draw=white] (4.866025,8.600000) -- (4.866025,9.200000);
	\draw[line width=3pt,draw=white] (4.866025,9.200000) -- (4.866025,9.400000);
	\draw[line width=3pt,draw=white] (4.866025,9.400000) -- (4.866025,10.000000);
	\draw[line width=3pt,draw=white] (4.866025,10.000000) -- (4.866025,10.200000);
	\draw[line width=3pt,draw=white] (4.866025,10.200000) -- (4.866025,10.800000);
	\draw[line width=3pt,draw=white] (4.866025,10.800000) -- (4.866025,11.000000);
	\draw[line width=3pt,draw=white] (4.866025,11.000000) -- (4.866025,11.600000);
	\draw[line width=3pt,draw=white] (4.866025,11.600000) -- (4.866025,11.800000);
	\draw[line width=3pt,draw=white] (4.866025,11.800000) -- (4.866025,12.400000);
	\draw[line width=3pt,draw=white] (4.866025,12.400000) -- (4.866025,12.600000);
	\draw[line width=3pt,draw=white] (4.866025,12.600000) -- (4.866025,13.200000);
	\draw[line width=3pt,draw=white] (4.866025,13.200000) -- (4.866025,13.400000);
	\draw[line width=1pt] (4.866025,0.500000) -- (4.866025,1.200000);
	\draw[line width=1pt] (4.866025,1.200000) -- (4.866025,1.400000);
	\draw[line width=1pt] (4.866025,1.400000) -- (4.866025,2.000000);
	\draw[line width=1pt] (4.866025,2.000000) -- (4.866025,2.200000);
	\draw[line width=1pt] (4.866025,2.200000) -- (4.866025,2.800000);
	\draw[line width=1pt] (4.866025,2.900000) -- (4.000000,2.400000);
	\draw[line width=1pt,fill=black] (4.866025,2.900000) circle (0.100000);
	\draw[line width=1pt,fill=white] (4.000000,2.400000) circle (0.200000);
	\draw[line width=1pt] (3.800000,2.400000) -- (4.200000,2.400000);
	\draw[line width=1pt] (4.000000,2.200000) -- (4.000000,2.600000);
	\draw[line width=1pt] (4.866025,3.000000) -- (4.866025,3.600000);
	\draw[line width=1pt] (4.866025,3.600000) -- (4.866025,3.800000);
	\draw[line width=1pt] (4.866025,3.800000) -- (4.866025,4.400000);
	\draw[line width=1pt] (4.866025,4.500000) -- (3.866025,4.500000);
	\draw[line width=1pt,fill=white] (4.866025,4.500000) circle (0.200000);
	\draw[line width=1pt] (4.666025,4.500000) -- (5.066025,4.500000);
	\draw[line width=1pt] (4.866025,4.300000) -- (4.866025,4.700000);
	\draw[line width=1pt,fill=black] (3.866025,4.500000) circle (0.100000);
	\draw[line width=1pt] (3.866025,4.600000) -- +(0.0,2.0pt);
	\draw[line width=1pt] (3.866025,4.400000) -- +(0.0,-2.0pt);
	\draw[line width=1pt] (4.866025,4.600000) -- (4.866025,5.200000);
	\draw[line width=1pt] (4.866025,5.300000) -- (5.732051,5.800000);
	\draw[line width=1pt,fill=black] (4.866025,5.300000) circle (0.100000);
	\draw[line width=1pt,fill=white] (5.732051,5.800000) circle (0.200000);
	\draw[line width=1pt] (5.532051,5.800000) -- (5.932051,5.800000);
	\draw[line width=1pt] (5.732051,5.600000) -- (5.732051,6.000000);
	\draw[line width=1pt] (4.866025,5.400000) -- (4.866025,6.000000);
	\draw[line width=1pt,fill=white] (4.766025,6.100000) -- (4.866025,6.200000) -- (4.966025,6.100000) -- (4.866025,6.000000) -- cycle;
	\draw[line width=1pt] (4.866025,6.200000) -- (4.866025,6.800000);
	\draw[line width=1pt,fill=white] (4.866025,6.900000) circle (0.100000);
	\draw[line width=1pt] (4.866025,7.000000) -- (4.866025,7.600000);
	\draw[line width=1pt] (4.866025,7.600000) -- (4.866025,7.800000);
	\draw[line width=1pt] (4.866025,7.800000) -- (4.866025,8.400000);
	\draw[line width=1pt] (4.866025,8.400000) -- (4.866025,8.600000);
	\draw[line width=1pt] (4.866025,8.600000) -- (4.866025,9.200000);
	\draw[line width=1pt] (4.866025,9.200000) -- (4.866025,9.400000);
	\draw[line width=1pt] (4.866025,9.400000) -- (4.866025,10.000000);
	\draw[line width=1pt] (4.866025,10.000000) -- (4.866025,10.200000);
	\draw[line width=1pt] (4.866025,10.200000) -- (4.866025,10.800000);
	\draw[line width=1pt] (4.866025,10.800000) -- (4.866025,11.000000);
	\draw[line width=1pt] (4.866025,11.000000) -- (4.866025,11.600000);
	\draw[line width=1pt] (4.866025,11.600000) -- (4.866025,11.800000);
	\draw[line width=1pt] (4.866025,11.800000) -- (4.866025,12.400000);
	\draw[line width=1pt] (4.866025,12.400000) -- (4.866025,12.600000);
	\draw[line width=1pt] (4.866025,12.600000) -- (4.866025,13.200000);
	\draw[line width=1pt] (4.866025,13.200000) -- (4.866025,13.400000);
	\draw[line width=3pt,draw=white] (3.000000,0.000000) -- (3.000000,0.700000);
	\draw[line width=3pt,draw=white] (3.000000,0.700000) -- (3.000000,0.900000);
	\draw[line width=3pt,draw=white] (3.000000,0.900000) -- (3.000000,1.500000);
	\draw[line width=3pt,draw=white] (3.000000,1.500000) -- (3.000000,1.700000);
	\draw[line width=3pt,draw=white] (3.000000,1.700000) -- (3.000000,2.300000);
	\draw[line width=3pt,draw=white] (3.000000,2.300000) -- (3.000000,2.500000);
	\draw[line width=3pt,draw=white] (3.000000,2.500000) -- (3.000000,3.100000);
	\draw[line width=3pt,draw=white,fill=white] (3.000000,3.200000) circle (0.100000);
	\draw[line width=3pt,draw=white] (3.000000,3.200000) -- (4.000000,3.200000);
	\draw[line width=3pt,draw=white,fill=white] (4.000000,3.200000) circle (0.200000);
	\draw[line width=3pt,draw=white] (3.800000,3.200000) -- (4.200000,3.200000);
	\draw[line width=3pt,draw=white] (4.000000,3.000000) -- (4.000000,3.400000);
	\draw[line width=3pt,draw=white] (3.000000,3.300000) -- (3.000000,3.900000);
	\draw[line width=3pt,draw=white,fill=white] (3.000000,4.000000) circle (0.100000);
	\draw[line width=3pt,draw=white] (3.000000,4.000000) -- (2.000000,4.000000);
	\draw[line width=3pt,draw=white,fill=white] (2.000000,4.000000) circle (0.200000);
	\draw[line width=3pt,draw=white] (1.800000,4.000000) -- (2.200000,4.000000);
	\draw[line width=3pt,draw=white] (2.000000,3.800000) -- (2.000000,4.200000);
	\draw[line width=3pt,draw=white] (3.000000,4.100000) -- (3.000000,4.700000);
	\draw[line width=3pt,draw=white,fill=white] (3.000000,4.800000) circle (0.200000);
	\draw[line width=3pt,draw=white] (2.800000,4.800000) -- (3.200000,4.800000);
	\draw[line width=3pt,draw=white] (3.000000,4.600000) -- (3.000000,5.000000);
	\draw[line width=3pt,draw=white] (3.000000,4.800000) -- (3.866025,5.300000);
	\draw[line width=3pt,draw=white,fill=white] (3.866025,5.300000) circle (0.100000);
	\draw[line width=3pt,draw=white] (3.000000,4.900000) -- (3.000000,5.500000);
	\draw[line width=3pt,fill=white,draw=white] (2.900000,5.600000) -- (3.000000,5.700000) -- (3.100000,5.600000) -- (3.000000,5.500000) -- cycle;
	\draw[line width=3pt,draw=white] (3.000000,5.700000) -- (3.000000,6.300000);
	\draw[line width=3pt,fill=white,draw=white] (3.000000,6.400000) circle (0.100000);
	\draw[line width=3pt,draw=white] (3.000000,6.500000) -- (3.000000,7.100000);
	\draw[line width=3pt,draw=white] (3.000000,7.100000) -- (3.000000,7.300000);
	\draw[line width=3pt,draw=white] (3.000000,7.300000) -- (3.000000,7.900000);
	\draw[line width=3pt,draw=white] (3.000000,7.900000) -- (3.000000,8.100000);
	\draw[line width=3pt,draw=white] (3.000000,8.100000) -- (3.000000,8.700000);
	\draw[line width=3pt,draw=white] (3.000000,8.700000) -- (3.000000,8.900000);
	\draw[line width=3pt,draw=white] (3.000000,8.900000) -- (3.000000,9.500000);
	\draw[line width=3pt,draw=white] (3.000000,9.500000) -- (3.000000,9.700000);
	\draw[line width=3pt,draw=white] (3.000000,9.700000) -- (3.000000,10.300000);
	\draw[line width=3pt,draw=white] (3.000000,10.300000) -- (3.000000,10.500000);
	\draw[line width=3pt,draw=white] (3.000000,10.500000) -- (3.000000,11.100000);
	\draw[line width=3pt,draw=white] (3.000000,11.100000) -- (3.000000,11.300000);
	\draw[line width=3pt,draw=white] (3.000000,11.300000) -- (3.000000,11.900000);
	\draw[line width=3pt,draw=white] (3.000000,11.900000) -- (3.000000,12.100000);
	\draw[line width=3pt,draw=white] (3.000000,12.100000) -- (3.000000,12.700000);
	\draw[line width=3pt,draw=white] (3.000000,12.700000) -- (3.000000,12.900000);
	\draw[line width=1pt] (3.000000,0.000000) -- (3.000000,0.700000);
	\draw[line width=1pt] (3.000000,0.700000) -- (3.000000,0.900000);
	\draw[line width=1pt] (3.000000,0.900000) -- (3.000000,1.500000);
	\draw[line width=1pt] (3.000000,1.500000) -- (3.000000,1.700000);
	\draw[line width=1pt] (3.000000,1.700000) -- (3.000000,2.300000);
	\draw[line width=1pt] (3.000000,2.300000) -- (3.000000,2.500000);
	\draw[line width=1pt] (3.000000,2.500000) -- (3.000000,3.100000);
	\draw[line width=1pt] (3.000000,3.200000) -- (4.000000,3.200000);
	\draw[line width=1pt,fill=black] (3.000000,3.200000) circle (0.100000);
	\draw[line width=1pt,fill=white] (4.000000,3.200000) circle (0.200000);
	\draw[line width=1pt] (3.800000,3.200000) -- (4.200000,3.200000);
	\draw[line width=1pt] (4.000000,3.000000) -- (4.000000,3.400000);
	\draw[line width=1pt] (3.000000,3.300000) -- (3.000000,3.900000);
	\draw[line width=1pt] (3.000000,4.000000) -- (2.000000,4.000000);
	\draw[line width=1pt,fill=black] (3.000000,4.000000) circle (0.100000);
	\draw[line width=1pt,fill=white] (2.000000,4.000000) circle (0.200000);
	\draw[line width=1pt] (1.800000,4.000000) -- (2.200000,4.000000);
	\draw[line width=1pt] (2.000000,3.800000) -- (2.000000,4.200000);
	\draw[line width=1pt] (3.000000,4.100000) -- (3.000000,4.700000);
	\draw[line width=1pt] (3.000000,4.800000) -- (3.866025,5.300000);
	\draw[line width=1pt,fill=white] (3.000000,4.800000) circle (0.200000);
	\draw[line width=1pt] (2.800000,4.800000) -- (3.200000,4.800000);
	\draw[line width=1pt] (3.000000,4.600000) -- (3.000000,5.000000);
	\draw[line width=1pt,fill=black] (3.866025,5.300000) circle (0.100000);
	\draw[line width=1pt] (3.866025,5.400000) -- +(0.0,2.0pt);
	\draw[line width=1pt] (3.866025,5.200000) -- +(0.0,-2.0pt);
	\draw[line width=1pt] (3.000000,4.900000) -- (3.000000,5.500000);
	\draw[line width=1pt,fill=white] (2.900000,5.600000) -- (3.000000,5.700000) -- (3.100000,5.600000) -- (3.000000,5.500000) -- cycle;
	\draw[line width=1pt] (3.000000,5.700000) -- (3.000000,6.300000);
	\draw[line width=1pt,fill=white] (3.000000,6.400000) circle (0.100000);
	\draw[line width=1pt] (3.000000,6.500000) -- (3.000000,7.100000);
	\draw[line width=1pt] (3.000000,7.100000) -- (3.000000,7.300000);
	\draw[line width=1pt] (3.000000,7.300000) -- (3.000000,7.900000);
	\draw[line width=1pt] (3.000000,7.900000) -- (3.000000,8.100000);
	\draw[line width=1pt] (3.000000,8.100000) -- (3.000000,8.700000);
	\draw[line width=1pt] (3.000000,8.700000) -- (3.000000,8.900000);
	\draw[line width=1pt] (3.000000,8.900000) -- (3.000000,9.500000);
	\draw[line width=1pt] (3.000000,9.500000) -- (3.000000,9.700000);
	\draw[line width=1pt] (3.000000,9.700000) -- (3.000000,10.300000);
	\draw[line width=1pt] (3.000000,10.300000) -- (3.000000,10.500000);
	\draw[line width=1pt] (3.000000,10.500000) -- (3.000000,11.100000);
	\draw[line width=1pt] (3.000000,11.100000) -- (3.000000,11.300000);
	\draw[line width=1pt] (3.000000,11.300000) -- (3.000000,11.900000);
	\draw[line width=1pt] (3.000000,11.900000) -- (3.000000,12.100000);
	\draw[line width=1pt] (3.000000,12.100000) -- (3.000000,12.700000);
	\draw[line width=1pt] (3.000000,12.700000) -- (3.000000,12.900000);
	\draw[line width=1pt,rounded corners,dashed] (1.366025,6.500000) -- (3.366025,6.500000) -- (3.366025,7.300000) -- (2.366025,7.300000) -- (2.366025,13.700000) -- (1.366025,13.700000) -- cycle;
	\draw[line width=1pt,rounded corners,dashed] (2.799038,7.150000) -- (1.933013,6.650000) -- (2.933013,6.650000) -- (2.066987,6.150000) -- (3.066987,6.150000) -- (3.933013,6.650000) -- (4.933013,6.650000) -- (5.799038,7.150000) -- (4.799038,7.150000) -- (5.665064,7.650000) -- (4.665064,7.650000) -- (3.799038,7.150000) -- cycle;
\end{tikzpicture}

%% file: pic_isotrimprimal2.tex
\begin{tikzpicture}[x=0.074272\linewidth,y=0.074272\linewidth]
	\draw[line width=1pt,fill=black] (0.000000,0.000000) circle (0.100000);
	\draw[line width=1pt] (0.779423,0.450000) -- (0.086603,0.050000);
	\draw[line width=1pt] (0.866025,0.500000) circle (0.100000);
	\draw[line width=1pt] (1.645448,0.950000) -- (0.952628,0.550000);
	\draw[line width=1pt,fill=black] (1.732051,1.000000) circle (0.100000);
	\draw[line width=1pt] (0.100000,0.000000) -- (0.900000,0.000000);
	\draw[line width=1pt] (1.000000,0.000000) circle (0.100000);
	\draw[line width=1pt,fill=black] (1.866025,0.500000) circle (0.100000);
	\draw[line width=1pt] (1.832051,1.000000) -- (2.632051,1.000000);
	\draw[line width=1pt] (2.732051,1.000000) circle (0.100000);
	\draw[line width=1pt] (1.100000,0.000000) -- (1.900000,0.000000);
	\draw[line width=1pt,fill=black] (2.000000,0.000000) circle (0.100000);
	\draw[line width=1pt] (2.779423,0.450000) -- (2.086603,0.050000);
	\draw[line width=1pt] (2.866025,0.500000) circle (0.100000);
	\draw[line width=1pt] (2.832051,1.000000) -- (3.632051,1.000000);
	\draw[line width=1pt] (3.645448,0.950000) -- (2.952628,0.550000);
	\draw[line width=1pt,fill=black] (3.732051,1.000000) circle (0.100000);
	\draw[line width=1pt] (2.100000,0.000000) -- (2.900000,0.000000);
	\draw[line width=1pt] (3.000000,0.000000) circle (0.100000);
	\draw[line width=1pt,fill=black] (3.866025,0.500000) circle (0.100000);
	\draw[line width=1pt] (3.832051,1.000000) -- (4.632051,1.000000);
	\draw[line width=1pt] (4.732051,1.000000) circle (0.100000);
	\draw[line width=1pt] (3.100000,0.000000) -- (3.900000,0.000000);
	\draw[line width=1pt,fill=black] (4.000000,0.000000) circle (0.100000);
	\draw[line width=1pt] (4.779423,0.450000) -- (4.086603,0.050000);
	\draw[line width=1pt] (4.866025,0.500000) circle (0.100000);
	\draw[line width=1pt] (4.832051,1.000000) -- (5.632051,1.000000);
	\draw[line width=1pt] (5.645448,0.950000) -- (4.952628,0.550000);
	\draw[line width=1pt,fill=black] (5.732051,1.000000) circle (0.100000);
	\draw[line width=3pt,draw=white] (1.866025,0.500000) -- (1.866025,1.200000);
	\draw[line width=3pt,draw=white] (1.866025,1.200000) -- (1.866025,1.400000);
	\draw[line width=3pt,draw=white] (1.866025,1.400000) -- (1.866025,2.000000);
	\draw[line width=3pt,fill=white,draw=white] (1.766025,2.000000) -- (1.766025,2.200000) -- (1.966025,2.200000) -- (1.966025,2.000000) -- cycle;
	\draw[line width=3pt,draw=white] (1.866025,2.200000) -- (1.866025,2.800000);
	\draw[line width=3pt,draw=white,fill=white] (1.866025,2.900000) circle (0.200000);
	\draw[line width=3pt,draw=white] (1.666025,2.900000) -- (2.066025,2.900000);
	\draw[line width=3pt,draw=white] (1.866025,2.700000) -- (1.866025,3.100000);
	\draw[line width=3pt,draw=white] (1.866025,2.900000) -- (2.732051,3.400000);
	\draw[line width=3pt,draw=white,fill=white] (2.732051,3.400000) circle (0.100000);
	\draw[line width=3pt,draw=white] (1.866025,3.000000) -- (1.866025,3.600000);
	\draw[line width=3pt,draw=white,fill=white] (1.866025,3.700000) circle (0.200000);
	\draw[line width=3pt,draw=white] (1.666025,3.700000) -- (2.066025,3.700000);
	\draw[line width=3pt,draw=white] (1.866025,3.500000) -- (1.866025,3.900000);
	\draw[line width=3pt,draw=white] (1.866025,3.700000) -- (0.866025,3.700000);
	\draw[line width=3pt,draw=white,fill=white] (0.866025,3.700000) circle (0.100000);
	\draw[line width=3pt,draw=white] (1.866025,3.800000) -- (1.866025,4.400000);
	\draw[line width=3pt,draw=white,fill=white] (1.866025,4.500000) circle (0.200000);
	\draw[line width=3pt,draw=white] (1.666025,4.500000) -- (2.066025,4.500000);
	\draw[line width=3pt,draw=white] (1.866025,4.300000) -- (1.866025,4.700000);
	\draw[line width=3pt,draw=white] (1.866025,4.500000) -- (2.866025,4.500000);
	\draw[line width=3pt,draw=white,fill=white] (2.866025,4.500000) circle (0.100000);
	\draw[line width=3pt,draw=white] (1.866025,4.600000) -- (1.866025,5.200000);
	\draw[line width=3pt,draw=white,fill=white] (1.866025,5.300000) circle (0.200000);
	\draw[line width=3pt,draw=white] (1.666025,5.300000) -- (2.066025,5.300000);
	\draw[line width=3pt,draw=white] (1.866025,5.100000) -- (1.866025,5.500000);
	\draw[line width=3pt,draw=white] (1.866025,5.300000) -- (1.000000,4.800000);
	\draw[line width=3pt,draw=white,fill=white] (1.000000,4.800000) circle (0.100000);
	\draw[line width=3pt,draw=white] (1.866025,5.400000) -- (1.866025,6.000000);
	\draw[line width=3pt,fill=white,draw=white] (1.866025,6.100000) circle (0.100000);
	\draw[line width=3pt,draw=white] (1.866025,6.200000) -- (1.866025,6.800000);
	\draw[line width=3pt,draw=white] (1.866025,6.800000) -- (1.866025,7.000000);
	\draw[line width=3pt,draw=white] (1.866025,7.000000) -- (1.866025,7.600000);
	\draw[line width=3pt,draw=white] (1.866025,7.600000) -- (1.866025,7.800000);
	\draw[line width=3pt,draw=white] (1.866025,7.800000) -- (1.866025,8.400000);
	\draw[line width=3pt,fill=white,draw=white] (1.766025,8.400000) -- (1.766025,8.600000) -- (1.966025,8.600000) -- (1.966025,8.400000) -- cycle;
	\draw[line width=3pt,draw=white] (1.866025,8.600000) -- (1.866025,9.200000);
	\draw[line width=3pt,draw=white,fill=white] (1.866025,9.300000) circle (0.200000);
	\draw[line width=3pt,draw=white] (1.666025,9.300000) -- (2.066025,9.300000);
	\draw[line width=3pt,draw=white] (1.866025,9.100000) -- (1.866025,9.500000);
	\draw[line width=3pt,draw=white] (1.866025,9.300000) -- (2.732051,9.800000);
	\draw[line width=3pt,draw=white,fill=white] (2.732051,9.800000) circle (0.100000);
	\draw[line width=3pt,draw=white] (1.866025,9.400000) -- (1.866025,10.000000);
	\draw[line width=3pt,draw=white,fill=white] (1.866025,10.100000) circle (0.200000);
	\draw[line width=3pt,draw=white] (1.666025,10.100000) -- (2.066025,10.100000);
	\draw[line width=3pt,draw=white] (1.866025,9.900000) -- (1.866025,10.300000);
	\draw[line width=3pt,draw=white] (1.866025,10.100000) -- (0.866025,10.100000);
	\draw[line width=3pt,draw=white,fill=white] (0.866025,10.100000) circle (0.100000);
	\draw[line width=3pt,draw=white] (1.866025,10.200000) -- (1.866025,10.800000);
	\draw[line width=3pt,draw=white,fill=white] (1.866025,10.900000) circle (0.200000);
	\draw[line width=3pt,draw=white] (1.666025,10.900000) -- (2.066025,10.900000);
	\draw[line width=3pt,draw=white] (1.866025,10.700000) -- (1.866025,11.100000);
	\draw[line width=3pt,draw=white] (1.866025,10.900000) -- (2.866025,10.900000);
	\draw[line width=3pt,draw=white,fill=white] (2.866025,10.900000) circle (0.100000);
	\draw[line width=3pt,draw=white] (1.866025,11.000000) -- (1.866025,11.600000);
	\draw[line width=3pt,draw=white,fill=white] (1.866025,11.700000) circle (0.200000);
	\draw[line width=3pt,draw=white] (1.666025,11.700000) -- (2.066025,11.700000);
	\draw[line width=3pt,draw=white] (1.866025,11.500000) -- (1.866025,11.900000);
	\draw[line width=3pt,draw=white] (1.866025,11.700000) -- (1.000000,11.200000);
	\draw[line width=3pt,draw=white,fill=white] (1.000000,11.200000) circle (0.100000);
	\draw[line width=3pt,draw=white] (1.866025,11.800000) -- (1.866025,12.400000);
	\draw[line width=3pt,fill=white,draw=white] (1.866025,12.500000) circle (0.100000);
	\draw[line width=3pt,draw=white] (1.866025,12.600000) -- (1.866025,13.200000);
	\draw[line width=3pt,draw=white] (1.866025,13.200000) -- (1.866025,13.400000);
	\draw[line width=1pt] (1.866025,0.500000) -- (1.866025,1.200000);
	\draw[line width=1pt] (1.866025,1.200000) -- (1.866025,1.400000);
	\draw[line width=1pt] (1.866025,1.400000) -- (1.866025,2.000000);
	\draw[line width=1pt,fill=white] (1.766025,2.000000) -- (1.766025,2.200000) -- (1.966025,2.200000) -- (1.966025,2.000000) -- cycle;
	\draw[line width=1pt] (1.866025,2.200000) -- (1.866025,2.800000);
	\draw[line width=1pt] (1.866025,2.900000) -- (2.732051,3.400000);
	\draw[line width=1pt,fill=white] (1.866025,2.900000) circle (0.200000);
	\draw[line width=1pt] (1.666025,2.900000) -- (2.066025,2.900000);
	\draw[line width=1pt] (1.866025,2.700000) -- (1.866025,3.100000);
	\draw[line width=1pt,fill=black] (2.732051,3.400000) circle (0.100000);
	\draw[line width=1pt] (1.866025,3.000000) -- (1.866025,3.600000);
	\draw[line width=1pt] (1.866025,3.700000) -- (0.866025,3.700000);
	\draw[line width=1pt,fill=white] (1.866025,3.700000) circle (0.200000);
	\draw[line width=1pt] (1.666025,3.700000) -- (2.066025,3.700000);
	\draw[line width=1pt] (1.866025,3.500000) -- (1.866025,3.900000);
	\draw[line width=1pt,fill=black] (0.866025,3.700000) circle (0.100000);
	\draw[line width=1pt] (1.866025,3.800000) -- (1.866025,4.400000);
	\draw[line width=1pt] (1.866025,4.500000) -- (2.866025,4.500000);
	\draw[line width=1pt,fill=white] (1.866025,4.500000) circle (0.200000);
	\draw[line width=1pt] (1.666025,4.500000) -- (2.066025,4.500000);
	\draw[line width=1pt] (1.866025,4.300000) -- (1.866025,4.700000);
	\draw[line width=1pt,fill=black] (2.866025,4.500000) circle (0.100000);
	\draw[line width=1pt] (2.866025,4.600000) -- +(0.0,2.0pt);
	\draw[line width=1pt] (2.866025,4.400000) -- +(0.0,-2.0pt);
	\draw[line width=1pt] (1.866025,4.600000) -- (1.866025,5.200000);
	\draw[line width=1pt] (1.866025,5.300000) -- (1.000000,4.800000);
	\draw[line width=1pt,fill=white] (1.866025,5.300000) circle (0.200000);
	\draw[line width=1pt] (1.666025,5.300000) -- (2.066025,5.300000);
	\draw[line width=1pt] (1.866025,5.100000) -- (1.866025,5.500000);
	\draw[line width=1pt,fill=black] (1.000000,4.800000) circle (0.100000);
	\draw[line width=1pt] (1.866025,5.400000) -- (1.866025,6.000000);
	\draw[line width=1pt,fill=white] (1.866025,6.100000) circle (0.100000);
	\draw[line width=1pt] (1.866025,6.200000) -- (1.866025,6.800000);
	\draw[line width=1pt] (1.866025,6.800000) -- (1.866025,7.000000);
	\draw[line width=1pt] (1.866025,7.000000) -- (1.866025,7.600000);
	\draw[line width=1pt] (1.866025,7.600000) -- (1.866025,7.800000);
	\draw[line width=1pt] (1.866025,7.800000) -- (1.866025,8.400000);
	\draw[line width=1pt,fill=white] (1.766025,8.400000) -- (1.766025,8.600000) -- (1.966025,8.600000) -- (1.966025,8.400000) -- cycle;
	\draw[line width=1pt] (1.866025,8.600000) -- (1.866025,9.200000);
	\draw[line width=1pt] (1.866025,9.300000) -- (2.732051,9.800000);
	\draw[line width=1pt,fill=white] (1.866025,9.300000) circle (0.200000);
	\draw[line width=1pt] (1.666025,9.300000) -- (2.066025,9.300000);
	\draw[line width=1pt] (1.866025,9.100000) -- (1.866025,9.500000);
	\draw[line width=1pt,fill=black] (2.732051,9.800000) circle (0.100000);
	\draw[line width=1pt] (1.866025,9.400000) -- (1.866025,10.000000);
	\draw[line width=1pt] (1.866025,10.100000) -- (0.866025,10.100000);
	\draw[line width=1pt,fill=white] (1.866025,10.100000) circle (0.200000);
	\draw[line width=1pt] (1.666025,10.100000) -- (2.066025,10.100000);
	\draw[line width=1pt] (1.866025,9.900000) -- (1.866025,10.300000);
	\draw[line width=1pt,fill=black] (0.866025,10.100000) circle (0.100000);
	\draw[line width=1pt] (1.866025,10.200000) -- (1.866025,10.800000);
	\draw[line width=1pt] (1.866025,10.900000) -- (2.866025,10.900000);
	\draw[line width=1pt,fill=white] (1.866025,10.900000) circle (0.200000);
	\draw[line width=1pt] (1.666025,10.900000) -- (2.066025,10.900000);
	\draw[line width=1pt] (1.866025,10.700000) -- (1.866025,11.100000);
	\draw[line width=1pt,fill=black] (2.866025,10.900000) circle (0.100000);
	\draw[line width=1pt] (2.866025,11.000000) -- +(0.0,2.0pt);
	\draw[line width=1pt] (2.866025,10.800000) -- +(0.0,-2.0pt);
	\draw[line width=1pt] (1.866025,11.000000) -- (1.866025,11.600000);
	\draw[line width=1pt] (1.866025,11.700000) -- (1.000000,11.200000);
	\draw[line width=1pt,fill=white] (1.866025,11.700000) circle (0.200000);
	\draw[line width=1pt] (1.666025,11.700000) -- (2.066025,11.700000);
	\draw[line width=1pt] (1.866025,11.500000) -- (1.866025,11.900000);
	\draw[line width=1pt,fill=black] (1.000000,11.200000) circle (0.100000);
	\draw[line width=1pt] (1.866025,11.800000) -- (1.866025,12.400000);
	\draw[line width=1pt,fill=white] (1.866025,12.500000) circle (0.100000);
	\draw[line width=1pt] (1.866025,12.600000) -- (1.866025,13.200000);
	\draw[line width=1pt] (1.866025,13.200000) -- (1.866025,13.400000);
	\draw[line width=3pt,draw=white] (2.866025,0.500000) -- (2.866025,1.200000);
	\draw[line width=3pt,draw=white] (2.866025,1.200000) -- (2.866025,1.400000);
	\draw[line width=3pt,draw=white] (2.866025,1.400000) -- (2.866025,2.000000);
	\draw[line width=3pt,draw=white] (2.866025,2.000000) -- (2.866025,2.200000);
	\draw[line width=3pt,draw=white] (2.866025,2.200000) -- (2.866025,2.800000);
	\draw[line width=3pt,draw=white,fill=white] (2.866025,2.900000) circle (0.200000);
	\draw[line width=3pt,draw=white] (2.666025,2.900000) -- (3.066025,2.900000);
	\draw[line width=3pt,draw=white] (2.866025,2.700000) -- (2.866025,3.100000);
	\draw[line width=3pt,draw=white] (2.866025,2.900000) -- (2.000000,2.400000);
	\draw[line width=3pt,draw=white,fill=white] (2.000000,2.400000) circle (0.100000);
	\draw[line width=3pt,draw=white] (2.866025,3.000000) -- (2.866025,3.600000);
	\draw[line width=3pt,draw=white,fill=white] (2.866025,3.700000) circle (0.100000);
	\draw[line width=3pt,draw=white] (2.866025,3.700000) -- (3.866025,3.700000);
	\draw[line width=3pt,draw=white,fill=white] (3.866025,3.700000) circle (0.200000);
	\draw[line width=3pt,draw=white] (3.666025,3.700000) -- (4.066025,3.700000);
	\draw[line width=3pt,draw=white] (3.866025,3.500000) -- (3.866025,3.900000);
	\draw[line width=3pt,draw=white] (2.866025,3.800000) -- (2.866025,4.400000);
	\draw[line width=3pt,draw=white,fill=white] (2.866025,4.500000) circle (0.100000);
	\draw[line width=3pt,draw=white] (2.866025,4.500000) -- (1.866025,4.500000);
	\draw[line width=3pt,draw=white,fill=white] (1.866025,4.500000) circle (0.200000);
	\draw[line width=3pt,draw=white] (1.666025,4.500000) -- (2.066025,4.500000);
	\draw[line width=3pt,draw=white] (1.866025,4.300000) -- (1.866025,4.700000);
	\draw[line width=3pt,draw=white] (2.866025,4.600000) -- (2.866025,5.200000);
	\draw[line width=3pt,draw=white,fill=white] (2.866025,5.300000) circle (0.200000);
	\draw[line width=3pt,draw=white] (2.666025,5.300000) -- (3.066025,5.300000);
	\draw[line width=3pt,draw=white] (2.866025,5.100000) -- (2.866025,5.500000);
	\draw[line width=3pt,draw=white] (2.866025,5.300000) -- (3.732051,5.800000);
	\draw[line width=3pt,draw=white,fill=white] (3.732051,5.800000) circle (0.100000);
	\draw[line width=3pt,draw=white] (2.866025,5.400000) -- (2.866025,6.000000);
	\draw[line width=3pt,draw=white] (2.866025,6.000000) -- (2.866025,6.200000);
	\draw[line width=3pt,draw=white] (2.866025,6.200000) -- (2.866025,6.800000);
	\draw[line width=3pt,draw=white] (2.866025,6.800000) -- (2.866025,7.000000);
	\draw[line width=3pt,draw=white] (2.866025,7.000000) -- (2.866025,7.600000);
	\draw[line width=3pt,draw=white] (2.866025,7.600000) -- (2.866025,7.800000);
	\draw[line width=3pt,draw=white] (2.866025,7.800000) -- (2.866025,8.400000);
	\draw[line width=3pt,draw=white] (2.866025,8.400000) -- (2.866025,8.600000);
	\draw[line width=3pt,draw=white] (2.866025,8.600000) -- (2.866025,9.200000);
	\draw[line width=3pt,draw=white,fill=white] (2.866025,9.300000) circle (0.200000);
	\draw[line width=3pt,draw=white] (2.666025,9.300000) -- (3.066025,9.300000);
	\draw[line width=3pt,draw=white] (2.866025,9.100000) -- (2.866025,9.500000);
	\draw[line width=3pt,draw=white] (2.866025,9.300000) -- (2.000000,8.800000);
	\draw[line width=3pt,draw=white,fill=white] (2.000000,8.800000) circle (0.100000);
	\draw[line width=3pt,draw=white] (2.866025,9.400000) -- (2.866025,10.000000);
	\draw[line width=3pt,draw=white] (2.866025,10.000000) -- (2.866025,10.200000);
	\draw[line width=3pt,draw=white] (2.866025,10.200000) -- (2.866025,10.800000);
	\draw[line width=3pt,draw=white,fill=white] (2.866025,10.900000) circle (0.100000);
	\draw[line width=3pt,draw=white] (2.866025,10.900000) -- (1.866025,10.900000);
	\draw[line width=3pt,draw=white,fill=white] (1.866025,10.900000) circle (0.200000);
	\draw[line width=3pt,draw=white] (1.666025,10.900000) -- (2.066025,10.900000);
	\draw[line width=3pt,draw=white] (1.866025,10.700000) -- (1.866025,11.100000);
	\draw[line width=3pt,draw=white] (2.866025,11.000000) -- (2.866025,11.600000);
	\draw[line width=3pt,draw=white,fill=white] (2.866025,11.700000) circle (0.200000);
	\draw[line width=3pt,draw=white] (2.666025,11.700000) -- (3.066025,11.700000);
	\draw[line width=3pt,draw=white] (2.866025,11.500000) -- (2.866025,11.900000);
	\draw[line width=3pt,draw=white] (2.866025,11.700000) -- (3.732051,12.200000);
	\draw[line width=3pt,draw=white,fill=white] (3.732051,12.200000) circle (0.100000);
	\draw[line width=3pt,draw=white] (2.866025,11.800000) -- (2.866025,12.400000);
	\draw[line width=3pt,draw=white] (2.866025,12.400000) -- (2.866025,12.600000);
	\draw[line width=3pt,draw=white] (2.866025,12.600000) -- (2.866025,13.200000);
	\draw[line width=3pt,draw=white] (2.866025,13.200000) -- (2.866025,13.400000);
	\draw[line width=1pt] (2.866025,0.500000) -- (2.866025,1.200000);
	\draw[line width=1pt] (2.866025,1.200000) -- (2.866025,1.400000);
	\draw[line width=1pt] (2.866025,1.400000) -- (2.866025,2.000000);
	\draw[line width=1pt] (2.866025,2.000000) -- (2.866025,2.200000);
	\draw[line width=1pt] (2.866025,2.200000) -- (2.866025,2.800000);
	\draw[line width=1pt] (2.866025,2.900000) -- (2.000000,2.400000);
	\draw[line width=1pt,fill=white] (2.866025,2.900000) circle (0.200000);
	\draw[line width=1pt] (2.666025,2.900000) -- (3.066025,2.900000);
	\draw[line width=1pt] (2.866025,2.700000) -- (2.866025,3.100000);
	\draw[line width=1pt,fill=black] (2.000000,2.400000) circle (0.100000);
	\draw[line width=1pt] (2.866025,3.000000) -- (2.866025,3.600000);
	\draw[line width=1pt] (2.866025,3.700000) -- (3.866025,3.700000);
	\draw[line width=1pt,fill=black] (2.866025,3.700000) circle (0.100000);
	\draw[line width=1pt,fill=white] (3.866025,3.700000) circle (0.200000);
	\draw[line width=1pt] (3.666025,3.700000) -- (4.066025,3.700000);
	\draw[line width=1pt] (3.866025,3.500000) -- (3.866025,3.900000);
	\draw[line width=1pt] (3.866025,3.900000) -- +(0.0,2.0pt);
	\draw[line width=1pt] (3.866025,3.500000) -- +(0.0,-2.0pt);
	\draw[line width=1pt] (2.866025,3.800000) -- (2.866025,4.400000);
	\draw[line width=1pt] (2.866025,4.500000) -- (1.866025,4.500000);
	\draw[line width=1pt,fill=black] (2.866025,4.500000) circle (0.100000);
	\draw[line width=1pt,fill=white] (1.866025,4.500000) circle (0.200000);
	\draw[line width=1pt] (1.666025,4.500000) -- (2.066025,4.500000);
	\draw[line width=1pt] (1.866025,4.300000) -- (1.866025,4.700000);
	\draw[line width=1pt] (1.866025,4.700000) -- +(0.0,2.0pt);
	\draw[line width=1pt] (1.866025,4.300000) -- +(0.0,-2.0pt);
	\draw[line width=1pt] (2.866025,4.600000) -- (2.866025,5.200000);
	\draw[line width=1pt] (2.866025,5.300000) -- (3.732051,5.800000);
	\draw[line width=1pt,fill=white] (2.866025,5.300000) circle (0.200000);
	\draw[line width=1pt] (2.666025,5.300000) -- (3.066025,5.300000);
	\draw[line width=1pt] (2.866025,5.100000) -- (2.866025,5.500000);
	\draw[line width=1pt,fill=black] (3.732051,5.800000) circle (0.100000);
	\draw[line width=1pt] (2.866025,5.400000) -- (2.866025,6.000000);
	\draw[line width=1pt] (2.866025,6.000000) -- (2.866025,6.200000);
	\draw[line width=1pt] (2.866025,6.200000) -- (2.866025,6.800000);
	\draw[line width=1pt] (2.866025,6.800000) -- (2.866025,7.000000);
	\draw[line width=1pt] (2.866025,7.000000) -- (2.866025,7.600000);
	\draw[line width=1pt] (2.866025,7.600000) -- (2.866025,7.800000);
	\draw[line width=1pt] (2.866025,7.800000) -- (2.866025,8.400000);
	\draw[line width=1pt] (2.866025,8.400000) -- (2.866025,8.600000);
	\draw[line width=1pt] (2.866025,8.600000) -- (2.866025,9.200000);
	\draw[line width=1pt] (2.866025,9.300000) -- (2.000000,8.800000);
	\draw[line width=1pt,fill=white] (2.866025,9.300000) circle (0.200000);
	\draw[line width=1pt] (2.666025,9.300000) -- (3.066025,9.300000);
	\draw[line width=1pt] (2.866025,9.100000) -- (2.866025,9.500000);
	\draw[line width=1pt,fill=black] (2.000000,8.800000) circle (0.100000);
	\draw[line width=1pt] (2.866025,9.400000) -- (2.866025,10.000000);
	\draw[line width=1pt] (2.866025,10.000000) -- (2.866025,10.200000);
	\draw[line width=1pt] (2.866025,10.200000) -- (2.866025,10.800000);
	\draw[line width=1pt] (2.866025,10.900000) -- (1.866025,10.900000);
	\draw[line width=1pt,fill=black] (2.866025,10.900000) circle (0.100000);
	\draw[line width=1pt,fill=white] (1.866025,10.900000) circle (0.200000);
	\draw[line width=1pt] (1.666025,10.900000) -- (2.066025,10.900000);
	\draw[line width=1pt] (1.866025,10.700000) -- (1.866025,11.100000);
	\draw[line width=1pt] (1.866025,11.100000) -- +(0.0,2.0pt);
	\draw[line width=1pt] (1.866025,10.700000) -- +(0.0,-2.0pt);
	\draw[line width=1pt] (2.866025,11.000000) -- (2.866025,11.600000);
	\draw[line width=1pt] (2.866025,11.700000) -- (3.732051,12.200000);
	\draw[line width=1pt,fill=white] (2.866025,11.700000) circle (0.200000);
	\draw[line width=1pt] (2.666025,11.700000) -- (3.066025,11.700000);
	\draw[line width=1pt] (2.866025,11.500000) -- (2.866025,11.900000);
	\draw[line width=1pt,fill=black] (3.732051,12.200000) circle (0.100000);
	\draw[line width=1pt] (2.866025,11.800000) -- (2.866025,12.400000);
	\draw[line width=1pt] (2.866025,12.400000) -- (2.866025,12.600000);
	\draw[line width=1pt] (2.866025,12.600000) -- (2.866025,13.200000);
	\draw[line width=1pt] (2.866025,13.200000) -- (2.866025,13.400000);
	\draw[line width=3pt,draw=white] (3.866025,0.500000) -- (3.866025,1.200000);
	\draw[line width=3pt,draw=white] (3.866025,1.200000) -- (3.866025,1.400000);
	\draw[line width=3pt,draw=white] (3.866025,1.400000) -- (3.866025,2.000000);
	\draw[line width=3pt,fill=white,draw=white] (3.766025,2.000000) -- (3.766025,2.200000) -- (3.966025,2.200000) -- (3.966025,2.000000) -- cycle;
	\draw[line width=3pt,draw=white] (3.866025,2.200000) -- (3.866025,2.800000);
	\draw[line width=3pt,draw=white,fill=white] (3.866025,2.900000) circle (0.200000);
	\draw[line width=3pt,draw=white] (3.666025,2.900000) -- (4.066025,2.900000);
	\draw[line width=3pt,draw=white] (3.866025,2.700000) -- (3.866025,3.100000);
	\draw[line width=3pt,draw=white] (3.866025,2.900000) -- (4.732051,3.400000);
	\draw[line width=3pt,draw=white,fill=white] (4.732051,3.400000) circle (0.100000);
	\draw[line width=3pt,draw=white] (3.866025,3.000000) -- (3.866025,3.600000);
	\draw[line width=3pt,draw=white,fill=white] (3.866025,3.700000) circle (0.200000);
	\draw[line width=3pt,draw=white] (3.666025,3.700000) -- (4.066025,3.700000);
	\draw[line width=3pt,draw=white] (3.866025,3.500000) -- (3.866025,3.900000);
	\draw[line width=3pt,draw=white] (3.866025,3.700000) -- (2.866025,3.700000);
	\draw[line width=3pt,draw=white,fill=white] (2.866025,3.700000) circle (0.100000);
	\draw[line width=3pt,draw=white] (3.866025,3.800000) -- (3.866025,4.400000);
	\draw[line width=3pt,draw=white,fill=white] (3.866025,4.500000) circle (0.200000);
	\draw[line width=3pt,draw=white] (3.666025,4.500000) -- (4.066025,4.500000);
	\draw[line width=3pt,draw=white] (3.866025,4.300000) -- (3.866025,4.700000);
	\draw[line width=3pt,draw=white] (3.866025,4.500000) -- (4.866025,4.500000);
	\draw[line width=3pt,draw=white,fill=white] (4.866025,4.500000) circle (0.100000);
	\draw[line width=3pt,draw=white] (3.866025,4.600000) -- (3.866025,5.200000);
	\draw[line width=3pt,draw=white,fill=white] (3.866025,5.300000) circle (0.200000);
	\draw[line width=3pt,draw=white] (3.666025,5.300000) -- (4.066025,5.300000);
	\draw[line width=3pt,draw=white] (3.866025,5.100000) -- (3.866025,5.500000);
	\draw[line width=3pt,draw=white] (3.866025,5.300000) -- (3.000000,4.800000);
	\draw[line width=3pt,draw=white,fill=white] (3.000000,4.800000) circle (0.100000);
	\draw[line width=3pt,draw=white] (3.866025,5.400000) -- (3.866025,6.000000);
	\draw[line width=3pt,fill=white,draw=white] (3.866025,6.100000) circle (0.100000);
	\draw[line width=3pt,draw=white] (3.866025,6.200000) -- (3.866025,6.800000);
	\draw[line width=3pt,draw=white] (3.866025,6.800000) -- (3.866025,7.000000);
	\draw[line width=3pt,draw=white] (3.866025,7.000000) -- (3.866025,7.600000);
	\draw[line width=3pt,draw=white] (3.866025,7.600000) -- (3.866025,7.800000);
	\draw[line width=3pt,draw=white] (3.866025,7.800000) -- (3.866025,8.400000);
	\draw[line width=3pt,draw=white] (3.866025,8.400000) -- (3.866025,8.600000);
	\draw[line width=3pt,draw=white] (3.866025,8.600000) -- (3.866025,9.200000);
	\draw[line width=3pt,draw=white] (3.866025,9.200000) -- (3.866025,9.400000);
	\draw[line width=3pt,draw=white] (3.866025,9.400000) -- (3.866025,10.000000);
	\draw[line width=3pt,draw=white] (3.866025,10.000000) -- (3.866025,10.200000);
	\draw[line width=3pt,draw=white] (3.866025,10.200000) -- (3.866025,10.800000);
	\draw[line width=3pt,draw=white] (3.866025,10.800000) -- (3.866025,11.000000);
	\draw[line width=3pt,draw=white] (3.866025,11.000000) -- (3.866025,11.600000);
	\draw[line width=3pt,draw=white] (3.866025,11.600000) -- (3.866025,11.800000);
	\draw[line width=3pt,draw=white] (3.866025,11.800000) -- (3.866025,12.400000);
	\draw[line width=3pt,draw=white] (3.866025,12.400000) -- (3.866025,12.600000);
	\draw[line width=3pt,draw=white] (3.866025,12.600000) -- (3.866025,13.200000);
	\draw[line width=3pt,draw=white] (3.866025,13.200000) -- (3.866025,13.400000);
	\draw[line width=1pt] (3.866025,0.500000) -- (3.866025,1.200000);
	\draw[line width=1pt] (3.866025,1.200000) -- (3.866025,1.400000);
	\draw[line width=1pt] (3.866025,1.400000) -- (3.866025,2.000000);
	\draw[line width=1pt,fill=white] (3.766025,2.000000) -- (3.766025,2.200000) -- (3.966025,2.200000) -- (3.966025,2.000000) -- cycle;
	\draw[line width=1pt] (3.866025,2.200000) -- (3.866025,2.800000);
	\draw[line width=1pt] (3.866025,2.900000) -- (4.732051,3.400000);
	\draw[line width=1pt,fill=white] (3.866025,2.900000) circle (0.200000);
	\draw[line width=1pt] (3.666025,2.900000) -- (4.066025,2.900000);
	\draw[line width=1pt] (3.866025,2.700000) -- (3.866025,3.100000);
	\draw[line width=1pt,fill=black] (4.732051,3.400000) circle (0.100000);
	\draw[line width=1pt] (3.866025,3.000000) -- (3.866025,3.600000);
	\draw[line width=1pt] (3.866025,3.700000) -- (2.866025,3.700000);
	\draw[line width=1pt,fill=white] (3.866025,3.700000) circle (0.200000);
	\draw[line width=1pt] (3.666025,3.700000) -- (4.066025,3.700000);
	\draw[line width=1pt] (3.866025,3.500000) -- (3.866025,3.900000);
	\draw[line width=1pt,fill=black] (2.866025,3.700000) circle (0.100000);
	\draw[line width=1pt] (2.866025,3.800000) -- +(0.0,2.0pt);
	\draw[line width=1pt] (2.866025,3.600000) -- +(0.0,-2.0pt);
	\draw[line width=1pt] (3.866025,3.800000) -- (3.866025,4.400000);
	\draw[line width=1pt] (3.866025,4.500000) -- (4.866025,4.500000);
	\draw[line width=1pt,fill=white] (3.866025,4.500000) circle (0.200000);
	\draw[line width=1pt] (3.666025,4.500000) -- (4.066025,4.500000);
	\draw[line width=1pt] (3.866025,4.300000) -- (3.866025,4.700000);
	\draw[line width=1pt,fill=black] (4.866025,4.500000) circle (0.100000);
	\draw[line width=1pt] (3.866025,4.600000) -- (3.866025,5.200000);
	\draw[line width=1pt] (3.866025,5.300000) -- (3.000000,4.800000);
	\draw[line width=1pt,fill=white] (3.866025,5.300000) circle (0.200000);
	\draw[line width=1pt] (3.666025,5.300000) -- (4.066025,5.300000);
	\draw[line width=1pt] (3.866025,5.100000) -- (3.866025,5.500000);
	\draw[line width=1pt,fill=black] (3.000000,4.800000) circle (0.100000);
	\draw[line width=1pt] (3.866025,5.400000) -- (3.866025,6.000000);
	\draw[line width=1pt,fill=white] (3.866025,6.100000) circle (0.100000);
	\draw[line width=1pt] (3.866025,6.200000) -- (3.866025,6.800000);
	\draw[line width=1pt] (3.866025,6.800000) -- (3.866025,7.000000);
	\draw[line width=1pt] (3.866025,7.000000) -- (3.866025,7.600000);
	\draw[line width=1pt] (3.866025,7.600000) -- (3.866025,7.800000);
	\draw[line width=1pt] (3.866025,7.800000) -- (3.866025,8.400000);
	\draw[line width=1pt] (3.866025,8.400000) -- (3.866025,8.600000);
	\draw[line width=1pt] (3.866025,8.600000) -- (3.866025,9.200000);
	\draw[line width=1pt] (3.866025,9.200000) -- (3.866025,9.400000);
	\draw[line width=1pt] (3.866025,9.400000) -- (3.866025,10.000000);
	\draw[line width=1pt] (3.866025,10.000000) -- (3.866025,10.200000);
	\draw[line width=1pt] (3.866025,10.200000) -- (3.866025,10.800000);
	\draw[line width=1pt] (3.866025,10.800000) -- (3.866025,11.000000);
	\draw[line width=1pt] (3.866025,11.000000) -- (3.866025,11.600000);
	\draw[line width=1pt] (3.866025,11.600000) -- (3.866025,11.800000);
	\draw[line width=1pt] (3.866025,11.800000) -- (3.866025,12.400000);
	\draw[line width=1pt] (3.866025,12.400000) -- (3.866025,12.600000);
	\draw[line width=1pt] (3.866025,12.600000) -- (3.866025,13.200000);
	\draw[line width=1pt] (3.866025,13.200000) -- (3.866025,13.400000);
	\draw[line width=1pt,rounded corners,dashed] (2.366025,5.700000) -- (2.366025,12.900000) -- (1.366025,12.900000) -- (1.366025,5.700000) -- cycle;
	\draw[line width=1pt,dotted] (3.866025,6.100000) circle (0.400000);
\end{tikzpicture}

%% file: pic_trimX.tex
\begin{tikzpicture}[x=0.064286\linewidth,y=0.064286\linewidth]
	\fill[gray!25] (2.000000,8.000000) -- (2.900000,8.000000) -- (2.000000,7.100000) -- cycle;
	\fill[gray!80] (3.000000,7.000000) -- (2.100000,7.000000) -- (3.000000,7.900000) -- cycle;
	\fill[gray!80] (3.000000,7.000000) -- (3.000000,6.100000) -- (2.100000,7.000000) -- cycle;
	\fill[gray!25] (2.000000,6.000000) -- (2.000000,6.900000) -- (2.900000,6.000000) -- cycle;
	\fill[gray!25] (2.000000,4.000000) -- (2.900000,4.000000) -- (2.000000,3.100000) -- cycle;
	\fill[gray!25] (4.000000,4.000000) -- (4.900000,4.000000) -- (4.000000,3.100000) -- cycle;
	\fill[gray!25] (4.000000,4.000000) -- (4.000000,3.100000) -- (3.100000,4.000000) -- cycle;
	\fill[gray!25] (6.000000,4.000000) -- (6.000000,3.100000) -- (5.100000,4.000000) -- cycle;
	\fill[gray!80] (3.000000,3.000000) -- (2.100000,3.000000) -- (3.000000,3.900000) -- cycle;
	\fill[gray!80] (3.000000,3.000000) -- (3.000000,3.900000) -- (3.900000,3.000000) -- cycle;
	\fill[gray!80] (3.000000,3.000000) -- (3.900000,3.000000) -- (3.000000,2.100000) -- cycle;
	\fill[gray!80] (3.000000,3.000000) -- (3.000000,2.100000) -- (2.100000,3.000000) -- cycle;
	\fill[gray!80] (5.000000,3.000000) -- (4.100000,3.000000) -- (5.000000,3.900000) -- cycle;
	\fill[gray!80] (5.000000,3.000000) -- (5.000000,3.900000) -- (5.900000,3.000000) -- cycle;
	\fill[gray!80] (5.000000,3.000000) -- (5.900000,3.000000) -- (5.000000,2.100000) -- cycle;
	\fill[gray!80] (5.000000,3.000000) -- (5.000000,2.100000) -- (4.100000,3.000000) -- cycle;
	\fill[gray!25] (2.000000,2.000000) -- (2.000000,2.900000) -- (2.900000,2.000000) -- cycle;
	\fill[gray!25] (4.000000,2.000000) -- (3.100000,2.000000) -- (4.000000,2.900000) -- cycle;
	\fill[gray!25] (4.000000,2.000000) -- (4.000000,2.900000) -- (4.900000,2.000000) -- cycle;
	\fill[gray!25] (6.000000,2.000000) -- (5.100000,2.000000) -- (6.000000,2.900000) -- cycle;
	\draw[line width=1pt,fill=black] (2.000000,8.000000) circle (0.100000);
	\draw[line width=1pt] (3.000000,8.000000) circle (0.100000);
	\draw[line width=1pt] (2.000000,7.000000) circle (0.100000);
	\draw[line width=1pt,fill=black] (3.000000,7.000000) circle (0.100000);
	\draw[line width=1pt,fill=black] (2.000000,6.000000) circle (0.100000);
	\draw[line width=1pt] (3.000000,6.000000) circle (0.100000);
	\draw[line width=1pt,fill=black] (2.000000,4.000000) circle (0.100000);
	\draw[line width=1pt] (3.000000,4.000000) circle (0.100000);
	\draw[line width=1pt,fill=black] (4.000000,4.000000) circle (0.100000);
	\draw (5.000000,4.000000) node [above right] {$M_X$};
	\draw[line width=1pt] (5.000000,4.000000) circle (0.100000);
	\draw[line width=1pt,fill=black] (6.000000,4.000000) circle (0.100000);
	\draw[line width=1pt] (2.000000,3.000000) circle (0.100000);
	\draw[line width=1pt,fill=black] (3.000000,3.000000) circle (0.100000);
	\draw (4.000000,3.000000) node [above right] {$M_X$};
	\draw[line width=1pt] (4.000000,3.000000) circle (0.100000);
	\draw[line width=1pt,fill=black] (5.000000,3.000000) circle (0.100000);
	\draw (6.000000,3.000000) node [above right] {$M_X$};
	\draw[line width=1pt] (6.000000,3.000000) circle (0.100000);
	\draw[line width=1pt,fill=black] (2.000000,2.000000) circle (0.100000);
	\draw[line width=1pt] (3.000000,2.000000) circle (0.100000);
	\draw[line width=1pt,fill=black] (4.000000,2.000000) circle (0.100000);
	\draw (5.000000,2.000000) node [above right] {$M_X$};
	\draw[line width=1pt] (5.000000,2.000000) circle (0.100000);
	\draw[line width=1pt,fill=black] (6.000000,2.000000) circle (0.100000);
\end{tikzpicture}

%% file: pic_trimX2.tex
\begin{tikzpicture}[x=0.064000\linewidth,y=0.064000\linewidth]
	\fill[gray!80] (1.000000,7.000000) -- (1.900000,7.000000) -- (1.000000,6.100000) -- cycle;
	\fill[gray!80] (3.000000,7.000000) -- (3.000000,6.100000) -- (2.100000,7.000000) -- cycle;
	\fill[gray!25] (2.000000,6.000000) -- (1.100000,6.000000) -- (2.000000,6.900000) -- cycle;
	\fill[gray!25] (2.000000,6.000000) -- (2.000000,6.900000) -- (2.900000,6.000000) -- cycle;
	\fill[gray!25] (2.000000,6.000000) -- (2.900000,6.000000) -- (2.000000,5.100000) -- cycle;
	\fill[gray!25] (2.000000,6.000000) -- (2.000000,5.100000) -- (1.100000,6.000000) -- cycle;
	\fill[gray!80] (1.000000,5.000000) -- (1.000000,5.900000) -- (1.900000,5.000000) -- cycle;
	\fill[gray!80] (3.000000,5.000000) -- (2.100000,5.000000) -- (3.000000,5.900000) -- cycle;
	\fill[gray!80] (1.000000,3.000000) -- (1.900000,3.000000) -- (1.000000,2.100000) -- cycle;
	\fill[gray!80] (3.000000,3.000000) -- (3.900000,3.000000) -- (3.000000,2.100000) -- cycle;
	\fill[gray!80] (3.000000,3.000000) -- (3.000000,2.100000) -- (2.100000,3.000000) -- cycle;
	\fill[gray!80] (5.000000,3.000000) -- (5.000000,2.100000) -- (4.100000,3.000000) -- cycle;
	\fill[gray!25] (2.000000,2.000000) -- (1.100000,2.000000) -- (2.000000,2.900000) -- cycle;
	\fill[gray!25] (2.000000,2.000000) -- (2.000000,2.900000) -- (2.900000,2.000000) -- cycle;
	\fill[gray!25] (2.000000,2.000000) -- (2.900000,2.000000) -- (2.000000,1.100000) -- cycle;
	\fill[gray!25] (2.000000,2.000000) -- (2.000000,1.100000) -- (1.100000,2.000000) -- cycle;
	\fill[gray!25] (4.000000,2.000000) -- (3.100000,2.000000) -- (4.000000,2.900000) -- cycle;
	\fill[gray!25] (4.000000,2.000000) -- (4.000000,2.900000) -- (4.900000,2.000000) -- cycle;
	\fill[gray!25] (4.000000,2.000000) -- (4.900000,2.000000) -- (4.000000,1.100000) -- cycle;
	\fill[gray!25] (4.000000,2.000000) -- (4.000000,1.100000) -- (3.100000,2.000000) -- cycle;
	\fill[gray!80] (1.000000,1.000000) -- (1.000000,1.900000) -- (1.900000,1.000000) -- cycle;
	\fill[gray!80] (3.000000,1.000000) -- (2.100000,1.000000) -- (3.000000,1.900000) -- cycle;
	\fill[gray!80] (3.000000,1.000000) -- (3.000000,1.900000) -- (3.900000,1.000000) -- cycle;
	\fill[gray!80] (5.000000,1.000000) -- (4.100000,1.000000) -- (5.000000,1.900000) -- cycle;
	\draw[line width=1pt,fill=black] (1.000000,7.000000) circle (0.100000);
	\draw[line width=1pt] (2.000000,7.000000) circle (0.100000);
	\draw[line width=1pt,fill=black] (3.000000,7.000000) circle (0.100000);
	\draw[line width=1pt] (1.000000,6.000000) circle (0.100000);
	\draw[line width=1pt,fill=black] (2.000000,6.000000) circle (0.100000);
	\draw[line width=1pt] (3.000000,6.000000) circle (0.100000);
	\draw[line width=1pt,fill=black] (1.000000,5.000000) circle (0.100000);
	\draw[line width=1pt] (2.000000,5.000000) circle (0.100000);
	\draw[line width=1pt,fill=black] (3.000000,5.000000) circle (0.100000);
	\draw[line width=1pt,fill=black] (1.000000,3.000000) circle (0.100000);
	\draw[line width=1pt] (2.000000,3.000000) circle (0.100000);
	\draw[line width=1pt,fill=black] (3.000000,3.000000) circle (0.100000);
	\draw (4.000000,3.000000) node [above right] {$M_X$};
	\draw[line width=1pt] (4.000000,3.000000) circle (0.100000);
	\draw[line width=1pt,fill=black] (5.000000,3.000000) circle (0.100000);
	\draw[line width=1pt] (1.000000,2.000000) circle (0.100000);
	\draw[line width=1pt,fill=black] (2.000000,2.000000) circle (0.100000);
	\draw[line width=1pt] (3.000000,2.000000) circle (0.100000);
	\draw[line width=1pt,fill=black] (4.000000,2.000000) circle (0.100000);
	\draw (5.000000,2.000000) node [above right] {$M_X$};
	\draw[line width=1pt] (5.000000,2.000000) circle (0.100000);
	\draw[line width=1pt,fill=black] (1.000000,1.000000) circle (0.100000);
	\draw[line width=1pt] (2.000000,1.000000) circle (0.100000);
	\draw[line width=1pt,fill=black] (3.000000,1.000000) circle (0.100000);
	\draw (4.000000,1.000000) node [above right] {$M_X$};
	\draw[line width=1pt] (4.000000,1.000000) circle (0.100000);
	\draw[line width=1pt,fill=black] (5.000000,1.000000) circle (0.100000);
\end{tikzpicture}

%% file: pic_swap.tex
\begin{tikzpicture}[x=0.090000\linewidth,y=0.090000\linewidth]
	\fill[gray!80] (3.000000,17.000000) -- (3.900000,17.000000) -- (3.000000,16.100000) -- cycle;
	\fill[gray!80] (5.000000,17.000000) -- (5.000000,16.100000) -- (4.100000,17.000000) -- cycle;
	\fill[gray!25] (4.000000,16.000000) -- (3.100000,16.000000) -- (4.000000,16.900000) -- cycle;
	\fill[gray!25] (4.000000,16.000000) -- (4.000000,16.900000) -- (4.900000,16.000000) -- cycle;
	\fill[gray!25] (4.000000,16.000000) -- (4.900000,16.000000) -- (4.000000,15.100000) -- cycle;
	\fill[gray!25] (4.000000,16.000000) -- (4.000000,15.100000) -- (3.100000,16.000000) -- cycle;
	\fill[gray!80] (3.000000,15.000000) -- (3.000000,15.900000) -- (3.900000,15.000000) -- cycle;
	\fill[gray!80] (3.000000,15.000000) -- (3.900000,15.000000) -- (3.000000,14.100000) -- cycle;
	\fill[gray!80] (5.000000,15.000000) -- (4.100000,15.000000) -- (5.000000,15.900000) -- cycle;
	\fill[gray!80] (5.000000,15.000000) -- (5.000000,14.100000) -- (4.100000,15.000000) -- cycle;
	\fill[gray!25] (4.000000,14.000000) -- (3.100000,14.000000) -- (4.000000,14.900000) -- cycle;
	\fill[gray!25] (4.000000,14.000000) -- (4.000000,14.900000) -- (4.900000,14.000000) -- cycle;
	\fill[gray!80] (5.000000,5.000000) -- (5.000000,4.100000) -- (4.100000,5.000000) -- cycle;
	\fill[gray!25] (4.000000,4.000000) -- (4.000000,4.900000) -- (4.900000,4.000000) -- cycle;
	\fill[gray!25] (4.000000,4.000000) -- (4.900000,4.000000) -- (4.000000,3.100000) -- cycle;
	\fill[gray!80] (5.000000,3.000000) -- (4.100000,3.000000) -- (5.000000,3.900000) -- cycle;
	\fill[gray!80] (5.000000,3.000000) -- (5.000000,2.100000) -- (4.100000,3.000000) -- cycle;
	\fill[gray!25] (4.000000,2.000000) -- (4.000000,2.900000) -- (4.900000,2.000000) -- cycle;
	\draw[line width=1pt,fill=black] (3.000000,17.000000) circle (0.100000);
	\draw[line width=1pt] (4.000000,17.000000) circle (0.100000);
	\draw[line width=1pt,fill=black] (5.000000,17.000000) circle (0.100000);
	\draw[line width=1pt] (3.000000,16.000000) circle (0.100000);
	\draw[line width=1pt,fill=black] (4.000000,16.000000) circle (0.100000);
	\draw[line width=1pt] (5.000000,16.000000) circle (0.100000);
	\draw[line width=1pt,fill=black] (3.000000,15.000000) circle (0.100000);
	\draw[line width=1pt] (4.000000,15.000000) circle (0.100000);
	\draw[line width=1pt,fill=black] (5.000000,15.000000) circle (0.100000);
	\draw[line width=1pt] (3.000000,14.000000) circle (0.100000);
	\draw[line width=1pt,fill=black] (4.000000,14.000000) circle (0.100000);
	\draw[line width=1pt] (5.000000,14.000000) circle (0.100000);
	\draw[line width=1pt,fill=black] (3.000000,11.000000) circle (0.100000);
	\draw[line width=1pt] (4.000000,11.000000) circle (0.100000);
	\draw[line width=1pt,fill=black] (5.000000,11.000000) circle (0.100000);
	\draw[line width=1pt] (3.000000,10.000000) circle (0.100000);
	\draw[line width=1pt,fill=black] (4.000000,10.000000) circle (0.100000);
	\draw[line width=1pt,->] (4.858579,10.141421) -- (4.141421,10.858579);
	\draw[line width=1pt] (5.000000,10.000000) circle (0.100000);
	\draw[line width=1pt,fill=black] (3.000000,9.000000) circle (0.100000);
	\draw[line width=1pt,->] (3.858579,9.141421) -- (3.141421,9.858579);
	\draw[line width=1pt] (4.000000,9.000000) circle (0.100000);
	\draw[line width=1pt,fill=black] (5.000000,9.000000) circle (0.100000);
	\draw[line width=1pt] (3.000000,8.000000) circle (0.100000);
	\draw[line width=1pt,fill=black] (4.000000,8.000000) circle (0.100000);
	\draw[line width=1pt,->] (4.858579,8.141421) -- (4.141421,8.858579);
	\draw[line width=1pt] (5.000000,8.000000) circle (0.100000);
	\draw (4.000000,5.000000) node [above right] {$H$};
	\draw[line width=1pt] (4.000000,5.000000) circle (0.100000);
	\draw[line width=1pt,fill=black] (5.000000,5.000000) circle (0.100000);
	\draw[line width=1pt,fill=black] (4.000000,4.000000) circle (0.100000);
	\draw (5.000000,4.000000) node [above right] {$H$};
	\draw[line width=1pt] (5.000000,4.000000) circle (0.100000);
	\draw (4.000000,3.000000) node [above right] {$H$};
	\draw[line width=1pt] (4.000000,3.000000) circle (0.100000);
	\draw[line width=1pt,fill=black] (5.000000,3.000000) circle (0.100000);
	\draw[line width=1pt,fill=black] (4.000000,2.000000) circle (0.100000);
	\draw (5.000000,2.000000) node [above right] {$H$};
	\draw[line width=1pt] (5.000000,2.000000) circle (0.100000);
\end{tikzpicture}

%% file: pic_isoswap.tex
\begin{tikzpicture}[x=0.080385\linewidth,y=0.080385\linewidth]
	\draw[line width=1pt] (0.000000,0.000000) circle (0.100000);
	\draw[line width=1pt] (0.779423,0.450000) -- (0.086603,0.050000);
	\draw[line width=1pt,fill=black] (0.866025,0.500000) circle (0.100000);
	\draw[line width=1pt] (1.645448,0.950000) -- (0.952628,0.550000);
	\draw[line width=1pt] (1.732051,1.000000) circle (0.100000);
	\draw[line width=1pt] (2.511474,1.450000) -- (1.818653,1.050000);
	\draw[line width=1pt,fill=black] (2.598076,1.500000) circle (0.100000);
	\draw[line width=1pt,fill=black] (1.000000,0.000000) circle (0.100000);
	\draw[line width=1pt] (0.966025,0.500000) -- (1.766025,0.500000);
	\draw[line width=1pt] (1.866025,0.500000) circle (0.100000);
	\draw[line width=1pt,fill=black] (2.732051,1.000000) circle (0.100000);
	\draw[line width=1pt] (2.698076,1.500000) -- (3.498076,1.500000);
	\draw[line width=1pt] (3.598076,1.500000) circle (0.100000);
	\draw[line width=1pt] (2.000000,0.000000) circle (0.100000);
	\draw[line width=1pt] (1.966025,0.500000) -- (2.766025,0.500000);
	\draw[line width=1pt] (2.779423,0.450000) -- (2.086603,0.050000);
	\draw[line width=1pt,fill=black] (2.866025,0.500000) circle (0.100000);
	\draw[line width=1pt] (3.645448,0.950000) -- (2.952628,0.550000);
	\draw[line width=1pt] (3.732051,1.000000) circle (0.100000);
	\draw[line width=1pt] (3.698076,1.500000) -- (4.498076,1.500000);
	\draw[line width=1pt] (4.511474,1.450000) -- (3.818653,1.050000);
	\draw[line width=1pt,fill=black] (4.598076,1.500000) circle (0.100000);
	\draw[line width=3pt,draw=white] (1.732051,1.000000) -- (1.732051,1.700000);
	\draw[line width=3pt,draw=white] (1.732051,1.700000) -- (1.732051,1.900000);
	\draw[line width=3pt,draw=white] (1.732051,1.900000) -- (1.732051,2.500000);
	\draw[line width=3pt,draw=white] (1.732051,2.500000) -- (1.732051,2.700000);
	\draw[line width=3pt,draw=white] (1.732051,2.700000) -- (1.732051,3.300000);
	\draw[line width=3pt,draw=white] (1.732051,3.300000) -- (1.732051,3.500000);
	\draw[line width=3pt,draw=white] (1.732051,3.500000) -- (1.732051,4.100000);
	\draw[line width=3pt,draw=white] (1.732051,4.100000) -- (1.732051,4.300000);
	\draw[line width=3pt,draw=white] (1.732051,4.300000) -- (1.732051,4.900000);
	\draw[line width=3pt,draw=white] (1.732051,4.900000) -- (1.732051,5.100000);
	\draw[line width=3pt,draw=white] (1.732051,5.100000) -- (1.732051,5.700000);
	\draw[line width=3pt,draw=white] (1.732051,5.700000) -- (1.732051,5.900000);
	\draw[line width=3pt,draw=white] (1.732051,5.900000) -- (1.732051,6.500000);
	\draw[line width=3pt,draw=white] (1.732051,6.500000) -- (1.732051,6.700000);
	\draw[line width=3pt,draw=white] (1.732051,6.700000) -- (1.732051,7.300000);
	\draw[line width=3pt,draw=white] (1.732051,7.300000) -- (1.732051,7.500000);
	\draw[line width=3pt,draw=white] (1.732051,7.500000) -- (1.732051,8.100000);
	\draw[line width=3pt,draw=white] (1.732051,8.100000) -- (1.732051,8.300000);
	\draw[line width=3pt,draw=white] (1.732051,8.300000) -- (1.732051,8.900000);
	\draw[line width=3pt,draw=white] (1.732051,8.800000) -- (1.732051,9.200000);
	\draw[line width=3pt,draw=white] (1.532051,8.800000) -- (1.932051,9.200000);
	\draw[line width=3pt,draw=white] (1.532051,9.200000) -- (1.932051,8.800000);
	\draw[line width=3pt,draw=white] (1.732051,9.000000) -- (2.732051,9.000000);
	\draw[line width=3pt,draw=white] (2.532051,8.800000) -- (2.932051,9.200000);
	\draw[line width=3pt,draw=white] (2.532051,9.200000) -- (2.932051,8.800000);
	\draw[line width=3pt,draw=white] (1.732051,9.100000) -- (1.732051,9.700000);
	\draw[line width=3pt,draw=white] (1.732051,9.700000) -- (1.732051,9.900000);
	\draw[line width=3pt,draw=white] (1.732051,9.900000) -- (1.732051,10.500000);
	\draw[line width=3pt,draw=white] (1.732051,10.500000) -- (1.732051,10.700000);
	\draw[line width=3pt,draw=white] (1.732051,10.700000) -- (1.732051,11.300000);
	\draw[line width=3pt,draw=white] (1.732051,11.300000) -- (1.732051,11.500000);
	\draw[line width=3pt,draw=white] (1.732051,11.500000) -- (1.732051,12.100000);
	\draw[line width=3pt,draw=white] (1.732051,12.100000) -- (1.732051,12.300000);
	\draw[line width=3pt,draw=white] (1.732051,12.300000) -- (1.732051,12.900000);
	\draw[line width=3pt,draw=white] (1.732051,12.900000) -- (1.732051,13.100000);
	\draw[line width=3pt,draw=white] (1.732051,13.100000) -- (1.732051,13.700000);
	\draw[line width=3pt,draw=white] (1.732051,13.700000) -- (1.732051,13.900000);
	\draw[line width=3pt,draw=white] (1.732051,13.900000) -- (1.732051,14.500000);
	\draw[line width=3pt,draw=white] (1.732051,14.500000) -- (1.732051,14.700000);
	\draw[line width=3pt,draw=white] (1.732051,14.700000) -- (1.732051,15.300000);
	\draw[line width=3pt,draw=white] (1.732051,15.300000) -- (1.732051,15.500000);
	\draw[line width=3pt,draw=white] (1.732051,15.500000) -- (1.732051,16.100000);
	\draw[line width=3pt,draw=white,fill=white] (1.732051,16.200000) circle (0.200000);
	\draw[line width=3pt,draw=white] (1.532051,16.200000) -- (1.932051,16.200000);
	\draw[line width=3pt,draw=white] (1.732051,16.000000) -- (1.732051,16.400000);
	\draw[line width=3pt,draw=white] (1.732051,16.200000) -- (0.866025,15.700000);
	\draw[line width=3pt,draw=white,fill=white] (0.866025,15.700000) circle (0.100000);
	\draw[line width=3pt,draw=white] (1.732051,16.300000) -- (1.732051,16.900000);
	\draw[line width=3pt,draw=white,fill=white] (1.732051,17.000000) circle (0.100000);
	\draw[line width=3pt,draw=white] (1.732051,17.000000) -- (2.732051,17.000000);
	\draw[line width=3pt,draw=white,fill=white] (2.732051,17.000000) circle (0.200000);
	\draw[line width=3pt,draw=white] (2.532051,17.000000) -- (2.932051,17.000000);
	\draw[line width=3pt,draw=white] (2.732051,16.800000) -- (2.732051,17.200000);
	\draw[line width=3pt,draw=white] (1.732051,17.100000) -- (1.732051,17.700000);
	\draw[line width=3pt,draw=white] (1.732051,17.700000) -- (1.732051,17.900000);
	\draw[line width=3pt,draw=white] (1.732051,17.900000) -- (1.732051,18.500000);
	\draw[line width=3pt,draw=white,fill=white] (1.732051,18.600000) circle (0.200000);
	\draw[line width=3pt,draw=white] (1.532051,18.600000) -- (1.932051,18.600000);
	\draw[line width=3pt,draw=white] (1.732051,18.400000) -- (1.732051,18.800000);
	\draw[line width=3pt,draw=white] (1.732051,18.600000) -- (2.598076,19.100000);
	\draw[line width=3pt,draw=white,fill=white] (2.598076,19.100000) circle (0.100000);
	\draw[line width=3pt,draw=white] (1.732051,18.700000) -- (1.732051,19.300000);
	\draw[line width=3pt,draw=white] (1.732051,19.300000) -- (1.732051,19.500000);
	\draw[line width=3pt,draw=white] (1.732051,19.500000) -- (1.732051,20.100000);
	\draw[line width=3pt,draw=white] (1.732051,20.100000) -- (1.732051,20.300000);
	\draw[line width=1pt] (1.732051,1.000000) -- (1.732051,1.700000);
	\draw[line width=1pt] (1.732051,1.700000) -- (1.732051,1.900000);
	\draw[line width=1pt] (1.732051,1.900000) -- (1.732051,2.500000);
	\draw[line width=1pt] (1.732051,2.500000) -- (1.732051,2.700000);
	\draw[line width=1pt] (1.732051,2.700000) -- (1.732051,3.300000);
	\draw[line width=1pt] (1.732051,3.300000) -- (1.732051,3.500000);
	\draw[line width=1pt] (1.732051,3.500000) -- (1.732051,4.100000);
	\draw[line width=1pt] (1.732051,4.100000) -- (1.732051,4.300000);
	\draw[line width=1pt] (1.732051,4.300000) -- (1.732051,4.900000);
	\draw[line width=1pt] (1.732051,4.900000) -- (1.732051,5.100000);
	\draw[line width=1pt] (1.732051,5.100000) -- (1.732051,5.700000);
	\draw[line width=1pt] (1.732051,5.700000) -- (1.732051,5.900000);
	\draw[line width=1pt] (1.732051,5.900000) -- (1.732051,6.500000);
	\draw[line width=1pt] (1.732051,6.500000) -- (1.732051,6.700000);
	\draw[line width=1pt] (1.732051,6.700000) -- (1.732051,7.300000);
	\draw[line width=1pt] (1.732051,7.300000) -- (1.732051,7.500000);
	\draw[line width=1pt] (1.732051,7.500000) -- (1.732051,8.100000);
	\draw[line width=1pt] (1.732051,8.100000) -- (1.732051,8.300000);
	\draw[line width=1pt] (1.732051,8.300000) -- (1.732051,8.900000);
	\draw[line width=1pt] (1.732051,8.800000) -- (1.732051,9.200000);
	\draw[line width=1pt] (1.532051,8.800000) -- (1.932051,9.200000);
	\draw[line width=1pt] (1.532051,9.200000) -- (1.932051,8.800000);
	\draw[line width=1pt] (1.732051,9.000000) -- (2.732051,9.000000);
	\draw[line width=1pt] (2.532051,8.800000) -- (2.932051,9.200000);
	\draw[line width=1pt] (2.532051,9.200000) -- (2.932051,8.800000);
	\draw[line width=1pt] (2.732051,8.800000) -- (2.732051,9.200000);
	\draw[line width=1pt] (1.732051,9.100000) -- (1.732051,9.700000);
	\draw[line width=1pt] (1.732051,9.700000) -- (1.732051,9.900000);
	\draw[line width=1pt] (1.732051,9.900000) -- (1.732051,10.500000);
	\draw[line width=1pt] (1.732051,10.500000) -- (1.732051,10.700000);
	\draw[line width=1pt] (1.732051,10.700000) -- (1.732051,11.300000);
	\draw[line width=1pt] (1.732051,11.300000) -- (1.732051,11.500000);
	\draw[line width=1pt] (1.732051,11.500000) -- (1.732051,12.100000);
	\draw[line width=1pt] (1.732051,12.100000) -- (1.732051,12.300000);
	\draw[line width=1pt] (1.732051,12.300000) -- (1.732051,12.900000);
	\draw[line width=1pt] (1.732051,12.900000) -- (1.732051,13.100000);
	\draw[line width=1pt] (1.732051,13.100000) -- (1.732051,13.700000);
	\draw[line width=1pt] (1.732051,13.700000) -- (1.732051,13.900000);
	\draw[line width=1pt] (1.732051,13.900000) -- (1.732051,14.500000);
	\draw[line width=1pt] (1.732051,14.500000) -- (1.732051,14.700000);
	\draw[line width=1pt] (1.732051,14.700000) -- (1.732051,15.300000);
	\draw[line width=1pt] (1.732051,15.300000) -- (1.732051,15.500000);
	\draw[line width=1pt] (1.732051,15.500000) -- (1.732051,16.100000);
	\draw[line width=1pt] (1.732051,16.200000) -- (0.866025,15.700000);
	\draw[line width=1pt,fill=white] (1.732051,16.200000) circle (0.200000);
	\draw[line width=1pt] (1.532051,16.200000) -- (1.932051,16.200000);
	\draw[line width=1pt] (1.732051,16.000000) -- (1.732051,16.400000);
	\draw[line width=1pt,fill=black] (0.866025,15.700000) circle (0.100000);
	\draw[line width=1pt] (0.866025,15.800000) -- +(0.0,2.0pt);
	\draw[line width=1pt] (0.866025,15.600000) -- +(0.0,-2.0pt);
	\draw[line width=1pt] (1.732051,16.300000) -- (1.732051,16.900000);
	\draw[line width=1pt] (1.732051,17.000000) -- (2.732051,17.000000);
	\draw[line width=1pt,fill=black] (1.732051,17.000000) circle (0.100000);
	\draw[line width=1pt,fill=white] (2.732051,17.000000) circle (0.200000);
	\draw[line width=1pt] (2.532051,17.000000) -- (2.932051,17.000000);
	\draw[line width=1pt] (2.732051,16.800000) -- (2.732051,17.200000);
	\draw[line width=1pt] (2.732051,17.200000) -- +(0.0,2.0pt);
	\draw[line width=1pt] (2.732051,16.800000) -- +(0.0,-2.0pt);
	\draw[line width=1pt] (1.732051,17.100000) -- (1.732051,17.700000);
	\draw[line width=1pt] (1.732051,17.700000) -- (1.732051,17.900000);
	\draw[line width=1pt] (1.732051,17.900000) -- (1.732051,18.500000);
	\draw[line width=1pt] (1.732051,18.600000) -- (2.598076,19.100000);
	\draw[line width=1pt,fill=white] (1.732051,18.600000) circle (0.200000);
	\draw[line width=1pt] (1.532051,18.600000) -- (1.932051,18.600000);
	\draw[line width=1pt] (1.732051,18.400000) -- (1.732051,18.800000);
	\draw[line width=1pt,fill=black] (2.598076,19.100000) circle (0.100000);
	\draw[line width=1pt] (1.732051,18.700000) -- (1.732051,19.300000);
	\draw[line width=1pt] (1.732051,19.300000) -- (1.732051,19.500000);
	\draw[line width=1pt] (1.732051,19.500000) -- (1.732051,20.100000);
	\draw[line width=1pt] (1.732051,20.100000) -- (1.732051,20.300000);
	\draw[line width=3pt,draw=white] (2.732051,1.000000) -- (2.732051,1.700000);
	\draw[line width=3pt,draw=white] (2.732051,1.700000) -- (2.732051,1.900000);
	\draw[line width=3pt,draw=white] (2.732051,1.900000) -- (2.732051,2.500000);
	\draw[line width=3pt,fill=white,draw=white] (2.632051,2.500000) -- (2.632051,2.700000) -- (2.832051,2.700000) -- (2.832051,2.500000) -- cycle;
	\draw[line width=3pt,draw=white] (2.732051,2.700000) -- (2.732051,3.300000);
	\draw[line width=3pt,draw=white,fill=white] (2.732051,3.400000) circle (0.200000);
	\draw[line width=3pt,draw=white] (2.532051,3.400000) -- (2.932051,3.400000);
	\draw[line width=3pt,draw=white] (2.732051,3.200000) -- (2.732051,3.600000);
	\draw[line width=3pt,draw=white] (2.732051,3.400000) -- (3.598076,3.900000);
	\draw[line width=3pt,draw=white,fill=white] (3.598076,3.900000) circle (0.100000);
	\draw[line width=3pt,draw=white] (2.732051,3.500000) -- (2.732051,4.100000);
	\draw[line width=3pt,draw=white] (2.732051,4.100000) -- (2.732051,4.300000);
	\draw[line width=3pt,draw=white] (2.732051,4.300000) -- (2.732051,4.900000);
	\draw[line width=3pt,draw=white,fill=white] (2.732051,5.000000) circle (0.200000);
	\draw[line width=3pt,draw=white] (2.532051,5.000000) -- (2.932051,5.000000);
	\draw[line width=3pt,draw=white] (2.732051,4.800000) -- (2.732051,5.200000);
	\draw[line width=3pt,draw=white] (2.732051,5.000000) -- (3.732051,5.000000);
	\draw[line width=3pt,draw=white,fill=white] (3.732051,5.000000) circle (0.100000);
	\draw[line width=3pt,draw=white] (2.732051,5.100000) -- (2.732051,5.700000);
	\draw[line width=3pt,draw=white,fill=white] (2.732051,5.800000) circle (0.200000);
	\draw[line width=3pt,draw=white] (2.532051,5.800000) -- (2.932051,5.800000);
	\draw[line width=3pt,draw=white] (2.732051,5.600000) -- (2.732051,6.000000);
	\draw[line width=3pt,draw=white] (2.732051,5.800000) -- (1.866025,5.300000);
	\draw[line width=3pt,draw=white,fill=white] (1.866025,5.300000) circle (0.100000);
	\draw[line width=3pt,draw=white] (2.732051,5.900000) -- (2.732051,6.500000);
	\draw[line width=3pt,fill=white,draw=white] (2.732051,6.600000) circle (0.100000);
	\draw[line width=3pt,draw=white] (2.732051,6.700000) -- (2.732051,7.300000);
	\draw[line width=3pt,draw=white] (2.732051,7.300000) -- (2.732051,7.500000);
	\draw[line width=3pt,draw=white] (2.732051,7.500000) -- (2.732051,8.100000);
	\draw[line width=3pt,draw=white] (2.732051,8.000000) -- (2.732051,8.400000);
	\draw[line width=3pt,draw=white] (2.532051,8.000000) -- (2.932051,8.400000);
	\draw[line width=3pt,draw=white] (2.532051,8.400000) -- (2.932051,8.000000);
	\draw[line width=3pt,draw=white] (2.732051,8.200000) -- (1.866025,7.700000);
	\draw[line width=3pt,draw=white] (1.666025,7.500000) -- (2.066025,7.900000);
	\draw[line width=3pt,draw=white] (1.666025,7.900000) -- (2.066025,7.500000);
	\draw[line width=3pt,draw=white] (2.732051,8.300000) -- (2.732051,8.900000);
	\draw[line width=3pt,draw=white] (2.732051,8.800000) -- (2.732051,9.200000);
	\draw[line width=3pt,draw=white] (2.532051,8.800000) -- (2.932051,9.200000);
	\draw[line width=3pt,draw=white] (2.532051,9.200000) -- (2.932051,8.800000);
	\draw[line width=3pt,draw=white] (2.732051,9.000000) -- (1.732051,9.000000);
	\draw[line width=3pt,draw=white] (1.532051,8.800000) -- (1.932051,9.200000);
	\draw[line width=3pt,draw=white] (1.532051,9.200000) -- (1.932051,8.800000);
	\draw[line width=3pt,draw=white] (2.732051,9.100000) -- (2.732051,9.700000);
	\draw[line width=3pt,draw=white] (2.732051,9.700000) -- (2.732051,9.900000);
	\draw[line width=3pt,draw=white] (2.732051,9.900000) -- (2.732051,10.500000);
	\draw[line width=3pt,draw=white] (2.732051,10.500000) -- (2.732051,10.700000);
	\draw[line width=3pt,draw=white] (2.732051,10.700000) -- (2.732051,11.300000);
	\draw[line width=3pt,draw=white] (2.732051,11.300000) -- (2.732051,11.500000);
	\draw[line width=3pt,draw=white] (2.732051,11.500000) -- (2.732051,12.100000);
	\draw[line width=3pt,draw=white] (2.732051,12.100000) -- (2.732051,12.300000);
	\draw[line width=3pt,draw=white] (2.732051,12.300000) -- (2.732051,12.900000);
	\draw[line width=3pt,draw=white] (2.732051,12.900000) -- (2.732051,13.100000);
	\draw[line width=3pt,draw=white] (2.732051,13.100000) -- (2.732051,13.700000);
	\draw[line width=3pt,draw=white] (2.732051,13.700000) -- (2.732051,13.900000);
	\draw[line width=3pt,draw=white] (2.732051,13.900000) -- (2.732051,14.500000);
	\draw[line width=3pt,draw=white] (2.732051,14.500000) -- (2.732051,14.700000);
	\draw[line width=3pt,draw=white] (2.732051,14.700000) -- (2.732051,15.300000);
	\draw[line width=3pt,fill=white,draw=white] (2.632051,15.300000) -- (2.632051,15.500000) -- (2.832051,15.500000) -- (2.832051,15.300000) -- cycle;
	\draw[line width=3pt,draw=white] (2.732051,15.500000) -- (2.732051,16.100000);
	\draw[line width=3pt,draw=white,fill=white] (2.732051,16.200000) circle (0.200000);
	\draw[line width=3pt,draw=white] (2.532051,16.200000) -- (2.932051,16.200000);
	\draw[line width=3pt,draw=white] (2.732051,16.000000) -- (2.732051,16.400000);
	\draw[line width=3pt,draw=white] (2.732051,16.200000) -- (3.598076,16.700000);
	\draw[line width=3pt,draw=white,fill=white] (3.598076,16.700000) circle (0.100000);
	\draw[line width=3pt,draw=white] (2.732051,16.300000) -- (2.732051,16.900000);
	\draw[line width=3pt,draw=white,fill=white] (2.732051,17.000000) circle (0.200000);
	\draw[line width=3pt,draw=white] (2.532051,17.000000) -- (2.932051,17.000000);
	\draw[line width=3pt,draw=white] (2.732051,16.800000) -- (2.732051,17.200000);
	\draw[line width=3pt,draw=white] (2.732051,17.000000) -- (1.732051,17.000000);
	\draw[line width=3pt,draw=white,fill=white] (1.732051,17.000000) circle (0.100000);
	\draw[line width=3pt,draw=white] (2.732051,17.100000) -- (2.732051,17.700000);
	\draw[line width=3pt,draw=white,fill=white] (2.732051,17.800000) circle (0.200000);
	\draw[line width=3pt,draw=white] (2.532051,17.800000) -- (2.932051,17.800000);
	\draw[line width=3pt,draw=white] (2.732051,17.600000) -- (2.732051,18.000000);
	\draw[line width=3pt,draw=white] (2.732051,17.800000) -- (3.732051,17.800000);
	\draw[line width=3pt,draw=white,fill=white] (3.732051,17.800000) circle (0.100000);
	\draw[line width=3pt,draw=white] (2.732051,17.900000) -- (2.732051,18.500000);
	\draw[line width=3pt,draw=white,fill=white] (2.732051,18.600000) circle (0.200000);
	\draw[line width=3pt,draw=white] (2.532051,18.600000) -- (2.932051,18.600000);
	\draw[line width=3pt,draw=white] (2.732051,18.400000) -- (2.732051,18.800000);
	\draw[line width=3pt,draw=white] (2.732051,18.600000) -- (1.866025,18.100000);
	\draw[line width=3pt,draw=white,fill=white] (1.866025,18.100000) circle (0.100000);
	\draw[line width=3pt,draw=white] (2.732051,18.700000) -- (2.732051,19.300000);
	\draw[line width=3pt,fill=white,draw=white] (2.732051,19.400000) circle (0.100000);
	\draw[line width=3pt,draw=white] (2.732051,19.500000) -- (2.732051,20.100000);
	\draw[line width=3pt,draw=white] (2.732051,20.100000) -- (2.732051,20.300000);
	\draw[line width=1pt] (2.732051,1.000000) -- (2.732051,1.700000);
	\draw[line width=1pt] (2.732051,1.700000) -- (2.732051,1.900000);
	\draw[line width=1pt] (2.732051,1.900000) -- (2.732051,2.500000);
	\draw[line width=1pt,fill=white] (2.632051,2.500000) -- (2.632051,2.700000) -- (2.832051,2.700000) -- (2.832051,2.500000) -- cycle;
	\draw[line width=1pt] (2.732051,2.700000) -- (2.732051,3.300000);
	\draw[line width=1pt] (2.732051,3.400000) -- (3.598076,3.900000);
	\draw[line width=1pt,fill=white] (2.732051,3.400000) circle (0.200000);
	\draw[line width=1pt] (2.532051,3.400000) -- (2.932051,3.400000);
	\draw[line width=1pt] (2.732051,3.200000) -- (2.732051,3.600000);
	\draw[line width=1pt,fill=black] (3.598076,3.900000) circle (0.100000);
	\draw[line width=1pt] (2.732051,3.500000) -- (2.732051,4.100000);
	\draw[line width=1pt] (2.732051,4.100000) -- (2.732051,4.300000);
	\draw[line width=1pt] (2.732051,4.300000) -- (2.732051,4.900000);
	\draw[line width=1pt] (2.732051,5.000000) -- (3.732051,5.000000);
	\draw[line width=1pt,fill=white] (2.732051,5.000000) circle (0.200000);
	\draw[line width=1pt] (2.532051,5.000000) -- (2.932051,5.000000);
	\draw[line width=1pt] (2.732051,4.800000) -- (2.732051,5.200000);
	\draw[line width=1pt,fill=black] (3.732051,5.000000) circle (0.100000);
	\draw[line width=1pt] (2.732051,5.100000) -- (2.732051,5.700000);
	\draw[line width=1pt] (2.732051,5.800000) -- (1.866025,5.300000);
	\draw[line width=1pt,fill=white] (2.732051,5.800000) circle (0.200000);
	\draw[line width=1pt] (2.532051,5.800000) -- (2.932051,5.800000);
	\draw[line width=1pt] (2.732051,5.600000) -- (2.732051,6.000000);
	\draw[line width=1pt,fill=black] (1.866025,5.300000) circle (0.100000);
	\draw[line width=1pt] (1.866025,5.400000) -- +(0.0,2.0pt);
	\draw[line width=1pt] (1.866025,5.200000) -- +(0.0,-2.0pt);
	\draw[line width=1pt] (2.732051,5.900000) -- (2.732051,6.500000);
	\draw[line width=1pt,fill=white] (2.732051,6.600000) circle (0.100000);
	\draw[line width=1pt] (2.732051,6.700000) -- (2.732051,7.300000);
	\draw[line width=1pt] (2.732051,7.300000) -- (2.732051,7.500000);
	\draw[line width=1pt] (2.732051,7.500000) -- (2.732051,8.100000);
	\draw[line width=1pt] (2.732051,8.000000) -- (2.732051,8.400000);
	\draw[line width=1pt] (2.532051,8.000000) -- (2.932051,8.400000);
	\draw[line width=1pt] (2.532051,8.400000) -- (2.932051,8.000000);
	\draw[line width=1pt] (2.732051,8.200000) -- (1.866025,7.700000);
	\draw[line width=1pt] (1.666025,7.500000) -- (2.066025,7.900000);
	\draw[line width=1pt] (1.666025,7.900000) -- (2.066025,7.500000);
	\draw[line width=1pt] (1.866025,7.500000) -- (1.866025,7.900000);
	\draw[line width=1pt] (2.732051,8.300000) -- (2.732051,8.900000);
	\draw[line width=1pt] (2.732051,8.800000) -- (2.732051,9.200000);
	\draw[line width=1pt] (2.532051,8.800000) -- (2.932051,9.200000);
	\draw[line width=1pt] (2.532051,9.200000) -- (2.932051,8.800000);
	\draw[line width=1pt] (2.732051,9.000000) -- (1.732051,9.000000);
	\draw[line width=1pt] (1.532051,8.800000) -- (1.932051,9.200000);
	\draw[line width=1pt] (1.532051,9.200000) -- (1.932051,8.800000);
	\draw[line width=1pt] (1.732051,8.800000) -- (1.732051,9.200000);
	\draw[line width=1pt] (2.732051,9.100000) -- (2.732051,9.700000);
	\draw[line width=1pt] (2.732051,9.700000) -- (2.732051,9.900000);
	\draw[line width=1pt] (2.732051,9.900000) -- (2.732051,10.500000);
	\draw[line width=1pt] (2.732051,10.500000) -- (2.732051,10.700000);
	\draw[line width=1pt] (2.732051,10.700000) -- (2.732051,11.300000);
	\draw[line width=1pt] (2.732051,11.300000) -- (2.732051,11.500000);
	\draw[line width=1pt] (2.732051,11.500000) -- (2.732051,12.100000);
	\draw[line width=1pt] (2.732051,12.100000) -- (2.732051,12.300000);
	\draw[line width=1pt] (2.732051,12.300000) -- (2.732051,12.900000);
	\draw[line width=1pt] (2.732051,12.900000) -- (2.732051,13.100000);
	\draw[line width=1pt] (2.732051,13.100000) -- (2.732051,13.700000);
	\draw[line width=1pt] (2.732051,13.700000) -- (2.732051,13.900000);
	\draw[line width=1pt] (2.732051,13.900000) -- (2.732051,14.500000);
	\draw[line width=1pt] (2.732051,14.500000) -- (2.732051,14.700000);
	\draw[line width=1pt] (2.732051,14.700000) -- (2.732051,15.300000);
	\draw[line width=1pt,fill=white] (2.632051,15.300000) -- (2.632051,15.500000) -- (2.832051,15.500000) -- (2.832051,15.300000) -- cycle;
	\draw[line width=1pt] (2.732051,15.500000) -- (2.732051,16.100000);
	\draw[line width=1pt] (2.732051,16.200000) -- (3.598076,16.700000);
	\draw[line width=1pt,fill=white] (2.732051,16.200000) circle (0.200000);
	\draw[line width=1pt] (2.532051,16.200000) -- (2.932051,16.200000);
	\draw[line width=1pt] (2.732051,16.000000) -- (2.732051,16.400000);
	\draw[line width=1pt,fill=black] (3.598076,16.700000) circle (0.100000);
	\draw[line width=1pt] (2.732051,16.300000) -- (2.732051,16.900000);
	\draw[line width=1pt] (2.732051,17.000000) -- (1.732051,17.000000);
	\draw[line width=1pt,fill=white] (2.732051,17.000000) circle (0.200000);
	\draw[line width=1pt] (2.532051,17.000000) -- (2.932051,17.000000);
	\draw[line width=1pt] (2.732051,16.800000) -- (2.732051,17.200000);
	\draw[line width=1pt,fill=black] (1.732051,17.000000) circle (0.100000);
	\draw[line width=1pt] (2.732051,17.100000) -- (2.732051,17.700000);
	\draw[line width=1pt] (2.732051,17.800000) -- (3.732051,17.800000);
	\draw[line width=1pt,fill=white] (2.732051,17.800000) circle (0.200000);
	\draw[line width=1pt] (2.532051,17.800000) -- (2.932051,17.800000);
	\draw[line width=1pt] (2.732051,17.600000) -- (2.732051,18.000000);
	\draw[line width=1pt,fill=black] (3.732051,17.800000) circle (0.100000);
	\draw[line width=1pt] (2.732051,17.900000) -- (2.732051,18.500000);
	\draw[line width=1pt] (2.732051,18.600000) -- (1.866025,18.100000);
	\draw[line width=1pt,fill=white] (2.732051,18.600000) circle (0.200000);
	\draw[line width=1pt] (2.532051,18.600000) -- (2.932051,18.600000);
	\draw[line width=1pt] (2.732051,18.400000) -- (2.732051,18.800000);
	\draw[line width=1pt,fill=black] (1.866025,18.100000) circle (0.100000);
	\draw[line width=1pt] (1.866025,18.200000) -- +(0.0,2.0pt);
	\draw[line width=1pt] (1.866025,18.000000) -- +(0.0,-2.0pt);
	\draw[line width=1pt] (2.732051,18.700000) -- (2.732051,19.300000);
	\draw[line width=1pt,fill=white] (2.732051,19.400000) circle (0.100000);
	\draw[line width=1pt] (2.732051,19.500000) -- (2.732051,20.100000);
	\draw[line width=1pt] (2.732051,20.100000) -- (2.732051,20.300000);
	\draw[line width=3pt,draw=white] (0.866025,0.500000) -- (0.866025,1.200000);
	\draw[line width=3pt,draw=white] (0.866025,1.200000) -- (0.866025,1.400000);
	\draw[line width=3pt,draw=white] (0.866025,1.400000) -- (0.866025,2.000000);
	\draw[line width=3pt,draw=white] (0.866025,2.000000) -- (0.866025,2.200000);
	\draw[line width=3pt,draw=white] (0.866025,2.200000) -- (0.866025,2.800000);
	\draw[line width=3pt,draw=white] (0.866025,2.800000) -- (0.866025,3.000000);
	\draw[line width=3pt,draw=white] (0.866025,3.000000) -- (0.866025,3.600000);
	\draw[line width=3pt,draw=white] (0.866025,3.600000) -- (0.866025,3.800000);
	\draw[line width=3pt,draw=white] (0.866025,3.800000) -- (0.866025,4.400000);
	\draw[line width=3pt,draw=white] (0.866025,4.400000) -- (0.866025,4.600000);
	\draw[line width=3pt,draw=white] (0.866025,4.600000) -- (0.866025,5.200000);
	\draw[line width=3pt,draw=white] (0.866025,5.200000) -- (0.866025,5.400000);
	\draw[line width=3pt,draw=white] (0.866025,5.400000) -- (0.866025,6.000000);
	\draw[line width=3pt,draw=white] (0.866025,6.000000) -- (0.866025,6.200000);
	\draw[line width=3pt,draw=white] (0.866025,6.200000) -- (0.866025,6.800000);
	\draw[line width=3pt,draw=white] (0.866025,6.800000) -- (0.866025,7.000000);
	\draw[line width=3pt,draw=white] (0.866025,7.000000) -- (0.866025,7.600000);
	\draw[line width=3pt,draw=white] (0.866025,7.600000) -- (0.866025,7.800000);
	\draw[line width=3pt,draw=white] (0.866025,7.800000) -- (0.866025,8.400000);
	\draw[line width=3pt,draw=white] (0.866025,8.400000) -- (0.866025,8.600000);
	\draw[line width=3pt,draw=white] (0.866025,8.600000) -- (0.866025,9.200000);
	\draw[line width=3pt,draw=white] (0.866025,9.200000) -- (0.866025,9.400000);
	\draw[line width=3pt,draw=white] (0.866025,9.400000) -- (0.866025,10.000000);
	\draw[line width=3pt,draw=white] (0.866025,10.000000) -- (0.866025,10.200000);
	\draw[line width=3pt,draw=white] (0.866025,10.200000) -- (0.866025,10.800000);
	\draw[line width=3pt,draw=white] (0.866025,10.800000) -- (0.866025,11.000000);
	\draw[line width=3pt,draw=white] (0.866025,11.000000) -- (0.866025,11.600000);
	\draw[line width=3pt,draw=white] (0.866025,11.600000) -- (0.866025,11.800000);
	\draw[line width=3pt,draw=white] (0.866025,11.800000) -- (0.866025,12.400000);
	\draw[line width=3pt,draw=white] (0.866025,12.400000) -- (0.866025,12.600000);
	\draw[line width=3pt,draw=white] (0.866025,12.600000) -- (0.866025,13.200000);
	\draw[line width=3pt,draw=white] (0.866025,13.200000) -- (0.866025,13.400000);
	\draw[line width=3pt,draw=white] (0.866025,13.400000) -- (0.866025,14.000000);
	\draw[line width=3pt,fill=white,draw=white] (0.766025,14.000000) -- (0.766025,14.200000) -- (0.966025,14.200000) -- (0.966025,14.000000) -- cycle;
	\draw[line width=3pt,draw=white] (0.866025,14.200000) -- (0.866025,14.800000);
	\draw[line width=3pt,fill=white,draw=white] (0.766025,14.900000) -- (0.866025,15.000000) -- (0.966025,14.900000) -- (0.866025,14.800000) -- cycle;
	\draw[line width=3pt,draw=white] (0.866025,15.000000) -- (0.866025,15.600000);
	\draw[line width=3pt,draw=white,fill=white] (0.866025,15.700000) circle (0.100000);
	\draw[line width=3pt,draw=white] (0.866025,15.700000) -- (1.732051,16.200000);
	\draw[line width=3pt,draw=white,fill=white] (1.732051,16.200000) circle (0.200000);
	\draw[line width=3pt,draw=white] (1.532051,16.200000) -- (1.932051,16.200000);
	\draw[line width=3pt,draw=white] (1.732051,16.000000) -- (1.732051,16.400000);
	\draw[line width=3pt,draw=white] (0.866025,15.800000) -- (0.866025,16.400000);
	\draw[line width=3pt,draw=white] (0.866025,16.400000) -- (0.866025,16.600000);
	\draw[line width=3pt,draw=white] (0.866025,16.600000) -- (0.866025,17.200000);
	\draw[line width=3pt,draw=white,fill=white] (0.866025,17.300000) circle (0.100000);
	\draw[line width=3pt,draw=white] (0.866025,17.300000) -- (1.866025,17.300000);
	\draw[line width=3pt,draw=white,fill=white] (1.866025,17.300000) circle (0.200000);
	\draw[line width=3pt,draw=white] (1.666025,17.300000) -- (2.066025,17.300000);
	\draw[line width=3pt,draw=white] (1.866025,17.100000) -- (1.866025,17.500000);
	\draw[line width=3pt,draw=white] (0.866025,17.400000) -- (0.866025,18.000000);
	\draw[line width=3pt,draw=white,fill=white] (0.866025,18.100000) circle (0.100000);
	\draw[line width=3pt,draw=white] (0.866025,18.100000) -- (0.000000,17.600000);
	\draw[line width=3pt,draw=white,fill=white] (0.000000,17.600000) circle (0.200000);
	\draw[line width=3pt,draw=white] (-0.200000,17.600000) -- (0.200000,17.600000);
	\draw[line width=3pt,draw=white] (0.000000,17.400000) -- (0.000000,17.800000);
	\draw[line width=3pt,draw=white] (0.866025,18.200000) -- (0.866025,18.800000);
	\draw[line width=3pt,fill=white,draw=white] (0.766025,18.900000) -- (0.866025,19.000000) -- (0.966025,18.900000) -- (0.866025,18.800000) -- cycle;
	\draw[line width=3pt,draw=white] (0.866025,19.000000) -- (0.866025,19.600000);
	\draw[line width=3pt,fill=white,draw=white] (0.866025,19.700000) circle (0.100000);
	\draw[line width=1pt] (0.866025,0.500000) -- (0.866025,1.200000);
	\draw[line width=1pt] (0.866025,1.200000) -- (0.866025,1.400000);
	\draw[line width=1pt] (0.866025,1.400000) -- (0.866025,2.000000);
	\draw[line width=1pt] (0.866025,2.000000) -- (0.866025,2.200000);
	\draw[line width=1pt] (0.866025,2.200000) -- (0.866025,2.800000);
	\draw[line width=1pt] (0.866025,2.800000) -- (0.866025,3.000000);
	\draw[line width=1pt] (0.866025,3.000000) -- (0.866025,3.600000);
	\draw[line width=1pt] (0.866025,3.600000) -- (0.866025,3.800000);
	\draw[line width=1pt] (0.866025,3.800000) -- (0.866025,4.400000);
	\draw[line width=1pt] (0.866025,4.400000) -- (0.866025,4.600000);
	\draw[line width=1pt] (0.866025,4.600000) -- (0.866025,5.200000);
	\draw[line width=1pt] (0.866025,5.200000) -- (0.866025,5.400000);
	\draw[line width=1pt] (0.866025,5.400000) -- (0.866025,6.000000);
	\draw[line width=1pt] (0.866025,6.000000) -- (0.866025,6.200000);
	\draw[line width=1pt] (0.866025,6.200000) -- (0.866025,6.800000);
	\draw[line width=1pt] (0.866025,6.800000) -- (0.866025,7.000000);
	\draw[line width=1pt] (0.866025,7.000000) -- (0.866025,7.600000);
	\draw[line width=1pt] (0.866025,7.600000) -- (0.866025,7.800000);
	\draw[line width=1pt] (0.866025,7.800000) -- (0.866025,8.400000);
	\draw[line width=1pt] (0.866025,8.400000) -- (0.866025,8.600000);
	\draw[line width=1pt] (0.866025,8.600000) -- (0.866025,9.200000);
	\draw[line width=1pt] (0.866025,9.200000) -- (0.866025,9.400000);
	\draw[line width=1pt] (0.866025,9.400000) -- (0.866025,10.000000);
	\draw[line width=1pt] (0.866025,10.000000) -- (0.866025,10.200000);
	\draw[line width=1pt] (0.866025,10.200000) -- (0.866025,10.800000);
	\draw[line width=1pt] (0.866025,10.800000) -- (0.866025,11.000000);
	\draw[line width=1pt] (0.866025,11.000000) -- (0.866025,11.600000);
	\draw[line width=1pt] (0.866025,11.600000) -- (0.866025,11.800000);
	\draw[line width=1pt] (0.866025,11.800000) -- (0.866025,12.400000);
	\draw[line width=1pt] (0.866025,12.400000) -- (0.866025,12.600000);
	\draw[line width=1pt] (0.866025,12.600000) -- (0.866025,13.200000);
	\draw[line width=1pt] (0.866025,13.200000) -- (0.866025,13.400000);
	\draw[line width=1pt] (0.866025,13.400000) -- (0.866025,14.000000);
	\draw[line width=1pt,fill=white] (0.766025,14.000000) -- (0.766025,14.200000) -- (0.966025,14.200000) -- (0.966025,14.000000) -- cycle;
	\draw[line width=1pt] (0.866025,14.200000) -- (0.866025,14.800000);
	\draw[line width=1pt,fill=white] (0.766025,14.900000) -- (0.866025,15.000000) -- (0.966025,14.900000) -- (0.866025,14.800000) -- cycle;
	\draw[line width=1pt] (0.866025,15.000000) -- (0.866025,15.600000);
	\draw[line width=1pt] (0.866025,15.700000) -- (1.732051,16.200000);
	\draw[line width=1pt,fill=black] (0.866025,15.700000) circle (0.100000);
	\draw[line width=1pt,fill=white] (1.732051,16.200000) circle (0.200000);
	\draw[line width=1pt] (1.532051,16.200000) -- (1.932051,16.200000);
	\draw[line width=1pt] (1.732051,16.000000) -- (1.732051,16.400000);
	\draw[line width=1pt] (1.732051,16.400000) -- +(0.0,2.0pt);
	\draw[line width=1pt] (1.732051,16.000000) -- +(0.0,-2.0pt);
	\draw[line width=1pt] (0.866025,15.800000) -- (0.866025,16.400000);
	\draw[line width=1pt] (0.866025,16.400000) -- (0.866025,16.600000);
	\draw[line width=1pt] (0.866025,16.600000) -- (0.866025,17.200000);
	\draw[line width=1pt] (0.866025,17.300000) -- (1.866025,17.300000);
	\draw[line width=1pt,fill=black] (0.866025,17.300000) circle (0.100000);
	\draw[line width=1pt,fill=white] (1.866025,17.300000) circle (0.200000);
	\draw[line width=1pt] (1.666025,17.300000) -- (2.066025,17.300000);
	\draw[line width=1pt] (1.866025,17.100000) -- (1.866025,17.500000);
	\draw[line width=1pt] (1.866025,17.500000) -- +(0.0,2.0pt);
	\draw[line width=1pt] (1.866025,17.100000) -- +(0.0,-2.0pt);
	\draw[line width=1pt] (0.866025,17.400000) -- (0.866025,18.000000);
	\draw[line width=1pt] (0.866025,18.100000) -- (0.000000,17.600000);
	\draw[line width=1pt,fill=black] (0.866025,18.100000) circle (0.100000);
	\draw[line width=1pt,fill=white] (0.000000,17.600000) circle (0.200000);
	\draw[line width=1pt] (-0.200000,17.600000) -- (0.200000,17.600000);
	\draw[line width=1pt] (0.000000,17.400000) -- (0.000000,17.800000);
	\draw[line width=1pt] (0.866025,18.200000) -- (0.866025,18.800000);
	\draw[line width=1pt,fill=white] (0.766025,18.900000) -- (0.866025,19.000000) -- (0.966025,18.900000) -- (0.866025,18.800000) -- cycle;
	\draw[line width=1pt] (0.866025,19.000000) -- (0.866025,19.600000);
	\draw[line width=1pt,fill=white] (0.866025,19.700000) circle (0.100000);
	\draw[line width=3pt,draw=white] (1.866025,0.500000) -- (1.866025,1.200000);
	\draw[line width=3pt,draw=white] (1.866025,1.200000) -- (1.866025,1.400000);
	\draw[line width=3pt,draw=white] (1.866025,1.400000) -- (1.866025,2.000000);
	\draw[line width=3pt,draw=white] (1.866025,2.000000) -- (1.866025,2.200000);
	\draw[line width=3pt,draw=white] (1.866025,2.200000) -- (1.866025,2.800000);
	\draw[line width=3pt,draw=white,fill=white] (1.866025,2.900000) circle (0.100000);
	\draw[line width=3pt,draw=white] (1.866025,2.900000) -- (1.000000,2.400000);
	\draw[line width=3pt,draw=white,fill=white] (1.000000,2.400000) circle (0.200000);
	\draw[line width=3pt,draw=white] (0.800000,2.400000) -- (1.200000,2.400000);
	\draw[line width=3pt,draw=white] (1.000000,2.200000) -- (1.000000,2.600000);
	\draw[line width=3pt,draw=white] (1.866025,3.000000) -- (1.866025,3.600000);
	\draw[line width=3pt,draw=white,fill=white] (1.866025,3.700000) circle (0.200000);
	\draw[line width=3pt,draw=white] (1.666025,3.700000) -- (2.066025,3.700000);
	\draw[line width=3pt,draw=white] (1.866025,3.500000) -- (1.866025,3.900000);
	\draw[line width=3pt,draw=white] (1.866025,3.700000) -- (2.866025,3.700000);
	\draw[line width=3pt,draw=white,fill=white] (2.866025,3.700000) circle (0.100000);
	\draw[line width=3pt,draw=white] (1.866025,3.800000) -- (1.866025,4.400000);
	\draw[line width=3pt,draw=white] (1.866025,4.400000) -- (1.866025,4.600000);
	\draw[line width=3pt,draw=white] (1.866025,4.600000) -- (1.866025,5.200000);
	\draw[line width=3pt,draw=white,fill=white] (1.866025,5.300000) circle (0.100000);
	\draw[line width=3pt,draw=white] (1.866025,5.300000) -- (2.732051,5.800000);
	\draw[line width=3pt,draw=white,fill=white] (2.732051,5.800000) circle (0.200000);
	\draw[line width=3pt,draw=white] (2.532051,5.800000) -- (2.932051,5.800000);
	\draw[line width=3pt,draw=white] (2.732051,5.600000) -- (2.732051,6.000000);
	\draw[line width=3pt,draw=white] (1.866025,5.400000) -- (1.866025,6.000000);
	\draw[line width=3pt,draw=white] (1.866025,6.000000) -- (1.866025,6.200000);
	\draw[line width=3pt,draw=white] (1.866025,6.200000) -- (1.866025,6.800000);
	\draw[line width=3pt,fill=white,draw=white] (1.766025,6.900000) -- (1.866025,7.000000) -- (1.966025,6.900000) -- (1.866025,6.800000) -- cycle;
	\draw[line width=3pt,draw=white] (1.866025,7.000000) -- (1.866025,7.600000);
	\draw[line width=3pt,draw=white] (1.866025,7.500000) -- (1.866025,7.900000);
	\draw[line width=3pt,draw=white] (1.666025,7.500000) -- (2.066025,7.900000);
	\draw[line width=3pt,draw=white] (1.666025,7.900000) -- (2.066025,7.500000);
	\draw[line width=3pt,draw=white] (1.866025,7.700000) -- (2.732051,8.200000);
	\draw[line width=3pt,draw=white] (2.532051,8.000000) -- (2.932051,8.400000);
	\draw[line width=3pt,draw=white] (2.532051,8.400000) -- (2.932051,8.000000);
	\draw[line width=3pt,draw=white] (1.866025,7.800000) -- (1.866025,8.400000);
	\draw[line width=3pt,draw=white] (1.866025,8.300000) -- (1.866025,8.700000);
	\draw[line width=3pt,draw=white] (1.666025,8.300000) -- (2.066025,8.700000);
	\draw[line width=3pt,draw=white] (1.666025,8.700000) -- (2.066025,8.300000);
	\draw[line width=3pt,draw=white] (1.866025,8.500000) -- (2.866025,8.500000);
	\draw[line width=3pt,draw=white] (2.666025,8.300000) -- (3.066025,8.700000);
	\draw[line width=3pt,draw=white] (2.666025,8.700000) -- (3.066025,8.300000);
	\draw[line width=3pt,draw=white] (1.866025,8.600000) -- (1.866025,9.200000);
	\draw[line width=3pt,draw=white] (1.866025,9.200000) -- (1.866025,9.400000);
	\draw[line width=3pt,draw=white] (1.866025,9.400000) -- (1.866025,10.000000);
	\draw[line width=3pt,draw=white] (1.866025,10.000000) -- (1.866025,10.200000);
	\draw[line width=3pt,draw=white] (1.866025,10.200000) -- (1.866025,10.800000);
	\draw[line width=3pt,draw=white] (1.866025,10.800000) -- (1.866025,11.000000);
	\draw[line width=3pt,draw=white] (1.866025,11.000000) -- (1.866025,11.600000);
	\draw[line width=3pt,draw=white] (1.866025,11.600000) -- (1.866025,11.800000);
	\draw[line width=3pt,draw=white] (1.866025,11.800000) -- (1.866025,12.400000);
	\draw[line width=3pt,draw=white] (1.866025,12.400000) -- (1.866025,12.600000);
	\draw[line width=3pt,draw=white] (1.866025,12.600000) -- (1.866025,13.200000);
	\draw[line width=3pt,draw=white] (1.866025,13.200000) -- (1.866025,13.400000);
	\draw[line width=3pt,draw=white] (1.866025,13.400000) -- (1.866025,14.000000);
	\draw[line width=3pt,draw=white] (1.866025,14.000000) -- (1.866025,14.200000);
	\draw[line width=3pt,draw=white] (1.866025,14.200000) -- (1.866025,14.800000);
	\draw[line width=3pt,draw=white] (1.866025,14.800000) -- (1.866025,15.000000);
	\draw[line width=3pt,draw=white] (1.866025,15.000000) -- (1.866025,15.600000);
	\draw[line width=3pt,draw=white,fill=white] (1.866025,15.700000) circle (0.100000);
	\draw[line width=3pt,draw=white] (1.866025,15.700000) -- (1.000000,15.200000);
	\draw[line width=3pt,draw=white,fill=white] (1.000000,15.200000) circle (0.200000);
	\draw[line width=3pt,draw=white] (0.800000,15.200000) -- (1.200000,15.200000);
	\draw[line width=3pt,draw=white] (1.000000,15.000000) -- (1.000000,15.400000);
	\draw[line width=3pt,draw=white] (1.866025,15.800000) -- (1.866025,16.400000);
	\draw[line width=3pt,draw=white,fill=white] (1.866025,16.500000) circle (0.200000);
	\draw[line width=3pt,draw=white] (1.666025,16.500000) -- (2.066025,16.500000);
	\draw[line width=3pt,draw=white] (1.866025,16.300000) -- (1.866025,16.700000);
	\draw[line width=3pt,draw=white] (1.866025,16.500000) -- (2.866025,16.500000);
	\draw[line width=3pt,draw=white,fill=white] (2.866025,16.500000) circle (0.100000);
	\draw[line width=3pt,draw=white] (1.866025,16.600000) -- (1.866025,17.200000);
	\draw[line width=3pt,draw=white,fill=white] (1.866025,17.300000) circle (0.200000);
	\draw[line width=3pt,draw=white] (1.666025,17.300000) -- (2.066025,17.300000);
	\draw[line width=3pt,draw=white] (1.866025,17.100000) -- (1.866025,17.500000);
	\draw[line width=3pt,draw=white] (1.866025,17.300000) -- (0.866025,17.300000);
	\draw[line width=3pt,draw=white,fill=white] (0.866025,17.300000) circle (0.100000);
	\draw[line width=3pt,draw=white] (1.866025,17.400000) -- (1.866025,18.000000);
	\draw[line width=3pt,draw=white,fill=white] (1.866025,18.100000) circle (0.100000);
	\draw[line width=3pt,draw=white] (1.866025,18.100000) -- (2.732051,18.600000);
	\draw[line width=3pt,draw=white,fill=white] (2.732051,18.600000) circle (0.200000);
	\draw[line width=3pt,draw=white] (2.532051,18.600000) -- (2.932051,18.600000);
	\draw[line width=3pt,draw=white] (2.732051,18.400000) -- (2.732051,18.800000);
	\draw[line width=3pt,draw=white] (1.866025,18.200000) -- (1.866025,18.800000);
	\draw[line width=3pt,draw=white] (1.866025,18.800000) -- (1.866025,19.000000);
	\draw[line width=3pt,draw=white] (1.866025,19.000000) -- (1.866025,19.600000);
	\draw[line width=3pt,draw=white] (1.866025,19.600000) -- (1.866025,19.800000);
	\draw[line width=1pt] (1.866025,0.500000) -- (1.866025,1.200000);
	\draw[line width=1pt] (1.866025,1.200000) -- (1.866025,1.400000);
	\draw[line width=1pt] (1.866025,1.400000) -- (1.866025,2.000000);
	\draw[line width=1pt] (1.866025,2.000000) -- (1.866025,2.200000);
	\draw[line width=1pt] (1.866025,2.200000) -- (1.866025,2.800000);
	\draw[line width=1pt] (1.866025,2.900000) -- (1.000000,2.400000);
	\draw[line width=1pt,fill=black] (1.866025,2.900000) circle (0.100000);
	\draw[line width=1pt,fill=white] (1.000000,2.400000) circle (0.200000);
	\draw[line width=1pt] (0.800000,2.400000) -- (1.200000,2.400000);
	\draw[line width=1pt] (1.000000,2.200000) -- (1.000000,2.600000);
	\draw[line width=1pt] (1.866025,3.000000) -- (1.866025,3.600000);
	\draw[line width=1pt] (1.866025,3.700000) -- (2.866025,3.700000);
	\draw[line width=1pt,fill=white] (1.866025,3.700000) circle (0.200000);
	\draw[line width=1pt] (1.666025,3.700000) -- (2.066025,3.700000);
	\draw[line width=1pt] (1.866025,3.500000) -- (1.866025,3.900000);
	\draw[line width=1pt,fill=black] (2.866025,3.700000) circle (0.100000);
	\draw[line width=1pt] (1.866025,3.800000) -- (1.866025,4.400000);
	\draw[line width=1pt] (1.866025,4.400000) -- (1.866025,4.600000);
	\draw[line width=1pt] (1.866025,4.600000) -- (1.866025,5.200000);
	\draw[line width=1pt] (1.866025,5.300000) -- (2.732051,5.800000);
	\draw[line width=1pt,fill=black] (1.866025,5.300000) circle (0.100000);
	\draw[line width=1pt,fill=white] (2.732051,5.800000) circle (0.200000);
	\draw[line width=1pt] (2.532051,5.800000) -- (2.932051,5.800000);
	\draw[line width=1pt] (2.732051,5.600000) -- (2.732051,6.000000);
	\draw[line width=1pt] (2.732051,6.000000) -- +(0.0,2.0pt);
	\draw[line width=1pt] (2.732051,5.600000) -- +(0.0,-2.0pt);
	\draw[line width=1pt] (1.866025,5.400000) -- (1.866025,6.000000);
	\draw[line width=1pt] (1.866025,6.000000) -- (1.866025,6.200000);
	\draw[line width=1pt] (1.866025,6.200000) -- (1.866025,6.800000);
	\draw[line width=1pt,fill=white] (1.766025,6.900000) -- (1.866025,7.000000) -- (1.966025,6.900000) -- (1.866025,6.800000) -- cycle;
	\draw[line width=1pt] (1.866025,7.000000) -- (1.866025,7.600000);
	\draw[line width=1pt] (1.866025,7.500000) -- (1.866025,7.900000);
	\draw[line width=1pt] (1.666025,7.500000) -- (2.066025,7.900000);
	\draw[line width=1pt] (1.666025,7.900000) -- (2.066025,7.500000);
	\draw[line width=1pt] (1.866025,7.700000) -- (2.732051,8.200000);
	\draw[line width=1pt] (2.532051,8.000000) -- (2.932051,8.400000);
	\draw[line width=1pt] (2.532051,8.400000) -- (2.932051,8.000000);
	\draw[line width=1pt] (2.732051,8.000000) -- (2.732051,8.400000);
	\draw[line width=1pt] (1.866025,7.800000) -- (1.866025,8.400000);
	\draw[line width=1pt] (1.866025,8.300000) -- (1.866025,8.700000);
	\draw[line width=1pt] (1.666025,8.300000) -- (2.066025,8.700000);
	\draw[line width=1pt] (1.666025,8.700000) -- (2.066025,8.300000);
	\draw[line width=1pt] (1.866025,8.500000) -- (2.866025,8.500000);
	\draw[line width=1pt] (2.666025,8.300000) -- (3.066025,8.700000);
	\draw[line width=1pt] (2.666025,8.700000) -- (3.066025,8.300000);
	\draw[line width=1pt] (1.866025,8.600000) -- (1.866025,9.200000);
	\draw[line width=1pt] (1.866025,9.200000) -- (1.866025,9.400000);
	\draw[line width=1pt] (1.866025,9.400000) -- (1.866025,10.000000);
	\draw[line width=1pt] (1.866025,10.000000) -- (1.866025,10.200000);
	\draw[line width=1pt] (1.866025,10.200000) -- (1.866025,10.800000);
	\draw[line width=1pt] (1.866025,10.800000) -- (1.866025,11.000000);
	\draw[line width=1pt] (1.866025,11.000000) -- (1.866025,11.600000);
	\draw[line width=1pt] (1.866025,11.600000) -- (1.866025,11.800000);
	\draw[line width=1pt] (1.866025,11.800000) -- (1.866025,12.400000);
	\draw[line width=1pt] (1.866025,12.400000) -- (1.866025,12.600000);
	\draw[line width=1pt] (1.866025,12.600000) -- (1.866025,13.200000);
	\draw[line width=1pt] (1.866025,13.200000) -- (1.866025,13.400000);
	\draw[line width=1pt] (1.866025,13.400000) -- (1.866025,14.000000);
	\draw[line width=1pt] (1.866025,14.000000) -- (1.866025,14.200000);
	\draw[line width=1pt] (1.866025,14.200000) -- (1.866025,14.800000);
	\draw[line width=1pt] (1.866025,14.800000) -- (1.866025,15.000000);
	\draw[line width=1pt] (1.866025,15.000000) -- (1.866025,15.600000);
	\draw[line width=1pt] (1.866025,15.700000) -- (1.000000,15.200000);
	\draw[line width=1pt,fill=black] (1.866025,15.700000) circle (0.100000);
	\draw[line width=1pt,fill=white] (1.000000,15.200000) circle (0.200000);
	\draw[line width=1pt] (0.800000,15.200000) -- (1.200000,15.200000);
	\draw[line width=1pt] (1.000000,15.000000) -- (1.000000,15.400000);
	\draw[line width=1pt] (1.866025,15.800000) -- (1.866025,16.400000);
	\draw[line width=1pt] (1.866025,16.500000) -- (2.866025,16.500000);
	\draw[line width=1pt,fill=white] (1.866025,16.500000) circle (0.200000);
	\draw[line width=1pt] (1.666025,16.500000) -- (2.066025,16.500000);
	\draw[line width=1pt] (1.866025,16.300000) -- (1.866025,16.700000);
	\draw[line width=1pt,fill=black] (2.866025,16.500000) circle (0.100000);
	\draw[line width=1pt] (1.866025,16.600000) -- (1.866025,17.200000);
	\draw[line width=1pt] (1.866025,17.300000) -- (0.866025,17.300000);
	\draw[line width=1pt,fill=white] (1.866025,17.300000) circle (0.200000);
	\draw[line width=1pt] (1.666025,17.300000) -- (2.066025,17.300000);
	\draw[line width=1pt] (1.866025,17.100000) -- (1.866025,17.500000);
	\draw[line width=1pt,fill=black] (0.866025,17.300000) circle (0.100000);
	\draw[line width=1pt] (0.866025,17.400000) -- +(0.0,2.0pt);
	\draw[line width=1pt] (0.866025,17.200000) -- +(0.0,-2.0pt);
	\draw[line width=1pt] (1.866025,17.400000) -- (1.866025,18.000000);
	\draw[line width=1pt] (1.866025,18.100000) -- (2.732051,18.600000);
	\draw[line width=1pt,fill=black] (1.866025,18.100000) circle (0.100000);
	\draw[line width=1pt,fill=white] (2.732051,18.600000) circle (0.200000);
	\draw[line width=1pt] (2.532051,18.600000) -- (2.932051,18.600000);
	\draw[line width=1pt] (2.732051,18.400000) -- (2.732051,18.800000);
	\draw[line width=1pt] (2.732051,18.800000) -- +(0.0,2.0pt);
	\draw[line width=1pt] (2.732051,18.400000) -- +(0.0,-2.0pt);
	\draw[line width=1pt] (1.866025,18.200000) -- (1.866025,18.800000);
	\draw[line width=1pt] (1.866025,18.800000) -- (1.866025,19.000000);
	\draw[line width=1pt] (1.866025,19.000000) -- (1.866025,19.600000);
	\draw[line width=1pt] (1.866025,19.600000) -- (1.866025,19.800000);
	\draw[line width=1pt,rounded corners,dashed] (3.232051,6.200000) -- (3.232051,19.800000) -- (2.232051,19.800000) -- (2.232051,6.200000) -- cycle;
	\draw[line width=1pt,dotted] (0.866025,19.700000) circle (0.400000);
\end{tikzpicture}

%% file: pic_swap2.tex
\begin{tikzpicture}[x=0.090000\linewidth,y=0.090000\linewidth]
	\fill[gray!80] (3.000000,17.000000) -- (3.000000,16.100000) -- (2.100000,17.000000) -- cycle;
	\fill[gray!25] (2.000000,16.000000) -- (2.000000,16.900000) -- (2.900000,16.000000) -- cycle;
	\fill[gray!25] (2.000000,16.000000) -- (2.900000,16.000000) -- (2.000000,15.100000) -- cycle;
	\fill[gray!80] (3.000000,15.000000) -- (2.100000,15.000000) -- (3.000000,15.900000) -- cycle;
	\fill[gray!80] (3.000000,15.000000) -- (3.000000,14.100000) -- (2.100000,15.000000) -- cycle;
	\fill[gray!25] (2.000000,14.000000) -- (2.000000,14.900000) -- (2.900000,14.000000) -- cycle;
	\fill[gray!80] (3.000000,5.000000) -- (3.900000,5.000000) -- (3.000000,4.100000) -- cycle;
	\fill[gray!80] (3.000000,5.000000) -- (3.000000,4.100000) -- (2.100000,5.000000) -- cycle;
	\fill[gray!25] (2.000000,4.000000) -- (2.000000,4.900000) -- (2.900000,4.000000) -- cycle;
	\fill[gray!25] (2.000000,4.000000) -- (2.900000,4.000000) -- (2.000000,3.100000) -- cycle;
	\fill[gray!25] (4.000000,4.000000) -- (3.100000,4.000000) -- (4.000000,4.900000) -- cycle;
	\fill[gray!25] (4.000000,4.000000) -- (4.000000,3.100000) -- (3.100000,4.000000) -- cycle;
	\fill[gray!80] (3.000000,3.000000) -- (2.100000,3.000000) -- (3.000000,3.900000) -- cycle;
	\fill[gray!80] (3.000000,3.000000) -- (3.000000,3.900000) -- (3.900000,3.000000) -- cycle;
	\fill[gray!80] (3.000000,3.000000) -- (3.900000,3.000000) -- (3.000000,2.100000) -- cycle;
	\fill[gray!80] (3.000000,3.000000) -- (3.000000,2.100000) -- (2.100000,3.000000) -- cycle;
	\fill[gray!25] (2.000000,2.000000) -- (2.000000,2.900000) -- (2.900000,2.000000) -- cycle;
	\fill[gray!25] (4.000000,2.000000) -- (3.100000,2.000000) -- (4.000000,2.900000) -- cycle;
	\draw[line width=1pt] (2.000000,17.000000) circle (0.100000);
	\draw[line width=1pt,fill=black] (3.000000,17.000000) circle (0.100000);
	\draw[line width=1pt,fill=black] (2.000000,16.000000) circle (0.100000);
	\draw[line width=1pt] (3.000000,16.000000) circle (0.100000);
	\draw[line width=1pt] (2.000000,15.000000) circle (0.100000);
	\draw[line width=1pt,fill=black] (3.000000,15.000000) circle (0.100000);
	\draw[line width=1pt,fill=black] (2.000000,14.000000) circle (0.100000);
	\draw[line width=1pt] (3.000000,14.000000) circle (0.100000);
	\draw[line width=1pt] (2.000000,11.000000) circle (0.100000);
	\draw[line width=1pt,fill=black] (3.000000,11.000000) circle (0.100000);
	\draw[line width=1pt] (4.000000,11.000000) circle (0.100000);
	\draw[line width=1pt,fill=black] (2.000000,10.000000) circle (0.100000);
	\draw[line width=1pt,->] (2.858579,10.141421) -- (2.141421,10.858579);
	\draw[line width=1pt] (3.000000,10.000000) circle (0.100000);
	\draw[line width=1pt] (2.000000,9.000000) circle (0.100000);
	\draw[line width=1pt,fill=black] (3.000000,9.000000) circle (0.100000);
	\draw[line width=1pt,->] (3.858579,9.141421) -- (3.141421,9.858579);
	\draw[line width=1pt] (4.000000,9.000000) circle (0.100000);
	\draw[line width=1pt,fill=black] (2.000000,8.000000) circle (0.100000);
	\draw[line width=1pt,->] (2.858579,8.141421) -- (2.141421,8.858579);
	\draw[line width=1pt] (3.000000,8.000000) circle (0.100000);
	\draw (2.000000,5.000000) node [above right] {$H$};
	\draw[line width=1pt] (2.000000,5.000000) circle (0.100000);
	\draw[line width=1pt,fill=black] (3.000000,5.000000) circle (0.100000);
	\draw (4.000000,5.000000) node [above right] {$H$};
	\draw[line width=1pt] (4.000000,5.000000) circle (0.100000);
	\draw[line width=1pt,fill=black] (2.000000,4.000000) circle (0.100000);
	\draw (3.000000,4.000000) node [above right] {$H$};
	\draw[line width=1pt] (3.000000,4.000000) circle (0.100000);
	\draw[line width=1pt,fill=black] (4.000000,4.000000) circle (0.100000);
	\draw (2.000000,3.000000) node [above right] {$H$};
	\draw[line width=1pt] (2.000000,3.000000) circle (0.100000);
	\draw[line width=1pt,fill=black] (3.000000,3.000000) circle (0.100000);
	\draw (4.000000,3.000000) node [above right] {$H$};
	\draw[line width=1pt] (4.000000,3.000000) circle (0.100000);
	\draw[line width=1pt,fill=black] (2.000000,2.000000) circle (0.100000);
	\draw (3.000000,2.000000) node [above right] {$H$};
	\draw[line width=1pt] (3.000000,2.000000) circle (0.100000);
	\draw[line width=1pt,fill=black] (4.000000,2.000000) circle (0.100000);
\end{tikzpicture}

%% file: pic_isoswap2.tex
\begin{tikzpicture}[x=0.080385\linewidth,y=0.080385\linewidth]
	\draw[line width=1pt,fill=black] (0.000000,0.000000) circle (0.100000);
	\draw[line width=1pt] (0.866025,0.500000) circle (0.100000);
	\draw[line width=1pt,fill=black] (1.732051,1.000000) circle (0.100000);
	\draw[line width=1pt] (2.598076,1.500000) circle (0.100000);
	\draw[line width=1pt] (1.000000,0.000000) circle (0.100000);
	\draw[line width=1pt] (0.966025,0.500000) -- (1.766025,0.500000);
	\draw[line width=1pt] (1.779423,0.450000) -- (1.086603,0.050000);
	\draw[line width=1pt,fill=black] (1.866025,0.500000) circle (0.100000);
	\draw[line width=1pt] (2.645448,0.950000) -- (1.952628,0.550000);
	\draw[line width=1pt] (2.732051,1.000000) circle (0.100000);
	\draw[line width=1pt] (2.698076,1.500000) -- (3.498076,1.500000);
	\draw[line width=1pt] (3.511474,1.450000) -- (2.818653,1.050000);
	\draw[line width=1pt,fill=black] (3.598076,1.500000) circle (0.100000);
	\draw[line width=1pt,fill=black] (2.000000,0.000000) circle (0.100000);
	\draw[line width=1pt] (1.966025,0.500000) -- (2.766025,0.500000);
	\draw[line width=1pt] (2.866025,0.500000) circle (0.100000);
	\draw[line width=1pt,fill=black] (3.732051,1.000000) circle (0.100000);
	\draw[line width=1pt] (3.698076,1.500000) -- (4.498076,1.500000);
	\draw[line width=1pt] (4.598076,1.500000) circle (0.100000);
	\draw[line width=3pt,draw=white] (2.732051,1.000000) -- (2.732051,1.700000);
	\draw[line width=3pt,draw=white] (2.732051,1.700000) -- (2.732051,1.900000);
	\draw[line width=3pt,draw=white] (2.732051,1.900000) -- (2.732051,2.500000);
	\draw[line width=3pt,draw=white] (2.732051,2.500000) -- (2.732051,2.700000);
	\draw[line width=3pt,draw=white] (2.732051,2.700000) -- (2.732051,3.300000);
	\draw[line width=3pt,draw=white,fill=white] (2.732051,3.400000) circle (0.200000);
	\draw[line width=3pt,draw=white] (2.532051,3.400000) -- (2.932051,3.400000);
	\draw[line width=3pt,draw=white] (2.732051,3.200000) -- (2.732051,3.600000);
	\draw[line width=3pt,draw=white] (2.732051,3.400000) -- (1.866025,2.900000);
	\draw[line width=3pt,draw=white,fill=white] (1.866025,2.900000) circle (0.100000);
	\draw[line width=3pt,draw=white] (2.732051,3.500000) -- (2.732051,4.100000);
	\draw[line width=3pt,draw=white,fill=white] (2.732051,4.200000) circle (0.100000);
	\draw[line width=3pt,draw=white] (2.732051,4.200000) -- (3.732051,4.200000);
	\draw[line width=3pt,draw=white,fill=white] (3.732051,4.200000) circle (0.200000);
	\draw[line width=3pt,draw=white] (3.532051,4.200000) -- (3.932051,4.200000);
	\draw[line width=3pt,draw=white] (3.732051,4.000000) -- (3.732051,4.400000);
	\draw[line width=3pt,draw=white] (2.732051,4.300000) -- (2.732051,4.900000);
	\draw[line width=3pt,draw=white,fill=white] (2.732051,5.000000) circle (0.100000);
	\draw[line width=3pt,draw=white] (2.732051,5.000000) -- (1.732051,5.000000);
	\draw[line width=3pt,draw=white,fill=white] (1.732051,5.000000) circle (0.200000);
	\draw[line width=3pt,draw=white] (1.532051,5.000000) -- (1.932051,5.000000);
	\draw[line width=3pt,draw=white] (1.732051,4.800000) -- (1.732051,5.200000);
	\draw[line width=3pt,draw=white] (2.732051,5.100000) -- (2.732051,5.700000);
	\draw[line width=3pt,draw=white,fill=white] (2.732051,5.800000) circle (0.200000);
	\draw[line width=3pt,draw=white] (2.532051,5.800000) -- (2.932051,5.800000);
	\draw[line width=3pt,draw=white] (2.732051,5.600000) -- (2.732051,6.000000);
	\draw[line width=3pt,draw=white] (2.732051,5.800000) -- (3.598076,6.300000);
	\draw[line width=3pt,draw=white,fill=white] (3.598076,6.300000) circle (0.100000);
	\draw[line width=3pt,draw=white] (2.732051,5.900000) -- (2.732051,6.500000);
	\draw[line width=3pt,draw=white] (2.732051,6.500000) -- (2.732051,6.700000);
	\draw[line width=3pt,draw=white] (2.732051,6.700000) -- (2.732051,7.300000);
	\draw[line width=3pt,fill=white,draw=white] (2.632051,7.400000) -- (2.732051,7.500000) -- (2.832051,7.400000) -- (2.732051,7.300000) -- cycle;
	\draw[line width=3pt,draw=white] (2.732051,7.500000) -- (2.732051,8.100000);
	\draw[line width=3pt,draw=white] (2.732051,8.000000) -- (2.732051,8.400000);
	\draw[line width=3pt,draw=white] (2.532051,8.000000) -- (2.932051,8.400000);
	\draw[line width=3pt,draw=white] (2.532051,8.400000) -- (2.932051,8.000000);
	\draw[line width=3pt,draw=white] (2.732051,8.200000) -- (3.598076,8.700000);
	\draw[line width=3pt,draw=white] (3.398076,8.500000) -- (3.798076,8.900000);
	\draw[line width=3pt,draw=white] (3.398076,8.900000) -- (3.798076,8.500000);
	\draw[line width=3pt,draw=white] (2.732051,8.300000) -- (2.732051,8.900000);
	\draw[line width=3pt,draw=white] (2.732051,8.800000) -- (2.732051,9.200000);
	\draw[line width=3pt,draw=white] (2.532051,8.800000) -- (2.932051,9.200000);
	\draw[line width=3pt,draw=white] (2.532051,9.200000) -- (2.932051,8.800000);
	\draw[line width=3pt,draw=white] (2.732051,9.000000) -- (3.732051,9.000000);
	\draw[line width=3pt,draw=white] (3.532051,8.800000) -- (3.932051,9.200000);
	\draw[line width=3pt,draw=white] (3.532051,9.200000) -- (3.932051,8.800000);
	\draw[line width=3pt,draw=white] (2.732051,9.100000) -- (2.732051,9.700000);
	\draw[line width=3pt,draw=white] (2.732051,9.700000) -- (2.732051,9.900000);
	\draw[line width=3pt,draw=white] (2.732051,9.900000) -- (2.732051,10.500000);
	\draw[line width=3pt,draw=white] (2.732051,10.500000) -- (2.732051,10.700000);
	\draw[line width=3pt,draw=white] (2.732051,10.700000) -- (2.732051,11.300000);
	\draw[line width=3pt,draw=white] (2.732051,11.300000) -- (2.732051,11.500000);
	\draw[line width=3pt,draw=white] (2.732051,11.500000) -- (2.732051,12.100000);
	\draw[line width=3pt,draw=white] (2.732051,12.100000) -- (2.732051,12.300000);
	\draw[line width=3pt,draw=white] (2.732051,12.300000) -- (2.732051,12.900000);
	\draw[line width=3pt,draw=white] (2.732051,12.900000) -- (2.732051,13.100000);
	\draw[line width=3pt,draw=white] (2.732051,13.100000) -- (2.732051,13.700000);
	\draw[line width=3pt,draw=white] (2.732051,13.700000) -- (2.732051,13.900000);
	\draw[line width=3pt,draw=white] (2.732051,13.900000) -- (2.732051,14.500000);
	\draw[line width=3pt,draw=white] (2.732051,14.500000) -- (2.732051,14.700000);
	\draw[line width=3pt,draw=white] (2.732051,14.700000) -- (2.732051,15.300000);
	\draw[line width=3pt,draw=white] (2.732051,15.300000) -- (2.732051,15.500000);
	\draw[line width=3pt,draw=white] (2.732051,15.500000) -- (2.732051,16.100000);
	\draw[line width=3pt,draw=white,fill=white] (2.732051,16.200000) circle (0.200000);
	\draw[line width=3pt,draw=white] (2.532051,16.200000) -- (2.932051,16.200000);
	\draw[line width=3pt,draw=white] (2.732051,16.000000) -- (2.732051,16.400000);
	\draw[line width=3pt,draw=white] (2.732051,16.200000) -- (1.866025,15.700000);
	\draw[line width=3pt,draw=white,fill=white] (1.866025,15.700000) circle (0.100000);
	\draw[line width=3pt,draw=white] (2.732051,16.300000) -- (2.732051,16.900000);
	\draw[line width=3pt,draw=white] (2.732051,16.900000) -- (2.732051,17.100000);
	\draw[line width=3pt,draw=white] (2.732051,17.100000) -- (2.732051,17.700000);
	\draw[line width=3pt,draw=white,fill=white] (2.732051,17.800000) circle (0.100000);
	\draw[line width=3pt,draw=white] (2.732051,17.800000) -- (1.732051,17.800000);
	\draw[line width=3pt,draw=white,fill=white] (1.732051,17.800000) circle (0.200000);
	\draw[line width=3pt,draw=white] (1.532051,17.800000) -- (1.932051,17.800000);
	\draw[line width=3pt,draw=white] (1.732051,17.600000) -- (1.732051,18.000000);
	\draw[line width=3pt,draw=white] (2.732051,17.900000) -- (2.732051,18.500000);
	\draw[line width=3pt,draw=white,fill=white] (2.732051,18.600000) circle (0.200000);
	\draw[line width=3pt,draw=white] (2.532051,18.600000) -- (2.932051,18.600000);
	\draw[line width=3pt,draw=white] (2.732051,18.400000) -- (2.732051,18.800000);
	\draw[line width=3pt,draw=white] (2.732051,18.600000) -- (3.598076,19.100000);
	\draw[line width=3pt,draw=white,fill=white] (3.598076,19.100000) circle (0.100000);
	\draw[line width=3pt,draw=white] (2.732051,18.700000) -- (2.732051,19.300000);
	\draw[line width=3pt,draw=white] (2.732051,19.300000) -- (2.732051,19.500000);
	\draw[line width=3pt,draw=white] (2.732051,19.500000) -- (2.732051,20.100000);
	\draw[line width=3pt,draw=white] (2.732051,20.100000) -- (2.732051,20.300000);
	\draw[line width=1pt] (2.732051,1.000000) -- (2.732051,1.700000);
	\draw[line width=1pt] (2.732051,1.700000) -- (2.732051,1.900000);
	\draw[line width=1pt] (2.732051,1.900000) -- (2.732051,2.500000);
	\draw[line width=1pt] (2.732051,2.500000) -- (2.732051,2.700000);
	\draw[line width=1pt] (2.732051,2.700000) -- (2.732051,3.300000);
	\draw[line width=1pt] (2.732051,3.400000) -- (1.866025,2.900000);
	\draw[line width=1pt,fill=white] (2.732051,3.400000) circle (0.200000);
	\draw[line width=1pt] (2.532051,3.400000) -- (2.932051,3.400000);
	\draw[line width=1pt] (2.732051,3.200000) -- (2.732051,3.600000);
	\draw[line width=1pt,fill=black] (1.866025,2.900000) circle (0.100000);
	\draw[line width=1pt] (1.866025,3.000000) -- +(0.0,2.0pt);
	\draw[line width=1pt] (1.866025,2.800000) -- +(0.0,-2.0pt);
	\draw[line width=1pt] (2.732051,3.500000) -- (2.732051,4.100000);
	\draw[line width=1pt] (2.732051,4.200000) -- (3.732051,4.200000);
	\draw[line width=1pt,fill=black] (2.732051,4.200000) circle (0.100000);
	\draw[line width=1pt,fill=white] (3.732051,4.200000) circle (0.200000);
	\draw[line width=1pt] (3.532051,4.200000) -- (3.932051,4.200000);
	\draw[line width=1pt] (3.732051,4.000000) -- (3.732051,4.400000);
	\draw[line width=1pt] (3.732051,4.400000) -- +(0.0,2.0pt);
	\draw[line width=1pt] (3.732051,4.000000) -- +(0.0,-2.0pt);
	\draw[line width=1pt] (2.732051,4.300000) -- (2.732051,4.900000);
	\draw[line width=1pt] (2.732051,5.000000) -- (1.732051,5.000000);
	\draw[line width=1pt,fill=black] (2.732051,5.000000) circle (0.100000);
	\draw[line width=1pt,fill=white] (1.732051,5.000000) circle (0.200000);
	\draw[line width=1pt] (1.532051,5.000000) -- (1.932051,5.000000);
	\draw[line width=1pt] (1.732051,4.800000) -- (1.732051,5.200000);
	\draw[line width=1pt] (2.732051,5.100000) -- (2.732051,5.700000);
	\draw[line width=1pt] (2.732051,5.800000) -- (3.598076,6.300000);
	\draw[line width=1pt,fill=white] (2.732051,5.800000) circle (0.200000);
	\draw[line width=1pt] (2.532051,5.800000) -- (2.932051,5.800000);
	\draw[line width=1pt] (2.732051,5.600000) -- (2.732051,6.000000);
	\draw[line width=1pt,fill=black] (3.598076,6.300000) circle (0.100000);
	\draw[line width=1pt] (2.732051,5.900000) -- (2.732051,6.500000);
	\draw[line width=1pt] (2.732051,6.500000) -- (2.732051,6.700000);
	\draw[line width=1pt] (2.732051,6.700000) -- (2.732051,7.300000);
	\draw[line width=1pt,fill=white] (2.632051,7.400000) -- (2.732051,7.500000) -- (2.832051,7.400000) -- (2.732051,7.300000) -- cycle;
	\draw[line width=1pt] (2.732051,7.500000) -- (2.732051,8.100000);
	\draw[line width=1pt] (2.732051,8.000000) -- (2.732051,8.400000);
	\draw[line width=1pt] (2.532051,8.000000) -- (2.932051,8.400000);
	\draw[line width=1pt] (2.532051,8.400000) -- (2.932051,8.000000);
	\draw[line width=1pt] (2.732051,8.200000) -- (3.598076,8.700000);
	\draw[line width=1pt] (3.398076,8.500000) -- (3.798076,8.900000);
	\draw[line width=1pt] (3.398076,8.900000) -- (3.798076,8.500000);
	\draw[line width=1pt] (2.732051,8.300000) -- (2.732051,8.900000);
	\draw[line width=1pt] (2.732051,8.800000) -- (2.732051,9.200000);
	\draw[line width=1pt] (2.532051,8.800000) -- (2.932051,9.200000);
	\draw[line width=1pt] (2.532051,9.200000) -- (2.932051,8.800000);
	\draw[line width=1pt] (2.732051,9.000000) -- (3.732051,9.000000);
	\draw[line width=1pt] (3.532051,8.800000) -- (3.932051,9.200000);
	\draw[line width=1pt] (3.532051,9.200000) -- (3.932051,8.800000);
	\draw[line width=1pt] (3.732051,8.800000) -- (3.732051,9.200000);
	\draw[line width=1pt] (2.732051,9.100000) -- (2.732051,9.700000);
	\draw[line width=1pt] (2.732051,9.700000) -- (2.732051,9.900000);
	\draw[line width=1pt] (2.732051,9.900000) -- (2.732051,10.500000);
	\draw[line width=1pt] (2.732051,10.500000) -- (2.732051,10.700000);
	\draw[line width=1pt] (2.732051,10.700000) -- (2.732051,11.300000);
	\draw[line width=1pt] (2.732051,11.300000) -- (2.732051,11.500000);
	\draw[line width=1pt] (2.732051,11.500000) -- (2.732051,12.100000);
	\draw[line width=1pt] (2.732051,12.100000) -- (2.732051,12.300000);
	\draw[line width=1pt] (2.732051,12.300000) -- (2.732051,12.900000);
	\draw[line width=1pt] (2.732051,12.900000) -- (2.732051,13.100000);
	\draw[line width=1pt] (2.732051,13.100000) -- (2.732051,13.700000);
	\draw[line width=1pt] (2.732051,13.700000) -- (2.732051,13.900000);
	\draw[line width=1pt] (2.732051,13.900000) -- (2.732051,14.500000);
	\draw[line width=1pt] (2.732051,14.500000) -- (2.732051,14.700000);
	\draw[line width=1pt] (2.732051,14.700000) -- (2.732051,15.300000);
	\draw[line width=1pt] (2.732051,15.300000) -- (2.732051,15.500000);
	\draw[line width=1pt] (2.732051,15.500000) -- (2.732051,16.100000);
	\draw[line width=1pt] (2.732051,16.200000) -- (1.866025,15.700000);
	\draw[line width=1pt,fill=white] (2.732051,16.200000) circle (0.200000);
	\draw[line width=1pt] (2.532051,16.200000) -- (2.932051,16.200000);
	\draw[line width=1pt] (2.732051,16.000000) -- (2.732051,16.400000);
	\draw[line width=1pt,fill=black] (1.866025,15.700000) circle (0.100000);
	\draw[line width=1pt] (1.866025,15.800000) -- +(0.0,2.0pt);
	\draw[line width=1pt] (1.866025,15.600000) -- +(0.0,-2.0pt);
	\draw[line width=1pt] (2.732051,16.300000) -- (2.732051,16.900000);
	\draw[line width=1pt] (2.732051,16.900000) -- (2.732051,17.100000);
	\draw[line width=1pt] (2.732051,17.100000) -- (2.732051,17.700000);
	\draw[line width=1pt] (2.732051,17.800000) -- (1.732051,17.800000);
	\draw[line width=1pt,fill=black] (2.732051,17.800000) circle (0.100000);
	\draw[line width=1pt,fill=white] (1.732051,17.800000) circle (0.200000);
	\draw[line width=1pt] (1.532051,17.800000) -- (1.932051,17.800000);
	\draw[line width=1pt] (1.732051,17.600000) -- (1.732051,18.000000);
	\draw[line width=1pt] (2.732051,17.900000) -- (2.732051,18.500000);
	\draw[line width=1pt] (2.732051,18.600000) -- (3.598076,19.100000);
	\draw[line width=1pt,fill=white] (2.732051,18.600000) circle (0.200000);
	\draw[line width=1pt] (2.532051,18.600000) -- (2.932051,18.600000);
	\draw[line width=1pt] (2.732051,18.400000) -- (2.732051,18.800000);
	\draw[line width=1pt,fill=black] (3.598076,19.100000) circle (0.100000);
	\draw[line width=1pt] (2.732051,18.700000) -- (2.732051,19.300000);
	\draw[line width=1pt] (2.732051,19.300000) -- (2.732051,19.500000);
	\draw[line width=1pt] (2.732051,19.500000) -- (2.732051,20.100000);
	\draw[line width=1pt] (2.732051,20.100000) -- (2.732051,20.300000);
	\draw[line width=3pt,draw=white] (3.732051,1.000000) -- (3.732051,1.700000);
	\draw[line width=3pt,draw=white] (3.732051,1.700000) -- (3.732051,1.900000);
	\draw[line width=3pt,draw=white] (3.732051,1.900000) -- (3.732051,2.500000);
	\draw[line width=3pt,fill=white,draw=white] (3.632051,2.500000) -- (3.632051,2.700000) -- (3.832051,2.700000) -- (3.832051,2.500000) -- cycle;
	\draw[line width=3pt,draw=white] (3.732051,2.700000) -- (3.732051,3.300000);
	\draw[line width=3pt,draw=white,fill=white] (3.732051,3.400000) circle (0.200000);
	\draw[line width=3pt,draw=white] (3.532051,3.400000) -- (3.932051,3.400000);
	\draw[line width=3pt,draw=white] (3.732051,3.200000) -- (3.732051,3.600000);
	\draw[line width=3pt,draw=white] (3.732051,3.400000) -- (4.598076,3.900000);
	\draw[line width=3pt,draw=white,fill=white] (4.598076,3.900000) circle (0.100000);
	\draw[line width=3pt,draw=white] (3.732051,3.500000) -- (3.732051,4.100000);
	\draw[line width=3pt,draw=white,fill=white] (3.732051,4.200000) circle (0.200000);
	\draw[line width=3pt,draw=white] (3.532051,4.200000) -- (3.932051,4.200000);
	\draw[line width=3pt,draw=white] (3.732051,4.000000) -- (3.732051,4.400000);
	\draw[line width=3pt,draw=white] (3.732051,4.200000) -- (2.732051,4.200000);
	\draw[line width=3pt,draw=white,fill=white] (2.732051,4.200000) circle (0.100000);
	\draw[line width=3pt,draw=white] (3.732051,4.300000) -- (3.732051,4.900000);
	\draw[line width=3pt,draw=white] (3.732051,4.900000) -- (3.732051,5.100000);
	\draw[line width=3pt,draw=white] (3.732051,5.100000) -- (3.732051,5.700000);
	\draw[line width=3pt,draw=white,fill=white] (3.732051,5.800000) circle (0.200000);
	\draw[line width=3pt,draw=white] (3.532051,5.800000) -- (3.932051,5.800000);
	\draw[line width=3pt,draw=white] (3.732051,5.600000) -- (3.732051,6.000000);
	\draw[line width=3pt,draw=white] (3.732051,5.800000) -- (2.866025,5.300000);
	\draw[line width=3pt,draw=white,fill=white] (2.866025,5.300000) circle (0.100000);
	\draw[line width=3pt,draw=white] (3.732051,5.900000) -- (3.732051,6.500000);
	\draw[line width=3pt,fill=white,draw=white] (3.732051,6.600000) circle (0.100000);
	\draw[line width=3pt,draw=white] (3.732051,6.700000) -- (3.732051,7.300000);
	\draw[line width=3pt,draw=white] (3.732051,7.300000) -- (3.732051,7.500000);
	\draw[line width=3pt,draw=white] (3.732051,7.500000) -- (3.732051,8.100000);
	\draw[line width=3pt,draw=white] (3.732051,8.000000) -- (3.732051,8.400000);
	\draw[line width=3pt,draw=white] (3.532051,8.000000) -- (3.932051,8.400000);
	\draw[line width=3pt,draw=white] (3.532051,8.400000) -- (3.932051,8.000000);
	\draw[line width=3pt,draw=white] (3.732051,8.200000) -- (2.866025,7.700000);
	\draw[line width=3pt,draw=white] (2.666025,7.500000) -- (3.066025,7.900000);
	\draw[line width=3pt,draw=white] (2.666025,7.900000) -- (3.066025,7.500000);
	\draw[line width=3pt,draw=white] (3.732051,8.300000) -- (3.732051,8.900000);
	\draw[line width=3pt,draw=white] (3.732051,8.800000) -- (3.732051,9.200000);
	\draw[line width=3pt,draw=white] (3.532051,8.800000) -- (3.932051,9.200000);
	\draw[line width=3pt,draw=white] (3.532051,9.200000) -- (3.932051,8.800000);
	\draw[line width=3pt,draw=white] (3.732051,9.000000) -- (2.732051,9.000000);
	\draw[line width=3pt,draw=white] (2.532051,8.800000) -- (2.932051,9.200000);
	\draw[line width=3pt,draw=white] (2.532051,9.200000) -- (2.932051,8.800000);
	\draw[line width=3pt,draw=white] (3.732051,9.100000) -- (3.732051,9.700000);
	\draw[line width=3pt,draw=white] (3.732051,9.700000) -- (3.732051,9.900000);
	\draw[line width=3pt,draw=white] (3.732051,9.900000) -- (3.732051,10.500000);
	\draw[line width=3pt,draw=white] (3.732051,10.500000) -- (3.732051,10.700000);
	\draw[line width=3pt,draw=white] (3.732051,10.700000) -- (3.732051,11.300000);
	\draw[line width=3pt,draw=white] (3.732051,11.300000) -- (3.732051,11.500000);
	\draw[line width=3pt,draw=white] (3.732051,11.500000) -- (3.732051,12.100000);
	\draw[line width=3pt,draw=white] (3.732051,12.100000) -- (3.732051,12.300000);
	\draw[line width=3pt,draw=white] (3.732051,12.300000) -- (3.732051,12.900000);
	\draw[line width=3pt,draw=white] (3.732051,12.900000) -- (3.732051,13.100000);
	\draw[line width=3pt,draw=white] (3.732051,13.100000) -- (3.732051,13.700000);
	\draw[line width=3pt,draw=white] (3.732051,13.700000) -- (3.732051,13.900000);
	\draw[line width=3pt,draw=white] (3.732051,13.900000) -- (3.732051,14.500000);
	\draw[line width=3pt,draw=white] (3.732051,14.500000) -- (3.732051,14.700000);
	\draw[line width=3pt,draw=white] (3.732051,14.700000) -- (3.732051,15.300000);
	\draw[line width=3pt,draw=white] (3.732051,15.300000) -- (3.732051,15.500000);
	\draw[line width=3pt,draw=white] (3.732051,15.500000) -- (3.732051,16.100000);
	\draw[line width=3pt,draw=white] (3.732051,16.100000) -- (3.732051,16.300000);
	\draw[line width=3pt,draw=white] (3.732051,16.300000) -- (3.732051,16.900000);
	\draw[line width=3pt,draw=white] (3.732051,16.900000) -- (3.732051,17.100000);
	\draw[line width=3pt,draw=white] (3.732051,17.100000) -- (3.732051,17.700000);
	\draw[line width=3pt,draw=white] (3.732051,17.700000) -- (3.732051,17.900000);
	\draw[line width=3pt,draw=white] (3.732051,17.900000) -- (3.732051,18.500000);
	\draw[line width=3pt,draw=white] (3.732051,18.500000) -- (3.732051,18.700000);
	\draw[line width=3pt,draw=white] (3.732051,18.700000) -- (3.732051,19.300000);
	\draw[line width=3pt,draw=white] (3.732051,19.300000) -- (3.732051,19.500000);
	\draw[line width=3pt,draw=white] (3.732051,19.500000) -- (3.732051,20.100000);
	\draw[line width=3pt,draw=white] (3.732051,20.100000) -- (3.732051,20.300000);
	\draw[line width=1pt] (3.732051,1.000000) -- (3.732051,1.700000);
	\draw[line width=1pt] (3.732051,1.700000) -- (3.732051,1.900000);
	\draw[line width=1pt] (3.732051,1.900000) -- (3.732051,2.500000);
	\draw[line width=1pt,fill=white] (3.632051,2.500000) -- (3.632051,2.700000) -- (3.832051,2.700000) -- (3.832051,2.500000) -- cycle;
	\draw[line width=1pt] (3.732051,2.700000) -- (3.732051,3.300000);
	\draw[line width=1pt] (3.732051,3.400000) -- (4.598076,3.900000);
	\draw[line width=1pt,fill=white] (3.732051,3.400000) circle (0.200000);
	\draw[line width=1pt] (3.532051,3.400000) -- (3.932051,3.400000);
	\draw[line width=1pt] (3.732051,3.200000) -- (3.732051,3.600000);
	\draw[line width=1pt,fill=black] (4.598076,3.900000) circle (0.100000);
	\draw[line width=1pt] (3.732051,3.500000) -- (3.732051,4.100000);
	\draw[line width=1pt] (3.732051,4.200000) -- (2.732051,4.200000);
	\draw[line width=1pt,fill=white] (3.732051,4.200000) circle (0.200000);
	\draw[line width=1pt] (3.532051,4.200000) -- (3.932051,4.200000);
	\draw[line width=1pt] (3.732051,4.000000) -- (3.732051,4.400000);
	\draw[line width=1pt,fill=black] (2.732051,4.200000) circle (0.100000);
	\draw[line width=1pt] (2.732051,4.300000) -- +(0.0,2.0pt);
	\draw[line width=1pt] (2.732051,4.100000) -- +(0.0,-2.0pt);
	\draw[line width=1pt] (3.732051,4.300000) -- (3.732051,4.900000);
	\draw[line width=1pt] (3.732051,4.900000) -- (3.732051,5.100000);
	\draw[line width=1pt] (3.732051,5.100000) -- (3.732051,5.700000);
	\draw[line width=1pt] (3.732051,5.800000) -- (2.866025,5.300000);
	\draw[line width=1pt,fill=white] (3.732051,5.800000) circle (0.200000);
	\draw[line width=1pt] (3.532051,5.800000) -- (3.932051,5.800000);
	\draw[line width=1pt] (3.732051,5.600000) -- (3.732051,6.000000);
	\draw[line width=1pt,fill=black] (2.866025,5.300000) circle (0.100000);
	\draw[line width=1pt] (2.866025,5.400000) -- +(0.0,2.0pt);
	\draw[line width=1pt] (2.866025,5.200000) -- +(0.0,-2.0pt);
	\draw[line width=1pt] (3.732051,5.900000) -- (3.732051,6.500000);
	\draw[line width=1pt,fill=white] (3.732051,6.600000) circle (0.100000);
	\draw[line width=1pt] (3.732051,6.700000) -- (3.732051,7.300000);
	\draw[line width=1pt] (3.732051,7.300000) -- (3.732051,7.500000);
	\draw[line width=1pt] (3.732051,7.500000) -- (3.732051,8.100000);
	\draw[line width=1pt] (3.732051,8.000000) -- (3.732051,8.400000);
	\draw[line width=1pt] (3.532051,8.000000) -- (3.932051,8.400000);
	\draw[line width=1pt] (3.532051,8.400000) -- (3.932051,8.000000);
	\draw[line width=1pt] (3.732051,8.200000) -- (2.866025,7.700000);
	\draw[line width=1pt] (2.666025,7.500000) -- (3.066025,7.900000);
	\draw[line width=1pt] (2.666025,7.900000) -- (3.066025,7.500000);
	\draw[line width=1pt] (2.866025,7.500000) -- (2.866025,7.900000);
	\draw[line width=1pt] (3.732051,8.300000) -- (3.732051,8.900000);
	\draw[line width=1pt] (3.732051,8.800000) -- (3.732051,9.200000);
	\draw[line width=1pt] (3.532051,8.800000) -- (3.932051,9.200000);
	\draw[line width=1pt] (3.532051,9.200000) -- (3.932051,8.800000);
	\draw[line width=1pt] (3.732051,9.000000) -- (2.732051,9.000000);
	\draw[line width=1pt] (2.532051,8.800000) -- (2.932051,9.200000);
	\draw[line width=1pt] (2.532051,9.200000) -- (2.932051,8.800000);
	\draw[line width=1pt] (2.732051,8.800000) -- (2.732051,9.200000);
	\draw[line width=1pt] (3.732051,9.100000) -- (3.732051,9.700000);
	\draw[line width=1pt] (3.732051,9.700000) -- (3.732051,9.900000);
	\draw[line width=1pt] (3.732051,9.900000) -- (3.732051,10.500000);
	\draw[line width=1pt] (3.732051,10.500000) -- (3.732051,10.700000);
	\draw[line width=1pt] (3.732051,10.700000) -- (3.732051,11.300000);
	\draw[line width=1pt] (3.732051,11.300000) -- (3.732051,11.500000);
	\draw[line width=1pt] (3.732051,11.500000) -- (3.732051,12.100000);
	\draw[line width=1pt] (3.732051,12.100000) -- (3.732051,12.300000);
	\draw[line width=1pt] (3.732051,12.300000) -- (3.732051,12.900000);
	\draw[line width=1pt] (3.732051,12.900000) -- (3.732051,13.100000);
	\draw[line width=1pt] (3.732051,13.100000) -- (3.732051,13.700000);
	\draw[line width=1pt] (3.732051,13.700000) -- (3.732051,13.900000);
	\draw[line width=1pt] (3.732051,13.900000) -- (3.732051,14.500000);
	\draw[line width=1pt] (3.732051,14.500000) -- (3.732051,14.700000);
	\draw[line width=1pt] (3.732051,14.700000) -- (3.732051,15.300000);
	\draw[line width=1pt] (3.732051,15.300000) -- (3.732051,15.500000);
	\draw[line width=1pt] (3.732051,15.500000) -- (3.732051,16.100000);
	\draw[line width=1pt] (3.732051,16.100000) -- (3.732051,16.300000);
	\draw[line width=1pt] (3.732051,16.300000) -- (3.732051,16.900000);
	\draw[line width=1pt] (3.732051,16.900000) -- (3.732051,17.100000);
	\draw[line width=1pt] (3.732051,17.100000) -- (3.732051,17.700000);
	\draw[line width=1pt] (3.732051,17.700000) -- (3.732051,17.900000);
	\draw[line width=1pt] (3.732051,17.900000) -- (3.732051,18.500000);
	\draw[line width=1pt] (3.732051,18.500000) -- (3.732051,18.700000);
	\draw[line width=1pt] (3.732051,18.700000) -- (3.732051,19.300000);
	\draw[line width=1pt] (3.732051,19.300000) -- (3.732051,19.500000);
	\draw[line width=1pt] (3.732051,19.500000) -- (3.732051,20.100000);
	\draw[line width=1pt] (3.732051,20.100000) -- (3.732051,20.300000);
	\draw[line width=3pt,draw=white] (1.866025,0.500000) -- (1.866025,1.200000);
	\draw[line width=3pt,fill=white,draw=white] (1.766025,1.200000) -- (1.766025,1.400000) -- (1.966025,1.400000) -- (1.966025,1.200000) -- cycle;
	\draw[line width=3pt,draw=white] (1.866025,1.400000) -- (1.866025,2.000000);
	\draw[line width=3pt,fill=white,draw=white] (1.766025,2.100000) -- (1.866025,2.200000) -- (1.966025,2.100000) -- (1.866025,2.000000) -- cycle;
	\draw[line width=3pt,draw=white] (1.866025,2.200000) -- (1.866025,2.800000);
	\draw[line width=3pt,draw=white,fill=white] (1.866025,2.900000) circle (0.100000);
	\draw[line width=3pt,draw=white] (1.866025,2.900000) -- (2.732051,3.400000);
	\draw[line width=3pt,draw=white,fill=white] (2.732051,3.400000) circle (0.200000);
	\draw[line width=3pt,draw=white] (2.532051,3.400000) -- (2.932051,3.400000);
	\draw[line width=3pt,draw=white] (2.732051,3.200000) -- (2.732051,3.600000);
	\draw[line width=3pt,draw=white] (1.866025,3.000000) -- (1.866025,3.600000);
	\draw[line width=3pt,draw=white,fill=white] (1.866025,3.700000) circle (0.100000);
	\draw[line width=3pt,draw=white] (1.866025,3.700000) -- (0.866025,3.700000);
	\draw[line width=3pt,draw=white,fill=white] (0.866025,3.700000) circle (0.200000);
	\draw[line width=3pt,draw=white] (0.666025,3.700000) -- (1.066025,3.700000);
	\draw[line width=3pt,draw=white] (0.866025,3.500000) -- (0.866025,3.900000);
	\draw[line width=3pt,draw=white] (1.866025,3.800000) -- (1.866025,4.400000);
	\draw[line width=3pt,draw=white,fill=white] (1.866025,4.500000) circle (0.100000);
	\draw[line width=3pt,draw=white] (1.866025,4.500000) -- (2.866025,4.500000);
	\draw[line width=3pt,draw=white,fill=white] (2.866025,4.500000) circle (0.200000);
	\draw[line width=3pt,draw=white] (2.666025,4.500000) -- (3.066025,4.500000);
	\draw[line width=3pt,draw=white] (2.866025,4.300000) -- (2.866025,4.700000);
	\draw[line width=3pt,draw=white] (1.866025,4.600000) -- (1.866025,5.200000);
	\draw[line width=3pt,draw=white,fill=white] (1.866025,5.300000) circle (0.100000);
	\draw[line width=3pt,draw=white] (1.866025,5.300000) -- (1.000000,4.800000);
	\draw[line width=3pt,draw=white,fill=white] (1.000000,4.800000) circle (0.200000);
	\draw[line width=3pt,draw=white] (0.800000,4.800000) -- (1.200000,4.800000);
	\draw[line width=3pt,draw=white] (1.000000,4.600000) -- (1.000000,5.000000);
	\draw[line width=3pt,draw=white] (1.866025,5.400000) -- (1.866025,6.000000);
	\draw[line width=3pt,fill=white,draw=white] (1.766025,6.100000) -- (1.866025,6.200000) -- (1.966025,6.100000) -- (1.866025,6.000000) -- cycle;
	\draw[line width=3pt,draw=white] (1.866025,6.200000) -- (1.866025,6.800000);
	\draw[line width=3pt,fill=white,draw=white] (1.866025,6.900000) circle (0.100000);
	\draw[line width=3pt,draw=white] (1.866025,7.000000) -- (1.866025,7.600000);
	\draw[line width=3pt,draw=white] (1.866025,7.500000) -- (1.866025,7.900000);
	\draw[line width=3pt,draw=white] (1.666025,7.500000) -- (2.066025,7.900000);
	\draw[line width=3pt,draw=white] (1.666025,7.900000) -- (2.066025,7.500000);
	\draw[line width=3pt,draw=white] (1.866025,7.700000) -- (1.000000,7.200000);
	\draw[line width=3pt,draw=white] (0.800000,7.000000) -- (1.200000,7.400000);
	\draw[line width=3pt,draw=white] (0.800000,7.400000) -- (1.200000,7.000000);
	\draw[line width=3pt,draw=white] (1.866025,7.800000) -- (1.866025,8.400000);
	\draw[line width=3pt,draw=white] (1.866025,8.300000) -- (1.866025,8.700000);
	\draw[line width=3pt,draw=white] (1.666025,8.300000) -- (2.066025,8.700000);
	\draw[line width=3pt,draw=white] (1.666025,8.700000) -- (2.066025,8.300000);
	\draw[line width=3pt,draw=white] (1.866025,8.500000) -- (0.866025,8.500000);
	\draw[line width=3pt,draw=white] (0.666025,8.300000) -- (1.066025,8.700000);
	\draw[line width=3pt,draw=white] (0.666025,8.700000) -- (1.066025,8.300000);
	\draw[line width=3pt,draw=white] (1.866025,8.600000) -- (1.866025,9.200000);
	\draw[line width=3pt,draw=white] (1.866025,9.200000) -- (1.866025,9.400000);
	\draw[line width=3pt,draw=white] (1.866025,9.400000) -- (1.866025,10.000000);
	\draw[line width=3pt,draw=white] (1.866025,10.000000) -- (1.866025,10.200000);
	\draw[line width=3pt,draw=white] (1.866025,10.200000) -- (1.866025,10.800000);
	\draw[line width=3pt,draw=white] (1.866025,10.800000) -- (1.866025,11.000000);
	\draw[line width=3pt,draw=white] (1.866025,11.000000) -- (1.866025,11.600000);
	\draw[line width=3pt,draw=white] (1.866025,11.600000) -- (1.866025,11.800000);
	\draw[line width=3pt,draw=white] (1.866025,11.800000) -- (1.866025,12.400000);
	\draw[line width=3pt,draw=white] (1.866025,12.400000) -- (1.866025,12.600000);
	\draw[line width=3pt,draw=white] (1.866025,12.600000) -- (1.866025,13.200000);
	\draw[line width=3pt,draw=white] (1.866025,13.200000) -- (1.866025,13.400000);
	\draw[line width=3pt,draw=white] (1.866025,13.400000) -- (1.866025,14.000000);
	\draw[line width=3pt,fill=white,draw=white] (1.766025,14.000000) -- (1.766025,14.200000) -- (1.966025,14.200000) -- (1.966025,14.000000) -- cycle;
	\draw[line width=3pt,draw=white] (1.866025,14.200000) -- (1.866025,14.800000);
	\draw[line width=3pt,fill=white,draw=white] (1.766025,14.900000) -- (1.866025,15.000000) -- (1.966025,14.900000) -- (1.866025,14.800000) -- cycle;
	\draw[line width=3pt,draw=white] (1.866025,15.000000) -- (1.866025,15.600000);
	\draw[line width=3pt,draw=white,fill=white] (1.866025,15.700000) circle (0.100000);
	\draw[line width=3pt,draw=white] (1.866025,15.700000) -- (2.732051,16.200000);
	\draw[line width=3pt,draw=white,fill=white] (2.732051,16.200000) circle (0.200000);
	\draw[line width=3pt,draw=white] (2.532051,16.200000) -- (2.932051,16.200000);
	\draw[line width=3pt,draw=white] (2.732051,16.000000) -- (2.732051,16.400000);
	\draw[line width=3pt,draw=white] (1.866025,15.800000) -- (1.866025,16.400000);
	\draw[line width=3pt,draw=white,fill=white] (1.866025,16.500000) circle (0.100000);
	\draw[line width=3pt,draw=white] (1.866025,16.500000) -- (0.866025,16.500000);
	\draw[line width=3pt,draw=white,fill=white] (0.866025,16.500000) circle (0.200000);
	\draw[line width=3pt,draw=white] (0.666025,16.500000) -- (1.066025,16.500000);
	\draw[line width=3pt,draw=white] (0.866025,16.300000) -- (0.866025,16.700000);
	\draw[line width=3pt,draw=white] (1.866025,16.600000) -- (1.866025,17.200000);
	\draw[line width=3pt,draw=white] (1.866025,17.200000) -- (1.866025,17.400000);
	\draw[line width=3pt,draw=white] (1.866025,17.400000) -- (1.866025,18.000000);
	\draw[line width=3pt,draw=white,fill=white] (1.866025,18.100000) circle (0.100000);
	\draw[line width=3pt,draw=white] (1.866025,18.100000) -- (1.000000,17.600000);
	\draw[line width=3pt,draw=white,fill=white] (1.000000,17.600000) circle (0.200000);
	\draw[line width=3pt,draw=white] (0.800000,17.600000) -- (1.200000,17.600000);
	\draw[line width=3pt,draw=white] (1.000000,17.400000) -- (1.000000,17.800000);
	\draw[line width=3pt,draw=white] (1.866025,18.200000) -- (1.866025,18.800000);
	\draw[line width=3pt,fill=white,draw=white] (1.766025,18.900000) -- (1.866025,19.000000) -- (1.966025,18.900000) -- (1.866025,18.800000) -- cycle;
	\draw[line width=3pt,draw=white] (1.866025,19.000000) -- (1.866025,19.600000);
	\draw[line width=3pt,fill=white,draw=white] (1.866025,19.700000) circle (0.100000);
	\draw[line width=1pt] (1.866025,0.500000) -- (1.866025,1.200000);
	\draw[line width=1pt,fill=white] (1.766025,1.200000) -- (1.766025,1.400000) -- (1.966025,1.400000) -- (1.966025,1.200000) -- cycle;
	\draw[line width=1pt] (1.866025,1.400000) -- (1.866025,2.000000);
	\draw[line width=1pt,fill=white] (1.766025,2.100000) -- (1.866025,2.200000) -- (1.966025,2.100000) -- (1.866025,2.000000) -- cycle;
	\draw[line width=1pt] (1.866025,2.200000) -- (1.866025,2.800000);
	\draw[line width=1pt] (1.866025,2.900000) -- (2.732051,3.400000);
	\draw[line width=1pt,fill=black] (1.866025,2.900000) circle (0.100000);
	\draw[line width=1pt,fill=white] (2.732051,3.400000) circle (0.200000);
	\draw[line width=1pt] (2.532051,3.400000) -- (2.932051,3.400000);
	\draw[line width=1pt] (2.732051,3.200000) -- (2.732051,3.600000);
	\draw[line width=1pt] (2.732051,3.600000) -- +(0.0,2.0pt);
	\draw[line width=1pt] (2.732051,3.200000) -- +(0.0,-2.0pt);
	\draw[line width=1pt] (1.866025,3.000000) -- (1.866025,3.600000);
	\draw[line width=1pt] (1.866025,3.700000) -- (0.866025,3.700000);
	\draw[line width=1pt,fill=black] (1.866025,3.700000) circle (0.100000);
	\draw[line width=1pt,fill=white] (0.866025,3.700000) circle (0.200000);
	\draw[line width=1pt] (0.666025,3.700000) -- (1.066025,3.700000);
	\draw[line width=1pt] (0.866025,3.500000) -- (0.866025,3.900000);
	\draw[line width=1pt] (1.866025,3.800000) -- (1.866025,4.400000);
	\draw[line width=1pt] (1.866025,4.500000) -- (2.866025,4.500000);
	\draw[line width=1pt,fill=black] (1.866025,4.500000) circle (0.100000);
	\draw[line width=1pt,fill=white] (2.866025,4.500000) circle (0.200000);
	\draw[line width=1pt] (2.666025,4.500000) -- (3.066025,4.500000);
	\draw[line width=1pt] (2.866025,4.300000) -- (2.866025,4.700000);
	\draw[line width=1pt] (2.866025,4.700000) -- +(0.0,2.0pt);
	\draw[line width=1pt] (2.866025,4.300000) -- +(0.0,-2.0pt);
	\draw[line width=1pt] (1.866025,4.600000) -- (1.866025,5.200000);
	\draw[line width=1pt] (1.866025,5.300000) -- (1.000000,4.800000);
	\draw[line width=1pt,fill=black] (1.866025,5.300000) circle (0.100000);
	\draw[line width=1pt,fill=white] (1.000000,4.800000) circle (0.200000);
	\draw[line width=1pt] (0.800000,4.800000) -- (1.200000,4.800000);
	\draw[line width=1pt] (1.000000,4.600000) -- (1.000000,5.000000);
	\draw[line width=1pt] (1.866025,5.400000) -- (1.866025,6.000000);
	\draw[line width=1pt,fill=white] (1.766025,6.100000) -- (1.866025,6.200000) -- (1.966025,6.100000) -- (1.866025,6.000000) -- cycle;
	\draw[line width=1pt] (1.866025,6.200000) -- (1.866025,6.800000);
	\draw[line width=1pt,fill=white] (1.866025,6.900000) circle (0.100000);
	\draw[line width=1pt] (1.866025,7.000000) -- (1.866025,7.600000);
	\draw[line width=1pt] (1.866025,7.500000) -- (1.866025,7.900000);
	\draw[line width=1pt] (1.666025,7.500000) -- (2.066025,7.900000);
	\draw[line width=1pt] (1.666025,7.900000) -- (2.066025,7.500000);
	\draw[line width=1pt] (1.866025,7.700000) -- (1.000000,7.200000);
	\draw[line width=1pt] (0.800000,7.000000) -- (1.200000,7.400000);
	\draw[line width=1pt] (0.800000,7.400000) -- (1.200000,7.000000);
	\draw[line width=1pt] (1.866025,7.800000) -- (1.866025,8.400000);
	\draw[line width=1pt] (1.866025,8.300000) -- (1.866025,8.700000);
	\draw[line width=1pt] (1.666025,8.300000) -- (2.066025,8.700000);
	\draw[line width=1pt] (1.666025,8.700000) -- (2.066025,8.300000);
	\draw[line width=1pt] (1.866025,8.500000) -- (0.866025,8.500000);
	\draw[line width=1pt] (0.666025,8.300000) -- (1.066025,8.700000);
	\draw[line width=1pt] (0.666025,8.700000) -- (1.066025,8.300000);
	\draw[line width=1pt] (1.866025,8.600000) -- (1.866025,9.200000);
	\draw[line width=1pt] (1.866025,9.200000) -- (1.866025,9.400000);
	\draw[line width=1pt] (1.866025,9.400000) -- (1.866025,10.000000);
	\draw[line width=1pt] (1.866025,10.000000) -- (1.866025,10.200000);
	\draw[line width=1pt] (1.866025,10.200000) -- (1.866025,10.800000);
	\draw[line width=1pt] (1.866025,10.800000) -- (1.866025,11.000000);
	\draw[line width=1pt] (1.866025,11.000000) -- (1.866025,11.600000);
	\draw[line width=1pt] (1.866025,11.600000) -- (1.866025,11.800000);
	\draw[line width=1pt] (1.866025,11.800000) -- (1.866025,12.400000);
	\draw[line width=1pt] (1.866025,12.400000) -- (1.866025,12.600000);
	\draw[line width=1pt] (1.866025,12.600000) -- (1.866025,13.200000);
	\draw[line width=1pt] (1.866025,13.200000) -- (1.866025,13.400000);
	\draw[line width=1pt] (1.866025,13.400000) -- (1.866025,14.000000);
	\draw[line width=1pt,fill=white] (1.766025,14.000000) -- (1.766025,14.200000) -- (1.966025,14.200000) -- (1.966025,14.000000) -- cycle;
	\draw[line width=1pt] (1.866025,14.200000) -- (1.866025,14.800000);
	\draw[line width=1pt,fill=white] (1.766025,14.900000) -- (1.866025,15.000000) -- (1.966025,14.900000) -- (1.866025,14.800000) -- cycle;
	\draw[line width=1pt] (1.866025,15.000000) -- (1.866025,15.600000);
	\draw[line width=1pt] (1.866025,15.700000) -- (2.732051,16.200000);
	\draw[line width=1pt,fill=black] (1.866025,15.700000) circle (0.100000);
	\draw[line width=1pt,fill=white] (2.732051,16.200000) circle (0.200000);
	\draw[line width=1pt] (2.532051,16.200000) -- (2.932051,16.200000);
	\draw[line width=1pt] (2.732051,16.000000) -- (2.732051,16.400000);
	\draw[line width=1pt] (2.732051,16.400000) -- +(0.0,2.0pt);
	\draw[line width=1pt] (2.732051,16.000000) -- +(0.0,-2.0pt);
	\draw[line width=1pt] (1.866025,15.800000) -- (1.866025,16.400000);
	\draw[line width=1pt] (1.866025,16.500000) -- (0.866025,16.500000);
	\draw[line width=1pt,fill=black] (1.866025,16.500000) circle (0.100000);
	\draw[line width=1pt,fill=white] (0.866025,16.500000) circle (0.200000);
	\draw[line width=1pt] (0.666025,16.500000) -- (1.066025,16.500000);
	\draw[line width=1pt] (0.866025,16.300000) -- (0.866025,16.700000);
	\draw[line width=1pt] (1.866025,16.600000) -- (1.866025,17.200000);
	\draw[line width=1pt] (1.866025,17.200000) -- (1.866025,17.400000);
	\draw[line width=1pt] (1.866025,17.400000) -- (1.866025,18.000000);
	\draw[line width=1pt] (1.866025,18.100000) -- (1.000000,17.600000);
	\draw[line width=1pt,fill=black] (1.866025,18.100000) circle (0.100000);
	\draw[line width=1pt,fill=white] (1.000000,17.600000) circle (0.200000);
	\draw[line width=1pt] (0.800000,17.600000) -- (1.200000,17.600000);
	\draw[line width=1pt] (1.000000,17.400000) -- (1.000000,17.800000);
	\draw[line width=1pt] (1.866025,18.200000) -- (1.866025,18.800000);
	\draw[line width=1pt,fill=white] (1.766025,18.900000) -- (1.866025,19.000000) -- (1.966025,18.900000) -- (1.866025,18.800000) -- cycle;
	\draw[line width=1pt] (1.866025,19.000000) -- (1.866025,19.600000);
	\draw[line width=1pt,fill=white] (1.866025,19.700000) circle (0.100000);
	\draw[line width=3pt,draw=white] (2.866025,0.500000) -- (2.866025,1.200000);
	\draw[line width=3pt,draw=white] (2.866025,1.200000) -- (2.866025,1.400000);
	\draw[line width=3pt,draw=white] (2.866025,1.400000) -- (2.866025,2.000000);
	\draw[line width=3pt,draw=white] (2.866025,2.000000) -- (2.866025,2.200000);
	\draw[line width=3pt,draw=white] (2.866025,2.200000) -- (2.866025,2.800000);
	\draw[line width=3pt,draw=white,fill=white] (2.866025,2.900000) circle (0.100000);
	\draw[line width=3pt,draw=white] (2.866025,2.900000) -- (2.000000,2.400000);
	\draw[line width=3pt,draw=white,fill=white] (2.000000,2.400000) circle (0.200000);
	\draw[line width=3pt,draw=white] (1.800000,2.400000) -- (2.200000,2.400000);
	\draw[line width=3pt,draw=white] (2.000000,2.200000) -- (2.000000,2.600000);
	\draw[line width=3pt,draw=white] (2.866025,3.000000) -- (2.866025,3.600000);
	\draw[line width=3pt,draw=white] (2.866025,3.600000) -- (2.866025,3.800000);
	\draw[line width=3pt,draw=white] (2.866025,3.800000) -- (2.866025,4.400000);
	\draw[line width=3pt,draw=white,fill=white] (2.866025,4.500000) circle (0.200000);
	\draw[line width=3pt,draw=white] (2.666025,4.500000) -- (3.066025,4.500000);
	\draw[line width=3pt,draw=white] (2.866025,4.300000) -- (2.866025,4.700000);
	\draw[line width=3pt,draw=white] (2.866025,4.500000) -- (1.866025,4.500000);
	\draw[line width=3pt,draw=white,fill=white] (1.866025,4.500000) circle (0.100000);
	\draw[line width=3pt,draw=white] (2.866025,4.600000) -- (2.866025,5.200000);
	\draw[line width=3pt,draw=white,fill=white] (2.866025,5.300000) circle (0.100000);
	\draw[line width=3pt,draw=white] (2.866025,5.300000) -- (3.732051,5.800000);
	\draw[line width=3pt,draw=white,fill=white] (3.732051,5.800000) circle (0.200000);
	\draw[line width=3pt,draw=white] (3.532051,5.800000) -- (3.932051,5.800000);
	\draw[line width=3pt,draw=white] (3.732051,5.600000) -- (3.732051,6.000000);
	\draw[line width=3pt,draw=white] (2.866025,5.400000) -- (2.866025,6.000000);
	\draw[line width=3pt,draw=white] (2.866025,6.000000) -- (2.866025,6.200000);
	\draw[line width=3pt,draw=white] (2.866025,6.200000) -- (2.866025,6.800000);
	\draw[line width=3pt,fill=white,draw=white] (2.766025,6.900000) -- (2.866025,7.000000) -- (2.966025,6.900000) -- (2.866025,6.800000) -- cycle;
	\draw[line width=3pt,draw=white] (2.866025,7.000000) -- (2.866025,7.600000);
	\draw[line width=3pt,draw=white] (2.866025,7.500000) -- (2.866025,7.900000);
	\draw[line width=3pt,draw=white] (2.666025,7.500000) -- (3.066025,7.900000);
	\draw[line width=3pt,draw=white] (2.666025,7.900000) -- (3.066025,7.500000);
	\draw[line width=3pt,draw=white] (2.866025,7.700000) -- (3.732051,8.200000);
	\draw[line width=3pt,draw=white] (3.532051,8.000000) -- (3.932051,8.400000);
	\draw[line width=3pt,draw=white] (3.532051,8.400000) -- (3.932051,8.000000);
	\draw[line width=3pt,draw=white] (2.866025,7.800000) -- (2.866025,8.400000);
	\draw[line width=3pt,draw=white] (2.866025,8.400000) -- (2.866025,8.600000);
	\draw[line width=3pt,draw=white] (2.866025,8.600000) -- (2.866025,9.200000);
	\draw[line width=3pt,draw=white] (2.866025,9.200000) -- (2.866025,9.400000);
	\draw[line width=3pt,draw=white] (2.866025,9.400000) -- (2.866025,10.000000);
	\draw[line width=3pt,draw=white] (2.866025,10.000000) -- (2.866025,10.200000);
	\draw[line width=3pt,draw=white] (2.866025,10.200000) -- (2.866025,10.800000);
	\draw[line width=3pt,draw=white] (2.866025,10.800000) -- (2.866025,11.000000);
	\draw[line width=3pt,draw=white] (2.866025,11.000000) -- (2.866025,11.600000);
	\draw[line width=3pt,draw=white] (2.866025,11.600000) -- (2.866025,11.800000);
	\draw[line width=3pt,draw=white] (2.866025,11.800000) -- (2.866025,12.400000);
	\draw[line width=3pt,draw=white] (2.866025,12.400000) -- (2.866025,12.600000);
	\draw[line width=3pt,draw=white] (2.866025,12.600000) -- (2.866025,13.200000);
	\draw[line width=3pt,draw=white] (2.866025,13.200000) -- (2.866025,13.400000);
	\draw[line width=3pt,draw=white] (2.866025,13.400000) -- (2.866025,14.000000);
	\draw[line width=3pt,draw=white] (2.866025,14.000000) -- (2.866025,14.200000);
	\draw[line width=3pt,draw=white] (2.866025,14.200000) -- (2.866025,14.800000);
	\draw[line width=3pt,draw=white] (2.866025,14.800000) -- (2.866025,15.000000);
	\draw[line width=3pt,draw=white] (2.866025,15.000000) -- (2.866025,15.600000);
	\draw[line width=3pt,draw=white] (2.866025,15.600000) -- (2.866025,15.800000);
	\draw[line width=3pt,draw=white] (2.866025,15.800000) -- (2.866025,16.400000);
	\draw[line width=3pt,draw=white] (2.866025,16.400000) -- (2.866025,16.600000);
	\draw[line width=3pt,draw=white] (2.866025,16.600000) -- (2.866025,17.200000);
	\draw[line width=3pt,draw=white] (2.866025,17.200000) -- (2.866025,17.400000);
	\draw[line width=3pt,draw=white] (2.866025,17.400000) -- (2.866025,18.000000);
	\draw[line width=3pt,draw=white] (2.866025,18.000000) -- (2.866025,18.200000);
	\draw[line width=3pt,draw=white] (2.866025,18.200000) -- (2.866025,18.800000);
	\draw[line width=3pt,draw=white] (2.866025,18.800000) -- (2.866025,19.000000);
	\draw[line width=3pt,draw=white] (2.866025,19.000000) -- (2.866025,19.600000);
	\draw[line width=3pt,draw=white] (2.866025,19.600000) -- (2.866025,19.800000);
	\draw[line width=1pt] (2.866025,0.500000) -- (2.866025,1.200000);
	\draw[line width=1pt] (2.866025,1.200000) -- (2.866025,1.400000);
	\draw[line width=1pt] (2.866025,1.400000) -- (2.866025,2.000000);
	\draw[line width=1pt] (2.866025,2.000000) -- (2.866025,2.200000);
	\draw[line width=1pt] (2.866025,2.200000) -- (2.866025,2.800000);
	\draw[line width=1pt] (2.866025,2.900000) -- (2.000000,2.400000);
	\draw[line width=1pt,fill=black] (2.866025,2.900000) circle (0.100000);
	\draw[line width=1pt,fill=white] (2.000000,2.400000) circle (0.200000);
	\draw[line width=1pt] (1.800000,2.400000) -- (2.200000,2.400000);
	\draw[line width=1pt] (2.000000,2.200000) -- (2.000000,2.600000);
	\draw[line width=1pt] (2.866025,3.000000) -- (2.866025,3.600000);
	\draw[line width=1pt] (2.866025,3.600000) -- (2.866025,3.800000);
	\draw[line width=1pt] (2.866025,3.800000) -- (2.866025,4.400000);
	\draw[line width=1pt] (2.866025,4.500000) -- (1.866025,4.500000);
	\draw[line width=1pt,fill=white] (2.866025,4.500000) circle (0.200000);
	\draw[line width=1pt] (2.666025,4.500000) -- (3.066025,4.500000);
	\draw[line width=1pt] (2.866025,4.300000) -- (2.866025,4.700000);
	\draw[line width=1pt,fill=black] (1.866025,4.500000) circle (0.100000);
	\draw[line width=1pt] (1.866025,4.600000) -- +(0.0,2.0pt);
	\draw[line width=1pt] (1.866025,4.400000) -- +(0.0,-2.0pt);
	\draw[line width=1pt] (2.866025,4.600000) -- (2.866025,5.200000);
	\draw[line width=1pt] (2.866025,5.300000) -- (3.732051,5.800000);
	\draw[line width=1pt,fill=black] (2.866025,5.300000) circle (0.100000);
	\draw[line width=1pt,fill=white] (3.732051,5.800000) circle (0.200000);
	\draw[line width=1pt] (3.532051,5.800000) -- (3.932051,5.800000);
	\draw[line width=1pt] (3.732051,5.600000) -- (3.732051,6.000000);
	\draw[line width=1pt] (3.732051,6.000000) -- +(0.0,2.0pt);
	\draw[line width=1pt] (3.732051,5.600000) -- +(0.0,-2.0pt);
	\draw[line width=1pt] (2.866025,5.400000) -- (2.866025,6.000000);
	\draw[line width=1pt] (2.866025,6.000000) -- (2.866025,6.200000);
	\draw[line width=1pt] (2.866025,6.200000) -- (2.866025,6.800000);
	\draw[line width=1pt,fill=white] (2.766025,6.900000) -- (2.866025,7.000000) -- (2.966025,6.900000) -- (2.866025,6.800000) -- cycle;
	\draw[line width=1pt] (2.866025,7.000000) -- (2.866025,7.600000);
	\draw[line width=1pt] (2.866025,7.500000) -- (2.866025,7.900000);
	\draw[line width=1pt] (2.666025,7.500000) -- (3.066025,7.900000);
	\draw[line width=1pt] (2.666025,7.900000) -- (3.066025,7.500000);
	\draw[line width=1pt] (2.866025,7.700000) -- (3.732051,8.200000);
	\draw[line width=1pt] (3.532051,8.000000) -- (3.932051,8.400000);
	\draw[line width=1pt] (3.532051,8.400000) -- (3.932051,8.000000);
	\draw[line width=1pt] (3.732051,8.000000) -- (3.732051,8.400000);
	\draw[line width=1pt] (2.866025,7.800000) -- (2.866025,8.400000);
	\draw[line width=1pt] (2.866025,8.400000) -- (2.866025,8.600000);
	\draw[line width=1pt] (2.866025,8.600000) -- (2.866025,9.200000);
	\draw[line width=1pt] (2.866025,9.200000) -- (2.866025,9.400000);
	\draw[line width=1pt] (2.866025,9.400000) -- (2.866025,10.000000);
	\draw[line width=1pt] (2.866025,10.000000) -- (2.866025,10.200000);
	\draw[line width=1pt] (2.866025,10.200000) -- (2.866025,10.800000);
	\draw[line width=1pt] (2.866025,10.800000) -- (2.866025,11.000000);
	\draw[line width=1pt] (2.866025,11.000000) -- (2.866025,11.600000);
	\draw[line width=1pt] (2.866025,11.600000) -- (2.866025,11.800000);
	\draw[line width=1pt] (2.866025,11.800000) -- (2.866025,12.400000);
	\draw[line width=1pt] (2.866025,12.400000) -- (2.866025,12.600000);
	\draw[line width=1pt] (2.866025,12.600000) -- (2.866025,13.200000);
	\draw[line width=1pt] (2.866025,13.200000) -- (2.866025,13.400000);
	\draw[line width=1pt] (2.866025,13.400000) -- (2.866025,14.000000);
	\draw[line width=1pt] (2.866025,14.000000) -- (2.866025,14.200000);
	\draw[line width=1pt] (2.866025,14.200000) -- (2.866025,14.800000);
	\draw[line width=1pt] (2.866025,14.800000) -- (2.866025,15.000000);
	\draw[line width=1pt] (2.866025,15.000000) -- (2.866025,15.600000);
	\draw[line width=1pt] (2.866025,15.600000) -- (2.866025,15.800000);
	\draw[line width=1pt] (2.866025,15.800000) -- (2.866025,16.400000);
	\draw[line width=1pt] (2.866025,16.400000) -- (2.866025,16.600000);
	\draw[line width=1pt] (2.866025,16.600000) -- (2.866025,17.200000);
	\draw[line width=1pt] (2.866025,17.200000) -- (2.866025,17.400000);
	\draw[line width=1pt] (2.866025,17.400000) -- (2.866025,18.000000);
	\draw[line width=1pt] (2.866025,18.000000) -- (2.866025,18.200000);
	\draw[line width=1pt] (2.866025,18.200000) -- (2.866025,18.800000);
	\draw[line width=1pt] (2.866025,18.800000) -- (2.866025,19.000000);
	\draw[line width=1pt] (2.866025,19.000000) -- (2.866025,19.600000);
	\draw[line width=1pt] (2.866025,19.600000) -- (2.866025,19.800000);
	\draw[line width=1pt,rounded corners,dashed] (2.366025,6.500000) -- (2.366025,20.100000) -- (1.366025,20.100000) -- (1.366025,6.500000) -- cycle;
	\draw[line width=1pt,dotted] (3.732051,6.600000) circle (0.400000);
\end{tikzpicture}

%% file: pic_d3_3.tex
\begin{tikzpicture}[x=0.1*\the\linewidth,y=0.1*\the\linewidth]
	\fill[gray!25] (3.000000,5.000000) -- (3.900000,5.000000) -- (3.000000,4.100000) -- cycle;
	\fill[gray!25] (3.000000,5.000000) -- (3.000000,4.100000) -- (2.100000,5.000000) -- cycle;
	\fill[gray!25] (5.000000,5.000000) -- (5.900000,5.000000) -- (5.000000,4.100000) -- cycle;
	\fill[gray!25] (5.000000,5.000000) -- (5.000000,4.100000) -- (4.100000,5.000000) -- cycle;
	\fill[gray!80] (2.000000,4.000000) -- (2.000000,4.900000) -- (2.900000,4.000000) -- cycle;
	\fill[gray!80] (2.000000,4.000000) -- (2.900000,4.000000) -- (2.000000,3.100000) -- cycle;
	\fill[gray!80] (4.000000,4.000000) -- (3.100000,4.000000) -- (4.000000,4.900000) -- cycle;
	\fill[gray!80] (4.000000,4.000000) -- (4.000000,4.900000) -- (4.900000,4.000000) -- cycle;
	\fill[gray!80] (4.000000,4.000000) -- (4.900000,4.000000) -- (4.000000,3.100000) -- cycle;
	\fill[gray!80] (4.000000,4.000000) -- (4.000000,3.100000) -- (3.100000,4.000000) -- cycle;
	\fill[gray!80] (6.000000,4.000000) -- (5.100000,4.000000) -- (6.000000,4.900000) -- cycle;
	\fill[gray!80] (6.000000,4.000000) -- (6.000000,3.100000) -- (5.100000,4.000000) -- cycle;
	\fill[gray!25] (3.000000,3.000000) -- (2.100000,3.000000) -- (3.000000,3.900000) -- cycle;
	\fill[gray!25] (3.000000,3.000000) -- (3.000000,3.900000) -- (3.900000,3.000000) -- cycle;
	\fill[gray!25] (3.000000,3.000000) -- (3.900000,3.000000) -- (3.000000,2.100000) -- cycle;
	\fill[gray!25] (3.000000,3.000000) -- (3.000000,2.100000) -- (2.100000,3.000000) -- cycle;
	\fill[gray!25] (5.000000,3.000000) -- (4.100000,3.000000) -- (5.000000,3.900000) -- cycle;
	\fill[gray!25] (5.000000,3.000000) -- (5.000000,3.900000) -- (5.900000,3.000000) -- cycle;
	\fill[gray!25] (5.000000,3.000000) -- (5.900000,3.000000) -- (5.000000,2.100000) -- cycle;
	\fill[gray!25] (5.000000,3.000000) -- (5.000000,2.100000) -- (4.100000,3.000000) -- cycle;
	\fill[gray!80] (2.000000,2.000000) -- (2.000000,2.900000) -- (2.900000,2.000000) -- cycle;
	\fill[gray!80] (2.000000,2.000000) -- (2.900000,2.000000) -- (2.000000,1.100000) -- cycle;
	\fill[gray!80] (4.000000,2.000000) -- (3.100000,2.000000) -- (4.000000,2.900000) -- cycle;
	\fill[gray!80] (4.000000,2.000000) -- (4.000000,2.900000) -- (4.900000,2.000000) -- cycle;
	\fill[gray!80] (4.000000,2.000000) -- (4.900000,2.000000) -- (4.000000,1.100000) -- cycle;
	\fill[gray!80] (4.000000,2.000000) -- (4.000000,1.100000) -- (3.100000,2.000000) -- cycle;
	\fill[gray!80] (6.000000,2.000000) -- (5.100000,2.000000) -- (6.000000,2.900000) -- cycle;
	\fill[gray!80] (6.000000,2.000000) -- (6.000000,1.100000) -- (5.100000,2.000000) -- cycle;
	\fill[gray!25] (3.000000,1.000000) -- (2.100000,1.000000) -- (3.000000,1.900000) -- cycle;
	\fill[gray!25] (3.000000,1.000000) -- (3.000000,1.900000) -- (3.900000,1.000000) -- cycle;
	\fill[gray!25] (5.000000,1.000000) -- (4.100000,1.000000) -- (5.000000,1.900000) -- cycle;
	\fill[gray!25] (5.000000,1.000000) -- (5.000000,1.900000) -- (5.900000,1.000000) -- cycle;
	\draw[line width=1pt] (2.000000,5.000000) circle (0.100000);
	\draw[line width=1pt,fill=black] (3.000000,5.000000) circle (0.100000);
	\draw[line width=1pt] (4.000000,5.000000) circle (0.100000);
	\draw[line width=1pt,fill=black] (5.000000,5.000000) circle (0.100000);
	\draw[line width=1pt] (6.000000,5.000000) circle (0.100000);
	\draw[line width=1pt,fill=black] (2.000000,4.000000) circle (0.100000);
	\draw[line width=1pt] (3.000000,4.000000) circle (0.100000);
	\draw[line width=1pt,fill=black] (4.000000,4.000000) circle (0.100000);
	\draw[line width=1pt] (5.000000,4.000000) circle (0.100000);
	\draw[line width=1pt,fill=black] (6.000000,4.000000) circle (0.100000);
	\draw[line width=1pt] (2.000000,3.000000) circle (0.100000);
	\draw[line width=1pt,fill=black] (3.000000,3.000000) circle (0.100000);
	\draw[line width=1pt] (4.000000,3.000000) circle (0.100000);
	\draw[line width=1pt,fill=black] (5.000000,3.000000) circle (0.100000);
	\draw[line width=1pt] (6.000000,3.000000) circle (0.100000);
	\draw[line width=1pt,fill=black] (2.000000,2.000000) circle (0.100000);
	\draw[line width=1pt] (3.000000,2.000000) circle (0.100000);
	\draw[line width=1pt,fill=black] (4.000000,2.000000) circle (0.100000);
	\draw[line width=1pt] (5.000000,2.000000) circle (0.100000);
	\draw[line width=1pt,fill=black] (6.000000,2.000000) circle (0.100000);
	\draw[line width=1pt] (2.000000,1.000000) circle (0.100000);
	\draw[line width=1pt,fill=black] (3.000000,1.000000) circle (0.100000);
	\draw[line width=1pt] (4.000000,1.000000) circle (0.100000);
	\draw[line width=1pt,fill=black] (5.000000,1.000000) circle (0.100000);
	\draw[line width=1pt] (6.000000,1.000000) circle (0.100000);

	\draw [line width=1pt,->] (2.2,4) -- (2.5,4) -- (2.5,3.5);
	\draw [line width=1pt,->] (4.2,4) -- (4.5,4) -- (4.5,3.5);
	\draw [line width=1pt,->] (6.2,4) -- (6.5,4) -- (6.5,3.5);

	\draw [line width=1pt,->] (2.2,2) -- (2.5,2) -- (2.5,1.5);
	\draw [line width=1pt,->] (4.2,2) -- (4.5,2) -- (4.5,1.5);
	\draw [line width=1pt,->] (6.2,2) -- (6.5,2) -- (6.5,1.5);

	\draw [line width=1pt,->,dashed] (2.8,5) -- (1.2,5);
	\draw [line width=1pt,->,dashed] (2.8,3) -- (1.2,3);
	\draw [line width=1pt,->,dashed] (2.8,1) -- (1.2,1);

	\draw [line width=1pt,->,dashed] (5.2,5) -- (6.8,5);
	\draw [line width=1pt,->,dashed] (5.2,3) -- (6.8,3);
	\draw [line width=1pt,->,dashed] (5.2,1) -- (6.8,1);

	\draw (2,1) node [above right] {$\ket{0}$};
	\draw (2,3) node [above right] {$\ket{0}$};
	\draw (2,5) node [above right] {$\ket{0}$};
	\draw (4,1) node [above right] {$\ket{0}$};
	\draw (4,3) node [above right] {$\ket{0}$};
	\draw (4,5) node [above right] {$\ket{0}$};
	\draw (6,1) node [above right] {$\ket{0}$};
	\draw (6,3) node [above right] {$\ket{0}$};
	\draw (6,5) node [above right] {$\ket{0}$};
	\draw (3,4) node [above right] {$\ket{0}$};
	\draw (3,2) node [above right] {$\ket{0}$};
	\draw (5,4) node [above right] {$\ket{0}$};
	\draw (5,2) node [above right] {$\ket{0}$};
\end{tikzpicture}

%% file: pic_d3_2.tex
\begin{tikzpicture}[x=0.1*\the\linewidth,y=0.1*\the\linewidth]
	\fill[gray!25] (3.000000,5.000000) -- (3.900000,5.000000) -- (3.000000,4.100000) -- cycle;
	\fill[gray!25] (3.000000,5.000000) -- (3.000000,4.100000) -- (2.100000,5.000000) -- cycle;
	\fill[gray!25] (5.000000,5.000000) -- (5.900000,5.000000) -- (5.000000,4.100000) -- cycle;
	\fill[gray!25] (5.000000,5.000000) -- (5.000000,4.100000) -- (4.100000,5.000000) -- cycle;
	\fill[gray!80] (2.000000,4.000000) -- (2.000000,4.900000) -- (2.900000,4.000000) -- cycle;
	\fill[gray!80] (2.000000,4.000000) -- (2.900000,4.000000) -- (2.000000,3.100000) -- cycle;
	\fill[gray!80] (4.000000,4.000000) -- (3.100000,4.000000) -- (4.000000,4.900000) -- cycle;
	\fill[gray!80] (4.000000,4.000000) -- (4.000000,4.900000) -- (4.900000,4.000000) -- cycle;
	\fill[gray!80] (4.000000,4.000000) -- (4.900000,4.000000) -- (4.000000,3.100000) -- cycle;
	\fill[gray!80] (4.000000,4.000000) -- (4.000000,3.100000) -- (3.100000,4.000000) -- cycle;
	\fill[gray!80] (6.000000,4.000000) -- (5.100000,4.000000) -- (6.000000,4.900000) -- cycle;
	\fill[gray!80] (6.000000,4.000000) -- (6.000000,3.100000) -- (5.100000,4.000000) -- cycle;
	\fill[gray!25] (3.000000,3.000000) -- (2.100000,3.000000) -- (3.000000,3.900000) -- cycle;
	\fill[gray!25] (3.000000,3.000000) -- (3.000000,3.900000) -- (3.900000,3.000000) -- cycle;
	\fill[gray!25] (3.000000,3.000000) -- (3.900000,3.000000) -- (3.000000,2.100000) -- cycle;
	\fill[gray!25] (3.000000,3.000000) -- (3.000000,2.100000) -- (2.100000,3.000000) -- cycle;
	\fill[gray!25] (5.000000,3.000000) -- (4.100000,3.000000) -- (5.000000,3.900000) -- cycle;
	\fill[gray!25] (5.000000,3.000000) -- (5.000000,3.900000) -- (5.900000,3.000000) -- cycle;
	\fill[gray!25] (5.000000,3.000000) -- (5.900000,3.000000) -- (5.000000,2.100000) -- cycle;
	\fill[gray!25] (5.000000,3.000000) -- (5.000000,2.100000) -- (4.100000,3.000000) -- cycle;
	\fill[gray!80] (2.000000,2.000000) -- (2.000000,2.900000) -- (2.900000,2.000000) -- cycle;
	\fill[gray!80] (2.000000,2.000000) -- (2.900000,2.000000) -- (2.000000,1.100000) -- cycle;
	\fill[gray!80] (4.000000,2.000000) -- (3.100000,2.000000) -- (4.000000,2.900000) -- cycle;
	\fill[gray!80] (4.000000,2.000000) -- (4.000000,2.900000) -- (4.900000,2.000000) -- cycle;
	\fill[gray!80] (4.000000,2.000000) -- (4.900000,2.000000) -- (4.000000,1.100000) -- cycle;
	\fill[gray!80] (4.000000,2.000000) -- (4.000000,1.100000) -- (3.100000,2.000000) -- cycle;
	\fill[gray!80] (6.000000,2.000000) -- (5.100000,2.000000) -- (6.000000,2.900000) -- cycle;
	\fill[gray!80] (6.000000,2.000000) -- (6.000000,1.100000) -- (5.100000,2.000000) -- cycle;
	\fill[gray!25] (3.000000,1.000000) -- (2.100000,1.000000) -- (3.000000,1.900000) -- cycle;
	\fill[gray!25] (3.000000,1.000000) -- (3.000000,1.900000) -- (3.900000,1.000000) -- cycle;
	\fill[gray!25] (5.000000,1.000000) -- (4.100000,1.000000) -- (5.000000,1.900000) -- cycle;
	\fill[gray!25] (5.000000,1.000000) -- (5.000000,1.900000) -- (5.900000,1.000000) -- cycle;
	\draw[line width=1pt] (2.000000,5.000000) circle (0.100000);
	\draw[line width=1pt,fill=black] (3.000000,5.000000) circle (0.100000);
	\draw[line width=1pt] (4.000000,5.000000) circle (0.100000);
	\draw[line width=1pt,fill=black] (5.000000,5.000000) circle (0.100000);
	\draw[line width=1pt] (6.000000,5.000000) circle (0.100000);
	\draw[line width=1pt,fill=black] (2.000000,4.000000) circle (0.100000);
	\draw[line width=1pt] (3.000000,4.000000) circle (0.100000);
	\draw[line width=1pt,fill=black] (4.000000,4.000000) circle (0.100000);
	\draw[line width=1pt] (5.000000,4.000000) circle (0.100000);
	\draw[line width=1pt,fill=black] (6.000000,4.000000) circle (0.100000);
	\draw[line width=1pt] (2.000000,3.000000) circle (0.100000);
	\draw[line width=1pt,fill=black] (3.000000,3.000000) circle (0.100000);
	\draw[line width=1pt] (4.000000,3.000000) circle (0.100000);
	\draw[line width=1pt,fill=black] (5.000000,3.000000) circle (0.100000);
	\draw[line width=1pt] (6.000000,3.000000) circle (0.100000);
	\draw[line width=1pt,fill=black] (2.000000,2.000000) circle (0.100000);
	\draw[line width=1pt] (3.000000,2.000000) circle (0.100000);
	\draw[line width=1pt,fill=black] (4.000000,2.000000) circle (0.100000);
	\draw[line width=1pt] (5.000000,2.000000) circle (0.100000);
	\draw[line width=1pt,fill=black] (6.000000,2.000000) circle (0.100000);
	\draw[line width=1pt] (2.000000,1.000000) circle (0.100000);
	\draw[line width=1pt,fill=black] (3.000000,1.000000) circle (0.100000);
	\draw[line width=1pt] (4.000000,1.000000) circle (0.100000);
	\draw[line width=1pt,fill=black] (5.000000,1.000000) circle (0.100000);
	\draw[line width=1pt] (6.000000,1.000000) circle (0.100000);

	\draw [line width=1pt,->] (2.2,4) -- (2.5,4) -- (2.5,3.5);
	\draw [line width=1pt,->] (4.2,4) -- (4.5,4) -- (4.5,3.5);
	\draw [line width=1pt,->] (6.2,4) -- (6.5,4) -- (6.5,3.5);

	\draw [line width=1pt,->] (2.2,2) -- (2.5,2) -- (2.5,1.5);
	\draw [line width=1pt,->] (4.2,2) -- (4.5,2) -- (4.5,1.5);
	\draw [line width=1pt,->] (6.2,2) -- (6.5,2) -- (6.5,1.5);

	\draw [line width=1pt,->,dashed] (2.8,5) -- (1.2,5);
	\draw [line width=1pt,->,dashed] (2.8,3) -- (1.2,3);
	\draw [line width=1pt,->,dashed] (2.8,1) -- (1.2,1);

	\draw [line width=1pt,->,dashed] (5.2,5) -- (6.8,5);
	\draw [line width=1pt,->,dashed] (5.2,3) -- (6.8,3);
	\draw [line width=1pt,->,dashed] (5.2,1) -- (6.8,1);
\end{tikzpicture}

%% file: pic_d3_1.tex
\begin{tikzpicture}[x=0.1*\the\linewidth,y=0.1*\the\linewidth]
	\fill[gray!25] (3.000000,5.000000) -- (3.900000,5.000000) -- (3.000000,4.100000) -- cycle;
	\fill[gray!25] (3.000000,5.000000) -- (3.000000,4.100000) -- (2.100000,5.000000) -- cycle;
	\fill[gray!25] (5.000000,5.000000) -- (5.900000,5.000000) -- (5.000000,4.100000) -- cycle;
	\fill[gray!25] (5.000000,5.000000) -- (5.000000,4.100000) -- (4.100000,5.000000) -- cycle;
	\fill[gray!80] (2.000000,4.000000) -- (2.000000,4.900000) -- (2.900000,4.000000) -- cycle;
	\fill[gray!80] (2.000000,4.000000) -- (2.900000,4.000000) -- (2.000000,3.100000) -- cycle;
	\fill[gray!80] (4.000000,4.000000) -- (3.100000,4.000000) -- (4.000000,4.900000) -- cycle;
	\fill[gray!80] (4.000000,4.000000) -- (4.000000,4.900000) -- (4.900000,4.000000) -- cycle;
	\fill[gray!80] (4.000000,4.000000) -- (4.900000,4.000000) -- (4.000000,3.100000) -- cycle;
	\fill[gray!80] (4.000000,4.000000) -- (4.000000,3.100000) -- (3.100000,4.000000) -- cycle;
	\fill[gray!80] (6.000000,4.000000) -- (5.100000,4.000000) -- (6.000000,4.900000) -- cycle;
	\fill[gray!80] (6.000000,4.000000) -- (6.000000,3.100000) -- (5.100000,4.000000) -- cycle;
	\fill[gray!25] (3.000000,3.000000) -- (2.100000,3.000000) -- (3.000000,3.900000) -- cycle;
	\fill[gray!25] (3.000000,3.000000) -- (3.000000,3.900000) -- (3.900000,3.000000) -- cycle;
	\fill[gray!25] (3.000000,3.000000) -- (3.900000,3.000000) -- (3.000000,2.100000) -- cycle;
	\fill[gray!25] (3.000000,3.000000) -- (3.000000,2.100000) -- (2.100000,3.000000) -- cycle;
	\fill[gray!25] (5.000000,3.000000) -- (4.100000,3.000000) -- (5.000000,3.900000) -- cycle;
	\fill[gray!25] (5.000000,3.000000) -- (5.000000,3.900000) -- (5.900000,3.000000) -- cycle;
	\fill[gray!25] (5.000000,3.000000) -- (5.900000,3.000000) -- (5.000000,2.100000) -- cycle;
	\fill[gray!25] (5.000000,3.000000) -- (5.000000,2.100000) -- (4.100000,3.000000) -- cycle;
	\fill[gray!80] (2.000000,2.000000) -- (2.000000,2.900000) -- (2.900000,2.000000) -- cycle;
	\fill[gray!80] (2.000000,2.000000) -- (2.900000,2.000000) -- (2.000000,1.100000) -- cycle;
	\fill[gray!80] (4.000000,2.000000) -- (3.100000,2.000000) -- (4.000000,2.900000) -- cycle;
	\fill[gray!80] (4.000000,2.000000) -- (4.000000,2.900000) -- (4.900000,2.000000) -- cycle;
	\fill[gray!80] (4.000000,2.000000) -- (4.900000,2.000000) -- (4.000000,1.100000) -- cycle;
	\fill[gray!80] (4.000000,2.000000) -- (4.000000,1.100000) -- (3.100000,2.000000) -- cycle;
	\fill[gray!80] (6.000000,2.000000) -- (5.100000,2.000000) -- (6.000000,2.900000) -- cycle;
	\fill[gray!80] (6.000000,2.000000) -- (6.000000,1.100000) -- (5.100000,2.000000) -- cycle;
	\fill[gray!25] (3.000000,1.000000) -- (2.100000,1.000000) -- (3.000000,1.900000) -- cycle;
	\fill[gray!25] (3.000000,1.000000) -- (3.000000,1.900000) -- (3.900000,1.000000) -- cycle;
	\fill[gray!25] (5.000000,1.000000) -- (4.100000,1.000000) -- (5.000000,1.900000) -- cycle;
	\fill[gray!25] (5.000000,1.000000) -- (5.000000,1.900000) -- (5.900000,1.000000) -- cycle;
	\draw[line width=1pt] (2.000000,5.000000) circle (0.100000);
	\draw[line width=1pt,fill=black] (3.000000,5.000000) circle (0.100000);
	\draw[line width=1pt] (4.000000,5.000000) circle (0.100000);
	\draw[line width=1pt,fill=black] (5.000000,5.000000) circle (0.100000);
	\draw[line width=1pt] (6.000000,5.000000) circle (0.100000);
	\draw[line width=1pt,fill=black] (2.000000,4.000000) circle (0.100000);
	\draw[line width=1pt] (3.000000,4.000000) circle (0.100000);
	\draw[line width=1pt,fill=black] (4.000000,4.000000) circle (0.100000);
	\draw[line width=1pt] (5.000000,4.000000) circle (0.100000);
	\draw[line width=1pt,fill=black] (6.000000,4.000000) circle (0.100000);
	\draw[line width=1pt] (2.000000,3.000000) circle (0.100000);
	\draw[line width=1pt,fill=black] (3.000000,3.000000) circle (0.100000);
	\draw[line width=1pt] (4.000000,3.000000) circle (0.100000);
	\draw[line width=1pt,fill=black] (5.000000,3.000000) circle (0.100000);
	\draw[line width=1pt] (6.000000,3.000000) circle (0.100000);
	\draw[line width=1pt,fill=black] (2.000000,2.000000) circle (0.100000);
	\draw[line width=1pt] (3.000000,2.000000) circle (0.100000);
	\draw[line width=1pt,fill=black] (4.000000,2.000000) circle (0.100000);
	\draw[line width=1pt] (5.000000,2.000000) circle (0.100000);
	\draw[line width=1pt,fill=black] (6.000000,2.000000) circle (0.100000);
	\draw[line width=1pt] (2.000000,1.000000) circle (0.100000);
	\draw[line width=1pt,fill=black] (3.000000,1.000000) circle (0.100000);
	\draw[line width=1pt] (4.000000,1.000000) circle (0.100000);
	\draw[line width=1pt,fill=black] (5.000000,1.000000) circle (0.100000);
	\draw[line width=1pt] (6.000000,1.000000) circle (0.100000);

	\draw [line width=1pt,->] (2,4.2) -- (2,5.8);
	\draw [line width=1pt,->] (4,4.2) -- (4,5.8);
	\draw [line width=1pt,->] (6,4.2) -- (6,5.8);

	\draw [line width=1pt,->] (2,1.8) -- (2,0.2);
	\draw [line width=1pt,->] (4,1.8) -- (4,0.2);
	\draw [line width=1pt,->] (6,1.8) -- (6,0.2);

	\draw [line width=1pt,->,dashed] (2.8,5) -- (1.2,5);
	\draw [line width=1pt,->,dashed] (2.8,3) -- (1.2,3);
	\draw [line width=1pt,->,dashed] (2.8,1) -- (1.2,1);

	\draw [line width=1pt,->,dashed] (5.2,5) -- (6.8,5);
	\draw [line width=1pt,->,dashed] (5.2,3) -- (6.8,3);
	\draw [line width=1pt,->,dashed] (5.2,1) -- (6.8,1);
\end{tikzpicture}

%% file: 002ex2.tex
\begin{tikzpicture}[x=0.030000\linewidth,y=0.030000\linewidth]
	\fill[gray!25] (4.000000,20.000000) -- (4.900000,20.000000) -- (4.000000,19.100000) -- cycle;
	\fill[gray!25] (4.000000,20.000000) -- (4.000000,19.100000) -- (3.100000,20.000000) -- cycle;
	\fill[gray!25] (6.000000,20.000000) -- (6.900000,20.000000) -- (6.000000,19.100000) -- cycle;
	\fill[gray!25] (6.000000,20.000000) -- (6.000000,19.100000) -- (5.100000,20.000000) -- cycle;
	\fill[gray!25] (10.000000,20.000000) -- (10.900000,20.000000) -- (10.000000,19.100000) -- cycle;
	\fill[gray!25] (10.000000,20.000000) -- (10.000000,19.100000) -- (9.100000,20.000000) -- cycle;
	\fill[gray!25] (12.000000,20.000000) -- (12.900000,20.000000) -- (12.000000,19.100000) -- cycle;
	\fill[gray!25] (12.000000,20.000000) -- (12.000000,19.100000) -- (11.100000,20.000000) -- cycle;
	\fill[gray!80] (3.000000,19.000000) -- (2.100000,19.000000) -- (3.000000,19.900000) -- cycle;
	\fill[gray!80] (3.000000,19.000000) -- (3.000000,19.900000) -- (3.900000,19.000000) -- cycle;
	\fill[gray!80] (3.000000,19.000000) -- (3.900000,19.000000) -- (3.000000,18.100000) -- cycle;
	\fill[gray!80] (3.000000,19.000000) -- (3.000000,18.100000) -- (2.100000,19.000000) -- cycle;
	\fill[gray!80] (5.000000,19.000000) -- (4.100000,19.000000) -- (5.000000,19.900000) -- cycle;
	\fill[gray!80] (5.000000,19.000000) -- (5.000000,19.900000) -- (5.900000,19.000000) -- cycle;
	\fill[gray!80] (5.000000,19.000000) -- (5.900000,19.000000) -- (5.000000,18.100000) -- cycle;
	\fill[gray!80] (5.000000,19.000000) -- (5.000000,18.100000) -- (4.100000,19.000000) -- cycle;
	\fill[gray!80] (7.000000,19.000000) -- (6.100000,19.000000) -- (7.000000,19.900000) -- cycle;
	\fill[gray!80] (7.000000,19.000000) -- (7.000000,19.900000) -- (7.900000,19.000000) -- cycle;
	\fill[gray!80] (7.000000,19.000000) -- (7.900000,19.000000) -- (7.000000,18.100000) -- cycle;
	\fill[gray!80] (7.000000,19.000000) -- (7.000000,18.100000) -- (6.100000,19.000000) -- cycle;
	\fill[gray!80] (9.000000,19.000000) -- (8.100000,19.000000) -- (9.000000,19.900000) -- cycle;
	\fill[gray!80] (9.000000,19.000000) -- (9.000000,19.900000) -- (9.900000,19.000000) -- cycle;
	\fill[gray!80] (9.000000,19.000000) -- (9.900000,19.000000) -- (9.000000,18.100000) -- cycle;
	\fill[gray!80] (9.000000,19.000000) -- (9.000000,18.100000) -- (8.100000,19.000000) -- cycle;
	\fill[gray!80] (11.000000,19.000000) -- (10.100000,19.000000) -- (11.000000,19.900000) -- cycle;
	\fill[gray!80] (11.000000,19.000000) -- (11.000000,19.900000) -- (11.900000,19.000000) -- cycle;
	\fill[gray!80] (11.000000,19.000000) -- (11.900000,19.000000) -- (11.000000,18.100000) -- cycle;
	\fill[gray!80] (11.000000,19.000000) -- (11.000000,18.100000) -- (10.100000,19.000000) -- cycle;
	\fill[gray!80] (13.000000,19.000000) -- (12.100000,19.000000) -- (13.000000,19.900000) -- cycle;
	\fill[gray!80] (13.000000,19.000000) -- (13.000000,19.900000) -- (13.900000,19.000000) -- cycle;
	\fill[gray!80] (13.000000,19.000000) -- (13.900000,19.000000) -- (13.000000,18.100000) -- cycle;
	\fill[gray!80] (13.000000,19.000000) -- (13.000000,18.100000) -- (12.100000,19.000000) -- cycle;
	\fill[gray!25] (2.000000,18.000000) -- (2.000000,18.900000) -- (2.900000,18.000000) -- cycle;
	\fill[gray!25] (2.000000,18.000000) -- (2.900000,18.000000) -- (2.000000,17.100000) -- cycle;
	\fill[gray!25] (4.000000,18.000000) -- (3.100000,18.000000) -- (4.000000,18.900000) -- cycle;
	\fill[gray!25] (4.000000,18.000000) -- (4.000000,18.900000) -- (4.900000,18.000000) -- cycle;
	\fill[gray!25] (4.000000,18.000000) -- (4.900000,18.000000) -- (4.000000,17.100000) -- cycle;
	\fill[gray!25] (4.000000,18.000000) -- (4.000000,17.100000) -- (3.100000,18.000000) -- cycle;
	\fill[gray!25] (6.000000,18.000000) -- (5.100000,18.000000) -- (6.000000,18.900000) -- cycle;
	\fill[gray!25] (6.000000,18.000000) -- (6.000000,18.900000) -- (6.900000,18.000000) -- cycle;
	\fill[gray!25] (6.000000,18.000000) -- (6.900000,18.000000) -- (6.000000,17.100000) -- cycle;
	\fill[gray!25] (6.000000,18.000000) -- (6.000000,17.100000) -- (5.100000,18.000000) -- cycle;
	\fill[gray!25] (8.000000,18.000000) -- (7.100000,18.000000) -- (8.000000,18.900000) -- cycle;
	\fill[gray!25] (8.000000,18.000000) -- (8.000000,18.900000) -- (8.900000,18.000000) -- cycle;
	\fill[gray!25] (8.000000,18.000000) -- (8.900000,18.000000) -- (8.000000,17.100000) -- cycle;
	\fill[gray!25] (8.000000,18.000000) -- (8.000000,17.100000) -- (7.100000,18.000000) -- cycle;
	\fill[gray!25] (10.000000,18.000000) -- (9.100000,18.000000) -- (10.000000,18.900000) -- cycle;
	\fill[gray!25] (10.000000,18.000000) -- (10.000000,18.900000) -- (10.900000,18.000000) -- cycle;
	\fill[gray!25] (10.000000,18.000000) -- (10.900000,18.000000) -- (10.000000,17.100000) -- cycle;
	\fill[gray!25] (10.000000,18.000000) -- (10.000000,17.100000) -- (9.100000,18.000000) -- cycle;
	\fill[gray!25] (12.000000,18.000000) -- (11.100000,18.000000) -- (12.000000,18.900000) -- cycle;
	\fill[gray!25] (12.000000,18.000000) -- (12.000000,18.900000) -- (12.900000,18.000000) -- cycle;
	\fill[gray!25] (12.000000,18.000000) -- (12.900000,18.000000) -- (12.000000,17.100000) -- cycle;
	\fill[gray!25] (12.000000,18.000000) -- (12.000000,17.100000) -- (11.100000,18.000000) -- cycle;
	\fill[gray!25] (14.000000,18.000000) -- (13.100000,18.000000) -- (14.000000,18.900000) -- cycle;
	\fill[gray!25] (14.000000,18.000000) -- (14.000000,17.100000) -- (13.100000,18.000000) -- cycle;
	\fill[gray!80] (3.000000,17.000000) -- (2.100000,17.000000) -- (3.000000,17.900000) -- cycle;
	\fill[gray!80] (3.000000,17.000000) -- (3.000000,17.900000) -- (3.900000,17.000000) -- cycle;
	\fill[gray!80] (3.000000,17.000000) -- (3.900000,17.000000) -- (3.000000,16.100000) -- cycle;
	\fill[gray!80] (3.000000,17.000000) -- (3.000000,16.100000) -- (2.100000,17.000000) -- cycle;
	\fill[gray!80] (5.000000,17.000000) -- (4.100000,17.000000) -- (5.000000,17.900000) -- cycle;
	\fill[gray!80] (5.000000,17.000000) -- (5.000000,17.900000) -- (5.900000,17.000000) -- cycle;
	\fill[gray!80] (5.000000,17.000000) -- (5.900000,17.000000) -- (5.000000,16.100000) -- cycle;
	\fill[gray!80] (5.000000,17.000000) -- (5.000000,16.100000) -- (4.100000,17.000000) -- cycle;
	\fill[gray!80] (7.000000,17.000000) -- (6.100000,17.000000) -- (7.000000,17.900000) -- cycle;
	\fill[gray!80] (7.000000,17.000000) -- (7.000000,17.900000) -- (7.900000,17.000000) -- cycle;
	\fill[gray!80] (7.000000,17.000000) -- (7.900000,17.000000) -- (7.000000,16.100000) -- cycle;
	\fill[gray!80] (7.000000,17.000000) -- (7.000000,16.100000) -- (6.100000,17.000000) -- cycle;
	\fill[gray!80] (9.000000,17.000000) -- (8.100000,17.000000) -- (9.000000,17.900000) -- cycle;
	\fill[gray!80] (9.000000,17.000000) -- (9.000000,17.900000) -- (9.900000,17.000000) -- cycle;
	\fill[gray!80] (9.000000,17.000000) -- (9.900000,17.000000) -- (9.000000,16.100000) -- cycle;
	\fill[gray!80] (9.000000,17.000000) -- (9.000000,16.100000) -- (8.100000,17.000000) -- cycle;
	\fill[gray!80] (11.000000,17.000000) -- (10.100000,17.000000) -- (11.000000,17.900000) -- cycle;
	\fill[gray!80] (11.000000,17.000000) -- (11.000000,17.900000) -- (11.900000,17.000000) -- cycle;
	\fill[gray!80] (11.000000,17.000000) -- (11.900000,17.000000) -- (11.000000,16.100000) -- cycle;
	\fill[gray!80] (11.000000,17.000000) -- (11.000000,16.100000) -- (10.100000,17.000000) -- cycle;
	\fill[gray!80] (13.000000,17.000000) -- (12.100000,17.000000) -- (13.000000,17.900000) -- cycle;
	\fill[gray!80] (13.000000,17.000000) -- (13.000000,17.900000) -- (13.900000,17.000000) -- cycle;
	\fill[gray!80] (13.000000,17.000000) -- (13.900000,17.000000) -- (13.000000,16.100000) -- cycle;
	\fill[gray!80] (13.000000,17.000000) -- (13.000000,16.100000) -- (12.100000,17.000000) -- cycle;
	\fill[gray!25] (2.000000,16.000000) -- (2.000000,16.900000) -- (2.900000,16.000000) -- cycle;
	\fill[gray!25] (2.000000,16.000000) -- (2.900000,16.000000) -- (2.000000,15.100000) -- cycle;
	\fill[gray!25] (4.000000,16.000000) -- (3.100000,16.000000) -- (4.000000,16.900000) -- cycle;
	\fill[gray!25] (4.000000,16.000000) -- (4.000000,16.900000) -- (4.900000,16.000000) -- cycle;
	\fill[gray!25] (6.000000,16.000000) -- (5.100000,16.000000) -- (6.000000,16.900000) -- cycle;
	\fill[gray!25] (6.000000,16.000000) -- (6.000000,16.900000) -- (6.900000,16.000000) -- cycle;
	\fill[gray!25] (8.000000,16.000000) -- (7.100000,16.000000) -- (8.000000,16.900000) -- cycle;
	\fill[gray!25] (8.000000,16.000000) -- (8.000000,16.900000) -- (8.900000,16.000000) -- cycle;
	\fill[gray!25] (10.000000,16.000000) -- (9.100000,16.000000) -- (10.000000,16.900000) -- cycle;
	\fill[gray!25] (10.000000,16.000000) -- (10.000000,16.900000) -- (10.900000,16.000000) -- cycle;
	\fill[gray!25] (12.000000,16.000000) -- (11.100000,16.000000) -- (12.000000,16.900000) -- cycle;
	\fill[gray!25] (12.000000,16.000000) -- (12.000000,16.900000) -- (12.900000,16.000000) -- cycle;
	\fill[gray!25] (12.000000,16.000000) -- (12.900000,16.000000) -- (12.000000,15.100000) -- cycle;
	\fill[gray!25] (12.000000,16.000000) -- (12.000000,15.100000) -- (11.100000,16.000000) -- cycle;
	\fill[gray!25] (14.000000,16.000000) -- (13.100000,16.000000) -- (14.000000,16.900000) -- cycle;
	\fill[gray!25] (14.000000,16.000000) -- (14.000000,15.100000) -- (13.100000,16.000000) -- cycle;
	\fill[gray!80] (13.000000,15.000000) -- (12.100000,15.000000) -- (13.000000,15.900000) -- cycle;
	\fill[gray!80] (13.000000,15.000000) -- (13.000000,15.900000) -- (13.900000,15.000000) -- cycle;
	\fill[gray!80] (13.000000,15.000000) -- (13.900000,15.000000) -- (13.000000,14.100000) -- cycle;
	\fill[gray!80] (13.000000,15.000000) -- (13.000000,14.100000) -- (12.100000,15.000000) -- cycle;
	\fill[gray!25] (4.000000,14.000000) -- (4.900000,14.000000) -- (4.000000,13.100000) -- cycle;
	\fill[gray!25] (6.000000,14.000000) -- (6.900000,14.000000) -- (6.000000,13.100000) -- cycle;
	\fill[gray!25] (6.000000,14.000000) -- (6.000000,13.100000) -- (5.100000,14.000000) -- cycle;
	\fill[gray!25] (10.000000,14.000000) -- (10.000000,13.100000) -- (9.100000,14.000000) -- cycle;
	\fill[gray!25] (12.000000,14.000000) -- (12.000000,14.900000) -- (12.900000,14.000000) -- cycle;
	\fill[gray!25] (12.000000,14.000000) -- (12.900000,14.000000) -- (12.000000,13.100000) -- cycle;
	\fill[gray!80] (5.000000,13.000000) -- (4.100000,13.000000) -- (5.000000,13.900000) -- cycle;
	\fill[gray!80] (5.000000,13.000000) -- (5.000000,13.900000) -- (5.900000,13.000000) -- cycle;
	\fill[gray!80] (5.000000,13.000000) -- (5.900000,13.000000) -- (5.000000,12.100000) -- cycle;
	\fill[gray!80] (5.000000,13.000000) -- (5.000000,12.100000) -- (4.100000,13.000000) -- cycle;
	\fill[gray!80] (7.000000,13.000000) -- (6.100000,13.000000) -- (7.000000,13.900000) -- cycle;
	\fill[gray!80] (7.000000,13.000000) -- (7.000000,13.900000) -- (7.900000,13.000000) -- cycle;
	\fill[gray!80] (7.000000,13.000000) -- (7.900000,13.000000) -- (7.000000,12.100000) -- cycle;
	\fill[gray!80] (7.000000,13.000000) -- (7.000000,12.100000) -- (6.100000,13.000000) -- cycle;
	\fill[gray!80] (9.000000,13.000000) -- (8.100000,13.000000) -- (9.000000,13.900000) -- cycle;
	\fill[gray!80] (9.000000,13.000000) -- (9.000000,13.900000) -- (9.900000,13.000000) -- cycle;
	\fill[gray!80] (9.000000,13.000000) -- (9.900000,13.000000) -- (9.000000,12.100000) -- cycle;
	\fill[gray!80] (9.000000,13.000000) -- (9.000000,12.100000) -- (8.100000,13.000000) -- cycle;
	\fill[gray!80] (13.000000,13.000000) -- (12.100000,13.000000) -- (13.000000,13.900000) -- cycle;
	\fill[gray!80] (13.000000,13.000000) -- (13.000000,13.900000) -- (13.900000,13.000000) -- cycle;
	\fill[gray!80] (13.000000,13.000000) -- (13.900000,13.000000) -- (13.000000,12.100000) -- cycle;
	\fill[gray!80] (13.000000,13.000000) -- (13.000000,12.100000) -- (12.100000,13.000000) -- cycle;
	\fill[gray!25] (4.000000,12.000000) -- (4.000000,12.900000) -- (4.900000,12.000000) -- cycle;
	\fill[gray!25] (4.000000,12.000000) -- (4.900000,12.000000) -- (4.000000,11.100000) -- cycle;
	\fill[gray!25] (6.000000,12.000000) -- (5.100000,12.000000) -- (6.000000,12.900000) -- cycle;
	\fill[gray!25] (6.000000,12.000000) -- (6.000000,12.900000) -- (6.900000,12.000000) -- cycle;
	\fill[gray!25] (6.000000,12.000000) -- (6.900000,12.000000) -- (6.000000,11.100000) -- cycle;
	\fill[gray!25] (6.000000,12.000000) -- (6.000000,11.100000) -- (5.100000,12.000000) -- cycle;
	\fill[gray!25] (8.000000,12.000000) -- (7.100000,12.000000) -- (8.000000,12.900000) -- cycle;
	\fill[gray!25] (8.000000,12.000000) -- (8.000000,12.900000) -- (8.900000,12.000000) -- cycle;
	\fill[gray!25] (8.000000,12.000000) -- (8.900000,12.000000) -- (8.000000,11.100000) -- cycle;
	\fill[gray!25] (8.000000,12.000000) -- (8.000000,11.100000) -- (7.100000,12.000000) -- cycle;
	\fill[gray!25] (10.000000,12.000000) -- (9.100000,12.000000) -- (10.000000,12.900000) -- cycle;
	\fill[gray!25] (10.000000,12.000000) -- (10.000000,11.100000) -- (9.100000,12.000000) -- cycle;
	\fill[gray!25] (12.000000,12.000000) -- (12.000000,12.900000) -- (12.900000,12.000000) -- cycle;
	\fill[gray!25] (12.000000,12.000000) -- (12.900000,12.000000) -- (12.000000,11.100000) -- cycle;
	\fill[gray!25] (14.000000,12.000000) -- (13.100000,12.000000) -- (14.000000,12.900000) -- cycle;
	\fill[gray!25] (14.000000,12.000000) -- (14.000000,11.100000) -- (13.100000,12.000000) -- cycle;
	\fill[gray!80] (5.000000,11.000000) -- (4.100000,11.000000) -- (5.000000,11.900000) -- cycle;
	\fill[gray!80] (5.000000,11.000000) -- (5.000000,11.900000) -- (5.900000,11.000000) -- cycle;
	\fill[gray!80] (5.000000,11.000000) -- (5.900000,11.000000) -- (5.000000,10.100000) -- cycle;
	\fill[gray!80] (5.000000,11.000000) -- (5.000000,10.100000) -- (4.100000,11.000000) -- cycle;
	\fill[gray!80] (7.000000,11.000000) -- (6.100000,11.000000) -- (7.000000,11.900000) -- cycle;
	\fill[gray!80] (7.000000,11.000000) -- (7.000000,11.900000) -- (7.900000,11.000000) -- cycle;
	\fill[gray!80] (7.000000,11.000000) -- (7.900000,11.000000) -- (7.000000,10.100000) -- cycle;
	\fill[gray!80] (7.000000,11.000000) -- (7.000000,10.100000) -- (6.100000,11.000000) -- cycle;
	\fill[gray!80] (9.000000,11.000000) -- (8.100000,11.000000) -- (9.000000,11.900000) -- cycle;
	\fill[gray!80] (9.000000,11.000000) -- (9.000000,11.900000) -- (9.900000,11.000000) -- cycle;
	\fill[gray!80] (9.000000,11.000000) -- (9.900000,11.000000) -- (9.000000,10.100000) -- cycle;
	\fill[gray!80] (9.000000,11.000000) -- (9.000000,10.100000) -- (8.100000,11.000000) -- cycle;
	\fill[gray!80] (13.000000,11.000000) -- (12.100000,11.000000) -- (13.000000,11.900000) -- cycle;
	\fill[gray!80] (13.000000,11.000000) -- (13.000000,11.900000) -- (13.900000,11.000000) -- cycle;
	\fill[gray!80] (13.000000,11.000000) -- (13.900000,11.000000) -- (13.000000,10.100000) -- cycle;
	\fill[gray!80] (13.000000,11.000000) -- (13.000000,10.100000) -- (12.100000,11.000000) -- cycle;
	\fill[gray!25] (4.000000,10.000000) -- (4.000000,10.900000) -- (4.900000,10.000000) -- cycle;
	\fill[gray!25] (4.000000,10.000000) -- (4.900000,10.000000) -- (4.000000,9.100000) -- cycle;
	\fill[gray!25] (6.000000,10.000000) -- (5.100000,10.000000) -- (6.000000,10.900000) -- cycle;
	\fill[gray!25] (6.000000,10.000000) -- (6.000000,10.900000) -- (6.900000,10.000000) -- cycle;
	\fill[gray!25] (6.000000,10.000000) -- (6.900000,10.000000) -- (6.000000,9.100000) -- cycle;
	\fill[gray!25] (6.000000,10.000000) -- (6.000000,9.100000) -- (5.100000,10.000000) -- cycle;
	\fill[gray!25] (8.000000,10.000000) -- (7.100000,10.000000) -- (8.000000,10.900000) -- cycle;
	\fill[gray!25] (8.000000,10.000000) -- (8.000000,10.900000) -- (8.900000,10.000000) -- cycle;
	\fill[gray!25] (8.000000,10.000000) -- (8.900000,10.000000) -- (8.000000,9.100000) -- cycle;
	\fill[gray!25] (8.000000,10.000000) -- (8.000000,9.100000) -- (7.100000,10.000000) -- cycle;
	\fill[gray!25] (10.000000,10.000000) -- (9.100000,10.000000) -- (10.000000,10.900000) -- cycle;
	\fill[gray!25] (10.000000,10.000000) -- (10.000000,9.100000) -- (9.100000,10.000000) -- cycle;
	\fill[gray!25] (12.000000,10.000000) -- (12.000000,10.900000) -- (12.900000,10.000000) -- cycle;
	\fill[gray!25] (12.000000,10.000000) -- (12.900000,10.000000) -- (12.000000,9.100000) -- cycle;
	\fill[gray!25] (14.000000,10.000000) -- (13.100000,10.000000) -- (14.000000,10.900000) -- cycle;
	\fill[gray!25] (14.000000,10.000000) -- (14.000000,9.100000) -- (13.100000,10.000000) -- cycle;
	\fill[gray!80] (5.000000,9.000000) -- (4.100000,9.000000) -- (5.000000,9.900000) -- cycle;
	\fill[gray!80] (5.000000,9.000000) -- (5.000000,9.900000) -- (5.900000,9.000000) -- cycle;
	\fill[gray!80] (5.000000,9.000000) -- (5.900000,9.000000) -- (5.000000,8.100000) -- cycle;
	\fill[gray!80] (5.000000,9.000000) -- (5.000000,8.100000) -- (4.100000,9.000000) -- cycle;
	\fill[gray!80] (7.000000,9.000000) -- (6.100000,9.000000) -- (7.000000,9.900000) -- cycle;
	\fill[gray!80] (7.000000,9.000000) -- (7.000000,9.900000) -- (7.900000,9.000000) -- cycle;
	\fill[gray!80] (7.000000,9.000000) -- (7.900000,9.000000) -- (7.000000,8.100000) -- cycle;
	\fill[gray!80] (7.000000,9.000000) -- (7.000000,8.100000) -- (6.100000,9.000000) -- cycle;
	\fill[gray!80] (9.000000,9.000000) -- (8.100000,9.000000) -- (9.000000,9.900000) -- cycle;
	\fill[gray!80] (9.000000,9.000000) -- (9.000000,9.900000) -- (9.900000,9.000000) -- cycle;
	\fill[gray!80] (9.000000,9.000000) -- (9.900000,9.000000) -- (9.000000,8.100000) -- cycle;
	\fill[gray!80] (9.000000,9.000000) -- (9.000000,8.100000) -- (8.100000,9.000000) -- cycle;
	\fill[gray!80] (13.000000,9.000000) -- (12.100000,9.000000) -- (13.000000,9.900000) -- cycle;
	\fill[gray!80] (13.000000,9.000000) -- (13.000000,9.900000) -- (13.900000,9.000000) -- cycle;
	\fill[gray!80] (13.000000,9.000000) -- (13.900000,9.000000) -- (13.000000,8.100000) -- cycle;
	\fill[gray!80] (13.000000,9.000000) -- (13.000000,8.100000) -- (12.100000,9.000000) -- cycle;
	\fill[gray!25] (4.000000,8.000000) -- (4.000000,8.900000) -- (4.900000,8.000000) -- cycle;
	\fill[gray!25] (6.000000,8.000000) -- (5.100000,8.000000) -- (6.000000,8.900000) -- cycle;
	\fill[gray!25] (6.000000,8.000000) -- (6.000000,8.900000) -- (6.900000,8.000000) -- cycle;
	\fill[gray!25] (10.000000,8.000000) -- (9.100000,8.000000) -- (10.000000,8.900000) -- cycle;
	\fill[gray!25] (12.000000,8.000000) -- (12.000000,8.900000) -- (12.900000,8.000000) -- cycle;
	\fill[gray!25] (12.000000,8.000000) -- (12.900000,8.000000) -- (12.000000,7.100000) -- cycle;
	\fill[gray!80] (13.000000,7.000000) -- (12.100000,7.000000) -- (13.000000,7.900000) -- cycle;
	\fill[gray!80] (13.000000,7.000000) -- (13.000000,7.900000) -- (13.900000,7.000000) -- cycle;
	\fill[gray!80] (13.000000,7.000000) -- (13.900000,7.000000) -- (13.000000,6.100000) -- cycle;
	\fill[gray!80] (13.000000,7.000000) -- (13.000000,6.100000) -- (12.100000,7.000000) -- cycle;
	\fill[gray!25] (2.000000,6.000000) -- (2.000000,6.900000) -- (2.900000,6.000000) -- cycle;
	\fill[gray!25] (2.000000,6.000000) -- (2.900000,6.000000) -- (2.000000,5.100000) -- cycle;
	\fill[gray!25] (4.000000,6.000000) -- (4.900000,6.000000) -- (4.000000,5.100000) -- cycle;
	\fill[gray!25] (4.000000,6.000000) -- (4.000000,5.100000) -- (3.100000,6.000000) -- cycle;
	\fill[gray!25] (6.000000,6.000000) -- (6.900000,6.000000) -- (6.000000,5.100000) -- cycle;
	\fill[gray!25] (6.000000,6.000000) -- (6.000000,5.100000) -- (5.100000,6.000000) -- cycle;
	\fill[gray!25] (8.000000,6.000000) -- (8.900000,6.000000) -- (8.000000,5.100000) -- cycle;
	\fill[gray!25] (8.000000,6.000000) -- (8.000000,5.100000) -- (7.100000,6.000000) -- cycle;
	\fill[gray!25] (10.000000,6.000000) -- (10.900000,6.000000) -- (10.000000,5.100000) -- cycle;
	\fill[gray!25] (10.000000,6.000000) -- (10.000000,5.100000) -- (9.100000,6.000000) -- cycle;
	\fill[gray!25] (12.000000,6.000000) -- (11.100000,6.000000) -- (12.000000,6.900000) -- cycle;
	\fill[gray!25] (12.000000,6.000000) -- (12.000000,6.900000) -- (12.900000,6.000000) -- cycle;
	\fill[gray!25] (12.000000,6.000000) -- (12.900000,6.000000) -- (12.000000,5.100000) -- cycle;
	\fill[gray!25] (12.000000,6.000000) -- (12.000000,5.100000) -- (11.100000,6.000000) -- cycle;
	\fill[gray!25] (14.000000,6.000000) -- (13.100000,6.000000) -- (14.000000,6.900000) -- cycle;
	\fill[gray!25] (14.000000,6.000000) -- (14.000000,5.100000) -- (13.100000,6.000000) -- cycle;
	\fill[gray!80] (3.000000,5.000000) -- (2.100000,5.000000) -- (3.000000,5.900000) -- cycle;
	\fill[gray!80] (3.000000,5.000000) -- (3.000000,5.900000) -- (3.900000,5.000000) -- cycle;
	\fill[gray!80] (3.000000,5.000000) -- (3.900000,5.000000) -- (3.000000,4.100000) -- cycle;
	\fill[gray!80] (3.000000,5.000000) -- (3.000000,4.100000) -- (2.100000,5.000000) -- cycle;
	\fill[gray!80] (5.000000,5.000000) -- (4.100000,5.000000) -- (5.000000,5.900000) -- cycle;
	\fill[gray!80] (5.000000,5.000000) -- (5.000000,5.900000) -- (5.900000,5.000000) -- cycle;
	\fill[gray!80] (5.000000,5.000000) -- (5.900000,5.000000) -- (5.000000,4.100000) -- cycle;
	\fill[gray!80] (5.000000,5.000000) -- (5.000000,4.100000) -- (4.100000,5.000000) -- cycle;
	\fill[gray!80] (7.000000,5.000000) -- (6.100000,5.000000) -- (7.000000,5.900000) -- cycle;
	\fill[gray!80] (7.000000,5.000000) -- (7.000000,5.900000) -- (7.900000,5.000000) -- cycle;
	\fill[gray!80] (7.000000,5.000000) -- (7.900000,5.000000) -- (7.000000,4.100000) -- cycle;
	\fill[gray!80] (7.000000,5.000000) -- (7.000000,4.100000) -- (6.100000,5.000000) -- cycle;
	\fill[gray!80] (9.000000,5.000000) -- (8.100000,5.000000) -- (9.000000,5.900000) -- cycle;
	\fill[gray!80] (9.000000,5.000000) -- (9.000000,5.900000) -- (9.900000,5.000000) -- cycle;
	\fill[gray!80] (9.000000,5.000000) -- (9.900000,5.000000) -- (9.000000,4.100000) -- cycle;
	\fill[gray!80] (9.000000,5.000000) -- (9.000000,4.100000) -- (8.100000,5.000000) -- cycle;
	\fill[gray!80] (11.000000,5.000000) -- (10.100000,5.000000) -- (11.000000,5.900000) -- cycle;
	\fill[gray!80] (11.000000,5.000000) -- (11.000000,5.900000) -- (11.900000,5.000000) -- cycle;
	\fill[gray!80] (11.000000,5.000000) -- (11.900000,5.000000) -- (11.000000,4.100000) -- cycle;
	\fill[gray!80] (11.000000,5.000000) -- (11.000000,4.100000) -- (10.100000,5.000000) -- cycle;
	\fill[gray!80] (13.000000,5.000000) -- (12.100000,5.000000) -- (13.000000,5.900000) -- cycle;
	\fill[gray!80] (13.000000,5.000000) -- (13.000000,5.900000) -- (13.900000,5.000000) -- cycle;
	\fill[gray!80] (13.000000,5.000000) -- (13.900000,5.000000) -- (13.000000,4.100000) -- cycle;
	\fill[gray!80] (13.000000,5.000000) -- (13.000000,4.100000) -- (12.100000,5.000000) -- cycle;
	\fill[gray!25] (2.000000,4.000000) -- (2.000000,4.900000) -- (2.900000,4.000000) -- cycle;
	\fill[gray!25] (2.000000,4.000000) -- (2.900000,4.000000) -- (2.000000,3.100000) -- cycle;
	\fill[gray!25] (4.000000,4.000000) -- (3.100000,4.000000) -- (4.000000,4.900000) -- cycle;
	\fill[gray!25] (4.000000,4.000000) -- (4.000000,4.900000) -- (4.900000,4.000000) -- cycle;
	\fill[gray!25] (4.000000,4.000000) -- (4.900000,4.000000) -- (4.000000,3.100000) -- cycle;
	\fill[gray!25] (4.000000,4.000000) -- (4.000000,3.100000) -- (3.100000,4.000000) -- cycle;
	\fill[gray!25] (6.000000,4.000000) -- (5.100000,4.000000) -- (6.000000,4.900000) -- cycle;
	\fill[gray!25] (6.000000,4.000000) -- (6.000000,4.900000) -- (6.900000,4.000000) -- cycle;
	\fill[gray!25] (6.000000,4.000000) -- (6.900000,4.000000) -- (6.000000,3.100000) -- cycle;
	\fill[gray!25] (6.000000,4.000000) -- (6.000000,3.100000) -- (5.100000,4.000000) -- cycle;
	\fill[gray!25] (8.000000,4.000000) -- (7.100000,4.000000) -- (8.000000,4.900000) -- cycle;
	\fill[gray!25] (8.000000,4.000000) -- (8.000000,4.900000) -- (8.900000,4.000000) -- cycle;
	\fill[gray!25] (8.000000,4.000000) -- (8.900000,4.000000) -- (8.000000,3.100000) -- cycle;
	\fill[gray!25] (8.000000,4.000000) -- (8.000000,3.100000) -- (7.100000,4.000000) -- cycle;
	\fill[gray!25] (10.000000,4.000000) -- (9.100000,4.000000) -- (10.000000,4.900000) -- cycle;
	\fill[gray!25] (10.000000,4.000000) -- (10.000000,4.900000) -- (10.900000,4.000000) -- cycle;
	\fill[gray!25] (10.000000,4.000000) -- (10.900000,4.000000) -- (10.000000,3.100000) -- cycle;
	\fill[gray!25] (10.000000,4.000000) -- (10.000000,3.100000) -- (9.100000,4.000000) -- cycle;
	\fill[gray!25] (12.000000,4.000000) -- (11.100000,4.000000) -- (12.000000,4.900000) -- cycle;
	\fill[gray!25] (12.000000,4.000000) -- (12.000000,4.900000) -- (12.900000,4.000000) -- cycle;
	\fill[gray!25] (12.000000,4.000000) -- (12.900000,4.000000) -- (12.000000,3.100000) -- cycle;
	\fill[gray!25] (12.000000,4.000000) -- (12.000000,3.100000) -- (11.100000,4.000000) -- cycle;
	\fill[gray!25] (14.000000,4.000000) -- (13.100000,4.000000) -- (14.000000,4.900000) -- cycle;
	\fill[gray!25] (14.000000,4.000000) -- (14.000000,3.100000) -- (13.100000,4.000000) -- cycle;
	\fill[gray!80] (3.000000,3.000000) -- (2.100000,3.000000) -- (3.000000,3.900000) -- cycle;
	\fill[gray!80] (3.000000,3.000000) -- (3.000000,3.900000) -- (3.900000,3.000000) -- cycle;
	\fill[gray!80] (3.000000,3.000000) -- (3.900000,3.000000) -- (3.000000,2.100000) -- cycle;
	\fill[gray!80] (3.000000,3.000000) -- (3.000000,2.100000) -- (2.100000,3.000000) -- cycle;
	\fill[gray!80] (5.000000,3.000000) -- (4.100000,3.000000) -- (5.000000,3.900000) -- cycle;
	\fill[gray!80] (5.000000,3.000000) -- (5.000000,3.900000) -- (5.900000,3.000000) -- cycle;
	\fill[gray!80] (5.000000,3.000000) -- (5.900000,3.000000) -- (5.000000,2.100000) -- cycle;
	\fill[gray!80] (5.000000,3.000000) -- (5.000000,2.100000) -- (4.100000,3.000000) -- cycle;
	\fill[gray!80] (7.000000,3.000000) -- (6.100000,3.000000) -- (7.000000,3.900000) -- cycle;
	\fill[gray!80] (7.000000,3.000000) -- (7.000000,3.900000) -- (7.900000,3.000000) -- cycle;
	\fill[gray!80] (7.000000,3.000000) -- (7.900000,3.000000) -- (7.000000,2.100000) -- cycle;
	\fill[gray!80] (7.000000,3.000000) -- (7.000000,2.100000) -- (6.100000,3.000000) -- cycle;
	\fill[gray!80] (9.000000,3.000000) -- (8.100000,3.000000) -- (9.000000,3.900000) -- cycle;
	\fill[gray!80] (9.000000,3.000000) -- (9.000000,3.900000) -- (9.900000,3.000000) -- cycle;
	\fill[gray!80] (9.000000,3.000000) -- (9.900000,3.000000) -- (9.000000,2.100000) -- cycle;
	\fill[gray!80] (9.000000,3.000000) -- (9.000000,2.100000) -- (8.100000,3.000000) -- cycle;
	\fill[gray!80] (11.000000,3.000000) -- (10.100000,3.000000) -- (11.000000,3.900000) -- cycle;
	\fill[gray!80] (11.000000,3.000000) -- (11.000000,3.900000) -- (11.900000,3.000000) -- cycle;
	\fill[gray!80] (11.000000,3.000000) -- (11.900000,3.000000) -- (11.000000,2.100000) -- cycle;
	\fill[gray!80] (11.000000,3.000000) -- (11.000000,2.100000) -- (10.100000,3.000000) -- cycle;
	\fill[gray!80] (13.000000,3.000000) -- (12.100000,3.000000) -- (13.000000,3.900000) -- cycle;
	\fill[gray!80] (13.000000,3.000000) -- (13.000000,3.900000) -- (13.900000,3.000000) -- cycle;
	\fill[gray!80] (13.000000,3.000000) -- (13.900000,3.000000) -- (13.000000,2.100000) -- cycle;
	\fill[gray!80] (13.000000,3.000000) -- (13.000000,2.100000) -- (12.100000,3.000000) -- cycle;
	\fill[gray!25] (4.000000,2.000000) -- (3.100000,2.000000) -- (4.000000,2.900000) -- cycle;
	\fill[gray!25] (4.000000,2.000000) -- (4.000000,2.900000) -- (4.900000,2.000000) -- cycle;
	\fill[gray!25] (6.000000,2.000000) -- (5.100000,2.000000) -- (6.000000,2.900000) -- cycle;
	\fill[gray!25] (6.000000,2.000000) -- (6.000000,2.900000) -- (6.900000,2.000000) -- cycle;
	\fill[gray!25] (10.000000,2.000000) -- (9.100000,2.000000) -- (10.000000,2.900000) -- cycle;
	\fill[gray!25] (10.000000,2.000000) -- (10.000000,2.900000) -- (10.900000,2.000000) -- cycle;
	\fill[gray!25] (12.000000,2.000000) -- (11.100000,2.000000) -- (12.000000,2.900000) -- cycle;
	\fill[gray!25] (12.000000,2.000000) -- (12.000000,2.900000) -- (12.900000,2.000000) -- cycle;
	\draw[line width=1pt] (3.000000,20.000000) circle (0.100000);
	\draw[line width=1pt,fill=black] (4.000000,20.000000) circle (0.100000);
	\draw[line width=1pt] (5.000000,20.000000) circle (0.100000);
	\draw[line width=1pt,fill=black] (6.000000,20.000000) circle (0.100000);
	\draw[line width=1pt] (7.000000,20.000000) circle (0.100000);
	\draw[line width=1pt] (9.000000,20.000000) circle (0.100000);
	\draw[line width=1pt,fill=black] (10.000000,20.000000) circle (0.100000);
	\draw[line width=1pt] (11.000000,20.000000) circle (0.100000);
	\draw[line width=1pt,fill=black] (12.000000,20.000000) circle (0.100000);
	\draw[line width=1pt] (13.000000,20.000000) circle (0.100000);
	\draw[line width=1pt] (2.000000,19.000000) circle (0.100000);
	\draw[line width=1pt,fill=black] (3.000000,19.000000) circle (0.100000);
	\draw[line width=1pt] (4.000000,19.000000) circle (0.100000);
	\draw[line width=1pt,fill=black] (5.000000,19.000000) circle (0.100000);
	\draw[line width=1pt] (6.000000,19.000000) circle (0.100000);
	\draw[line width=1pt,fill=black] (7.000000,19.000000) circle (0.100000);
	\draw[line width=1pt] (8.000000,19.000000) circle (0.100000);
	\draw[line width=1pt,fill=black] (9.000000,19.000000) circle (0.100000);
	\draw[line width=1pt] (10.000000,19.000000) circle (0.100000);
	\draw[line width=1pt,fill=black] (11.000000,19.000000) circle (0.100000);
	\draw[line width=1pt] (12.000000,19.000000) circle (0.100000);
	\draw[line width=1pt,fill=black] (13.000000,19.000000) circle (0.100000);
	\draw[line width=1pt] (14.000000,19.000000) circle (0.100000);
	\draw[line width=1pt,fill=black] (2.000000,18.000000) circle (0.100000);
	\draw[line width=1pt] (3.000000,18.000000) circle (0.100000);
	\draw[line width=1pt,fill=black] (4.000000,18.000000) circle (0.100000);
	\draw[line width=1pt] (5.000000,18.000000) circle (0.100000);
	\draw[line width=1pt,fill=black] (6.000000,18.000000) circle (0.100000);
	\draw[line width=1pt] (7.000000,18.000000) circle (0.100000);
	\draw[line width=1pt,fill=black] (8.000000,18.000000) circle (0.100000);
	\draw[line width=1pt] (9.000000,18.000000) circle (0.100000);
	\draw[line width=1pt,fill=black] (10.000000,18.000000) circle (0.100000);
	\draw[line width=1pt] (11.000000,18.000000) circle (0.100000);
	\draw[line width=1pt,fill=black] (12.000000,18.000000) circle (0.100000);
	\draw[line width=1pt] (13.000000,18.000000) circle (0.100000);
	\draw[line width=1pt,fill=black] (14.000000,18.000000) circle (0.100000);
	\draw[line width=1pt] (2.000000,17.000000) circle (0.100000);
	\draw[line width=1pt,fill=black] (3.000000,17.000000) circle (0.100000);
	\draw[line width=1pt] (4.000000,17.000000) circle (0.100000);
	\draw[line width=1pt,fill=black] (5.000000,17.000000) circle (0.100000);
	\draw[line width=1pt] (6.000000,17.000000) circle (0.100000);
	\draw[line width=1pt,fill=black] (7.000000,17.000000) circle (0.100000);
	\draw[line width=1pt] (8.000000,17.000000) circle (0.100000);
	\draw[line width=1pt,fill=black] (9.000000,17.000000) circle (0.100000);
	\draw[line width=1pt] (10.000000,17.000000) circle (0.100000);
	\draw[line width=1pt,fill=black] (11.000000,17.000000) circle (0.100000);
	\draw[line width=1pt] (12.000000,17.000000) circle (0.100000);
	\draw[line width=1pt,fill=black] (13.000000,17.000000) circle (0.100000);
	\draw[line width=1pt] (14.000000,17.000000) circle (0.100000);
	\draw[line width=1pt,fill=black] (2.000000,16.000000) circle (0.100000);
	\draw[line width=1pt] (3.000000,16.000000) circle (0.100000);
	\draw[line width=1pt,fill=black] (4.000000,16.000000) circle (0.100000);
	\draw[line width=1pt] (5.000000,16.000000) circle (0.100000);
	\draw[line width=1pt,fill=black] (6.000000,16.000000) circle (0.100000);
	\draw[line width=1pt] (7.000000,16.000000) circle (0.100000);
	\draw[line width=1pt,fill=black] (8.000000,16.000000) circle (0.100000);
	\draw[line width=1pt] (9.000000,16.000000) circle (0.100000);
	\draw[line width=1pt,fill=black] (10.000000,16.000000) circle (0.100000);
	\draw[line width=1pt] (11.000000,16.000000) circle (0.100000);
	\draw[line width=1pt,fill=black] (12.000000,16.000000) circle (0.100000);
	\draw[line width=1pt] (13.000000,16.000000) circle (0.100000);
	\draw[line width=1pt,fill=black] (14.000000,16.000000) circle (0.100000);
	\draw[line width=1pt] (2.000000,15.000000) circle (0.100000);
	\draw[line width=1pt] (12.000000,15.000000) circle (0.100000);
	\draw[line width=1pt,fill=black] (13.000000,15.000000) circle (0.100000);
	\draw[line width=1pt] (14.000000,15.000000) circle (0.100000);
	\draw[line width=1pt,fill=black] (4.000000,14.000000) circle (0.100000);
	\draw (5.000000,14.000000) node [above right] {$M_Z$};
	\draw[line width=1pt] (5.000000,14.000000) circle (0.100000);
	\draw[line width=1pt,fill=black] (6.000000,14.000000) circle (0.100000);
	\draw (7.000000,14.000000) node [above right] {$M_X$};
	\draw[line width=1pt] (7.000000,14.000000) circle (0.100000);
	\draw (9.000000,14.000000) node [above right] {$M_X$};
	\draw[line width=1pt] (9.000000,14.000000) circle (0.100000);
	\draw[line width=1pt,fill=black] (10.000000,14.000000) circle (0.100000);
	\draw[line width=1pt,fill=black] (12.000000,14.000000) circle (0.100000);
	\draw[line width=1pt] (13.000000,14.000000) circle (0.100000);
	\draw[line width=1pt] (2.000000,13.000000) circle (0.100000);
	\draw (4.000000,13.000000) node [above right] {$M_Z$};
	\draw[line width=1pt] (4.000000,13.000000) circle (0.100000);
	\draw[line width=1pt,fill=black] (5.000000,13.000000) circle (0.100000);
	\draw[line width=1pt] (6.000000,13.000000) circle (0.100000);
	\draw[line width=1pt,fill=black] (7.000000,13.000000) circle (0.100000);
	\draw[line width=1pt] (8.000000,13.000000) circle (0.100000);
	\draw[line width=1pt,fill=black] (9.000000,13.000000) circle (0.100000);
	\draw[line width=1pt] (10.000000,13.000000) circle (0.100000);
	\draw[line width=1pt] (12.000000,13.000000) circle (0.100000);
	\draw[line width=1pt,fill=black] (13.000000,13.000000) circle (0.100000);
	\draw[line width=1pt] (14.000000,13.000000) circle (0.100000);
	\draw[line width=1pt,fill=black] (2.000000,12.000000) circle (0.100000);
	\draw[line width=1pt,fill=black] (4.000000,12.000000) circle (0.100000);
	\draw (5.000000,12.000000) node [above right] {$M_Z$};
	\draw[line width=1pt] (5.000000,12.000000) circle (0.100000);
	\draw[line width=1pt,fill=black] (6.000000,12.000000) circle (0.100000);
	\draw[line width=1pt] (7.000000,12.000000) circle (0.100000);
	\draw[line width=1pt,fill=black] (8.000000,12.000000) circle (0.100000);
	\draw[line width=1pt] (9.000000,12.000000) circle (0.100000);
	\draw[line width=1pt,fill=black] (10.000000,12.000000) circle (0.100000);
	\draw[line width=1pt,fill=black] (12.000000,12.000000) circle (0.100000);
	\draw[line width=1pt] (13.000000,12.000000) circle (0.100000);
	\draw[line width=1pt,fill=black] (14.000000,12.000000) circle (0.100000);
	\draw[line width=1pt] (2.000000,11.000000) circle (0.100000);
	\draw (4.000000,11.000000) node [above right] {$M_Z$};
	\draw[line width=1pt] (4.000000,11.000000) circle (0.100000);
	\draw[line width=1pt,fill=black] (5.000000,11.000000) circle (0.100000);
	\draw[line width=1pt] (6.000000,11.000000) circle (0.100000);
	\draw[line width=1pt,fill=black] (7.000000,11.000000) circle (0.100000);
	\draw[line width=1pt] (8.000000,11.000000) circle (0.100000);
	\draw[line width=1pt,fill=black] (9.000000,11.000000) circle (0.100000);
	\draw[line width=1pt] (10.000000,11.000000) circle (0.100000);
	\draw[line width=1pt] (12.000000,11.000000) circle (0.100000);
	\draw[line width=1pt,fill=black] (13.000000,11.000000) circle (0.100000);
	\draw[line width=1pt] (14.000000,11.000000) circle (0.100000);
	\draw[line width=1pt,fill=black] (2.000000,10.000000) circle (0.100000);
	\draw[line width=1pt,fill=black] (4.000000,10.000000) circle (0.100000);
	\draw (5.000000,10.000000) node [above right] {$M_Z$};
	\draw[line width=1pt] (5.000000,10.000000) circle (0.100000);
	\draw[line width=1pt,fill=black] (6.000000,10.000000) circle (0.100000);
	\draw[line width=1pt] (7.000000,10.000000) circle (0.100000);
	\draw[line width=1pt,fill=black] (8.000000,10.000000) circle (0.100000);
	\draw[line width=1pt] (9.000000,10.000000) circle (0.100000);
	\draw[line width=1pt,fill=black] (10.000000,10.000000) circle (0.100000);
	\draw[line width=1pt,fill=black] (12.000000,10.000000) circle (0.100000);
	\draw[line width=1pt] (13.000000,10.000000) circle (0.100000);
	\draw[line width=1pt,fill=black] (14.000000,10.000000) circle (0.100000);
	\draw[line width=1pt] (2.000000,9.000000) circle (0.100000);
	\draw (4.000000,9.000000) node [above right] {$M_Z$};
	\draw[line width=1pt] (4.000000,9.000000) circle (0.100000);
	\draw[line width=1pt,fill=black] (5.000000,9.000000) circle (0.100000);
	\draw[line width=1pt] (6.000000,9.000000) circle (0.100000);
	\draw[line width=1pt,fill=black] (7.000000,9.000000) circle (0.100000);
	\draw[line width=1pt] (8.000000,9.000000) circle (0.100000);
	\draw[line width=1pt,fill=black] (9.000000,9.000000) circle (0.100000);
	\draw[line width=1pt] (10.000000,9.000000) circle (0.100000);
	\draw[line width=1pt] (12.000000,9.000000) circle (0.100000);
	\draw[line width=1pt,fill=black] (13.000000,9.000000) circle (0.100000);
	\draw[line width=1pt] (14.000000,9.000000) circle (0.100000);
	\draw[line width=1pt,fill=black] (4.000000,8.000000) circle (0.100000);
	\draw (5.000000,8.000000) node [above right] {$M_Z$};
	\draw[line width=1pt] (5.000000,8.000000) circle (0.100000);
	\draw[line width=1pt,fill=black] (6.000000,8.000000) circle (0.100000);
	\draw (7.000000,8.000000) node [above right] {$M_X$};
	\draw[line width=1pt] (7.000000,8.000000) circle (0.100000);
	\draw (9.000000,8.000000) node [above right] {$M_X$};
	\draw[line width=1pt] (9.000000,8.000000) circle (0.100000);
	\draw[line width=1pt,fill=black] (10.000000,8.000000) circle (0.100000);
	\draw[line width=1pt,fill=black] (12.000000,8.000000) circle (0.100000);
	\draw[line width=1pt] (13.000000,8.000000) circle (0.100000);
	\draw[line width=1pt] (2.000000,7.000000) circle (0.100000);
	\draw[line width=1pt] (12.000000,7.000000) circle (0.100000);
	\draw[line width=1pt,fill=black] (13.000000,7.000000) circle (0.100000);
	\draw[line width=1pt] (14.000000,7.000000) circle (0.100000);
	\draw[line width=1pt,fill=black] (2.000000,6.000000) circle (0.100000);
	\draw[line width=1pt] (3.000000,6.000000) circle (0.100000);
	\draw[line width=1pt,fill=black] (4.000000,6.000000) circle (0.100000);
	\draw[line width=1pt] (5.000000,6.000000) circle (0.100000);
	\draw[line width=1pt,fill=black] (6.000000,6.000000) circle (0.100000);
	\draw[line width=1pt] (7.000000,6.000000) circle (0.100000);
	\draw[line width=1pt,fill=black] (8.000000,6.000000) circle (0.100000);
	\draw[line width=1pt] (9.000000,6.000000) circle (0.100000);
	\draw[line width=1pt,fill=black] (10.000000,6.000000) circle (0.100000);
	\draw[line width=1pt] (11.000000,6.000000) circle (0.100000);
	\draw[line width=1pt,fill=black] (12.000000,6.000000) circle (0.100000);
	\draw[line width=1pt] (13.000000,6.000000) circle (0.100000);
	\draw[line width=1pt,fill=black] (14.000000,6.000000) circle (0.100000);
	\draw[line width=1pt] (2.000000,5.000000) circle (0.100000);
	\draw[line width=1pt,fill=black] (3.000000,5.000000) circle (0.100000);
	\draw[line width=1pt] (4.000000,5.000000) circle (0.100000);
	\draw[line width=1pt,fill=black] (5.000000,5.000000) circle (0.100000);
	\draw[line width=1pt] (6.000000,5.000000) circle (0.100000);
	\draw[line width=1pt,fill=black] (7.000000,5.000000) circle (0.100000);
	\draw[line width=1pt] (8.000000,5.000000) circle (0.100000);
	\draw[line width=1pt,fill=black] (9.000000,5.000000) circle (0.100000);
	\draw[line width=1pt] (10.000000,5.000000) circle (0.100000);
	\draw[line width=1pt,fill=black] (11.000000,5.000000) circle (0.100000);
	\draw[line width=1pt] (12.000000,5.000000) circle (0.100000);
	\draw[line width=1pt,fill=black] (13.000000,5.000000) circle (0.100000);
	\draw[line width=1pt] (14.000000,5.000000) circle (0.100000);
	\draw[line width=1pt,fill=black] (2.000000,4.000000) circle (0.100000);
	\draw[line width=1pt] (3.000000,4.000000) circle (0.100000);
	\draw[line width=1pt,fill=black] (4.000000,4.000000) circle (0.100000);
	\draw[line width=1pt] (5.000000,4.000000) circle (0.100000);
	\draw[line width=1pt,fill=black] (6.000000,4.000000) circle (0.100000);
	\draw[line width=1pt] (7.000000,4.000000) circle (0.100000);
	\draw[line width=1pt,fill=black] (8.000000,4.000000) circle (0.100000);
	\draw[line width=1pt] (9.000000,4.000000) circle (0.100000);
	\draw[line width=1pt,fill=black] (10.000000,4.000000) circle (0.100000);
	\draw[line width=1pt] (11.000000,4.000000) circle (0.100000);
	\draw[line width=1pt,fill=black] (12.000000,4.000000) circle (0.100000);
	\draw[line width=1pt] (13.000000,4.000000) circle (0.100000);
	\draw[line width=1pt,fill=black] (14.000000,4.000000) circle (0.100000);
	\draw[line width=1pt] (2.000000,3.000000) circle (0.100000);
	\draw[line width=1pt,fill=black] (3.000000,3.000000) circle (0.100000);
	\draw[line width=1pt] (4.000000,3.000000) circle (0.100000);
	\draw[line width=1pt,fill=black] (5.000000,3.000000) circle (0.100000);
	\draw[line width=1pt] (6.000000,3.000000) circle (0.100000);
	\draw[line width=1pt,fill=black] (7.000000,3.000000) circle (0.100000);
	\draw[line width=1pt] (8.000000,3.000000) circle (0.100000);
	\draw[line width=1pt,fill=black] (9.000000,3.000000) circle (0.100000);
	\draw[line width=1pt] (10.000000,3.000000) circle (0.100000);
	\draw[line width=1pt,fill=black] (11.000000,3.000000) circle (0.100000);
	\draw[line width=1pt] (12.000000,3.000000) circle (0.100000);
	\draw[line width=1pt,fill=black] (13.000000,3.000000) circle (0.100000);
	\draw[line width=1pt] (14.000000,3.000000) circle (0.100000);
	\draw[line width=1pt] (3.000000,2.000000) circle (0.100000);
	\draw[line width=1pt,fill=black] (4.000000,2.000000) circle (0.100000);
	\draw[line width=1pt] (5.000000,2.000000) circle (0.100000);
	\draw[line width=1pt,fill=black] (6.000000,2.000000) circle (0.100000);
	\draw[line width=1pt] (7.000000,2.000000) circle (0.100000);
	\draw[line width=1pt] (9.000000,2.000000) circle (0.100000);
	\draw[line width=1pt,fill=black] (10.000000,2.000000) circle (0.100000);
	\draw[line width=1pt] (11.000000,2.000000) circle (0.100000);
	\draw[line width=1pt,fill=black] (12.000000,2.000000) circle (0.100000);
	\draw[line width=1pt] (13.000000,2.000000) circle (0.100000);
\end{tikzpicture}

%% file: 003ex2.tex
\begin{tikzpicture}[x=0.030000\linewidth,y=0.030000\linewidth]
	\fill[gray!25] (4.000000,20.000000) -- (4.900000,20.000000) -- (4.000000,19.100000) -- cycle;
	\fill[gray!25] (4.000000,20.000000) -- (4.000000,19.100000) -- (3.100000,20.000000) -- cycle;
	\fill[gray!25] (6.000000,20.000000) -- (6.900000,20.000000) -- (6.000000,19.100000) -- cycle;
	\fill[gray!25] (6.000000,20.000000) -- (6.000000,19.100000) -- (5.100000,20.000000) -- cycle;
	\fill[gray!25] (10.000000,20.000000) -- (10.900000,20.000000) -- (10.000000,19.100000) -- cycle;
	\fill[gray!25] (10.000000,20.000000) -- (10.000000,19.100000) -- (9.100000,20.000000) -- cycle;
	\fill[gray!25] (12.000000,20.000000) -- (12.900000,20.000000) -- (12.000000,19.100000) -- cycle;
	\fill[gray!25] (12.000000,20.000000) -- (12.000000,19.100000) -- (11.100000,20.000000) -- cycle;
	\fill[gray!80] (3.000000,19.000000) -- (2.100000,19.000000) -- (3.000000,19.900000) -- cycle;
	\fill[gray!80] (3.000000,19.000000) -- (3.000000,19.900000) -- (3.900000,19.000000) -- cycle;
	\fill[gray!80] (3.000000,19.000000) -- (3.900000,19.000000) -- (3.000000,18.100000) -- cycle;
	\fill[gray!80] (3.000000,19.000000) -- (3.000000,18.100000) -- (2.100000,19.000000) -- cycle;
	\fill[gray!80] (5.000000,19.000000) -- (4.100000,19.000000) -- (5.000000,19.900000) -- cycle;
	\fill[gray!80] (5.000000,19.000000) -- (5.000000,19.900000) -- (5.900000,19.000000) -- cycle;
	\fill[gray!80] (5.000000,19.000000) -- (5.900000,19.000000) -- (5.000000,18.100000) -- cycle;
	\fill[gray!80] (5.000000,19.000000) -- (5.000000,18.100000) -- (4.100000,19.000000) -- cycle;
	\fill[gray!80] (7.000000,19.000000) -- (6.100000,19.000000) -- (7.000000,19.900000) -- cycle;
	\fill[gray!80] (7.000000,19.000000) -- (7.000000,19.900000) -- (7.900000,19.000000) -- cycle;
	\fill[gray!80] (7.000000,19.000000) -- (7.900000,19.000000) -- (7.000000,18.100000) -- cycle;
	\fill[gray!80] (7.000000,19.000000) -- (7.000000,18.100000) -- (6.100000,19.000000) -- cycle;
	\fill[gray!80] (9.000000,19.000000) -- (8.100000,19.000000) -- (9.000000,19.900000) -- cycle;
	\fill[gray!80] (9.000000,19.000000) -- (9.000000,19.900000) -- (9.900000,19.000000) -- cycle;
	\fill[gray!80] (9.000000,19.000000) -- (9.900000,19.000000) -- (9.000000,18.100000) -- cycle;
	\fill[gray!80] (9.000000,19.000000) -- (9.000000,18.100000) -- (8.100000,19.000000) -- cycle;
	\fill[gray!80] (11.000000,19.000000) -- (10.100000,19.000000) -- (11.000000,19.900000) -- cycle;
	\fill[gray!80] (11.000000,19.000000) -- (11.000000,19.900000) -- (11.900000,19.000000) -- cycle;
	\fill[gray!80] (11.000000,19.000000) -- (11.900000,19.000000) -- (11.000000,18.100000) -- cycle;
	\fill[gray!80] (11.000000,19.000000) -- (11.000000,18.100000) -- (10.100000,19.000000) -- cycle;
	\fill[gray!80] (13.000000,19.000000) -- (12.100000,19.000000) -- (13.000000,19.900000) -- cycle;
	\fill[gray!80] (13.000000,19.000000) -- (13.000000,19.900000) -- (13.900000,19.000000) -- cycle;
	\fill[gray!80] (13.000000,19.000000) -- (13.900000,19.000000) -- (13.000000,18.100000) -- cycle;
	\fill[gray!80] (13.000000,19.000000) -- (13.000000,18.100000) -- (12.100000,19.000000) -- cycle;
	\fill[gray!25] (2.000000,18.000000) -- (2.000000,18.900000) -- (2.900000,18.000000) -- cycle;
	\fill[gray!25] (2.000000,18.000000) -- (2.900000,18.000000) -- (2.000000,17.100000) -- cycle;
	\fill[gray!25] (4.000000,18.000000) -- (3.100000,18.000000) -- (4.000000,18.900000) -- cycle;
	\fill[gray!25] (4.000000,18.000000) -- (4.000000,18.900000) -- (4.900000,18.000000) -- cycle;
	\fill[gray!25] (4.000000,18.000000) -- (4.900000,18.000000) -- (4.000000,17.100000) -- cycle;
	\fill[gray!25] (4.000000,18.000000) -- (4.000000,17.100000) -- (3.100000,18.000000) -- cycle;
	\fill[gray!25] (6.000000,18.000000) -- (5.100000,18.000000) -- (6.000000,18.900000) -- cycle;
	\fill[gray!25] (6.000000,18.000000) -- (6.000000,18.900000) -- (6.900000,18.000000) -- cycle;
	\fill[gray!25] (6.000000,18.000000) -- (6.900000,18.000000) -- (6.000000,17.100000) -- cycle;
	\fill[gray!25] (6.000000,18.000000) -- (6.000000,17.100000) -- (5.100000,18.000000) -- cycle;
	\fill[gray!25] (8.000000,18.000000) -- (7.100000,18.000000) -- (8.000000,18.900000) -- cycle;
	\fill[gray!25] (8.000000,18.000000) -- (8.000000,18.900000) -- (8.900000,18.000000) -- cycle;
	\fill[gray!25] (8.000000,18.000000) -- (8.900000,18.000000) -- (8.000000,17.100000) -- cycle;
	\fill[gray!25] (8.000000,18.000000) -- (8.000000,17.100000) -- (7.100000,18.000000) -- cycle;
	\fill[gray!25] (10.000000,18.000000) -- (9.100000,18.000000) -- (10.000000,18.900000) -- cycle;
	\fill[gray!25] (10.000000,18.000000) -- (10.000000,18.900000) -- (10.900000,18.000000) -- cycle;
	\fill[gray!25] (10.000000,18.000000) -- (10.900000,18.000000) -- (10.000000,17.100000) -- cycle;
	\fill[gray!25] (10.000000,18.000000) -- (10.000000,17.100000) -- (9.100000,18.000000) -- cycle;
	\fill[gray!25] (12.000000,18.000000) -- (11.100000,18.000000) -- (12.000000,18.900000) -- cycle;
	\fill[gray!25] (12.000000,18.000000) -- (12.000000,18.900000) -- (12.900000,18.000000) -- cycle;
	\fill[gray!25] (12.000000,18.000000) -- (12.900000,18.000000) -- (12.000000,17.100000) -- cycle;
	\fill[gray!25] (12.000000,18.000000) -- (12.000000,17.100000) -- (11.100000,18.000000) -- cycle;
	\fill[gray!25] (14.000000,18.000000) -- (13.100000,18.000000) -- (14.000000,18.900000) -- cycle;
	\fill[gray!25] (14.000000,18.000000) -- (14.000000,17.100000) -- (13.100000,18.000000) -- cycle;
	\fill[gray!80] (3.000000,17.000000) -- (2.100000,17.000000) -- (3.000000,17.900000) -- cycle;
	\fill[gray!80] (3.000000,17.000000) -- (3.000000,17.900000) -- (3.900000,17.000000) -- cycle;
	\fill[gray!80] (3.000000,17.000000) -- (3.900000,17.000000) -- (3.000000,16.100000) -- cycle;
	\fill[gray!80] (3.000000,17.000000) -- (3.000000,16.100000) -- (2.100000,17.000000) -- cycle;
	\fill[gray!80] (5.000000,17.000000) -- (4.100000,17.000000) -- (5.000000,17.900000) -- cycle;
	\fill[gray!80] (5.000000,17.000000) -- (5.000000,17.900000) -- (5.900000,17.000000) -- cycle;
	\fill[gray!80] (5.000000,17.000000) -- (5.900000,17.000000) -- (5.000000,16.100000) -- cycle;
	\fill[gray!80] (5.000000,17.000000) -- (5.000000,16.100000) -- (4.100000,17.000000) -- cycle;
	\fill[gray!80] (7.000000,17.000000) -- (6.100000,17.000000) -- (7.000000,17.900000) -- cycle;
	\fill[gray!80] (7.000000,17.000000) -- (7.000000,17.900000) -- (7.900000,17.000000) -- cycle;
	\fill[gray!80] (7.000000,17.000000) -- (7.900000,17.000000) -- (7.000000,16.100000) -- cycle;
	\fill[gray!80] (7.000000,17.000000) -- (7.000000,16.100000) -- (6.100000,17.000000) -- cycle;
	\fill[gray!80] (9.000000,17.000000) -- (8.100000,17.000000) -- (9.000000,17.900000) -- cycle;
	\fill[gray!80] (9.000000,17.000000) -- (9.000000,17.900000) -- (9.900000,17.000000) -- cycle;
	\fill[gray!80] (9.000000,17.000000) -- (9.900000,17.000000) -- (9.000000,16.100000) -- cycle;
	\fill[gray!80] (9.000000,17.000000) -- (9.000000,16.100000) -- (8.100000,17.000000) -- cycle;
	\fill[gray!80] (11.000000,17.000000) -- (10.100000,17.000000) -- (11.000000,17.900000) -- cycle;
	\fill[gray!80] (11.000000,17.000000) -- (11.000000,17.900000) -- (11.900000,17.000000) -- cycle;
	\fill[gray!80] (11.000000,17.000000) -- (11.900000,17.000000) -- (11.000000,16.100000) -- cycle;
	\fill[gray!80] (11.000000,17.000000) -- (11.000000,16.100000) -- (10.100000,17.000000) -- cycle;
	\fill[gray!80] (13.000000,17.000000) -- (12.100000,17.000000) -- (13.000000,17.900000) -- cycle;
	\fill[gray!80] (13.000000,17.000000) -- (13.000000,17.900000) -- (13.900000,17.000000) -- cycle;
	\fill[gray!80] (13.000000,17.000000) -- (13.900000,17.000000) -- (13.000000,16.100000) -- cycle;
	\fill[gray!80] (13.000000,17.000000) -- (13.000000,16.100000) -- (12.100000,17.000000) -- cycle;
	\fill[gray!25] (2.000000,16.000000) -- (2.000000,16.900000) -- (2.900000,16.000000) -- cycle;
	\fill[gray!25] (2.000000,16.000000) -- (2.900000,16.000000) -- (2.000000,15.100000) -- cycle;
	\fill[gray!25] (4.000000,16.000000) -- (3.100000,16.000000) -- (4.000000,16.900000) -- cycle;
	\fill[gray!25] (4.000000,16.000000) -- (4.000000,16.900000) -- (4.900000,16.000000) -- cycle;
	\fill[gray!25] (6.000000,16.000000) -- (5.100000,16.000000) -- (6.000000,16.900000) -- cycle;
	\fill[gray!25] (6.000000,16.000000) -- (6.000000,16.900000) -- (6.900000,16.000000) -- cycle;
	\fill[gray!25] (8.000000,16.000000) -- (7.100000,16.000000) -- (8.000000,16.900000) -- cycle;
	\fill[gray!25] (8.000000,16.000000) -- (8.000000,16.900000) -- (8.900000,16.000000) -- cycle;
	\fill[gray!25] (10.000000,16.000000) -- (9.100000,16.000000) -- (10.000000,16.900000) -- cycle;
	\fill[gray!25] (10.000000,16.000000) -- (10.000000,16.900000) -- (10.900000,16.000000) -- cycle;
	\fill[gray!25] (12.000000,16.000000) -- (11.100000,16.000000) -- (12.000000,16.900000) -- cycle;
	\fill[gray!25] (12.000000,16.000000) -- (12.000000,16.900000) -- (12.900000,16.000000) -- cycle;
	\fill[gray!25] (12.000000,16.000000) -- (12.900000,16.000000) -- (12.000000,15.100000) -- cycle;
	\fill[gray!25] (12.000000,16.000000) -- (12.000000,15.100000) -- (11.100000,16.000000) -- cycle;
	\fill[gray!25] (14.000000,16.000000) -- (13.100000,16.000000) -- (14.000000,16.900000) -- cycle;
	\fill[gray!25] (14.000000,16.000000) -- (14.000000,15.100000) -- (13.100000,16.000000) -- cycle;
	\fill[gray!80] (13.000000,15.000000) -- (12.100000,15.000000) -- (13.000000,15.900000) -- cycle;
	\fill[gray!80] (13.000000,15.000000) -- (13.000000,15.900000) -- (13.900000,15.000000) -- cycle;
	\fill[gray!80] (13.000000,15.000000) -- (13.900000,15.000000) -- (13.000000,14.100000) -- cycle;
	\fill[gray!80] (13.000000,15.000000) -- (13.000000,14.100000) -- (12.100000,15.000000) -- cycle;
	\fill[gray!25] (12.000000,14.000000) -- (12.000000,14.900000) -- (12.900000,14.000000) -- cycle;
	\fill[gray!25] (12.000000,14.000000) -- (12.900000,14.000000) -- (12.000000,13.100000) -- cycle;
	\fill[gray!80] (7.000000,13.000000) -- (7.900000,13.000000) -- (7.000000,12.100000) -- cycle;
	\fill[gray!80] (7.000000,13.000000) -- (7.000000,12.100000) -- (6.100000,13.000000) -- cycle;
	\fill[gray!80] (9.000000,13.000000) -- (9.900000,13.000000) -- (9.000000,12.100000) -- cycle;
	\fill[gray!80] (9.000000,13.000000) -- (9.000000,12.100000) -- (8.100000,13.000000) -- cycle;
	\fill[gray!80] (13.000000,13.000000) -- (12.100000,13.000000) -- (13.000000,13.900000) -- cycle;
	\fill[gray!80] (13.000000,13.000000) -- (13.000000,13.900000) -- (13.900000,13.000000) -- cycle;
	\fill[gray!80] (13.000000,13.000000) -- (13.900000,13.000000) -- (13.000000,12.100000) -- cycle;
	\fill[gray!80] (13.000000,13.000000) -- (13.000000,12.100000) -- (12.100000,13.000000) -- cycle;
	\fill[gray!25] (6.000000,12.000000) -- (6.000000,12.900000) -- (6.900000,12.000000) -- cycle;
	\fill[gray!25] (6.000000,12.000000) -- (6.900000,12.000000) -- (6.000000,11.100000) -- cycle;
	\fill[gray!25] (8.000000,12.000000) -- (7.100000,12.000000) -- (8.000000,12.900000) -- cycle;
	\fill[gray!25] (8.000000,12.000000) -- (8.000000,12.900000) -- (8.900000,12.000000) -- cycle;
	\fill[gray!25] (8.000000,12.000000) -- (8.900000,12.000000) -- (8.000000,11.100000) -- cycle;
	\fill[gray!25] (8.000000,12.000000) -- (8.000000,11.100000) -- (7.100000,12.000000) -- cycle;
	\fill[gray!25] (10.000000,12.000000) -- (9.100000,12.000000) -- (10.000000,12.900000) -- cycle;
	\fill[gray!25] (10.000000,12.000000) -- (10.000000,11.100000) -- (9.100000,12.000000) -- cycle;
	\fill[gray!25] (12.000000,12.000000) -- (12.000000,12.900000) -- (12.900000,12.000000) -- cycle;
	\fill[gray!25] (12.000000,12.000000) -- (12.900000,12.000000) -- (12.000000,11.100000) -- cycle;
	\fill[gray!25] (14.000000,12.000000) -- (13.100000,12.000000) -- (14.000000,12.900000) -- cycle;
	\fill[gray!25] (14.000000,12.000000) -- (14.000000,11.100000) -- (13.100000,12.000000) -- cycle;
	\fill[gray!80] (7.000000,11.000000) -- (6.100000,11.000000) -- (7.000000,11.900000) -- cycle;
	\fill[gray!80] (7.000000,11.000000) -- (7.000000,11.900000) -- (7.900000,11.000000) -- cycle;
	\fill[gray!80] (7.000000,11.000000) -- (7.900000,11.000000) -- (7.000000,10.100000) -- cycle;
	\fill[gray!80] (7.000000,11.000000) -- (7.000000,10.100000) -- (6.100000,11.000000) -- cycle;
	\fill[gray!80] (9.000000,11.000000) -- (8.100000,11.000000) -- (9.000000,11.900000) -- cycle;
	\fill[gray!80] (9.000000,11.000000) -- (9.000000,11.900000) -- (9.900000,11.000000) -- cycle;
	\fill[gray!80] (9.000000,11.000000) -- (9.900000,11.000000) -- (9.000000,10.100000) -- cycle;
	\fill[gray!80] (9.000000,11.000000) -- (9.000000,10.100000) -- (8.100000,11.000000) -- cycle;
	\fill[gray!80] (13.000000,11.000000) -- (12.100000,11.000000) -- (13.000000,11.900000) -- cycle;
	\fill[gray!80] (13.000000,11.000000) -- (13.000000,11.900000) -- (13.900000,11.000000) -- cycle;
	\fill[gray!80] (13.000000,11.000000) -- (13.900000,11.000000) -- (13.000000,10.100000) -- cycle;
	\fill[gray!80] (13.000000,11.000000) -- (13.000000,10.100000) -- (12.100000,11.000000) -- cycle;
	\fill[gray!25] (6.000000,10.000000) -- (6.000000,10.900000) -- (6.900000,10.000000) -- cycle;
	\fill[gray!25] (6.000000,10.000000) -- (6.900000,10.000000) -- (6.000000,9.100000) -- cycle;
	\fill[gray!25] (8.000000,10.000000) -- (7.100000,10.000000) -- (8.000000,10.900000) -- cycle;
	\fill[gray!25] (8.000000,10.000000) -- (8.000000,10.900000) -- (8.900000,10.000000) -- cycle;
	\fill[gray!25] (8.000000,10.000000) -- (8.900000,10.000000) -- (8.000000,9.100000) -- cycle;
	\fill[gray!25] (8.000000,10.000000) -- (8.000000,9.100000) -- (7.100000,10.000000) -- cycle;
	\fill[gray!25] (10.000000,10.000000) -- (9.100000,10.000000) -- (10.000000,10.900000) -- cycle;
	\fill[gray!25] (10.000000,10.000000) -- (10.000000,9.100000) -- (9.100000,10.000000) -- cycle;
	\fill[gray!25] (12.000000,10.000000) -- (12.000000,10.900000) -- (12.900000,10.000000) -- cycle;
	\fill[gray!25] (12.000000,10.000000) -- (12.900000,10.000000) -- (12.000000,9.100000) -- cycle;
	\fill[gray!25] (14.000000,10.000000) -- (13.100000,10.000000) -- (14.000000,10.900000) -- cycle;
	\fill[gray!25] (14.000000,10.000000) -- (14.000000,9.100000) -- (13.100000,10.000000) -- cycle;
	\fill[gray!80] (7.000000,9.000000) -- (6.100000,9.000000) -- (7.000000,9.900000) -- cycle;
	\fill[gray!80] (7.000000,9.000000) -- (7.000000,9.900000) -- (7.900000,9.000000) -- cycle;
	\fill[gray!80] (9.000000,9.000000) -- (8.100000,9.000000) -- (9.000000,9.900000) -- cycle;
	\fill[gray!80] (9.000000,9.000000) -- (9.000000,9.900000) -- (9.900000,9.000000) -- cycle;
	\fill[gray!80] (13.000000,9.000000) -- (12.100000,9.000000) -- (13.000000,9.900000) -- cycle;
	\fill[gray!80] (13.000000,9.000000) -- (13.000000,9.900000) -- (13.900000,9.000000) -- cycle;
	\fill[gray!80] (13.000000,9.000000) -- (13.900000,9.000000) -- (13.000000,8.100000) -- cycle;
	\fill[gray!80] (13.000000,9.000000) -- (13.000000,8.100000) -- (12.100000,9.000000) -- cycle;
	\fill[gray!25] (12.000000,8.000000) -- (12.000000,8.900000) -- (12.900000,8.000000) -- cycle;
	\fill[gray!25] (12.000000,8.000000) -- (12.900000,8.000000) -- (12.000000,7.100000) -- cycle;
	\fill[gray!80] (13.000000,7.000000) -- (12.100000,7.000000) -- (13.000000,7.900000) -- cycle;
	\fill[gray!80] (13.000000,7.000000) -- (13.000000,7.900000) -- (13.900000,7.000000) -- cycle;
	\fill[gray!80] (13.000000,7.000000) -- (13.900000,7.000000) -- (13.000000,6.100000) -- cycle;
	\fill[gray!80] (13.000000,7.000000) -- (13.000000,6.100000) -- (12.100000,7.000000) -- cycle;
	\fill[gray!25] (2.000000,6.000000) -- (2.000000,6.900000) -- (2.900000,6.000000) -- cycle;
	\fill[gray!25] (2.000000,6.000000) -- (2.900000,6.000000) -- (2.000000,5.100000) -- cycle;
	\fill[gray!25] (4.000000,6.000000) -- (4.900000,6.000000) -- (4.000000,5.100000) -- cycle;
	\fill[gray!25] (4.000000,6.000000) -- (4.000000,5.100000) -- (3.100000,6.000000) -- cycle;
	\fill[gray!25] (6.000000,6.000000) -- (6.900000,6.000000) -- (6.000000,5.100000) -- cycle;
	\fill[gray!25] (6.000000,6.000000) -- (6.000000,5.100000) -- (5.100000,6.000000) -- cycle;
	\fill[gray!25] (8.000000,6.000000) -- (8.900000,6.000000) -- (8.000000,5.100000) -- cycle;
	\fill[gray!25] (8.000000,6.000000) -- (8.000000,5.100000) -- (7.100000,6.000000) -- cycle;
	\fill[gray!25] (10.000000,6.000000) -- (10.900000,6.000000) -- (10.000000,5.100000) -- cycle;
	\fill[gray!25] (10.000000,6.000000) -- (10.000000,5.100000) -- (9.100000,6.000000) -- cycle;
	\fill[gray!25] (12.000000,6.000000) -- (11.100000,6.000000) -- (12.000000,6.900000) -- cycle;
	\fill[gray!25] (12.000000,6.000000) -- (12.000000,6.900000) -- (12.900000,6.000000) -- cycle;
	\fill[gray!25] (12.000000,6.000000) -- (12.900000,6.000000) -- (12.000000,5.100000) -- cycle;
	\fill[gray!25] (12.000000,6.000000) -- (12.000000,5.100000) -- (11.100000,6.000000) -- cycle;
	\fill[gray!25] (14.000000,6.000000) -- (13.100000,6.000000) -- (14.000000,6.900000) -- cycle;
	\fill[gray!25] (14.000000,6.000000) -- (14.000000,5.100000) -- (13.100000,6.000000) -- cycle;
	\fill[gray!80] (3.000000,5.000000) -- (2.100000,5.000000) -- (3.000000,5.900000) -- cycle;
	\fill[gray!80] (3.000000,5.000000) -- (3.000000,5.900000) -- (3.900000,5.000000) -- cycle;
	\fill[gray!80] (3.000000,5.000000) -- (3.900000,5.000000) -- (3.000000,4.100000) -- cycle;
	\fill[gray!80] (3.000000,5.000000) -- (3.000000,4.100000) -- (2.100000,5.000000) -- cycle;
	\fill[gray!80] (5.000000,5.000000) -- (4.100000,5.000000) -- (5.000000,5.900000) -- cycle;
	\fill[gray!80] (5.000000,5.000000) -- (5.000000,5.900000) -- (5.900000,5.000000) -- cycle;
	\fill[gray!80] (5.000000,5.000000) -- (5.900000,5.000000) -- (5.000000,4.100000) -- cycle;
	\fill[gray!80] (5.000000,5.000000) -- (5.000000,4.100000) -- (4.100000,5.000000) -- cycle;
	\fill[gray!80] (7.000000,5.000000) -- (6.100000,5.000000) -- (7.000000,5.900000) -- cycle;
	\fill[gray!80] (7.000000,5.000000) -- (7.000000,5.900000) -- (7.900000,5.000000) -- cycle;
	\fill[gray!80] (7.000000,5.000000) -- (7.900000,5.000000) -- (7.000000,4.100000) -- cycle;
	\fill[gray!80] (7.000000,5.000000) -- (7.000000,4.100000) -- (6.100000,5.000000) -- cycle;
	\fill[gray!80] (9.000000,5.000000) -- (8.100000,5.000000) -- (9.000000,5.900000) -- cycle;
	\fill[gray!80] (9.000000,5.000000) -- (9.000000,5.900000) -- (9.900000,5.000000) -- cycle;
	\fill[gray!80] (9.000000,5.000000) -- (9.900000,5.000000) -- (9.000000,4.100000) -- cycle;
	\fill[gray!80] (9.000000,5.000000) -- (9.000000,4.100000) -- (8.100000,5.000000) -- cycle;
	\fill[gray!80] (11.000000,5.000000) -- (10.100000,5.000000) -- (11.000000,5.900000) -- cycle;
	\fill[gray!80] (11.000000,5.000000) -- (11.000000,5.900000) -- (11.900000,5.000000) -- cycle;
	\fill[gray!80] (11.000000,5.000000) -- (11.900000,5.000000) -- (11.000000,4.100000) -- cycle;
	\fill[gray!80] (11.000000,5.000000) -- (11.000000,4.100000) -- (10.100000,5.000000) -- cycle;
	\fill[gray!80] (13.000000,5.000000) -- (12.100000,5.000000) -- (13.000000,5.900000) -- cycle;
	\fill[gray!80] (13.000000,5.000000) -- (13.000000,5.900000) -- (13.900000,5.000000) -- cycle;
	\fill[gray!80] (13.000000,5.000000) -- (13.900000,5.000000) -- (13.000000,4.100000) -- cycle;
	\fill[gray!80] (13.000000,5.000000) -- (13.000000,4.100000) -- (12.100000,5.000000) -- cycle;
	\fill[gray!25] (2.000000,4.000000) -- (2.000000,4.900000) -- (2.900000,4.000000) -- cycle;
	\fill[gray!25] (2.000000,4.000000) -- (2.900000,4.000000) -- (2.000000,3.100000) -- cycle;
	\fill[gray!25] (4.000000,4.000000) -- (3.100000,4.000000) -- (4.000000,4.900000) -- cycle;
	\fill[gray!25] (4.000000,4.000000) -- (4.000000,4.900000) -- (4.900000,4.000000) -- cycle;
	\fill[gray!25] (4.000000,4.000000) -- (4.900000,4.000000) -- (4.000000,3.100000) -- cycle;
	\fill[gray!25] (4.000000,4.000000) -- (4.000000,3.100000) -- (3.100000,4.000000) -- cycle;
	\fill[gray!25] (6.000000,4.000000) -- (5.100000,4.000000) -- (6.000000,4.900000) -- cycle;
	\fill[gray!25] (6.000000,4.000000) -- (6.000000,4.900000) -- (6.900000,4.000000) -- cycle;
	\fill[gray!25] (6.000000,4.000000) -- (6.900000,4.000000) -- (6.000000,3.100000) -- cycle;
	\fill[gray!25] (6.000000,4.000000) -- (6.000000,3.100000) -- (5.100000,4.000000) -- cycle;
	\fill[gray!25] (8.000000,4.000000) -- (7.100000,4.000000) -- (8.000000,4.900000) -- cycle;
	\fill[gray!25] (8.000000,4.000000) -- (8.000000,4.900000) -- (8.900000,4.000000) -- cycle;
	\fill[gray!25] (8.000000,4.000000) -- (8.900000,4.000000) -- (8.000000,3.100000) -- cycle;
	\fill[gray!25] (8.000000,4.000000) -- (8.000000,3.100000) -- (7.100000,4.000000) -- cycle;
	\fill[gray!25] (10.000000,4.000000) -- (9.100000,4.000000) -- (10.000000,4.900000) -- cycle;
	\fill[gray!25] (10.000000,4.000000) -- (10.000000,4.900000) -- (10.900000,4.000000) -- cycle;
	\fill[gray!25] (10.000000,4.000000) -- (10.900000,4.000000) -- (10.000000,3.100000) -- cycle;
	\fill[gray!25] (10.000000,4.000000) -- (10.000000,3.100000) -- (9.100000,4.000000) -- cycle;
	\fill[gray!25] (12.000000,4.000000) -- (11.100000,4.000000) -- (12.000000,4.900000) -- cycle;
	\fill[gray!25] (12.000000,4.000000) -- (12.000000,4.900000) -- (12.900000,4.000000) -- cycle;
	\fill[gray!25] (12.000000,4.000000) -- (12.900000,4.000000) -- (12.000000,3.100000) -- cycle;
	\fill[gray!25] (12.000000,4.000000) -- (12.000000,3.100000) -- (11.100000,4.000000) -- cycle;
	\fill[gray!25] (14.000000,4.000000) -- (13.100000,4.000000) -- (14.000000,4.900000) -- cycle;
	\fill[gray!25] (14.000000,4.000000) -- (14.000000,3.100000) -- (13.100000,4.000000) -- cycle;
	\fill[gray!80] (3.000000,3.000000) -- (2.100000,3.000000) -- (3.000000,3.900000) -- cycle;
	\fill[gray!80] (3.000000,3.000000) -- (3.000000,3.900000) -- (3.900000,3.000000) -- cycle;
	\fill[gray!80] (3.000000,3.000000) -- (3.900000,3.000000) -- (3.000000,2.100000) -- cycle;
	\fill[gray!80] (3.000000,3.000000) -- (3.000000,2.100000) -- (2.100000,3.000000) -- cycle;
	\fill[gray!80] (5.000000,3.000000) -- (4.100000,3.000000) -- (5.000000,3.900000) -- cycle;
	\fill[gray!80] (5.000000,3.000000) -- (5.000000,3.900000) -- (5.900000,3.000000) -- cycle;
	\fill[gray!80] (5.000000,3.000000) -- (5.900000,3.000000) -- (5.000000,2.100000) -- cycle;
	\fill[gray!80] (5.000000,3.000000) -- (5.000000,2.100000) -- (4.100000,3.000000) -- cycle;
	\fill[gray!80] (7.000000,3.000000) -- (6.100000,3.000000) -- (7.000000,3.900000) -- cycle;
	\fill[gray!80] (7.000000,3.000000) -- (7.000000,3.900000) -- (7.900000,3.000000) -- cycle;
	\fill[gray!80] (7.000000,3.000000) -- (7.900000,3.000000) -- (7.000000,2.100000) -- cycle;
	\fill[gray!80] (7.000000,3.000000) -- (7.000000,2.100000) -- (6.100000,3.000000) -- cycle;
	\fill[gray!80] (9.000000,3.000000) -- (8.100000,3.000000) -- (9.000000,3.900000) -- cycle;
	\fill[gray!80] (9.000000,3.000000) -- (9.000000,3.900000) -- (9.900000,3.000000) -- cycle;
	\fill[gray!80] (9.000000,3.000000) -- (9.900000,3.000000) -- (9.000000,2.100000) -- cycle;
	\fill[gray!80] (9.000000,3.000000) -- (9.000000,2.100000) -- (8.100000,3.000000) -- cycle;
	\fill[gray!80] (11.000000,3.000000) -- (10.100000,3.000000) -- (11.000000,3.900000) -- cycle;
	\fill[gray!80] (11.000000,3.000000) -- (11.000000,3.900000) -- (11.900000,3.000000) -- cycle;
	\fill[gray!80] (11.000000,3.000000) -- (11.900000,3.000000) -- (11.000000,2.100000) -- cycle;
	\fill[gray!80] (11.000000,3.000000) -- (11.000000,2.100000) -- (10.100000,3.000000) -- cycle;
	\fill[gray!80] (13.000000,3.000000) -- (12.100000,3.000000) -- (13.000000,3.900000) -- cycle;
	\fill[gray!80] (13.000000,3.000000) -- (13.000000,3.900000) -- (13.900000,3.000000) -- cycle;
	\fill[gray!80] (13.000000,3.000000) -- (13.900000,3.000000) -- (13.000000,2.100000) -- cycle;
	\fill[gray!80] (13.000000,3.000000) -- (13.000000,2.100000) -- (12.100000,3.000000) -- cycle;
	\fill[gray!25] (4.000000,2.000000) -- (3.100000,2.000000) -- (4.000000,2.900000) -- cycle;
	\fill[gray!25] (4.000000,2.000000) -- (4.000000,2.900000) -- (4.900000,2.000000) -- cycle;
	\fill[gray!25] (6.000000,2.000000) -- (5.100000,2.000000) -- (6.000000,2.900000) -- cycle;
	\fill[gray!25] (6.000000,2.000000) -- (6.000000,2.900000) -- (6.900000,2.000000) -- cycle;
	\fill[gray!25] (10.000000,2.000000) -- (9.100000,2.000000) -- (10.000000,2.900000) -- cycle;
	\fill[gray!25] (10.000000,2.000000) -- (10.000000,2.900000) -- (10.900000,2.000000) -- cycle;
	\fill[gray!25] (12.000000,2.000000) -- (11.100000,2.000000) -- (12.000000,2.900000) -- cycle;
	\fill[gray!25] (12.000000,2.000000) -- (12.000000,2.900000) -- (12.900000,2.000000) -- cycle;
	\draw[line width=1pt] (3.000000,20.000000) circle (0.100000);
	\draw[line width=1pt,fill=black] (4.000000,20.000000) circle (0.100000);
	\draw[line width=1pt] (5.000000,20.000000) circle (0.100000);
	\draw[line width=1pt,fill=black] (6.000000,20.000000) circle (0.100000);
	\draw[line width=1pt] (7.000000,20.000000) circle (0.100000);
	\draw[line width=1pt] (9.000000,20.000000) circle (0.100000);
	\draw[line width=1pt,fill=black] (10.000000,20.000000) circle (0.100000);
	\draw[line width=1pt] (11.000000,20.000000) circle (0.100000);
	\draw[line width=1pt,fill=black] (12.000000,20.000000) circle (0.100000);
	\draw[line width=1pt] (13.000000,20.000000) circle (0.100000);
	\draw[line width=1pt] (2.000000,19.000000) circle (0.100000);
	\draw[line width=1pt,fill=black] (3.000000,19.000000) circle (0.100000);
	\draw[line width=1pt] (4.000000,19.000000) circle (0.100000);
	\draw[line width=1pt,fill=black] (5.000000,19.000000) circle (0.100000);
	\draw[line width=1pt] (6.000000,19.000000) circle (0.100000);
	\draw[line width=1pt,fill=black] (7.000000,19.000000) circle (0.100000);
	\draw[line width=1pt] (8.000000,19.000000) circle (0.100000);
	\draw[line width=1pt,fill=black] (9.000000,19.000000) circle (0.100000);
	\draw[line width=1pt] (10.000000,19.000000) circle (0.100000);
	\draw[line width=1pt,fill=black] (11.000000,19.000000) circle (0.100000);
	\draw[line width=1pt] (12.000000,19.000000) circle (0.100000);
	\draw[line width=1pt,fill=black] (13.000000,19.000000) circle (0.100000);
	\draw[line width=1pt] (14.000000,19.000000) circle (0.100000);
	\draw[line width=1pt,fill=black] (2.000000,18.000000) circle (0.100000);
	\draw[line width=1pt] (3.000000,18.000000) circle (0.100000);
	\draw[line width=1pt,fill=black] (4.000000,18.000000) circle (0.100000);
	\draw[line width=1pt] (5.000000,18.000000) circle (0.100000);
	\draw[line width=1pt,fill=black] (6.000000,18.000000) circle (0.100000);
	\draw[line width=1pt] (7.000000,18.000000) circle (0.100000);
	\draw[line width=1pt,fill=black] (8.000000,18.000000) circle (0.100000);
	\draw[line width=1pt] (9.000000,18.000000) circle (0.100000);
	\draw[line width=1pt,fill=black] (10.000000,18.000000) circle (0.100000);
	\draw[line width=1pt] (11.000000,18.000000) circle (0.100000);
	\draw[line width=1pt,fill=black] (12.000000,18.000000) circle (0.100000);
	\draw[line width=1pt] (13.000000,18.000000) circle (0.100000);
	\draw[line width=1pt,fill=black] (14.000000,18.000000) circle (0.100000);
	\draw[line width=1pt] (2.000000,17.000000) circle (0.100000);
	\draw[line width=1pt,fill=black] (3.000000,17.000000) circle (0.100000);
	\draw[line width=1pt] (4.000000,17.000000) circle (0.100000);
	\draw[line width=1pt,fill=black] (5.000000,17.000000) circle (0.100000);
	\draw[line width=1pt] (6.000000,17.000000) circle (0.100000);
	\draw[line width=1pt,fill=black] (7.000000,17.000000) circle (0.100000);
	\draw[line width=1pt] (8.000000,17.000000) circle (0.100000);
	\draw[line width=1pt,fill=black] (9.000000,17.000000) circle (0.100000);
	\draw[line width=1pt] (10.000000,17.000000) circle (0.100000);
	\draw[line width=1pt,fill=black] (11.000000,17.000000) circle (0.100000);
	\draw[line width=1pt] (12.000000,17.000000) circle (0.100000);
	\draw[line width=1pt,fill=black] (13.000000,17.000000) circle (0.100000);
	\draw[line width=1pt] (14.000000,17.000000) circle (0.100000);
	\draw[line width=1pt,fill=black] (2.000000,16.000000) circle (0.100000);
	\draw[line width=1pt] (3.000000,16.000000) circle (0.100000);
	\draw[line width=1pt,fill=black] (4.000000,16.000000) circle (0.100000);
	\draw[line width=1pt] (5.000000,16.000000) circle (0.100000);
	\draw[line width=1pt,fill=black] (6.000000,16.000000) circle (0.100000);
	\draw[line width=1pt] (7.000000,16.000000) circle (0.100000);
	\draw[line width=1pt,fill=black] (8.000000,16.000000) circle (0.100000);
	\draw[line width=1pt] (9.000000,16.000000) circle (0.100000);
	\draw[line width=1pt,fill=black] (10.000000,16.000000) circle (0.100000);
	\draw[line width=1pt] (11.000000,16.000000) circle (0.100000);
	\draw[line width=1pt,fill=black] (12.000000,16.000000) circle (0.100000);
	\draw[line width=1pt] (13.000000,16.000000) circle (0.100000);
	\draw[line width=1pt,fill=black] (14.000000,16.000000) circle (0.100000);
	\draw[line width=1pt] (2.000000,15.000000) circle (0.100000);
	\draw[line width=1pt] (12.000000,15.000000) circle (0.100000);
	\draw[line width=1pt,fill=black] (13.000000,15.000000) circle (0.100000);
	\draw[line width=1pt] (14.000000,15.000000) circle (0.100000);
	\draw[line width=1pt,fill=black] (12.000000,14.000000) circle (0.100000);
	\draw[line width=1pt] (13.000000,14.000000) circle (0.100000);
	\draw[line width=1pt] (2.000000,13.000000) circle (0.100000);
	\draw (6.000000,13.000000) node [above right] {$H$};
	\draw[line width=1pt] (6.000000,13.000000) circle (0.100000);
	\draw[line width=1pt,fill=black] (7.000000,13.000000) circle (0.100000);
	\draw (8.000000,13.000000) node [above right] {$H$};
	\draw[line width=1pt] (8.000000,13.000000) circle (0.100000);
	\draw[line width=1pt,fill=black] (9.000000,13.000000) circle (0.100000);
	\draw (10.000000,13.000000) node [above right] {$H$};
	\draw[line width=1pt] (10.000000,13.000000) circle (0.100000);
	\draw[line width=1pt] (12.000000,13.000000) circle (0.100000);
	\draw[line width=1pt,fill=black] (13.000000,13.000000) circle (0.100000);
	\draw[line width=1pt] (14.000000,13.000000) circle (0.100000);
	\draw[line width=1pt,fill=black] (2.000000,12.000000) circle (0.100000);
	\draw[line width=1pt,fill=black] (6.000000,12.000000) circle (0.100000);
	\draw (7.000000,12.000000) node [above right] {$H$};
	\draw[line width=1pt] (7.000000,12.000000) circle (0.100000);
	\draw[line width=1pt,fill=black] (8.000000,12.000000) circle (0.100000);
	\draw (9.000000,12.000000) node [above right] {$H$};
	\draw[line width=1pt] (9.000000,12.000000) circle (0.100000);
	\draw[line width=1pt,fill=black] (10.000000,12.000000) circle (0.100000);
	\draw[line width=1pt,fill=black] (12.000000,12.000000) circle (0.100000);
	\draw[line width=1pt] (13.000000,12.000000) circle (0.100000);
	\draw[line width=1pt,fill=black] (14.000000,12.000000) circle (0.100000);
	\draw[line width=1pt] (2.000000,11.000000) circle (0.100000);
	\draw (6.000000,11.000000) node [above right] {$H$};
	\draw[line width=1pt] (6.000000,11.000000) circle (0.100000);
	\draw[line width=1pt,fill=black] (7.000000,11.000000) circle (0.100000);
	\draw (8.000000,11.000000) node [above right] {$H$};
	\draw[line width=1pt] (8.000000,11.000000) circle (0.100000);
	\draw[line width=1pt,fill=black] (9.000000,11.000000) circle (0.100000);
	\draw (10.000000,11.000000) node [above right] {$H$};
	\draw[line width=1pt] (10.000000,11.000000) circle (0.100000);
	\draw[line width=1pt] (12.000000,11.000000) circle (0.100000);
	\draw[line width=1pt,fill=black] (13.000000,11.000000) circle (0.100000);
	\draw[line width=1pt] (14.000000,11.000000) circle (0.100000);
	\draw[line width=1pt,fill=black] (2.000000,10.000000) circle (0.100000);
	\draw[line width=1pt,fill=black] (6.000000,10.000000) circle (0.100000);
	\draw (7.000000,10.000000) node [above right] {$H$};
	\draw[line width=1pt] (7.000000,10.000000) circle (0.100000);
	\draw[line width=1pt,fill=black] (8.000000,10.000000) circle (0.100000);
	\draw (9.000000,10.000000) node [above right] {$H$};
	\draw[line width=1pt] (9.000000,10.000000) circle (0.100000);
	\draw[line width=1pt,fill=black] (10.000000,10.000000) circle (0.100000);
	\draw[line width=1pt,fill=black] (12.000000,10.000000) circle (0.100000);
	\draw[line width=1pt] (13.000000,10.000000) circle (0.100000);
	\draw[line width=1pt,fill=black] (14.000000,10.000000) circle (0.100000);
	\draw[line width=1pt] (2.000000,9.000000) circle (0.100000);
	\draw (6.000000,9.000000) node [above right] {$H$};
	\draw[line width=1pt] (6.000000,9.000000) circle (0.100000);
	\draw[line width=1pt,fill=black] (7.000000,9.000000) circle (0.100000);
	\draw (8.000000,9.000000) node [above right] {$H$};
	\draw[line width=1pt] (8.000000,9.000000) circle (0.100000);
	\draw[line width=1pt,fill=black] (9.000000,9.000000) circle (0.100000);
	\draw (10.000000,9.000000) node [above right] {$H$};
	\draw[line width=1pt] (10.000000,9.000000) circle (0.100000);
	\draw[line width=1pt] (12.000000,9.000000) circle (0.100000);
	\draw[line width=1pt,fill=black] (13.000000,9.000000) circle (0.100000);
	\draw[line width=1pt] (14.000000,9.000000) circle (0.100000);
	\draw[line width=1pt,fill=black] (12.000000,8.000000) circle (0.100000);
	\draw[line width=1pt] (13.000000,8.000000) circle (0.100000);
	\draw[line width=1pt] (2.000000,7.000000) circle (0.100000);
	\draw[line width=1pt] (12.000000,7.000000) circle (0.100000);
	\draw[line width=1pt,fill=black] (13.000000,7.000000) circle (0.100000);
	\draw[line width=1pt] (14.000000,7.000000) circle (0.100000);
	\draw[line width=1pt,fill=black] (2.000000,6.000000) circle (0.100000);
	\draw[line width=1pt] (3.000000,6.000000) circle (0.100000);
	\draw[line width=1pt,fill=black] (4.000000,6.000000) circle (0.100000);
	\draw[line width=1pt] (5.000000,6.000000) circle (0.100000);
	\draw[line width=1pt,fill=black] (6.000000,6.000000) circle (0.100000);
	\draw[line width=1pt] (7.000000,6.000000) circle (0.100000);
	\draw[line width=1pt,fill=black] (8.000000,6.000000) circle (0.100000);
	\draw[line width=1pt] (9.000000,6.000000) circle (0.100000);
	\draw[line width=1pt,fill=black] (10.000000,6.000000) circle (0.100000);
	\draw[line width=1pt] (11.000000,6.000000) circle (0.100000);
	\draw[line width=1pt,fill=black] (12.000000,6.000000) circle (0.100000);
	\draw[line width=1pt] (13.000000,6.000000) circle (0.100000);
	\draw[line width=1pt,fill=black] (14.000000,6.000000) circle (0.100000);
	\draw[line width=1pt] (2.000000,5.000000) circle (0.100000);
	\draw[line width=1pt,fill=black] (3.000000,5.000000) circle (0.100000);
	\draw[line width=1pt] (4.000000,5.000000) circle (0.100000);
	\draw[line width=1pt,fill=black] (5.000000,5.000000) circle (0.100000);
	\draw[line width=1pt] (6.000000,5.000000) circle (0.100000);
	\draw[line width=1pt,fill=black] (7.000000,5.000000) circle (0.100000);
	\draw[line width=1pt] (8.000000,5.000000) circle (0.100000);
	\draw[line width=1pt,fill=black] (9.000000,5.000000) circle (0.100000);
	\draw[line width=1pt] (10.000000,5.000000) circle (0.100000);
	\draw[line width=1pt,fill=black] (11.000000,5.000000) circle (0.100000);
	\draw[line width=1pt] (12.000000,5.000000) circle (0.100000);
	\draw[line width=1pt,fill=black] (13.000000,5.000000) circle (0.100000);
	\draw[line width=1pt] (14.000000,5.000000) circle (0.100000);
	\draw[line width=1pt,fill=black] (2.000000,4.000000) circle (0.100000);
	\draw[line width=1pt] (3.000000,4.000000) circle (0.100000);
	\draw[line width=1pt,fill=black] (4.000000,4.000000) circle (0.100000);
	\draw[line width=1pt] (5.000000,4.000000) circle (0.100000);
	\draw[line width=1pt,fill=black] (6.000000,4.000000) circle (0.100000);
	\draw[line width=1pt] (7.000000,4.000000) circle (0.100000);
	\draw[line width=1pt,fill=black] (8.000000,4.000000) circle (0.100000);
	\draw[line width=1pt] (9.000000,4.000000) circle (0.100000);
	\draw[line width=1pt,fill=black] (10.000000,4.000000) circle (0.100000);
	\draw[line width=1pt] (11.000000,4.000000) circle (0.100000);
	\draw[line width=1pt,fill=black] (12.000000,4.000000) circle (0.100000);
	\draw[line width=1pt] (13.000000,4.000000) circle (0.100000);
	\draw[line width=1pt,fill=black] (14.000000,4.000000) circle (0.100000);
	\draw[line width=1pt] (2.000000,3.000000) circle (0.100000);
	\draw[line width=1pt,fill=black] (3.000000,3.000000) circle (0.100000);
	\draw[line width=1pt] (4.000000,3.000000) circle (0.100000);
	\draw[line width=1pt,fill=black] (5.000000,3.000000) circle (0.100000);
	\draw[line width=1pt] (6.000000,3.000000) circle (0.100000);
	\draw[line width=1pt,fill=black] (7.000000,3.000000) circle (0.100000);
	\draw[line width=1pt] (8.000000,3.000000) circle (0.100000);
	\draw[line width=1pt,fill=black] (9.000000,3.000000) circle (0.100000);
	\draw[line width=1pt] (10.000000,3.000000) circle (0.100000);
	\draw[line width=1pt,fill=black] (11.000000,3.000000) circle (0.100000);
	\draw[line width=1pt] (12.000000,3.000000) circle (0.100000);
	\draw[line width=1pt,fill=black] (13.000000,3.000000) circle (0.100000);
	\draw[line width=1pt] (14.000000,3.000000) circle (0.100000);
	\draw[line width=1pt] (3.000000,2.000000) circle (0.100000);
	\draw[line width=1pt,fill=black] (4.000000,2.000000) circle (0.100000);
	\draw[line width=1pt] (5.000000,2.000000) circle (0.100000);
	\draw[line width=1pt,fill=black] (6.000000,2.000000) circle (0.100000);
	\draw[line width=1pt] (7.000000,2.000000) circle (0.100000);
	\draw[line width=1pt] (9.000000,2.000000) circle (0.100000);
	\draw[line width=1pt,fill=black] (10.000000,2.000000) circle (0.100000);
	\draw[line width=1pt] (11.000000,2.000000) circle (0.100000);
	\draw[line width=1pt,fill=black] (12.000000,2.000000) circle (0.100000);
	\draw[line width=1pt] (13.000000,2.000000) circle (0.100000);
\end{tikzpicture}

%% file: 004ex2.tex
\begin{tikzpicture}[x=0.030000\linewidth,y=0.030000\linewidth]
	\fill[gray!25] (4.000000,20.000000) -- (4.900000,20.000000) -- (4.000000,19.100000) -- cycle;
	\fill[gray!25] (4.000000,20.000000) -- (4.000000,19.100000) -- (3.100000,20.000000) -- cycle;
	\fill[gray!25] (6.000000,20.000000) -- (6.900000,20.000000) -- (6.000000,19.100000) -- cycle;
	\fill[gray!25] (6.000000,20.000000) -- (6.000000,19.100000) -- (5.100000,20.000000) -- cycle;
	\fill[gray!25] (10.000000,20.000000) -- (10.900000,20.000000) -- (10.000000,19.100000) -- cycle;
	\fill[gray!25] (10.000000,20.000000) -- (10.000000,19.100000) -- (9.100000,20.000000) -- cycle;
	\fill[gray!25] (12.000000,20.000000) -- (12.900000,20.000000) -- (12.000000,19.100000) -- cycle;
	\fill[gray!25] (12.000000,20.000000) -- (12.000000,19.100000) -- (11.100000,20.000000) -- cycle;
	\fill[gray!80] (3.000000,19.000000) -- (2.100000,19.000000) -- (3.000000,19.900000) -- cycle;
	\fill[gray!80] (3.000000,19.000000) -- (3.000000,19.900000) -- (3.900000,19.000000) -- cycle;
	\fill[gray!80] (3.000000,19.000000) -- (3.900000,19.000000) -- (3.000000,18.100000) -- cycle;
	\fill[gray!80] (3.000000,19.000000) -- (3.000000,18.100000) -- (2.100000,19.000000) -- cycle;
	\fill[gray!80] (5.000000,19.000000) -- (4.100000,19.000000) -- (5.000000,19.900000) -- cycle;
	\fill[gray!80] (5.000000,19.000000) -- (5.000000,19.900000) -- (5.900000,19.000000) -- cycle;
	\fill[gray!80] (5.000000,19.000000) -- (5.900000,19.000000) -- (5.000000,18.100000) -- cycle;
	\fill[gray!80] (5.000000,19.000000) -- (5.000000,18.100000) -- (4.100000,19.000000) -- cycle;
	\fill[gray!80] (7.000000,19.000000) -- (6.100000,19.000000) -- (7.000000,19.900000) -- cycle;
	\fill[gray!80] (7.000000,19.000000) -- (7.000000,19.900000) -- (7.900000,19.000000) -- cycle;
	\fill[gray!80] (7.000000,19.000000) -- (7.900000,19.000000) -- (7.000000,18.100000) -- cycle;
	\fill[gray!80] (7.000000,19.000000) -- (7.000000,18.100000) -- (6.100000,19.000000) -- cycle;
	\fill[gray!80] (9.000000,19.000000) -- (8.100000,19.000000) -- (9.000000,19.900000) -- cycle;
	\fill[gray!80] (9.000000,19.000000) -- (9.000000,19.900000) -- (9.900000,19.000000) -- cycle;
	\fill[gray!80] (9.000000,19.000000) -- (9.900000,19.000000) -- (9.000000,18.100000) -- cycle;
	\fill[gray!80] (9.000000,19.000000) -- (9.000000,18.100000) -- (8.100000,19.000000) -- cycle;
	\fill[gray!80] (11.000000,19.000000) -- (10.100000,19.000000) -- (11.000000,19.900000) -- cycle;
	\fill[gray!80] (11.000000,19.000000) -- (11.000000,19.900000) -- (11.900000,19.000000) -- cycle;
	\fill[gray!80] (11.000000,19.000000) -- (11.900000,19.000000) -- (11.000000,18.100000) -- cycle;
	\fill[gray!80] (11.000000,19.000000) -- (11.000000,18.100000) -- (10.100000,19.000000) -- cycle;
	\fill[gray!80] (13.000000,19.000000) -- (12.100000,19.000000) -- (13.000000,19.900000) -- cycle;
	\fill[gray!80] (13.000000,19.000000) -- (13.000000,19.900000) -- (13.900000,19.000000) -- cycle;
	\fill[gray!80] (13.000000,19.000000) -- (13.900000,19.000000) -- (13.000000,18.100000) -- cycle;
	\fill[gray!80] (13.000000,19.000000) -- (13.000000,18.100000) -- (12.100000,19.000000) -- cycle;
	\fill[gray!25] (2.000000,18.000000) -- (2.000000,18.900000) -- (2.900000,18.000000) -- cycle;
	\fill[gray!25] (2.000000,18.000000) -- (2.900000,18.000000) -- (2.000000,17.100000) -- cycle;
	\fill[gray!25] (4.000000,18.000000) -- (3.100000,18.000000) -- (4.000000,18.900000) -- cycle;
	\fill[gray!25] (4.000000,18.000000) -- (4.000000,18.900000) -- (4.900000,18.000000) -- cycle;
	\fill[gray!25] (4.000000,18.000000) -- (4.900000,18.000000) -- (4.000000,17.100000) -- cycle;
	\fill[gray!25] (4.000000,18.000000) -- (4.000000,17.100000) -- (3.100000,18.000000) -- cycle;
	\fill[gray!25] (6.000000,18.000000) -- (5.100000,18.000000) -- (6.000000,18.900000) -- cycle;
	\fill[gray!25] (6.000000,18.000000) -- (6.000000,18.900000) -- (6.900000,18.000000) -- cycle;
	\fill[gray!25] (6.000000,18.000000) -- (6.900000,18.000000) -- (6.000000,17.100000) -- cycle;
	\fill[gray!25] (6.000000,18.000000) -- (6.000000,17.100000) -- (5.100000,18.000000) -- cycle;
	\fill[gray!25] (8.000000,18.000000) -- (7.100000,18.000000) -- (8.000000,18.900000) -- cycle;
	\fill[gray!25] (8.000000,18.000000) -- (8.000000,18.900000) -- (8.900000,18.000000) -- cycle;
	\fill[gray!25] (8.000000,18.000000) -- (8.900000,18.000000) -- (8.000000,17.100000) -- cycle;
	\fill[gray!25] (8.000000,18.000000) -- (8.000000,17.100000) -- (7.100000,18.000000) -- cycle;
	\fill[gray!25] (10.000000,18.000000) -- (9.100000,18.000000) -- (10.000000,18.900000) -- cycle;
	\fill[gray!25] (10.000000,18.000000) -- (10.000000,18.900000) -- (10.900000,18.000000) -- cycle;
	\fill[gray!25] (10.000000,18.000000) -- (10.900000,18.000000) -- (10.000000,17.100000) -- cycle;
	\fill[gray!25] (10.000000,18.000000) -- (10.000000,17.100000) -- (9.100000,18.000000) -- cycle;
	\fill[gray!25] (12.000000,18.000000) -- (11.100000,18.000000) -- (12.000000,18.900000) -- cycle;
	\fill[gray!25] (12.000000,18.000000) -- (12.000000,18.900000) -- (12.900000,18.000000) -- cycle;
	\fill[gray!25] (12.000000,18.000000) -- (12.900000,18.000000) -- (12.000000,17.100000) -- cycle;
	\fill[gray!25] (12.000000,18.000000) -- (12.000000,17.100000) -- (11.100000,18.000000) -- cycle;
	\fill[gray!25] (14.000000,18.000000) -- (13.100000,18.000000) -- (14.000000,18.900000) -- cycle;
	\fill[gray!25] (14.000000,18.000000) -- (14.000000,17.100000) -- (13.100000,18.000000) -- cycle;
	\fill[gray!80] (3.000000,17.000000) -- (2.100000,17.000000) -- (3.000000,17.900000) -- cycle;
	\fill[gray!80] (3.000000,17.000000) -- (3.000000,17.900000) -- (3.900000,17.000000) -- cycle;
	\fill[gray!80] (3.000000,17.000000) -- (3.900000,17.000000) -- (3.000000,16.100000) -- cycle;
	\fill[gray!80] (3.000000,17.000000) -- (3.000000,16.100000) -- (2.100000,17.000000) -- cycle;
	\fill[gray!80] (5.000000,17.000000) -- (4.100000,17.000000) -- (5.000000,17.900000) -- cycle;
	\fill[gray!80] (5.000000,17.000000) -- (5.000000,17.900000) -- (5.900000,17.000000) -- cycle;
	\fill[gray!80] (5.000000,17.000000) -- (5.900000,17.000000) -- (5.000000,16.100000) -- cycle;
	\fill[gray!80] (5.000000,17.000000) -- (5.000000,16.100000) -- (4.100000,17.000000) -- cycle;
	\fill[gray!80] (7.000000,17.000000) -- (6.100000,17.000000) -- (7.000000,17.900000) -- cycle;
	\fill[gray!80] (7.000000,17.000000) -- (7.000000,17.900000) -- (7.900000,17.000000) -- cycle;
	\fill[gray!80] (7.000000,17.000000) -- (7.900000,17.000000) -- (7.000000,16.100000) -- cycle;
	\fill[gray!80] (7.000000,17.000000) -- (7.000000,16.100000) -- (6.100000,17.000000) -- cycle;
	\fill[gray!80] (9.000000,17.000000) -- (8.100000,17.000000) -- (9.000000,17.900000) -- cycle;
	\fill[gray!80] (9.000000,17.000000) -- (9.000000,17.900000) -- (9.900000,17.000000) -- cycle;
	\fill[gray!80] (9.000000,17.000000) -- (9.900000,17.000000) -- (9.000000,16.100000) -- cycle;
	\fill[gray!80] (9.000000,17.000000) -- (9.000000,16.100000) -- (8.100000,17.000000) -- cycle;
	\fill[gray!80] (11.000000,17.000000) -- (10.100000,17.000000) -- (11.000000,17.900000) -- cycle;
	\fill[gray!80] (11.000000,17.000000) -- (11.000000,17.900000) -- (11.900000,17.000000) -- cycle;
	\fill[gray!80] (11.000000,17.000000) -- (11.900000,17.000000) -- (11.000000,16.100000) -- cycle;
	\fill[gray!80] (11.000000,17.000000) -- (11.000000,16.100000) -- (10.100000,17.000000) -- cycle;
	\fill[gray!80] (13.000000,17.000000) -- (12.100000,17.000000) -- (13.000000,17.900000) -- cycle;
	\fill[gray!80] (13.000000,17.000000) -- (13.000000,17.900000) -- (13.900000,17.000000) -- cycle;
	\fill[gray!80] (13.000000,17.000000) -- (13.900000,17.000000) -- (13.000000,16.100000) -- cycle;
	\fill[gray!80] (13.000000,17.000000) -- (13.000000,16.100000) -- (12.100000,17.000000) -- cycle;
	\fill[gray!25] (2.000000,16.000000) -- (2.000000,16.900000) -- (2.900000,16.000000) -- cycle;
	\fill[gray!25] (2.000000,16.000000) -- (2.900000,16.000000) -- (2.000000,15.100000) -- cycle;
	\fill[gray!25] (4.000000,16.000000) -- (3.100000,16.000000) -- (4.000000,16.900000) -- cycle;
	\fill[gray!25] (4.000000,16.000000) -- (4.000000,16.900000) -- (4.900000,16.000000) -- cycle;
	\fill[gray!25] (6.000000,16.000000) -- (5.100000,16.000000) -- (6.000000,16.900000) -- cycle;
	\fill[gray!25] (6.000000,16.000000) -- (6.000000,16.900000) -- (6.900000,16.000000) -- cycle;
	\fill[gray!25] (8.000000,16.000000) -- (7.100000,16.000000) -- (8.000000,16.900000) -- cycle;
	\fill[gray!25] (8.000000,16.000000) -- (8.000000,16.900000) -- (8.900000,16.000000) -- cycle;
	\fill[gray!25] (10.000000,16.000000) -- (9.100000,16.000000) -- (10.000000,16.900000) -- cycle;
	\fill[gray!25] (10.000000,16.000000) -- (10.000000,16.900000) -- (10.900000,16.000000) -- cycle;
	\fill[gray!25] (12.000000,16.000000) -- (11.100000,16.000000) -- (12.000000,16.900000) -- cycle;
	\fill[gray!25] (12.000000,16.000000) -- (12.000000,16.900000) -- (12.900000,16.000000) -- cycle;
	\fill[gray!25] (12.000000,16.000000) -- (12.900000,16.000000) -- (12.000000,15.100000) -- cycle;
	\fill[gray!25] (12.000000,16.000000) -- (12.000000,15.100000) -- (11.100000,16.000000) -- cycle;
	\fill[gray!25] (14.000000,16.000000) -- (13.100000,16.000000) -- (14.000000,16.900000) -- cycle;
	\fill[gray!25] (14.000000,16.000000) -- (14.000000,15.100000) -- (13.100000,16.000000) -- cycle;
	\fill[gray!80] (13.000000,15.000000) -- (12.100000,15.000000) -- (13.000000,15.900000) -- cycle;
	\fill[gray!80] (13.000000,15.000000) -- (13.000000,15.900000) -- (13.900000,15.000000) -- cycle;
	\fill[gray!80] (13.000000,15.000000) -- (13.900000,15.000000) -- (13.000000,14.100000) -- cycle;
	\fill[gray!80] (13.000000,15.000000) -- (13.000000,14.100000) -- (12.100000,15.000000) -- cycle;
	\fill[gray!25] (12.000000,14.000000) -- (12.000000,14.900000) -- (12.900000,14.000000) -- cycle;
	\fill[gray!25] (12.000000,14.000000) -- (12.900000,14.000000) -- (12.000000,13.100000) -- cycle;
	\fill[gray!80] (13.000000,13.000000) -- (12.100000,13.000000) -- (13.000000,13.900000) -- cycle;
	\fill[gray!80] (13.000000,13.000000) -- (13.000000,13.900000) -- (13.900000,13.000000) -- cycle;
	\fill[gray!80] (13.000000,13.000000) -- (13.900000,13.000000) -- (13.000000,12.100000) -- cycle;
	\fill[gray!80] (13.000000,13.000000) -- (13.000000,12.100000) -- (12.100000,13.000000) -- cycle;
	\fill[gray!25] (12.000000,12.000000) -- (12.000000,12.900000) -- (12.900000,12.000000) -- cycle;
	\fill[gray!25] (12.000000,12.000000) -- (12.900000,12.000000) -- (12.000000,11.100000) -- cycle;
	\fill[gray!25] (14.000000,12.000000) -- (13.100000,12.000000) -- (14.000000,12.900000) -- cycle;
	\fill[gray!25] (14.000000,12.000000) -- (14.000000,11.100000) -- (13.100000,12.000000) -- cycle;
	\fill[gray!80] (13.000000,11.000000) -- (12.100000,11.000000) -- (13.000000,11.900000) -- cycle;
	\fill[gray!80] (13.000000,11.000000) -- (13.000000,11.900000) -- (13.900000,11.000000) -- cycle;
	\fill[gray!80] (13.000000,11.000000) -- (13.900000,11.000000) -- (13.000000,10.100000) -- cycle;
	\fill[gray!80] (13.000000,11.000000) -- (13.000000,10.100000) -- (12.100000,11.000000) -- cycle;
	\fill[gray!25] (12.000000,10.000000) -- (12.000000,10.900000) -- (12.900000,10.000000) -- cycle;
	\fill[gray!25] (12.000000,10.000000) -- (12.900000,10.000000) -- (12.000000,9.100000) -- cycle;
	\fill[gray!25] (14.000000,10.000000) -- (13.100000,10.000000) -- (14.000000,10.900000) -- cycle;
	\fill[gray!25] (14.000000,10.000000) -- (14.000000,9.100000) -- (13.100000,10.000000) -- cycle;
	\fill[gray!80] (13.000000,9.000000) -- (12.100000,9.000000) -- (13.000000,9.900000) -- cycle;
	\fill[gray!80] (13.000000,9.000000) -- (13.000000,9.900000) -- (13.900000,9.000000) -- cycle;
	\fill[gray!80] (13.000000,9.000000) -- (13.900000,9.000000) -- (13.000000,8.100000) -- cycle;
	\fill[gray!80] (13.000000,9.000000) -- (13.000000,8.100000) -- (12.100000,9.000000) -- cycle;
	\fill[gray!25] (12.000000,8.000000) -- (12.000000,8.900000) -- (12.900000,8.000000) -- cycle;
	\fill[gray!25] (12.000000,8.000000) -- (12.900000,8.000000) -- (12.000000,7.100000) -- cycle;
	\fill[gray!80] (13.000000,7.000000) -- (12.100000,7.000000) -- (13.000000,7.900000) -- cycle;
	\fill[gray!80] (13.000000,7.000000) -- (13.000000,7.900000) -- (13.900000,7.000000) -- cycle;
	\fill[gray!80] (13.000000,7.000000) -- (13.900000,7.000000) -- (13.000000,6.100000) -- cycle;
	\fill[gray!80] (13.000000,7.000000) -- (13.000000,6.100000) -- (12.100000,7.000000) -- cycle;
	\fill[gray!25] (2.000000,6.000000) -- (2.000000,6.900000) -- (2.900000,6.000000) -- cycle;
	\fill[gray!25] (2.000000,6.000000) -- (2.900000,6.000000) -- (2.000000,5.100000) -- cycle;
	\fill[gray!25] (4.000000,6.000000) -- (4.900000,6.000000) -- (4.000000,5.100000) -- cycle;
	\fill[gray!25] (4.000000,6.000000) -- (4.000000,5.100000) -- (3.100000,6.000000) -- cycle;
	\fill[gray!25] (6.000000,6.000000) -- (6.900000,6.000000) -- (6.000000,5.100000) -- cycle;
	\fill[gray!25] (6.000000,6.000000) -- (6.000000,5.100000) -- (5.100000,6.000000) -- cycle;
	\fill[gray!25] (8.000000,6.000000) -- (8.900000,6.000000) -- (8.000000,5.100000) -- cycle;
	\fill[gray!25] (8.000000,6.000000) -- (8.000000,5.100000) -- (7.100000,6.000000) -- cycle;
	\fill[gray!25] (10.000000,6.000000) -- (10.900000,6.000000) -- (10.000000,5.100000) -- cycle;
	\fill[gray!25] (10.000000,6.000000) -- (10.000000,5.100000) -- (9.100000,6.000000) -- cycle;
	\fill[gray!25] (12.000000,6.000000) -- (11.100000,6.000000) -- (12.000000,6.900000) -- cycle;
	\fill[gray!25] (12.000000,6.000000) -- (12.000000,6.900000) -- (12.900000,6.000000) -- cycle;
	\fill[gray!25] (12.000000,6.000000) -- (12.900000,6.000000) -- (12.000000,5.100000) -- cycle;
	\fill[gray!25] (12.000000,6.000000) -- (12.000000,5.100000) -- (11.100000,6.000000) -- cycle;
	\fill[gray!25] (14.000000,6.000000) -- (13.100000,6.000000) -- (14.000000,6.900000) -- cycle;
	\fill[gray!25] (14.000000,6.000000) -- (14.000000,5.100000) -- (13.100000,6.000000) -- cycle;
	\fill[gray!80] (3.000000,5.000000) -- (2.100000,5.000000) -- (3.000000,5.900000) -- cycle;
	\fill[gray!80] (3.000000,5.000000) -- (3.000000,5.900000) -- (3.900000,5.000000) -- cycle;
	\fill[gray!80] (3.000000,5.000000) -- (3.900000,5.000000) -- (3.000000,4.100000) -- cycle;
	\fill[gray!80] (3.000000,5.000000) -- (3.000000,4.100000) -- (2.100000,5.000000) -- cycle;
	\fill[gray!80] (5.000000,5.000000) -- (4.100000,5.000000) -- (5.000000,5.900000) -- cycle;
	\fill[gray!80] (5.000000,5.000000) -- (5.000000,5.900000) -- (5.900000,5.000000) -- cycle;
	\fill[gray!80] (5.000000,5.000000) -- (5.900000,5.000000) -- (5.000000,4.100000) -- cycle;
	\fill[gray!80] (5.000000,5.000000) -- (5.000000,4.100000) -- (4.100000,5.000000) -- cycle;
	\fill[gray!80] (7.000000,5.000000) -- (6.100000,5.000000) -- (7.000000,5.900000) -- cycle;
	\fill[gray!80] (7.000000,5.000000) -- (7.000000,5.900000) -- (7.900000,5.000000) -- cycle;
	\fill[gray!80] (7.000000,5.000000) -- (7.900000,5.000000) -- (7.000000,4.100000) -- cycle;
	\fill[gray!80] (7.000000,5.000000) -- (7.000000,4.100000) -- (6.100000,5.000000) -- cycle;
	\fill[gray!80] (9.000000,5.000000) -- (8.100000,5.000000) -- (9.000000,5.900000) -- cycle;
	\fill[gray!80] (9.000000,5.000000) -- (9.000000,5.900000) -- (9.900000,5.000000) -- cycle;
	\fill[gray!80] (9.000000,5.000000) -- (9.900000,5.000000) -- (9.000000,4.100000) -- cycle;
	\fill[gray!80] (9.000000,5.000000) -- (9.000000,4.100000) -- (8.100000,5.000000) -- cycle;
	\fill[gray!80] (11.000000,5.000000) -- (10.100000,5.000000) -- (11.000000,5.900000) -- cycle;
	\fill[gray!80] (11.000000,5.000000) -- (11.000000,5.900000) -- (11.900000,5.000000) -- cycle;
	\fill[gray!80] (11.000000,5.000000) -- (11.900000,5.000000) -- (11.000000,4.100000) -- cycle;
	\fill[gray!80] (11.000000,5.000000) -- (11.000000,4.100000) -- (10.100000,5.000000) -- cycle;
	\fill[gray!80] (13.000000,5.000000) -- (12.100000,5.000000) -- (13.000000,5.900000) -- cycle;
	\fill[gray!80] (13.000000,5.000000) -- (13.000000,5.900000) -- (13.900000,5.000000) -- cycle;
	\fill[gray!80] (13.000000,5.000000) -- (13.900000,5.000000) -- (13.000000,4.100000) -- cycle;
	\fill[gray!80] (13.000000,5.000000) -- (13.000000,4.100000) -- (12.100000,5.000000) -- cycle;
	\fill[gray!25] (2.000000,4.000000) -- (2.000000,4.900000) -- (2.900000,4.000000) -- cycle;
	\fill[gray!25] (2.000000,4.000000) -- (2.900000,4.000000) -- (2.000000,3.100000) -- cycle;
	\fill[gray!25] (4.000000,4.000000) -- (3.100000,4.000000) -- (4.000000,4.900000) -- cycle;
	\fill[gray!25] (4.000000,4.000000) -- (4.000000,4.900000) -- (4.900000,4.000000) -- cycle;
	\fill[gray!25] (4.000000,4.000000) -- (4.900000,4.000000) -- (4.000000,3.100000) -- cycle;
	\fill[gray!25] (4.000000,4.000000) -- (4.000000,3.100000) -- (3.100000,4.000000) -- cycle;
	\fill[gray!25] (6.000000,4.000000) -- (5.100000,4.000000) -- (6.000000,4.900000) -- cycle;
	\fill[gray!25] (6.000000,4.000000) -- (6.000000,4.900000) -- (6.900000,4.000000) -- cycle;
	\fill[gray!25] (6.000000,4.000000) -- (6.900000,4.000000) -- (6.000000,3.100000) -- cycle;
	\fill[gray!25] (6.000000,4.000000) -- (6.000000,3.100000) -- (5.100000,4.000000) -- cycle;
	\fill[gray!25] (8.000000,4.000000) -- (7.100000,4.000000) -- (8.000000,4.900000) -- cycle;
	\fill[gray!25] (8.000000,4.000000) -- (8.000000,4.900000) -- (8.900000,4.000000) -- cycle;
	\fill[gray!25] (8.000000,4.000000) -- (8.900000,4.000000) -- (8.000000,3.100000) -- cycle;
	\fill[gray!25] (8.000000,4.000000) -- (8.000000,3.100000) -- (7.100000,4.000000) -- cycle;
	\fill[gray!25] (10.000000,4.000000) -- (9.100000,4.000000) -- (10.000000,4.900000) -- cycle;
	\fill[gray!25] (10.000000,4.000000) -- (10.000000,4.900000) -- (10.900000,4.000000) -- cycle;
	\fill[gray!25] (10.000000,4.000000) -- (10.900000,4.000000) -- (10.000000,3.100000) -- cycle;
	\fill[gray!25] (10.000000,4.000000) -- (10.000000,3.100000) -- (9.100000,4.000000) -- cycle;
	\fill[gray!25] (12.000000,4.000000) -- (11.100000,4.000000) -- (12.000000,4.900000) -- cycle;
	\fill[gray!25] (12.000000,4.000000) -- (12.000000,4.900000) -- (12.900000,4.000000) -- cycle;
	\fill[gray!25] (12.000000,4.000000) -- (12.900000,4.000000) -- (12.000000,3.100000) -- cycle;
	\fill[gray!25] (12.000000,4.000000) -- (12.000000,3.100000) -- (11.100000,4.000000) -- cycle;
	\fill[gray!25] (14.000000,4.000000) -- (13.100000,4.000000) -- (14.000000,4.900000) -- cycle;
	\fill[gray!25] (14.000000,4.000000) -- (14.000000,3.100000) -- (13.100000,4.000000) -- cycle;
	\fill[gray!80] (3.000000,3.000000) -- (2.100000,3.000000) -- (3.000000,3.900000) -- cycle;
	\fill[gray!80] (3.000000,3.000000) -- (3.000000,3.900000) -- (3.900000,3.000000) -- cycle;
	\fill[gray!80] (3.000000,3.000000) -- (3.900000,3.000000) -- (3.000000,2.100000) -- cycle;
	\fill[gray!80] (3.000000,3.000000) -- (3.000000,2.100000) -- (2.100000,3.000000) -- cycle;
	\fill[gray!80] (5.000000,3.000000) -- (4.100000,3.000000) -- (5.000000,3.900000) -- cycle;
	\fill[gray!80] (5.000000,3.000000) -- (5.000000,3.900000) -- (5.900000,3.000000) -- cycle;
	\fill[gray!80] (5.000000,3.000000) -- (5.900000,3.000000) -- (5.000000,2.100000) -- cycle;
	\fill[gray!80] (5.000000,3.000000) -- (5.000000,2.100000) -- (4.100000,3.000000) -- cycle;
	\fill[gray!80] (7.000000,3.000000) -- (6.100000,3.000000) -- (7.000000,3.900000) -- cycle;
	\fill[gray!80] (7.000000,3.000000) -- (7.000000,3.900000) -- (7.900000,3.000000) -- cycle;
	\fill[gray!80] (7.000000,3.000000) -- (7.900000,3.000000) -- (7.000000,2.100000) -- cycle;
	\fill[gray!80] (7.000000,3.000000) -- (7.000000,2.100000) -- (6.100000,3.000000) -- cycle;
	\fill[gray!80] (9.000000,3.000000) -- (8.100000,3.000000) -- (9.000000,3.900000) -- cycle;
	\fill[gray!80] (9.000000,3.000000) -- (9.000000,3.900000) -- (9.900000,3.000000) -- cycle;
	\fill[gray!80] (9.000000,3.000000) -- (9.900000,3.000000) -- (9.000000,2.100000) -- cycle;
	\fill[gray!80] (9.000000,3.000000) -- (9.000000,2.100000) -- (8.100000,3.000000) -- cycle;
	\fill[gray!80] (11.000000,3.000000) -- (10.100000,3.000000) -- (11.000000,3.900000) -- cycle;
	\fill[gray!80] (11.000000,3.000000) -- (11.000000,3.900000) -- (11.900000,3.000000) -- cycle;
	\fill[gray!80] (11.000000,3.000000) -- (11.900000,3.000000) -- (11.000000,2.100000) -- cycle;
	\fill[gray!80] (11.000000,3.000000) -- (11.000000,2.100000) -- (10.100000,3.000000) -- cycle;
	\fill[gray!80] (13.000000,3.000000) -- (12.100000,3.000000) -- (13.000000,3.900000) -- cycle;
	\fill[gray!80] (13.000000,3.000000) -- (13.000000,3.900000) -- (13.900000,3.000000) -- cycle;
	\fill[gray!80] (13.000000,3.000000) -- (13.900000,3.000000) -- (13.000000,2.100000) -- cycle;
	\fill[gray!80] (13.000000,3.000000) -- (13.000000,2.100000) -- (12.100000,3.000000) -- cycle;
	\fill[gray!25] (4.000000,2.000000) -- (3.100000,2.000000) -- (4.000000,2.900000) -- cycle;
	\fill[gray!25] (4.000000,2.000000) -- (4.000000,2.900000) -- (4.900000,2.000000) -- cycle;
	\fill[gray!25] (6.000000,2.000000) -- (5.100000,2.000000) -- (6.000000,2.900000) -- cycle;
	\fill[gray!25] (6.000000,2.000000) -- (6.000000,2.900000) -- (6.900000,2.000000) -- cycle;
	\fill[gray!25] (10.000000,2.000000) -- (9.100000,2.000000) -- (10.000000,2.900000) -- cycle;
	\fill[gray!25] (10.000000,2.000000) -- (10.000000,2.900000) -- (10.900000,2.000000) -- cycle;
	\fill[gray!25] (12.000000,2.000000) -- (11.100000,2.000000) -- (12.000000,2.900000) -- cycle;
	\fill[gray!25] (12.000000,2.000000) -- (12.000000,2.900000) -- (12.900000,2.000000) -- cycle;
	\draw[line width=1pt] (3.000000,20.000000) circle (0.100000);
	\draw[line width=1pt,fill=black] (4.000000,20.000000) circle (0.100000);
	\draw[line width=1pt] (5.000000,20.000000) circle (0.100000);
	\draw[line width=1pt,fill=black] (6.000000,20.000000) circle (0.100000);
	\draw[line width=1pt] (7.000000,20.000000) circle (0.100000);
	\draw[line width=1pt] (9.000000,20.000000) circle (0.100000);
	\draw[line width=1pt,fill=black] (10.000000,20.000000) circle (0.100000);
	\draw[line width=1pt] (11.000000,20.000000) circle (0.100000);
	\draw[line width=1pt,fill=black] (12.000000,20.000000) circle (0.100000);
	\draw[line width=1pt] (13.000000,20.000000) circle (0.100000);
	\draw[line width=1pt] (2.000000,19.000000) circle (0.100000);
	\draw[line width=1pt,fill=black] (3.000000,19.000000) circle (0.100000);
	\draw[line width=1pt] (4.000000,19.000000) circle (0.100000);
	\draw[line width=1pt,fill=black] (5.000000,19.000000) circle (0.100000);
	\draw[line width=1pt] (6.000000,19.000000) circle (0.100000);
	\draw[line width=1pt,fill=black] (7.000000,19.000000) circle (0.100000);
	\draw[line width=1pt] (8.000000,19.000000) circle (0.100000);
	\draw[line width=1pt,fill=black] (9.000000,19.000000) circle (0.100000);
	\draw[line width=1pt] (10.000000,19.000000) circle (0.100000);
	\draw[line width=1pt,fill=black] (11.000000,19.000000) circle (0.100000);
	\draw[line width=1pt] (12.000000,19.000000) circle (0.100000);
	\draw[line width=1pt,fill=black] (13.000000,19.000000) circle (0.100000);
	\draw[line width=1pt] (14.000000,19.000000) circle (0.100000);
	\draw[line width=1pt,fill=black] (2.000000,18.000000) circle (0.100000);
	\draw[line width=1pt] (3.000000,18.000000) circle (0.100000);
	\draw[line width=1pt,fill=black] (4.000000,18.000000) circle (0.100000);
	\draw[line width=1pt] (5.000000,18.000000) circle (0.100000);
	\draw[line width=1pt,fill=black] (6.000000,18.000000) circle (0.100000);
	\draw[line width=1pt] (7.000000,18.000000) circle (0.100000);
	\draw[line width=1pt,fill=black] (8.000000,18.000000) circle (0.100000);
	\draw[line width=1pt] (9.000000,18.000000) circle (0.100000);
	\draw[line width=1pt,fill=black] (10.000000,18.000000) circle (0.100000);
	\draw[line width=1pt] (11.000000,18.000000) circle (0.100000);
	\draw[line width=1pt,fill=black] (12.000000,18.000000) circle (0.100000);
	\draw[line width=1pt] (13.000000,18.000000) circle (0.100000);
	\draw[line width=1pt,fill=black] (14.000000,18.000000) circle (0.100000);
	\draw[line width=1pt] (2.000000,17.000000) circle (0.100000);
	\draw[line width=1pt,fill=black] (3.000000,17.000000) circle (0.100000);
	\draw[line width=1pt] (4.000000,17.000000) circle (0.100000);
	\draw[line width=1pt,fill=black] (5.000000,17.000000) circle (0.100000);
	\draw[line width=1pt] (6.000000,17.000000) circle (0.100000);
	\draw[line width=1pt,fill=black] (7.000000,17.000000) circle (0.100000);
	\draw[line width=1pt] (8.000000,17.000000) circle (0.100000);
	\draw[line width=1pt,fill=black] (9.000000,17.000000) circle (0.100000);
	\draw[line width=1pt] (10.000000,17.000000) circle (0.100000);
	\draw[line width=1pt,fill=black] (11.000000,17.000000) circle (0.100000);
	\draw[line width=1pt] (12.000000,17.000000) circle (0.100000);
	\draw[line width=1pt,fill=black] (13.000000,17.000000) circle (0.100000);
	\draw[line width=1pt] (14.000000,17.000000) circle (0.100000);
	\draw[line width=1pt,fill=black] (2.000000,16.000000) circle (0.100000);
	\draw[line width=1pt] (3.000000,16.000000) circle (0.100000);
	\draw[line width=1pt,fill=black] (4.000000,16.000000) circle (0.100000);
	\draw[line width=1pt] (5.000000,16.000000) circle (0.100000);
	\draw[line width=1pt,fill=black] (6.000000,16.000000) circle (0.100000);
	\draw[line width=1pt] (7.000000,16.000000) circle (0.100000);
	\draw[line width=1pt,fill=black] (8.000000,16.000000) circle (0.100000);
	\draw[line width=1pt] (9.000000,16.000000) circle (0.100000);
	\draw[line width=1pt,fill=black] (10.000000,16.000000) circle (0.100000);
	\draw[line width=1pt] (11.000000,16.000000) circle (0.100000);
	\draw[line width=1pt,fill=black] (12.000000,16.000000) circle (0.100000);
	\draw[line width=1pt] (13.000000,16.000000) circle (0.100000);
	\draw[line width=1pt,fill=black] (14.000000,16.000000) circle (0.100000);
	\draw[line width=1pt] (2.000000,15.000000) circle (0.100000);
	\draw[line width=1pt] (12.000000,15.000000) circle (0.100000);
	\draw[line width=1pt,fill=black] (13.000000,15.000000) circle (0.100000);
	\draw[line width=1pt] (14.000000,15.000000) circle (0.100000);
	\draw[line width=1pt] (5.000000,14.000000) circle (0.100000);
	\draw[line width=1pt,fill=black] (6.000000,14.000000) circle (0.100000);
	\draw[line width=1pt] (7.000000,14.000000) circle (0.100000);
	\draw[line width=1pt,fill=black] (8.000000,14.000000) circle (0.100000);
	\draw[line width=1pt] (9.000000,14.000000) circle (0.100000);
	\draw[line width=1pt,fill=black] (12.000000,14.000000) circle (0.100000);
	\draw[line width=1pt] (13.000000,14.000000) circle (0.100000);
	\draw[line width=1pt] (2.000000,13.000000) circle (0.100000);
	\draw[line width=1pt,fill=black] (5.000000,13.000000) circle (0.100000);
	\draw[line width=1pt,->] (5.858579,13.141421) -- (5.141421,13.858579);
	\draw[line width=1pt] (6.000000,13.000000) circle (0.100000);
	\draw[line width=1pt,fill=black] (7.000000,13.000000) circle (0.100000);
	\draw[line width=1pt,->] (7.858579,13.141421) -- (7.141421,13.858579);
	\draw[line width=1pt] (8.000000,13.000000) circle (0.100000);
	\draw[line width=1pt,fill=black] (9.000000,13.000000) circle (0.100000);
	\draw[line width=1pt,->] (9.858579,13.141421) -- (9.141421,13.858579);
	\draw[line width=1pt] (10.000000,13.000000) circle (0.100000);
	\draw[line width=1pt] (12.000000,13.000000) circle (0.100000);
	\draw[line width=1pt,fill=black] (13.000000,13.000000) circle (0.100000);
	\draw[line width=1pt] (14.000000,13.000000) circle (0.100000);
	\draw[line width=1pt,fill=black] (2.000000,12.000000) circle (0.100000);
	\draw[line width=1pt] (5.000000,12.000000) circle (0.100000);
	\draw[line width=1pt,fill=black] (6.000000,12.000000) circle (0.100000);
	\draw[line width=1pt,->] (6.858579,12.141421) -- (6.141421,12.858579);
	\draw[line width=1pt] (7.000000,12.000000) circle (0.100000);
	\draw[line width=1pt,fill=black] (8.000000,12.000000) circle (0.100000);
	\draw[line width=1pt,->] (8.858579,12.141421) -- (8.141421,12.858579);
	\draw[line width=1pt] (9.000000,12.000000) circle (0.100000);
	\draw[line width=1pt,fill=black] (12.000000,12.000000) circle (0.100000);
	\draw[line width=1pt] (13.000000,12.000000) circle (0.100000);
	\draw[line width=1pt,fill=black] (14.000000,12.000000) circle (0.100000);
	\draw[line width=1pt] (2.000000,11.000000) circle (0.100000);
	\draw[line width=1pt,fill=black] (5.000000,11.000000) circle (0.100000);
	\draw[line width=1pt,->] (5.858579,11.141421) -- (5.141421,11.858579);
	\draw[line width=1pt] (6.000000,11.000000) circle (0.100000);
	\draw[line width=1pt,fill=black] (7.000000,11.000000) circle (0.100000);
	\draw[line width=1pt,->] (7.858579,11.141421) -- (7.141421,11.858579);
	\draw[line width=1pt] (8.000000,11.000000) circle (0.100000);
	\draw[line width=1pt,fill=black] (9.000000,11.000000) circle (0.100000);
	\draw[line width=1pt,->] (9.858579,11.141421) -- (9.141421,11.858579);
	\draw[line width=1pt] (10.000000,11.000000) circle (0.100000);
	\draw[line width=1pt] (12.000000,11.000000) circle (0.100000);
	\draw[line width=1pt,fill=black] (13.000000,11.000000) circle (0.100000);
	\draw[line width=1pt] (14.000000,11.000000) circle (0.100000);
	\draw[line width=1pt,fill=black] (2.000000,10.000000) circle (0.100000);
	\draw[line width=1pt] (5.000000,10.000000) circle (0.100000);
	\draw[line width=1pt,fill=black] (6.000000,10.000000) circle (0.100000);
	\draw[line width=1pt,->] (6.858579,10.141421) -- (6.141421,10.858579);
	\draw[line width=1pt] (7.000000,10.000000) circle (0.100000);
	\draw[line width=1pt,fill=black] (8.000000,10.000000) circle (0.100000);
	\draw[line width=1pt,->] (8.858579,10.141421) -- (8.141421,10.858579);
	\draw[line width=1pt] (9.000000,10.000000) circle (0.100000);
	\draw[line width=1pt,fill=black] (12.000000,10.000000) circle (0.100000);
	\draw[line width=1pt] (13.000000,10.000000) circle (0.100000);
	\draw[line width=1pt,fill=black] (14.000000,10.000000) circle (0.100000);
	\draw[line width=1pt] (2.000000,9.000000) circle (0.100000);
	\draw[line width=1pt,->] (5.858579,9.141421) -- (5.141421,9.858579);
	\draw[line width=1pt] (6.000000,9.000000) circle (0.100000);
	\draw[line width=1pt,->] (7.858579,9.141421) -- (7.141421,9.858579);
	\draw[line width=1pt] (8.000000,9.000000) circle (0.100000);
	\draw[line width=1pt,->] (9.858579,9.141421) -- (9.141421,9.858579);
	\draw[line width=1pt] (10.000000,9.000000) circle (0.100000);
	\draw[line width=1pt] (12.000000,9.000000) circle (0.100000);
	\draw[line width=1pt,fill=black] (13.000000,9.000000) circle (0.100000);
	\draw[line width=1pt] (14.000000,9.000000) circle (0.100000);
	\draw[line width=1pt,fill=black] (12.000000,8.000000) circle (0.100000);
	\draw[line width=1pt] (13.000000,8.000000) circle (0.100000);
	\draw[line width=1pt] (2.000000,7.000000) circle (0.100000);
	\draw[line width=1pt] (12.000000,7.000000) circle (0.100000);
	\draw[line width=1pt,fill=black] (13.000000,7.000000) circle (0.100000);
	\draw[line width=1pt] (14.000000,7.000000) circle (0.100000);
	\draw[line width=1pt,fill=black] (2.000000,6.000000) circle (0.100000);
	\draw[line width=1pt] (3.000000,6.000000) circle (0.100000);
	\draw[line width=1pt,fill=black] (4.000000,6.000000) circle (0.100000);
	\draw[line width=1pt] (5.000000,6.000000) circle (0.100000);
	\draw[line width=1pt,fill=black] (6.000000,6.000000) circle (0.100000);
	\draw[line width=1pt] (7.000000,6.000000) circle (0.100000);
	\draw[line width=1pt,fill=black] (8.000000,6.000000) circle (0.100000);
	\draw[line width=1pt] (9.000000,6.000000) circle (0.100000);
	\draw[line width=1pt,fill=black] (10.000000,6.000000) circle (0.100000);
	\draw[line width=1pt] (11.000000,6.000000) circle (0.100000);
	\draw[line width=1pt,fill=black] (12.000000,6.000000) circle (0.100000);
	\draw[line width=1pt] (13.000000,6.000000) circle (0.100000);
	\draw[line width=1pt,fill=black] (14.000000,6.000000) circle (0.100000);
	\draw[line width=1pt] (2.000000,5.000000) circle (0.100000);
	\draw[line width=1pt,fill=black] (3.000000,5.000000) circle (0.100000);
	\draw[line width=1pt] (4.000000,5.000000) circle (0.100000);
	\draw[line width=1pt,fill=black] (5.000000,5.000000) circle (0.100000);
	\draw[line width=1pt] (6.000000,5.000000) circle (0.100000);
	\draw[line width=1pt,fill=black] (7.000000,5.000000) circle (0.100000);
	\draw[line width=1pt] (8.000000,5.000000) circle (0.100000);
	\draw[line width=1pt,fill=black] (9.000000,5.000000) circle (0.100000);
	\draw[line width=1pt] (10.000000,5.000000) circle (0.100000);
	\draw[line width=1pt,fill=black] (11.000000,5.000000) circle (0.100000);
	\draw[line width=1pt] (12.000000,5.000000) circle (0.100000);
	\draw[line width=1pt,fill=black] (13.000000,5.000000) circle (0.100000);
	\draw[line width=1pt] (14.000000,5.000000) circle (0.100000);
	\draw[line width=1pt,fill=black] (2.000000,4.000000) circle (0.100000);
	\draw[line width=1pt] (3.000000,4.000000) circle (0.100000);
	\draw[line width=1pt,fill=black] (4.000000,4.000000) circle (0.100000);
	\draw[line width=1pt] (5.000000,4.000000) circle (0.100000);
	\draw[line width=1pt,fill=black] (6.000000,4.000000) circle (0.100000);
	\draw[line width=1pt] (7.000000,4.000000) circle (0.100000);
	\draw[line width=1pt,fill=black] (8.000000,4.000000) circle (0.100000);
	\draw[line width=1pt] (9.000000,4.000000) circle (0.100000);
	\draw[line width=1pt,fill=black] (10.000000,4.000000) circle (0.100000);
	\draw[line width=1pt] (11.000000,4.000000) circle (0.100000);
	\draw[line width=1pt,fill=black] (12.000000,4.000000) circle (0.100000);
	\draw[line width=1pt] (13.000000,4.000000) circle (0.100000);
	\draw[line width=1pt,fill=black] (14.000000,4.000000) circle (0.100000);
	\draw[line width=1pt] (2.000000,3.000000) circle (0.100000);
	\draw[line width=1pt,fill=black] (3.000000,3.000000) circle (0.100000);
	\draw[line width=1pt] (4.000000,3.000000) circle (0.100000);
	\draw[line width=1pt,fill=black] (5.000000,3.000000) circle (0.100000);
	\draw[line width=1pt] (6.000000,3.000000) circle (0.100000);
	\draw[line width=1pt,fill=black] (7.000000,3.000000) circle (0.100000);
	\draw[line width=1pt] (8.000000,3.000000) circle (0.100000);
	\draw[line width=1pt,fill=black] (9.000000,3.000000) circle (0.100000);
	\draw[line width=1pt] (10.000000,3.000000) circle (0.100000);
	\draw[line width=1pt,fill=black] (11.000000,3.000000) circle (0.100000);
	\draw[line width=1pt] (12.000000,3.000000) circle (0.100000);
	\draw[line width=1pt,fill=black] (13.000000,3.000000) circle (0.100000);
	\draw[line width=1pt] (14.000000,3.000000) circle (0.100000);
	\draw[line width=1pt] (3.000000,2.000000) circle (0.100000);
	\draw[line width=1pt,fill=black] (4.000000,2.000000) circle (0.100000);
	\draw[line width=1pt] (5.000000,2.000000) circle (0.100000);
	\draw[line width=1pt,fill=black] (6.000000,2.000000) circle (0.100000);
	\draw[line width=1pt] (7.000000,2.000000) circle (0.100000);
	\draw[line width=1pt] (9.000000,2.000000) circle (0.100000);
	\draw[line width=1pt,fill=black] (10.000000,2.000000) circle (0.100000);
	\draw[line width=1pt] (11.000000,2.000000) circle (0.100000);
	\draw[line width=1pt,fill=black] (12.000000,2.000000) circle (0.100000);
	\draw[line width=1pt] (13.000000,2.000000) circle (0.100000);
\end{tikzpicture}

%% file: 005ex2.tex
\begin{tikzpicture}[x=0.030000\linewidth,y=0.030000\linewidth]
	\fill[gray!25] (4.000000,20.000000) -- (4.900000,20.000000) -- (4.000000,19.100000) -- cycle;
	\fill[gray!25] (4.000000,20.000000) -- (4.000000,19.100000) -- (3.100000,20.000000) -- cycle;
	\fill[gray!25] (6.000000,20.000000) -- (6.900000,20.000000) -- (6.000000,19.100000) -- cycle;
	\fill[gray!25] (6.000000,20.000000) -- (6.000000,19.100000) -- (5.100000,20.000000) -- cycle;
	\fill[gray!25] (10.000000,20.000000) -- (10.900000,20.000000) -- (10.000000,19.100000) -- cycle;
	\fill[gray!25] (10.000000,20.000000) -- (10.000000,19.100000) -- (9.100000,20.000000) -- cycle;
	\fill[gray!25] (12.000000,20.000000) -- (12.900000,20.000000) -- (12.000000,19.100000) -- cycle;
	\fill[gray!25] (12.000000,20.000000) -- (12.000000,19.100000) -- (11.100000,20.000000) -- cycle;
	\fill[gray!80] (3.000000,19.000000) -- (2.100000,19.000000) -- (3.000000,19.900000) -- cycle;
	\fill[gray!80] (3.000000,19.000000) -- (3.000000,19.900000) -- (3.900000,19.000000) -- cycle;
	\fill[gray!80] (3.000000,19.000000) -- (3.900000,19.000000) -- (3.000000,18.100000) -- cycle;
	\fill[gray!80] (3.000000,19.000000) -- (3.000000,18.100000) -- (2.100000,19.000000) -- cycle;
	\fill[gray!80] (5.000000,19.000000) -- (4.100000,19.000000) -- (5.000000,19.900000) -- cycle;
	\fill[gray!80] (5.000000,19.000000) -- (5.000000,19.900000) -- (5.900000,19.000000) -- cycle;
	\fill[gray!80] (5.000000,19.000000) -- (5.900000,19.000000) -- (5.000000,18.100000) -- cycle;
	\fill[gray!80] (5.000000,19.000000) -- (5.000000,18.100000) -- (4.100000,19.000000) -- cycle;
	\fill[gray!80] (7.000000,19.000000) -- (6.100000,19.000000) -- (7.000000,19.900000) -- cycle;
	\fill[gray!80] (7.000000,19.000000) -- (7.000000,19.900000) -- (7.900000,19.000000) -- cycle;
	\fill[gray!80] (7.000000,19.000000) -- (7.900000,19.000000) -- (7.000000,18.100000) -- cycle;
	\fill[gray!80] (7.000000,19.000000) -- (7.000000,18.100000) -- (6.100000,19.000000) -- cycle;
	\fill[gray!80] (9.000000,19.000000) -- (8.100000,19.000000) -- (9.000000,19.900000) -- cycle;
	\fill[gray!80] (9.000000,19.000000) -- (9.000000,19.900000) -- (9.900000,19.000000) -- cycle;
	\fill[gray!80] (9.000000,19.000000) -- (9.900000,19.000000) -- (9.000000,18.100000) -- cycle;
	\fill[gray!80] (9.000000,19.000000) -- (9.000000,18.100000) -- (8.100000,19.000000) -- cycle;
	\fill[gray!80] (11.000000,19.000000) -- (10.100000,19.000000) -- (11.000000,19.900000) -- cycle;
	\fill[gray!80] (11.000000,19.000000) -- (11.000000,19.900000) -- (11.900000,19.000000) -- cycle;
	\fill[gray!80] (11.000000,19.000000) -- (11.900000,19.000000) -- (11.000000,18.100000) -- cycle;
	\fill[gray!80] (11.000000,19.000000) -- (11.000000,18.100000) -- (10.100000,19.000000) -- cycle;
	\fill[gray!80] (13.000000,19.000000) -- (12.100000,19.000000) -- (13.000000,19.900000) -- cycle;
	\fill[gray!80] (13.000000,19.000000) -- (13.000000,19.900000) -- (13.900000,19.000000) -- cycle;
	\fill[gray!80] (13.000000,19.000000) -- (13.900000,19.000000) -- (13.000000,18.100000) -- cycle;
	\fill[gray!80] (13.000000,19.000000) -- (13.000000,18.100000) -- (12.100000,19.000000) -- cycle;
	\fill[gray!25] (2.000000,18.000000) -- (2.000000,18.900000) -- (2.900000,18.000000) -- cycle;
	\fill[gray!25] (2.000000,18.000000) -- (2.900000,18.000000) -- (2.000000,17.100000) -- cycle;
	\fill[gray!25] (4.000000,18.000000) -- (3.100000,18.000000) -- (4.000000,18.900000) -- cycle;
	\fill[gray!25] (4.000000,18.000000) -- (4.000000,18.900000) -- (4.900000,18.000000) -- cycle;
	\fill[gray!25] (4.000000,18.000000) -- (4.900000,18.000000) -- (4.000000,17.100000) -- cycle;
	\fill[gray!25] (4.000000,18.000000) -- (4.000000,17.100000) -- (3.100000,18.000000) -- cycle;
	\fill[gray!25] (6.000000,18.000000) -- (5.100000,18.000000) -- (6.000000,18.900000) -- cycle;
	\fill[gray!25] (6.000000,18.000000) -- (6.000000,18.900000) -- (6.900000,18.000000) -- cycle;
	\fill[gray!25] (6.000000,18.000000) -- (6.900000,18.000000) -- (6.000000,17.100000) -- cycle;
	\fill[gray!25] (6.000000,18.000000) -- (6.000000,17.100000) -- (5.100000,18.000000) -- cycle;
	\fill[gray!25] (8.000000,18.000000) -- (7.100000,18.000000) -- (8.000000,18.900000) -- cycle;
	\fill[gray!25] (8.000000,18.000000) -- (8.000000,18.900000) -- (8.900000,18.000000) -- cycle;
	\fill[gray!25] (8.000000,18.000000) -- (8.900000,18.000000) -- (8.000000,17.100000) -- cycle;
	\fill[gray!25] (8.000000,18.000000) -- (8.000000,17.100000) -- (7.100000,18.000000) -- cycle;
	\fill[gray!25] (10.000000,18.000000) -- (9.100000,18.000000) -- (10.000000,18.900000) -- cycle;
	\fill[gray!25] (10.000000,18.000000) -- (10.000000,18.900000) -- (10.900000,18.000000) -- cycle;
	\fill[gray!25] (10.000000,18.000000) -- (10.900000,18.000000) -- (10.000000,17.100000) -- cycle;
	\fill[gray!25] (10.000000,18.000000) -- (10.000000,17.100000) -- (9.100000,18.000000) -- cycle;
	\fill[gray!25] (12.000000,18.000000) -- (11.100000,18.000000) -- (12.000000,18.900000) -- cycle;
	\fill[gray!25] (12.000000,18.000000) -- (12.000000,18.900000) -- (12.900000,18.000000) -- cycle;
	\fill[gray!25] (12.000000,18.000000) -- (12.900000,18.000000) -- (12.000000,17.100000) -- cycle;
	\fill[gray!25] (12.000000,18.000000) -- (12.000000,17.100000) -- (11.100000,18.000000) -- cycle;
	\fill[gray!25] (14.000000,18.000000) -- (13.100000,18.000000) -- (14.000000,18.900000) -- cycle;
	\fill[gray!25] (14.000000,18.000000) -- (14.000000,17.100000) -- (13.100000,18.000000) -- cycle;
	\fill[gray!80] (3.000000,17.000000) -- (2.100000,17.000000) -- (3.000000,17.900000) -- cycle;
	\fill[gray!80] (3.000000,17.000000) -- (3.000000,17.900000) -- (3.900000,17.000000) -- cycle;
	\fill[gray!80] (3.000000,17.000000) -- (3.900000,17.000000) -- (3.000000,16.100000) -- cycle;
	\fill[gray!80] (3.000000,17.000000) -- (3.000000,16.100000) -- (2.100000,17.000000) -- cycle;
	\fill[gray!80] (5.000000,17.000000) -- (4.100000,17.000000) -- (5.000000,17.900000) -- cycle;
	\fill[gray!80] (5.000000,17.000000) -- (5.000000,17.900000) -- (5.900000,17.000000) -- cycle;
	\fill[gray!80] (5.000000,17.000000) -- (5.900000,17.000000) -- (5.000000,16.100000) -- cycle;
	\fill[gray!80] (5.000000,17.000000) -- (5.000000,16.100000) -- (4.100000,17.000000) -- cycle;
	\fill[gray!80] (7.000000,17.000000) -- (6.100000,17.000000) -- (7.000000,17.900000) -- cycle;
	\fill[gray!80] (7.000000,17.000000) -- (7.000000,17.900000) -- (7.900000,17.000000) -- cycle;
	\fill[gray!80] (7.000000,17.000000) -- (7.900000,17.000000) -- (7.000000,16.100000) -- cycle;
	\fill[gray!80] (7.000000,17.000000) -- (7.000000,16.100000) -- (6.100000,17.000000) -- cycle;
	\fill[gray!80] (9.000000,17.000000) -- (8.100000,17.000000) -- (9.000000,17.900000) -- cycle;
	\fill[gray!80] (9.000000,17.000000) -- (9.000000,17.900000) -- (9.900000,17.000000) -- cycle;
	\fill[gray!80] (9.000000,17.000000) -- (9.900000,17.000000) -- (9.000000,16.100000) -- cycle;
	\fill[gray!80] (9.000000,17.000000) -- (9.000000,16.100000) -- (8.100000,17.000000) -- cycle;
	\fill[gray!80] (11.000000,17.000000) -- (10.100000,17.000000) -- (11.000000,17.900000) -- cycle;
	\fill[gray!80] (11.000000,17.000000) -- (11.000000,17.900000) -- (11.900000,17.000000) -- cycle;
	\fill[gray!80] (11.000000,17.000000) -- (11.900000,17.000000) -- (11.000000,16.100000) -- cycle;
	\fill[gray!80] (11.000000,17.000000) -- (11.000000,16.100000) -- (10.100000,17.000000) -- cycle;
	\fill[gray!80] (13.000000,17.000000) -- (12.100000,17.000000) -- (13.000000,17.900000) -- cycle;
	\fill[gray!80] (13.000000,17.000000) -- (13.000000,17.900000) -- (13.900000,17.000000) -- cycle;
	\fill[gray!80] (13.000000,17.000000) -- (13.900000,17.000000) -- (13.000000,16.100000) -- cycle;
	\fill[gray!80] (13.000000,17.000000) -- (13.000000,16.100000) -- (12.100000,17.000000) -- cycle;
	\fill[gray!25] (2.000000,16.000000) -- (2.000000,16.900000) -- (2.900000,16.000000) -- cycle;
	\fill[gray!25] (2.000000,16.000000) -- (2.900000,16.000000) -- (2.000000,15.100000) -- cycle;
	\fill[gray!25] (4.000000,16.000000) -- (3.100000,16.000000) -- (4.000000,16.900000) -- cycle;
	\fill[gray!25] (4.000000,16.000000) -- (4.000000,16.900000) -- (4.900000,16.000000) -- cycle;
	\fill[gray!25] (6.000000,16.000000) -- (5.100000,16.000000) -- (6.000000,16.900000) -- cycle;
	\fill[gray!25] (6.000000,16.000000) -- (6.000000,16.900000) -- (6.900000,16.000000) -- cycle;
	\fill[gray!25] (8.000000,16.000000) -- (7.100000,16.000000) -- (8.000000,16.900000) -- cycle;
	\fill[gray!25] (8.000000,16.000000) -- (8.000000,16.900000) -- (8.900000,16.000000) -- cycle;
	\fill[gray!25] (10.000000,16.000000) -- (9.100000,16.000000) -- (10.000000,16.900000) -- cycle;
	\fill[gray!25] (10.000000,16.000000) -- (10.000000,16.900000) -- (10.900000,16.000000) -- cycle;
	\fill[gray!25] (12.000000,16.000000) -- (11.100000,16.000000) -- (12.000000,16.900000) -- cycle;
	\fill[gray!25] (12.000000,16.000000) -- (12.000000,16.900000) -- (12.900000,16.000000) -- cycle;
	\fill[gray!25] (12.000000,16.000000) -- (12.900000,16.000000) -- (12.000000,15.100000) -- cycle;
	\fill[gray!25] (12.000000,16.000000) -- (12.000000,15.100000) -- (11.100000,16.000000) -- cycle;
	\fill[gray!25] (14.000000,16.000000) -- (13.100000,16.000000) -- (14.000000,16.900000) -- cycle;
	\fill[gray!25] (14.000000,16.000000) -- (14.000000,15.100000) -- (13.100000,16.000000) -- cycle;
	\fill[gray!80] (13.000000,15.000000) -- (12.100000,15.000000) -- (13.000000,15.900000) -- cycle;
	\fill[gray!80] (13.000000,15.000000) -- (13.000000,15.900000) -- (13.900000,15.000000) -- cycle;
	\fill[gray!80] (13.000000,15.000000) -- (13.900000,15.000000) -- (13.000000,14.100000) -- cycle;
	\fill[gray!80] (13.000000,15.000000) -- (13.000000,14.100000) -- (12.100000,15.000000) -- cycle;
	\fill[gray!25] (4.000000,14.000000) -- (4.900000,14.000000) -- (4.000000,13.100000) -- cycle;
	\fill[gray!25] (6.000000,14.000000) -- (6.900000,14.000000) -- (6.000000,13.100000) -- cycle;
	\fill[gray!25] (6.000000,14.000000) -- (6.000000,13.100000) -- (5.100000,14.000000) -- cycle;
	\fill[gray!25] (8.000000,14.000000) -- (8.900000,14.000000) -- (8.000000,13.100000) -- cycle;
	\fill[gray!25] (8.000000,14.000000) -- (8.000000,13.100000) -- (7.100000,14.000000) -- cycle;
	\fill[gray!25] (10.000000,14.000000) -- (10.000000,13.100000) -- (9.100000,14.000000) -- cycle;
	\fill[gray!25] (12.000000,14.000000) -- (12.000000,14.900000) -- (12.900000,14.000000) -- cycle;
	\fill[gray!25] (12.000000,14.000000) -- (12.900000,14.000000) -- (12.000000,13.100000) -- cycle;
	\fill[gray!80] (5.000000,13.000000) -- (4.100000,13.000000) -- (5.000000,13.900000) -- cycle;
	\fill[gray!80] (5.000000,13.000000) -- (5.000000,13.900000) -- (5.900000,13.000000) -- cycle;
	\fill[gray!80] (5.000000,13.000000) -- (5.900000,13.000000) -- (5.000000,12.100000) -- cycle;
	\fill[gray!80] (5.000000,13.000000) -- (5.000000,12.100000) -- (4.100000,13.000000) -- cycle;
	\fill[gray!80] (7.000000,13.000000) -- (6.100000,13.000000) -- (7.000000,13.900000) -- cycle;
	\fill[gray!80] (7.000000,13.000000) -- (7.000000,13.900000) -- (7.900000,13.000000) -- cycle;
	\fill[gray!80] (7.000000,13.000000) -- (7.900000,13.000000) -- (7.000000,12.100000) -- cycle;
	\fill[gray!80] (7.000000,13.000000) -- (7.000000,12.100000) -- (6.100000,13.000000) -- cycle;
	\fill[gray!80] (9.000000,13.000000) -- (8.100000,13.000000) -- (9.000000,13.900000) -- cycle;
	\fill[gray!80] (9.000000,13.000000) -- (9.000000,13.900000) -- (9.900000,13.000000) -- cycle;
	\fill[gray!80] (9.000000,13.000000) -- (9.900000,13.000000) -- (9.000000,12.100000) -- cycle;
	\fill[gray!80] (9.000000,13.000000) -- (9.000000,12.100000) -- (8.100000,13.000000) -- cycle;
	\fill[gray!80] (13.000000,13.000000) -- (12.100000,13.000000) -- (13.000000,13.900000) -- cycle;
	\fill[gray!80] (13.000000,13.000000) -- (13.000000,13.900000) -- (13.900000,13.000000) -- cycle;
	\fill[gray!80] (13.000000,13.000000) -- (13.900000,13.000000) -- (13.000000,12.100000) -- cycle;
	\fill[gray!80] (13.000000,13.000000) -- (13.000000,12.100000) -- (12.100000,13.000000) -- cycle;
	\fill[gray!25] (6.000000,12.000000) -- (5.100000,12.000000) -- (6.000000,12.900000) -- cycle;
	\fill[gray!25] (6.000000,12.000000) -- (6.000000,12.900000) -- (6.900000,12.000000) -- cycle;
	\fill[gray!25] (6.000000,12.000000) -- (6.900000,12.000000) -- (6.000000,11.100000) -- cycle;
	\fill[gray!25] (6.000000,12.000000) -- (6.000000,11.100000) -- (5.100000,12.000000) -- cycle;
	\fill[gray!25] (8.000000,12.000000) -- (7.100000,12.000000) -- (8.000000,12.900000) -- cycle;
	\fill[gray!25] (8.000000,12.000000) -- (8.000000,12.900000) -- (8.900000,12.000000) -- cycle;
	\fill[gray!25] (8.000000,12.000000) -- (8.900000,12.000000) -- (8.000000,11.100000) -- cycle;
	\fill[gray!25] (8.000000,12.000000) -- (8.000000,11.100000) -- (7.100000,12.000000) -- cycle;
	\fill[gray!25] (12.000000,12.000000) -- (12.000000,12.900000) -- (12.900000,12.000000) -- cycle;
	\fill[gray!25] (12.000000,12.000000) -- (12.900000,12.000000) -- (12.000000,11.100000) -- cycle;
	\fill[gray!25] (14.000000,12.000000) -- (13.100000,12.000000) -- (14.000000,12.900000) -- cycle;
	\fill[gray!25] (14.000000,12.000000) -- (14.000000,11.100000) -- (13.100000,12.000000) -- cycle;
	\fill[gray!80] (5.000000,11.000000) -- (4.100000,11.000000) -- (5.000000,11.900000) -- cycle;
	\fill[gray!80] (5.000000,11.000000) -- (5.000000,11.900000) -- (5.900000,11.000000) -- cycle;
	\fill[gray!80] (5.000000,11.000000) -- (5.900000,11.000000) -- (5.000000,10.100000) -- cycle;
	\fill[gray!80] (5.000000,11.000000) -- (5.000000,10.100000) -- (4.100000,11.000000) -- cycle;
	\fill[gray!80] (7.000000,11.000000) -- (6.100000,11.000000) -- (7.000000,11.900000) -- cycle;
	\fill[gray!80] (7.000000,11.000000) -- (7.000000,11.900000) -- (7.900000,11.000000) -- cycle;
	\fill[gray!80] (7.000000,11.000000) -- (7.900000,11.000000) -- (7.000000,10.100000) -- cycle;
	\fill[gray!80] (7.000000,11.000000) -- (7.000000,10.100000) -- (6.100000,11.000000) -- cycle;
	\fill[gray!80] (9.000000,11.000000) -- (8.100000,11.000000) -- (9.000000,11.900000) -- cycle;
	\fill[gray!80] (9.000000,11.000000) -- (9.000000,11.900000) -- (9.900000,11.000000) -- cycle;
	\fill[gray!80] (9.000000,11.000000) -- (9.900000,11.000000) -- (9.000000,10.100000) -- cycle;
	\fill[gray!80] (9.000000,11.000000) -- (9.000000,10.100000) -- (8.100000,11.000000) -- cycle;
	\fill[gray!80] (13.000000,11.000000) -- (12.100000,11.000000) -- (13.000000,11.900000) -- cycle;
	\fill[gray!80] (13.000000,11.000000) -- (13.000000,11.900000) -- (13.900000,11.000000) -- cycle;
	\fill[gray!80] (13.000000,11.000000) -- (13.900000,11.000000) -- (13.000000,10.100000) -- cycle;
	\fill[gray!80] (13.000000,11.000000) -- (13.000000,10.100000) -- (12.100000,11.000000) -- cycle;
	\fill[gray!25] (4.000000,10.000000) -- (4.000000,10.900000) -- (4.900000,10.000000) -- cycle;
	\fill[gray!25] (4.000000,10.000000) -- (4.900000,10.000000) -- (4.000000,9.100000) -- cycle;
	\fill[gray!25] (6.000000,10.000000) -- (5.100000,10.000000) -- (6.000000,10.900000) -- cycle;
	\fill[gray!25] (6.000000,10.000000) -- (6.000000,10.900000) -- (6.900000,10.000000) -- cycle;
	\fill[gray!25] (6.000000,10.000000) -- (6.900000,10.000000) -- (6.000000,9.100000) -- cycle;
	\fill[gray!25] (6.000000,10.000000) -- (6.000000,9.100000) -- (5.100000,10.000000) -- cycle;
	\fill[gray!25] (8.000000,10.000000) -- (7.100000,10.000000) -- (8.000000,10.900000) -- cycle;
	\fill[gray!25] (8.000000,10.000000) -- (8.000000,10.900000) -- (8.900000,10.000000) -- cycle;
	\fill[gray!25] (8.000000,10.000000) -- (8.900000,10.000000) -- (8.000000,9.100000) -- cycle;
	\fill[gray!25] (8.000000,10.000000) -- (8.000000,9.100000) -- (7.100000,10.000000) -- cycle;
	\fill[gray!25] (10.000000,10.000000) -- (9.100000,10.000000) -- (10.000000,10.900000) -- cycle;
	\fill[gray!25] (10.000000,10.000000) -- (10.000000,9.100000) -- (9.100000,10.000000) -- cycle;
	\fill[gray!25] (12.000000,10.000000) -- (12.000000,10.900000) -- (12.900000,10.000000) -- cycle;
	\fill[gray!25] (12.000000,10.000000) -- (12.900000,10.000000) -- (12.000000,9.100000) -- cycle;
	\fill[gray!25] (14.000000,10.000000) -- (13.100000,10.000000) -- (14.000000,10.900000) -- cycle;
	\fill[gray!25] (14.000000,10.000000) -- (14.000000,9.100000) -- (13.100000,10.000000) -- cycle;
	\fill[gray!80] (5.000000,9.000000) -- (4.100000,9.000000) -- (5.000000,9.900000) -- cycle;
	\fill[gray!80] (5.000000,9.000000) -- (5.000000,9.900000) -- (5.900000,9.000000) -- cycle;
	\fill[gray!80] (5.000000,9.000000) -- (5.900000,9.000000) -- (5.000000,8.100000) -- cycle;
	\fill[gray!80] (5.000000,9.000000) -- (5.000000,8.100000) -- (4.100000,9.000000) -- cycle;
	\fill[gray!80] (7.000000,9.000000) -- (6.100000,9.000000) -- (7.000000,9.900000) -- cycle;
	\fill[gray!80] (7.000000,9.000000) -- (7.000000,9.900000) -- (7.900000,9.000000) -- cycle;
	\fill[gray!80] (7.000000,9.000000) -- (7.900000,9.000000) -- (7.000000,8.100000) -- cycle;
	\fill[gray!80] (7.000000,9.000000) -- (7.000000,8.100000) -- (6.100000,9.000000) -- cycle;
	\fill[gray!80] (9.000000,9.000000) -- (8.100000,9.000000) -- (9.000000,9.900000) -- cycle;
	\fill[gray!80] (9.000000,9.000000) -- (9.000000,9.900000) -- (9.900000,9.000000) -- cycle;
	\fill[gray!80] (9.000000,9.000000) -- (9.900000,9.000000) -- (9.000000,8.100000) -- cycle;
	\fill[gray!80] (9.000000,9.000000) -- (9.000000,8.100000) -- (8.100000,9.000000) -- cycle;
	\fill[gray!80] (13.000000,9.000000) -- (12.100000,9.000000) -- (13.000000,9.900000) -- cycle;
	\fill[gray!80] (13.000000,9.000000) -- (13.000000,9.900000) -- (13.900000,9.000000) -- cycle;
	\fill[gray!80] (13.000000,9.000000) -- (13.900000,9.000000) -- (13.000000,8.100000) -- cycle;
	\fill[gray!80] (13.000000,9.000000) -- (13.000000,8.100000) -- (12.100000,9.000000) -- cycle;
	\fill[gray!25] (4.000000,8.000000) -- (4.000000,8.900000) -- (4.900000,8.000000) -- cycle;
	\fill[gray!25] (6.000000,8.000000) -- (5.100000,8.000000) -- (6.000000,8.900000) -- cycle;
	\fill[gray!25] (6.000000,8.000000) -- (6.000000,8.900000) -- (6.900000,8.000000) -- cycle;
	\fill[gray!25] (8.000000,8.000000) -- (7.100000,8.000000) -- (8.000000,8.900000) -- cycle;
	\fill[gray!25] (8.000000,8.000000) -- (8.000000,8.900000) -- (8.900000,8.000000) -- cycle;
	\fill[gray!25] (10.000000,8.000000) -- (9.100000,8.000000) -- (10.000000,8.900000) -- cycle;
	\fill[gray!25] (12.000000,8.000000) -- (12.000000,8.900000) -- (12.900000,8.000000) -- cycle;
	\fill[gray!25] (12.000000,8.000000) -- (12.900000,8.000000) -- (12.000000,7.100000) -- cycle;
	\fill[gray!80] (13.000000,7.000000) -- (12.100000,7.000000) -- (13.000000,7.900000) -- cycle;
	\fill[gray!80] (13.000000,7.000000) -- (13.000000,7.900000) -- (13.900000,7.000000) -- cycle;
	\fill[gray!80] (13.000000,7.000000) -- (13.900000,7.000000) -- (13.000000,6.100000) -- cycle;
	\fill[gray!80] (13.000000,7.000000) -- (13.000000,6.100000) -- (12.100000,7.000000) -- cycle;
	\fill[gray!25] (2.000000,6.000000) -- (2.000000,6.900000) -- (2.900000,6.000000) -- cycle;
	\fill[gray!25] (2.000000,6.000000) -- (2.900000,6.000000) -- (2.000000,5.100000) -- cycle;
	\fill[gray!25] (4.000000,6.000000) -- (4.900000,6.000000) -- (4.000000,5.100000) -- cycle;
	\fill[gray!25] (4.000000,6.000000) -- (4.000000,5.100000) -- (3.100000,6.000000) -- cycle;
	\fill[gray!25] (6.000000,6.000000) -- (6.900000,6.000000) -- (6.000000,5.100000) -- cycle;
	\fill[gray!25] (6.000000,6.000000) -- (6.000000,5.100000) -- (5.100000,6.000000) -- cycle;
	\fill[gray!25] (8.000000,6.000000) -- (8.900000,6.000000) -- (8.000000,5.100000) -- cycle;
	\fill[gray!25] (8.000000,6.000000) -- (8.000000,5.100000) -- (7.100000,6.000000) -- cycle;
	\fill[gray!25] (10.000000,6.000000) -- (10.900000,6.000000) -- (10.000000,5.100000) -- cycle;
	\fill[gray!25] (10.000000,6.000000) -- (10.000000,5.100000) -- (9.100000,6.000000) -- cycle;
	\fill[gray!25] (12.000000,6.000000) -- (11.100000,6.000000) -- (12.000000,6.900000) -- cycle;
	\fill[gray!25] (12.000000,6.000000) -- (12.000000,6.900000) -- (12.900000,6.000000) -- cycle;
	\fill[gray!25] (12.000000,6.000000) -- (12.900000,6.000000) -- (12.000000,5.100000) -- cycle;
	\fill[gray!25] (12.000000,6.000000) -- (12.000000,5.100000) -- (11.100000,6.000000) -- cycle;
	\fill[gray!25] (14.000000,6.000000) -- (13.100000,6.000000) -- (14.000000,6.900000) -- cycle;
	\fill[gray!25] (14.000000,6.000000) -- (14.000000,5.100000) -- (13.100000,6.000000) -- cycle;
	\fill[gray!80] (3.000000,5.000000) -- (2.100000,5.000000) -- (3.000000,5.900000) -- cycle;
	\fill[gray!80] (3.000000,5.000000) -- (3.000000,5.900000) -- (3.900000,5.000000) -- cycle;
	\fill[gray!80] (3.000000,5.000000) -- (3.900000,5.000000) -- (3.000000,4.100000) -- cycle;
	\fill[gray!80] (3.000000,5.000000) -- (3.000000,4.100000) -- (2.100000,5.000000) -- cycle;
	\fill[gray!80] (5.000000,5.000000) -- (4.100000,5.000000) -- (5.000000,5.900000) -- cycle;
	\fill[gray!80] (5.000000,5.000000) -- (5.000000,5.900000) -- (5.900000,5.000000) -- cycle;
	\fill[gray!80] (5.000000,5.000000) -- (5.900000,5.000000) -- (5.000000,4.100000) -- cycle;
	\fill[gray!80] (5.000000,5.000000) -- (5.000000,4.100000) -- (4.100000,5.000000) -- cycle;
	\fill[gray!80] (7.000000,5.000000) -- (6.100000,5.000000) -- (7.000000,5.900000) -- cycle;
	\fill[gray!80] (7.000000,5.000000) -- (7.000000,5.900000) -- (7.900000,5.000000) -- cycle;
	\fill[gray!80] (7.000000,5.000000) -- (7.900000,5.000000) -- (7.000000,4.100000) -- cycle;
	\fill[gray!80] (7.000000,5.000000) -- (7.000000,4.100000) -- (6.100000,5.000000) -- cycle;
	\fill[gray!80] (9.000000,5.000000) -- (8.100000,5.000000) -- (9.000000,5.900000) -- cycle;
	\fill[gray!80] (9.000000,5.000000) -- (9.000000,5.900000) -- (9.900000,5.000000) -- cycle;
	\fill[gray!80] (9.000000,5.000000) -- (9.900000,5.000000) -- (9.000000,4.100000) -- cycle;
	\fill[gray!80] (9.000000,5.000000) -- (9.000000,4.100000) -- (8.100000,5.000000) -- cycle;
	\fill[gray!80] (11.000000,5.000000) -- (10.100000,5.000000) -- (11.000000,5.900000) -- cycle;
	\fill[gray!80] (11.000000,5.000000) -- (11.000000,5.900000) -- (11.900000,5.000000) -- cycle;
	\fill[gray!80] (11.000000,5.000000) -- (11.900000,5.000000) -- (11.000000,4.100000) -- cycle;
	\fill[gray!80] (11.000000,5.000000) -- (11.000000,4.100000) -- (10.100000,5.000000) -- cycle;
	\fill[gray!80] (13.000000,5.000000) -- (12.100000,5.000000) -- (13.000000,5.900000) -- cycle;
	\fill[gray!80] (13.000000,5.000000) -- (13.000000,5.900000) -- (13.900000,5.000000) -- cycle;
	\fill[gray!80] (13.000000,5.000000) -- (13.900000,5.000000) -- (13.000000,4.100000) -- cycle;
	\fill[gray!80] (13.000000,5.000000) -- (13.000000,4.100000) -- (12.100000,5.000000) -- cycle;
	\fill[gray!25] (2.000000,4.000000) -- (2.000000,4.900000) -- (2.900000,4.000000) -- cycle;
	\fill[gray!25] (2.000000,4.000000) -- (2.900000,4.000000) -- (2.000000,3.100000) -- cycle;
	\fill[gray!25] (4.000000,4.000000) -- (3.100000,4.000000) -- (4.000000,4.900000) -- cycle;
	\fill[gray!25] (4.000000,4.000000) -- (4.000000,4.900000) -- (4.900000,4.000000) -- cycle;
	\fill[gray!25] (4.000000,4.000000) -- (4.900000,4.000000) -- (4.000000,3.100000) -- cycle;
	\fill[gray!25] (4.000000,4.000000) -- (4.000000,3.100000) -- (3.100000,4.000000) -- cycle;
	\fill[gray!25] (6.000000,4.000000) -- (5.100000,4.000000) -- (6.000000,4.900000) -- cycle;
	\fill[gray!25] (6.000000,4.000000) -- (6.000000,4.900000) -- (6.900000,4.000000) -- cycle;
	\fill[gray!25] (6.000000,4.000000) -- (6.900000,4.000000) -- (6.000000,3.100000) -- cycle;
	\fill[gray!25] (6.000000,4.000000) -- (6.000000,3.100000) -- (5.100000,4.000000) -- cycle;
	\fill[gray!25] (8.000000,4.000000) -- (7.100000,4.000000) -- (8.000000,4.900000) -- cycle;
	\fill[gray!25] (8.000000,4.000000) -- (8.000000,4.900000) -- (8.900000,4.000000) -- cycle;
	\fill[gray!25] (8.000000,4.000000) -- (8.900000,4.000000) -- (8.000000,3.100000) -- cycle;
	\fill[gray!25] (8.000000,4.000000) -- (8.000000,3.100000) -- (7.100000,4.000000) -- cycle;
	\fill[gray!25] (10.000000,4.000000) -- (9.100000,4.000000) -- (10.000000,4.900000) -- cycle;
	\fill[gray!25] (10.000000,4.000000) -- (10.000000,4.900000) -- (10.900000,4.000000) -- cycle;
	\fill[gray!25] (10.000000,4.000000) -- (10.900000,4.000000) -- (10.000000,3.100000) -- cycle;
	\fill[gray!25] (10.000000,4.000000) -- (10.000000,3.100000) -- (9.100000,4.000000) -- cycle;
	\fill[gray!25] (12.000000,4.000000) -- (11.100000,4.000000) -- (12.000000,4.900000) -- cycle;
	\fill[gray!25] (12.000000,4.000000) -- (12.000000,4.900000) -- (12.900000,4.000000) -- cycle;
	\fill[gray!25] (12.000000,4.000000) -- (12.900000,4.000000) -- (12.000000,3.100000) -- cycle;
	\fill[gray!25] (12.000000,4.000000) -- (12.000000,3.100000) -- (11.100000,4.000000) -- cycle;
	\fill[gray!25] (14.000000,4.000000) -- (13.100000,4.000000) -- (14.000000,4.900000) -- cycle;
	\fill[gray!25] (14.000000,4.000000) -- (14.000000,3.100000) -- (13.100000,4.000000) -- cycle;
	\fill[gray!80] (3.000000,3.000000) -- (2.100000,3.000000) -- (3.000000,3.900000) -- cycle;
	\fill[gray!80] (3.000000,3.000000) -- (3.000000,3.900000) -- (3.900000,3.000000) -- cycle;
	\fill[gray!80] (3.000000,3.000000) -- (3.900000,3.000000) -- (3.000000,2.100000) -- cycle;
	\fill[gray!80] (3.000000,3.000000) -- (3.000000,2.100000) -- (2.100000,3.000000) -- cycle;
	\fill[gray!80] (5.000000,3.000000) -- (4.100000,3.000000) -- (5.000000,3.900000) -- cycle;
	\fill[gray!80] (5.000000,3.000000) -- (5.000000,3.900000) -- (5.900000,3.000000) -- cycle;
	\fill[gray!80] (5.000000,3.000000) -- (5.900000,3.000000) -- (5.000000,2.100000) -- cycle;
	\fill[gray!80] (5.000000,3.000000) -- (5.000000,2.100000) -- (4.100000,3.000000) -- cycle;
	\fill[gray!80] (7.000000,3.000000) -- (6.100000,3.000000) -- (7.000000,3.900000) -- cycle;
	\fill[gray!80] (7.000000,3.000000) -- (7.000000,3.900000) -- (7.900000,3.000000) -- cycle;
	\fill[gray!80] (7.000000,3.000000) -- (7.900000,3.000000) -- (7.000000,2.100000) -- cycle;
	\fill[gray!80] (7.000000,3.000000) -- (7.000000,2.100000) -- (6.100000,3.000000) -- cycle;
	\fill[gray!80] (9.000000,3.000000) -- (8.100000,3.000000) -- (9.000000,3.900000) -- cycle;
	\fill[gray!80] (9.000000,3.000000) -- (9.000000,3.900000) -- (9.900000,3.000000) -- cycle;
	\fill[gray!80] (9.000000,3.000000) -- (9.900000,3.000000) -- (9.000000,2.100000) -- cycle;
	\fill[gray!80] (9.000000,3.000000) -- (9.000000,2.100000) -- (8.100000,3.000000) -- cycle;
	\fill[gray!80] (11.000000,3.000000) -- (10.100000,3.000000) -- (11.000000,3.900000) -- cycle;
	\fill[gray!80] (11.000000,3.000000) -- (11.000000,3.900000) -- (11.900000,3.000000) -- cycle;
	\fill[gray!80] (11.000000,3.000000) -- (11.900000,3.000000) -- (11.000000,2.100000) -- cycle;
	\fill[gray!80] (11.000000,3.000000) -- (11.000000,2.100000) -- (10.100000,3.000000) -- cycle;
	\fill[gray!80] (13.000000,3.000000) -- (12.100000,3.000000) -- (13.000000,3.900000) -- cycle;
	\fill[gray!80] (13.000000,3.000000) -- (13.000000,3.900000) -- (13.900000,3.000000) -- cycle;
	\fill[gray!80] (13.000000,3.000000) -- (13.900000,3.000000) -- (13.000000,2.100000) -- cycle;
	\fill[gray!80] (13.000000,3.000000) -- (13.000000,2.100000) -- (12.100000,3.000000) -- cycle;
	\fill[gray!25] (4.000000,2.000000) -- (3.100000,2.000000) -- (4.000000,2.900000) -- cycle;
	\fill[gray!25] (4.000000,2.000000) -- (4.000000,2.900000) -- (4.900000,2.000000) -- cycle;
	\fill[gray!25] (6.000000,2.000000) -- (5.100000,2.000000) -- (6.000000,2.900000) -- cycle;
	\fill[gray!25] (6.000000,2.000000) -- (6.000000,2.900000) -- (6.900000,2.000000) -- cycle;
	\fill[gray!25] (10.000000,2.000000) -- (9.100000,2.000000) -- (10.000000,2.900000) -- cycle;
	\fill[gray!25] (10.000000,2.000000) -- (10.000000,2.900000) -- (10.900000,2.000000) -- cycle;
	\fill[gray!25] (12.000000,2.000000) -- (11.100000,2.000000) -- (12.000000,2.900000) -- cycle;
	\fill[gray!25] (12.000000,2.000000) -- (12.000000,2.900000) -- (12.900000,2.000000) -- cycle;
	\draw[line width=1pt] (3.000000,20.000000) circle (0.100000);
	\draw[line width=1pt,fill=black] (4.000000,20.000000) circle (0.100000);
	\draw[line width=1pt] (5.000000,20.000000) circle (0.100000);
	\draw[line width=1pt,fill=black] (6.000000,20.000000) circle (0.100000);
	\draw[line width=1pt] (7.000000,20.000000) circle (0.100000);
	\draw[line width=1pt] (9.000000,20.000000) circle (0.100000);
	\draw[line width=1pt,fill=black] (10.000000,20.000000) circle (0.100000);
	\draw[line width=1pt] (11.000000,20.000000) circle (0.100000);
	\draw[line width=1pt,fill=black] (12.000000,20.000000) circle (0.100000);
	\draw[line width=1pt] (13.000000,20.000000) circle (0.100000);
	\draw[line width=1pt] (2.000000,19.000000) circle (0.100000);
	\draw[line width=1pt,fill=black] (3.000000,19.000000) circle (0.100000);
	\draw[line width=1pt] (4.000000,19.000000) circle (0.100000);
	\draw[line width=1pt,fill=black] (5.000000,19.000000) circle (0.100000);
	\draw[line width=1pt] (6.000000,19.000000) circle (0.100000);
	\draw[line width=1pt,fill=black] (7.000000,19.000000) circle (0.100000);
	\draw[line width=1pt] (8.000000,19.000000) circle (0.100000);
	\draw[line width=1pt,fill=black] (9.000000,19.000000) circle (0.100000);
	\draw[line width=1pt] (10.000000,19.000000) circle (0.100000);
	\draw[line width=1pt,fill=black] (11.000000,19.000000) circle (0.100000);
	\draw[line width=1pt] (12.000000,19.000000) circle (0.100000);
	\draw[line width=1pt,fill=black] (13.000000,19.000000) circle (0.100000);
	\draw[line width=1pt] (14.000000,19.000000) circle (0.100000);
	\draw[line width=1pt,fill=black] (2.000000,18.000000) circle (0.100000);
	\draw[line width=1pt] (3.000000,18.000000) circle (0.100000);
	\draw[line width=1pt,fill=black] (4.000000,18.000000) circle (0.100000);
	\draw[line width=1pt] (5.000000,18.000000) circle (0.100000);
	\draw[line width=1pt,fill=black] (6.000000,18.000000) circle (0.100000);
	\draw[line width=1pt] (7.000000,18.000000) circle (0.100000);
	\draw[line width=1pt,fill=black] (8.000000,18.000000) circle (0.100000);
	\draw[line width=1pt] (9.000000,18.000000) circle (0.100000);
	\draw[line width=1pt,fill=black] (10.000000,18.000000) circle (0.100000);
	\draw[line width=1pt] (11.000000,18.000000) circle (0.100000);
	\draw[line width=1pt,fill=black] (12.000000,18.000000) circle (0.100000);
	\draw[line width=1pt] (13.000000,18.000000) circle (0.100000);
	\draw[line width=1pt,fill=black] (14.000000,18.000000) circle (0.100000);
	\draw[line width=1pt] (2.000000,17.000000) circle (0.100000);
	\draw[line width=1pt,fill=black] (3.000000,17.000000) circle (0.100000);
	\draw[line width=1pt] (4.000000,17.000000) circle (0.100000);
	\draw[line width=1pt,fill=black] (5.000000,17.000000) circle (0.100000);
	\draw[line width=1pt] (6.000000,17.000000) circle (0.100000);
	\draw[line width=1pt,fill=black] (7.000000,17.000000) circle (0.100000);
	\draw[line width=1pt] (8.000000,17.000000) circle (0.100000);
	\draw[line width=1pt,fill=black] (9.000000,17.000000) circle (0.100000);
	\draw[line width=1pt] (10.000000,17.000000) circle (0.100000);
	\draw[line width=1pt,fill=black] (11.000000,17.000000) circle (0.100000);
	\draw[line width=1pt] (12.000000,17.000000) circle (0.100000);
	\draw[line width=1pt,fill=black] (13.000000,17.000000) circle (0.100000);
	\draw[line width=1pt] (14.000000,17.000000) circle (0.100000);
	\draw[line width=1pt,fill=black] (2.000000,16.000000) circle (0.100000);
	\draw[line width=1pt] (3.000000,16.000000) circle (0.100000);
	\draw[line width=1pt,fill=black] (4.000000,16.000000) circle (0.100000);
	\draw[line width=1pt] (5.000000,16.000000) circle (0.100000);
	\draw[line width=1pt,fill=black] (6.000000,16.000000) circle (0.100000);
	\draw[line width=1pt] (7.000000,16.000000) circle (0.100000);
	\draw[line width=1pt,fill=black] (8.000000,16.000000) circle (0.100000);
	\draw[line width=1pt] (9.000000,16.000000) circle (0.100000);
	\draw[line width=1pt,fill=black] (10.000000,16.000000) circle (0.100000);
	\draw[line width=1pt] (11.000000,16.000000) circle (0.100000);
	\draw[line width=1pt,fill=black] (12.000000,16.000000) circle (0.100000);
	\draw[line width=1pt] (13.000000,16.000000) circle (0.100000);
	\draw[line width=1pt,fill=black] (14.000000,16.000000) circle (0.100000);
	\draw[line width=1pt] (2.000000,15.000000) circle (0.100000);
	\draw[line width=1pt] (12.000000,15.000000) circle (0.100000);
	\draw[line width=1pt,fill=black] (13.000000,15.000000) circle (0.100000);
	\draw[line width=1pt] (14.000000,15.000000) circle (0.100000);
	\draw[line width=1pt,fill=black] (4.000000,14.000000) circle (0.100000);
	\draw[line width=1pt] (5.000000,14.000000) circle (0.100000);
	\draw[line width=1pt,fill=black] (6.000000,14.000000) circle (0.100000);
	\draw[line width=1pt] (7.000000,14.000000) circle (0.100000);
	\draw[line width=1pt,fill=black] (8.000000,14.000000) circle (0.100000);
	\draw[line width=1pt] (9.000000,14.000000) circle (0.100000);
	\draw[line width=1pt,fill=black] (10.000000,14.000000) circle (0.100000);
	\draw[line width=1pt,fill=black] (12.000000,14.000000) circle (0.100000);
	\draw[line width=1pt] (13.000000,14.000000) circle (0.100000);
	\draw[line width=1pt] (2.000000,13.000000) circle (0.100000);
	\draw (4.000000,13.000000) node [above right] {$|+\rangle$};
grestore
	\draw[line width=1pt] (4.000000,13.000000) circle (0.100000);
	\draw[line width=1pt,fill=black] (5.000000,13.000000) circle (0.100000);
	\draw[line width=1pt] (6.000000,13.000000) circle (0.100000);
	\draw[line width=1pt,fill=black] (7.000000,13.000000) circle (0.100000);
	\draw[line width=1pt] (8.000000,13.000000) circle (0.100000);
	\draw[line width=1pt,fill=black] (9.000000,13.000000) circle (0.100000);
	\draw (10.000000,13.000000) node [above right] {$|+\rangle$};
grestore
	\draw[line width=1pt] (10.000000,13.000000) circle (0.100000);
	\draw[line width=1pt] (12.000000,13.000000) circle (0.100000);
	\draw[line width=1pt,fill=black] (13.000000,13.000000) circle (0.100000);
	\draw[line width=1pt] (14.000000,13.000000) circle (0.100000);
	\draw[line width=1pt,fill=black] (2.000000,12.000000) circle (0.100000);
	\draw[line width=1pt] (5.000000,12.000000) circle (0.100000);
	\draw[line width=1pt,fill=black] (6.000000,12.000000) circle (0.100000);
	\draw[line width=1pt] (7.000000,12.000000) circle (0.100000);
	\draw[line width=1pt,fill=black] (8.000000,12.000000) circle (0.100000);
	\draw[line width=1pt] (9.000000,12.000000) circle (0.100000);
	\draw[line width=1pt,fill=black] (12.000000,12.000000) circle (0.100000);
	\draw[line width=1pt] (13.000000,12.000000) circle (0.100000);
	\draw[line width=1pt,fill=black] (14.000000,12.000000) circle (0.100000);
	\draw[line width=1pt] (2.000000,11.000000) circle (0.100000);
	\draw (4.000000,11.000000) node [above right] {$|+\rangle$};
grestore
	\draw[line width=1pt] (4.000000,11.000000) circle (0.100000);
	\draw[line width=1pt,fill=black] (5.000000,11.000000) circle (0.100000);
	\draw[line width=1pt] (6.000000,11.000000) circle (0.100000);
	\draw[line width=1pt,fill=black] (7.000000,11.000000) circle (0.100000);
	\draw[line width=1pt] (8.000000,11.000000) circle (0.100000);
	\draw[line width=1pt,fill=black] (9.000000,11.000000) circle (0.100000);
	\draw (10.000000,11.000000) node [above right] {$|+\rangle$};
grestore
	\draw[line width=1pt] (10.000000,11.000000) circle (0.100000);
	\draw[line width=1pt] (12.000000,11.000000) circle (0.100000);
	\draw[line width=1pt,fill=black] (13.000000,11.000000) circle (0.100000);
	\draw[line width=1pt] (14.000000,11.000000) circle (0.100000);
	\draw[line width=1pt,fill=black] (2.000000,10.000000) circle (0.100000);
	\draw[line width=1pt,fill=black] (4.000000,10.000000) circle (0.100000);
	\draw[line width=1pt] (5.000000,10.000000) circle (0.100000);
	\draw[line width=1pt,fill=black] (6.000000,10.000000) circle (0.100000);
	\draw[line width=1pt] (7.000000,10.000000) circle (0.100000);
	\draw[line width=1pt,fill=black] (8.000000,10.000000) circle (0.100000);
	\draw[line width=1pt] (9.000000,10.000000) circle (0.100000);
	\draw[line width=1pt,fill=black] (10.000000,10.000000) circle (0.100000);
	\draw[line width=1pt,fill=black] (12.000000,10.000000) circle (0.100000);
	\draw[line width=1pt] (13.000000,10.000000) circle (0.100000);
	\draw[line width=1pt,fill=black] (14.000000,10.000000) circle (0.100000);
	\draw[line width=1pt] (2.000000,9.000000) circle (0.100000);
	\draw (4.000000,9.000000) node [above right] {$|0\rangle$};
grestore
	\draw[line width=1pt] (4.000000,9.000000) circle (0.100000);
	\draw[line width=1pt,fill=black] (5.000000,9.000000) circle (0.100000);
	\draw (6.000000,9.000000) node [above right] {$|0\rangle$};
grestore
	\draw[line width=1pt] (6.000000,9.000000) circle (0.100000);
	\draw[line width=1pt,fill=black] (7.000000,9.000000) circle (0.100000);
	\draw (8.000000,9.000000) node [above right] {$|0\rangle$};
grestore
	\draw[line width=1pt] (8.000000,9.000000) circle (0.100000);
	\draw[line width=1pt,fill=black] (9.000000,9.000000) circle (0.100000);
	\draw (10.000000,9.000000) node [above right] {$|0\rangle$};
grestore
	\draw[line width=1pt] (10.000000,9.000000) circle (0.100000);
	\draw[line width=1pt] (12.000000,9.000000) circle (0.100000);
	\draw[line width=1pt,fill=black] (13.000000,9.000000) circle (0.100000);
	\draw[line width=1pt] (14.000000,9.000000) circle (0.100000);
	\draw[line width=1pt,fill=black] (4.000000,8.000000) circle (0.100000);
	\draw (5.000000,8.000000) node [above right] {$|0\rangle$};
grestore
	\draw[line width=1pt] (5.000000,8.000000) circle (0.100000);
	\draw[line width=1pt,fill=black] (6.000000,8.000000) circle (0.100000);
	\draw (7.000000,8.000000) node [above right] {$|0\rangle$};
grestore
	\draw[line width=1pt] (7.000000,8.000000) circle (0.100000);
	\draw[line width=1pt,fill=black] (8.000000,8.000000) circle (0.100000);
	\draw (9.000000,8.000000) node [above right] {$|0\rangle$};
grestore
	\draw[line width=1pt] (9.000000,8.000000) circle (0.100000);
	\draw[line width=1pt,fill=black] (10.000000,8.000000) circle (0.100000);
	\draw[line width=1pt,fill=black] (12.000000,8.000000) circle (0.100000);
	\draw[line width=1pt] (13.000000,8.000000) circle (0.100000);
	\draw[line width=1pt] (2.000000,7.000000) circle (0.100000);
	\draw[line width=1pt] (12.000000,7.000000) circle (0.100000);
	\draw[line width=1pt,fill=black] (13.000000,7.000000) circle (0.100000);
	\draw[line width=1pt] (14.000000,7.000000) circle (0.100000);
	\draw[line width=1pt,fill=black] (2.000000,6.000000) circle (0.100000);
	\draw[line width=1pt] (3.000000,6.000000) circle (0.100000);
	\draw[line width=1pt,fill=black] (4.000000,6.000000) circle (0.100000);
	\draw[line width=1pt] (5.000000,6.000000) circle (0.100000);
	\draw[line width=1pt,fill=black] (6.000000,6.000000) circle (0.100000);
	\draw[line width=1pt] (7.000000,6.000000) circle (0.100000);
	\draw[line width=1pt,fill=black] (8.000000,6.000000) circle (0.100000);
	\draw[line width=1pt] (9.000000,6.000000) circle (0.100000);
	\draw[line width=1pt,fill=black] (10.000000,6.000000) circle (0.100000);
	\draw[line width=1pt] (11.000000,6.000000) circle (0.100000);
	\draw[line width=1pt,fill=black] (12.000000,6.000000) circle (0.100000);
	\draw[line width=1pt] (13.000000,6.000000) circle (0.100000);
	\draw[line width=1pt,fill=black] (14.000000,6.000000) circle (0.100000);
	\draw[line width=1pt] (2.000000,5.000000) circle (0.100000);
	\draw[line width=1pt,fill=black] (3.000000,5.000000) circle (0.100000);
	\draw[line width=1pt] (4.000000,5.000000) circle (0.100000);
	\draw[line width=1pt,fill=black] (5.000000,5.000000) circle (0.100000);
	\draw[line width=1pt] (6.000000,5.000000) circle (0.100000);
	\draw[line width=1pt,fill=black] (7.000000,5.000000) circle (0.100000);
	\draw[line width=1pt] (8.000000,5.000000) circle (0.100000);
	\draw[line width=1pt,fill=black] (9.000000,5.000000) circle (0.100000);
	\draw[line width=1pt] (10.000000,5.000000) circle (0.100000);
	\draw[line width=1pt,fill=black] (11.000000,5.000000) circle (0.100000);
	\draw[line width=1pt] (12.000000,5.000000) circle (0.100000);
	\draw[line width=1pt,fill=black] (13.000000,5.000000) circle (0.100000);
	\draw[line width=1pt] (14.000000,5.000000) circle (0.100000);
	\draw[line width=1pt,fill=black] (2.000000,4.000000) circle (0.100000);
	\draw[line width=1pt] (3.000000,4.000000) circle (0.100000);
	\draw[line width=1pt,fill=black] (4.000000,4.000000) circle (0.100000);
	\draw[line width=1pt] (5.000000,4.000000) circle (0.100000);
	\draw[line width=1pt,fill=black] (6.000000,4.000000) circle (0.100000);
	\draw[line width=1pt] (7.000000,4.000000) circle (0.100000);
	\draw[line width=1pt,fill=black] (8.000000,4.000000) circle (0.100000);
	\draw[line width=1pt] (9.000000,4.000000) circle (0.100000);
	\draw[line width=1pt,fill=black] (10.000000,4.000000) circle (0.100000);
	\draw[line width=1pt] (11.000000,4.000000) circle (0.100000);
	\draw[line width=1pt,fill=black] (12.000000,4.000000) circle (0.100000);
	\draw[line width=1pt] (13.000000,4.000000) circle (0.100000);
	\draw[line width=1pt,fill=black] (14.000000,4.000000) circle (0.100000);
	\draw[line width=1pt] (2.000000,3.000000) circle (0.100000);
	\draw[line width=1pt,fill=black] (3.000000,3.000000) circle (0.100000);
	\draw[line width=1pt] (4.000000,3.000000) circle (0.100000);
	\draw[line width=1pt,fill=black] (5.000000,3.000000) circle (0.100000);
	\draw[line width=1pt] (6.000000,3.000000) circle (0.100000);
	\draw[line width=1pt,fill=black] (7.000000,3.000000) circle (0.100000);
	\draw[line width=1pt] (8.000000,3.000000) circle (0.100000);
	\draw[line width=1pt,fill=black] (9.000000,3.000000) circle (0.100000);
	\draw[line width=1pt] (10.000000,3.000000) circle (0.100000);
	\draw[line width=1pt,fill=black] (11.000000,3.000000) circle (0.100000);
	\draw[line width=1pt] (12.000000,3.000000) circle (0.100000);
	\draw[line width=1pt,fill=black] (13.000000,3.000000) circle (0.100000);
	\draw[line width=1pt] (14.000000,3.000000) circle (0.100000);
	\draw[line width=1pt] (3.000000,2.000000) circle (0.100000);
	\draw[line width=1pt,fill=black] (4.000000,2.000000) circle (0.100000);
	\draw[line width=1pt] (5.000000,2.000000) circle (0.100000);
	\draw[line width=1pt,fill=black] (6.000000,2.000000) circle (0.100000);
	\draw[line width=1pt] (7.000000,2.000000) circle (0.100000);
	\draw[line width=1pt] (9.000000,2.000000) circle (0.100000);
	\draw[line width=1pt,fill=black] (10.000000,2.000000) circle (0.100000);
	\draw[line width=1pt] (11.000000,2.000000) circle (0.100000);
	\draw[line width=1pt,fill=black] (12.000000,2.000000) circle (0.100000);
	\draw[line width=1pt] (13.000000,2.000000) circle (0.100000);
\end{tikzpicture}

%% file: 006ex2.tex
\begin{tikzpicture}[x=0.030000\linewidth,y=0.030000\linewidth]
	\fill[gray!25] (4.000000,20.000000) -- (4.900000,20.000000) -- (4.000000,19.100000) -- cycle;
	\fill[gray!25] (4.000000,20.000000) -- (4.000000,19.100000) -- (3.100000,20.000000) -- cycle;
	\fill[gray!25] (6.000000,20.000000) -- (6.900000,20.000000) -- (6.000000,19.100000) -- cycle;
	\fill[gray!25] (6.000000,20.000000) -- (6.000000,19.100000) -- (5.100000,20.000000) -- cycle;
	\fill[gray!25] (10.000000,20.000000) -- (10.900000,20.000000) -- (10.000000,19.100000) -- cycle;
	\fill[gray!25] (10.000000,20.000000) -- (10.000000,19.100000) -- (9.100000,20.000000) -- cycle;
	\fill[gray!25] (12.000000,20.000000) -- (12.900000,20.000000) -- (12.000000,19.100000) -- cycle;
	\fill[gray!25] (12.000000,20.000000) -- (12.000000,19.100000) -- (11.100000,20.000000) -- cycle;
	\fill[gray!80] (3.000000,19.000000) -- (2.100000,19.000000) -- (3.000000,19.900000) -- cycle;
	\fill[gray!80] (3.000000,19.000000) -- (3.000000,19.900000) -- (3.900000,19.000000) -- cycle;
	\fill[gray!80] (3.000000,19.000000) -- (3.900000,19.000000) -- (3.000000,18.100000) -- cycle;
	\fill[gray!80] (3.000000,19.000000) -- (3.000000,18.100000) -- (2.100000,19.000000) -- cycle;
	\fill[gray!80] (5.000000,19.000000) -- (4.100000,19.000000) -- (5.000000,19.900000) -- cycle;
	\fill[gray!80] (5.000000,19.000000) -- (5.000000,19.900000) -- (5.900000,19.000000) -- cycle;
	\fill[gray!80] (5.000000,19.000000) -- (5.900000,19.000000) -- (5.000000,18.100000) -- cycle;
	\fill[gray!80] (5.000000,19.000000) -- (5.000000,18.100000) -- (4.100000,19.000000) -- cycle;
	\fill[gray!80] (7.000000,19.000000) -- (6.100000,19.000000) -- (7.000000,19.900000) -- cycle;
	\fill[gray!80] (7.000000,19.000000) -- (7.000000,19.900000) -- (7.900000,19.000000) -- cycle;
	\fill[gray!80] (7.000000,19.000000) -- (7.900000,19.000000) -- (7.000000,18.100000) -- cycle;
	\fill[gray!80] (7.000000,19.000000) -- (7.000000,18.100000) -- (6.100000,19.000000) -- cycle;
	\fill[gray!80] (9.000000,19.000000) -- (8.100000,19.000000) -- (9.000000,19.900000) -- cycle;
	\fill[gray!80] (9.000000,19.000000) -- (9.000000,19.900000) -- (9.900000,19.000000) -- cycle;
	\fill[gray!80] (9.000000,19.000000) -- (9.900000,19.000000) -- (9.000000,18.100000) -- cycle;
	\fill[gray!80] (9.000000,19.000000) -- (9.000000,18.100000) -- (8.100000,19.000000) -- cycle;
	\fill[gray!80] (11.000000,19.000000) -- (10.100000,19.000000) -- (11.000000,19.900000) -- cycle;
	\fill[gray!80] (11.000000,19.000000) -- (11.000000,19.900000) -- (11.900000,19.000000) -- cycle;
	\fill[gray!80] (11.000000,19.000000) -- (11.900000,19.000000) -- (11.000000,18.100000) -- cycle;
	\fill[gray!80] (11.000000,19.000000) -- (11.000000,18.100000) -- (10.100000,19.000000) -- cycle;
	\fill[gray!80] (13.000000,19.000000) -- (12.100000,19.000000) -- (13.000000,19.900000) -- cycle;
	\fill[gray!80] (13.000000,19.000000) -- (13.000000,19.900000) -- (13.900000,19.000000) -- cycle;
	\fill[gray!80] (13.000000,19.000000) -- (13.900000,19.000000) -- (13.000000,18.100000) -- cycle;
	\fill[gray!80] (13.000000,19.000000) -- (13.000000,18.100000) -- (12.100000,19.000000) -- cycle;
	\fill[gray!25] (2.000000,18.000000) -- (2.000000,18.900000) -- (2.900000,18.000000) -- cycle;
	\fill[gray!25] (2.000000,18.000000) -- (2.900000,18.000000) -- (2.000000,17.100000) -- cycle;
	\fill[gray!25] (4.000000,18.000000) -- (3.100000,18.000000) -- (4.000000,18.900000) -- cycle;
	\fill[gray!25] (4.000000,18.000000) -- (4.000000,18.900000) -- (4.900000,18.000000) -- cycle;
	\fill[gray!25] (4.000000,18.000000) -- (4.900000,18.000000) -- (4.000000,17.100000) -- cycle;
	\fill[gray!25] (4.000000,18.000000) -- (4.000000,17.100000) -- (3.100000,18.000000) -- cycle;
	\fill[gray!25] (6.000000,18.000000) -- (5.100000,18.000000) -- (6.000000,18.900000) -- cycle;
	\fill[gray!25] (6.000000,18.000000) -- (6.000000,18.900000) -- (6.900000,18.000000) -- cycle;
	\fill[gray!25] (6.000000,18.000000) -- (6.900000,18.000000) -- (6.000000,17.100000) -- cycle;
	\fill[gray!25] (6.000000,18.000000) -- (6.000000,17.100000) -- (5.100000,18.000000) -- cycle;
	\fill[gray!25] (8.000000,18.000000) -- (7.100000,18.000000) -- (8.000000,18.900000) -- cycle;
	\fill[gray!25] (8.000000,18.000000) -- (8.000000,18.900000) -- (8.900000,18.000000) -- cycle;
	\fill[gray!25] (8.000000,18.000000) -- (8.900000,18.000000) -- (8.000000,17.100000) -- cycle;
	\fill[gray!25] (8.000000,18.000000) -- (8.000000,17.100000) -- (7.100000,18.000000) -- cycle;
	\fill[gray!25] (10.000000,18.000000) -- (9.100000,18.000000) -- (10.000000,18.900000) -- cycle;
	\fill[gray!25] (10.000000,18.000000) -- (10.000000,18.900000) -- (10.900000,18.000000) -- cycle;
	\fill[gray!25] (10.000000,18.000000) -- (10.900000,18.000000) -- (10.000000,17.100000) -- cycle;
	\fill[gray!25] (10.000000,18.000000) -- (10.000000,17.100000) -- (9.100000,18.000000) -- cycle;
	\fill[gray!25] (12.000000,18.000000) -- (11.100000,18.000000) -- (12.000000,18.900000) -- cycle;
	\fill[gray!25] (12.000000,18.000000) -- (12.000000,18.900000) -- (12.900000,18.000000) -- cycle;
	\fill[gray!25] (12.000000,18.000000) -- (12.900000,18.000000) -- (12.000000,17.100000) -- cycle;
	\fill[gray!25] (12.000000,18.000000) -- (12.000000,17.100000) -- (11.100000,18.000000) -- cycle;
	\fill[gray!25] (14.000000,18.000000) -- (13.100000,18.000000) -- (14.000000,18.900000) -- cycle;
	\fill[gray!25] (14.000000,18.000000) -- (14.000000,17.100000) -- (13.100000,18.000000) -- cycle;
	\fill[gray!80] (3.000000,17.000000) -- (2.100000,17.000000) -- (3.000000,17.900000) -- cycle;
	\fill[gray!80] (3.000000,17.000000) -- (3.000000,17.900000) -- (3.900000,17.000000) -- cycle;
	\fill[gray!80] (3.000000,17.000000) -- (3.900000,17.000000) -- (3.000000,16.100000) -- cycle;
	\fill[gray!80] (3.000000,17.000000) -- (3.000000,16.100000) -- (2.100000,17.000000) -- cycle;
	\fill[gray!80] (5.000000,17.000000) -- (4.100000,17.000000) -- (5.000000,17.900000) -- cycle;
	\fill[gray!80] (5.000000,17.000000) -- (5.000000,17.900000) -- (5.900000,17.000000) -- cycle;
	\fill[gray!80] (5.000000,17.000000) -- (5.900000,17.000000) -- (5.000000,16.100000) -- cycle;
	\fill[gray!80] (5.000000,17.000000) -- (5.000000,16.100000) -- (4.100000,17.000000) -- cycle;
	\fill[gray!80] (7.000000,17.000000) -- (6.100000,17.000000) -- (7.000000,17.900000) -- cycle;
	\fill[gray!80] (7.000000,17.000000) -- (7.000000,17.900000) -- (7.900000,17.000000) -- cycle;
	\fill[gray!80] (7.000000,17.000000) -- (7.900000,17.000000) -- (7.000000,16.100000) -- cycle;
	\fill[gray!80] (7.000000,17.000000) -- (7.000000,16.100000) -- (6.100000,17.000000) -- cycle;
	\fill[gray!80] (9.000000,17.000000) -- (8.100000,17.000000) -- (9.000000,17.900000) -- cycle;
	\fill[gray!80] (9.000000,17.000000) -- (9.000000,17.900000) -- (9.900000,17.000000) -- cycle;
	\fill[gray!80] (9.000000,17.000000) -- (9.900000,17.000000) -- (9.000000,16.100000) -- cycle;
	\fill[gray!80] (9.000000,17.000000) -- (9.000000,16.100000) -- (8.100000,17.000000) -- cycle;
	\fill[gray!80] (11.000000,17.000000) -- (10.100000,17.000000) -- (11.000000,17.900000) -- cycle;
	\fill[gray!80] (11.000000,17.000000) -- (11.000000,17.900000) -- (11.900000,17.000000) -- cycle;
	\fill[gray!80] (11.000000,17.000000) -- (11.900000,17.000000) -- (11.000000,16.100000) -- cycle;
	\fill[gray!80] (11.000000,17.000000) -- (11.000000,16.100000) -- (10.100000,17.000000) -- cycle;
	\fill[gray!80] (13.000000,17.000000) -- (12.100000,17.000000) -- (13.000000,17.900000) -- cycle;
	\fill[gray!80] (13.000000,17.000000) -- (13.000000,17.900000) -- (13.900000,17.000000) -- cycle;
	\fill[gray!80] (13.000000,17.000000) -- (13.900000,17.000000) -- (13.000000,16.100000) -- cycle;
	\fill[gray!80] (13.000000,17.000000) -- (13.000000,16.100000) -- (12.100000,17.000000) -- cycle;
	\fill[gray!25] (2.000000,16.000000) -- (2.000000,16.900000) -- (2.900000,16.000000) -- cycle;
	\fill[gray!25] (2.000000,16.000000) -- (2.900000,16.000000) -- (2.000000,15.100000) -- cycle;
	\fill[gray!25] (4.000000,16.000000) -- (3.100000,16.000000) -- (4.000000,16.900000) -- cycle;
	\fill[gray!25] (4.000000,16.000000) -- (4.000000,16.900000) -- (4.900000,16.000000) -- cycle;
	\fill[gray!25] (6.000000,16.000000) -- (5.100000,16.000000) -- (6.000000,16.900000) -- cycle;
	\fill[gray!25] (6.000000,16.000000) -- (6.000000,16.900000) -- (6.900000,16.000000) -- cycle;
	\fill[gray!25] (8.000000,16.000000) -- (7.100000,16.000000) -- (8.000000,16.900000) -- cycle;
	\fill[gray!25] (8.000000,16.000000) -- (8.000000,16.900000) -- (8.900000,16.000000) -- cycle;
	\fill[gray!25] (10.000000,16.000000) -- (9.100000,16.000000) -- (10.000000,16.900000) -- cycle;
	\fill[gray!25] (10.000000,16.000000) -- (10.000000,16.900000) -- (10.900000,16.000000) -- cycle;
	\fill[gray!25] (12.000000,16.000000) -- (11.100000,16.000000) -- (12.000000,16.900000) -- cycle;
	\fill[gray!25] (12.000000,16.000000) -- (12.000000,16.900000) -- (12.900000,16.000000) -- cycle;
	\fill[gray!25] (12.000000,16.000000) -- (12.900000,16.000000) -- (12.000000,15.100000) -- cycle;
	\fill[gray!25] (12.000000,16.000000) -- (12.000000,15.100000) -- (11.100000,16.000000) -- cycle;
	\fill[gray!25] (14.000000,16.000000) -- (13.100000,16.000000) -- (14.000000,16.900000) -- cycle;
	\fill[gray!25] (14.000000,16.000000) -- (14.000000,15.100000) -- (13.100000,16.000000) -- cycle;
	\fill[gray!80] (13.000000,15.000000) -- (12.100000,15.000000) -- (13.000000,15.900000) -- cycle;
	\fill[gray!80] (13.000000,15.000000) -- (13.000000,15.900000) -- (13.900000,15.000000) -- cycle;
	\fill[gray!80] (13.000000,15.000000) -- (13.900000,15.000000) -- (13.000000,14.100000) -- cycle;
	\fill[gray!80] (13.000000,15.000000) -- (13.000000,14.100000) -- (12.100000,15.000000) -- cycle;
	\fill[gray!25] (4.000000,14.000000) -- (4.900000,14.000000) -- (4.000000,13.100000) -- cycle;
	\fill[gray!25] (6.000000,14.000000) -- (6.900000,14.000000) -- (6.000000,13.100000) -- cycle;
	\fill[gray!25] (6.000000,14.000000) -- (6.000000,13.100000) -- (5.100000,14.000000) -- cycle;
	\fill[gray!25] (8.000000,14.000000) -- (8.900000,14.000000) -- (8.000000,13.100000) -- cycle;
	\fill[gray!25] (8.000000,14.000000) -- (8.000000,13.100000) -- (7.100000,14.000000) -- cycle;
	\fill[gray!25] (10.000000,14.000000) -- (10.000000,13.100000) -- (9.100000,14.000000) -- cycle;
	\fill[gray!25] (12.000000,14.000000) -- (12.000000,14.900000) -- (12.900000,14.000000) -- cycle;
	\fill[gray!25] (12.000000,14.000000) -- (12.900000,14.000000) -- (12.000000,13.100000) -- cycle;
	\fill[gray!80] (5.000000,13.000000) -- (4.100000,13.000000) -- (5.000000,13.900000) -- cycle;
	\fill[gray!80] (5.000000,13.000000) -- (5.000000,13.900000) -- (5.900000,13.000000) -- cycle;
	\fill[gray!80] (5.000000,13.000000) -- (5.900000,13.000000) -- (5.000000,12.100000) -- cycle;
	\fill[gray!80] (5.000000,13.000000) -- (5.000000,12.100000) -- (4.100000,13.000000) -- cycle;
	\fill[gray!80] (7.000000,13.000000) -- (6.100000,13.000000) -- (7.000000,13.900000) -- cycle;
	\fill[gray!80] (7.000000,13.000000) -- (7.000000,13.900000) -- (7.900000,13.000000) -- cycle;
	\fill[gray!80] (7.000000,13.000000) -- (7.900000,13.000000) -- (7.000000,12.100000) -- cycle;
	\fill[gray!80] (7.000000,13.000000) -- (7.000000,12.100000) -- (6.100000,13.000000) -- cycle;
	\fill[gray!80] (9.000000,13.000000) -- (8.100000,13.000000) -- (9.000000,13.900000) -- cycle;
	\fill[gray!80] (9.000000,13.000000) -- (9.000000,13.900000) -- (9.900000,13.000000) -- cycle;
	\fill[gray!80] (9.000000,13.000000) -- (9.900000,13.000000) -- (9.000000,12.100000) -- cycle;
	\fill[gray!80] (9.000000,13.000000) -- (9.000000,12.100000) -- (8.100000,13.000000) -- cycle;
	\fill[gray!80] (13.000000,13.000000) -- (12.100000,13.000000) -- (13.000000,13.900000) -- cycle;
	\fill[gray!80] (13.000000,13.000000) -- (13.000000,13.900000) -- (13.900000,13.000000) -- cycle;
	\fill[gray!80] (13.000000,13.000000) -- (13.900000,13.000000) -- (13.000000,12.100000) -- cycle;
	\fill[gray!80] (13.000000,13.000000) -- (13.000000,12.100000) -- (12.100000,13.000000) -- cycle;
	\fill[gray!25] (6.000000,12.000000) -- (5.100000,12.000000) -- (6.000000,12.900000) -- cycle;
	\fill[gray!25] (6.000000,12.000000) -- (6.000000,12.900000) -- (6.900000,12.000000) -- cycle;
	\fill[gray!25] (6.000000,12.000000) -- (6.900000,12.000000) -- (6.000000,11.100000) -- cycle;
	\fill[gray!25] (6.000000,12.000000) -- (6.000000,11.100000) -- (5.100000,12.000000) -- cycle;
	\fill[gray!25] (8.000000,12.000000) -- (7.100000,12.000000) -- (8.000000,12.900000) -- cycle;
	\fill[gray!25] (8.000000,12.000000) -- (8.000000,12.900000) -- (8.900000,12.000000) -- cycle;
	\fill[gray!25] (8.000000,12.000000) -- (8.900000,12.000000) -- (8.000000,11.100000) -- cycle;
	\fill[gray!25] (8.000000,12.000000) -- (8.000000,11.100000) -- (7.100000,12.000000) -- cycle;
	\fill[gray!25] (12.000000,12.000000) -- (12.000000,12.900000) -- (12.900000,12.000000) -- cycle;
	\fill[gray!25] (12.000000,12.000000) -- (12.900000,12.000000) -- (12.000000,11.100000) -- cycle;
	\fill[gray!25] (14.000000,12.000000) -- (13.100000,12.000000) -- (14.000000,12.900000) -- cycle;
	\fill[gray!25] (14.000000,12.000000) -- (14.000000,11.100000) -- (13.100000,12.000000) -- cycle;
	\fill[gray!80] (5.000000,11.000000) -- (4.100000,11.000000) -- (5.000000,11.900000) -- cycle;
	\fill[gray!80] (5.000000,11.000000) -- (5.000000,11.900000) -- (5.900000,11.000000) -- cycle;
	\fill[gray!80] (5.000000,11.000000) -- (5.900000,11.000000) -- (5.000000,10.100000) -- cycle;
	\fill[gray!80] (5.000000,11.000000) -- (5.000000,10.100000) -- (4.100000,11.000000) -- cycle;
	\fill[gray!80] (7.000000,11.000000) -- (6.100000,11.000000) -- (7.000000,11.900000) -- cycle;
	\fill[gray!80] (7.000000,11.000000) -- (7.000000,11.900000) -- (7.900000,11.000000) -- cycle;
	\fill[gray!80] (7.000000,11.000000) -- (7.900000,11.000000) -- (7.000000,10.100000) -- cycle;
	\fill[gray!80] (7.000000,11.000000) -- (7.000000,10.100000) -- (6.100000,11.000000) -- cycle;
	\fill[gray!80] (9.000000,11.000000) -- (8.100000,11.000000) -- (9.000000,11.900000) -- cycle;
	\fill[gray!80] (9.000000,11.000000) -- (9.000000,11.900000) -- (9.900000,11.000000) -- cycle;
	\fill[gray!80] (9.000000,11.000000) -- (9.900000,11.000000) -- (9.000000,10.100000) -- cycle;
	\fill[gray!80] (9.000000,11.000000) -- (9.000000,10.100000) -- (8.100000,11.000000) -- cycle;
	\fill[gray!80] (13.000000,11.000000) -- (12.100000,11.000000) -- (13.000000,11.900000) -- cycle;
	\fill[gray!80] (13.000000,11.000000) -- (13.000000,11.900000) -- (13.900000,11.000000) -- cycle;
	\fill[gray!80] (13.000000,11.000000) -- (13.900000,11.000000) -- (13.000000,10.100000) -- cycle;
	\fill[gray!80] (13.000000,11.000000) -- (13.000000,10.100000) -- (12.100000,11.000000) -- cycle;
	\fill[gray!25] (4.000000,10.000000) -- (4.000000,10.900000) -- (4.900000,10.000000) -- cycle;
	\fill[gray!25] (4.000000,10.000000) -- (4.900000,10.000000) -- (4.000000,9.100000) -- cycle;
	\fill[gray!25] (6.000000,10.000000) -- (5.100000,10.000000) -- (6.000000,10.900000) -- cycle;
	\fill[gray!25] (6.000000,10.000000) -- (6.000000,10.900000) -- (6.900000,10.000000) -- cycle;
	\fill[gray!25] (6.000000,10.000000) -- (6.900000,10.000000) -- (6.000000,9.100000) -- cycle;
	\fill[gray!25] (6.000000,10.000000) -- (6.000000,9.100000) -- (5.100000,10.000000) -- cycle;
	\fill[gray!25] (8.000000,10.000000) -- (7.100000,10.000000) -- (8.000000,10.900000) -- cycle;
	\fill[gray!25] (8.000000,10.000000) -- (8.000000,10.900000) -- (8.900000,10.000000) -- cycle;
	\fill[gray!25] (8.000000,10.000000) -- (8.900000,10.000000) -- (8.000000,9.100000) -- cycle;
	\fill[gray!25] (8.000000,10.000000) -- (8.000000,9.100000) -- (7.100000,10.000000) -- cycle;
	\fill[gray!25] (10.000000,10.000000) -- (9.100000,10.000000) -- (10.000000,10.900000) -- cycle;
	\fill[gray!25] (10.000000,10.000000) -- (10.000000,9.100000) -- (9.100000,10.000000) -- cycle;
	\fill[gray!25] (12.000000,10.000000) -- (12.000000,10.900000) -- (12.900000,10.000000) -- cycle;
	\fill[gray!25] (12.000000,10.000000) -- (12.900000,10.000000) -- (12.000000,9.100000) -- cycle;
	\fill[gray!25] (14.000000,10.000000) -- (13.100000,10.000000) -- (14.000000,10.900000) -- cycle;
	\fill[gray!25] (14.000000,10.000000) -- (14.000000,9.100000) -- (13.100000,10.000000) -- cycle;
	\fill[gray!80] (5.000000,9.000000) -- (4.100000,9.000000) -- (5.000000,9.900000) -- cycle;
	\fill[gray!80] (5.000000,9.000000) -- (5.000000,9.900000) -- (5.900000,9.000000) -- cycle;
	\fill[gray!80] (5.000000,9.000000) -- (5.900000,9.000000) -- (5.000000,8.100000) -- cycle;
	\fill[gray!80] (5.000000,9.000000) -- (5.000000,8.100000) -- (4.100000,9.000000) -- cycle;
	\fill[gray!80] (7.000000,9.000000) -- (6.100000,9.000000) -- (7.000000,9.900000) -- cycle;
	\fill[gray!80] (7.000000,9.000000) -- (7.000000,9.900000) -- (7.900000,9.000000) -- cycle;
	\fill[gray!80] (7.000000,9.000000) -- (7.900000,9.000000) -- (7.000000,8.100000) -- cycle;
	\fill[gray!80] (7.000000,9.000000) -- (7.000000,8.100000) -- (6.100000,9.000000) -- cycle;
	\fill[gray!80] (9.000000,9.000000) -- (8.100000,9.000000) -- (9.000000,9.900000) -- cycle;
	\fill[gray!80] (9.000000,9.000000) -- (9.000000,9.900000) -- (9.900000,9.000000) -- cycle;
	\fill[gray!80] (9.000000,9.000000) -- (9.900000,9.000000) -- (9.000000,8.100000) -- cycle;
	\fill[gray!80] (9.000000,9.000000) -- (9.000000,8.100000) -- (8.100000,9.000000) -- cycle;
	\fill[gray!80] (13.000000,9.000000) -- (12.100000,9.000000) -- (13.000000,9.900000) -- cycle;
	\fill[gray!80] (13.000000,9.000000) -- (13.000000,9.900000) -- (13.900000,9.000000) -- cycle;
	\fill[gray!80] (13.000000,9.000000) -- (13.900000,9.000000) -- (13.000000,8.100000) -- cycle;
	\fill[gray!80] (13.000000,9.000000) -- (13.000000,8.100000) -- (12.100000,9.000000) -- cycle;
	\fill[gray!25] (4.000000,8.000000) -- (4.000000,8.900000) -- (4.900000,8.000000) -- cycle;
	\fill[gray!25] (6.000000,8.000000) -- (5.100000,8.000000) -- (6.000000,8.900000) -- cycle;
	\fill[gray!25] (6.000000,8.000000) -- (6.000000,8.900000) -- (6.900000,8.000000) -- cycle;
	\fill[gray!25] (8.000000,8.000000) -- (7.100000,8.000000) -- (8.000000,8.900000) -- cycle;
	\fill[gray!25] (8.000000,8.000000) -- (8.000000,8.900000) -- (8.900000,8.000000) -- cycle;
	\fill[gray!25] (10.000000,8.000000) -- (9.100000,8.000000) -- (10.000000,8.900000) -- cycle;
	\fill[gray!25] (12.000000,8.000000) -- (12.000000,8.900000) -- (12.900000,8.000000) -- cycle;
	\fill[gray!25] (12.000000,8.000000) -- (12.900000,8.000000) -- (12.000000,7.100000) -- cycle;
	\fill[gray!80] (13.000000,7.000000) -- (12.100000,7.000000) -- (13.000000,7.900000) -- cycle;
	\fill[gray!80] (13.000000,7.000000) -- (13.000000,7.900000) -- (13.900000,7.000000) -- cycle;
	\fill[gray!80] (13.000000,7.000000) -- (13.900000,7.000000) -- (13.000000,6.100000) -- cycle;
	\fill[gray!80] (13.000000,7.000000) -- (13.000000,6.100000) -- (12.100000,7.000000) -- cycle;
	\fill[gray!25] (2.000000,6.000000) -- (2.000000,6.900000) -- (2.900000,6.000000) -- cycle;
	\fill[gray!25] (2.000000,6.000000) -- (2.900000,6.000000) -- (2.000000,5.100000) -- cycle;
	\fill[gray!25] (4.000000,6.000000) -- (4.900000,6.000000) -- (4.000000,5.100000) -- cycle;
	\fill[gray!25] (4.000000,6.000000) -- (4.000000,5.100000) -- (3.100000,6.000000) -- cycle;
	\fill[gray!25] (6.000000,6.000000) -- (6.900000,6.000000) -- (6.000000,5.100000) -- cycle;
	\fill[gray!25] (6.000000,6.000000) -- (6.000000,5.100000) -- (5.100000,6.000000) -- cycle;
	\fill[gray!25] (8.000000,6.000000) -- (8.900000,6.000000) -- (8.000000,5.100000) -- cycle;
	\fill[gray!25] (8.000000,6.000000) -- (8.000000,5.100000) -- (7.100000,6.000000) -- cycle;
	\fill[gray!25] (10.000000,6.000000) -- (10.900000,6.000000) -- (10.000000,5.100000) -- cycle;
	\fill[gray!25] (10.000000,6.000000) -- (10.000000,5.100000) -- (9.100000,6.000000) -- cycle;
	\fill[gray!25] (12.000000,6.000000) -- (11.100000,6.000000) -- (12.000000,6.900000) -- cycle;
	\fill[gray!25] (12.000000,6.000000) -- (12.000000,6.900000) -- (12.900000,6.000000) -- cycle;
	\fill[gray!25] (12.000000,6.000000) -- (12.900000,6.000000) -- (12.000000,5.100000) -- cycle;
	\fill[gray!25] (12.000000,6.000000) -- (12.000000,5.100000) -- (11.100000,6.000000) -- cycle;
	\fill[gray!25] (14.000000,6.000000) -- (13.100000,6.000000) -- (14.000000,6.900000) -- cycle;
	\fill[gray!25] (14.000000,6.000000) -- (14.000000,5.100000) -- (13.100000,6.000000) -- cycle;
	\fill[gray!80] (3.000000,5.000000) -- (2.100000,5.000000) -- (3.000000,5.900000) -- cycle;
	\fill[gray!80] (3.000000,5.000000) -- (3.000000,5.900000) -- (3.900000,5.000000) -- cycle;
	\fill[gray!80] (3.000000,5.000000) -- (3.900000,5.000000) -- (3.000000,4.100000) -- cycle;
	\fill[gray!80] (3.000000,5.000000) -- (3.000000,4.100000) -- (2.100000,5.000000) -- cycle;
	\fill[gray!80] (5.000000,5.000000) -- (4.100000,5.000000) -- (5.000000,5.900000) -- cycle;
	\fill[gray!80] (5.000000,5.000000) -- (5.000000,5.900000) -- (5.900000,5.000000) -- cycle;
	\fill[gray!80] (5.000000,5.000000) -- (5.900000,5.000000) -- (5.000000,4.100000) -- cycle;
	\fill[gray!80] (5.000000,5.000000) -- (5.000000,4.100000) -- (4.100000,5.000000) -- cycle;
	\fill[gray!80] (7.000000,5.000000) -- (6.100000,5.000000) -- (7.000000,5.900000) -- cycle;
	\fill[gray!80] (7.000000,5.000000) -- (7.000000,5.900000) -- (7.900000,5.000000) -- cycle;
	\fill[gray!80] (7.000000,5.000000) -- (7.900000,5.000000) -- (7.000000,4.100000) -- cycle;
	\fill[gray!80] (7.000000,5.000000) -- (7.000000,4.100000) -- (6.100000,5.000000) -- cycle;
	\fill[gray!80] (9.000000,5.000000) -- (8.100000,5.000000) -- (9.000000,5.900000) -- cycle;
	\fill[gray!80] (9.000000,5.000000) -- (9.000000,5.900000) -- (9.900000,5.000000) -- cycle;
	\fill[gray!80] (9.000000,5.000000) -- (9.900000,5.000000) -- (9.000000,4.100000) -- cycle;
	\fill[gray!80] (9.000000,5.000000) -- (9.000000,4.100000) -- (8.100000,5.000000) -- cycle;
	\fill[gray!80] (11.000000,5.000000) -- (10.100000,5.000000) -- (11.000000,5.900000) -- cycle;
	\fill[gray!80] (11.000000,5.000000) -- (11.000000,5.900000) -- (11.900000,5.000000) -- cycle;
	\fill[gray!80] (11.000000,5.000000) -- (11.900000,5.000000) -- (11.000000,4.100000) -- cycle;
	\fill[gray!80] (11.000000,5.000000) -- (11.000000,4.100000) -- (10.100000,5.000000) -- cycle;
	\fill[gray!80] (13.000000,5.000000) -- (12.100000,5.000000) -- (13.000000,5.900000) -- cycle;
	\fill[gray!80] (13.000000,5.000000) -- (13.000000,5.900000) -- (13.900000,5.000000) -- cycle;
	\fill[gray!80] (13.000000,5.000000) -- (13.900000,5.000000) -- (13.000000,4.100000) -- cycle;
	\fill[gray!80] (13.000000,5.000000) -- (13.000000,4.100000) -- (12.100000,5.000000) -- cycle;
	\fill[gray!25] (2.000000,4.000000) -- (2.000000,4.900000) -- (2.900000,4.000000) -- cycle;
	\fill[gray!25] (2.000000,4.000000) -- (2.900000,4.000000) -- (2.000000,3.100000) -- cycle;
	\fill[gray!25] (4.000000,4.000000) -- (3.100000,4.000000) -- (4.000000,4.900000) -- cycle;
	\fill[gray!25] (4.000000,4.000000) -- (4.000000,4.900000) -- (4.900000,4.000000) -- cycle;
	\fill[gray!25] (4.000000,4.000000) -- (4.900000,4.000000) -- (4.000000,3.100000) -- cycle;
	\fill[gray!25] (4.000000,4.000000) -- (4.000000,3.100000) -- (3.100000,4.000000) -- cycle;
	\fill[gray!25] (6.000000,4.000000) -- (5.100000,4.000000) -- (6.000000,4.900000) -- cycle;
	\fill[gray!25] (6.000000,4.000000) -- (6.000000,4.900000) -- (6.900000,4.000000) -- cycle;
	\fill[gray!25] (6.000000,4.000000) -- (6.900000,4.000000) -- (6.000000,3.100000) -- cycle;
	\fill[gray!25] (6.000000,4.000000) -- (6.000000,3.100000) -- (5.100000,4.000000) -- cycle;
	\fill[gray!25] (8.000000,4.000000) -- (7.100000,4.000000) -- (8.000000,4.900000) -- cycle;
	\fill[gray!25] (8.000000,4.000000) -- (8.000000,4.900000) -- (8.900000,4.000000) -- cycle;
	\fill[gray!25] (8.000000,4.000000) -- (8.900000,4.000000) -- (8.000000,3.100000) -- cycle;
	\fill[gray!25] (8.000000,4.000000) -- (8.000000,3.100000) -- (7.100000,4.000000) -- cycle;
	\fill[gray!25] (10.000000,4.000000) -- (9.100000,4.000000) -- (10.000000,4.900000) -- cycle;
	\fill[gray!25] (10.000000,4.000000) -- (10.000000,4.900000) -- (10.900000,4.000000) -- cycle;
	\fill[gray!25] (10.000000,4.000000) -- (10.900000,4.000000) -- (10.000000,3.100000) -- cycle;
	\fill[gray!25] (10.000000,4.000000) -- (10.000000,3.100000) -- (9.100000,4.000000) -- cycle;
	\fill[gray!25] (12.000000,4.000000) -- (11.100000,4.000000) -- (12.000000,4.900000) -- cycle;
	\fill[gray!25] (12.000000,4.000000) -- (12.000000,4.900000) -- (12.900000,4.000000) -- cycle;
	\fill[gray!25] (12.000000,4.000000) -- (12.900000,4.000000) -- (12.000000,3.100000) -- cycle;
	\fill[gray!25] (12.000000,4.000000) -- (12.000000,3.100000) -- (11.100000,4.000000) -- cycle;
	\fill[gray!25] (14.000000,4.000000) -- (13.100000,4.000000) -- (14.000000,4.900000) -- cycle;
	\fill[gray!25] (14.000000,4.000000) -- (14.000000,3.100000) -- (13.100000,4.000000) -- cycle;
	\fill[gray!80] (3.000000,3.000000) -- (2.100000,3.000000) -- (3.000000,3.900000) -- cycle;
	\fill[gray!80] (3.000000,3.000000) -- (3.000000,3.900000) -- (3.900000,3.000000) -- cycle;
	\fill[gray!80] (3.000000,3.000000) -- (3.900000,3.000000) -- (3.000000,2.100000) -- cycle;
	\fill[gray!80] (3.000000,3.000000) -- (3.000000,2.100000) -- (2.100000,3.000000) -- cycle;
	\fill[gray!80] (5.000000,3.000000) -- (4.100000,3.000000) -- (5.000000,3.900000) -- cycle;
	\fill[gray!80] (5.000000,3.000000) -- (5.000000,3.900000) -- (5.900000,3.000000) -- cycle;
	\fill[gray!80] (5.000000,3.000000) -- (5.900000,3.000000) -- (5.000000,2.100000) -- cycle;
	\fill[gray!80] (5.000000,3.000000) -- (5.000000,2.100000) -- (4.100000,3.000000) -- cycle;
	\fill[gray!80] (7.000000,3.000000) -- (6.100000,3.000000) -- (7.000000,3.900000) -- cycle;
	\fill[gray!80] (7.000000,3.000000) -- (7.000000,3.900000) -- (7.900000,3.000000) -- cycle;
	\fill[gray!80] (7.000000,3.000000) -- (7.900000,3.000000) -- (7.000000,2.100000) -- cycle;
	\fill[gray!80] (7.000000,3.000000) -- (7.000000,2.100000) -- (6.100000,3.000000) -- cycle;
	\fill[gray!80] (9.000000,3.000000) -- (8.100000,3.000000) -- (9.000000,3.900000) -- cycle;
	\fill[gray!80] (9.000000,3.000000) -- (9.000000,3.900000) -- (9.900000,3.000000) -- cycle;
	\fill[gray!80] (9.000000,3.000000) -- (9.900000,3.000000) -- (9.000000,2.100000) -- cycle;
	\fill[gray!80] (9.000000,3.000000) -- (9.000000,2.100000) -- (8.100000,3.000000) -- cycle;
	\fill[gray!80] (11.000000,3.000000) -- (10.100000,3.000000) -- (11.000000,3.900000) -- cycle;
	\fill[gray!80] (11.000000,3.000000) -- (11.000000,3.900000) -- (11.900000,3.000000) -- cycle;
	\fill[gray!80] (11.000000,3.000000) -- (11.900000,3.000000) -- (11.000000,2.100000) -- cycle;
	\fill[gray!80] (11.000000,3.000000) -- (11.000000,2.100000) -- (10.100000,3.000000) -- cycle;
	\fill[gray!80] (13.000000,3.000000) -- (12.100000,3.000000) -- (13.000000,3.900000) -- cycle;
	\fill[gray!80] (13.000000,3.000000) -- (13.000000,3.900000) -- (13.900000,3.000000) -- cycle;
	\fill[gray!80] (13.000000,3.000000) -- (13.900000,3.000000) -- (13.000000,2.100000) -- cycle;
	\fill[gray!80] (13.000000,3.000000) -- (13.000000,2.100000) -- (12.100000,3.000000) -- cycle;
	\fill[gray!25] (4.000000,2.000000) -- (3.100000,2.000000) -- (4.000000,2.900000) -- cycle;
	\fill[gray!25] (4.000000,2.000000) -- (4.000000,2.900000) -- (4.900000,2.000000) -- cycle;
	\fill[gray!25] (6.000000,2.000000) -- (5.100000,2.000000) -- (6.000000,2.900000) -- cycle;
	\fill[gray!25] (6.000000,2.000000) -- (6.000000,2.900000) -- (6.900000,2.000000) -- cycle;
	\fill[gray!25] (10.000000,2.000000) -- (9.100000,2.000000) -- (10.000000,2.900000) -- cycle;
	\fill[gray!25] (10.000000,2.000000) -- (10.000000,2.900000) -- (10.900000,2.000000) -- cycle;
	\fill[gray!25] (12.000000,2.000000) -- (11.100000,2.000000) -- (12.000000,2.900000) -- cycle;
	\fill[gray!25] (12.000000,2.000000) -- (12.000000,2.900000) -- (12.900000,2.000000) -- cycle;
	\draw[line width=1pt] (3.000000,20.000000) circle (0.100000);
	\draw[line width=1pt,fill=black] (4.000000,20.000000) circle (0.100000);
	\draw[line width=1pt] (5.000000,20.000000) circle (0.100000);
	\draw[line width=1pt,fill=black] (6.000000,20.000000) circle (0.100000);
	\draw[line width=1pt] (7.000000,20.000000) circle (0.100000);
	\draw[line width=1pt] (9.000000,20.000000) circle (0.100000);
	\draw[line width=1pt,fill=black] (10.000000,20.000000) circle (0.100000);
	\draw[line width=1pt] (11.000000,20.000000) circle (0.100000);
	\draw[line width=1pt,fill=black] (12.000000,20.000000) circle (0.100000);
	\draw[line width=1pt] (13.000000,20.000000) circle (0.100000);
	\draw[line width=1pt] (2.000000,19.000000) circle (0.100000);
	\draw[line width=1pt,fill=black] (3.000000,19.000000) circle (0.100000);
	\draw[line width=1pt] (4.000000,19.000000) circle (0.100000);
	\draw[line width=1pt,fill=black] (5.000000,19.000000) circle (0.100000);
	\draw[line width=1pt] (6.000000,19.000000) circle (0.100000);
	\draw[line width=1pt,fill=black] (7.000000,19.000000) circle (0.100000);
	\draw[line width=1pt] (8.000000,19.000000) circle (0.100000);
	\draw[line width=1pt,fill=black] (9.000000,19.000000) circle (0.100000);
	\draw[line width=1pt] (10.000000,19.000000) circle (0.100000);
	\draw[line width=1pt,fill=black] (11.000000,19.000000) circle (0.100000);
	\draw[line width=1pt] (12.000000,19.000000) circle (0.100000);
	\draw[line width=1pt,fill=black] (13.000000,19.000000) circle (0.100000);
	\draw[line width=1pt] (14.000000,19.000000) circle (0.100000);
	\draw[line width=1pt,fill=black] (2.000000,18.000000) circle (0.100000);
	\draw[line width=1pt] (3.000000,18.000000) circle (0.100000);
	\draw[line width=1pt,fill=black] (4.000000,18.000000) circle (0.100000);
	\draw[line width=1pt] (5.000000,18.000000) circle (0.100000);
	\draw[line width=1pt,fill=black] (6.000000,18.000000) circle (0.100000);
	\draw[line width=1pt] (7.000000,18.000000) circle (0.100000);
	\draw[line width=1pt,fill=black] (8.000000,18.000000) circle (0.100000);
	\draw[line width=1pt] (9.000000,18.000000) circle (0.100000);
	\draw[line width=1pt,fill=black] (10.000000,18.000000) circle (0.100000);
	\draw[line width=1pt] (11.000000,18.000000) circle (0.100000);
	\draw[line width=1pt,fill=black] (12.000000,18.000000) circle (0.100000);
	\draw[line width=1pt] (13.000000,18.000000) circle (0.100000);
	\draw[line width=1pt,fill=black] (14.000000,18.000000) circle (0.100000);
	\draw[line width=1pt] (2.000000,17.000000) circle (0.100000);
	\draw[line width=1pt,fill=black] (3.000000,17.000000) circle (0.100000);
	\draw[line width=1pt] (4.000000,17.000000) circle (0.100000);
	\draw[line width=1pt,fill=black] (5.000000,17.000000) circle (0.100000);
	\draw[line width=1pt] (6.000000,17.000000) circle (0.100000);
	\draw[line width=1pt,fill=black] (7.000000,17.000000) circle (0.100000);
	\draw[line width=1pt] (8.000000,17.000000) circle (0.100000);
	\draw[line width=1pt,fill=black] (9.000000,17.000000) circle (0.100000);
	\draw[line width=1pt] (10.000000,17.000000) circle (0.100000);
	\draw[line width=1pt,fill=black] (11.000000,17.000000) circle (0.100000);
	\draw[line width=1pt] (12.000000,17.000000) circle (0.100000);
	\draw[line width=1pt,fill=black] (13.000000,17.000000) circle (0.100000);
	\draw[line width=1pt] (14.000000,17.000000) circle (0.100000);
	\draw[line width=1pt,fill=black] (2.000000,16.000000) circle (0.100000);
	\draw[line width=1pt] (3.000000,16.000000) circle (0.100000);
	\draw[line width=1pt,fill=black] (4.000000,16.000000) circle (0.100000);
	\draw[line width=1pt] (5.000000,16.000000) circle (0.100000);
	\draw[line width=1pt,fill=black] (6.000000,16.000000) circle (0.100000);
	\draw[line width=1pt] (7.000000,16.000000) circle (0.100000);
	\draw[line width=1pt,fill=black] (8.000000,16.000000) circle (0.100000);
	\draw[line width=1pt] (9.000000,16.000000) circle (0.100000);
	\draw[line width=1pt,fill=black] (10.000000,16.000000) circle (0.100000);
	\draw[line width=1pt] (11.000000,16.000000) circle (0.100000);
	\draw[line width=1pt,fill=black] (12.000000,16.000000) circle (0.100000);
	\draw[line width=1pt] (13.000000,16.000000) circle (0.100000);
	\draw[line width=1pt,fill=black] (14.000000,16.000000) circle (0.100000);
	\draw[line width=1pt] (2.000000,15.000000) circle (0.100000);
	\draw[line width=1pt] (12.000000,15.000000) circle (0.100000);
	\draw[line width=1pt,fill=black] (13.000000,15.000000) circle (0.100000);
	\draw[line width=1pt] (14.000000,15.000000) circle (0.100000);
	\draw[line width=1pt,fill=black] (4.000000,14.000000) circle (0.100000);
	\draw[line width=1pt] (5.000000,14.000000) circle (0.100000);
	\draw[line width=1pt,fill=black] (6.000000,14.000000) circle (0.100000);
	\draw[line width=1pt] (7.000000,14.000000) circle (0.100000);
	\draw[line width=1pt,fill=black] (8.000000,14.000000) circle (0.100000);
	\draw[line width=1pt] (9.000000,14.000000) circle (0.100000);
	\draw[line width=1pt,fill=black] (10.000000,14.000000) circle (0.100000);
	\draw[line width=1pt,fill=black] (12.000000,14.000000) circle (0.100000);
	\draw[line width=1pt] (13.000000,14.000000) circle (0.100000);
	\draw[line width=1pt] (2.000000,13.000000) circle (0.100000);
	\draw[line width=1pt] (4.000000,13.000000) circle (0.100000);
	\draw[line width=1pt,fill=black] (5.000000,13.000000) circle (0.100000);
	\draw[line width=1pt] (6.000000,13.000000) circle (0.100000);
	\draw[line width=1pt,fill=black] (7.000000,13.000000) circle (0.100000);
	\draw[line width=1pt] (8.000000,13.000000) circle (0.100000);
	\draw[line width=1pt,fill=black] (9.000000,13.000000) circle (0.100000);
	\draw[line width=1pt] (10.000000,13.000000) circle (0.100000);
	\draw[line width=1pt] (12.000000,13.000000) circle (0.100000);
	\draw[line width=1pt,fill=black] (13.000000,13.000000) circle (0.100000);
	\draw[line width=1pt] (14.000000,13.000000) circle (0.100000);
	\draw[line width=1pt,fill=black] (2.000000,12.000000) circle (0.100000);
	\draw[line width=1pt] (5.000000,12.000000) circle (0.100000);
	\draw[line width=1pt,fill=black] (6.000000,12.000000) circle (0.100000);
	\draw[line width=1pt] (7.000000,12.000000) circle (0.100000);
	\draw[line width=1pt,fill=black] (8.000000,12.000000) circle (0.100000);
	\draw[line width=1pt] (9.000000,12.000000) circle (0.100000);
	\draw[line width=1pt,fill=black] (12.000000,12.000000) circle (0.100000);
	\draw[line width=1pt] (13.000000,12.000000) circle (0.100000);
	\draw[line width=1pt,fill=black] (14.000000,12.000000) circle (0.100000);
	\draw[line width=1pt] (2.000000,11.000000) circle (0.100000);
	\draw[line width=1pt] (4.000000,11.000000) circle (0.100000);
	\draw[line width=1pt,fill=black] (5.000000,11.000000) circle (0.100000);
	\draw[line width=1pt] (6.000000,11.000000) circle (0.100000);
	\draw[line width=1pt,fill=black] (7.000000,11.000000) circle (0.100000);
	\draw[line width=1pt] (8.000000,11.000000) circle (0.100000);
	\draw[line width=1pt,fill=black] (9.000000,11.000000) circle (0.100000);
	\draw (10.000000,11.000000) node [above right] {$M_X$};
	\draw[line width=1pt] (10.000000,11.000000) circle (0.100000);
	\draw[line width=1pt] (12.000000,11.000000) circle (0.100000);
	\draw[line width=1pt,fill=black] (13.000000,11.000000) circle (0.100000);
	\draw[line width=1pt] (14.000000,11.000000) circle (0.100000);
	\draw[line width=1pt,fill=black] (2.000000,10.000000) circle (0.100000);
	\draw[line width=1pt,fill=black] (4.000000,10.000000) circle (0.100000);
	\draw[line width=1pt] (5.000000,10.000000) circle (0.100000);
	\draw[line width=1pt,fill=black] (6.000000,10.000000) circle (0.100000);
	\draw[line width=1pt] (7.000000,10.000000) circle (0.100000);
	\draw[line width=1pt,fill=black] (8.000000,10.000000) circle (0.100000);
	\draw[line width=1pt] (9.000000,10.000000) circle (0.100000);
	\draw[line width=1pt,fill=black] (10.000000,10.000000) circle (0.100000);
	\draw[line width=1pt,fill=black] (12.000000,10.000000) circle (0.100000);
	\draw[line width=1pt] (13.000000,10.000000) circle (0.100000);
	\draw[line width=1pt,fill=black] (14.000000,10.000000) circle (0.100000);
	\draw[line width=1pt] (2.000000,9.000000) circle (0.100000);
	\draw[line width=1pt] (4.000000,9.000000) circle (0.100000);
	\draw[line width=1pt,fill=black] (5.000000,9.000000) circle (0.100000);
	\draw[line width=1pt] (6.000000,9.000000) circle (0.100000);
	\draw[line width=1pt,fill=black] (7.000000,9.000000) circle (0.100000);
	\draw[line width=1pt] (8.000000,9.000000) circle (0.100000);
	\draw[line width=1pt,fill=black] (9.000000,9.000000) circle (0.100000);
	\draw (10.000000,9.000000) node [above right] {$M_X$};
	\draw[line width=1pt] (10.000000,9.000000) circle (0.100000);
	\draw[line width=1pt] (12.000000,9.000000) circle (0.100000);
	\draw[line width=1pt,fill=black] (13.000000,9.000000) circle (0.100000);
	\draw[line width=1pt] (14.000000,9.000000) circle (0.100000);
	\draw[line width=1pt,fill=black] (4.000000,8.000000) circle (0.100000);
	\draw[line width=1pt] (5.000000,8.000000) circle (0.100000);
	\draw[line width=1pt,fill=black] (6.000000,8.000000) circle (0.100000);
	\draw[line width=1pt] (7.000000,8.000000) circle (0.100000);
	\draw[line width=1pt,fill=black] (8.000000,8.000000) circle (0.100000);
	\draw (9.000000,8.000000) node [above right] {$M_X$};
	\draw[line width=1pt] (9.000000,8.000000) circle (0.100000);
	\draw[line width=1pt,fill=black] (10.000000,8.000000) circle (0.100000);
	\draw[line width=1pt,fill=black] (12.000000,8.000000) circle (0.100000);
	\draw[line width=1pt] (13.000000,8.000000) circle (0.100000);
	\draw[line width=1pt] (2.000000,7.000000) circle (0.100000);
	\draw[line width=1pt] (12.000000,7.000000) circle (0.100000);
	\draw[line width=1pt,fill=black] (13.000000,7.000000) circle (0.100000);
	\draw[line width=1pt] (14.000000,7.000000) circle (0.100000);
	\draw[line width=1pt,fill=black] (2.000000,6.000000) circle (0.100000);
	\draw[line width=1pt] (3.000000,6.000000) circle (0.100000);
	\draw[line width=1pt,fill=black] (4.000000,6.000000) circle (0.100000);
	\draw[line width=1pt] (5.000000,6.000000) circle (0.100000);
	\draw[line width=1pt,fill=black] (6.000000,6.000000) circle (0.100000);
	\draw[line width=1pt] (7.000000,6.000000) circle (0.100000);
	\draw[line width=1pt,fill=black] (8.000000,6.000000) circle (0.100000);
	\draw[line width=1pt] (9.000000,6.000000) circle (0.100000);
	\draw[line width=1pt,fill=black] (10.000000,6.000000) circle (0.100000);
	\draw[line width=1pt] (11.000000,6.000000) circle (0.100000);
	\draw[line width=1pt,fill=black] (12.000000,6.000000) circle (0.100000);
	\draw[line width=1pt] (13.000000,6.000000) circle (0.100000);
	\draw[line width=1pt,fill=black] (14.000000,6.000000) circle (0.100000);
	\draw[line width=1pt] (2.000000,5.000000) circle (0.100000);
	\draw[line width=1pt,fill=black] (3.000000,5.000000) circle (0.100000);
	\draw[line width=1pt] (4.000000,5.000000) circle (0.100000);
	\draw[line width=1pt,fill=black] (5.000000,5.000000) circle (0.100000);
	\draw[line width=1pt] (6.000000,5.000000) circle (0.100000);
	\draw[line width=1pt,fill=black] (7.000000,5.000000) circle (0.100000);
	\draw[line width=1pt] (8.000000,5.000000) circle (0.100000);
	\draw[line width=1pt,fill=black] (9.000000,5.000000) circle (0.100000);
	\draw[line width=1pt] (10.000000,5.000000) circle (0.100000);
	\draw[line width=1pt,fill=black] (11.000000,5.000000) circle (0.100000);
	\draw[line width=1pt] (12.000000,5.000000) circle (0.100000);
	\draw[line width=1pt,fill=black] (13.000000,5.000000) circle (0.100000);
	\draw[line width=1pt] (14.000000,5.000000) circle (0.100000);
	\draw[line width=1pt,fill=black] (2.000000,4.000000) circle (0.100000);
	\draw[line width=1pt] (3.000000,4.000000) circle (0.100000);
	\draw[line width=1pt,fill=black] (4.000000,4.000000) circle (0.100000);
	\draw[line width=1pt] (5.000000,4.000000) circle (0.100000);
	\draw[line width=1pt,fill=black] (6.000000,4.000000) circle (0.100000);
	\draw[line width=1pt] (7.000000,4.000000) circle (0.100000);
	\draw[line width=1pt,fill=black] (8.000000,4.000000) circle (0.100000);
	\draw[line width=1pt] (9.000000,4.000000) circle (0.100000);
	\draw[line width=1pt,fill=black] (10.000000,4.000000) circle (0.100000);
	\draw[line width=1pt] (11.000000,4.000000) circle (0.100000);
	\draw[line width=1pt,fill=black] (12.000000,4.000000) circle (0.100000);
	\draw[line width=1pt] (13.000000,4.000000) circle (0.100000);
	\draw[line width=1pt,fill=black] (14.000000,4.000000) circle (0.100000);
	\draw[line width=1pt] (2.000000,3.000000) circle (0.100000);
	\draw[line width=1pt,fill=black] (3.000000,3.000000) circle (0.100000);
	\draw[line width=1pt] (4.000000,3.000000) circle (0.100000);
	\draw[line width=1pt,fill=black] (5.000000,3.000000) circle (0.100000);
	\draw[line width=1pt] (6.000000,3.000000) circle (0.100000);
	\draw[line width=1pt,fill=black] (7.000000,3.000000) circle (0.100000);
	\draw[line width=1pt] (8.000000,3.000000) circle (0.100000);
	\draw[line width=1pt,fill=black] (9.000000,3.000000) circle (0.100000);
	\draw[line width=1pt] (10.000000,3.000000) circle (0.100000);
	\draw[line width=1pt,fill=black] (11.000000,3.000000) circle (0.100000);
	\draw[line width=1pt] (12.000000,3.000000) circle (0.100000);
	\draw[line width=1pt,fill=black] (13.000000,3.000000) circle (0.100000);
	\draw[line width=1pt] (14.000000,3.000000) circle (0.100000);
	\draw[line width=1pt] (3.000000,2.000000) circle (0.100000);
	\draw[line width=1pt,fill=black] (4.000000,2.000000) circle (0.100000);
	\draw[line width=1pt] (5.000000,2.000000) circle (0.100000);
	\draw[line width=1pt,fill=black] (6.000000,2.000000) circle (0.100000);
	\draw[line width=1pt] (7.000000,2.000000) circle (0.100000);
	\draw[line width=1pt] (9.000000,2.000000) circle (0.100000);
	\draw[line width=1pt,fill=black] (10.000000,2.000000) circle (0.100000);
	\draw[line width=1pt] (11.000000,2.000000) circle (0.100000);
	\draw[line width=1pt,fill=black] (12.000000,2.000000) circle (0.100000);
	\draw[line width=1pt] (13.000000,2.000000) circle (0.100000);
\end{tikzpicture}

%% file: 007ex2.tex
\begin{tikzpicture}[x=0.030000\linewidth,y=0.030000\linewidth]
	\fill[gray!25] (4.000000,20.000000) -- (4.900000,20.000000) -- (4.000000,19.100000) -- cycle;
	\fill[gray!25] (4.000000,20.000000) -- (4.000000,19.100000) -- (3.100000,20.000000) -- cycle;
	\fill[gray!25] (6.000000,20.000000) -- (6.900000,20.000000) -- (6.000000,19.100000) -- cycle;
	\fill[gray!25] (6.000000,20.000000) -- (6.000000,19.100000) -- (5.100000,20.000000) -- cycle;
	\fill[gray!25] (10.000000,20.000000) -- (10.900000,20.000000) -- (10.000000,19.100000) -- cycle;
	\fill[gray!25] (10.000000,20.000000) -- (10.000000,19.100000) -- (9.100000,20.000000) -- cycle;
	\fill[gray!25] (12.000000,20.000000) -- (12.900000,20.000000) -- (12.000000,19.100000) -- cycle;
	\fill[gray!25] (12.000000,20.000000) -- (12.000000,19.100000) -- (11.100000,20.000000) -- cycle;
	\fill[gray!80] (3.000000,19.000000) -- (2.100000,19.000000) -- (3.000000,19.900000) -- cycle;
	\fill[gray!80] (3.000000,19.000000) -- (3.000000,19.900000) -- (3.900000,19.000000) -- cycle;
	\fill[gray!80] (3.000000,19.000000) -- (3.900000,19.000000) -- (3.000000,18.100000) -- cycle;
	\fill[gray!80] (3.000000,19.000000) -- (3.000000,18.100000) -- (2.100000,19.000000) -- cycle;
	\fill[gray!80] (5.000000,19.000000) -- (4.100000,19.000000) -- (5.000000,19.900000) -- cycle;
	\fill[gray!80] (5.000000,19.000000) -- (5.000000,19.900000) -- (5.900000,19.000000) -- cycle;
	\fill[gray!80] (5.000000,19.000000) -- (5.900000,19.000000) -- (5.000000,18.100000) -- cycle;
	\fill[gray!80] (5.000000,19.000000) -- (5.000000,18.100000) -- (4.100000,19.000000) -- cycle;
	\fill[gray!80] (7.000000,19.000000) -- (6.100000,19.000000) -- (7.000000,19.900000) -- cycle;
	\fill[gray!80] (7.000000,19.000000) -- (7.000000,19.900000) -- (7.900000,19.000000) -- cycle;
	\fill[gray!80] (7.000000,19.000000) -- (7.900000,19.000000) -- (7.000000,18.100000) -- cycle;
	\fill[gray!80] (7.000000,19.000000) -- (7.000000,18.100000) -- (6.100000,19.000000) -- cycle;
	\fill[gray!80] (9.000000,19.000000) -- (8.100000,19.000000) -- (9.000000,19.900000) -- cycle;
	\fill[gray!80] (9.000000,19.000000) -- (9.000000,19.900000) -- (9.900000,19.000000) -- cycle;
	\fill[gray!80] (9.000000,19.000000) -- (9.900000,19.000000) -- (9.000000,18.100000) -- cycle;
	\fill[gray!80] (9.000000,19.000000) -- (9.000000,18.100000) -- (8.100000,19.000000) -- cycle;
	\fill[gray!80] (11.000000,19.000000) -- (10.100000,19.000000) -- (11.000000,19.900000) -- cycle;
	\fill[gray!80] (11.000000,19.000000) -- (11.000000,19.900000) -- (11.900000,19.000000) -- cycle;
	\fill[gray!80] (11.000000,19.000000) -- (11.900000,19.000000) -- (11.000000,18.100000) -- cycle;
	\fill[gray!80] (11.000000,19.000000) -- (11.000000,18.100000) -- (10.100000,19.000000) -- cycle;
	\fill[gray!80] (13.000000,19.000000) -- (12.100000,19.000000) -- (13.000000,19.900000) -- cycle;
	\fill[gray!80] (13.000000,19.000000) -- (13.000000,19.900000) -- (13.900000,19.000000) -- cycle;
	\fill[gray!80] (13.000000,19.000000) -- (13.900000,19.000000) -- (13.000000,18.100000) -- cycle;
	\fill[gray!80] (13.000000,19.000000) -- (13.000000,18.100000) -- (12.100000,19.000000) -- cycle;
	\fill[gray!25] (2.000000,18.000000) -- (2.000000,18.900000) -- (2.900000,18.000000) -- cycle;
	\fill[gray!25] (2.000000,18.000000) -- (2.900000,18.000000) -- (2.000000,17.100000) -- cycle;
	\fill[gray!25] (4.000000,18.000000) -- (3.100000,18.000000) -- (4.000000,18.900000) -- cycle;
	\fill[gray!25] (4.000000,18.000000) -- (4.000000,18.900000) -- (4.900000,18.000000) -- cycle;
	\fill[gray!25] (4.000000,18.000000) -- (4.900000,18.000000) -- (4.000000,17.100000) -- cycle;
	\fill[gray!25] (4.000000,18.000000) -- (4.000000,17.100000) -- (3.100000,18.000000) -- cycle;
	\fill[gray!25] (6.000000,18.000000) -- (5.100000,18.000000) -- (6.000000,18.900000) -- cycle;
	\fill[gray!25] (6.000000,18.000000) -- (6.000000,18.900000) -- (6.900000,18.000000) -- cycle;
	\fill[gray!25] (6.000000,18.000000) -- (6.900000,18.000000) -- (6.000000,17.100000) -- cycle;
	\fill[gray!25] (6.000000,18.000000) -- (6.000000,17.100000) -- (5.100000,18.000000) -- cycle;
	\fill[gray!25] (8.000000,18.000000) -- (7.100000,18.000000) -- (8.000000,18.900000) -- cycle;
	\fill[gray!25] (8.000000,18.000000) -- (8.000000,18.900000) -- (8.900000,18.000000) -- cycle;
	\fill[gray!25] (8.000000,18.000000) -- (8.900000,18.000000) -- (8.000000,17.100000) -- cycle;
	\fill[gray!25] (8.000000,18.000000) -- (8.000000,17.100000) -- (7.100000,18.000000) -- cycle;
	\fill[gray!25] (10.000000,18.000000) -- (9.100000,18.000000) -- (10.000000,18.900000) -- cycle;
	\fill[gray!25] (10.000000,18.000000) -- (10.000000,18.900000) -- (10.900000,18.000000) -- cycle;
	\fill[gray!25] (10.000000,18.000000) -- (10.900000,18.000000) -- (10.000000,17.100000) -- cycle;
	\fill[gray!25] (10.000000,18.000000) -- (10.000000,17.100000) -- (9.100000,18.000000) -- cycle;
	\fill[gray!25] (12.000000,18.000000) -- (11.100000,18.000000) -- (12.000000,18.900000) -- cycle;
	\fill[gray!25] (12.000000,18.000000) -- (12.000000,18.900000) -- (12.900000,18.000000) -- cycle;
	\fill[gray!25] (12.000000,18.000000) -- (12.900000,18.000000) -- (12.000000,17.100000) -- cycle;
	\fill[gray!25] (12.000000,18.000000) -- (12.000000,17.100000) -- (11.100000,18.000000) -- cycle;
	\fill[gray!25] (14.000000,18.000000) -- (13.100000,18.000000) -- (14.000000,18.900000) -- cycle;
	\fill[gray!25] (14.000000,18.000000) -- (14.000000,17.100000) -- (13.100000,18.000000) -- cycle;
	\fill[gray!80] (3.000000,17.000000) -- (2.100000,17.000000) -- (3.000000,17.900000) -- cycle;
	\fill[gray!80] (3.000000,17.000000) -- (3.000000,17.900000) -- (3.900000,17.000000) -- cycle;
	\fill[gray!80] (3.000000,17.000000) -- (3.900000,17.000000) -- (3.000000,16.100000) -- cycle;
	\fill[gray!80] (3.000000,17.000000) -- (3.000000,16.100000) -- (2.100000,17.000000) -- cycle;
	\fill[gray!80] (5.000000,17.000000) -- (4.100000,17.000000) -- (5.000000,17.900000) -- cycle;
	\fill[gray!80] (5.000000,17.000000) -- (5.000000,17.900000) -- (5.900000,17.000000) -- cycle;
	\fill[gray!80] (5.000000,17.000000) -- (5.900000,17.000000) -- (5.000000,16.100000) -- cycle;
	\fill[gray!80] (5.000000,17.000000) -- (5.000000,16.100000) -- (4.100000,17.000000) -- cycle;
	\fill[gray!80] (7.000000,17.000000) -- (6.100000,17.000000) -- (7.000000,17.900000) -- cycle;
	\fill[gray!80] (7.000000,17.000000) -- (7.000000,17.900000) -- (7.900000,17.000000) -- cycle;
	\fill[gray!80] (7.000000,17.000000) -- (7.900000,17.000000) -- (7.000000,16.100000) -- cycle;
	\fill[gray!80] (7.000000,17.000000) -- (7.000000,16.100000) -- (6.100000,17.000000) -- cycle;
	\fill[gray!80] (9.000000,17.000000) -- (8.100000,17.000000) -- (9.000000,17.900000) -- cycle;
	\fill[gray!80] (9.000000,17.000000) -- (9.000000,17.900000) -- (9.900000,17.000000) -- cycle;
	\fill[gray!80] (9.000000,17.000000) -- (9.900000,17.000000) -- (9.000000,16.100000) -- cycle;
	\fill[gray!80] (9.000000,17.000000) -- (9.000000,16.100000) -- (8.100000,17.000000) -- cycle;
	\fill[gray!80] (11.000000,17.000000) -- (10.100000,17.000000) -- (11.000000,17.900000) -- cycle;
	\fill[gray!80] (11.000000,17.000000) -- (11.000000,17.900000) -- (11.900000,17.000000) -- cycle;
	\fill[gray!80] (11.000000,17.000000) -- (11.900000,17.000000) -- (11.000000,16.100000) -- cycle;
	\fill[gray!80] (11.000000,17.000000) -- (11.000000,16.100000) -- (10.100000,17.000000) -- cycle;
	\fill[gray!80] (13.000000,17.000000) -- (12.100000,17.000000) -- (13.000000,17.900000) -- cycle;
	\fill[gray!80] (13.000000,17.000000) -- (13.000000,17.900000) -- (13.900000,17.000000) -- cycle;
	\fill[gray!80] (13.000000,17.000000) -- (13.900000,17.000000) -- (13.000000,16.100000) -- cycle;
	\fill[gray!80] (13.000000,17.000000) -- (13.000000,16.100000) -- (12.100000,17.000000) -- cycle;
	\fill[gray!25] (2.000000,16.000000) -- (2.000000,16.900000) -- (2.900000,16.000000) -- cycle;
	\fill[gray!25] (2.000000,16.000000) -- (2.900000,16.000000) -- (2.000000,15.100000) -- cycle;
	\fill[gray!25] (4.000000,16.000000) -- (3.100000,16.000000) -- (4.000000,16.900000) -- cycle;
	\fill[gray!25] (4.000000,16.000000) -- (4.000000,16.900000) -- (4.900000,16.000000) -- cycle;
	\fill[gray!25] (6.000000,16.000000) -- (5.100000,16.000000) -- (6.000000,16.900000) -- cycle;
	\fill[gray!25] (6.000000,16.000000) -- (6.000000,16.900000) -- (6.900000,16.000000) -- cycle;
	\fill[gray!25] (8.000000,16.000000) -- (7.100000,16.000000) -- (8.000000,16.900000) -- cycle;
	\fill[gray!25] (8.000000,16.000000) -- (8.000000,16.900000) -- (8.900000,16.000000) -- cycle;
	\fill[gray!25] (10.000000,16.000000) -- (9.100000,16.000000) -- (10.000000,16.900000) -- cycle;
	\fill[gray!25] (10.000000,16.000000) -- (10.000000,16.900000) -- (10.900000,16.000000) -- cycle;
	\fill[gray!25] (12.000000,16.000000) -- (11.100000,16.000000) -- (12.000000,16.900000) -- cycle;
	\fill[gray!25] (12.000000,16.000000) -- (12.000000,16.900000) -- (12.900000,16.000000) -- cycle;
	\fill[gray!25] (12.000000,16.000000) -- (12.900000,16.000000) -- (12.000000,15.100000) -- cycle;
	\fill[gray!25] (12.000000,16.000000) -- (12.000000,15.100000) -- (11.100000,16.000000) -- cycle;
	\fill[gray!25] (14.000000,16.000000) -- (13.100000,16.000000) -- (14.000000,16.900000) -- cycle;
	\fill[gray!25] (14.000000,16.000000) -- (14.000000,15.100000) -- (13.100000,16.000000) -- cycle;
	\fill[gray!80] (13.000000,15.000000) -- (12.100000,15.000000) -- (13.000000,15.900000) -- cycle;
	\fill[gray!80] (13.000000,15.000000) -- (13.000000,15.900000) -- (13.900000,15.000000) -- cycle;
	\fill[gray!80] (13.000000,15.000000) -- (13.900000,15.000000) -- (13.000000,14.100000) -- cycle;
	\fill[gray!80] (13.000000,15.000000) -- (13.000000,14.100000) -- (12.100000,15.000000) -- cycle;
	\fill[gray!25] (4.000000,14.000000) -- (4.900000,14.000000) -- (4.000000,13.100000) -- cycle;
	\fill[gray!25] (6.000000,14.000000) -- (6.900000,14.000000) -- (6.000000,13.100000) -- cycle;
	\fill[gray!25] (6.000000,14.000000) -- (6.000000,13.100000) -- (5.100000,14.000000) -- cycle;
	\fill[gray!25] (8.000000,14.000000) -- (8.900000,14.000000) -- (8.000000,13.100000) -- cycle;
	\fill[gray!25] (8.000000,14.000000) -- (8.000000,13.100000) -- (7.100000,14.000000) -- cycle;
	\fill[gray!25] (10.000000,14.000000) -- (10.000000,13.100000) -- (9.100000,14.000000) -- cycle;
	\fill[gray!25] (12.000000,14.000000) -- (12.000000,14.900000) -- (12.900000,14.000000) -- cycle;
	\fill[gray!25] (12.000000,14.000000) -- (12.900000,14.000000) -- (12.000000,13.100000) -- cycle;
	\fill[gray!80] (5.000000,13.000000) -- (4.100000,13.000000) -- (5.000000,13.900000) -- cycle;
	\fill[gray!80] (5.000000,13.000000) -- (5.000000,13.900000) -- (5.900000,13.000000) -- cycle;
	\fill[gray!80] (5.000000,13.000000) -- (5.900000,13.000000) -- (5.000000,12.100000) -- cycle;
	\fill[gray!80] (5.000000,13.000000) -- (5.000000,12.100000) -- (4.100000,13.000000) -- cycle;
	\fill[gray!80] (7.000000,13.000000) -- (6.100000,13.000000) -- (7.000000,13.900000) -- cycle;
	\fill[gray!80] (7.000000,13.000000) -- (7.000000,13.900000) -- (7.900000,13.000000) -- cycle;
	\fill[gray!80] (7.000000,13.000000) -- (7.900000,13.000000) -- (7.000000,12.100000) -- cycle;
	\fill[gray!80] (7.000000,13.000000) -- (7.000000,12.100000) -- (6.100000,13.000000) -- cycle;
	\fill[gray!80] (9.000000,13.000000) -- (8.100000,13.000000) -- (9.000000,13.900000) -- cycle;
	\fill[gray!80] (9.000000,13.000000) -- (9.000000,13.900000) -- (9.900000,13.000000) -- cycle;
	\fill[gray!80] (9.000000,13.000000) -- (9.900000,13.000000) -- (9.000000,12.100000) -- cycle;
	\fill[gray!80] (9.000000,13.000000) -- (9.000000,12.100000) -- (8.100000,13.000000) -- cycle;
	\fill[gray!80] (13.000000,13.000000) -- (12.100000,13.000000) -- (13.000000,13.900000) -- cycle;
	\fill[gray!80] (13.000000,13.000000) -- (13.000000,13.900000) -- (13.900000,13.000000) -- cycle;
	\fill[gray!80] (13.000000,13.000000) -- (13.900000,13.000000) -- (13.000000,12.100000) -- cycle;
	\fill[gray!80] (13.000000,13.000000) -- (13.000000,12.100000) -- (12.100000,13.000000) -- cycle;
	\fill[gray!25] (6.000000,12.000000) -- (5.100000,12.000000) -- (6.000000,12.900000) -- cycle;
	\fill[gray!25] (6.000000,12.000000) -- (6.000000,12.900000) -- (6.900000,12.000000) -- cycle;
	\fill[gray!25] (6.000000,12.000000) -- (6.900000,12.000000) -- (6.000000,11.100000) -- cycle;
	\fill[gray!25] (6.000000,12.000000) -- (6.000000,11.100000) -- (5.100000,12.000000) -- cycle;
	\fill[gray!25] (8.000000,12.000000) -- (7.100000,12.000000) -- (8.000000,12.900000) -- cycle;
	\fill[gray!25] (8.000000,12.000000) -- (8.000000,12.900000) -- (8.900000,12.000000) -- cycle;
	\fill[gray!25] (8.000000,12.000000) -- (8.900000,12.000000) -- (8.000000,11.100000) -- cycle;
	\fill[gray!25] (8.000000,12.000000) -- (8.000000,11.100000) -- (7.100000,12.000000) -- cycle;
	\fill[gray!25] (10.000000,12.000000) -- (9.100000,12.000000) -- (10.000000,12.900000) -- cycle;
	\fill[gray!25] (10.000000,12.000000) -- (10.000000,11.100000) -- (9.100000,12.000000) -- cycle;
	\fill[gray!25] (12.000000,12.000000) -- (12.000000,12.900000) -- (12.900000,12.000000) -- cycle;
	\fill[gray!25] (12.000000,12.000000) -- (12.900000,12.000000) -- (12.000000,11.100000) -- cycle;
	\fill[gray!25] (14.000000,12.000000) -- (13.100000,12.000000) -- (14.000000,12.900000) -- cycle;
	\fill[gray!25] (14.000000,12.000000) -- (14.000000,11.100000) -- (13.100000,12.000000) -- cycle;
	\fill[gray!80] (5.000000,11.000000) -- (4.100000,11.000000) -- (5.000000,11.900000) -- cycle;
	\fill[gray!80] (5.000000,11.000000) -- (5.000000,11.900000) -- (5.900000,11.000000) -- cycle;
	\fill[gray!80] (5.000000,11.000000) -- (5.900000,11.000000) -- (5.000000,10.100000) -- cycle;
	\fill[gray!80] (5.000000,11.000000) -- (5.000000,10.100000) -- (4.100000,11.000000) -- cycle;
	\fill[gray!80] (7.000000,11.000000) -- (6.100000,11.000000) -- (7.000000,11.900000) -- cycle;
	\fill[gray!80] (7.000000,11.000000) -- (7.000000,11.900000) -- (7.900000,11.000000) -- cycle;
	\fill[gray!80] (7.000000,11.000000) -- (7.900000,11.000000) -- (7.000000,10.100000) -- cycle;
	\fill[gray!80] (7.000000,11.000000) -- (7.000000,10.100000) -- (6.100000,11.000000) -- cycle;
	\fill[gray!80] (9.000000,11.000000) -- (8.100000,11.000000) -- (9.000000,11.900000) -- cycle;
	\fill[gray!80] (9.000000,11.000000) -- (9.000000,11.900000) -- (9.900000,11.000000) -- cycle;
	\fill[gray!80] (9.000000,11.000000) -- (9.900000,11.000000) -- (9.000000,10.100000) -- cycle;
	\fill[gray!80] (9.000000,11.000000) -- (9.000000,10.100000) -- (8.100000,11.000000) -- cycle;
	\fill[gray!80] (13.000000,11.000000) -- (12.100000,11.000000) -- (13.000000,11.900000) -- cycle;
	\fill[gray!80] (13.000000,11.000000) -- (13.000000,11.900000) -- (13.900000,11.000000) -- cycle;
	\fill[gray!80] (13.000000,11.000000) -- (13.900000,11.000000) -- (13.000000,10.100000) -- cycle;
	\fill[gray!80] (13.000000,11.000000) -- (13.000000,10.100000) -- (12.100000,11.000000) -- cycle;
	\fill[gray!25] (4.000000,10.000000) -- (4.000000,10.900000) -- (4.900000,10.000000) -- cycle;
	\fill[gray!25] (4.000000,10.000000) -- (4.900000,10.000000) -- (4.000000,9.100000) -- cycle;
	\fill[gray!25] (6.000000,10.000000) -- (5.100000,10.000000) -- (6.000000,10.900000) -- cycle;
	\fill[gray!25] (6.000000,10.000000) -- (6.000000,10.900000) -- (6.900000,10.000000) -- cycle;
	\fill[gray!25] (6.000000,10.000000) -- (6.900000,10.000000) -- (6.000000,9.100000) -- cycle;
	\fill[gray!25] (6.000000,10.000000) -- (6.000000,9.100000) -- (5.100000,10.000000) -- cycle;
	\fill[gray!25] (8.000000,10.000000) -- (7.100000,10.000000) -- (8.000000,10.900000) -- cycle;
	\fill[gray!25] (8.000000,10.000000) -- (8.000000,10.900000) -- (8.900000,10.000000) -- cycle;
	\fill[gray!25] (8.000000,10.000000) -- (8.900000,10.000000) -- (8.000000,9.100000) -- cycle;
	\fill[gray!25] (8.000000,10.000000) -- (8.000000,9.100000) -- (7.100000,10.000000) -- cycle;
	\fill[gray!25] (10.000000,10.000000) -- (9.100000,10.000000) -- (10.000000,10.900000) -- cycle;
	\fill[gray!25] (10.000000,10.000000) -- (10.000000,9.100000) -- (9.100000,10.000000) -- cycle;
	\fill[gray!25] (12.000000,10.000000) -- (12.000000,10.900000) -- (12.900000,10.000000) -- cycle;
	\fill[gray!25] (12.000000,10.000000) -- (12.900000,10.000000) -- (12.000000,9.100000) -- cycle;
	\fill[gray!25] (14.000000,10.000000) -- (13.100000,10.000000) -- (14.000000,10.900000) -- cycle;
	\fill[gray!25] (14.000000,10.000000) -- (14.000000,9.100000) -- (13.100000,10.000000) -- cycle;
	\fill[gray!80] (5.000000,9.000000) -- (4.100000,9.000000) -- (5.000000,9.900000) -- cycle;
	\fill[gray!80] (5.000000,9.000000) -- (5.000000,9.900000) -- (5.900000,9.000000) -- cycle;
	\fill[gray!80] (5.000000,9.000000) -- (5.900000,9.000000) -- (5.000000,8.100000) -- cycle;
	\fill[gray!80] (5.000000,9.000000) -- (5.000000,8.100000) -- (4.100000,9.000000) -- cycle;
	\fill[gray!80] (7.000000,9.000000) -- (6.100000,9.000000) -- (7.000000,9.900000) -- cycle;
	\fill[gray!80] (7.000000,9.000000) -- (7.000000,9.900000) -- (7.900000,9.000000) -- cycle;
	\fill[gray!80] (7.000000,9.000000) -- (7.900000,9.000000) -- (7.000000,8.100000) -- cycle;
	\fill[gray!80] (7.000000,9.000000) -- (7.000000,8.100000) -- (6.100000,9.000000) -- cycle;
	\fill[gray!80] (9.000000,9.000000) -- (8.100000,9.000000) -- (9.000000,9.900000) -- cycle;
	\fill[gray!80] (9.000000,9.000000) -- (9.000000,9.900000) -- (9.900000,9.000000) -- cycle;
	\fill[gray!80] (9.000000,9.000000) -- (9.900000,9.000000) -- (9.000000,8.100000) -- cycle;
	\fill[gray!80] (9.000000,9.000000) -- (9.000000,8.100000) -- (8.100000,9.000000) -- cycle;
	\fill[gray!80] (13.000000,9.000000) -- (12.100000,9.000000) -- (13.000000,9.900000) -- cycle;
	\fill[gray!80] (13.000000,9.000000) -- (13.000000,9.900000) -- (13.900000,9.000000) -- cycle;
	\fill[gray!80] (13.000000,9.000000) -- (13.900000,9.000000) -- (13.000000,8.100000) -- cycle;
	\fill[gray!80] (13.000000,9.000000) -- (13.000000,8.100000) -- (12.100000,9.000000) -- cycle;
	\fill[gray!25] (4.000000,8.000000) -- (4.000000,8.900000) -- (4.900000,8.000000) -- cycle;
	\fill[gray!25] (6.000000,8.000000) -- (5.100000,8.000000) -- (6.000000,8.900000) -- cycle;
	\fill[gray!25] (6.000000,8.000000) -- (6.000000,8.900000) -- (6.900000,8.000000) -- cycle;
	\fill[gray!25] (10.000000,8.000000) -- (9.100000,8.000000) -- (10.000000,8.900000) -- cycle;
	\fill[gray!25] (12.000000,8.000000) -- (12.000000,8.900000) -- (12.900000,8.000000) -- cycle;
	\fill[gray!25] (12.000000,8.000000) -- (12.900000,8.000000) -- (12.000000,7.100000) -- cycle;
	\fill[gray!80] (13.000000,7.000000) -- (12.100000,7.000000) -- (13.000000,7.900000) -- cycle;
	\fill[gray!80] (13.000000,7.000000) -- (13.000000,7.900000) -- (13.900000,7.000000) -- cycle;
	\fill[gray!80] (13.000000,7.000000) -- (13.900000,7.000000) -- (13.000000,6.100000) -- cycle;
	\fill[gray!80] (13.000000,7.000000) -- (13.000000,6.100000) -- (12.100000,7.000000) -- cycle;
	\fill[gray!25] (2.000000,6.000000) -- (2.000000,6.900000) -- (2.900000,6.000000) -- cycle;
	\fill[gray!25] (2.000000,6.000000) -- (2.900000,6.000000) -- (2.000000,5.100000) -- cycle;
	\fill[gray!25] (4.000000,6.000000) -- (4.900000,6.000000) -- (4.000000,5.100000) -- cycle;
	\fill[gray!25] (4.000000,6.000000) -- (4.000000,5.100000) -- (3.100000,6.000000) -- cycle;
	\fill[gray!25] (6.000000,6.000000) -- (6.900000,6.000000) -- (6.000000,5.100000) -- cycle;
	\fill[gray!25] (6.000000,6.000000) -- (6.000000,5.100000) -- (5.100000,6.000000) -- cycle;
	\fill[gray!25] (8.000000,6.000000) -- (8.900000,6.000000) -- (8.000000,5.100000) -- cycle;
	\fill[gray!25] (8.000000,6.000000) -- (8.000000,5.100000) -- (7.100000,6.000000) -- cycle;
	\fill[gray!25] (10.000000,6.000000) -- (10.900000,6.000000) -- (10.000000,5.100000) -- cycle;
	\fill[gray!25] (10.000000,6.000000) -- (10.000000,5.100000) -- (9.100000,6.000000) -- cycle;
	\fill[gray!25] (12.000000,6.000000) -- (11.100000,6.000000) -- (12.000000,6.900000) -- cycle;
	\fill[gray!25] (12.000000,6.000000) -- (12.000000,6.900000) -- (12.900000,6.000000) -- cycle;
	\fill[gray!25] (12.000000,6.000000) -- (12.900000,6.000000) -- (12.000000,5.100000) -- cycle;
	\fill[gray!25] (12.000000,6.000000) -- (12.000000,5.100000) -- (11.100000,6.000000) -- cycle;
	\fill[gray!25] (14.000000,6.000000) -- (13.100000,6.000000) -- (14.000000,6.900000) -- cycle;
	\fill[gray!25] (14.000000,6.000000) -- (14.000000,5.100000) -- (13.100000,6.000000) -- cycle;
	\fill[gray!80] (3.000000,5.000000) -- (2.100000,5.000000) -- (3.000000,5.900000) -- cycle;
	\fill[gray!80] (3.000000,5.000000) -- (3.000000,5.900000) -- (3.900000,5.000000) -- cycle;
	\fill[gray!80] (3.000000,5.000000) -- (3.900000,5.000000) -- (3.000000,4.100000) -- cycle;
	\fill[gray!80] (3.000000,5.000000) -- (3.000000,4.100000) -- (2.100000,5.000000) -- cycle;
	\fill[gray!80] (5.000000,5.000000) -- (4.100000,5.000000) -- (5.000000,5.900000) -- cycle;
	\fill[gray!80] (5.000000,5.000000) -- (5.000000,5.900000) -- (5.900000,5.000000) -- cycle;
	\fill[gray!80] (5.000000,5.000000) -- (5.900000,5.000000) -- (5.000000,4.100000) -- cycle;
	\fill[gray!80] (5.000000,5.000000) -- (5.000000,4.100000) -- (4.100000,5.000000) -- cycle;
	\fill[gray!80] (7.000000,5.000000) -- (6.100000,5.000000) -- (7.000000,5.900000) -- cycle;
	\fill[gray!80] (7.000000,5.000000) -- (7.000000,5.900000) -- (7.900000,5.000000) -- cycle;
	\fill[gray!80] (7.000000,5.000000) -- (7.900000,5.000000) -- (7.000000,4.100000) -- cycle;
	\fill[gray!80] (7.000000,5.000000) -- (7.000000,4.100000) -- (6.100000,5.000000) -- cycle;
	\fill[gray!80] (9.000000,5.000000) -- (8.100000,5.000000) -- (9.000000,5.900000) -- cycle;
	\fill[gray!80] (9.000000,5.000000) -- (9.000000,5.900000) -- (9.900000,5.000000) -- cycle;
	\fill[gray!80] (9.000000,5.000000) -- (9.900000,5.000000) -- (9.000000,4.100000) -- cycle;
	\fill[gray!80] (9.000000,5.000000) -- (9.000000,4.100000) -- (8.100000,5.000000) -- cycle;
	\fill[gray!80] (11.000000,5.000000) -- (10.100000,5.000000) -- (11.000000,5.900000) -- cycle;
	\fill[gray!80] (11.000000,5.000000) -- (11.000000,5.900000) -- (11.900000,5.000000) -- cycle;
	\fill[gray!80] (11.000000,5.000000) -- (11.900000,5.000000) -- (11.000000,4.100000) -- cycle;
	\fill[gray!80] (11.000000,5.000000) -- (11.000000,4.100000) -- (10.100000,5.000000) -- cycle;
	\fill[gray!80] (13.000000,5.000000) -- (12.100000,5.000000) -- (13.000000,5.900000) -- cycle;
	\fill[gray!80] (13.000000,5.000000) -- (13.000000,5.900000) -- (13.900000,5.000000) -- cycle;
	\fill[gray!80] (13.000000,5.000000) -- (13.900000,5.000000) -- (13.000000,4.100000) -- cycle;
	\fill[gray!80] (13.000000,5.000000) -- (13.000000,4.100000) -- (12.100000,5.000000) -- cycle;
	\fill[gray!25] (2.000000,4.000000) -- (2.000000,4.900000) -- (2.900000,4.000000) -- cycle;
	\fill[gray!25] (2.000000,4.000000) -- (2.900000,4.000000) -- (2.000000,3.100000) -- cycle;
	\fill[gray!25] (4.000000,4.000000) -- (3.100000,4.000000) -- (4.000000,4.900000) -- cycle;
	\fill[gray!25] (4.000000,4.000000) -- (4.000000,4.900000) -- (4.900000,4.000000) -- cycle;
	\fill[gray!25] (4.000000,4.000000) -- (4.900000,4.000000) -- (4.000000,3.100000) -- cycle;
	\fill[gray!25] (4.000000,4.000000) -- (4.000000,3.100000) -- (3.100000,4.000000) -- cycle;
	\fill[gray!25] (6.000000,4.000000) -- (5.100000,4.000000) -- (6.000000,4.900000) -- cycle;
	\fill[gray!25] (6.000000,4.000000) -- (6.000000,4.900000) -- (6.900000,4.000000) -- cycle;
	\fill[gray!25] (6.000000,4.000000) -- (6.900000,4.000000) -- (6.000000,3.100000) -- cycle;
	\fill[gray!25] (6.000000,4.000000) -- (6.000000,3.100000) -- (5.100000,4.000000) -- cycle;
	\fill[gray!25] (8.000000,4.000000) -- (7.100000,4.000000) -- (8.000000,4.900000) -- cycle;
	\fill[gray!25] (8.000000,4.000000) -- (8.000000,4.900000) -- (8.900000,4.000000) -- cycle;
	\fill[gray!25] (8.000000,4.000000) -- (8.900000,4.000000) -- (8.000000,3.100000) -- cycle;
	\fill[gray!25] (8.000000,4.000000) -- (8.000000,3.100000) -- (7.100000,4.000000) -- cycle;
	\fill[gray!25] (10.000000,4.000000) -- (9.100000,4.000000) -- (10.000000,4.900000) -- cycle;
	\fill[gray!25] (10.000000,4.000000) -- (10.000000,4.900000) -- (10.900000,4.000000) -- cycle;
	\fill[gray!25] (10.000000,4.000000) -- (10.900000,4.000000) -- (10.000000,3.100000) -- cycle;
	\fill[gray!25] (10.000000,4.000000) -- (10.000000,3.100000) -- (9.100000,4.000000) -- cycle;
	\fill[gray!25] (12.000000,4.000000) -- (11.100000,4.000000) -- (12.000000,4.900000) -- cycle;
	\fill[gray!25] (12.000000,4.000000) -- (12.000000,4.900000) -- (12.900000,4.000000) -- cycle;
	\fill[gray!25] (12.000000,4.000000) -- (12.900000,4.000000) -- (12.000000,3.100000) -- cycle;
	\fill[gray!25] (12.000000,4.000000) -- (12.000000,3.100000) -- (11.100000,4.000000) -- cycle;
	\fill[gray!25] (14.000000,4.000000) -- (13.100000,4.000000) -- (14.000000,4.900000) -- cycle;
	\fill[gray!25] (14.000000,4.000000) -- (14.000000,3.100000) -- (13.100000,4.000000) -- cycle;
	\fill[gray!80] (3.000000,3.000000) -- (2.100000,3.000000) -- (3.000000,3.900000) -- cycle;
	\fill[gray!80] (3.000000,3.000000) -- (3.000000,3.900000) -- (3.900000,3.000000) -- cycle;
	\fill[gray!80] (3.000000,3.000000) -- (3.900000,3.000000) -- (3.000000,2.100000) -- cycle;
	\fill[gray!80] (3.000000,3.000000) -- (3.000000,2.100000) -- (2.100000,3.000000) -- cycle;
	\fill[gray!80] (5.000000,3.000000) -- (4.100000,3.000000) -- (5.000000,3.900000) -- cycle;
	\fill[gray!80] (5.000000,3.000000) -- (5.000000,3.900000) -- (5.900000,3.000000) -- cycle;
	\fill[gray!80] (5.000000,3.000000) -- (5.900000,3.000000) -- (5.000000,2.100000) -- cycle;
	\fill[gray!80] (5.000000,3.000000) -- (5.000000,2.100000) -- (4.100000,3.000000) -- cycle;
	\fill[gray!80] (7.000000,3.000000) -- (6.100000,3.000000) -- (7.000000,3.900000) -- cycle;
	\fill[gray!80] (7.000000,3.000000) -- (7.000000,3.900000) -- (7.900000,3.000000) -- cycle;
	\fill[gray!80] (7.000000,3.000000) -- (7.900000,3.000000) -- (7.000000,2.100000) -- cycle;
	\fill[gray!80] (7.000000,3.000000) -- (7.000000,2.100000) -- (6.100000,3.000000) -- cycle;
	\fill[gray!80] (9.000000,3.000000) -- (8.100000,3.000000) -- (9.000000,3.900000) -- cycle;
	\fill[gray!80] (9.000000,3.000000) -- (9.000000,3.900000) -- (9.900000,3.000000) -- cycle;
	\fill[gray!80] (9.000000,3.000000) -- (9.900000,3.000000) -- (9.000000,2.100000) -- cycle;
	\fill[gray!80] (9.000000,3.000000) -- (9.000000,2.100000) -- (8.100000,3.000000) -- cycle;
	\fill[gray!80] (11.000000,3.000000) -- (10.100000,3.000000) -- (11.000000,3.900000) -- cycle;
	\fill[gray!80] (11.000000,3.000000) -- (11.000000,3.900000) -- (11.900000,3.000000) -- cycle;
	\fill[gray!80] (11.000000,3.000000) -- (11.900000,3.000000) -- (11.000000,2.100000) -- cycle;
	\fill[gray!80] (11.000000,3.000000) -- (11.000000,2.100000) -- (10.100000,3.000000) -- cycle;
	\fill[gray!80] (13.000000,3.000000) -- (12.100000,3.000000) -- (13.000000,3.900000) -- cycle;
	\fill[gray!80] (13.000000,3.000000) -- (13.000000,3.900000) -- (13.900000,3.000000) -- cycle;
	\fill[gray!80] (13.000000,3.000000) -- (13.900000,3.000000) -- (13.000000,2.100000) -- cycle;
	\fill[gray!80] (13.000000,3.000000) -- (13.000000,2.100000) -- (12.100000,3.000000) -- cycle;
	\fill[gray!25] (4.000000,2.000000) -- (3.100000,2.000000) -- (4.000000,2.900000) -- cycle;
	\fill[gray!25] (4.000000,2.000000) -- (4.000000,2.900000) -- (4.900000,2.000000) -- cycle;
	\fill[gray!25] (6.000000,2.000000) -- (5.100000,2.000000) -- (6.000000,2.900000) -- cycle;
	\fill[gray!25] (6.000000,2.000000) -- (6.000000,2.900000) -- (6.900000,2.000000) -- cycle;
	\fill[gray!25] (10.000000,2.000000) -- (9.100000,2.000000) -- (10.000000,2.900000) -- cycle;
	\fill[gray!25] (10.000000,2.000000) -- (10.000000,2.900000) -- (10.900000,2.000000) -- cycle;
	\fill[gray!25] (12.000000,2.000000) -- (11.100000,2.000000) -- (12.000000,2.900000) -- cycle;
	\fill[gray!25] (12.000000,2.000000) -- (12.000000,2.900000) -- (12.900000,2.000000) -- cycle;
	\draw[line width=1pt] (3.000000,20.000000) circle (0.100000);
	\draw[line width=1pt,fill=black] (4.000000,20.000000) circle (0.100000);
	\draw[line width=1pt] (5.000000,20.000000) circle (0.100000);
	\draw[line width=1pt,fill=black] (6.000000,20.000000) circle (0.100000);
	\draw[line width=1pt] (7.000000,20.000000) circle (0.100000);
	\draw[line width=1pt] (9.000000,20.000000) circle (0.100000);
	\draw[line width=1pt,fill=black] (10.000000,20.000000) circle (0.100000);
	\draw[line width=1pt] (11.000000,20.000000) circle (0.100000);
	\draw[line width=1pt,fill=black] (12.000000,20.000000) circle (0.100000);
	\draw[line width=1pt] (13.000000,20.000000) circle (0.100000);
	\draw[line width=1pt] (2.000000,19.000000) circle (0.100000);
	\draw[line width=1pt,fill=black] (3.000000,19.000000) circle (0.100000);
	\draw[line width=1pt] (4.000000,19.000000) circle (0.100000);
	\draw[line width=1pt,fill=black] (5.000000,19.000000) circle (0.100000);
	\draw[line width=1pt] (6.000000,19.000000) circle (0.100000);
	\draw[line width=1pt,fill=black] (7.000000,19.000000) circle (0.100000);
	\draw[line width=1pt] (8.000000,19.000000) circle (0.100000);
	\draw[line width=1pt,fill=black] (9.000000,19.000000) circle (0.100000);
	\draw[line width=1pt] (10.000000,19.000000) circle (0.100000);
	\draw[line width=1pt,fill=black] (11.000000,19.000000) circle (0.100000);
	\draw[line width=1pt] (12.000000,19.000000) circle (0.100000);
	\draw[line width=1pt,fill=black] (13.000000,19.000000) circle (0.100000);
	\draw[line width=1pt] (14.000000,19.000000) circle (0.100000);
	\draw[line width=1pt,fill=black] (2.000000,18.000000) circle (0.100000);
	\draw[line width=1pt] (3.000000,18.000000) circle (0.100000);
	\draw[line width=1pt,fill=black] (4.000000,18.000000) circle (0.100000);
	\draw[line width=1pt] (5.000000,18.000000) circle (0.100000);
	\draw[line width=1pt,fill=black] (6.000000,18.000000) circle (0.100000);
	\draw[line width=1pt] (7.000000,18.000000) circle (0.100000);
	\draw[line width=1pt,fill=black] (8.000000,18.000000) circle (0.100000);
	\draw[line width=1pt] (9.000000,18.000000) circle (0.100000);
	\draw[line width=1pt,fill=black] (10.000000,18.000000) circle (0.100000);
	\draw[line width=1pt] (11.000000,18.000000) circle (0.100000);
	\draw[line width=1pt,fill=black] (12.000000,18.000000) circle (0.100000);
	\draw[line width=1pt] (13.000000,18.000000) circle (0.100000);
	\draw[line width=1pt,fill=black] (14.000000,18.000000) circle (0.100000);
	\draw[line width=1pt] (2.000000,17.000000) circle (0.100000);
	\draw[line width=1pt,fill=black] (3.000000,17.000000) circle (0.100000);
	\draw[line width=1pt] (4.000000,17.000000) circle (0.100000);
	\draw[line width=1pt,fill=black] (5.000000,17.000000) circle (0.100000);
	\draw[line width=1pt] (6.000000,17.000000) circle (0.100000);
	\draw[line width=1pt,fill=black] (7.000000,17.000000) circle (0.100000);
	\draw[line width=1pt] (8.000000,17.000000) circle (0.100000);
	\draw[line width=1pt,fill=black] (9.000000,17.000000) circle (0.100000);
	\draw[line width=1pt] (10.000000,17.000000) circle (0.100000);
	\draw[line width=1pt,fill=black] (11.000000,17.000000) circle (0.100000);
	\draw[line width=1pt] (12.000000,17.000000) circle (0.100000);
	\draw[line width=1pt,fill=black] (13.000000,17.000000) circle (0.100000);
	\draw[line width=1pt] (14.000000,17.000000) circle (0.100000);
	\draw[line width=1pt,fill=black] (2.000000,16.000000) circle (0.100000);
	\draw[line width=1pt] (3.000000,16.000000) circle (0.100000);
	\draw[line width=1pt,fill=black] (4.000000,16.000000) circle (0.100000);
	\draw[line width=1pt] (5.000000,16.000000) circle (0.100000);
	\draw[line width=1pt,fill=black] (6.000000,16.000000) circle (0.100000);
	\draw[line width=1pt] (7.000000,16.000000) circle (0.100000);
	\draw[line width=1pt,fill=black] (8.000000,16.000000) circle (0.100000);
	\draw[line width=1pt] (9.000000,16.000000) circle (0.100000);
	\draw[line width=1pt,fill=black] (10.000000,16.000000) circle (0.100000);
	\draw[line width=1pt] (11.000000,16.000000) circle (0.100000);
	\draw[line width=1pt,fill=black] (12.000000,16.000000) circle (0.100000);
	\draw[line width=1pt] (13.000000,16.000000) circle (0.100000);
	\draw[line width=1pt,fill=black] (14.000000,16.000000) circle (0.100000);
	\draw[line width=1pt] (2.000000,15.000000) circle (0.100000);
	\draw[line width=1pt] (12.000000,15.000000) circle (0.100000);
	\draw[line width=1pt,fill=black] (13.000000,15.000000) circle (0.100000);
	\draw[line width=1pt] (14.000000,15.000000) circle (0.100000);
	\draw[line width=1pt,fill=black] (4.000000,14.000000) circle (0.100000);
	\draw[line width=1pt] (5.000000,14.000000) circle (0.100000);
	\draw[line width=1pt,fill=black] (6.000000,14.000000) circle (0.100000);
	\draw[line width=1pt] (7.000000,14.000000) circle (0.100000);
	\draw[line width=1pt,fill=black] (8.000000,14.000000) circle (0.100000);
	\draw[line width=1pt] (9.000000,14.000000) circle (0.100000);
	\draw[line width=1pt,fill=black] (10.000000,14.000000) circle (0.100000);
	\draw[line width=1pt,fill=black] (12.000000,14.000000) circle (0.100000);
	\draw[line width=1pt] (13.000000,14.000000) circle (0.100000);
	\draw[line width=1pt] (2.000000,13.000000) circle (0.100000);
	\draw[line width=1pt] (4.000000,13.000000) circle (0.100000);
	\draw[line width=1pt,fill=black] (5.000000,13.000000) circle (0.100000);
	\draw[line width=1pt] (6.000000,13.000000) circle (0.100000);
	\draw[line width=1pt,fill=black] (7.000000,13.000000) circle (0.100000);
	\draw[line width=1pt] (8.000000,13.000000) circle (0.100000);
	\draw[line width=1pt,fill=black] (9.000000,13.000000) circle (0.100000);
	\draw[line width=1pt] (10.000000,13.000000) circle (0.100000);
	\draw[line width=1pt] (12.000000,13.000000) circle (0.100000);
	\draw[line width=1pt,fill=black] (13.000000,13.000000) circle (0.100000);
	\draw[line width=1pt] (14.000000,13.000000) circle (0.100000);
	\draw[line width=1pt,fill=black] (2.000000,12.000000) circle (0.100000);
	\draw[line width=1pt] (5.000000,12.000000) circle (0.100000);
	\draw[line width=1pt,fill=black] (6.000000,12.000000) circle (0.100000);
	\draw[line width=1pt] (7.000000,12.000000) circle (0.100000);
	\draw[line width=1pt,fill=black] (8.000000,12.000000) circle (0.100000);
	\draw[line width=1pt] (9.000000,12.000000) circle (0.100000);
	\draw[line width=1pt,fill=black] (10.000000,12.000000) circle (0.100000);
	\draw[line width=1pt,fill=black] (12.000000,12.000000) circle (0.100000);
	\draw[line width=1pt] (13.000000,12.000000) circle (0.100000);
	\draw[line width=1pt,fill=black] (14.000000,12.000000) circle (0.100000);
	\draw[line width=1pt] (2.000000,11.000000) circle (0.100000);
	\draw[line width=1pt] (4.000000,11.000000) circle (0.100000);
	\draw[line width=1pt,fill=black] (5.000000,11.000000) circle (0.100000);
	\draw[line width=1pt] (6.000000,11.000000) circle (0.100000);
	\draw[line width=1pt,fill=black] (7.000000,11.000000) circle (0.100000);
	\draw[line width=1pt] (8.000000,11.000000) circle (0.100000);
	\draw[line width=1pt,fill=black] (9.000000,11.000000) circle (0.100000);
	\draw (10.000000,11.000000) node [above right] {$|+\rangle$};
grestore
	\draw[line width=1pt] (10.000000,11.000000) circle (0.100000);
	\draw[line width=1pt] (12.000000,11.000000) circle (0.100000);
	\draw[line width=1pt,fill=black] (13.000000,11.000000) circle (0.100000);
	\draw[line width=1pt] (14.000000,11.000000) circle (0.100000);
	\draw[line width=1pt,fill=black] (2.000000,10.000000) circle (0.100000);
	\draw[line width=1pt,fill=black] (4.000000,10.000000) circle (0.100000);
	\draw[line width=1pt] (5.000000,10.000000) circle (0.100000);
	\draw[line width=1pt,fill=black] (6.000000,10.000000) circle (0.100000);
	\draw[line width=1pt] (7.000000,10.000000) circle (0.100000);
	\draw[line width=1pt,fill=black] (8.000000,10.000000) circle (0.100000);
	\draw[line width=1pt] (9.000000,10.000000) circle (0.100000);
	\draw[line width=1pt,fill=black] (10.000000,10.000000) circle (0.100000);
	\draw[line width=1pt,fill=black] (12.000000,10.000000) circle (0.100000);
	\draw[line width=1pt] (13.000000,10.000000) circle (0.100000);
	\draw[line width=1pt,fill=black] (14.000000,10.000000) circle (0.100000);
	\draw[line width=1pt] (2.000000,9.000000) circle (0.100000);
	\draw[line width=1pt] (4.000000,9.000000) circle (0.100000);
	\draw[line width=1pt,fill=black] (5.000000,9.000000) circle (0.100000);
	\draw[line width=1pt] (6.000000,9.000000) circle (0.100000);
	\draw[line width=1pt,fill=black] (7.000000,9.000000) circle (0.100000);
	\draw[line width=1pt] (8.000000,9.000000) circle (0.100000);
	\draw[line width=1pt,fill=black] (9.000000,9.000000) circle (0.100000);
	\draw (10.000000,9.000000) node [above right] {$|+\rangle$};
grestore
	\draw[line width=1pt] (10.000000,9.000000) circle (0.100000);
	\draw[line width=1pt] (12.000000,9.000000) circle (0.100000);
	\draw[line width=1pt,fill=black] (13.000000,9.000000) circle (0.100000);
	\draw[line width=1pt] (14.000000,9.000000) circle (0.100000);
	\draw[line width=1pt,fill=black] (4.000000,8.000000) circle (0.100000);
	\draw[line width=1pt] (5.000000,8.000000) circle (0.100000);
	\draw[line width=1pt,fill=black] (6.000000,8.000000) circle (0.100000);
	\draw[line width=1pt] (7.000000,8.000000) circle (0.100000);
	\draw (9.000000,8.000000) node [above right] {$|+\rangle$};
grestore
	\draw[line width=1pt] (9.000000,8.000000) circle (0.100000);
	\draw[line width=1pt,fill=black] (10.000000,8.000000) circle (0.100000);
	\draw[line width=1pt,fill=black] (12.000000,8.000000) circle (0.100000);
	\draw[line width=1pt] (13.000000,8.000000) circle (0.100000);
	\draw[line width=1pt] (2.000000,7.000000) circle (0.100000);
	\draw[line width=1pt] (12.000000,7.000000) circle (0.100000);
	\draw[line width=1pt,fill=black] (13.000000,7.000000) circle (0.100000);
	\draw[line width=1pt] (14.000000,7.000000) circle (0.100000);
	\draw[line width=1pt,fill=black] (2.000000,6.000000) circle (0.100000);
	\draw[line width=1pt] (3.000000,6.000000) circle (0.100000);
	\draw[line width=1pt,fill=black] (4.000000,6.000000) circle (0.100000);
	\draw[line width=1pt] (5.000000,6.000000) circle (0.100000);
	\draw[line width=1pt,fill=black] (6.000000,6.000000) circle (0.100000);
	\draw[line width=1pt] (7.000000,6.000000) circle (0.100000);
	\draw[line width=1pt,fill=black] (8.000000,6.000000) circle (0.100000);
	\draw[line width=1pt] (9.000000,6.000000) circle (0.100000);
	\draw[line width=1pt,fill=black] (10.000000,6.000000) circle (0.100000);
	\draw[line width=1pt] (11.000000,6.000000) circle (0.100000);
	\draw[line width=1pt,fill=black] (12.000000,6.000000) circle (0.100000);
	\draw[line width=1pt] (13.000000,6.000000) circle (0.100000);
	\draw[line width=1pt,fill=black] (14.000000,6.000000) circle (0.100000);
	\draw[line width=1pt] (2.000000,5.000000) circle (0.100000);
	\draw[line width=1pt,fill=black] (3.000000,5.000000) circle (0.100000);
	\draw[line width=1pt] (4.000000,5.000000) circle (0.100000);
	\draw[line width=1pt,fill=black] (5.000000,5.000000) circle (0.100000);
	\draw[line width=1pt] (6.000000,5.000000) circle (0.100000);
	\draw[line width=1pt,fill=black] (7.000000,5.000000) circle (0.100000);
	\draw[line width=1pt] (8.000000,5.000000) circle (0.100000);
	\draw[line width=1pt,fill=black] (9.000000,5.000000) circle (0.100000);
	\draw[line width=1pt] (10.000000,5.000000) circle (0.100000);
	\draw[line width=1pt,fill=black] (11.000000,5.000000) circle (0.100000);
	\draw[line width=1pt] (12.000000,5.000000) circle (0.100000);
	\draw[line width=1pt,fill=black] (13.000000,5.000000) circle (0.100000);
	\draw[line width=1pt] (14.000000,5.000000) circle (0.100000);
	\draw[line width=1pt,fill=black] (2.000000,4.000000) circle (0.100000);
	\draw[line width=1pt] (3.000000,4.000000) circle (0.100000);
	\draw[line width=1pt,fill=black] (4.000000,4.000000) circle (0.100000);
	\draw[line width=1pt] (5.000000,4.000000) circle (0.100000);
	\draw[line width=1pt,fill=black] (6.000000,4.000000) circle (0.100000);
	\draw[line width=1pt] (7.000000,4.000000) circle (0.100000);
	\draw[line width=1pt,fill=black] (8.000000,4.000000) circle (0.100000);
	\draw[line width=1pt] (9.000000,4.000000) circle (0.100000);
	\draw[line width=1pt,fill=black] (10.000000,4.000000) circle (0.100000);
	\draw[line width=1pt] (11.000000,4.000000) circle (0.100000);
	\draw[line width=1pt,fill=black] (12.000000,4.000000) circle (0.100000);
	\draw[line width=1pt] (13.000000,4.000000) circle (0.100000);
	\draw[line width=1pt,fill=black] (14.000000,4.000000) circle (0.100000);
	\draw[line width=1pt] (2.000000,3.000000) circle (0.100000);
	\draw[line width=1pt,fill=black] (3.000000,3.000000) circle (0.100000);
	\draw[line width=1pt] (4.000000,3.000000) circle (0.100000);
	\draw[line width=1pt,fill=black] (5.000000,3.000000) circle (0.100000);
	\draw[line width=1pt] (6.000000,3.000000) circle (0.100000);
	\draw[line width=1pt,fill=black] (7.000000,3.000000) circle (0.100000);
	\draw[line width=1pt] (8.000000,3.000000) circle (0.100000);
	\draw[line width=1pt,fill=black] (9.000000,3.000000) circle (0.100000);
	\draw[line width=1pt] (10.000000,3.000000) circle (0.100000);
	\draw[line width=1pt,fill=black] (11.000000,3.000000) circle (0.100000);
	\draw[line width=1pt] (12.000000,3.000000) circle (0.100000);
	\draw[line width=1pt,fill=black] (13.000000,3.000000) circle (0.100000);
	\draw[line width=1pt] (14.000000,3.000000) circle (0.100000);
	\draw[line width=1pt] (3.000000,2.000000) circle (0.100000);
	\draw[line width=1pt,fill=black] (4.000000,2.000000) circle (0.100000);
	\draw[line width=1pt] (5.000000,2.000000) circle (0.100000);
	\draw[line width=1pt,fill=black] (6.000000,2.000000) circle (0.100000);
	\draw[line width=1pt] (7.000000,2.000000) circle (0.100000);
	\draw[line width=1pt] (9.000000,2.000000) circle (0.100000);
	\draw[line width=1pt,fill=black] (10.000000,2.000000) circle (0.100000);
	\draw[line width=1pt] (11.000000,2.000000) circle (0.100000);
	\draw[line width=1pt,fill=black] (12.000000,2.000000) circle (0.100000);
	\draw[line width=1pt] (13.000000,2.000000) circle (0.100000);
\end{tikzpicture}

%% file: 008ex2.tex
\begin{tikzpicture}[x=0.030000\linewidth,y=0.030000\linewidth]
	\fill[gray!25] (4.000000,20.000000) -- (4.900000,20.000000) -- (4.000000,19.100000) -- cycle;
	\fill[gray!25] (4.000000,20.000000) -- (4.000000,19.100000) -- (3.100000,20.000000) -- cycle;
	\fill[gray!25] (6.000000,20.000000) -- (6.900000,20.000000) -- (6.000000,19.100000) -- cycle;
	\fill[gray!25] (6.000000,20.000000) -- (6.000000,19.100000) -- (5.100000,20.000000) -- cycle;
	\fill[gray!25] (10.000000,20.000000) -- (10.900000,20.000000) -- (10.000000,19.100000) -- cycle;
	\fill[gray!25] (10.000000,20.000000) -- (10.000000,19.100000) -- (9.100000,20.000000) -- cycle;
	\fill[gray!25] (12.000000,20.000000) -- (12.900000,20.000000) -- (12.000000,19.100000) -- cycle;
	\fill[gray!25] (12.000000,20.000000) -- (12.000000,19.100000) -- (11.100000,20.000000) -- cycle;
	\fill[gray!80] (3.000000,19.000000) -- (2.100000,19.000000) -- (3.000000,19.900000) -- cycle;
	\fill[gray!80] (3.000000,19.000000) -- (3.000000,19.900000) -- (3.900000,19.000000) -- cycle;
	\fill[gray!80] (3.000000,19.000000) -- (3.900000,19.000000) -- (3.000000,18.100000) -- cycle;
	\fill[gray!80] (3.000000,19.000000) -- (3.000000,18.100000) -- (2.100000,19.000000) -- cycle;
	\fill[gray!80] (5.000000,19.000000) -- (4.100000,19.000000) -- (5.000000,19.900000) -- cycle;
	\fill[gray!80] (5.000000,19.000000) -- (5.000000,19.900000) -- (5.900000,19.000000) -- cycle;
	\fill[gray!80] (5.000000,19.000000) -- (5.900000,19.000000) -- (5.000000,18.100000) -- cycle;
	\fill[gray!80] (5.000000,19.000000) -- (5.000000,18.100000) -- (4.100000,19.000000) -- cycle;
	\fill[gray!80] (7.000000,19.000000) -- (6.100000,19.000000) -- (7.000000,19.900000) -- cycle;
	\fill[gray!80] (7.000000,19.000000) -- (7.000000,19.900000) -- (7.900000,19.000000) -- cycle;
	\fill[gray!80] (7.000000,19.000000) -- (7.900000,19.000000) -- (7.000000,18.100000) -- cycle;
	\fill[gray!80] (7.000000,19.000000) -- (7.000000,18.100000) -- (6.100000,19.000000) -- cycle;
	\fill[gray!80] (9.000000,19.000000) -- (8.100000,19.000000) -- (9.000000,19.900000) -- cycle;
	\fill[gray!80] (9.000000,19.000000) -- (9.000000,19.900000) -- (9.900000,19.000000) -- cycle;
	\fill[gray!80] (9.000000,19.000000) -- (9.900000,19.000000) -- (9.000000,18.100000) -- cycle;
	\fill[gray!80] (9.000000,19.000000) -- (9.000000,18.100000) -- (8.100000,19.000000) -- cycle;
	\fill[gray!80] (11.000000,19.000000) -- (10.100000,19.000000) -- (11.000000,19.900000) -- cycle;
	\fill[gray!80] (11.000000,19.000000) -- (11.000000,19.900000) -- (11.900000,19.000000) -- cycle;
	\fill[gray!80] (11.000000,19.000000) -- (11.900000,19.000000) -- (11.000000,18.100000) -- cycle;
	\fill[gray!80] (11.000000,19.000000) -- (11.000000,18.100000) -- (10.100000,19.000000) -- cycle;
	\fill[gray!80] (13.000000,19.000000) -- (12.100000,19.000000) -- (13.000000,19.900000) -- cycle;
	\fill[gray!80] (13.000000,19.000000) -- (13.000000,19.900000) -- (13.900000,19.000000) -- cycle;
	\fill[gray!80] (13.000000,19.000000) -- (13.900000,19.000000) -- (13.000000,18.100000) -- cycle;
	\fill[gray!80] (13.000000,19.000000) -- (13.000000,18.100000) -- (12.100000,19.000000) -- cycle;
	\fill[gray!25] (2.000000,18.000000) -- (2.000000,18.900000) -- (2.900000,18.000000) -- cycle;
	\fill[gray!25] (2.000000,18.000000) -- (2.900000,18.000000) -- (2.000000,17.100000) -- cycle;
	\fill[gray!25] (4.000000,18.000000) -- (3.100000,18.000000) -- (4.000000,18.900000) -- cycle;
	\fill[gray!25] (4.000000,18.000000) -- (4.000000,18.900000) -- (4.900000,18.000000) -- cycle;
	\fill[gray!25] (4.000000,18.000000) -- (4.900000,18.000000) -- (4.000000,17.100000) -- cycle;
	\fill[gray!25] (4.000000,18.000000) -- (4.000000,17.100000) -- (3.100000,18.000000) -- cycle;
	\fill[gray!25] (6.000000,18.000000) -- (5.100000,18.000000) -- (6.000000,18.900000) -- cycle;
	\fill[gray!25] (6.000000,18.000000) -- (6.000000,18.900000) -- (6.900000,18.000000) -- cycle;
	\fill[gray!25] (6.000000,18.000000) -- (6.900000,18.000000) -- (6.000000,17.100000) -- cycle;
	\fill[gray!25] (6.000000,18.000000) -- (6.000000,17.100000) -- (5.100000,18.000000) -- cycle;
	\fill[gray!25] (8.000000,18.000000) -- (7.100000,18.000000) -- (8.000000,18.900000) -- cycle;
	\fill[gray!25] (8.000000,18.000000) -- (8.000000,18.900000) -- (8.900000,18.000000) -- cycle;
	\fill[gray!25] (8.000000,18.000000) -- (8.900000,18.000000) -- (8.000000,17.100000) -- cycle;
	\fill[gray!25] (8.000000,18.000000) -- (8.000000,17.100000) -- (7.100000,18.000000) -- cycle;
	\fill[gray!25] (10.000000,18.000000) -- (9.100000,18.000000) -- (10.000000,18.900000) -- cycle;
	\fill[gray!25] (10.000000,18.000000) -- (10.000000,18.900000) -- (10.900000,18.000000) -- cycle;
	\fill[gray!25] (10.000000,18.000000) -- (10.900000,18.000000) -- (10.000000,17.100000) -- cycle;
	\fill[gray!25] (10.000000,18.000000) -- (10.000000,17.100000) -- (9.100000,18.000000) -- cycle;
	\fill[gray!25] (12.000000,18.000000) -- (11.100000,18.000000) -- (12.000000,18.900000) -- cycle;
	\fill[gray!25] (12.000000,18.000000) -- (12.000000,18.900000) -- (12.900000,18.000000) -- cycle;
	\fill[gray!25] (12.000000,18.000000) -- (12.900000,18.000000) -- (12.000000,17.100000) -- cycle;
	\fill[gray!25] (12.000000,18.000000) -- (12.000000,17.100000) -- (11.100000,18.000000) -- cycle;
	\fill[gray!25] (14.000000,18.000000) -- (13.100000,18.000000) -- (14.000000,18.900000) -- cycle;
	\fill[gray!25] (14.000000,18.000000) -- (14.000000,17.100000) -- (13.100000,18.000000) -- cycle;
	\fill[gray!80] (3.000000,17.000000) -- (2.100000,17.000000) -- (3.000000,17.900000) -- cycle;
	\fill[gray!80] (3.000000,17.000000) -- (3.000000,17.900000) -- (3.900000,17.000000) -- cycle;
	\fill[gray!80] (3.000000,17.000000) -- (3.900000,17.000000) -- (3.000000,16.100000) -- cycle;
	\fill[gray!80] (3.000000,17.000000) -- (3.000000,16.100000) -- (2.100000,17.000000) -- cycle;
	\fill[gray!80] (5.000000,17.000000) -- (4.100000,17.000000) -- (5.000000,17.900000) -- cycle;
	\fill[gray!80] (5.000000,17.000000) -- (5.000000,17.900000) -- (5.900000,17.000000) -- cycle;
	\fill[gray!80] (5.000000,17.000000) -- (5.900000,17.000000) -- (5.000000,16.100000) -- cycle;
	\fill[gray!80] (5.000000,17.000000) -- (5.000000,16.100000) -- (4.100000,17.000000) -- cycle;
	\fill[gray!80] (7.000000,17.000000) -- (6.100000,17.000000) -- (7.000000,17.900000) -- cycle;
	\fill[gray!80] (7.000000,17.000000) -- (7.000000,17.900000) -- (7.900000,17.000000) -- cycle;
	\fill[gray!80] (7.000000,17.000000) -- (7.900000,17.000000) -- (7.000000,16.100000) -- cycle;
	\fill[gray!80] (7.000000,17.000000) -- (7.000000,16.100000) -- (6.100000,17.000000) -- cycle;
	\fill[gray!80] (9.000000,17.000000) -- (8.100000,17.000000) -- (9.000000,17.900000) -- cycle;
	\fill[gray!80] (9.000000,17.000000) -- (9.000000,17.900000) -- (9.900000,17.000000) -- cycle;
	\fill[gray!80] (9.000000,17.000000) -- (9.900000,17.000000) -- (9.000000,16.100000) -- cycle;
	\fill[gray!80] (9.000000,17.000000) -- (9.000000,16.100000) -- (8.100000,17.000000) -- cycle;
	\fill[gray!80] (11.000000,17.000000) -- (10.100000,17.000000) -- (11.000000,17.900000) -- cycle;
	\fill[gray!80] (11.000000,17.000000) -- (11.000000,17.900000) -- (11.900000,17.000000) -- cycle;
	\fill[gray!80] (11.000000,17.000000) -- (11.900000,17.000000) -- (11.000000,16.100000) -- cycle;
	\fill[gray!80] (11.000000,17.000000) -- (11.000000,16.100000) -- (10.100000,17.000000) -- cycle;
	\fill[gray!80] (13.000000,17.000000) -- (12.100000,17.000000) -- (13.000000,17.900000) -- cycle;
	\fill[gray!80] (13.000000,17.000000) -- (13.000000,17.900000) -- (13.900000,17.000000) -- cycle;
	\fill[gray!80] (13.000000,17.000000) -- (13.900000,17.000000) -- (13.000000,16.100000) -- cycle;
	\fill[gray!80] (13.000000,17.000000) -- (13.000000,16.100000) -- (12.100000,17.000000) -- cycle;
	\fill[gray!25] (2.000000,16.000000) -- (2.000000,16.900000) -- (2.900000,16.000000) -- cycle;
	\fill[gray!25] (2.000000,16.000000) -- (2.900000,16.000000) -- (2.000000,15.100000) -- cycle;
	\fill[gray!25] (4.000000,16.000000) -- (3.100000,16.000000) -- (4.000000,16.900000) -- cycle;
	\fill[gray!25] (4.000000,16.000000) -- (4.000000,16.900000) -- (4.900000,16.000000) -- cycle;
	\fill[gray!25] (6.000000,16.000000) -- (5.100000,16.000000) -- (6.000000,16.900000) -- cycle;
	\fill[gray!25] (6.000000,16.000000) -- (6.000000,16.900000) -- (6.900000,16.000000) -- cycle;
	\fill[gray!25] (8.000000,16.000000) -- (7.100000,16.000000) -- (8.000000,16.900000) -- cycle;
	\fill[gray!25] (8.000000,16.000000) -- (8.000000,16.900000) -- (8.900000,16.000000) -- cycle;
	\fill[gray!25] (10.000000,16.000000) -- (9.100000,16.000000) -- (10.000000,16.900000) -- cycle;
	\fill[gray!25] (10.000000,16.000000) -- (10.000000,16.900000) -- (10.900000,16.000000) -- cycle;
	\fill[gray!25] (12.000000,16.000000) -- (11.100000,16.000000) -- (12.000000,16.900000) -- cycle;
	\fill[gray!25] (12.000000,16.000000) -- (12.000000,16.900000) -- (12.900000,16.000000) -- cycle;
	\fill[gray!25] (12.000000,16.000000) -- (12.900000,16.000000) -- (12.000000,15.100000) -- cycle;
	\fill[gray!25] (12.000000,16.000000) -- (12.000000,15.100000) -- (11.100000,16.000000) -- cycle;
	\fill[gray!25] (14.000000,16.000000) -- (13.100000,16.000000) -- (14.000000,16.900000) -- cycle;
	\fill[gray!25] (14.000000,16.000000) -- (14.000000,15.100000) -- (13.100000,16.000000) -- cycle;
	\fill[gray!80] (13.000000,15.000000) -- (12.100000,15.000000) -- (13.000000,15.900000) -- cycle;
	\fill[gray!80] (13.000000,15.000000) -- (13.000000,15.900000) -- (13.900000,15.000000) -- cycle;
	\fill[gray!80] (13.000000,15.000000) -- (13.900000,15.000000) -- (13.000000,14.100000) -- cycle;
	\fill[gray!80] (13.000000,15.000000) -- (13.000000,14.100000) -- (12.100000,15.000000) -- cycle;
	\fill[gray!25] (4.000000,14.000000) -- (4.900000,14.000000) -- (4.000000,13.100000) -- cycle;
	\fill[gray!25] (6.000000,14.000000) -- (6.900000,14.000000) -- (6.000000,13.100000) -- cycle;
	\fill[gray!25] (6.000000,14.000000) -- (6.000000,13.100000) -- (5.100000,14.000000) -- cycle;
	\fill[gray!25] (8.000000,14.000000) -- (8.900000,14.000000) -- (8.000000,13.100000) -- cycle;
	\fill[gray!25] (8.000000,14.000000) -- (8.000000,13.100000) -- (7.100000,14.000000) -- cycle;
	\fill[gray!25] (10.000000,14.000000) -- (10.000000,13.100000) -- (9.100000,14.000000) -- cycle;
	\fill[gray!25] (12.000000,14.000000) -- (12.000000,14.900000) -- (12.900000,14.000000) -- cycle;
	\fill[gray!25] (12.000000,14.000000) -- (12.900000,14.000000) -- (12.000000,13.100000) -- cycle;
	\fill[gray!80] (5.000000,13.000000) -- (4.100000,13.000000) -- (5.000000,13.900000) -- cycle;
	\fill[gray!80] (5.000000,13.000000) -- (5.000000,13.900000) -- (5.900000,13.000000) -- cycle;
	\fill[gray!80] (5.000000,13.000000) -- (5.900000,13.000000) -- (5.000000,12.100000) -- cycle;
	\fill[gray!80] (5.000000,13.000000) -- (5.000000,12.100000) -- (4.100000,13.000000) -- cycle;
	\fill[gray!80] (7.000000,13.000000) -- (6.100000,13.000000) -- (7.000000,13.900000) -- cycle;
	\fill[gray!80] (7.000000,13.000000) -- (7.000000,13.900000) -- (7.900000,13.000000) -- cycle;
	\fill[gray!80] (7.000000,13.000000) -- (7.900000,13.000000) -- (7.000000,12.100000) -- cycle;
	\fill[gray!80] (7.000000,13.000000) -- (7.000000,12.100000) -- (6.100000,13.000000) -- cycle;
	\fill[gray!80] (9.000000,13.000000) -- (8.100000,13.000000) -- (9.000000,13.900000) -- cycle;
	\fill[gray!80] (9.000000,13.000000) -- (9.000000,13.900000) -- (9.900000,13.000000) -- cycle;
	\fill[gray!80] (9.000000,13.000000) -- (9.900000,13.000000) -- (9.000000,12.100000) -- cycle;
	\fill[gray!80] (9.000000,13.000000) -- (9.000000,12.100000) -- (8.100000,13.000000) -- cycle;
	\fill[gray!80] (13.000000,13.000000) -- (12.100000,13.000000) -- (13.000000,13.900000) -- cycle;
	\fill[gray!80] (13.000000,13.000000) -- (13.000000,13.900000) -- (13.900000,13.000000) -- cycle;
	\fill[gray!80] (13.000000,13.000000) -- (13.900000,13.000000) -- (13.000000,12.100000) -- cycle;
	\fill[gray!80] (13.000000,13.000000) -- (13.000000,12.100000) -- (12.100000,13.000000) -- cycle;
	\fill[gray!25] (6.000000,12.000000) -- (5.100000,12.000000) -- (6.000000,12.900000) -- cycle;
	\fill[gray!25] (6.000000,12.000000) -- (6.000000,12.900000) -- (6.900000,12.000000) -- cycle;
	\fill[gray!25] (6.000000,12.000000) -- (6.900000,12.000000) -- (6.000000,11.100000) -- cycle;
	\fill[gray!25] (6.000000,12.000000) -- (6.000000,11.100000) -- (5.100000,12.000000) -- cycle;
	\fill[gray!25] (8.000000,12.000000) -- (7.100000,12.000000) -- (8.000000,12.900000) -- cycle;
	\fill[gray!25] (8.000000,12.000000) -- (8.000000,12.900000) -- (8.900000,12.000000) -- cycle;
	\fill[gray!25] (8.000000,12.000000) -- (8.900000,12.000000) -- (8.000000,11.100000) -- cycle;
	\fill[gray!25] (8.000000,12.000000) -- (8.000000,11.100000) -- (7.100000,12.000000) -- cycle;
	\fill[gray!25] (10.000000,12.000000) -- (9.100000,12.000000) -- (10.000000,12.900000) -- cycle;
	\fill[gray!25] (10.000000,12.000000) -- (10.000000,11.100000) -- (9.100000,12.000000) -- cycle;
	\fill[gray!25] (12.000000,12.000000) -- (12.000000,12.900000) -- (12.900000,12.000000) -- cycle;
	\fill[gray!25] (12.000000,12.000000) -- (12.900000,12.000000) -- (12.000000,11.100000) -- cycle;
	\fill[gray!25] (14.000000,12.000000) -- (13.100000,12.000000) -- (14.000000,12.900000) -- cycle;
	\fill[gray!25] (14.000000,12.000000) -- (14.000000,11.100000) -- (13.100000,12.000000) -- cycle;
	\fill[gray!80] (5.000000,11.000000) -- (4.100000,11.000000) -- (5.000000,11.900000) -- cycle;
	\fill[gray!80] (5.000000,11.000000) -- (5.000000,11.900000) -- (5.900000,11.000000) -- cycle;
	\fill[gray!80] (5.000000,11.000000) -- (5.900000,11.000000) -- (5.000000,10.100000) -- cycle;
	\fill[gray!80] (5.000000,11.000000) -- (5.000000,10.100000) -- (4.100000,11.000000) -- cycle;
	\fill[gray!80] (7.000000,11.000000) -- (6.100000,11.000000) -- (7.000000,11.900000) -- cycle;
	\fill[gray!80] (7.000000,11.000000) -- (7.000000,11.900000) -- (7.900000,11.000000) -- cycle;
	\fill[gray!80] (7.000000,11.000000) -- (7.900000,11.000000) -- (7.000000,10.100000) -- cycle;
	\fill[gray!80] (7.000000,11.000000) -- (7.000000,10.100000) -- (6.100000,11.000000) -- cycle;
	\fill[gray!80] (9.000000,11.000000) -- (8.100000,11.000000) -- (9.000000,11.900000) -- cycle;
	\fill[gray!80] (9.000000,11.000000) -- (9.000000,11.900000) -- (9.900000,11.000000) -- cycle;
	\fill[gray!80] (9.000000,11.000000) -- (9.900000,11.000000) -- (9.000000,10.100000) -- cycle;
	\fill[gray!80] (9.000000,11.000000) -- (9.000000,10.100000) -- (8.100000,11.000000) -- cycle;
	\fill[gray!80] (13.000000,11.000000) -- (12.100000,11.000000) -- (13.000000,11.900000) -- cycle;
	\fill[gray!80] (13.000000,11.000000) -- (13.000000,11.900000) -- (13.900000,11.000000) -- cycle;
	\fill[gray!80] (13.000000,11.000000) -- (13.900000,11.000000) -- (13.000000,10.100000) -- cycle;
	\fill[gray!80] (13.000000,11.000000) -- (13.000000,10.100000) -- (12.100000,11.000000) -- cycle;
	\fill[gray!25] (4.000000,10.000000) -- (4.000000,10.900000) -- (4.900000,10.000000) -- cycle;
	\fill[gray!25] (4.000000,10.000000) -- (4.900000,10.000000) -- (4.000000,9.100000) -- cycle;
	\fill[gray!25] (6.000000,10.000000) -- (5.100000,10.000000) -- (6.000000,10.900000) -- cycle;
	\fill[gray!25] (6.000000,10.000000) -- (6.000000,10.900000) -- (6.900000,10.000000) -- cycle;
	\fill[gray!25] (6.000000,10.000000) -- (6.900000,10.000000) -- (6.000000,9.100000) -- cycle;
	\fill[gray!25] (6.000000,10.000000) -- (6.000000,9.100000) -- (5.100000,10.000000) -- cycle;
	\fill[gray!25] (8.000000,10.000000) -- (7.100000,10.000000) -- (8.000000,10.900000) -- cycle;
	\fill[gray!25] (8.000000,10.000000) -- (8.000000,10.900000) -- (8.900000,10.000000) -- cycle;
	\fill[gray!25] (8.000000,10.000000) -- (8.900000,10.000000) -- (8.000000,9.100000) -- cycle;
	\fill[gray!25] (8.000000,10.000000) -- (8.000000,9.100000) -- (7.100000,10.000000) -- cycle;
	\fill[gray!25] (10.000000,10.000000) -- (9.100000,10.000000) -- (10.000000,10.900000) -- cycle;
	\fill[gray!25] (10.000000,10.000000) -- (10.000000,9.100000) -- (9.100000,10.000000) -- cycle;
	\fill[gray!25] (12.000000,10.000000) -- (12.000000,10.900000) -- (12.900000,10.000000) -- cycle;
	\fill[gray!25] (12.000000,10.000000) -- (12.900000,10.000000) -- (12.000000,9.100000) -- cycle;
	\fill[gray!25] (14.000000,10.000000) -- (13.100000,10.000000) -- (14.000000,10.900000) -- cycle;
	\fill[gray!25] (14.000000,10.000000) -- (14.000000,9.100000) -- (13.100000,10.000000) -- cycle;
	\fill[gray!80] (5.000000,9.000000) -- (4.100000,9.000000) -- (5.000000,9.900000) -- cycle;
	\fill[gray!80] (5.000000,9.000000) -- (5.000000,9.900000) -- (5.900000,9.000000) -- cycle;
	\fill[gray!80] (5.000000,9.000000) -- (5.900000,9.000000) -- (5.000000,8.100000) -- cycle;
	\fill[gray!80] (5.000000,9.000000) -- (5.000000,8.100000) -- (4.100000,9.000000) -- cycle;
	\fill[gray!80] (7.000000,9.000000) -- (6.100000,9.000000) -- (7.000000,9.900000) -- cycle;
	\fill[gray!80] (7.000000,9.000000) -- (7.000000,9.900000) -- (7.900000,9.000000) -- cycle;
	\fill[gray!80] (7.000000,9.000000) -- (7.900000,9.000000) -- (7.000000,8.100000) -- cycle;
	\fill[gray!80] (7.000000,9.000000) -- (7.000000,8.100000) -- (6.100000,9.000000) -- cycle;
	\fill[gray!80] (9.000000,9.000000) -- (8.100000,9.000000) -- (9.000000,9.900000) -- cycle;
	\fill[gray!80] (9.000000,9.000000) -- (9.000000,9.900000) -- (9.900000,9.000000) -- cycle;
	\fill[gray!80] (9.000000,9.000000) -- (9.900000,9.000000) -- (9.000000,8.100000) -- cycle;
	\fill[gray!80] (9.000000,9.000000) -- (9.000000,8.100000) -- (8.100000,9.000000) -- cycle;
	\fill[gray!80] (13.000000,9.000000) -- (12.100000,9.000000) -- (13.000000,9.900000) -- cycle;
	\fill[gray!80] (13.000000,9.000000) -- (13.000000,9.900000) -- (13.900000,9.000000) -- cycle;
	\fill[gray!80] (13.000000,9.000000) -- (13.900000,9.000000) -- (13.000000,8.100000) -- cycle;
	\fill[gray!80] (13.000000,9.000000) -- (13.000000,8.100000) -- (12.100000,9.000000) -- cycle;
	\fill[gray!25] (4.000000,8.000000) -- (4.000000,8.900000) -- (4.900000,8.000000) -- cycle;
	\fill[gray!25] (6.000000,8.000000) -- (5.100000,8.000000) -- (6.000000,8.900000) -- cycle;
	\fill[gray!25] (6.000000,8.000000) -- (6.000000,8.900000) -- (6.900000,8.000000) -- cycle;
	\fill[gray!25] (10.000000,8.000000) -- (9.100000,8.000000) -- (10.000000,8.900000) -- cycle;
	\fill[gray!25] (12.000000,8.000000) -- (12.000000,8.900000) -- (12.900000,8.000000) -- cycle;
	\fill[gray!25] (12.000000,8.000000) -- (12.900000,8.000000) -- (12.000000,7.100000) -- cycle;
	\fill[gray!80] (13.000000,7.000000) -- (12.100000,7.000000) -- (13.000000,7.900000) -- cycle;
	\fill[gray!80] (13.000000,7.000000) -- (13.000000,7.900000) -- (13.900000,7.000000) -- cycle;
	\fill[gray!80] (13.000000,7.000000) -- (13.900000,7.000000) -- (13.000000,6.100000) -- cycle;
	\fill[gray!80] (13.000000,7.000000) -- (13.000000,6.100000) -- (12.100000,7.000000) -- cycle;
	\fill[gray!25] (2.000000,6.000000) -- (2.000000,6.900000) -- (2.900000,6.000000) -- cycle;
	\fill[gray!25] (2.000000,6.000000) -- (2.900000,6.000000) -- (2.000000,5.100000) -- cycle;
	\fill[gray!25] (4.000000,6.000000) -- (4.900000,6.000000) -- (4.000000,5.100000) -- cycle;
	\fill[gray!25] (4.000000,6.000000) -- (4.000000,5.100000) -- (3.100000,6.000000) -- cycle;
	\fill[gray!25] (6.000000,6.000000) -- (6.900000,6.000000) -- (6.000000,5.100000) -- cycle;
	\fill[gray!25] (6.000000,6.000000) -- (6.000000,5.100000) -- (5.100000,6.000000) -- cycle;
	\fill[gray!25] (8.000000,6.000000) -- (8.900000,6.000000) -- (8.000000,5.100000) -- cycle;
	\fill[gray!25] (8.000000,6.000000) -- (8.000000,5.100000) -- (7.100000,6.000000) -- cycle;
	\fill[gray!25] (10.000000,6.000000) -- (10.900000,6.000000) -- (10.000000,5.100000) -- cycle;
	\fill[gray!25] (10.000000,6.000000) -- (10.000000,5.100000) -- (9.100000,6.000000) -- cycle;
	\fill[gray!25] (12.000000,6.000000) -- (11.100000,6.000000) -- (12.000000,6.900000) -- cycle;
	\fill[gray!25] (12.000000,6.000000) -- (12.000000,6.900000) -- (12.900000,6.000000) -- cycle;
	\fill[gray!25] (12.000000,6.000000) -- (12.900000,6.000000) -- (12.000000,5.100000) -- cycle;
	\fill[gray!25] (12.000000,6.000000) -- (12.000000,5.100000) -- (11.100000,6.000000) -- cycle;
	\fill[gray!25] (14.000000,6.000000) -- (13.100000,6.000000) -- (14.000000,6.900000) -- cycle;
	\fill[gray!25] (14.000000,6.000000) -- (14.000000,5.100000) -- (13.100000,6.000000) -- cycle;
	\fill[gray!80] (3.000000,5.000000) -- (2.100000,5.000000) -- (3.000000,5.900000) -- cycle;
	\fill[gray!80] (3.000000,5.000000) -- (3.000000,5.900000) -- (3.900000,5.000000) -- cycle;
	\fill[gray!80] (3.000000,5.000000) -- (3.900000,5.000000) -- (3.000000,4.100000) -- cycle;
	\fill[gray!80] (3.000000,5.000000) -- (3.000000,4.100000) -- (2.100000,5.000000) -- cycle;
	\fill[gray!80] (5.000000,5.000000) -- (4.100000,5.000000) -- (5.000000,5.900000) -- cycle;
	\fill[gray!80] (5.000000,5.000000) -- (5.000000,5.900000) -- (5.900000,5.000000) -- cycle;
	\fill[gray!80] (5.000000,5.000000) -- (5.900000,5.000000) -- (5.000000,4.100000) -- cycle;
	\fill[gray!80] (5.000000,5.000000) -- (5.000000,4.100000) -- (4.100000,5.000000) -- cycle;
	\fill[gray!80] (7.000000,5.000000) -- (6.100000,5.000000) -- (7.000000,5.900000) -- cycle;
	\fill[gray!80] (7.000000,5.000000) -- (7.000000,5.900000) -- (7.900000,5.000000) -- cycle;
	\fill[gray!80] (7.000000,5.000000) -- (7.900000,5.000000) -- (7.000000,4.100000) -- cycle;
	\fill[gray!80] (7.000000,5.000000) -- (7.000000,4.100000) -- (6.100000,5.000000) -- cycle;
	\fill[gray!80] (9.000000,5.000000) -- (8.100000,5.000000) -- (9.000000,5.900000) -- cycle;
	\fill[gray!80] (9.000000,5.000000) -- (9.000000,5.900000) -- (9.900000,5.000000) -- cycle;
	\fill[gray!80] (9.000000,5.000000) -- (9.900000,5.000000) -- (9.000000,4.100000) -- cycle;
	\fill[gray!80] (9.000000,5.000000) -- (9.000000,4.100000) -- (8.100000,5.000000) -- cycle;
	\fill[gray!80] (11.000000,5.000000) -- (10.100000,5.000000) -- (11.000000,5.900000) -- cycle;
	\fill[gray!80] (11.000000,5.000000) -- (11.000000,5.900000) -- (11.900000,5.000000) -- cycle;
	\fill[gray!80] (11.000000,5.000000) -- (11.900000,5.000000) -- (11.000000,4.100000) -- cycle;
	\fill[gray!80] (11.000000,5.000000) -- (11.000000,4.100000) -- (10.100000,5.000000) -- cycle;
	\fill[gray!80] (13.000000,5.000000) -- (12.100000,5.000000) -- (13.000000,5.900000) -- cycle;
	\fill[gray!80] (13.000000,5.000000) -- (13.000000,5.900000) -- (13.900000,5.000000) -- cycle;
	\fill[gray!80] (13.000000,5.000000) -- (13.900000,5.000000) -- (13.000000,4.100000) -- cycle;
	\fill[gray!80] (13.000000,5.000000) -- (13.000000,4.100000) -- (12.100000,5.000000) -- cycle;
	\fill[gray!25] (2.000000,4.000000) -- (2.000000,4.900000) -- (2.900000,4.000000) -- cycle;
	\fill[gray!25] (2.000000,4.000000) -- (2.900000,4.000000) -- (2.000000,3.100000) -- cycle;
	\fill[gray!25] (4.000000,4.000000) -- (3.100000,4.000000) -- (4.000000,4.900000) -- cycle;
	\fill[gray!25] (4.000000,4.000000) -- (4.000000,4.900000) -- (4.900000,4.000000) -- cycle;
	\fill[gray!25] (4.000000,4.000000) -- (4.900000,4.000000) -- (4.000000,3.100000) -- cycle;
	\fill[gray!25] (4.000000,4.000000) -- (4.000000,3.100000) -- (3.100000,4.000000) -- cycle;
	\fill[gray!25] (6.000000,4.000000) -- (5.100000,4.000000) -- (6.000000,4.900000) -- cycle;
	\fill[gray!25] (6.000000,4.000000) -- (6.000000,4.900000) -- (6.900000,4.000000) -- cycle;
	\fill[gray!25] (6.000000,4.000000) -- (6.900000,4.000000) -- (6.000000,3.100000) -- cycle;
	\fill[gray!25] (6.000000,4.000000) -- (6.000000,3.100000) -- (5.100000,4.000000) -- cycle;
	\fill[gray!25] (8.000000,4.000000) -- (7.100000,4.000000) -- (8.000000,4.900000) -- cycle;
	\fill[gray!25] (8.000000,4.000000) -- (8.000000,4.900000) -- (8.900000,4.000000) -- cycle;
	\fill[gray!25] (8.000000,4.000000) -- (8.900000,4.000000) -- (8.000000,3.100000) -- cycle;
	\fill[gray!25] (8.000000,4.000000) -- (8.000000,3.100000) -- (7.100000,4.000000) -- cycle;
	\fill[gray!25] (10.000000,4.000000) -- (9.100000,4.000000) -- (10.000000,4.900000) -- cycle;
	\fill[gray!25] (10.000000,4.000000) -- (10.000000,4.900000) -- (10.900000,4.000000) -- cycle;
	\fill[gray!25] (10.000000,4.000000) -- (10.900000,4.000000) -- (10.000000,3.100000) -- cycle;
	\fill[gray!25] (10.000000,4.000000) -- (10.000000,3.100000) -- (9.100000,4.000000) -- cycle;
	\fill[gray!25] (12.000000,4.000000) -- (11.100000,4.000000) -- (12.000000,4.900000) -- cycle;
	\fill[gray!25] (12.000000,4.000000) -- (12.000000,4.900000) -- (12.900000,4.000000) -- cycle;
	\fill[gray!25] (12.000000,4.000000) -- (12.900000,4.000000) -- (12.000000,3.100000) -- cycle;
	\fill[gray!25] (12.000000,4.000000) -- (12.000000,3.100000) -- (11.100000,4.000000) -- cycle;
	\fill[gray!25] (14.000000,4.000000) -- (13.100000,4.000000) -- (14.000000,4.900000) -- cycle;
	\fill[gray!25] (14.000000,4.000000) -- (14.000000,3.100000) -- (13.100000,4.000000) -- cycle;
	\fill[gray!80] (3.000000,3.000000) -- (2.100000,3.000000) -- (3.000000,3.900000) -- cycle;
	\fill[gray!80] (3.000000,3.000000) -- (3.000000,3.900000) -- (3.900000,3.000000) -- cycle;
	\fill[gray!80] (3.000000,3.000000) -- (3.900000,3.000000) -- (3.000000,2.100000) -- cycle;
	\fill[gray!80] (3.000000,3.000000) -- (3.000000,2.100000) -- (2.100000,3.000000) -- cycle;
	\fill[gray!80] (5.000000,3.000000) -- (4.100000,3.000000) -- (5.000000,3.900000) -- cycle;
	\fill[gray!80] (5.000000,3.000000) -- (5.000000,3.900000) -- (5.900000,3.000000) -- cycle;
	\fill[gray!80] (5.000000,3.000000) -- (5.900000,3.000000) -- (5.000000,2.100000) -- cycle;
	\fill[gray!80] (5.000000,3.000000) -- (5.000000,2.100000) -- (4.100000,3.000000) -- cycle;
	\fill[gray!80] (7.000000,3.000000) -- (6.100000,3.000000) -- (7.000000,3.900000) -- cycle;
	\fill[gray!80] (7.000000,3.000000) -- (7.000000,3.900000) -- (7.900000,3.000000) -- cycle;
	\fill[gray!80] (7.000000,3.000000) -- (7.900000,3.000000) -- (7.000000,2.100000) -- cycle;
	\fill[gray!80] (7.000000,3.000000) -- (7.000000,2.100000) -- (6.100000,3.000000) -- cycle;
	\fill[gray!80] (9.000000,3.000000) -- (8.100000,3.000000) -- (9.000000,3.900000) -- cycle;
	\fill[gray!80] (9.000000,3.000000) -- (9.000000,3.900000) -- (9.900000,3.000000) -- cycle;
	\fill[gray!80] (9.000000,3.000000) -- (9.900000,3.000000) -- (9.000000,2.100000) -- cycle;
	\fill[gray!80] (9.000000,3.000000) -- (9.000000,2.100000) -- (8.100000,3.000000) -- cycle;
	\fill[gray!80] (11.000000,3.000000) -- (10.100000,3.000000) -- (11.000000,3.900000) -- cycle;
	\fill[gray!80] (11.000000,3.000000) -- (11.000000,3.900000) -- (11.900000,3.000000) -- cycle;
	\fill[gray!80] (11.000000,3.000000) -- (11.900000,3.000000) -- (11.000000,2.100000) -- cycle;
	\fill[gray!80] (11.000000,3.000000) -- (11.000000,2.100000) -- (10.100000,3.000000) -- cycle;
	\fill[gray!80] (13.000000,3.000000) -- (12.100000,3.000000) -- (13.000000,3.900000) -- cycle;
	\fill[gray!80] (13.000000,3.000000) -- (13.000000,3.900000) -- (13.900000,3.000000) -- cycle;
	\fill[gray!80] (13.000000,3.000000) -- (13.900000,3.000000) -- (13.000000,2.100000) -- cycle;
	\fill[gray!80] (13.000000,3.000000) -- (13.000000,2.100000) -- (12.100000,3.000000) -- cycle;
	\fill[gray!25] (4.000000,2.000000) -- (3.100000,2.000000) -- (4.000000,2.900000) -- cycle;
	\fill[gray!25] (4.000000,2.000000) -- (4.000000,2.900000) -- (4.900000,2.000000) -- cycle;
	\fill[gray!25] (6.000000,2.000000) -- (5.100000,2.000000) -- (6.000000,2.900000) -- cycle;
	\fill[gray!25] (6.000000,2.000000) -- (6.000000,2.900000) -- (6.900000,2.000000) -- cycle;
	\fill[gray!25] (10.000000,2.000000) -- (9.100000,2.000000) -- (10.000000,2.900000) -- cycle;
	\fill[gray!25] (10.000000,2.000000) -- (10.000000,2.900000) -- (10.900000,2.000000) -- cycle;
	\fill[gray!25] (12.000000,2.000000) -- (11.100000,2.000000) -- (12.000000,2.900000) -- cycle;
	\fill[gray!25] (12.000000,2.000000) -- (12.000000,2.900000) -- (12.900000,2.000000) -- cycle;
	\draw[line width=1pt] (3.000000,20.000000) circle (0.100000);
	\draw[line width=1pt,fill=black] (4.000000,20.000000) circle (0.100000);
	\draw[line width=1pt] (5.000000,20.000000) circle (0.100000);
	\draw[line width=1pt,fill=black] (6.000000,20.000000) circle (0.100000);
	\draw[line width=1pt] (7.000000,20.000000) circle (0.100000);
	\draw[line width=1pt] (9.000000,20.000000) circle (0.100000);
	\draw[line width=1pt,fill=black] (10.000000,20.000000) circle (0.100000);
	\draw[line width=1pt] (11.000000,20.000000) circle (0.100000);
	\draw[line width=1pt,fill=black] (12.000000,20.000000) circle (0.100000);
	\draw[line width=1pt] (13.000000,20.000000) circle (0.100000);
	\draw[line width=1pt] (2.000000,19.000000) circle (0.100000);
	\draw[line width=1pt,fill=black] (3.000000,19.000000) circle (0.100000);
	\draw[line width=1pt] (4.000000,19.000000) circle (0.100000);
	\draw[line width=1pt,fill=black] (5.000000,19.000000) circle (0.100000);
	\draw[line width=1pt] (6.000000,19.000000) circle (0.100000);
	\draw[line width=1pt,fill=black] (7.000000,19.000000) circle (0.100000);
	\draw[line width=1pt] (8.000000,19.000000) circle (0.100000);
	\draw[line width=1pt,fill=black] (9.000000,19.000000) circle (0.100000);
	\draw[line width=1pt] (10.000000,19.000000) circle (0.100000);
	\draw[line width=1pt,fill=black] (11.000000,19.000000) circle (0.100000);
	\draw[line width=1pt] (12.000000,19.000000) circle (0.100000);
	\draw[line width=1pt,fill=black] (13.000000,19.000000) circle (0.100000);
	\draw[line width=1pt] (14.000000,19.000000) circle (0.100000);
	\draw[line width=1pt,fill=black] (2.000000,18.000000) circle (0.100000);
	\draw[line width=1pt] (3.000000,18.000000) circle (0.100000);
	\draw[line width=1pt,fill=black] (4.000000,18.000000) circle (0.100000);
	\draw[line width=1pt] (5.000000,18.000000) circle (0.100000);
	\draw[line width=1pt,fill=black] (6.000000,18.000000) circle (0.100000);
	\draw[line width=1pt] (7.000000,18.000000) circle (0.100000);
	\draw[line width=1pt,fill=black] (8.000000,18.000000) circle (0.100000);
	\draw[line width=1pt] (9.000000,18.000000) circle (0.100000);
	\draw[line width=1pt,fill=black] (10.000000,18.000000) circle (0.100000);
	\draw[line width=1pt] (11.000000,18.000000) circle (0.100000);
	\draw[line width=1pt,fill=black] (12.000000,18.000000) circle (0.100000);
	\draw[line width=1pt] (13.000000,18.000000) circle (0.100000);
	\draw[line width=1pt,fill=black] (14.000000,18.000000) circle (0.100000);
	\draw[line width=1pt] (2.000000,17.000000) circle (0.100000);
	\draw[line width=1pt,fill=black] (3.000000,17.000000) circle (0.100000);
	\draw[line width=1pt] (4.000000,17.000000) circle (0.100000);
	\draw[line width=1pt,fill=black] (5.000000,17.000000) circle (0.100000);
	\draw[line width=1pt] (6.000000,17.000000) circle (0.100000);
	\draw[line width=1pt,fill=black] (7.000000,17.000000) circle (0.100000);
	\draw[line width=1pt] (8.000000,17.000000) circle (0.100000);
	\draw[line width=1pt,fill=black] (9.000000,17.000000) circle (0.100000);
	\draw[line width=1pt] (10.000000,17.000000) circle (0.100000);
	\draw[line width=1pt,fill=black] (11.000000,17.000000) circle (0.100000);
	\draw[line width=1pt] (12.000000,17.000000) circle (0.100000);
	\draw[line width=1pt,fill=black] (13.000000,17.000000) circle (0.100000);
	\draw[line width=1pt] (14.000000,17.000000) circle (0.100000);
	\draw[line width=1pt,fill=black] (2.000000,16.000000) circle (0.100000);
	\draw[line width=1pt] (3.000000,16.000000) circle (0.100000);
	\draw[line width=1pt,fill=black] (4.000000,16.000000) circle (0.100000);
	\draw[line width=1pt] (5.000000,16.000000) circle (0.100000);
	\draw[line width=1pt,fill=black] (6.000000,16.000000) circle (0.100000);
	\draw[line width=1pt] (7.000000,16.000000) circle (0.100000);
	\draw[line width=1pt,fill=black] (8.000000,16.000000) circle (0.100000);
	\draw[line width=1pt] (9.000000,16.000000) circle (0.100000);
	\draw[line width=1pt,fill=black] (10.000000,16.000000) circle (0.100000);
	\draw[line width=1pt] (11.000000,16.000000) circle (0.100000);
	\draw[line width=1pt,fill=black] (12.000000,16.000000) circle (0.100000);
	\draw[line width=1pt] (13.000000,16.000000) circle (0.100000);
	\draw[line width=1pt,fill=black] (14.000000,16.000000) circle (0.100000);
	\draw[line width=1pt] (2.000000,15.000000) circle (0.100000);
	\draw[line width=1pt] (12.000000,15.000000) circle (0.100000);
	\draw[line width=1pt,fill=black] (13.000000,15.000000) circle (0.100000);
	\draw[line width=1pt] (14.000000,15.000000) circle (0.100000);
	\draw[line width=1pt,fill=black] (4.000000,14.000000) circle (0.100000);
	\draw (5.000000,14.000000) node [above right] {$M_X$};
	\draw[line width=1pt] (5.000000,14.000000) circle (0.100000);
	\draw[line width=1pt,fill=black] (6.000000,14.000000) circle (0.100000);
	\draw (7.000000,14.000000) node [above right] {$M_X$};
	\draw[line width=1pt] (7.000000,14.000000) circle (0.100000);
	\draw[line width=1pt,fill=black] (8.000000,14.000000) circle (0.100000);
	\draw[line width=1pt] (9.000000,14.000000) circle (0.100000);
	\draw[line width=1pt,fill=black] (10.000000,14.000000) circle (0.100000);
	\draw[line width=1pt,fill=black] (12.000000,14.000000) circle (0.100000);
	\draw[line width=1pt] (13.000000,14.000000) circle (0.100000);
	\draw[line width=1pt] (2.000000,13.000000) circle (0.100000);
	\draw (4.000000,13.000000) node [above right] {$M_X$};
	\draw[line width=1pt] (4.000000,13.000000) circle (0.100000);
	\draw[line width=1pt,fill=black] (5.000000,13.000000) circle (0.100000);
	\draw[line width=1pt] (6.000000,13.000000) circle (0.100000);
	\draw[line width=1pt,fill=black] (7.000000,13.000000) circle (0.100000);
	\draw[line width=1pt] (8.000000,13.000000) circle (0.100000);
	\draw[line width=1pt,fill=black] (9.000000,13.000000) circle (0.100000);
	\draw[line width=1pt] (10.000000,13.000000) circle (0.100000);
	\draw[line width=1pt] (12.000000,13.000000) circle (0.100000);
	\draw[line width=1pt,fill=black] (13.000000,13.000000) circle (0.100000);
	\draw[line width=1pt] (14.000000,13.000000) circle (0.100000);
	\draw[line width=1pt,fill=black] (2.000000,12.000000) circle (0.100000);
	\draw[line width=1pt] (5.000000,12.000000) circle (0.100000);
	\draw[line width=1pt,fill=black] (6.000000,12.000000) circle (0.100000);
	\draw[line width=1pt] (7.000000,12.000000) circle (0.100000);
	\draw[line width=1pt,fill=black] (8.000000,12.000000) circle (0.100000);
	\draw[line width=1pt] (9.000000,12.000000) circle (0.100000);
	\draw[line width=1pt,fill=black] (10.000000,12.000000) circle (0.100000);
	\draw[line width=1pt,fill=black] (12.000000,12.000000) circle (0.100000);
	\draw[line width=1pt] (13.000000,12.000000) circle (0.100000);
	\draw[line width=1pt,fill=black] (14.000000,12.000000) circle (0.100000);
	\draw[line width=1pt] (2.000000,11.000000) circle (0.100000);
	\draw[line width=1pt] (4.000000,11.000000) circle (0.100000);
	\draw[line width=1pt,fill=black] (5.000000,11.000000) circle (0.100000);
	\draw[line width=1pt] (6.000000,11.000000) circle (0.100000);
	\draw[line width=1pt,fill=black] (7.000000,11.000000) circle (0.100000);
	\draw[line width=1pt] (8.000000,11.000000) circle (0.100000);
	\draw[line width=1pt,fill=black] (9.000000,11.000000) circle (0.100000);
	\draw[line width=1pt] (10.000000,11.000000) circle (0.100000);
	\draw[line width=1pt] (12.000000,11.000000) circle (0.100000);
	\draw[line width=1pt,fill=black] (13.000000,11.000000) circle (0.100000);
	\draw[line width=1pt] (14.000000,11.000000) circle (0.100000);
	\draw[line width=1pt,fill=black] (2.000000,10.000000) circle (0.100000);
	\draw[line width=1pt,fill=black] (4.000000,10.000000) circle (0.100000);
	\draw[line width=1pt] (5.000000,10.000000) circle (0.100000);
	\draw[line width=1pt,fill=black] (6.000000,10.000000) circle (0.100000);
	\draw[line width=1pt] (7.000000,10.000000) circle (0.100000);
	\draw[line width=1pt,fill=black] (8.000000,10.000000) circle (0.100000);
	\draw[line width=1pt] (9.000000,10.000000) circle (0.100000);
	\draw[line width=1pt,fill=black] (10.000000,10.000000) circle (0.100000);
	\draw[line width=1pt,fill=black] (12.000000,10.000000) circle (0.100000);
	\draw[line width=1pt] (13.000000,10.000000) circle (0.100000);
	\draw[line width=1pt,fill=black] (14.000000,10.000000) circle (0.100000);
	\draw[line width=1pt] (2.000000,9.000000) circle (0.100000);
	\draw[line width=1pt] (4.000000,9.000000) circle (0.100000);
	\draw[line width=1pt,fill=black] (5.000000,9.000000) circle (0.100000);
	\draw[line width=1pt] (6.000000,9.000000) circle (0.100000);
	\draw[line width=1pt,fill=black] (7.000000,9.000000) circle (0.100000);
	\draw[line width=1pt] (8.000000,9.000000) circle (0.100000);
	\draw[line width=1pt,fill=black] (9.000000,9.000000) circle (0.100000);
	\draw[line width=1pt] (10.000000,9.000000) circle (0.100000);
	\draw[line width=1pt] (12.000000,9.000000) circle (0.100000);
	\draw[line width=1pt,fill=black] (13.000000,9.000000) circle (0.100000);
	\draw[line width=1pt] (14.000000,9.000000) circle (0.100000);
	\draw[line width=1pt,fill=black] (4.000000,8.000000) circle (0.100000);
	\draw[line width=1pt] (5.000000,8.000000) circle (0.100000);
	\draw[line width=1pt,fill=black] (6.000000,8.000000) circle (0.100000);
	\draw[line width=1pt] (7.000000,8.000000) circle (0.100000);
	\draw[line width=1pt] (9.000000,8.000000) circle (0.100000);
	\draw[line width=1pt,fill=black] (10.000000,8.000000) circle (0.100000);
	\draw[line width=1pt,fill=black] (12.000000,8.000000) circle (0.100000);
	\draw[line width=1pt] (13.000000,8.000000) circle (0.100000);
	\draw[line width=1pt] (2.000000,7.000000) circle (0.100000);
	\draw[line width=1pt] (12.000000,7.000000) circle (0.100000);
	\draw[line width=1pt,fill=black] (13.000000,7.000000) circle (0.100000);
	\draw[line width=1pt] (14.000000,7.000000) circle (0.100000);
	\draw[line width=1pt,fill=black] (2.000000,6.000000) circle (0.100000);
	\draw[line width=1pt] (3.000000,6.000000) circle (0.100000);
	\draw[line width=1pt,fill=black] (4.000000,6.000000) circle (0.100000);
	\draw[line width=1pt] (5.000000,6.000000) circle (0.100000);
	\draw[line width=1pt,fill=black] (6.000000,6.000000) circle (0.100000);
	\draw[line width=1pt] (7.000000,6.000000) circle (0.100000);
	\draw[line width=1pt,fill=black] (8.000000,6.000000) circle (0.100000);
	\draw[line width=1pt] (9.000000,6.000000) circle (0.100000);
	\draw[line width=1pt,fill=black] (10.000000,6.000000) circle (0.100000);
	\draw[line width=1pt] (11.000000,6.000000) circle (0.100000);
	\draw[line width=1pt,fill=black] (12.000000,6.000000) circle (0.100000);
	\draw[line width=1pt] (13.000000,6.000000) circle (0.100000);
	\draw[line width=1pt,fill=black] (14.000000,6.000000) circle (0.100000);
	\draw[line width=1pt] (2.000000,5.000000) circle (0.100000);
	\draw[line width=1pt,fill=black] (3.000000,5.000000) circle (0.100000);
	\draw[line width=1pt] (4.000000,5.000000) circle (0.100000);
	\draw[line width=1pt,fill=black] (5.000000,5.000000) circle (0.100000);
	\draw[line width=1pt] (6.000000,5.000000) circle (0.100000);
	\draw[line width=1pt,fill=black] (7.000000,5.000000) circle (0.100000);
	\draw[line width=1pt] (8.000000,5.000000) circle (0.100000);
	\draw[line width=1pt,fill=black] (9.000000,5.000000) circle (0.100000);
	\draw[line width=1pt] (10.000000,5.000000) circle (0.100000);
	\draw[line width=1pt,fill=black] (11.000000,5.000000) circle (0.100000);
	\draw[line width=1pt] (12.000000,5.000000) circle (0.100000);
	\draw[line width=1pt,fill=black] (13.000000,5.000000) circle (0.100000);
	\draw[line width=1pt] (14.000000,5.000000) circle (0.100000);
	\draw[line width=1pt,fill=black] (2.000000,4.000000) circle (0.100000);
	\draw[line width=1pt] (3.000000,4.000000) circle (0.100000);
	\draw[line width=1pt,fill=black] (4.000000,4.000000) circle (0.100000);
	\draw[line width=1pt] (5.000000,4.000000) circle (0.100000);
	\draw[line width=1pt,fill=black] (6.000000,4.000000) circle (0.100000);
	\draw[line width=1pt] (7.000000,4.000000) circle (0.100000);
	\draw[line width=1pt,fill=black] (8.000000,4.000000) circle (0.100000);
	\draw[line width=1pt] (9.000000,4.000000) circle (0.100000);
	\draw[line width=1pt,fill=black] (10.000000,4.000000) circle (0.100000);
	\draw[line width=1pt] (11.000000,4.000000) circle (0.100000);
	\draw[line width=1pt,fill=black] (12.000000,4.000000) circle (0.100000);
	\draw[line width=1pt] (13.000000,4.000000) circle (0.100000);
	\draw[line width=1pt,fill=black] (14.000000,4.000000) circle (0.100000);
	\draw[line width=1pt] (2.000000,3.000000) circle (0.100000);
	\draw[line width=1pt,fill=black] (3.000000,3.000000) circle (0.100000);
	\draw[line width=1pt] (4.000000,3.000000) circle (0.100000);
	\draw[line width=1pt,fill=black] (5.000000,3.000000) circle (0.100000);
	\draw[line width=1pt] (6.000000,3.000000) circle (0.100000);
	\draw[line width=1pt,fill=black] (7.000000,3.000000) circle (0.100000);
	\draw[line width=1pt] (8.000000,3.000000) circle (0.100000);
	\draw[line width=1pt,fill=black] (9.000000,3.000000) circle (0.100000);
	\draw[line width=1pt] (10.000000,3.000000) circle (0.100000);
	\draw[line width=1pt,fill=black] (11.000000,3.000000) circle (0.100000);
	\draw[line width=1pt] (12.000000,3.000000) circle (0.100000);
	\draw[line width=1pt,fill=black] (13.000000,3.000000) circle (0.100000);
	\draw[line width=1pt] (14.000000,3.000000) circle (0.100000);
	\draw[line width=1pt] (3.000000,2.000000) circle (0.100000);
	\draw[line width=1pt,fill=black] (4.000000,2.000000) circle (0.100000);
	\draw[line width=1pt] (5.000000,2.000000) circle (0.100000);
	\draw[line width=1pt,fill=black] (6.000000,2.000000) circle (0.100000);
	\draw[line width=1pt] (7.000000,2.000000) circle (0.100000);
	\draw[line width=1pt] (9.000000,2.000000) circle (0.100000);
	\draw[line width=1pt,fill=black] (10.000000,2.000000) circle (0.100000);
	\draw[line width=1pt] (11.000000,2.000000) circle (0.100000);
	\draw[line width=1pt,fill=black] (12.000000,2.000000) circle (0.100000);
	\draw[line width=1pt] (13.000000,2.000000) circle (0.100000);
\end{tikzpicture}

%% file: 009ex2.tex
\begin{tikzpicture}[x=0.030000\linewidth,y=0.030000\linewidth]
	\fill[gray!25] (4.000000,20.000000) -- (4.900000,20.000000) -- (4.000000,19.100000) -- cycle;
	\fill[gray!25] (4.000000,20.000000) -- (4.000000,19.100000) -- (3.100000,20.000000) -- cycle;
	\fill[gray!25] (6.000000,20.000000) -- (6.900000,20.000000) -- (6.000000,19.100000) -- cycle;
	\fill[gray!25] (6.000000,20.000000) -- (6.000000,19.100000) -- (5.100000,20.000000) -- cycle;
	\fill[gray!25] (10.000000,20.000000) -- (10.900000,20.000000) -- (10.000000,19.100000) -- cycle;
	\fill[gray!25] (10.000000,20.000000) -- (10.000000,19.100000) -- (9.100000,20.000000) -- cycle;
	\fill[gray!25] (12.000000,20.000000) -- (12.900000,20.000000) -- (12.000000,19.100000) -- cycle;
	\fill[gray!25] (12.000000,20.000000) -- (12.000000,19.100000) -- (11.100000,20.000000) -- cycle;
	\fill[gray!80] (3.000000,19.000000) -- (2.100000,19.000000) -- (3.000000,19.900000) -- cycle;
	\fill[gray!80] (3.000000,19.000000) -- (3.000000,19.900000) -- (3.900000,19.000000) -- cycle;
	\fill[gray!80] (3.000000,19.000000) -- (3.900000,19.000000) -- (3.000000,18.100000) -- cycle;
	\fill[gray!80] (3.000000,19.000000) -- (3.000000,18.100000) -- (2.100000,19.000000) -- cycle;
	\fill[gray!80] (5.000000,19.000000) -- (4.100000,19.000000) -- (5.000000,19.900000) -- cycle;
	\fill[gray!80] (5.000000,19.000000) -- (5.000000,19.900000) -- (5.900000,19.000000) -- cycle;
	\fill[gray!80] (5.000000,19.000000) -- (5.900000,19.000000) -- (5.000000,18.100000) -- cycle;
	\fill[gray!80] (5.000000,19.000000) -- (5.000000,18.100000) -- (4.100000,19.000000) -- cycle;
	\fill[gray!80] (7.000000,19.000000) -- (6.100000,19.000000) -- (7.000000,19.900000) -- cycle;
	\fill[gray!80] (7.000000,19.000000) -- (7.000000,19.900000) -- (7.900000,19.000000) -- cycle;
	\fill[gray!80] (7.000000,19.000000) -- (7.900000,19.000000) -- (7.000000,18.100000) -- cycle;
	\fill[gray!80] (7.000000,19.000000) -- (7.000000,18.100000) -- (6.100000,19.000000) -- cycle;
	\fill[gray!80] (9.000000,19.000000) -- (8.100000,19.000000) -- (9.000000,19.900000) -- cycle;
	\fill[gray!80] (9.000000,19.000000) -- (9.000000,19.900000) -- (9.900000,19.000000) -- cycle;
	\fill[gray!80] (9.000000,19.000000) -- (9.900000,19.000000) -- (9.000000,18.100000) -- cycle;
	\fill[gray!80] (9.000000,19.000000) -- (9.000000,18.100000) -- (8.100000,19.000000) -- cycle;
	\fill[gray!80] (11.000000,19.000000) -- (10.100000,19.000000) -- (11.000000,19.900000) -- cycle;
	\fill[gray!80] (11.000000,19.000000) -- (11.000000,19.900000) -- (11.900000,19.000000) -- cycle;
	\fill[gray!80] (11.000000,19.000000) -- (11.900000,19.000000) -- (11.000000,18.100000) -- cycle;
	\fill[gray!80] (11.000000,19.000000) -- (11.000000,18.100000) -- (10.100000,19.000000) -- cycle;
	\fill[gray!80] (13.000000,19.000000) -- (12.100000,19.000000) -- (13.000000,19.900000) -- cycle;
	\fill[gray!80] (13.000000,19.000000) -- (13.000000,19.900000) -- (13.900000,19.000000) -- cycle;
	\fill[gray!80] (13.000000,19.000000) -- (13.900000,19.000000) -- (13.000000,18.100000) -- cycle;
	\fill[gray!80] (13.000000,19.000000) -- (13.000000,18.100000) -- (12.100000,19.000000) -- cycle;
	\fill[gray!25] (2.000000,18.000000) -- (2.000000,18.900000) -- (2.900000,18.000000) -- cycle;
	\fill[gray!25] (2.000000,18.000000) -- (2.900000,18.000000) -- (2.000000,17.100000) -- cycle;
	\fill[gray!25] (4.000000,18.000000) -- (3.100000,18.000000) -- (4.000000,18.900000) -- cycle;
	\fill[gray!25] (4.000000,18.000000) -- (4.000000,18.900000) -- (4.900000,18.000000) -- cycle;
	\fill[gray!25] (4.000000,18.000000) -- (4.900000,18.000000) -- (4.000000,17.100000) -- cycle;
	\fill[gray!25] (4.000000,18.000000) -- (4.000000,17.100000) -- (3.100000,18.000000) -- cycle;
	\fill[gray!25] (6.000000,18.000000) -- (5.100000,18.000000) -- (6.000000,18.900000) -- cycle;
	\fill[gray!25] (6.000000,18.000000) -- (6.000000,18.900000) -- (6.900000,18.000000) -- cycle;
	\fill[gray!25] (6.000000,18.000000) -- (6.900000,18.000000) -- (6.000000,17.100000) -- cycle;
	\fill[gray!25] (6.000000,18.000000) -- (6.000000,17.100000) -- (5.100000,18.000000) -- cycle;
	\fill[gray!25] (8.000000,18.000000) -- (7.100000,18.000000) -- (8.000000,18.900000) -- cycle;
	\fill[gray!25] (8.000000,18.000000) -- (8.000000,18.900000) -- (8.900000,18.000000) -- cycle;
	\fill[gray!25] (8.000000,18.000000) -- (8.900000,18.000000) -- (8.000000,17.100000) -- cycle;
	\fill[gray!25] (8.000000,18.000000) -- (8.000000,17.100000) -- (7.100000,18.000000) -- cycle;
	\fill[gray!25] (10.000000,18.000000) -- (9.100000,18.000000) -- (10.000000,18.900000) -- cycle;
	\fill[gray!25] (10.000000,18.000000) -- (10.000000,18.900000) -- (10.900000,18.000000) -- cycle;
	\fill[gray!25] (10.000000,18.000000) -- (10.900000,18.000000) -- (10.000000,17.100000) -- cycle;
	\fill[gray!25] (10.000000,18.000000) -- (10.000000,17.100000) -- (9.100000,18.000000) -- cycle;
	\fill[gray!25] (12.000000,18.000000) -- (11.100000,18.000000) -- (12.000000,18.900000) -- cycle;
	\fill[gray!25] (12.000000,18.000000) -- (12.000000,18.900000) -- (12.900000,18.000000) -- cycle;
	\fill[gray!25] (12.000000,18.000000) -- (12.900000,18.000000) -- (12.000000,17.100000) -- cycle;
	\fill[gray!25] (12.000000,18.000000) -- (12.000000,17.100000) -- (11.100000,18.000000) -- cycle;
	\fill[gray!25] (14.000000,18.000000) -- (13.100000,18.000000) -- (14.000000,18.900000) -- cycle;
	\fill[gray!25] (14.000000,18.000000) -- (14.000000,17.100000) -- (13.100000,18.000000) -- cycle;
	\fill[gray!80] (3.000000,17.000000) -- (2.100000,17.000000) -- (3.000000,17.900000) -- cycle;
	\fill[gray!80] (3.000000,17.000000) -- (3.000000,17.900000) -- (3.900000,17.000000) -- cycle;
	\fill[gray!80] (3.000000,17.000000) -- (3.900000,17.000000) -- (3.000000,16.100000) -- cycle;
	\fill[gray!80] (3.000000,17.000000) -- (3.000000,16.100000) -- (2.100000,17.000000) -- cycle;
	\fill[gray!80] (5.000000,17.000000) -- (4.100000,17.000000) -- (5.000000,17.900000) -- cycle;
	\fill[gray!80] (5.000000,17.000000) -- (5.000000,17.900000) -- (5.900000,17.000000) -- cycle;
	\fill[gray!80] (5.000000,17.000000) -- (5.900000,17.000000) -- (5.000000,16.100000) -- cycle;
	\fill[gray!80] (5.000000,17.000000) -- (5.000000,16.100000) -- (4.100000,17.000000) -- cycle;
	\fill[gray!80] (7.000000,17.000000) -- (6.100000,17.000000) -- (7.000000,17.900000) -- cycle;
	\fill[gray!80] (7.000000,17.000000) -- (7.000000,17.900000) -- (7.900000,17.000000) -- cycle;
	\fill[gray!80] (7.000000,17.000000) -- (7.900000,17.000000) -- (7.000000,16.100000) -- cycle;
	\fill[gray!80] (7.000000,17.000000) -- (7.000000,16.100000) -- (6.100000,17.000000) -- cycle;
	\fill[gray!80] (9.000000,17.000000) -- (8.100000,17.000000) -- (9.000000,17.900000) -- cycle;
	\fill[gray!80] (9.000000,17.000000) -- (9.000000,17.900000) -- (9.900000,17.000000) -- cycle;
	\fill[gray!80] (9.000000,17.000000) -- (9.900000,17.000000) -- (9.000000,16.100000) -- cycle;
	\fill[gray!80] (9.000000,17.000000) -- (9.000000,16.100000) -- (8.100000,17.000000) -- cycle;
	\fill[gray!80] (11.000000,17.000000) -- (10.100000,17.000000) -- (11.000000,17.900000) -- cycle;
	\fill[gray!80] (11.000000,17.000000) -- (11.000000,17.900000) -- (11.900000,17.000000) -- cycle;
	\fill[gray!80] (11.000000,17.000000) -- (11.900000,17.000000) -- (11.000000,16.100000) -- cycle;
	\fill[gray!80] (11.000000,17.000000) -- (11.000000,16.100000) -- (10.100000,17.000000) -- cycle;
	\fill[gray!80] (13.000000,17.000000) -- (12.100000,17.000000) -- (13.000000,17.900000) -- cycle;
	\fill[gray!80] (13.000000,17.000000) -- (13.000000,17.900000) -- (13.900000,17.000000) -- cycle;
	\fill[gray!80] (13.000000,17.000000) -- (13.900000,17.000000) -- (13.000000,16.100000) -- cycle;
	\fill[gray!80] (13.000000,17.000000) -- (13.000000,16.100000) -- (12.100000,17.000000) -- cycle;
	\fill[gray!25] (2.000000,16.000000) -- (2.000000,16.900000) -- (2.900000,16.000000) -- cycle;
	\fill[gray!25] (2.000000,16.000000) -- (2.900000,16.000000) -- (2.000000,15.100000) -- cycle;
	\fill[gray!25] (4.000000,16.000000) -- (3.100000,16.000000) -- (4.000000,16.900000) -- cycle;
	\fill[gray!25] (4.000000,16.000000) -- (4.000000,16.900000) -- (4.900000,16.000000) -- cycle;
	\fill[gray!25] (6.000000,16.000000) -- (5.100000,16.000000) -- (6.000000,16.900000) -- cycle;
	\fill[gray!25] (6.000000,16.000000) -- (6.000000,16.900000) -- (6.900000,16.000000) -- cycle;
	\fill[gray!25] (8.000000,16.000000) -- (7.100000,16.000000) -- (8.000000,16.900000) -- cycle;
	\fill[gray!25] (8.000000,16.000000) -- (8.000000,16.900000) -- (8.900000,16.000000) -- cycle;
	\fill[gray!25] (10.000000,16.000000) -- (9.100000,16.000000) -- (10.000000,16.900000) -- cycle;
	\fill[gray!25] (10.000000,16.000000) -- (10.000000,16.900000) -- (10.900000,16.000000) -- cycle;
	\fill[gray!25] (12.000000,16.000000) -- (11.100000,16.000000) -- (12.000000,16.900000) -- cycle;
	\fill[gray!25] (12.000000,16.000000) -- (12.000000,16.900000) -- (12.900000,16.000000) -- cycle;
	\fill[gray!25] (12.000000,16.000000) -- (12.900000,16.000000) -- (12.000000,15.100000) -- cycle;
	\fill[gray!25] (12.000000,16.000000) -- (12.000000,15.100000) -- (11.100000,16.000000) -- cycle;
	\fill[gray!25] (14.000000,16.000000) -- (13.100000,16.000000) -- (14.000000,16.900000) -- cycle;
	\fill[gray!25] (14.000000,16.000000) -- (14.000000,15.100000) -- (13.100000,16.000000) -- cycle;
	\fill[gray!80] (13.000000,15.000000) -- (12.100000,15.000000) -- (13.000000,15.900000) -- cycle;
	\fill[gray!80] (13.000000,15.000000) -- (13.000000,15.900000) -- (13.900000,15.000000) -- cycle;
	\fill[gray!80] (13.000000,15.000000) -- (13.900000,15.000000) -- (13.000000,14.100000) -- cycle;
	\fill[gray!80] (13.000000,15.000000) -- (13.000000,14.100000) -- (12.100000,15.000000) -- cycle;
	\fill[gray!25] (4.000000,14.000000) -- (4.900000,14.000000) -- (4.000000,13.100000) -- cycle;
	\fill[gray!25] (6.000000,14.000000) -- (6.900000,14.000000) -- (6.000000,13.100000) -- cycle;
	\fill[gray!25] (6.000000,14.000000) -- (6.000000,13.100000) -- (5.100000,14.000000) -- cycle;
	\fill[gray!25] (10.000000,14.000000) -- (10.000000,13.100000) -- (9.100000,14.000000) -- cycle;
	\fill[gray!25] (12.000000,14.000000) -- (12.000000,14.900000) -- (12.900000,14.000000) -- cycle;
	\fill[gray!25] (12.000000,14.000000) -- (12.900000,14.000000) -- (12.000000,13.100000) -- cycle;
	\fill[gray!80] (5.000000,13.000000) -- (4.100000,13.000000) -- (5.000000,13.900000) -- cycle;
	\fill[gray!80] (5.000000,13.000000) -- (5.000000,13.900000) -- (5.900000,13.000000) -- cycle;
	\fill[gray!80] (5.000000,13.000000) -- (5.900000,13.000000) -- (5.000000,12.100000) -- cycle;
	\fill[gray!80] (5.000000,13.000000) -- (5.000000,12.100000) -- (4.100000,13.000000) -- cycle;
	\fill[gray!80] (7.000000,13.000000) -- (6.100000,13.000000) -- (7.000000,13.900000) -- cycle;
	\fill[gray!80] (7.000000,13.000000) -- (7.000000,13.900000) -- (7.900000,13.000000) -- cycle;
	\fill[gray!80] (7.000000,13.000000) -- (7.900000,13.000000) -- (7.000000,12.100000) -- cycle;
	\fill[gray!80] (7.000000,13.000000) -- (7.000000,12.100000) -- (6.100000,13.000000) -- cycle;
	\fill[gray!80] (9.000000,13.000000) -- (8.100000,13.000000) -- (9.000000,13.900000) -- cycle;
	\fill[gray!80] (9.000000,13.000000) -- (9.000000,13.900000) -- (9.900000,13.000000) -- cycle;
	\fill[gray!80] (9.000000,13.000000) -- (9.900000,13.000000) -- (9.000000,12.100000) -- cycle;
	\fill[gray!80] (9.000000,13.000000) -- (9.000000,12.100000) -- (8.100000,13.000000) -- cycle;
	\fill[gray!80] (13.000000,13.000000) -- (12.100000,13.000000) -- (13.000000,13.900000) -- cycle;
	\fill[gray!80] (13.000000,13.000000) -- (13.000000,13.900000) -- (13.900000,13.000000) -- cycle;
	\fill[gray!80] (13.000000,13.000000) -- (13.900000,13.000000) -- (13.000000,12.100000) -- cycle;
	\fill[gray!80] (13.000000,13.000000) -- (13.000000,12.100000) -- (12.100000,13.000000) -- cycle;
	\fill[gray!25] (4.000000,12.000000) -- (4.000000,12.900000) -- (4.900000,12.000000) -- cycle;
	\fill[gray!25] (4.000000,12.000000) -- (4.900000,12.000000) -- (4.000000,11.100000) -- cycle;
	\fill[gray!25] (6.000000,12.000000) -- (5.100000,12.000000) -- (6.000000,12.900000) -- cycle;
	\fill[gray!25] (6.000000,12.000000) -- (6.000000,12.900000) -- (6.900000,12.000000) -- cycle;
	\fill[gray!25] (6.000000,12.000000) -- (6.900000,12.000000) -- (6.000000,11.100000) -- cycle;
	\fill[gray!25] (6.000000,12.000000) -- (6.000000,11.100000) -- (5.100000,12.000000) -- cycle;
	\fill[gray!25] (8.000000,12.000000) -- (7.100000,12.000000) -- (8.000000,12.900000) -- cycle;
	\fill[gray!25] (8.000000,12.000000) -- (8.000000,12.900000) -- (8.900000,12.000000) -- cycle;
	\fill[gray!25] (8.000000,12.000000) -- (8.900000,12.000000) -- (8.000000,11.100000) -- cycle;
	\fill[gray!25] (8.000000,12.000000) -- (8.000000,11.100000) -- (7.100000,12.000000) -- cycle;
	\fill[gray!25] (10.000000,12.000000) -- (9.100000,12.000000) -- (10.000000,12.900000) -- cycle;
	\fill[gray!25] (10.000000,12.000000) -- (10.000000,11.100000) -- (9.100000,12.000000) -- cycle;
	\fill[gray!25] (12.000000,12.000000) -- (12.000000,12.900000) -- (12.900000,12.000000) -- cycle;
	\fill[gray!25] (12.000000,12.000000) -- (12.900000,12.000000) -- (12.000000,11.100000) -- cycle;
	\fill[gray!25] (14.000000,12.000000) -- (13.100000,12.000000) -- (14.000000,12.900000) -- cycle;
	\fill[gray!25] (14.000000,12.000000) -- (14.000000,11.100000) -- (13.100000,12.000000) -- cycle;
	\fill[gray!80] (5.000000,11.000000) -- (4.100000,11.000000) -- (5.000000,11.900000) -- cycle;
	\fill[gray!80] (5.000000,11.000000) -- (5.000000,11.900000) -- (5.900000,11.000000) -- cycle;
	\fill[gray!80] (5.000000,11.000000) -- (5.900000,11.000000) -- (5.000000,10.100000) -- cycle;
	\fill[gray!80] (5.000000,11.000000) -- (5.000000,10.100000) -- (4.100000,11.000000) -- cycle;
	\fill[gray!80] (7.000000,11.000000) -- (6.100000,11.000000) -- (7.000000,11.900000) -- cycle;
	\fill[gray!80] (7.000000,11.000000) -- (7.000000,11.900000) -- (7.900000,11.000000) -- cycle;
	\fill[gray!80] (7.000000,11.000000) -- (7.900000,11.000000) -- (7.000000,10.100000) -- cycle;
	\fill[gray!80] (7.000000,11.000000) -- (7.000000,10.100000) -- (6.100000,11.000000) -- cycle;
	\fill[gray!80] (9.000000,11.000000) -- (8.100000,11.000000) -- (9.000000,11.900000) -- cycle;
	\fill[gray!80] (9.000000,11.000000) -- (9.000000,11.900000) -- (9.900000,11.000000) -- cycle;
	\fill[gray!80] (9.000000,11.000000) -- (9.900000,11.000000) -- (9.000000,10.100000) -- cycle;
	\fill[gray!80] (9.000000,11.000000) -- (9.000000,10.100000) -- (8.100000,11.000000) -- cycle;
	\fill[gray!80] (13.000000,11.000000) -- (12.100000,11.000000) -- (13.000000,11.900000) -- cycle;
	\fill[gray!80] (13.000000,11.000000) -- (13.000000,11.900000) -- (13.900000,11.000000) -- cycle;
	\fill[gray!80] (13.000000,11.000000) -- (13.900000,11.000000) -- (13.000000,10.100000) -- cycle;
	\fill[gray!80] (13.000000,11.000000) -- (13.000000,10.100000) -- (12.100000,11.000000) -- cycle;
	\fill[gray!25] (4.000000,10.000000) -- (4.000000,10.900000) -- (4.900000,10.000000) -- cycle;
	\fill[gray!25] (4.000000,10.000000) -- (4.900000,10.000000) -- (4.000000,9.100000) -- cycle;
	\fill[gray!25] (6.000000,10.000000) -- (5.100000,10.000000) -- (6.000000,10.900000) -- cycle;
	\fill[gray!25] (6.000000,10.000000) -- (6.000000,10.900000) -- (6.900000,10.000000) -- cycle;
	\fill[gray!25] (6.000000,10.000000) -- (6.900000,10.000000) -- (6.000000,9.100000) -- cycle;
	\fill[gray!25] (6.000000,10.000000) -- (6.000000,9.100000) -- (5.100000,10.000000) -- cycle;
	\fill[gray!25] (8.000000,10.000000) -- (7.100000,10.000000) -- (8.000000,10.900000) -- cycle;
	\fill[gray!25] (8.000000,10.000000) -- (8.000000,10.900000) -- (8.900000,10.000000) -- cycle;
	\fill[gray!25] (8.000000,10.000000) -- (8.900000,10.000000) -- (8.000000,9.100000) -- cycle;
	\fill[gray!25] (8.000000,10.000000) -- (8.000000,9.100000) -- (7.100000,10.000000) -- cycle;
	\fill[gray!25] (10.000000,10.000000) -- (9.100000,10.000000) -- (10.000000,10.900000) -- cycle;
	\fill[gray!25] (10.000000,10.000000) -- (10.000000,9.100000) -- (9.100000,10.000000) -- cycle;
	\fill[gray!25] (12.000000,10.000000) -- (12.000000,10.900000) -- (12.900000,10.000000) -- cycle;
	\fill[gray!25] (12.000000,10.000000) -- (12.900000,10.000000) -- (12.000000,9.100000) -- cycle;
	\fill[gray!25] (14.000000,10.000000) -- (13.100000,10.000000) -- (14.000000,10.900000) -- cycle;
	\fill[gray!25] (14.000000,10.000000) -- (14.000000,9.100000) -- (13.100000,10.000000) -- cycle;
	\fill[gray!80] (5.000000,9.000000) -- (4.100000,9.000000) -- (5.000000,9.900000) -- cycle;
	\fill[gray!80] (5.000000,9.000000) -- (5.000000,9.900000) -- (5.900000,9.000000) -- cycle;
	\fill[gray!80] (5.000000,9.000000) -- (5.900000,9.000000) -- (5.000000,8.100000) -- cycle;
	\fill[gray!80] (5.000000,9.000000) -- (5.000000,8.100000) -- (4.100000,9.000000) -- cycle;
	\fill[gray!80] (7.000000,9.000000) -- (6.100000,9.000000) -- (7.000000,9.900000) -- cycle;
	\fill[gray!80] (7.000000,9.000000) -- (7.000000,9.900000) -- (7.900000,9.000000) -- cycle;
	\fill[gray!80] (7.000000,9.000000) -- (7.900000,9.000000) -- (7.000000,8.100000) -- cycle;
	\fill[gray!80] (7.000000,9.000000) -- (7.000000,8.100000) -- (6.100000,9.000000) -- cycle;
	\fill[gray!80] (9.000000,9.000000) -- (8.100000,9.000000) -- (9.000000,9.900000) -- cycle;
	\fill[gray!80] (9.000000,9.000000) -- (9.000000,9.900000) -- (9.900000,9.000000) -- cycle;
	\fill[gray!80] (9.000000,9.000000) -- (9.900000,9.000000) -- (9.000000,8.100000) -- cycle;
	\fill[gray!80] (9.000000,9.000000) -- (9.000000,8.100000) -- (8.100000,9.000000) -- cycle;
	\fill[gray!80] (13.000000,9.000000) -- (12.100000,9.000000) -- (13.000000,9.900000) -- cycle;
	\fill[gray!80] (13.000000,9.000000) -- (13.000000,9.900000) -- (13.900000,9.000000) -- cycle;
	\fill[gray!80] (13.000000,9.000000) -- (13.900000,9.000000) -- (13.000000,8.100000) -- cycle;
	\fill[gray!80] (13.000000,9.000000) -- (13.000000,8.100000) -- (12.100000,9.000000) -- cycle;
	\fill[gray!25] (4.000000,8.000000) -- (4.000000,8.900000) -- (4.900000,8.000000) -- cycle;
	\fill[gray!25] (6.000000,8.000000) -- (5.100000,8.000000) -- (6.000000,8.900000) -- cycle;
	\fill[gray!25] (6.000000,8.000000) -- (6.000000,8.900000) -- (6.900000,8.000000) -- cycle;
	\fill[gray!25] (10.000000,8.000000) -- (9.100000,8.000000) -- (10.000000,8.900000) -- cycle;
	\fill[gray!25] (12.000000,8.000000) -- (12.000000,8.900000) -- (12.900000,8.000000) -- cycle;
	\fill[gray!25] (12.000000,8.000000) -- (12.900000,8.000000) -- (12.000000,7.100000) -- cycle;
	\fill[gray!80] (13.000000,7.000000) -- (12.100000,7.000000) -- (13.000000,7.900000) -- cycle;
	\fill[gray!80] (13.000000,7.000000) -- (13.000000,7.900000) -- (13.900000,7.000000) -- cycle;
	\fill[gray!80] (13.000000,7.000000) -- (13.900000,7.000000) -- (13.000000,6.100000) -- cycle;
	\fill[gray!80] (13.000000,7.000000) -- (13.000000,6.100000) -- (12.100000,7.000000) -- cycle;
	\fill[gray!25] (2.000000,6.000000) -- (2.000000,6.900000) -- (2.900000,6.000000) -- cycle;
	\fill[gray!25] (2.000000,6.000000) -- (2.900000,6.000000) -- (2.000000,5.100000) -- cycle;
	\fill[gray!25] (4.000000,6.000000) -- (4.900000,6.000000) -- (4.000000,5.100000) -- cycle;
	\fill[gray!25] (4.000000,6.000000) -- (4.000000,5.100000) -- (3.100000,6.000000) -- cycle;
	\fill[gray!25] (6.000000,6.000000) -- (6.900000,6.000000) -- (6.000000,5.100000) -- cycle;
	\fill[gray!25] (6.000000,6.000000) -- (6.000000,5.100000) -- (5.100000,6.000000) -- cycle;
	\fill[gray!25] (8.000000,6.000000) -- (8.900000,6.000000) -- (8.000000,5.100000) -- cycle;
	\fill[gray!25] (8.000000,6.000000) -- (8.000000,5.100000) -- (7.100000,6.000000) -- cycle;
	\fill[gray!25] (10.000000,6.000000) -- (10.900000,6.000000) -- (10.000000,5.100000) -- cycle;
	\fill[gray!25] (10.000000,6.000000) -- (10.000000,5.100000) -- (9.100000,6.000000) -- cycle;
	\fill[gray!25] (12.000000,6.000000) -- (11.100000,6.000000) -- (12.000000,6.900000) -- cycle;
	\fill[gray!25] (12.000000,6.000000) -- (12.000000,6.900000) -- (12.900000,6.000000) -- cycle;
	\fill[gray!25] (12.000000,6.000000) -- (12.900000,6.000000) -- (12.000000,5.100000) -- cycle;
	\fill[gray!25] (12.000000,6.000000) -- (12.000000,5.100000) -- (11.100000,6.000000) -- cycle;
	\fill[gray!25] (14.000000,6.000000) -- (13.100000,6.000000) -- (14.000000,6.900000) -- cycle;
	\fill[gray!25] (14.000000,6.000000) -- (14.000000,5.100000) -- (13.100000,6.000000) -- cycle;
	\fill[gray!80] (3.000000,5.000000) -- (2.100000,5.000000) -- (3.000000,5.900000) -- cycle;
	\fill[gray!80] (3.000000,5.000000) -- (3.000000,5.900000) -- (3.900000,5.000000) -- cycle;
	\fill[gray!80] (3.000000,5.000000) -- (3.900000,5.000000) -- (3.000000,4.100000) -- cycle;
	\fill[gray!80] (3.000000,5.000000) -- (3.000000,4.100000) -- (2.100000,5.000000) -- cycle;
	\fill[gray!80] (5.000000,5.000000) -- (4.100000,5.000000) -- (5.000000,5.900000) -- cycle;
	\fill[gray!80] (5.000000,5.000000) -- (5.000000,5.900000) -- (5.900000,5.000000) -- cycle;
	\fill[gray!80] (5.000000,5.000000) -- (5.900000,5.000000) -- (5.000000,4.100000) -- cycle;
	\fill[gray!80] (5.000000,5.000000) -- (5.000000,4.100000) -- (4.100000,5.000000) -- cycle;
	\fill[gray!80] (7.000000,5.000000) -- (6.100000,5.000000) -- (7.000000,5.900000) -- cycle;
	\fill[gray!80] (7.000000,5.000000) -- (7.000000,5.900000) -- (7.900000,5.000000) -- cycle;
	\fill[gray!80] (7.000000,5.000000) -- (7.900000,5.000000) -- (7.000000,4.100000) -- cycle;
	\fill[gray!80] (7.000000,5.000000) -- (7.000000,4.100000) -- (6.100000,5.000000) -- cycle;
	\fill[gray!80] (9.000000,5.000000) -- (8.100000,5.000000) -- (9.000000,5.900000) -- cycle;
	\fill[gray!80] (9.000000,5.000000) -- (9.000000,5.900000) -- (9.900000,5.000000) -- cycle;
	\fill[gray!80] (9.000000,5.000000) -- (9.900000,5.000000) -- (9.000000,4.100000) -- cycle;
	\fill[gray!80] (9.000000,5.000000) -- (9.000000,4.100000) -- (8.100000,5.000000) -- cycle;
	\fill[gray!80] (11.000000,5.000000) -- (10.100000,5.000000) -- (11.000000,5.900000) -- cycle;
	\fill[gray!80] (11.000000,5.000000) -- (11.000000,5.900000) -- (11.900000,5.000000) -- cycle;
	\fill[gray!80] (11.000000,5.000000) -- (11.900000,5.000000) -- (11.000000,4.100000) -- cycle;
	\fill[gray!80] (11.000000,5.000000) -- (11.000000,4.100000) -- (10.100000,5.000000) -- cycle;
	\fill[gray!80] (13.000000,5.000000) -- (12.100000,5.000000) -- (13.000000,5.900000) -- cycle;
	\fill[gray!80] (13.000000,5.000000) -- (13.000000,5.900000) -- (13.900000,5.000000) -- cycle;
	\fill[gray!80] (13.000000,5.000000) -- (13.900000,5.000000) -- (13.000000,4.100000) -- cycle;
	\fill[gray!80] (13.000000,5.000000) -- (13.000000,4.100000) -- (12.100000,5.000000) -- cycle;
	\fill[gray!25] (2.000000,4.000000) -- (2.000000,4.900000) -- (2.900000,4.000000) -- cycle;
	\fill[gray!25] (2.000000,4.000000) -- (2.900000,4.000000) -- (2.000000,3.100000) -- cycle;
	\fill[gray!25] (4.000000,4.000000) -- (3.100000,4.000000) -- (4.000000,4.900000) -- cycle;
	\fill[gray!25] (4.000000,4.000000) -- (4.000000,4.900000) -- (4.900000,4.000000) -- cycle;
	\fill[gray!25] (4.000000,4.000000) -- (4.900000,4.000000) -- (4.000000,3.100000) -- cycle;
	\fill[gray!25] (4.000000,4.000000) -- (4.000000,3.100000) -- (3.100000,4.000000) -- cycle;
	\fill[gray!25] (6.000000,4.000000) -- (5.100000,4.000000) -- (6.000000,4.900000) -- cycle;
	\fill[gray!25] (6.000000,4.000000) -- (6.000000,4.900000) -- (6.900000,4.000000) -- cycle;
	\fill[gray!25] (6.000000,4.000000) -- (6.900000,4.000000) -- (6.000000,3.100000) -- cycle;
	\fill[gray!25] (6.000000,4.000000) -- (6.000000,3.100000) -- (5.100000,4.000000) -- cycle;
	\fill[gray!25] (8.000000,4.000000) -- (7.100000,4.000000) -- (8.000000,4.900000) -- cycle;
	\fill[gray!25] (8.000000,4.000000) -- (8.000000,4.900000) -- (8.900000,4.000000) -- cycle;
	\fill[gray!25] (8.000000,4.000000) -- (8.900000,4.000000) -- (8.000000,3.100000) -- cycle;
	\fill[gray!25] (8.000000,4.000000) -- (8.000000,3.100000) -- (7.100000,4.000000) -- cycle;
	\fill[gray!25] (10.000000,4.000000) -- (9.100000,4.000000) -- (10.000000,4.900000) -- cycle;
	\fill[gray!25] (10.000000,4.000000) -- (10.000000,4.900000) -- (10.900000,4.000000) -- cycle;
	\fill[gray!25] (10.000000,4.000000) -- (10.900000,4.000000) -- (10.000000,3.100000) -- cycle;
	\fill[gray!25] (10.000000,4.000000) -- (10.000000,3.100000) -- (9.100000,4.000000) -- cycle;
	\fill[gray!25] (12.000000,4.000000) -- (11.100000,4.000000) -- (12.000000,4.900000) -- cycle;
	\fill[gray!25] (12.000000,4.000000) -- (12.000000,4.900000) -- (12.900000,4.000000) -- cycle;
	\fill[gray!25] (12.000000,4.000000) -- (12.900000,4.000000) -- (12.000000,3.100000) -- cycle;
	\fill[gray!25] (12.000000,4.000000) -- (12.000000,3.100000) -- (11.100000,4.000000) -- cycle;
	\fill[gray!25] (14.000000,4.000000) -- (13.100000,4.000000) -- (14.000000,4.900000) -- cycle;
	\fill[gray!25] (14.000000,4.000000) -- (14.000000,3.100000) -- (13.100000,4.000000) -- cycle;
	\fill[gray!80] (3.000000,3.000000) -- (2.100000,3.000000) -- (3.000000,3.900000) -- cycle;
	\fill[gray!80] (3.000000,3.000000) -- (3.000000,3.900000) -- (3.900000,3.000000) -- cycle;
	\fill[gray!80] (3.000000,3.000000) -- (3.900000,3.000000) -- (3.000000,2.100000) -- cycle;
	\fill[gray!80] (3.000000,3.000000) -- (3.000000,2.100000) -- (2.100000,3.000000) -- cycle;
	\fill[gray!80] (5.000000,3.000000) -- (4.100000,3.000000) -- (5.000000,3.900000) -- cycle;
	\fill[gray!80] (5.000000,3.000000) -- (5.000000,3.900000) -- (5.900000,3.000000) -- cycle;
	\fill[gray!80] (5.000000,3.000000) -- (5.900000,3.000000) -- (5.000000,2.100000) -- cycle;
	\fill[gray!80] (5.000000,3.000000) -- (5.000000,2.100000) -- (4.100000,3.000000) -- cycle;
	\fill[gray!80] (7.000000,3.000000) -- (6.100000,3.000000) -- (7.000000,3.900000) -- cycle;
	\fill[gray!80] (7.000000,3.000000) -- (7.000000,3.900000) -- (7.900000,3.000000) -- cycle;
	\fill[gray!80] (7.000000,3.000000) -- (7.900000,3.000000) -- (7.000000,2.100000) -- cycle;
	\fill[gray!80] (7.000000,3.000000) -- (7.000000,2.100000) -- (6.100000,3.000000) -- cycle;
	\fill[gray!80] (9.000000,3.000000) -- (8.100000,3.000000) -- (9.000000,3.900000) -- cycle;
	\fill[gray!80] (9.000000,3.000000) -- (9.000000,3.900000) -- (9.900000,3.000000) -- cycle;
	\fill[gray!80] (9.000000,3.000000) -- (9.900000,3.000000) -- (9.000000,2.100000) -- cycle;
	\fill[gray!80] (9.000000,3.000000) -- (9.000000,2.100000) -- (8.100000,3.000000) -- cycle;
	\fill[gray!80] (11.000000,3.000000) -- (10.100000,3.000000) -- (11.000000,3.900000) -- cycle;
	\fill[gray!80] (11.000000,3.000000) -- (11.000000,3.900000) -- (11.900000,3.000000) -- cycle;
	\fill[gray!80] (11.000000,3.000000) -- (11.900000,3.000000) -- (11.000000,2.100000) -- cycle;
	\fill[gray!80] (11.000000,3.000000) -- (11.000000,2.100000) -- (10.100000,3.000000) -- cycle;
	\fill[gray!80] (13.000000,3.000000) -- (12.100000,3.000000) -- (13.000000,3.900000) -- cycle;
	\fill[gray!80] (13.000000,3.000000) -- (13.000000,3.900000) -- (13.900000,3.000000) -- cycle;
	\fill[gray!80] (13.000000,3.000000) -- (13.900000,3.000000) -- (13.000000,2.100000) -- cycle;
	\fill[gray!80] (13.000000,3.000000) -- (13.000000,2.100000) -- (12.100000,3.000000) -- cycle;
	\fill[gray!25] (4.000000,2.000000) -- (3.100000,2.000000) -- (4.000000,2.900000) -- cycle;
	\fill[gray!25] (4.000000,2.000000) -- (4.000000,2.900000) -- (4.900000,2.000000) -- cycle;
	\fill[gray!25] (6.000000,2.000000) -- (5.100000,2.000000) -- (6.000000,2.900000) -- cycle;
	\fill[gray!25] (6.000000,2.000000) -- (6.000000,2.900000) -- (6.900000,2.000000) -- cycle;
	\fill[gray!25] (10.000000,2.000000) -- (9.100000,2.000000) -- (10.000000,2.900000) -- cycle;
	\fill[gray!25] (10.000000,2.000000) -- (10.000000,2.900000) -- (10.900000,2.000000) -- cycle;
	\fill[gray!25] (12.000000,2.000000) -- (11.100000,2.000000) -- (12.000000,2.900000) -- cycle;
	\fill[gray!25] (12.000000,2.000000) -- (12.000000,2.900000) -- (12.900000,2.000000) -- cycle;
	\draw[line width=1pt] (3.000000,20.000000) circle (0.100000);
	\draw[line width=1pt,fill=black] (4.000000,20.000000) circle (0.100000);
	\draw[line width=1pt] (5.000000,20.000000) circle (0.100000);
	\draw[line width=1pt,fill=black] (6.000000,20.000000) circle (0.100000);
	\draw[line width=1pt] (7.000000,20.000000) circle (0.100000);
	\draw[line width=1pt] (9.000000,20.000000) circle (0.100000);
	\draw[line width=1pt,fill=black] (10.000000,20.000000) circle (0.100000);
	\draw[line width=1pt] (11.000000,20.000000) circle (0.100000);
	\draw[line width=1pt,fill=black] (12.000000,20.000000) circle (0.100000);
	\draw[line width=1pt] (13.000000,20.000000) circle (0.100000);
	\draw[line width=1pt] (2.000000,19.000000) circle (0.100000);
	\draw[line width=1pt,fill=black] (3.000000,19.000000) circle (0.100000);
	\draw[line width=1pt] (4.000000,19.000000) circle (0.100000);
	\draw[line width=1pt,fill=black] (5.000000,19.000000) circle (0.100000);
	\draw[line width=1pt] (6.000000,19.000000) circle (0.100000);
	\draw[line width=1pt,fill=black] (7.000000,19.000000) circle (0.100000);
	\draw[line width=1pt] (8.000000,19.000000) circle (0.100000);
	\draw[line width=1pt,fill=black] (9.000000,19.000000) circle (0.100000);
	\draw[line width=1pt] (10.000000,19.000000) circle (0.100000);
	\draw[line width=1pt,fill=black] (11.000000,19.000000) circle (0.100000);
	\draw[line width=1pt] (12.000000,19.000000) circle (0.100000);
	\draw[line width=1pt,fill=black] (13.000000,19.000000) circle (0.100000);
	\draw[line width=1pt] (14.000000,19.000000) circle (0.100000);
	\draw[line width=1pt,fill=black] (2.000000,18.000000) circle (0.100000);
	\draw[line width=1pt] (3.000000,18.000000) circle (0.100000);
	\draw[line width=1pt,fill=black] (4.000000,18.000000) circle (0.100000);
	\draw[line width=1pt] (5.000000,18.000000) circle (0.100000);
	\draw[line width=1pt,fill=black] (6.000000,18.000000) circle (0.100000);
	\draw[line width=1pt] (7.000000,18.000000) circle (0.100000);
	\draw[line width=1pt,fill=black] (8.000000,18.000000) circle (0.100000);
	\draw[line width=1pt] (9.000000,18.000000) circle (0.100000);
	\draw[line width=1pt,fill=black] (10.000000,18.000000) circle (0.100000);
	\draw[line width=1pt] (11.000000,18.000000) circle (0.100000);
	\draw[line width=1pt,fill=black] (12.000000,18.000000) circle (0.100000);
	\draw[line width=1pt] (13.000000,18.000000) circle (0.100000);
	\draw[line width=1pt,fill=black] (14.000000,18.000000) circle (0.100000);
	\draw[line width=1pt] (2.000000,17.000000) circle (0.100000);
	\draw[line width=1pt,fill=black] (3.000000,17.000000) circle (0.100000);
	\draw[line width=1pt] (4.000000,17.000000) circle (0.100000);
	\draw[line width=1pt,fill=black] (5.000000,17.000000) circle (0.100000);
	\draw[line width=1pt] (6.000000,17.000000) circle (0.100000);
	\draw[line width=1pt,fill=black] (7.000000,17.000000) circle (0.100000);
	\draw[line width=1pt] (8.000000,17.000000) circle (0.100000);
	\draw[line width=1pt,fill=black] (9.000000,17.000000) circle (0.100000);
	\draw[line width=1pt] (10.000000,17.000000) circle (0.100000);
	\draw[line width=1pt,fill=black] (11.000000,17.000000) circle (0.100000);
	\draw[line width=1pt] (12.000000,17.000000) circle (0.100000);
	\draw[line width=1pt,fill=black] (13.000000,17.000000) circle (0.100000);
	\draw[line width=1pt] (14.000000,17.000000) circle (0.100000);
	\draw[line width=1pt,fill=black] (2.000000,16.000000) circle (0.100000);
	\draw[line width=1pt] (3.000000,16.000000) circle (0.100000);
	\draw[line width=1pt,fill=black] (4.000000,16.000000) circle (0.100000);
	\draw[line width=1pt] (5.000000,16.000000) circle (0.100000);
	\draw[line width=1pt,fill=black] (6.000000,16.000000) circle (0.100000);
	\draw[line width=1pt] (7.000000,16.000000) circle (0.100000);
	\draw[line width=1pt,fill=black] (8.000000,16.000000) circle (0.100000);
	\draw[line width=1pt] (9.000000,16.000000) circle (0.100000);
	\draw[line width=1pt,fill=black] (10.000000,16.000000) circle (0.100000);
	\draw[line width=1pt] (11.000000,16.000000) circle (0.100000);
	\draw[line width=1pt,fill=black] (12.000000,16.000000) circle (0.100000);
	\draw[line width=1pt] (13.000000,16.000000) circle (0.100000);
	\draw[line width=1pt,fill=black] (14.000000,16.000000) circle (0.100000);
	\draw[line width=1pt] (2.000000,15.000000) circle (0.100000);
	\draw[line width=1pt] (12.000000,15.000000) circle (0.100000);
	\draw[line width=1pt,fill=black] (13.000000,15.000000) circle (0.100000);
	\draw[line width=1pt] (14.000000,15.000000) circle (0.100000);
	\draw[line width=1pt,fill=black] (4.000000,14.000000) circle (0.100000);
	\draw (5.000000,14.000000) node [above right] {$|+\rangle$};
grestore
	\draw[line width=1pt] (5.000000,14.000000) circle (0.100000);
	\draw[line width=1pt,fill=black] (6.000000,14.000000) circle (0.100000);
	\draw (7.000000,14.000000) node [above right] {$|+\rangle$};
grestore
	\draw[line width=1pt] (7.000000,14.000000) circle (0.100000);
	\draw[line width=1pt] (9.000000,14.000000) circle (0.100000);
	\draw[line width=1pt,fill=black] (10.000000,14.000000) circle (0.100000);
	\draw[line width=1pt,fill=black] (12.000000,14.000000) circle (0.100000);
	\draw[line width=1pt] (13.000000,14.000000) circle (0.100000);
	\draw[line width=1pt] (2.000000,13.000000) circle (0.100000);
	\draw (4.000000,13.000000) node [above right] {$|+\rangle$};
grestore
	\draw[line width=1pt] (4.000000,13.000000) circle (0.100000);
	\draw[line width=1pt,fill=black] (5.000000,13.000000) circle (0.100000);
	\draw[line width=1pt] (6.000000,13.000000) circle (0.100000);
	\draw[line width=1pt,fill=black] (7.000000,13.000000) circle (0.100000);
	\draw[line width=1pt] (8.000000,13.000000) circle (0.100000);
	\draw[line width=1pt,fill=black] (9.000000,13.000000) circle (0.100000);
	\draw[line width=1pt] (10.000000,13.000000) circle (0.100000);
	\draw[line width=1pt] (12.000000,13.000000) circle (0.100000);
	\draw[line width=1pt,fill=black] (13.000000,13.000000) circle (0.100000);
	\draw[line width=1pt] (14.000000,13.000000) circle (0.100000);
	\draw[line width=1pt,fill=black] (2.000000,12.000000) circle (0.100000);
	\draw[line width=1pt,fill=black] (4.000000,12.000000) circle (0.100000);
	\draw[line width=1pt] (5.000000,12.000000) circle (0.100000);
	\draw[line width=1pt,fill=black] (6.000000,12.000000) circle (0.100000);
	\draw[line width=1pt] (7.000000,12.000000) circle (0.100000);
	\draw[line width=1pt,fill=black] (8.000000,12.000000) circle (0.100000);
	\draw[line width=1pt] (9.000000,12.000000) circle (0.100000);
	\draw[line width=1pt,fill=black] (10.000000,12.000000) circle (0.100000);
	\draw[line width=1pt,fill=black] (12.000000,12.000000) circle (0.100000);
	\draw[line width=1pt] (13.000000,12.000000) circle (0.100000);
	\draw[line width=1pt,fill=black] (14.000000,12.000000) circle (0.100000);
	\draw[line width=1pt] (2.000000,11.000000) circle (0.100000);
	\draw[line width=1pt] (4.000000,11.000000) circle (0.100000);
	\draw[line width=1pt,fill=black] (5.000000,11.000000) circle (0.100000);
	\draw[line width=1pt] (6.000000,11.000000) circle (0.100000);
	\draw[line width=1pt,fill=black] (7.000000,11.000000) circle (0.100000);
	\draw[line width=1pt] (8.000000,11.000000) circle (0.100000);
	\draw[line width=1pt,fill=black] (9.000000,11.000000) circle (0.100000);
	\draw[line width=1pt] (10.000000,11.000000) circle (0.100000);
	\draw[line width=1pt] (12.000000,11.000000) circle (0.100000);
	\draw[line width=1pt,fill=black] (13.000000,11.000000) circle (0.100000);
	\draw[line width=1pt] (14.000000,11.000000) circle (0.100000);
	\draw[line width=1pt,fill=black] (2.000000,10.000000) circle (0.100000);
	\draw[line width=1pt,fill=black] (4.000000,10.000000) circle (0.100000);
	\draw[line width=1pt] (5.000000,10.000000) circle (0.100000);
	\draw[line width=1pt,fill=black] (6.000000,10.000000) circle (0.100000);
	\draw[line width=1pt] (7.000000,10.000000) circle (0.100000);
	\draw[line width=1pt,fill=black] (8.000000,10.000000) circle (0.100000);
	\draw[line width=1pt] (9.000000,10.000000) circle (0.100000);
	\draw[line width=1pt,fill=black] (10.000000,10.000000) circle (0.100000);
	\draw[line width=1pt,fill=black] (12.000000,10.000000) circle (0.100000);
	\draw[line width=1pt] (13.000000,10.000000) circle (0.100000);
	\draw[line width=1pt,fill=black] (14.000000,10.000000) circle (0.100000);
	\draw[line width=1pt] (2.000000,9.000000) circle (0.100000);
	\draw[line width=1pt] (4.000000,9.000000) circle (0.100000);
	\draw[line width=1pt,fill=black] (5.000000,9.000000) circle (0.100000);
	\draw[line width=1pt] (6.000000,9.000000) circle (0.100000);
	\draw[line width=1pt,fill=black] (7.000000,9.000000) circle (0.100000);
	\draw[line width=1pt] (8.000000,9.000000) circle (0.100000);
	\draw[line width=1pt,fill=black] (9.000000,9.000000) circle (0.100000);
	\draw[line width=1pt] (10.000000,9.000000) circle (0.100000);
	\draw[line width=1pt] (12.000000,9.000000) circle (0.100000);
	\draw[line width=1pt,fill=black] (13.000000,9.000000) circle (0.100000);
	\draw[line width=1pt] (14.000000,9.000000) circle (0.100000);
	\draw[line width=1pt,fill=black] (4.000000,8.000000) circle (0.100000);
	\draw[line width=1pt] (5.000000,8.000000) circle (0.100000);
	\draw[line width=1pt,fill=black] (6.000000,8.000000) circle (0.100000);
	\draw[line width=1pt] (7.000000,8.000000) circle (0.100000);
	\draw[line width=1pt] (9.000000,8.000000) circle (0.100000);
	\draw[line width=1pt,fill=black] (10.000000,8.000000) circle (0.100000);
	\draw[line width=1pt,fill=black] (12.000000,8.000000) circle (0.100000);
	\draw[line width=1pt] (13.000000,8.000000) circle (0.100000);
	\draw[line width=1pt] (2.000000,7.000000) circle (0.100000);
	\draw[line width=1pt] (12.000000,7.000000) circle (0.100000);
	\draw[line width=1pt,fill=black] (13.000000,7.000000) circle (0.100000);
	\draw[line width=1pt] (14.000000,7.000000) circle (0.100000);
	\draw[line width=1pt,fill=black] (2.000000,6.000000) circle (0.100000);
	\draw[line width=1pt] (3.000000,6.000000) circle (0.100000);
	\draw[line width=1pt,fill=black] (4.000000,6.000000) circle (0.100000);
	\draw[line width=1pt] (5.000000,6.000000) circle (0.100000);
	\draw[line width=1pt,fill=black] (6.000000,6.000000) circle (0.100000);
	\draw[line width=1pt] (7.000000,6.000000) circle (0.100000);
	\draw[line width=1pt,fill=black] (8.000000,6.000000) circle (0.100000);
	\draw[line width=1pt] (9.000000,6.000000) circle (0.100000);
	\draw[line width=1pt,fill=black] (10.000000,6.000000) circle (0.100000);
	\draw[line width=1pt] (11.000000,6.000000) circle (0.100000);
	\draw[line width=1pt,fill=black] (12.000000,6.000000) circle (0.100000);
	\draw[line width=1pt] (13.000000,6.000000) circle (0.100000);
	\draw[line width=1pt,fill=black] (14.000000,6.000000) circle (0.100000);
	\draw[line width=1pt] (2.000000,5.000000) circle (0.100000);
	\draw[line width=1pt,fill=black] (3.000000,5.000000) circle (0.100000);
	\draw[line width=1pt] (4.000000,5.000000) circle (0.100000);
	\draw[line width=1pt,fill=black] (5.000000,5.000000) circle (0.100000);
	\draw[line width=1pt] (6.000000,5.000000) circle (0.100000);
	\draw[line width=1pt,fill=black] (7.000000,5.000000) circle (0.100000);
	\draw[line width=1pt] (8.000000,5.000000) circle (0.100000);
	\draw[line width=1pt,fill=black] (9.000000,5.000000) circle (0.100000);
	\draw[line width=1pt] (10.000000,5.000000) circle (0.100000);
	\draw[line width=1pt,fill=black] (11.000000,5.000000) circle (0.100000);
	\draw[line width=1pt] (12.000000,5.000000) circle (0.100000);
	\draw[line width=1pt,fill=black] (13.000000,5.000000) circle (0.100000);
	\draw[line width=1pt] (14.000000,5.000000) circle (0.100000);
	\draw[line width=1pt,fill=black] (2.000000,4.000000) circle (0.100000);
	\draw[line width=1pt] (3.000000,4.000000) circle (0.100000);
	\draw[line width=1pt,fill=black] (4.000000,4.000000) circle (0.100000);
	\draw[line width=1pt] (5.000000,4.000000) circle (0.100000);
	\draw[line width=1pt,fill=black] (6.000000,4.000000) circle (0.100000);
	\draw[line width=1pt] (7.000000,4.000000) circle (0.100000);
	\draw[line width=1pt,fill=black] (8.000000,4.000000) circle (0.100000);
	\draw[line width=1pt] (9.000000,4.000000) circle (0.100000);
	\draw[line width=1pt,fill=black] (10.000000,4.000000) circle (0.100000);
	\draw[line width=1pt] (11.000000,4.000000) circle (0.100000);
	\draw[line width=1pt,fill=black] (12.000000,4.000000) circle (0.100000);
	\draw[line width=1pt] (13.000000,4.000000) circle (0.100000);
	\draw[line width=1pt,fill=black] (14.000000,4.000000) circle (0.100000);
	\draw[line width=1pt] (2.000000,3.000000) circle (0.100000);
	\draw[line width=1pt,fill=black] (3.000000,3.000000) circle (0.100000);
	\draw[line width=1pt] (4.000000,3.000000) circle (0.100000);
	\draw[line width=1pt,fill=black] (5.000000,3.000000) circle (0.100000);
	\draw[line width=1pt] (6.000000,3.000000) circle (0.100000);
	\draw[line width=1pt,fill=black] (7.000000,3.000000) circle (0.100000);
	\draw[line width=1pt] (8.000000,3.000000) circle (0.100000);
	\draw[line width=1pt,fill=black] (9.000000,3.000000) circle (0.100000);
	\draw[line width=1pt] (10.000000,3.000000) circle (0.100000);
	\draw[line width=1pt,fill=black] (11.000000,3.000000) circle (0.100000);
	\draw[line width=1pt] (12.000000,3.000000) circle (0.100000);
	\draw[line width=1pt,fill=black] (13.000000,3.000000) circle (0.100000);
	\draw[line width=1pt] (14.000000,3.000000) circle (0.100000);
	\draw[line width=1pt] (3.000000,2.000000) circle (0.100000);
	\draw[line width=1pt,fill=black] (4.000000,2.000000) circle (0.100000);
	\draw[line width=1pt] (5.000000,2.000000) circle (0.100000);
	\draw[line width=1pt,fill=black] (6.000000,2.000000) circle (0.100000);
	\draw[line width=1pt] (7.000000,2.000000) circle (0.100000);
	\draw[line width=1pt] (9.000000,2.000000) circle (0.100000);
	\draw[line width=1pt,fill=black] (10.000000,2.000000) circle (0.100000);
	\draw[line width=1pt] (11.000000,2.000000) circle (0.100000);
	\draw[line width=1pt,fill=black] (12.000000,2.000000) circle (0.100000);
	\draw[line width=1pt] (13.000000,2.000000) circle (0.100000);
\end{tikzpicture}